\documentclass[aps,preprint,superscriptaddress,twoside,nofootinbib,11pt]{revtex4-1}
\usepackage{graphicx} % Include figure files 
\usepackage{dcolumn} % Align table columns on decimal point 
\usepackage{bm} % bold math 
\usepackage{hyperref} % add hypertext capabilities 
\usepackage[mathlines]{lineno} % Enable numbering of text and display math 
\usepackage[usenames]{color} % color for proofread
\usepackage{multirow}
\usepackage{amsmath}
\usepackage{amssymb} 
\usepackage{bm}
\usepackage{hyperref}
\usepackage{bmpsize}
\usepackage{amsopn}
\usepackage[caption=false]{subfig}
\usepackage{filecontents}
\usepackage{xspace}
\usepackage{tabularx}

\RequirePackage{lineno}

\usepackage{array}

\usepackage{hyperref}
\usepackage{chngcntr}

\counterwithin{section}{part}

\begin{document}

\newcommand{\numu}{\ensuremath{\nu_{\mu}}}                   
\newcommand{\numubar}{\ensuremath{\overline{\nu}_{\mu}}}                   
\newcommand{\nue}{\ensuremath{\nu_{e}}}                   
\newcommand{\nuebar}{\ensuremath{\overline{\nu}_{e}}}                   
\newcommand{\enurec}{\ensuremath{E_\nu^\mathrm{rec}}}     
\newcommand{\deltacp}{\ensuremath{\delta_{CP}} }

\newcommand{\Nova}{NO{$\nu$A} }
\newcommand{\sthetatt}{\ensuremath{\sin^2\theta_{23}\ }}

\newcommand{\hksingletank}{1TankHD}

\newcolumntype{L}[1]{>{\raggedright\let\newline\\\arraybackslash\hspace{0pt}}m{#1}}
\newcolumntype{C}[1]{>{\centering\let\newline\\\arraybackslash\hspace{0pt}}m{#1}}
\newcolumntype{R}[1]{>{\raggedleft\let\newline\\\arraybackslash\hspace{0pt}}m{#1}}

\vspace{2cm}
\begin{figure}[htbp]
  \begin{center}
    \includegraphics[scale=0.28]{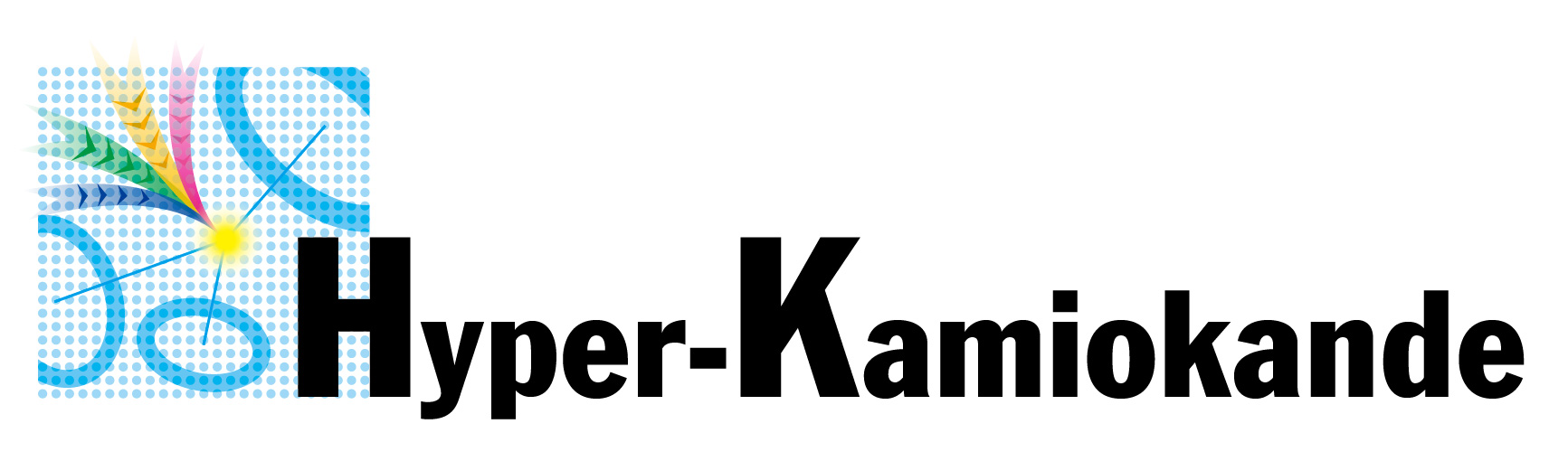}
  \end{center}
\end{figure}

\vspace{2cm}
\begin{center}
{\Huge \bf Design Report \\
(Dated: \today} )%\\
\end{center}
      \newpage



\newcommand{\BOSTON}{\affiliation{Boston University, Department of Physics, Boston, Massachusetts, U.S.A.}}
\newcommand{\UBC}{\affiliation{University of British Columbia, Department of Physics and Astronomy, Vancouver, British Columbia, Canada }}
\newcommand{\UCDAVIS}{\affiliation{University of California, Davis, Department of Physics, Davis, California, U.S.A.}}
\newcommand{\UCI}{\affiliation{University of California, Irvine, Department of Physics and Astronomy, Irvine, California, U.S.A.}}
\newcommand{\CSU}{\affiliation{California State University, Department of Physics, Carson, California, U.S.A.}}
\newcommand{\SACLAY}{\affiliation{IRFU, CEA Saclay, Gif-sur-Yvette, France}}
\newcommand{\CHONNAM}{\affiliation{Chonnam National University, Department of Physics, Gwangju, Korea}}
\newcommand{\DONGSHIN}{\affiliation{Dongshin University, Department of Physics, Naju, Korea}}
\newcommand{\LLR}{\affiliation{Ecole Polytechnique, IN2P3-CNRS, Laboratoire Leprince-Ringuet, Palaiseau, France}}
\newcommand{\LPNHE}{\affiliation{Laboratoire de Physique Nucleaire et de Hautes Energies, UPMC and Universite Paris-Diderot and CNRS/IN2P3, Paris, France}}
\newcommand{\EDINBURGH}{\affiliation{University of Edinburgh, School of Physics and Astronomy, Edinburgh, United Kingdom}}
\newcommand{\GENEVA}{\affiliation{University of Geneva, Section de Physique, DPNC, Geneva, Switzerland}}
\newcommand{\GIST}{\affiliation{GIST College, Gwangju Institute of Science and Technology, Gwangju 500-712, Korea}}
\newcommand{\HAWAII}{\affiliation{University of Hawaii, Department of Physics and Astronomy, Honolulu, Hawaii, U.S.A.}}
\newcommand{\IMPERIAL}{\affiliation{Imperial College London, Department of Physics, London, United Kingdom}}
\newcommand{\BARI}{\affiliation{INFN Sezione di Bari and Universit`a e Politecnico di Bari,Bari Italy}}
\newcommand{\NAPOLI}{\affiliation{INFN Sezione di Napoli and Universit\`a di Napoli, Dipartimento di Fisica, Napoli, Italy}}
\newcommand{\PADOVA}{\affiliation{INFN Sezione di Padova and Universit\`a di Padova, Dipartimento di Fisica, Padova, Italy}}
\newcommand{\ROME}{\affiliation{INFN Sezione di Roma, Universita' La Sapienza, Dipartimento di Fisica, Roma, Italy}}
\newcommand{\INR}{\affiliation{Institute for Nuclear Research of the Russian Academy of Sciences, Moscow, Russia}}
\newcommand{\ISU}{\affiliation{Iowa State University, Department of Physics and Astronomy, Ames, Iowa, U.S.A.}}
\newcommand{\KEK}{\affiliation{High Energy Accelerator Research Organization (KEK), Tsukuba, Ibaraki, Japan}}
\newcommand{\KOBE}{\affiliation{Kobe University, Department of Physics, Kobe, Japan}}
\newcommand{\KYOTO}{\affiliation{Kyoto University, Department of Physics, Kyoto, Japan}}
\newcommand{\YITP}{\affiliation{Kyoto University, Yukawa Institute for Theoretical Physics, Kyoto, Japan}}
\newcommand{\LNF}{\affiliation{Laboratori Nazionali di Frascati, Frascati, Italy}}
\newcommand{\LANCASTER}{\affiliation{Lancaster University, Physics Department, Lancaster, United Kingdom}}
\newcommand{\LIVERPOOL}{\affiliation{University of Liverpool, Department of Physics, Liverpool, United Kingdom}}
\newcommand{\LANL}{\affiliation{Los Alamos National Laboratory, New Mexico, U.S.A.}}
\newcommand{\LSU}{\affiliation{Louisiana State University, Department of Physics and Astronomy, Baton Rouge, Louisiana, U.S.A. }}
\newcommand{\MADRID}{\affiliation{University Autonoma Madrid, Department of Theoretical Physics, Madrid, Spain}}
\newcommand{\MADRIDIFT}{\affiliation{Instituto de F\'{\i}sica Te\'orica, UAM/CSIC, Madrid, Spain}}
\newcommand{\MSU}{\affiliation{Michigan State University, Department of Physics and Astronomy,  East Lansing, Michigan, U.S.A.}}
\newcommand{\MIYAGI}{\affiliation{Miyagi University of Education, Department of Physics, Sendai, Japan}}
\newcommand{\NAGOYA}{\affiliation{Nagoya University, Graduate School of Science, Nagoya, Japan}}
\newcommand{\KMI}{\affiliation{Nagoya University, Kobayashi-Maskawa Institute for the Origin of Particles and the Universe, Nagoya, Japan}}
\newcommand{\STELAB}{\affiliation{Nagoya University, Institute for Space-Earth Environmental Research, Nagoya, Japan}}
\newcommand{\NCBJ}{\affiliation{National Centre for Nuclear Research, Warsaw, Poland}}
\newcommand{\OKAYAMA}{\affiliation{Okayama University, Department of Physics, Okayama, Japan}}
\newcommand{\OCU}{\affiliation{Osaka City University, Department of Physics, Osaka, Japan}}
\newcommand{\OXFORD}{\affiliation{Oxford University, Department of Physics, Oxford, United Kingdom}}
\newcommand{\PENN}{\affiliation{Pennsylvania State University, Department of Physics, University Park, PA 16802, U.S.A.}}
\newcommand{\PITTSBURGH}{\affiliation{University of Pittsburgh, Department of Physics and Astronomy, Pittsburgh, Pennsylvania, U.S.A.}}
\newcommand{\RAL}{\affiliation{STFC, Rutherford Appleton Laboratory, Harwell Oxford, and Daresbury Laboratory, Warrington, United Kingdom}}
\newcommand{\REGINA}{\affiliation{University of Regina, Department of Physics, Regina, Saskatchewan, Canada}}
\newcommand{\RHUL}{\affiliation{Royal Holloway University of London, Department of Physics, Egham, Surrey, United Kingdom}}
\newcommand{\RIO}{\affiliation{Pontif{\'\i}cia Universidade Cat{\'o}lica do Rio de Janeiro, Departamento de F\'{\i}sica, Rio de Janeiro, Brazil}}
\newcommand{\ROCHESTER}{\affiliation{University of Rochester, Department of Physics and Astronomy, Rochester, New York, U.S.A.}}
\newcommand{\QMUL}{\affiliation{Queen Mary University of London, School of Physics and Astronomy, London, United Kingdom}}
\newcommand{\SHEFFIELD}{\affiliation{University of Sheffield, Department of Physics and Astronomy, Sheffield, United Kingdom}}
\newcommand{\SNU}{\affiliation{Seoul National University, Department of Physics, Seoul, Korea}}
\newcommand{\SEOYEONG}{\affiliation{Seoyeong University, Department of Fire Safety, Gwangju, Korea }}
\newcommand{\STONYBROOK}{\affiliation{State University of New York at Stony Brook, Department of Physics and Astronomy, Stony Brook, New York, U.S.A.}}
\newcommand{\SKKU}{\affiliation{Sungkyunkwan University, Department of Physics, Suwon, Korea}}
\newcommand{\TOHOKU}{\affiliation{Research Center for Neutrino Science, Tohoku University, Sendai, Japan}}
\newcommand{\ERI}{\affiliation{University of Tokyo, Earthquake Research Institute, Tokyo, Japan}}
\newcommand{\KAMIOKA}{\affiliation{University of Tokyo, Institute for Cosmic Ray Research, Kamioka Observatory, Kamioka, Japan}}
\newcommand{\RCCN}{\affiliation{University of Tokyo, Institute for Cosmic Ray Research, Research Center for Cosmic Neutrinos, Kashiwa, Japan}}
\newcommand{\IPMU}{\affiliation{University of Tokyo, Kavli Institute for the Physics and Mathematics of the Universe (WPI), Todai Institutes for Advanced Study, Kashiwa, Chiba, Japan}}
\newcommand{\TOKYO}{\affiliation{University of Tokyo, Department of Physics, Tokyo, Japan}}
\newcommand{\TITECH}{\affiliation{Tokyo Institute of Technology, Department of Physics, Tokyo, Japan}}
\newcommand{\TRIUMF }{\affiliation{TRIUMF, Vancouver, British Columbia, Canada}}
\newcommand{\TORONTO}{\affiliation{University of Toronto, Department of Physics, Toronto, Ontario, Canada}}
\newcommand{\WARSAW}{\affiliation{University of Warsaw, Faculty of Physics, Warsaw, Poland}}
\newcommand{\WUT}{\affiliation{Warsaw University of Technology, Institute of Radioelectronics and Multimedia Technology, Warsaw, Poland}}
\newcommand{\WARWICK}{\affiliation{University of Warwick, Department of Physics, Coventry, United Kingdom}}
\newcommand{\WASHINGTON}{\affiliation{University of Washington, Department of Physics, Seattle, Washington, U.S.A.}}
\newcommand{\WINNIPEG}{\affiliation{University of Winnipeg, Department of Physics, Winnipeg, Manitoba, Canada}}
\newcommand{\VT}{\affiliation{Virginia Tech, Center for Neutrino Physics, Blacksburg, Virginia, U.S.A.}}
\newcommand{\WROCLAW}{\affiliation{Wroclaw University, Faculty of Physics and Astronomy, Wroclaw, Poland}}
\newcommand{\YEREVAN}{\affiliation{Yerevan Institute for Theoretical Physics and Modeling, Halabian Str. 34; Yerevan 0036, Armenia}}
\newcommand{\YORK}{\affiliation{York University, Department of Physics and Astronomy, Toronto, Ontario, Canada}}
\newcommand{\KYIV}{\affiliation{Kyiv National University, Department of Nuclear  Physics, Kyiv, Ukraine}}
\newcommand{\YOKOHAMA}{\affiliation{Yokohama National University, Faculty of Engineering, Yokohama, Japan}}
\newcommand{\TUS}{\affiliation{Tokyo University of Science, Department of Physics, Chiba, Japan}}
\newcommand{\STOCKHOLM}{\affiliation{Stockholm University, Oskar Klein Centre and Dept. of Physics,  Stockholm, Sweden}}

\BOSTON
\UBC
\UCDAVIS
\UCI
\CSU
\SACLAY
\CHONNAM
\DONGSHIN
\LLR
\EDINBURGH
\GENEVA
\GIST
\HAWAII
\IMPERIAL
\BARI
\NAPOLI
\PADOVA
\ROME
\INR
\ISU
\KEK
\KOBE
\KYOTO
\LPNHE
\LNF
\LANCASTER
\LIVERPOOL
\LANL
\LSU
\MADRID
\MADRIDIFT
\MSU
\MIYAGI
\NAGOYA
\KMI
\STELAB
\NCBJ
\OKAYAMA
\OCU
\OXFORD
\PENN
\PITTSBURGH
\RAL
\REGINA
\RHUL
\RIO
\ROCHESTER
\QMUL
\SHEFFIELD
\SNU
\SEOYEONG
\STOCKHOLM
\STONYBROOK
\SKKU
\TOHOKU
\ERI
\KAMIOKA
\RCCN
\IPMU
\TOKYO
\TITECH
\TRIUMF
\TORONTO
\TUS
\WARSAW
\WUT
\WARWICK
\WASHINGTON
\WINNIPEG
\VT
\WROCLAW
\YEREVAN
\YORK
\KYIV
\YOKOHAMA

\author{K.~Abe}\KAMIOKA\IPMU
\author{Ke.~Abe}\KOBE
\author{H.~Aihara}\IPMU\TOKYO
\author{A.~Aimi}\PADOVA
\author{R.~Akutsu}\RCCN
\author{C.~Andreopoulos}\LIVERPOOL\RAL
\author{I.~Anghel}\ISU
\author{L.H.V.~Anthony}\LIVERPOOL
\author{M.~Antonova}\INR
\author{Y.~Ashida}\KYOTO
% ...

\collaboration{Hyper-Kamiokande proto-collaboration}


\begin{abstract}

Hyper-Kamiokande is a next generation underground water Cherenkov
detector, based on the highly successful Super-Kamiokande
experiment. It will serve as a far detector, 295\,km away, of a long
baseline neutrino experiment for the upgraded J-PARC
beam. It will also be a detector capable of observing - far beyond the
sensitivity of the Super-Kamiokande detector - proton decay,
atmospheric neutrinos, and neutrinos from astronomical sources.

The detector is much larger than Super-Kamiokande and
presents new experimental challenges that are addressed in this
report, where a full overview of the cavern and detector design R\&D
is given. This is also supported by a description of the upgraded beam
and near detector suite. Based on the design of the experiment the
expected sensitivity for both beam and atmospheric neutrinos, proton
decays, solar and astrophysical neutrinos, non standard physics,
etc. is shown. 
\end{abstract}

\maketitle

\tableofcontents

\clearpage
\color{black}
\section*{List of the acronyms}
We introduce here the acronyms used throughout the document:
\begin{itemize}
    \setlength{\itemsep}{0pt}%
    \setlength{\parskip}{0pt}%
\item AD: Avalanche Diode
\item B\&L: Box-and-Line dynode
\item BSM: Beyond the Standard Model
\item CC: charged currents
\item CCSNe: Core-Collapse Supernovae 
\item CCQE: charge current quasi-elastic
\item CE: Collection Efficiency
\item CPL: Concrete Protective Liner
\item DAQ: Data Acquisition
\item DR: Design Report
\item DT: deuterium-tritium
\item EBU: Event Building Unit
\item ECal: ND280 Electromagnetic Calorimeter
\item FC: Fully Contained
\item FCFV: Fully Contained in Fiducial Volume 
\item FGD: Fine Grained Detector
\item FRP: Fiber Reinforced Plastics
\item FV: Fiducial Volume
\item GUT: Grand Unified Theory
\item HDPE: High Density PolyEthylene
\item HK: Hyper-Kamiokande
\item HPD: Hybrid Photodetector
\item HPTPC: High Pressure Time Projection Chamber
\item HQE: High Quantum Efficiency
\item Hyper-K: Hyper-Kamiokande
\item IBC: International Board Representatives 
\item IBD: Inverse Beta Decay
\item ID: Inner Detector
\item INGRID: Interactive Neutrino GRID
\item ISC: International Steering Committee
\item IWCD: Intermediate Water Cherenkov Detector
\item LAPPD: Large Area Picosecond PhotoDetector 
\item LBNE: Long Baseline Neutrino Experiment
\item LAr: Liquid Argon calorimeter
\item LD: Laser Diode
\item LLDPE:Linear Low-Density PolyEthylene
\item LV: Lorentz Violation
\item MC: Monte Carlo
\item MLF: Material Science Facility
\item mPMT: Multi-channel Optical Module
\item MR: Main Ring synchrotron
\item NC: neutral currents
\item ND280: Near Detector 280m
\item NF: Nano Filter
\item OD: Outer Detector
\item PC: Partially Contained
\item PE: Photo Electron
\item PS: Power Supply
\item PTF: Photosensor Testing Facility
\item MH: neutrino mass hierarchy
\item QA: quality assurance
\item RBU: Readout Buffer Unit
\item RO: Reverse Osmosis
\item RCS: Rapid Cycling Synchrotron
\item SK: Super-Kamiokande
\item SM: Standard Model
\item Super-K: Super-Kamiokande
\item SUS: Stainless Steel (or Steel Use Stainless)
\item TITUS: Tokai Intermediate Tank for Unoscillated Spectrum
\item TPU: Trigger Processing Unit
\item TS: Target Station
\item UF: Ultra Filter
\item UPW: Ultra Purified Water
\item WAGASCI: Water Grid And SCIntillator detector
\item WC: Water Cherenkov
\end{itemize}

\clearpage
\color{black}
\graphicspath{{summary/figures/}}

\section*{Executive summary}

On the strength of a double Nobel prize winning experiment
(Super)Kamiokande and an extremely successful long baseline neutrino
programme, the third generation Water Cherenkov detector,
Hyper-Kamiokande, is being developed by an international collaboration
as a leading worldwide experiment based in Japan.

It will address the biggest unsolved questions in physics through a
multi-decade physics programme that will start in the middle of the
next decade.

The Hyper-Kamiokande detector will be hosted in the Tochibora mine,
about 295\,km away from the J-PARC proton accelerator research complex
in Tokai, Japan.

The currently existing accelerator will be steadily upgraded to reach
a MW beam by the start of the experiment.
A suite of near detectors will be vital to constrain the beam for
neutrino oscillation measurements. They will be a combination of
upgraded and new detectors at a distance ranging from 280\,m to
1-2\,km from the neutrino target.

A new cavern will be excavated at the Tochibora mine to host the
detector. The corresponding infrastructure will be built.  The
experiment will be the largest underground water Cherenkov detector in
the world and will be instrumented with new technology photosensors,
faster and with higher quantum efficiency than the ones in
Super-Kamiokande. Pressure tests demonstrate that they will be able to
support the pressure due to the massive tank.

The science that will be developed will be able to shape the future
theoretical framework and generations of experiments.
Hyper-Kamiokande will be able to measure with the highest precision
the leptonic CP violation that could explain the baryon asymmetry in
the Universe. The experiment also has a demonstrated excellent capability to
search for proton decay, providing a significant improvement in
discovery sensitivity over current searches for the proton lifetime.
The atmospheric neutrinos will allow to determine the neutrino mass
ordering and, together with the beam, able to precisely test the
three-flavour neutrino oscillation paradigm and search for new
phenomena. A strong astrophysical programme will be carried out at the
experiment that will also allow to measure precisely solar neutrino
oscillation.  A set of other main physics searches is planned, like
indirect dark matter.

In summary, a new experiment, based on the experience and facilities of
the already existing Super-Kamiokande and long baseline neutrino
experiment as T2K, is being developed by the international physics
community to provide a wide and groundbreaking multi-decade physics
programme from the middle of the next decade (see Table~\ref{tab:intro:phys}).

\begin{table}[hbtp]
  \caption{Expected sensitivities of the Hyper-Kamiokande experiment assuming 1 tank for 10 years.} 	
  \label{tab:intro:phys}
  \begin{center}
    \begin{tabular}{lll} \hline \hline
      Physics Target & Sensitivity & Conditions \\
      \hline \hline
      Neutrino study w/ J-PARC $\nu$~~ && 1.3\,MW $\times$ $10^8$ sec\\
      $-$ $CP$ phase precision & $<23^\circ$ & @ $\sin^22\theta_{13}=0.1$, mass hierarchy known \\
      $-$ $CPV$ discovery coverage & 76\% (3\,$\sigma$), 57\% ($5\,\sigma$) & @ $\sin^22\theta_{13}=0.1$, mass hierarchy known \\
      $-$ $\sin^2\theta_{23}$ & $\pm 0.017$ & 1$\sigma$ @ $\sin^2\theta_{23}=0.5$ \\
      \hline
      Atmospheric neutrino study && 10 years observation\\
      $-$ MH determination & $> 2.2\,\sigma$ CL & @ $\sin^2\theta_{23}>0.4$ \\
      $-$ $\theta_{23}$ octant determination & $> 3\,\sigma$ CL & @  $|\theta_{23} - 45^{\circ}| > 4^{\circ}$ \\\hline
      \hline
      Atmospheric and Beam Combination && 10 years observation\\
      $-$ MH determination & $> 3.8\,\sigma$ CL & @ $\sin^2\theta_{23}>0.4$ \\
      $-$ $\theta_{23}$ octant determination & $> 3\,\sigma$ CL & @  $|\theta_{23} - 45^{\circ}| > 2.3^{\circ}$ \\\hline
     Nucleon Decay Searches && 1.9 Mton$\cdot$year exposure \\
      $-$ $p\rightarrow e^+ + \pi^0$ & $7.8 \times 10^{34}$ yrs (90\% CL UL) &\\
                     & $6.3 \times 10^{34}$ yrs ($3\,\sigma$ discovery) &\\
      $-$ $p\rightarrow \bar{\nu} + K^+$ & $3.2 \times 10^{34}$ yrs (90\% CL UL) &\\
                     & $2.0 \times 10^{34}$ yrs ($3\,\sigma$ discovery) &\\ 
      \hline
      Astrophysical neutrino sources && \\
      $-$ $^8$B $\nu$ from Sun & 130 $\nu$'s / day & 4.5\,MeV threshold (visible
                                                     energy) w/ osc.\\
      $-$ Supernova burst $\nu$ & 54,000$-$90,000 $\nu$'s & @ Galactic center (10 kpc)\\ 
                     & $\sim$10 $\nu$'s & @ M31 (Andromeda galaxy) \\ 
      $-$ Supernova relic $\nu$ & 70 $\nu$'s / 10 years & 10$-$30\,MeV, 4.2$\sigma$ non-zero significance  \\
      $-$ WIMP annihilation in the Earth & & 10 years observation\\
      ~~($\sigma_{SD}$: WIMP-proton spin & $\sigma_{SD}=10^{-40}$cm$^2$ & @ $M_{\rm WIMP}=10$\,GeV, $\chi\chi\rightarrow b\bar b$ dominant\\
      ~~~~dependent cross section)& $\sigma_{SD}=10^{-44}$cm$^2$ & @ $M_{\rm WIMP}=50$\,GeV, $\chi\chi\rightarrow \tau^+ \tau^-$ dominant\\
      \hline \hline
    \end{tabular}
  \end{center}
\end{table}

\clearpage
\color{black}
\part{Introduction}
\graphicspath{{introduction/figures}}

\section{Introduction}
\label{section:intro}

Recent advances in experimental particle physics have yielded
fascinating insights into the inner workings of the smallest-scale
phenomena.  In 2012, the last missing piece of the standard model (SM)
of elementary particles, the Higgs boson, was finally observed by the
ATLAS and CMS experiments at the Large Hadron Collider (LHC) in
CERN~\cite{Aad:2012tfa,Chatrchyan:2012xdj}. The SM is highly
successful in explaining experimental data, however our current
ability to describe nature from a fundamental physics point of view is
far from satisfactory, most significantly the fact that neutrino mass
cannot be incorporated, and so we need beyond the standard model (BSM)
physics. 

The Nobel Prize in 2002 was awarded for the detection of cosmic
neutrinos (in particular the ones coming from supernova) in Kamiokande
and for the pioneering solar neutrino experiment at the Homestake
mine. More recently, the 2015 Nobel Prize was awarded for the
discovery of neutrino oscillations using data taken by the
Super-Kamiokande (Super-K) and the Sudbury Neutrino Observatory collaborations,
which has the very profound implication that neutrinos have non-zero
but very tiny masses.

Building on the expertise gained from the past and current
experiments, Kamiokande and Super-Kamiokande, Hyper-Kamiokande
(Hyper-K) is a natural progression for the highly successful
Japanese-hosted neutrino program.

\begin{figure}[htbp]
  \begin{center}
  \begin{tabular}{cc}
    \includegraphics[width=0.8\textwidth]{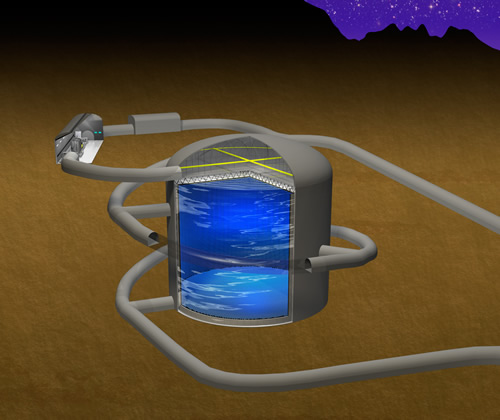}
  \end{tabular}
  \caption{Illustration of the Hyper-Kamiokande first cylindrical tank in Japan.}
  \label{fig:hk-perspective-JTank}
  \end{center}
\end{figure}

Hyper-Kamiokande is a next-generation, large-scale water Cherenkov
neutrino detector. A dedicated task force determined the optimal tank
design to be two cylindrical detectors that are 60\,m in height and
74\,m in diameter with 40\% photocoverage, where a staging between the
first and second tank is considered. We first focus on building the
first tank in Japan, see Fig.~\ref{fig:hk-perspective-JTank} for the
drawing.

Candidate sites for the Hyper-K experiment were selected such that
neutrinos generated in the J-PARC accelerator facility in Tokai, Japan
can be measured in the detector. J-PARC will operate a 750\,kW beam in
the near future, and has a long-term projection to operate with 1300\,kW
of beam power. Near detectors placed close to the J-PARC beam line
will determine the information about the neutrinos coming from the
beam, thus allowing for the extraction of oscillation parameters from
the Hyper-K detector. The ND280 detector suite, which has been used
successfully by the T2K experiment, could be upgraded to further
improve the measurement of neutrino cross section and flux. The
WAGASCI detector is a new concept under development that would have a
larger angular acceptance and a larger mass ratio of water (and thus
making the properties more similar to the Hyper-K detector) than the
ND280 design. Intermediate detectors, placed 1-2 km from the J-PARC
beam line, would measure the beam properties directly on a water
target. Details of the beam, as well as the near and intermediate
detectors, can be found in Section~\ref{section:jparc}.

Hyper-K is a truly international proto-collaboration with over 70
participating institutions from Armenia, Brazil, Canada, France,
Italy, Korea, Poland, Russia, Spain, Sweden, Switzerland, Ukraine, the United
Kingdom and the United States, in addition to Japan.

Hyper-K will be a multipurpose neutrino detector with a rich physics
program that aims to address some of the most significant questions
facing particle physicists today. Oscillation studies from
accelerator, atmospheric and solar neutrinos will refine the neutrino
mixing angles and mass squared difference parameters and will aim to
make the first observation of asymmetries in neutrino and antineutrino
oscillations arising from a CP-violating phase, shedding light on
one of the most promising explanations for the matter-antimatter
asymmetry in the Universe. The search for nucleon decays will probe
one of the key tenets of Grand Unified Theories. In the case of a
nearby supernova, Hyper-K will observe an unprecedented number of
neutrino events, providing much needed experimental results to
researchers seeking to understand the mechanism of the
explosion. Finally, the detection of astrophysical neutrinos from
sources such as dark matter annihilation, gamma ray burst jets, and
pulsar winds could further our understanding of some of the most
spectacular, and least understood, phenomena in the Universe. These
topics will be discussed further in Section~\ref{section:physics}.

This design report is organized as follows. There are a total of five
parts.  The remainder of this Part~\ref{section:intro} outlines the
theoretical framework for the physics topics contained in this report
and discusses the relationships between Hyper-K and other large-scale
neutrino experiments. Part~\ref{part:experimentalconfiguration}
describes the experimental configuration where
Section~\ref{section:jparc} describes the J-PARC neutrino beam line
and near detector facility; Section~\ref{section:design} discusses the
technical details of the experimental design and
Section~\ref{section:software} details the software packages that will
be utilized by the Hyper-K experiment. A discussion of pertinent
radioactive backgrounds is contained in
Section~\ref{section:background}. Part~\ref{section:physics} explains
the physics capabilities for
Hyper-K. Part~\ref{section:secon-detector-korea} introduces a possible
second tank in Korea.  
The last part, Part~\ref{section:appendices}, is the Appendix with
details on the liner sheet tests (A) and a
description of a possible second tank in Japan at Hakamagoshi (B).

\subsection{Neutrino oscillations}

Neutrino oscillations, discovered by the Super-Kamiokande (Super-K)
experiment in 1998~\cite{Fukuda:1998mi}, implies that neutrinos have
nonzero masses and flavor mixing, providing one of the most convincing
experimental proofs known today for the existence of physics beyond
the Standard Model (BSM).
Indeed neutrino oscillation has been established as a very powerful
tool to probe extremely small neutrino masses (or their differences)
as well as lepton flavor mixing.

Throughout this design report, unless stated otherwise, we consider
the standard three flavor neutrino framework.  The 3$\times3$ unitary
matrix $U$ which describes the mixing of neutrinos~\cite{Maki:1962mu}
(that is often referred to as the Pontecorvo-Maki-Nakagawa-Sakata
(PMNS) or Maki-Nakagawa-Sakata
(MNS)~\cite{Pontecorvo:1967fh,Maki:1962mu} matrix) relates the flavor
and mass eigenstates of neutrinos as
\begin{eqnarray} 
\nu_\alpha  = \sum_{i=1}^3 U_{\alpha i} \nu_i
\ \ (\alpha = e, \mu, \tau),
\end{eqnarray} 
where $\nu_\alpha  (\alpha = e, \mu, \tau)$ 
and $\nu_i (i = 1,2,3)$ 
denote neutrino fields with definite flavor and mass, respectively. 

Using the standard parameterization, found, e.g. in
Ref.~\cite{Agashe:2014kda}, $U$ can be expressed as,
\begin{eqnarray}
\hskip -0.5cm
U 
& = &
\left(
\begin{array}{ccc}
1 & 0   & 0 \\
0 & c_{23} & s_{23} \\
0 & -s_{23} & c_{23} \\
\end{array}
\right)
\left(
\begin{array}{ccc}
c_{13} & 0   & s_{13}e^{-i\deltacp} \\
0 & 1 & 0 \\
-s_{13}e^{i\deltacp} & 0 & c_{13} \\
\end{array}
\right)
\left(
\begin{array}{ccc}
c_{12} & s_{12}  & 0 \\
-s_{12} & c_{12} & 0 \\
0 & 0 & 1 \\
\end{array}
\right) 
\left(
\begin{array}{ccc}
1 & 0 & 0 \\
0 & e^{i\frac{\alpha_{21}}{2}} & 0 \\
0 & 0 & e^{i\frac{\alpha_{31}}{2}} \\
\end{array}
\right)
\label{eq:mixing}
\end{eqnarray}
where $c_{ij} \equiv \cos\theta_{ij}$, $s_{ij} \equiv \sin\theta_{ij}$, 
and $\deltacp$ --- often called the Dirac $CP$ phase ---,  
is the Kobayashi-Maskawa type $CP$ phase~\cite{Kobayashi:1973fv} 
in the lepton sector. 
On the other hand, the two phases, $\alpha_{21}$ and $\alpha_{31}$,
--- often called Majorana $CP$ phases --- exist only if neutrinos are
of Majorana type~\cite{Schechter:1980gr, Bilenky:1980cx,Doi:1980yb}.
While the Majorana $CP$ phases can not be observed in neutrino
oscillation, they can be probed by lepton number violating processes
such as neutrinoless double beta ($0\nu \beta\beta$) decay.

In the standard three neutrino flavor framework, only two mass squared
differences, $\Delta m^2_{21}$ and $\Delta m^2_{31}$, for example, are
independent.  Here, the definition of mass squared differences is
$\Delta m^2_{ij}$ $\equiv$ $m^2_i - m^2_j$. Therefore, for a given
energy and baseline, there are six independent parameters that
describe neutrino oscillations: three mixing angles, one $CP$ phase,
and two mass squared differences.
Among these six parameters, $\theta_{12}$ and $\Delta m^2_{21}$ have
been measured by solar~\cite{Ahmad:2002jz,Ahmad:2001an,Abe:2010hy} and
reactor~\cite{Eguchi:2002dm,Araki:2004mb,Abe:2008aa} neutrino
experiments.  The parameters $\theta_{23}$ and $|\Delta m^2_{32}|$
(only its absolute value) have been measured by
atmospheric~\cite{Ashie:2005ik,Ashie:2004mr} and
accelerator~\cite{Ahn:2006zza,Adamson:2011ig,Abe:2012gx,Abe:2014ugx}
neutrino experiments.
In the last few years, \(\theta_{13}\) has also been measured by
accelerator~\cite{Abe:2011sj,Adamson:2011qu,Abe:2013xua,Abe:2013hdq}
and reactor experiments~\cite{Abe:2011fz,Ahn:2012nd,An:2012eh,
  An:2013zwz,Abe:2014lus}.
Remarkably, the Super-K detector has successfully measured all of
these mixing parameters, apart from the $CP$ phase and the sign of
$\Delta m^2_{32}$.
The current best-measured values of the mixing parameters are listed
in ~\cite{Agashe:2014kda}, where the mass hierarchy and $CP$ phase are still unknown 
though there are some weak preferences by the current 
neutrino data as will be mentioned later in this section. 

By studying neutrino oscillation behaviour, Hyper-K is expected to
improve the current bounds obtained by Super-K for various
non-standard neutrino properties, such as the possible presence of
sterile neutrinos~\cite{Abe:2014gda}, non-standard interactions of
neutrinos with matter~\cite{Mitsuka:2011ty}, or violation of Lorentz
invariance~\cite{Abe:2014wla}.

\subsubsection{Mass Hierarchy}

The positive or negative sign of $\Delta m^2_{32}$ (or equivalently
that of $\Delta m^2_{31}$) corresponds, respectively, to the case of
normal ($m_2 < m_3$) or inverted ($m_3 < m_2$) mass hierarchy
(ordering).  From a theoretical point of view, it is of great interest
to know the mass hierarchy to understand or obtain clues about how the
neutrino masses and mixing are generated (see
e.g. \cite{Mohapatra:2006gs} for a review).  Also the mass hierarchy
has a significant impact on the observation of the $0\nu \beta \beta $
decay for the case where neutrinos are Majorana particles.  If the
mass hierarchy is inverted, a positive signal of $0\nu\beta\beta$
is expected in future experiments if the current sensitivity on the effective
Majorana mass can be improved by about one order of magnitude beyond
the current limit.

In the $\nu_\mu \to \nu_e$ appearance channel, its oscillation
probability at around the first oscillation maximum, $O(L/E_\nu) \sim
1$, tends to be enhanced (suppressed) if the mass hierarchy is normal
(inverted) due to the matter effect or the so called
Mikheev-Smirnov-Wolfenstein (MSW)
effect~\cite{Mikheev:1986gs,Wolfenstein:1977ue} as we will see in Part
\ref{section:physics}.  For the antineutrino channel, $\bar{\nu}_\mu
\to \bar{\nu}_e$, the effect become opposite, namely, the normal
(inverted) mass hierarchy tends to suppress (enhance) the appearance
probability.  The longer the baseline ($L$), larger the effect of such
enhancement or suppression.  Therefore, in principle, the mass
hierarchy can be determined by measuring the oscillation probability
provided that the matter effect is sufficiently large.  This is the
most familiar way to determine the mass hierarchy in neutrino
oscillation which can be done using accelerator or atmospheric
neutrinos.

Independently from this method, it is also possible to determine the
mass hierarchy by observing the small interference effects caused by
$\Delta m^2_{31}$ and $\Delta m^2_{32}$ in the medium baseline ($L
\sim 50$ km) reactor neutrino oscillation experiment as first
discussed in \cite{Petcov:2001sy}.  The proposed projects such as
JUNO~\cite{An:2015jdp} and RENO-50~\cite{Kim:2014rfa} aim to determine
the mass hierarchy by this method.
Furthermore, in principle, it is possible to determine the mass
hierarchy by comparing the absolute values of the effective mass
squared differences determined by reactor ($\bar{\nu}_e$
disappearance) and accelerator ($\nu_\mu$ disappearance) with high
precision ~\cite{deGouvea:2005hk,Nunokawa:2005nx}.

It is expected by the time Hyper-K will start its operation,
around the year 2025, the mass hierarchy could be determined at $\sim$
(3-4)$\sigma$ or more by combining the future data coming from the
ongoing experiments such as NOvA, T2K and reactor experiments, Daya
Bay~\cite{Guo:2007ug}, RENO~\cite{Ahn:2010vy}, Double
Chooz~\cite{Ardellier:2006mn}, and proposed future experiments such as
JUNO~\cite{An:2015jdp}, RENO-50~\cite{Kim:2014rfa},
ICAL~\cite{Ahmed:2015jtv}, PINGU~\cite{Aartsen:2014oha}, and
ORCA~\cite{Katz:2014tta} where the last three projects will use
atmospheric neutrinos to determine the mass hierarchy.

\subsubsection{CP Violation}

The magnitude of the charge-parity ($CP$) violation in neutrino
oscillation can be characterized by the difference of neutrino
oscillation probabilities between neutrino and anti-neutrino channels
~\cite{Barger:1980jm,Pakvasa:1980bz}.

The current data coming from T2K~\cite{Abe:2015awa} and
NOvA~\cite{NOvA-talk-nufact15}, when combined with the result of the
reactor $\theta_{13}$ measurement, prefer the value around
$\delta_{CP} \sim - \pi/2$ (or equivalently, $\delta_{CP} \sim
3\pi/2$) for both mass hierarchies though the statistical significance
is still small.
Interestingly, the Super-K atmospheric neutrino data also prefers
similar $\delta_{CP}$ values with a similar statistical
significance~\cite{Wendell:2014dka}.

If $CP$ is maximally violated ($|\sin \delta_{CP} | \sim 1$), $CP$
violation ($\sin \delta_{CP} \ne 0$) could be established at $\sim$
(2-3)$\sigma$ CL by combining the future data coming from T2K and NOvA
as well as with data coming from the reactor $\theta_{13}$
measurements.

In Hyper-K the neutrino oscillation parameters will be measured using
two neutrino sources which can provide complementary information.
Both atmospheric neutrinos, where neutrino oscillations were first
confirmed by Super-K, and a long baseline neutrino beam, where
electron neutrino appearance was first observed by T2K, will be
employed.

With a total exposure of 1.3~MW $\times$ 10$^8$ sec integrated proton
beam power (corresponding to $2.7\times10^{22}$ protons on target
with a 30~GeV proton beam) to a $2.5$-degree off-axis neutrino beam,
it is expected that the leptonic $CP$ phase $\deltacp$ can be
determined to better than 23 degrees for all possible values of
$\deltacp$, and $CP$ violation can be established with a statistical
significance of more than $3\sigma$ ($5\sigma$) for $76\%$
($57\%$) of the $\deltacp$ parameter space.

\begin{figure}
   \centering
   \includegraphics[width=0.9\textwidth]
                   {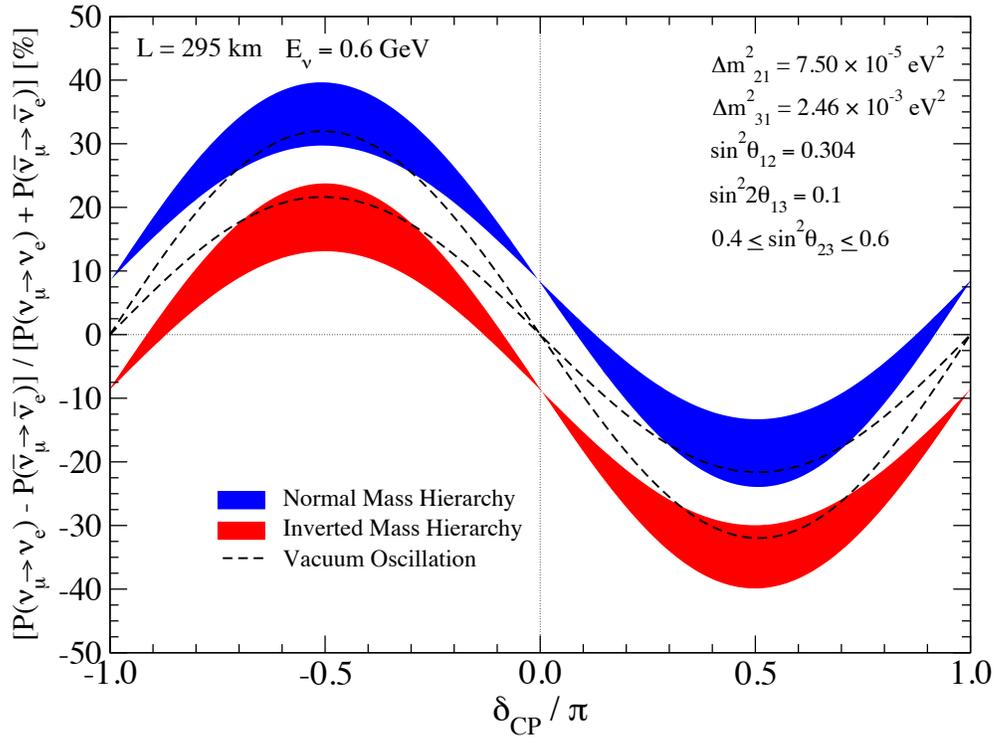} 
   \caption{The effect of neutrino mass hierarchy and $CP$-violation ($\delta_{CP}$) on
     the neutrino/antineutrino detection probability, for a specific
     set of neutrino mixing parameters, neutrino energy(E$_\nu$), and
     propagation length (L).}
   \label{fig:hierarchy_CP}
\end{figure}

Figure~\ref{fig:hierarchy_CP} shows how both CP-violation and mass
hierarchy affect the difference between $\nu_\mu \to \nu_e$ detection
probability relative to $\bar{\nu}_\mu \to \bar{\nu}_e$ detection
probability for a given set of
neutrino parameters. 

\subsection{Astrophysical neutrino observations \label{subsection:intro/astrophysics}}

 Hyper-K is also capable of observing neutrinos from various
 astrophysical objects.  One main advantage of the detector is that
 its energy threshold can be set as low as several MeV; this enables
 us to reconstruct neutrinos from the Sun and supernovae on an
 event-by-event basis.

The Sun is an abundant and nearby source of neutrinos. Recently,
Super-K showed the first indication of the terrestrial matter effects
on $^8$B solar neutrino oscillations ~\cite{sk4-daynight}. This was a
direct confirmation of the MSW model
\cite{Wolfenstein:1977ue,Mikheyev:1985zz,Mikheyev:1986zz} predictions
for neutrino interactions with matter, which is also used to describe
neutrino behaviour as it travels through the Sun. Furthermore,
terrestrial matter effects hint at an intriguing possibility of using
atmospheric and long baseline neutrinos to measure mass hierarchy and
$CP$ phase as both these parameters affect how neutrinos interact with
matter. Hyper-K hopes to measure terrestrial matter effects with
higher precision to better understand neutrino oscillation behaviour
in the presence of matter. This also might resolve the $\sim 2 \sigma$
tension between the current best fit values of $\Delta m^2_{21}$ from
solar and reactor neutrino experiments, which is thought to be due to
solar neutrino interactions in matter. Additionally, there are several
physics goals for the solar neutrino observations in Hyper-K, such as
long and short time variation of the $^8$B flux, the first measurement
of $hep$ neutrinos, and precise measurement of solar neutrino energy
spectrum.

Computational simulations of core-collapse supernovae (CCSN) have
failed to successfully reproduce explosions for more than 40 years.
However, thanks to the recent advances in modeling techniques and the
growth of available computation power, multi-dimensional (2D and 3D)
simulations can now produce successful explosions \cite{Bruenn:13,
  Melson:15a, Lentz:15, Takiwaki:14}.  Nevertheless, there are still
some puzzles, such as the finding that the total explosion energy of
the available multi-dimensional models is small compared to the
SN1987A observation.  Furthermore, the available 3D models are
generally less energetic (or unsuccessful) compared with the more
extensively simulated 2D models \cite{Couch:14, Tamborra:13,
  Lentz:15,Takiwaki:14}.  Clearly, details of the supernova explosion
mechanism are still lacking.  High statistics observations of
neutrinos from a CCSN (along with gravitational waves) are the only
way to obtain precious inside information on the dynamics of the CCSN
central engine and the explosion mechanism
\cite{OConnor:13,Tamborra:13}.  If a CCSN explosion were to take place
near the center of our Galaxy, Hyper-K would observe as many as
tens of thousands of neutrino interactions (see Section~\ref{sec:supernova}).
Furthermore, Hyper-K will have the ability to precisely determine the
arrival time of supernova neutrinos, which will help contribute to the
understandings of both neutrino and CCSN properties. For example, by
comparing of the number of $\nu_e$ and $\bar{\nu}_e$ during the CCSN
neutronization burst (first $\sim$10\,msec) we will be able to
determine the neutrino mass hierarchy (see
Section~\ref{sec:supernova}). High frequency timing will also provide
experimental evidence of the multidimensional dynamics thought to be
crucial in the CCSN explosion mechanism \cite{Tamborra:13}. A large
target volume like that of Hyper-K is also required to observe
neutrinos from CCSN explosions in nearby ($\sim$ few Mpc) galaxies.
In this volume, CCSNe occur every few years \cite{Ando:2005ka}.
Meanwhile, while waiting for a nearby explosion to occur, the
continuous flux of relic supernova neutrinos from all past CCSN
explosions in the observable universe will guarantee a steady
accumulation of valuable astrophysical data.

Thanks to its good low energy performance for upward-going muons,
Hyper-K has a larger effective area for upward-going muons below 30
GeV than do cubic kilometer-scale neutrino telescopes.  Additionally,
fully contained events in Hyper-K have energy, direction, and flavor
reconstruction and resolutions as good as those in Super-K.  This high
performance will be useful for further background suppression or
studies of source properties.  For example, the detector is extremely
sensitive to the energy range of neutrinos from annihilations of light
(below 100 GeV) WIMP dark matter, a region which is suggested by
recent direct dark matter search experiments. Hyper-K can search for
dark matter WIMPs by looking for neutrinos created in pair
annihilation from trapped dark matter in the Galactic centre or the
centre of the Sun. Atmospheric neutrinos are a background to this WIMP
search, so spacial cuts are made to determine if there is an excess of
neutrinos coming from the Galactic centre or the Sun. Hyper-K will
have the ability to detect both $\nu_{e}$ and $\nu_{\mu}$ components
of the signal, making it more sensitive to this type of analysis.

The detection of neutrinos from solar flares is another astrophysics
goal for Hyper-K.  This will give us important information about the
mechanism of the particle acceleration at work in solar flares.  There
have been some estimations of the number of expected
neutrinos. Although it has large uncertainties, about 20 neutrinos
will be observed at Hyper-K during a solar flare as large as the one
in 20 January 2005.
Hyper-K also has the potential to see neutrinos from astrophysical
sources such as magnetars, pulsar wind nebulae, active galactic
nuclei, and gamma ray bursts. The large target volume of Hyper-K,
combined with the potential for these sources to emit neutrinos with
energy at the GeV-TeV scale, could make Hyper-K an interesting
experiment for observing these neutrinos. As with dark matter
searches, the most significant background for detecting neutrinos for
these astrophysical sources are atmospheric neutrinos. Spacial, and in
some cases temporal, cuts need to be utilized to disentangle the
astrophysical neutrino signal from the atmospheric neutrino
background.

\subsection{Nucleon decay searches\label{subsection:intro/nucleondecay}}

The stability of everyday matter motivated Weyl, Stueckelberg, Wigner,
and other early quantum physicists to introduce a conserved quantity,
baryon number, to explain the observed and unobserved particle
reactions. Baryon number violation is believed to have played an
important role during the formation of the universe, and comprises one
of the famous Sakharov Conditions to explain the baryon asymmetry of
the universe. Proton decay and the decay of bound neutrons are
observable consequences of the violation of baryon number.

The Standard Model Lagrangian explicitly conserves baryon number,
although anomalous quantum effects do violate baryon number at an
unobservably tiny level. Nevertheless, there are reasons to believe
that the Standard Model is part of a more expansive theory. Baryon
number violation is a generic prediction of Grand Unified Theories
(GUTs) that combine quarks and leptons and include interactions that
allow their transition from one to the other. These theories are well
motivated by observations such as the equality of the sum of quark and
lepton charges, the convergence of the running gauge couplings at an
energy scale of about $10^{16}$ GeV, and frequently have mechanisms to
generate neutrino mass. If new forces carrying particles have masses
at this GUT scale, the lifetime of the proton will be in excess of
$10^{30}$ years, where past, present, and future proton decay
experiments must search.

\begin{figure}
   \centering
   \includegraphics[width=0.9\textwidth]
                   {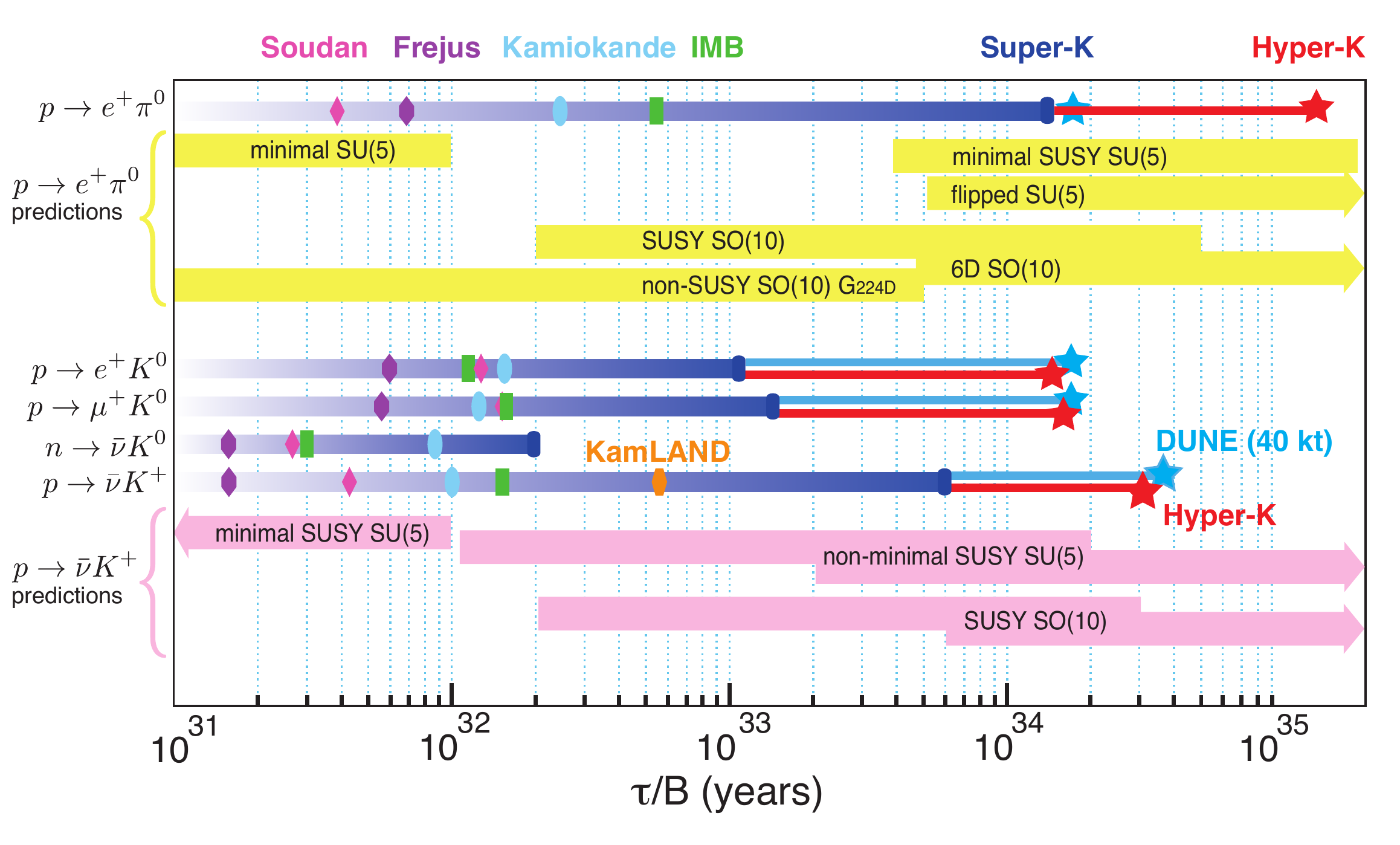} 
   \caption{A comparison of historical experimental limits on the rate
     of nucleon decay for several key modes to indicative ranges of
     theoretical prediction. Included in the
     figure are projected limits for Hyper-Kamiokande and DUNE based
     on 10 years of exposure.}
   \label{fig:limits-compared-theory}
\end{figure}

Baryon number violation has never been experimentally observed and
lifetime limits, mainly by Super-Kamiokande, greatly restrict
allowable Grand Unified theories and other interactions of interest to
model building theorists. In Fig.~\ref{fig:limits-compared-theory}, we
show 90\% CL lifetime limits by Super-Kamiokande and earlier
experiments compared with representative lifetime ranges predicted by
various GUTs. We also show the improvement in lifetime limits expected
for 10 years of Hyper-Kamiokande exposure. The complementary
experiment DUNE, assumed to be a 40 kt liquid argon time projection
chamber (LArTPC) is also a sensitive to nucleon decay. Due to its
smaller mass compared to Hyper-K, it is competitive mainly in modes
with distinctive final state tracks such as those involving kaons.

\begin{figure}
   \centering
   \includegraphics[width=0.9\textwidth]
                   {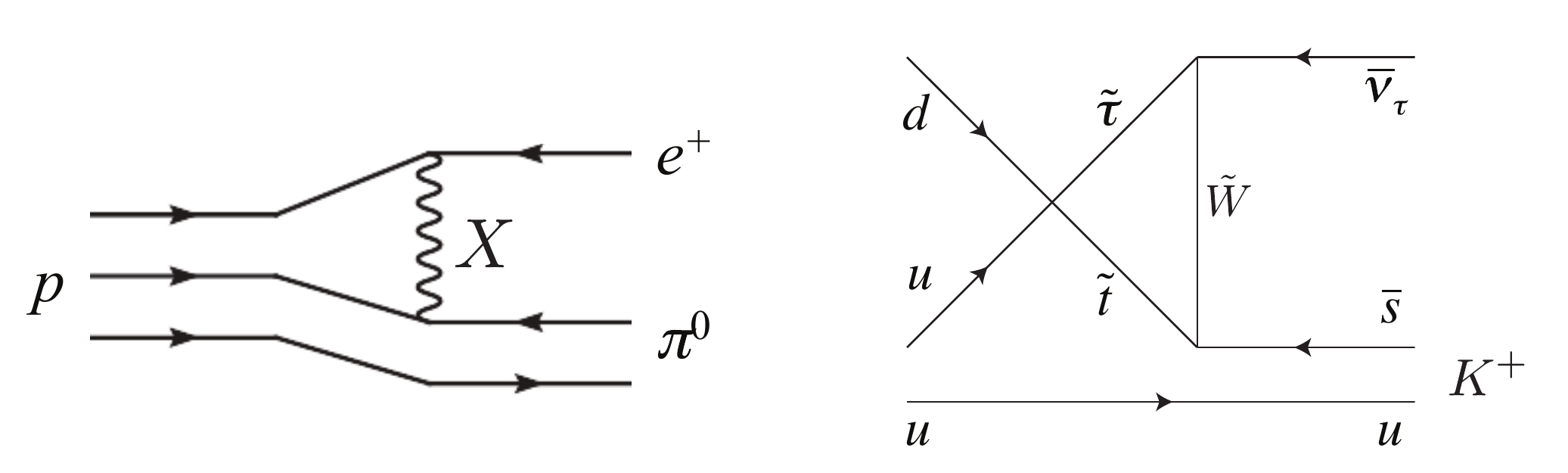} 
   \caption{Two sample Feynman diagrams that could be responsible for
     proton decay. The left diagram is a $d = 6$ interaction mediated
     by a heavy gauge boson, $X$, with mass at the GUT scale. The
     right diagram contains superymmetric particles and a $d = 5$
     operator that is predicted in many SUSY GUTs.}
   \label{fig:diagrams-ek}
\end{figure}

The message the reader should conclude from this figure is that 10
years of Hyper-K exposure is sensitive to lifetimes that are commonly
predicted by modern grand unified theories. The key decay channel $p
\rightarrow e^+\pi^0$ has been emphasized, because it is dominant in a
number of models, and represents a nearly model independent reaction
mediated by the exchange of a new heavy gauge boson with a mass at the
GUT scale. The other key channels involve kaons, wherein a final state
containing second generation quarks are generic predictions of GUTs
that include supersymmetry. Example Feynman diagrams are shown in
Fig.~\ref{fig:diagrams-ek}.

Generally, nucleon decay may occur through multiple channels and
ideally, experiments would reveal information about the underlying GUT
by measuring branching ratios.  It is a strength of Hyper-K that it is
sensitive to a wide range of nucleon decay channels, however the few
shown here are sufficient to discuss the details of the search for
nucleon decay by Hyper-Kamiokande later in this document.

Practically, because of the stringent limits from more than
300\,kt$\cdot$y of Super-K running, the next generation experiments
will have to concentrate on the discovery of nucleon decay, perhaps by
one or a small number of events. The predictions are uncertain to two
or three orders of magnitude, and one should not expect a negative
search to definitively rule out the idea of GUTs. To excel in the
search for proton decay, Hyper-Kamiokande requires the largest mass
that is affordable in combination with sufficient instrumentation to
minimize experimental background.

\subsection{Synergies between Hyper-K and other neutrino experiments}

This section will focus on understanding how Hyper-K will fit into the
context of the global neutrino community.  This includes currently
operating experiments such as T2K and Super-K, as well as future
experiments like DUNE.

\subsubsection{T2K}

T2K \cite{Abe:2011sj} is a currently-operating experiment which uses
Super-K to measure neutrinos produced in the J-PARC beam line. Hyper-K
will use much of the existing infrastructure used by T2K, particularly
the beam line and near detectors. Hyper-K will also benefit from any
improved data analysis techniques developed for T2K. Several important
T2K upgrades and improvements are planned for the coming years, and
this will have a direct impact on improved Hyper-K performance.

\begin{itemize}
	\item
          \textbf{Near detector improvements}: The T2K experiment uses
          the ND280 near detector suite. Future analysis improvements
          in the ND280 detector aim to reduce the cross section and
          flux uncertainties. Hardware upgrades, particularly to the
          time projection chamber component has also been
          proposed. The reader should refer to
          Section~\ref{subsection:nd280} for the full details of the
          current and future status of the ND280 detector.
	\item
          \textbf{Increased beam power}: J-PARC is planning an upgrade
          of the proton drivers in the neutrino beam. The near-term
          goal is to improve the beam power from 365\,kW to
          750\,kW. After the proton driver upgrade, beam power is
          projected to reach 1300\,kW. See Section~\ref{section:jparc}
          for more details.
	\item
          \textbf{Better data analysis techniques}: T2K demonstrated
          in its publications about $\nu_e$ appearance that
          aspects of the data analysis such $\pi^0$ rejection can be
          improved. Other improvements to the data analysis technique
          are under development, including $\nu_e$ detection
          efficiency, precision of the vertex
          determination (which could allow for an increased fiducial
          volume), and an improvement in $\pi/\mu$ separation.
\end{itemize}

In addition to benefiting directly from the upgraded hardware and
analysis techniques, Hyper-K will also benefit from the expertise
gained through implementing these upgrades. Furthermore, these
upgrades can serve as a test bed for new near detector designs that
have been proposed for Hyper-K (see Section~\ref{section:jparc}).

\subsubsection{Super-K}

In June 2015, the Super-Kamiokande Collaboration approved the SK-Gd
project.  This project is an upgrade of the detector's capabilities,
achieved by dissolving 0.2\% gadolinium sulfate into Super-K's water
in order to enhance detection efficiency of neutrons from neutrino
interactions. One of the main motivations of SK-Gd is to discover
supernova relic neutrinos (SRN), the diffuse flux of neutrinos emitted
by all supernovae since the beginning of the universe.  SRN primarily
interact in Super-K via inverse beta decay (IBD). Therefore,
following the prompt detection of a positron, the accompanying IBD
neutron can be identified in SK-Gd by a delayed gamma cascade, the
result of the neutron's capture on gadolinium.  As a result of this
positive identification of true IBD events, a much improved separation
between signal and background can be achieved.

As Super-K will be the first example of gadolinium loading in a
large-scale water Cherenkov detector, this will be a template for any
future possibility of loading gadolinium into Hyper-K. In addition to
determining the physics performance of gadolinium-loaded water,
Hyper-K will also benefit from the extensive research done to optimize
the water purification system, as well as the tests for material
compatibility that was required to upgrade the Super-K detector.

\subsubsection{DUNE}
The Deep Underground Neutrino Experiment (DUNE), formerly LBNE
\cite{LBNE}, is a 40 kilotonne liquid argon neutrino experiment that
is projected to begin taking data around the same time as
Hyper-K. Because DUNE will use a different target material than
Hyper-K (liquid argon rather than water), many complementary
measurements can be made, including nucleon decay measurements (as
described in Section~\ref{subsection:intro/nucleondecay}) and
supernova neutrino detection.

As mentioned in Section~\ref{subsection:intro/astrophysics},
information about the neutrino signature from supernovae is much
sought after, and Hyper-K and DUNE will each add to the overall
picture. The primary reaction channel for these neutrinos in Hyper-K
is the inverse beta decay channel, in which only electron
antineutrinos will take part. In DUNE, the reaction channel will be
the charged-current reaction on $^{40}$Ar, which measures electron
neutrinos. Taken together, these measurements will be able to
determine the relative abundance of neutrinos to
antineutrinos. Furthermore, DUNE will be able to better determine some
features of the neutrino spectrum which are dominated by the electron neutrino
signal, such as the neutronization burst that occurs during early
times, while Hyper-K will better measure features where there is an
antineutrino signal, such as the accretion and cooling phases that
occur at late times.

Due to the fact that the baseline between the accelerator facility and Hyper-K will be
shorter than the proposed baseline for the DUNE experiment, the two
experiments will have some complementarity in the information they can
extract from their accelerator programs. The longer baseline to the
DUNE experiment
means their measurement will be more affected by matter effects, which
will give them more sensitivity to the mass
hierarchy. The shorter baseline of Hyper-K experiment means less
sensitivity to matter effects, which should lead to an increased
sensitivity to the measurement of the CP-violation phase. This is further
described in Section~\ref{sec:cp}.

\clearpage
\color{black}
\part{Experimental Configuration} \label{part:experimentalconfiguration}
\section{J-PARC neutrino beam facility} \label{section:jparc}
The accelerator neutrinos detected by Hyper-K are produced at the
Japan Proton Accelerator Research Complex (J-PARC)~\cite{JPARCTDR}.
The proton accelerator chain, neutrino beamline and near detectors are
located within J-PARC.  Proposed intermediate detectors would be
located near the J-PARC site at a distance of 1-2\,km from the
production target. This section describes the proton accelerator
chain, neutrino beamline, near detectors and proposed intermediate
detectors.  In each case, the current configuration and future
upgrades are described.

\subsection{Neutrino beam and near detectors in long baseline oscillation measurements \label{sec:beam_and_nd}}
The neutrino beam is produced by colliding 30\,GeV protons
extracted from the J-PARC accelerator chain with a 91\,cm long graphite
target.  Three magnetic horns focus secondary charged particles that
are produced in the proton-target collisions.  The polarity of the
horns' currents determine which charge is focused and defocused,
allowing for the creation of neutrino or antineutrino enhanced beams.
The secondary particles are allowed to decay in a 96\,m long decay
volume.  The dominant source of neutrinos is the decay of $\pi^{\pm}$.
Most $\mu^{\pm}$ are stopped in the absorber located at the end of the
decay volume before they decay, and $\nu_{e}(\bar{\nu_{e}})$ from
$\mu^{\pm}$ decays contribute less than 1\% to the total neutrino flux
at at the peak energy.

\begin{figure}
\centering
\includegraphics[width=0.45\textwidth]{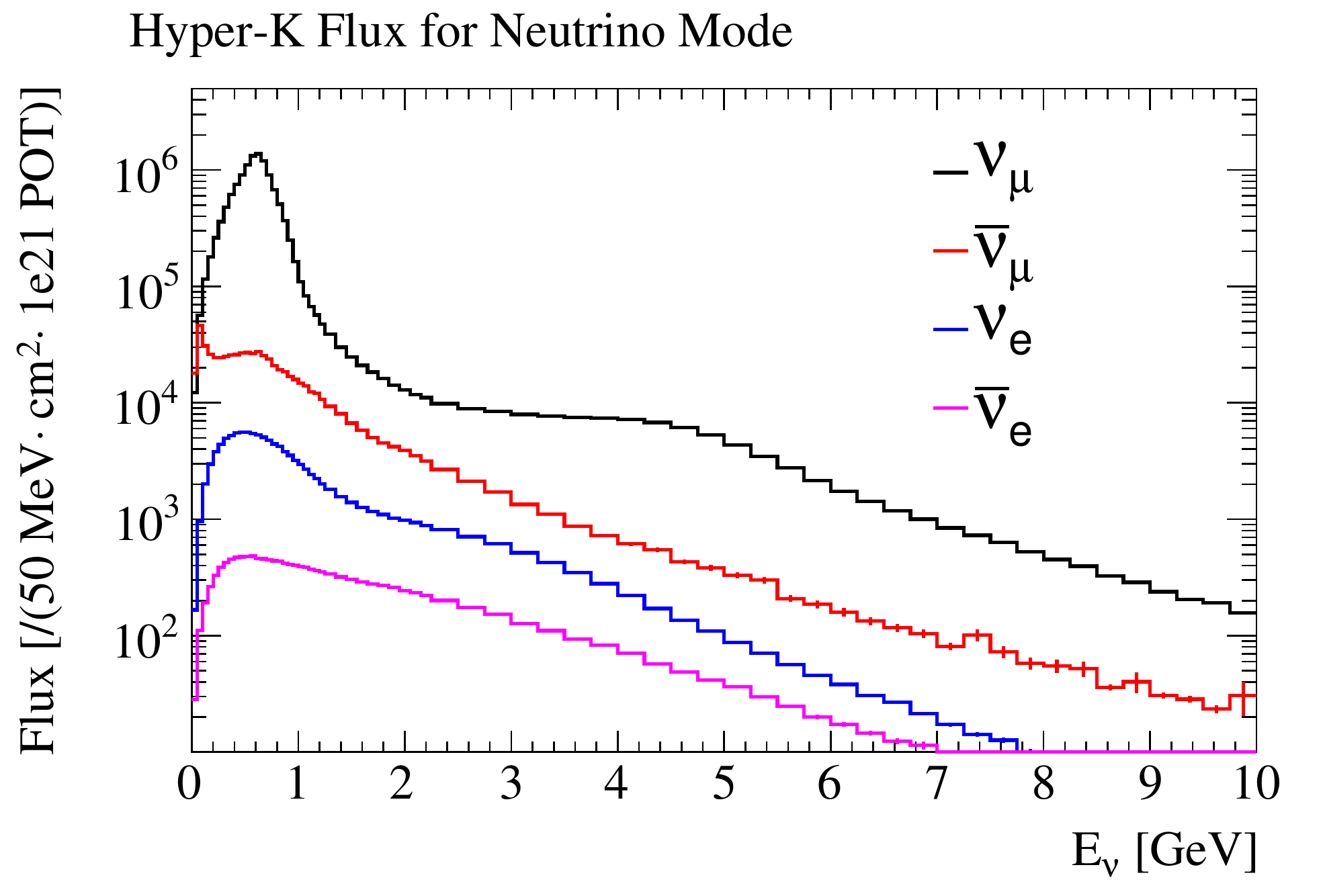}
\includegraphics[width=0.45\textwidth]{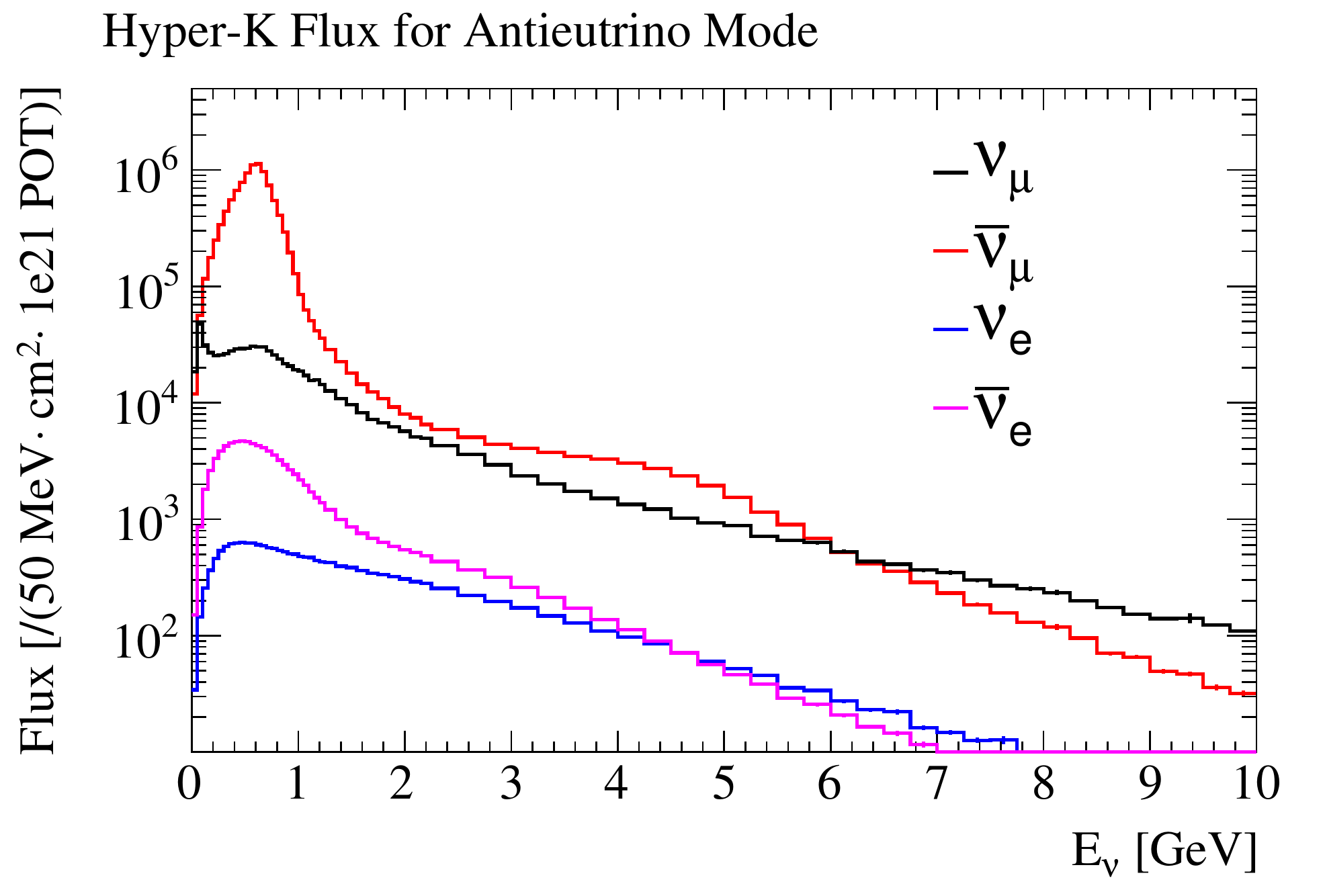}
\caption{The neutrino spectra at Hyper-K for the neutrino enhanced
  (left) and antineutrino enhanced (right) horn current polarities
  with the absolute horn current set to 320\,kA.
\label{fig:hk_fluxes}}
\end{figure}

The J-PARC beam is aimed 2.5$^{\circ}$ away from the Super-K and
Hyper-K detectors to take advantage of the pion decay kinematics to
produce a narrow band beam~\cite{Beavis-BNL-52459} with a spectrum
peaked at 600\,MeV, at the first oscillation maximum for a baseline of
295\,km. Fig.~\ref{fig:hk_fluxes} shows the calculated energy dependent
neutrino fluxes in the absence of neutrino oscillations impinging on
Hyper-K for 320\,kA horn currents in both horn polarities.  Neglecting
oscillations, neutrino detectors located near the neutrino source
observe a similar neutrino spectrum to the far detector spectrum, but
the peak of the spectrum is broader since the beam appears as a line
source for near detectors, compared to a point source for far
detectors.

\begin{figure}
\centering
\includegraphics[width=0.45\textwidth]{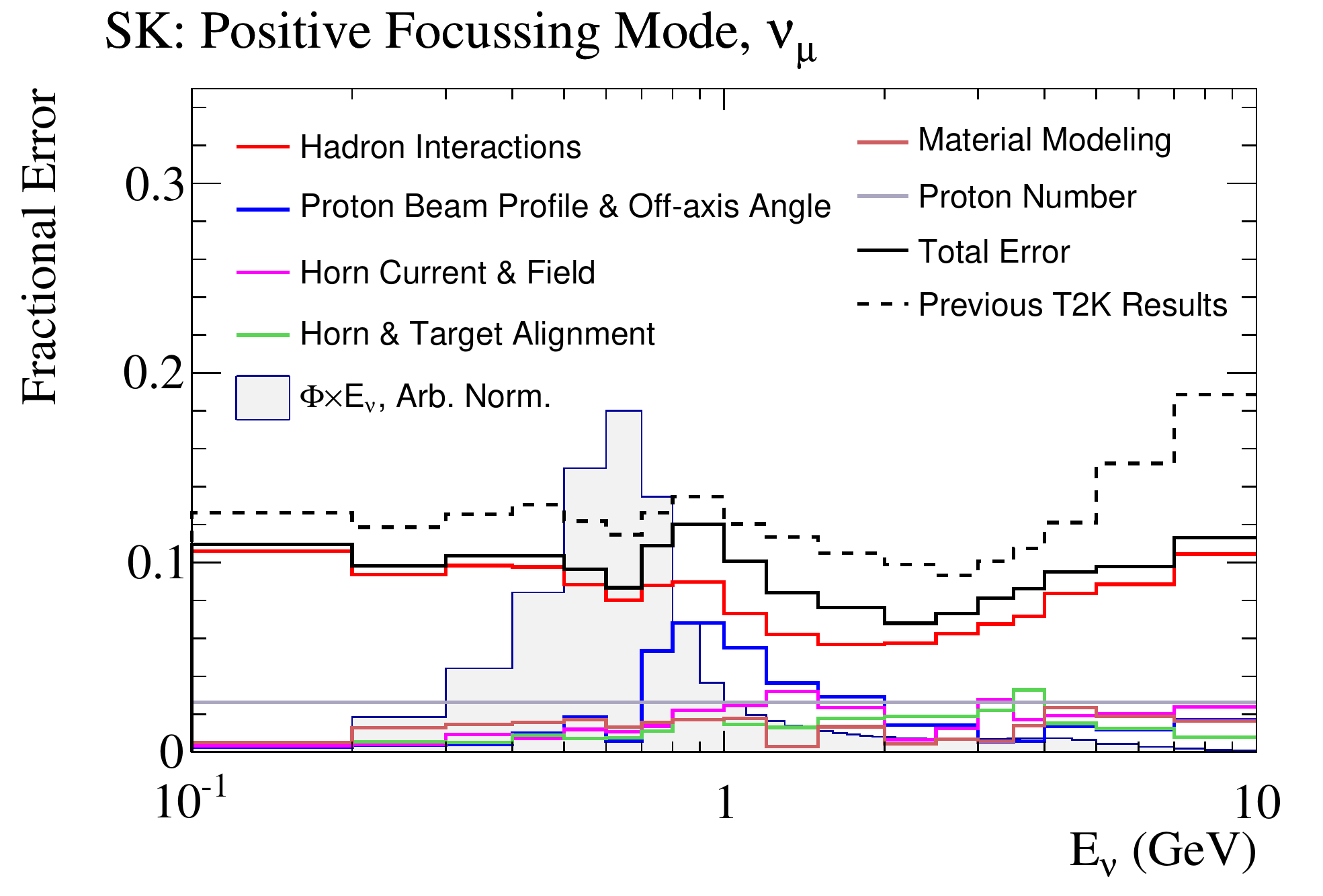}
\includegraphics[width=0.45\textwidth]{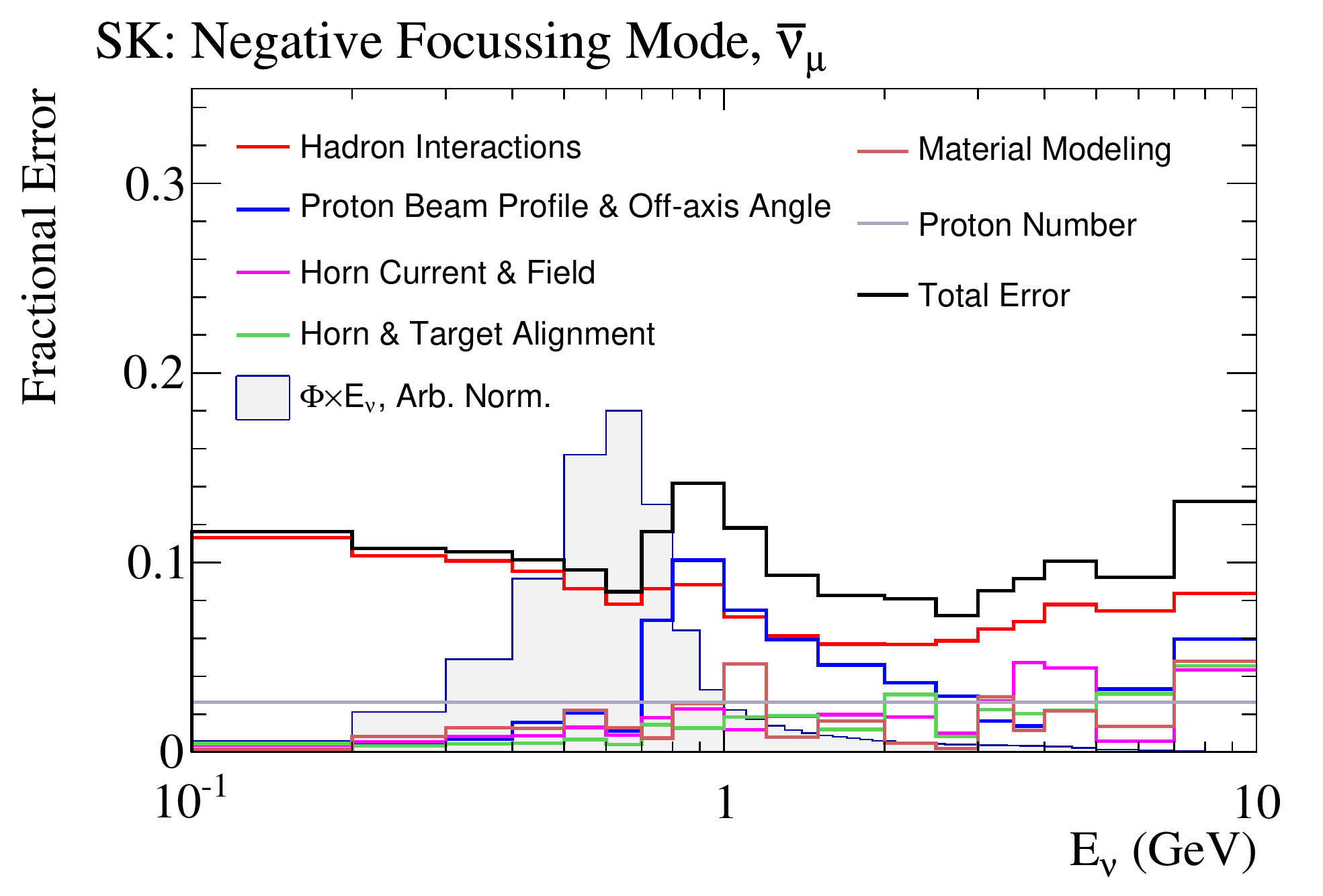}
\caption{The uncertainties on the T2K flux calculation at Super-K for neutrino enhanced (left) and 
antineutrino enhanced (right) beams.
\label{fig:t2k_flux_errors}}
\end{figure}

The neutrino flux is calculated using a data-driven simulation that
employs primary proton beam measurements, hadron production
measurements, beamline element alignment measurements and horn current
and field measurements~\cite{Abe:2012av}.  The dominant uncertainty on
the flux calculation arises from modeling of hadron production in the
graphite target and surrounding material.  To minimize the hadron
production uncertainties, the NA61/SHINE
experiment~\cite{Abgrall:2014xwa} has measured particle production
with 30\,GeV protons incident on a thin (4\% of an interaction length)
target~\cite{Abgrall:2011ae,Abgrall:2011ts}, and a replica T2K
target~\cite{Abgrall:2012pp}. The thin target data have been used in
the T2K flux calculation, and a 10\% uncertainty on the flux
calculation has been achieved, as shown in
Fig.~\ref{fig:t2k_flux_errors}.  Much of the remaining uncertainty
arises from the modeling of secondary particle re-interactions inside
the target.  Preliminary work suggests that the hadron production
uncertainty can be reduced to $\sim5\%$ by using the NA61/SHINE
measurement of the particle multiplicities exiting the T2K replica
target~\cite{Hasler:2039148}.  In the context of Hyper-K, the thin
target data from NA61/SHINE are applicable to the flux calculation,
and the replica target data may also be used if the target geometry
does not change significantly.  If the target geometry or material 
are changed for Hyper-K, then new hadron production measurements will
be necessary.  

The near neutrino detectors of T2K are located 280\,m from the pion
production target and they include the INGRID on-axis
detector~\cite{Otani:2010zza} and the ND280 off-axis
detector~\cite{Assylbekov:2011sh,Allan:2013ofa,Aoki:2012mf,Amaudruz:2012agx,Abgrall:2010hi}.
The INGRID detector is used primarily to measure the beam direction and neutrino yield,
while ND280 measurements provide constraints on the neutrino flux and
interaction models that are used to predicted the event rate at the
far detector after oscillations. Measurements with the ND280 detector
are used for both dedicated neutrino cross-section measurements and
event rate constraints that are used directly in neutrino oscillation
measurements.  For neutrino cross-section measurements, the neutrino
flux is derived from the previously described flux calculation and the
neutrino cross-section is inferred from the event rate and particle
kinematics measured with the ND280 detector.  The cross-section
measurements provide constraints on the building of models of
neutrino-nucleus interactions that are ultimately used in oscillation
measurements.  For the oscillation measurements themselves, nuisance
parameters are introduced to describe the uncertainty on the neutrino
flux and interaction models.  A fit to a subset of the ND280 data
simultaneously constrains the flux and interaction model nuisance
parameters, and the predicted event rate and uncertainty at the far
detector are updated~\cite{Abe:2015awa}.  The beam direction
measurement, neutrino cross section measurements and direct
constraints on the neutrino event rate for oscillation measurements
are all necessary for long baseline oscillations measurements at
Hyper-K.

\subsection{The J-PARC accelerator chain }
The J-PARC accelerator cascade~\cite{JPARCTDR} 
consists of a normal-conducting LINAC as an injection 
system, a Rapid Cycling Synchrotron (RCS), and a Main Ring synchrotron (MR). 
H$^-$ ion beams, with a peak current of 50 mA and pulse width of 500 $\mu$s, 
are accelerated to 400 MeV by the LINAC. 
Conversion into a proton beam is achieved by charge-stripping foils at
injection into the RCS ring, which accumulates and accelerates two
proton beam bunches up to 3 GeV at a repetition rate of 25 Hz. Most of
the bunches are extracted to the Materials and Life science Facility
(MLF) to generate intense neutron/muon beams.  The beam power of RCS
extraction is rated at 1\,MW.  With a prescribed repetition cycle, four
successive beam pulses are injected from the RCS into the MR at 40 ms
(= 25 Hz) intervals to form eight bunches in a cycle, and
accelerated up to 30 GeV.  In fast extraction (FX) mode operation, the
circulating proton beam bunches are extracted within a single turn
into the neutrino primary beamline by a kicker/septum magnet system.

\begin{table}[t]
 \caption{Main Ring rated parameters for fast extraction, with numbers
   achieved as of December 2017. The columns show (left to right): the
   currently achieved operation parameters, the original design
   parameters, the projected parameters after the MR RF and magnet
   power supply upgrade, and the projected parameters for the maximum
   beam power achievable after the upgrade. }
 \label{jparc:MRFXpara}
 \begin{center}
   \begin{tabular}{lcccc}
     \hline \hline
     Parameter   & Achieved & Original  & Doubled rep-rate & Long-term Projection \\
     \hline
     Circumference          & \multicolumn{4}{c}{1,567.5\,m } \\
     Beam kinetic energy    & 30\,GeV  & 50\,GeV & 30\,GeV & 30\,GeV \\
     Beam intensity         & $2.45\times 10^{14}$\,ppp
     & $3.3\times 10^{14}$\,ppp & $2.0\times 10^{14}$\,ppp & $3.2\times 10^{14}$\,ppp  \\
     ~                      & $3.1\times 10^{13}$\,ppb
     & $4.1\times 10^{13}$\,ppb & $2.5\times 10^{13}$\,ppb & $4.0\times 10^{13}$\,ppb \\
     $[$ RCS equivalent power $]$  & $[$ 575\,kW $]$
     &   $[$ 1\,MW $]$      & $[$ 610\,kW $]$ &  $[$ 1\,MW $]$  \\
     Harmonic number        & \multicolumn{4}{c}{9} \\
     Bunch number           & \multicolumn{4}{c}{8~/~spill} \\
     Spill width            & \multicolumn{4}{c}{$\sim$~5\,$\mu$s} \\
     Bunch full width at extraction  & $\sim$50\,ns & -- & $\sim$50\,ns & $\sim$50\,ns  \\
     Maximum RF voltage     & 280\,kV  & 280\,kV & 560\,kV & 560\,kV  \\
     Repetition period      & 2.48\,sec & 3.52\,sec & 1.32\,sec & 1.16\,sec \\
      \hline
     Beam power             & 485\,kW\footnote{As of 2018} & 750\,kW & 750\,kW & 1340\,kW  \\
     \hline
     \hline
   \end{tabular}
 \end{center}
\end{table}

\begin {figure}[htbp]
  \begin{center}
    \includegraphics[width=0.8\textwidth]
     {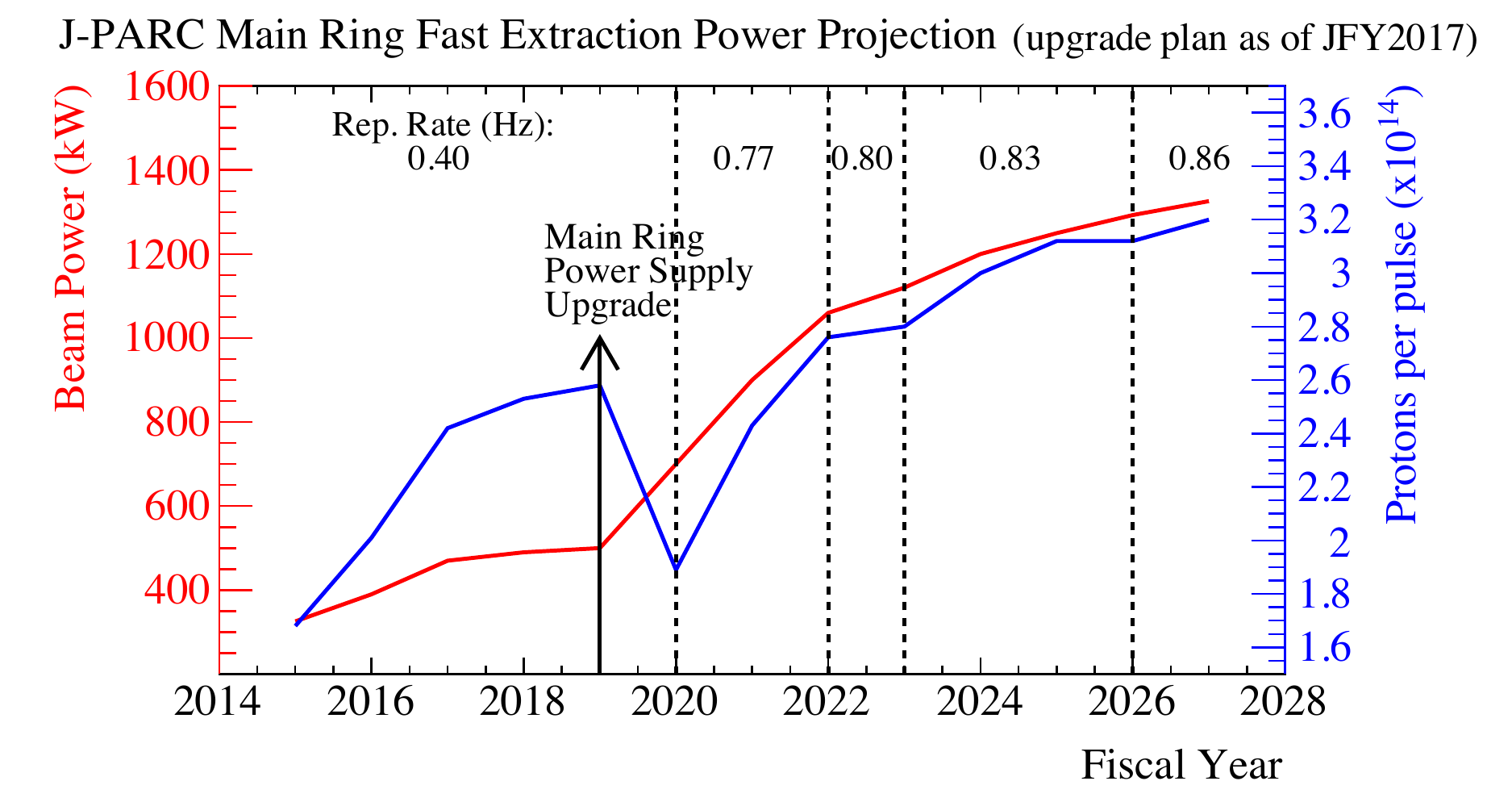}
    \caption{The projected Main Ring fast extraction performance up to 2028, including the beam power, the protons per pulse, and the repitition rate.}
    \label{fig:power_proj}
  \end{center}
\end {figure}

In the MR FX mode operation,  a beam intensity of 2.45$\times$10$^{14}$ proton-per-pulse (ppp) 
has been achieved, corresponding to $\sim$485\,kW beam power (as of 2018).
The accelerator team is following 
a concrete upgrade scenario~\cite{jparc-midterm-new}
to reach the design power of 750\,kW in forthcoming years, 
with a typical planned parameter set as listed in Table~\ref{jparc:MRFXpara}.
This will double the current repetition rate by 
(i) replacing the magnet power supplies, 
(ii) replacing the RF system, and 
(iii) upgrading injection/extraction devices. 
Based on high intensity studies of the current accelerator performance,
it is expected that 1-1.3\,MW beam power can be achieved after these upgrades~\cite{jparc-status-upgrade,jparc-plan-2026}.
The projected beam performance up to 2028 is shown in Fig.~\ref{fig:power_proj}.
For operation larger than 2\,MW beam power, conceptual design studies are now
underway~\cite{jparc-longterm-new}, and they include approaches such as raising the RCS top energy, enlarging the MR
aperture, or inserting an “emittance-damping” ring between the RCS and MR.

\subsection{Neutrino beamline}
\begin {figure}[htbp]
  \begin{center}
    \includegraphics[width=0.8\textwidth]
     {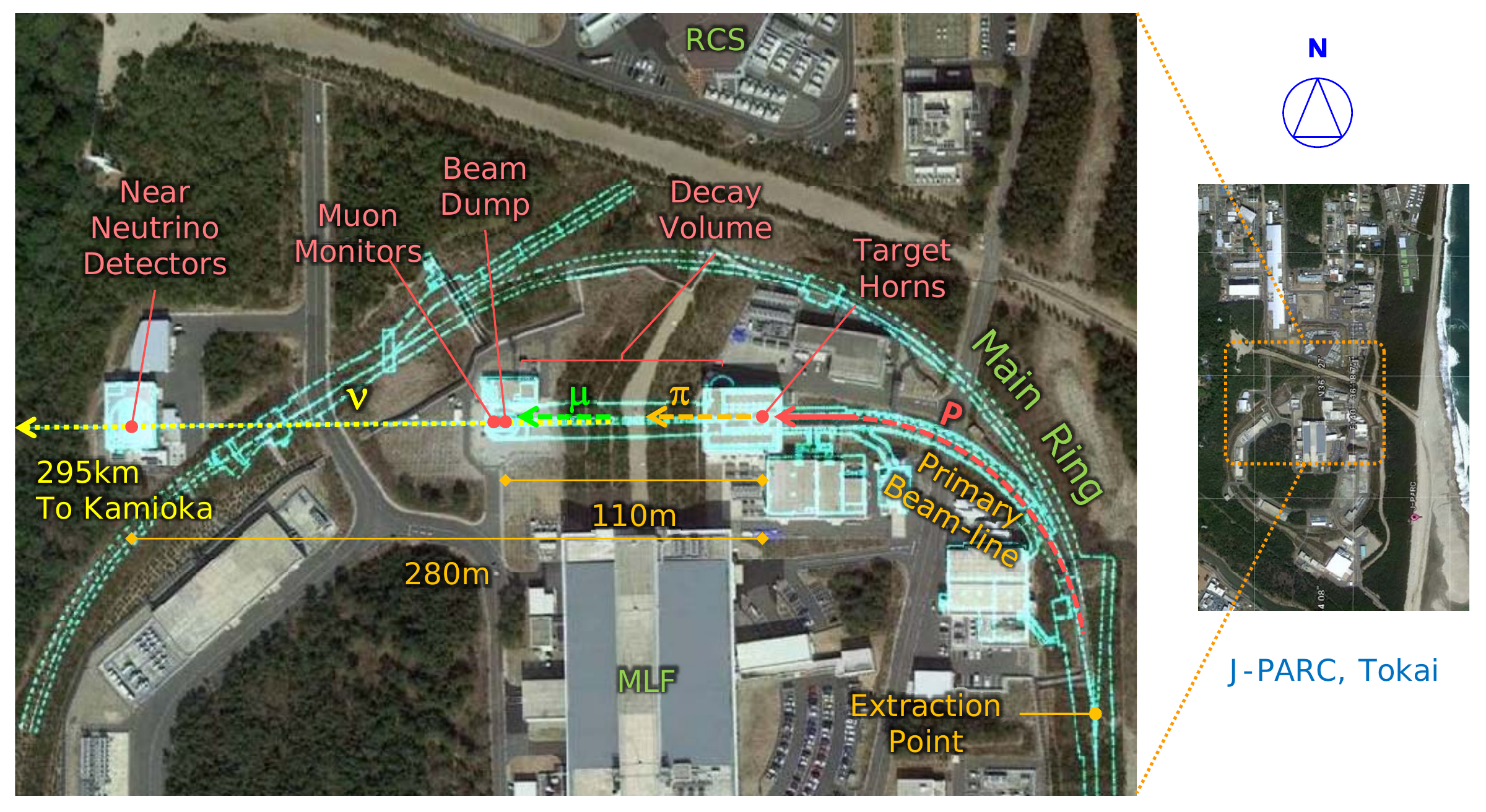}
    \caption{The neutrino experimental facility 
             (neutrino beamline) at J-PARC.}
    \label{fig:beamline}
  \end{center}
\end {figure}
Fig.~\ref{fig:beamline} shows an overview of the neutrino experimental
facility~\cite{Abe:2011ks}. The primary beamline guides the extracted
proton beam to a production target/pion-focusing horn system in a
target station (TS). The pions decay into muons and neutrinos during
their flight in a 96 m-long decay volume. A graphite beam dump is
installed at the end of the decay volume, and muon monitors downstream
of the beam dump monitor the muon profile.  The beam is aimed
2.5$^{\circ}$ off-axis~\cite{OffAxisBeam} from the direction to
Super-K and the beamline has the capability to vary the off-axis angle
between 2.0$^\circ$ to 2.5$^\circ$.
The centreline of the beamline extends 295 km to the west, 
passing midway between Tochibora and Mozumi, so that both 
sites have identical off-axis angles.  
\begin {figure}[htbp]
  \begin{center}
    \includegraphics[width=0.8\textwidth]
     {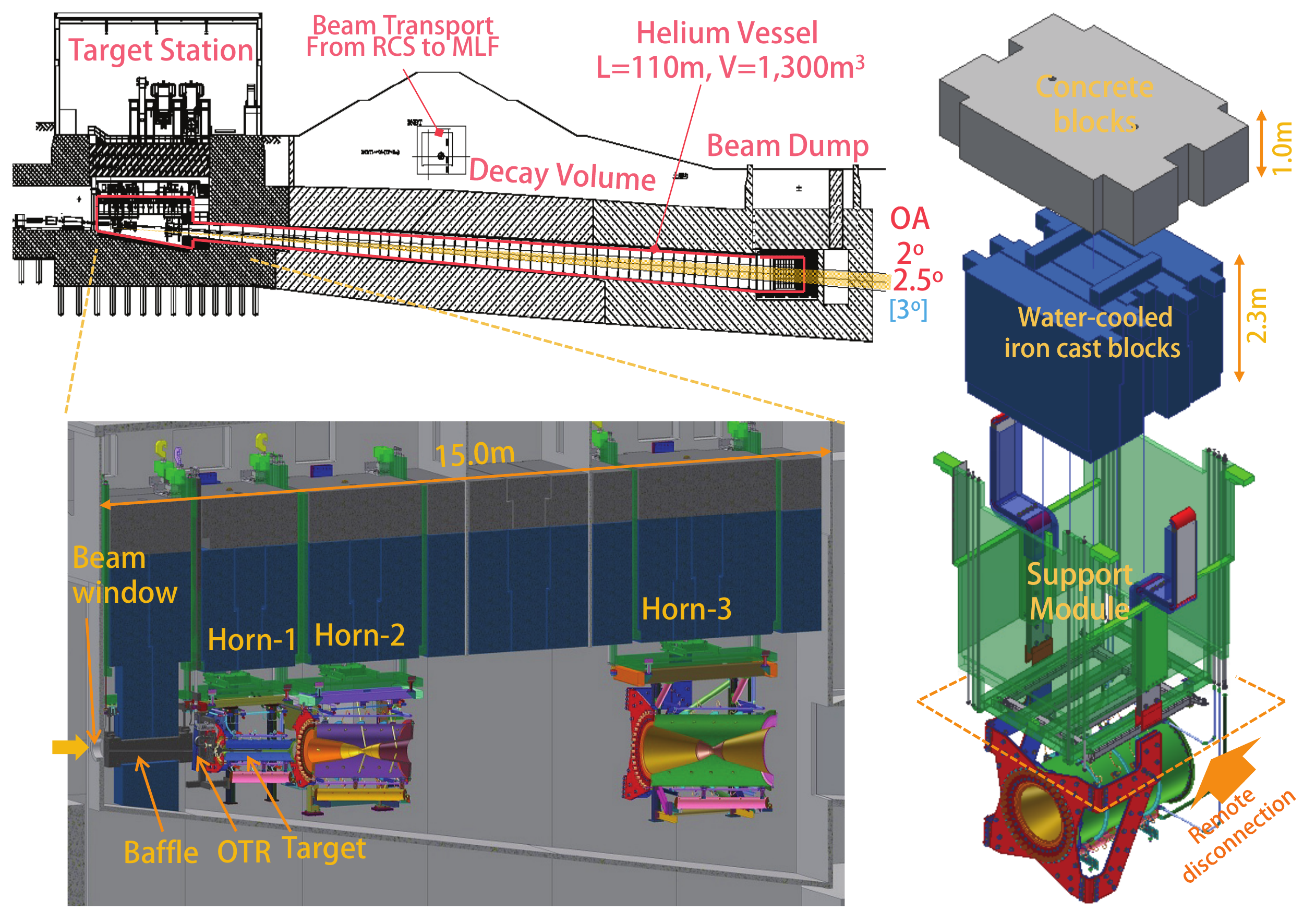}
    \caption{(Left) Side view of the secondary beamline, 
             with a close up of the target station helium vessel.
             (Right) A schematic view of a support module and shield blocks 
             for horn-3. If a horn fails, the horn together with its 
             support module is transferred remotely to a purpose-built 
             maintenance area, disconnected from the support module 
             and replaced. 
             }
    \label{fig:secondaryBL}
  \end{center}
\end {figure}

\subsubsection{Secondary beamline}
The secondary beamline consists of the beamline from the TS entrance
to the muon monitors.  Fig.~\ref{fig:secondaryBL} shows a cross
section of the secondary beamline, and a close-up of the TS helium
vessel.  The secondary beamline components and their capability to
accept high power beam are reviewed here.

A helium cooled, double skin titanium alloy beam window separates the
helium environment in the TS vessel ($\sim$1 atm pressure) from the
vacuum of the primary beamline.  The proton beam collides with a
helium-cooled graphite production target that is inserted within the
bore of the first of a three-horn pion-focusing system.  At 750\,kW
operation, a $\sim$20\,kW heat load is generated in the target.  The
neutrino production target and the beam window are designed for 750\,kW
operation with 3.3$\times$10$^{14}$ ppp (equivalent to RCS 1\,MW
operation) and 2.1\,sec cycle. In the target, the pulsed beam generates
an instantaneous temperature rise per pulse of 200 C$^\circ$ and a
thermal stress wave of magnitude 7 MPa.  Given the tensile strength,
the safety factor is $\sim$5.  Although the tensile strength and
safety factor will be reduced by cyclic fatigue, radiation damage and
oxidization of the graphite, a lifetime of 2$-$5 years is expected.

The horns are suspended from the lid of the TS helium vessel.  Each
horn comprises two co-axial cylindrical conductors which carry up to a
320 kA pulsed current. This generates a peak toroidal magnetic field
of 2.1\,Tesla which focuses one sign of pions.  The heat load generated
in the inner conductors by secondary particles and by joule heating is
removed by water spray cooling.  So far the horns were operated with a
250 kA pulsed current and a minimum repetition cycle of 2.48 sec.  To
operate the horns at a doubled repetition rate of $\sim$1 Hz requires
new individual power supplies for each horn utilizing an energy
recovery scheme and low inductance/resistance striplines.  These
upgrades will reduce the charging voltage/risk of failure, and, as
another benefit, increase the pulsed current to 320 kA.  The horn-1
water-spray cooling system has sufficient capacity to keep the
conductor below the required 80$^\circ$C at up to 2\,MW.

All secondary beamline components become highly radioactive during
operation and replacements require handling by a remotely controlled
overhead crane in the target station.  Failed targets can be replaced
within horn-1 using a bespoke target installation and exchange
mechanism.

Both the decay volume and the beam dump dissipate $\sim$1/3 of the
total beam power, respectively.  The steel walls of the decay volume
and the graphite blocks of the hadron absorber (core of the beam dump)
are water cooled and both are designed to accept 3$\sim$4\,MW beam
power since neither can be upgraded nor maintained after irradiation
by the beam.

Considerable experience has been gained on the path to achieving
475\,kW beam power operation, and the beamline group is promoting
upgrades to realize 750\,kW operation and to expand the facilities for the
treatment of activated water.  Table~\ref{jparc:BLupgrade} gives a
summary of acceptable beam power and/or achievable parameters for each
beamline component~\cite{IFW-nu750kW,IFW-numultiMW}, for both the
current configuration and after the proposed upgrades in forthcoming
years.

\begin{table}[t]
 \caption{Acceptable beam power and achievable parameters
   for each beamline component after proposed upgrades.
   Limitations as of December 2017 are also given in
   parentheses.}
 \label{jparc:BLupgrade}
 \begin{center}
   \begin{tabular}{lcc}
     \hline \hline
     Component   & \multicolumn{2}{c}
     {Acceptable beam power or achievable parameter}  \\
     \hline
     Target      & \multicolumn{2}{c}{3.3$\times$10$^{14}$ ppp } \\
     Beam window & \multicolumn{2}{c}{3.3$\times$10$^{14}$ ppp } \\
     Horn        &   ~  & ~ \\
     \multicolumn{1}{c}{cooling for conductors} &
     \multicolumn{2}{c}{2 MW} \\
     \multicolumn{1}{c}{stripline cooling}
     & ( 750 kW $\rightarrow$) & $\sim$3 MW \\
     \multicolumn{1}{c}{hydrogen production}
     & ( 1 MW $\rightarrow$) & $\sim$2 MW \\
     \multicolumn{1}{c}{power supply} & ( 250 kA $\rightarrow$) & 320
kA  \\
     ~              &  ( 0.4 Hz $\rightarrow$)  &  1 Hz   \\
     Decay volume   &  \multicolumn{2}{c}{4 MW} \\
     Hadron absorber (beam dump)  &  \multicolumn{2}{c}{3 MW} \\
     \multicolumn{1}{c}{water-cooling facilities}
      & ( 750 kW $\rightarrow$) & $\sim$2 MW \\
     Radiation shielding  & ( 750 kW $\rightarrow$) & 4 MW \\
     Radioactive cooling water treatment
     &  ( 600 kW $\rightarrow$) & $\sim$1.3 MW \\
     \hline \hline
   \end{tabular}
 \end{center}
\end{table}
%
%=============================%
%
\subsection{Near detector complex\label{sec:NDcomplex}}
The accelerator neutrino event rate observed at Hyper-K depends on the
oscillation probability, neutrino flux, neutrino interaction
cross-section, detection efficiency, and the detector fiducial mass of
Hyper-K.  To extract estimates of the oscillation parameters from
data, one must model the neutrino flux, cross-section and detection
efficiency with sufficient precision.  In the case of the neutrino
cross-section, the model must describe the exclusive differential
cross-section that includes the dependence on the incident neutrino
energy, $E_{\nu}$, the kinematics of the outgoing lepton, momentum
$p_{l}$ and scattering angle $\theta_{l}$, and the kinematics of final
state hadrons and photons.  In our case, the neutrino energy is
inferred from the lepton kinematics, while the reconstruction
efficiencies depend on the hadronic final state as well.

The near detectors measure the neutrino interaction rates close enough to the neutrino
production point so that oscillation effects are negligible.  The prediction of 
event rates at Hyper-K for a given set of oscillation parmeters will be precisely
constrained by event rates measured in the near detector and the flux simulation
based on hadron production data from NA61/SHINE and other hadron production experiments.
Our approach to using near detector data will build on the experience of
T2K while considering new near detectors that address
limitations in reducing neutrino cross section modelling uncertainties
with the current T2K near detector suite.

The near detectors should be capable of measuring the signal and
background processes relevant for neutrino oscillation measurements
made using the accelerator produced neutrinos.  The processes include:
\begin{itemize}
\item The charged current interactions with no detected final state
  pion (CC0$\pi$) that are the signal channel for the oscillation
  measurements in Hyper-K.
\item The intrinsic electron neutrino component of the beam from muon
  and kaon decays, which is a background for the electron
  (anti)neutrino appearance signal.
\item The neutral current interactions with $\pi^{0}$ production
  (NC$\pi^{0}$) that are a background for the electron (anti)neutrino
  appearance signal.
\item The wrong-sign charged current processes (neutrinos in the
  antineutrino beam and vise versa) which are a background in the CP
  violation measurement.
\end{itemize}
In addition to measuring these processes, the near detectors should be
designed to maximize the cancellation of systematic uncertainties when
extrapolating from measured event rates in the near detector to
predict the event rate at Hyper-K.  Hence, the near detector should be
able to make measurements with the same angular acceptance (4$\pi$)
and target nuclei (H$_2$O) as Hyper-K.  Another source of uncertainty
in the extrapolation is the difference between the near and far
detector neutrino spectra due to oscillations, which can amplify
systematic errors related to the modeling of the relationship between
the final state lepton kinematics and the incident neutrino
energy~\cite{Martini:2012fa,Lalakulich:2012hs,Martini:2012uc}.  The
near detectors should be able to sufficiently constrain the modeling
of the dependence of lepton kinematics on neutrino energy over the
relevant neutrino energy range.

The near detectors can also be used to constrain important neutrino
interaction modes for atmospheric neutrino and nucleon decay
measurements at Hyper-K.  For example, Hyper-K may use neutron
captures on Gd or H to statistically separate neutrinos and
antineutrinos in the atmospheric measurements, or to reject
atmospheric neutrino backgrounds in the nucleon decay measurements.
The neutron multiplicities produced in the interactions of neutrinos
and antineutrinos with energy of $\mathcal{O}($1 GeV$)$ can be
measured in the near detectors.  The dominant sources of uncertainty
in the determination of the mass hierarchy and $\theta_{23}$ quadrant
with atmospheric neutrinos are uncertainties in the neutrino to
anti-neutrino cross section ratio for both CCQE and single pion
production modes, the axial vector nucleon form factor, the
neutrino-tau cross section, and the DIS cross section model.
Near detector measurements that would constrain these uncertainties
for interactions on water, have the potential to significantly improve
the sensitivity of these atmospheric neutrino measurements.
The near detectors can also be used to measure the interaction modes for nucleon decay
backgrounds, including the CC$\pi^{0}$ background to the $e^{+}(\mu^{+})\pi^{0}$ mode and the kaon production background to the $K^{+}\nu$ mode.

To summarize, the near and intermediate detectors for Hyper-K should cover the full momentum and angular acceptance of the far detector, include homogenous H$_2$O targets to make precision measurements on H$_2$O, have sign selection capability to measure wrong sign backgrounds, be capable of directly measuring intrinsic $\nu_e,\bar{\nu_e}$ and NC$\pi^0$ backgrounds, be capable of reconstructing exclusive final states with low particle thresholds, be capable of measuring the $\nu_e,\bar{\nu_e}$ cross sections, be capable of measuring the final state neutron multiplicities and be capable of measurements at multiple off-axis angles with varying peak neutrino energies.  We have not identified a single detector technology that can achieve all of these capabilities.  In the minimum configuration, it is necessary to have both a magnetized tracking detector and kiloton scale water Cherenkov detector.  The magnetized tracking detector may provide full momentum and angular acceptance, sign selection, reconstruction of exclusive final states with low particle thresholds, measurement of the $\nu_e,\bar{\nu_e}$ rates and off-axis spanning measurements.  The water Cherenkov detector may provide full angular acceptance, a homogenous water target, direct measurement of the intrinsic $\nu_e,\bar{\nu_e}$  and NC$\pi^0$ backgrounds, measurement of the $\nu_e,\bar{\nu_e}$ cross sections, measurements of final state neutron multiplicities and the off-axis spanning measurements.  Hence, in the following text we present the upgrade for the current ND280 magnetized tracking detector, new tracking detector technologies, and intermediate water Cherenkov detectors as potential components of the Hyper-K near and intermediate detector suite.

    \subsubsection{The ND280 Detector Suite}
\begin {figure}[htbp]
  \begin{center}
    \includegraphics[width=0.85\textwidth]{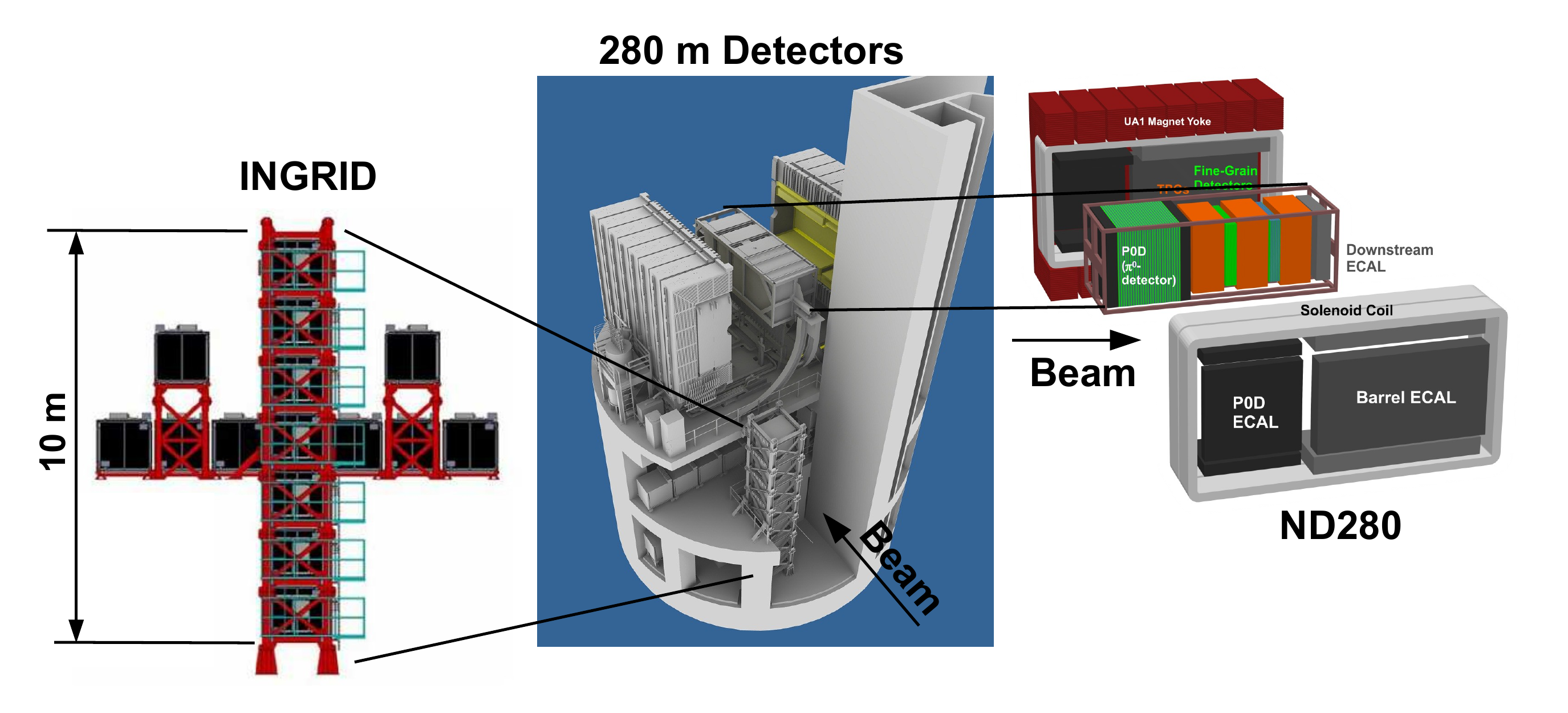}
    \caption{The ND280 detector complex (center) including the INGRID (left) and ND280 (right) detectors.}
    \label{fig:nd280}
  \end{center}
\end {figure}

    The T2K ND280 detector suite, shown in Fig.~\ref{fig:nd280}, comprises two detectors: INGRID~\cite{Otani:2010zza}, which consists of 16 iron-scintillator modules in a cross pattern centered on the neutrino beam axis, and ND280, a multi-component detector at an angle of 2.5$^{\circ}$ from the beam direction.  
The primary purpose of the INGRID detector is to constrain the neutrino beam direction, whilst the off-axis detector is used to characterise the various neutrino interaction rates before oscillations.

The ND280 off-axis detector is composed of an inner tracking region surrounded by an upstream Pi-zero detector~\cite{Assylbekov:2011sh}, electromagnetic calorimeters~\cite{Allan:2013ofa} and side muon range detectors~\cite{Aoki:2012mf}, all of which are held inside the UA1 magnet.
The 0.2~T magnetic field allows for momentum measurement and sign selection of charged particles, and is important for operation in antineutrino mode where the neutrino background is large.
The tracking region is composed of three time projection chambers (TPCs)~\cite{Abgrall:2010hi} separated by two fine grained detectors (FGDs)~\cite{Amaudruz:2012agx}, the second of which contains passive water layers to allow for neutrino interaction rate measurements on oxygen.
The FGDs are the neutrino target whilst the TPCs provide precise momentum measurements, particle identification and sign selection.

T2K has successfully applied a method of fitting to ND280 data with parameterized models of the neutrino flux and interaction cross-sections, as described in~\cite{Abe:2015awa}.
Using the ND280 measurements, the systematic uncertainties (estimated assuming specific models) on neutrino interaction processes constrained by ND280 have been reduced to $\sim$3\% on the Super-K (SK) predicted event rates, as shown in Table~\ref{tab:current_uncertainties}.

\begin{table}[htbp]
    \caption{Current systematic uncertainty contributions to the T2K oscillation measurements taken from 
    \cite{PhysRevLett.118.151801} and \cite{abe-arxiv1704}
    } 
    \centering
    \begin{tabular}{lC{2.5cm}C{2.5cm}C{2.5cm}C{2.5cm}}
        \hline \hline
            Source of uncertainty & $\nu_{\mu}~CC$ & $\nu_{\textrm{e}}~CC$  & $\bar{\nu_{\mu}}~CC$ & $\bar{\nu_{\textrm{e}}}~CC$\\
        \hline
            Flux and common cross sections & ~ & ~ & ~ & ~\\
            (w/o ND280 constraint) & 10.8\% & 10.9\%& 11.9\%& 12.4\% \\
            (w/ ND280 constraint) & 2.8\% & 2.9\% & 3.3\% & 3.2\%\\
        \hline
            Unconstrained cross sections & 0.8\% & 3.0\% & 0.8\% & 3.3\%\\
        \hline
            SK & 3.9\% &  2.4\% & 3.3\% &3.1\%\\
            FSI + SI(+ PN) & 1.5\% & 2.5\% & 2.1\% & 2.5\%\\
        \hline
            Total & ~ & ~ & ~ & ~\\
        \hline
            (w/o ND280 constraint) & 11.9\% & 12.2\% & 13.0\% & 13.4\%\\
            (w/ ND280 constraint) & 5.1\% & 5.4\% & 5.2\% & 6.2\%\\
        \hline \hline
    \end{tabular}
    \label{tab:current_uncertainties}
\end{table}%

\noindent
{\it Current limitations and future analysis improvements}

As shown in Table~\ref{tab:current_uncertainties}, the uncertainty in the estimated neutrino flux numbers at Super-K are 5--6\%. Without the near detector constraints, the uncertainties would be more than double this, but with the flux and cross-section uncertainties contributing roughly half the total uncertainty, tighter near detector constraints are still necessary to achieve the 1--2\% systematic target for Hyper-K physics measurements.  

\begin{figure}[tbh]
    \begin{center}
      \includegraphics[width=0.5\linewidth]{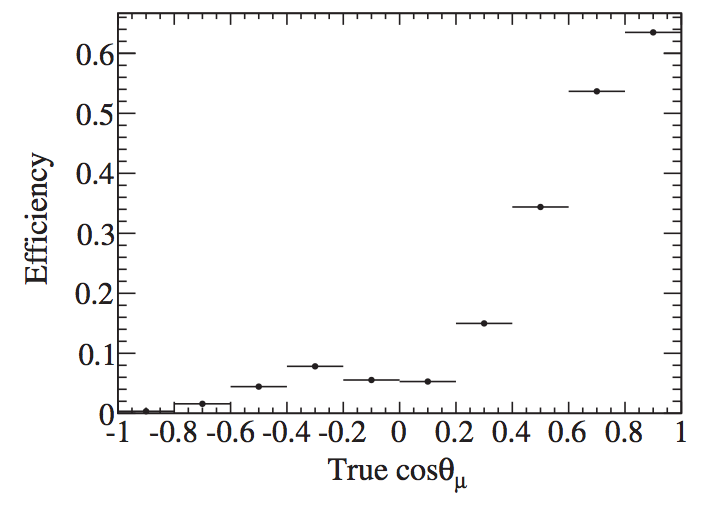} \\
    \end{center}
    \caption{The angular efficiency of CC events with no pions from the ND280 measurement of the muon neutrino charged-current interactions on C$_8$H$_8$ without pions in the final state~\cite{Abe:2016tmq}.
    }
    \label{fig:nd280_ang_eff}
\end{figure}

Cross section model parameters unconstrained by the near detector  (``Unconstrained cross sections'' in Table~\ref{tab:current_uncertainties}) are more significant for the $\nu_e$ and $\bar{\nu_e}$ rates due to the uncertainty on the $\nu_e$ and $\bar{\nu_e}$ cross sections relative to $\nu_{\mu}$ and $\bar{\nu_{\mu}}$.  Reduction of these uncertainties with direct measurements is a challenge due to the relative small fraction of intrinsic $\nu_e$ and $\bar{\nu_e}$ in the beam before oscillations.
Separate selections with interactions in FGD1 (all scintillator) and FGD2 (40\% water, 60\% scintillator by mass) help to constrain cross-sections on water.   Compared to a measurement on a pure water target, additional systematic and statistical errors are introduced on the FGD2 water measurements due to the uncertainties on the subtraction of interactions on the scintillator in FGD2.
Another limitation of the current ND280 analysis is the difference in acceptance between ND280 and SK.
These differences in accessible phase space mean that ND280 and SK are sensitive to different parts of the neutrino interaction model, which introduces additional uncertainty when performing oscillation analyses.
The current ND280 analysis of CC events with no pions has an efficiency that strongly depends on the angle of the muon trajectory relative to the Z axis, which is roughly aligned with the neutrino beam direction.  Fig.~\ref{fig:nd280_ang_eff} shows the angular dependence of the efficiency for events in FGD1, which is $>$50\% when the cosine of the angle is $>$0.6~\cite{Abe:2016tmq}.  
Future analysis improvements will provide some acceptance of events at greater angles and with backward-going muons, but will not provide the same coverage as a true $4\pi$ detector, such as SK. Other potential analysis improvements include the use of pion kinematic information in addition to the muon kinematic information of the ND280 selected events which could help reduce potential reconstruction bias in the neutrino energy.

A fundamental limitation of near detectors in oscillation analyses is that the near and far detectors do not sample the same neutrino spectrum 
due to the neutrino oscillations.  
Since the observed final state particle distributions are produced by different spectra in the near and far detectors,
the interaction models are necessary to extrapolate the rate observed with the near detector spectrum to the rate expected with the far detector spectrum.
This extrapolation is sensitive to the modeling of the observed final state particle kinematics for each incident neutrino energy.  For water Cherenkov 
detectors such as Hyper-K, the relation of the final state lepton kinematics and neutrino energy for events with only a visible lepton is critical since
the neutrino spectrum is infered from the lepton kinematics, and the signal sample is chosen by the presence of a single lepton candidate ring.  
  If this relationship
is not properly modeled, systematic biases will arise in the extrapolation.  This effect has been studied for both muon neutrino disappearance and 
electron neutrino appearance measurements and found to be a potentially limiting source of systematic uncertainty for the oscillation measurements~\cite{Martini:2012fa,Martini:2012uc,Lalakulich:2012hs,Coloma:2013tba}.  Measurements of neutrino interactions from multiple neutrino spectra peaked at different energies
may be used to constrain and improve the models to the level necessary for Hyper-K.  This may be done at different experiments or with a single experiment
by using the off-axis dependence of the neutrino beam.  Alternatively, measurements at many off-axis angles
may also be used to directly predict the expected event rate for the oscillated spectrum, as discussed in Section~\ref{sec:nuprism}.  The models
may also be improved with better measurements of exclusive final hadronic states.

In summary, we expect some decrease in the uncertainty (current value $\sim$3--4\%) in the event rate prediction from the flux and cross section model from future analysis improvements at ND280 but upgrades to the ND280 hardware, as discussed in the next section could significantly increase sensitivity to interactions on water and those with leptons at large angles thus providing new samples to further drive down the uncertainties. However, there are fundamental limitations to the ND280 upgrade potential, due to size and space constraints within the magnet, so we also need to consider alternative $4\pi$ water cerenkov detection as discussed in the following sections.

\noindent
{\it ND280 Hardware Upgrade possibilities}

The upgrade of the ND280 detector has become an official T2K project since February 2017.
Its goal is to reduce the total systematic uncertainties on the neutrino event rate prediction at the far detector,
in the presence of oscillations, to better than 4\%
by improving the acceptance of particles produced at high angles or backward-going.
The basic idea is to cover almost all the $4\pi$ angular region thanks to 
a neutrino target detector rotated along the neutrino beam direction sandwiched between two horizontal TPCs.

\begin{figure}[tbh]
    \begin{center}
      \includegraphics[width=0.6\linewidth]{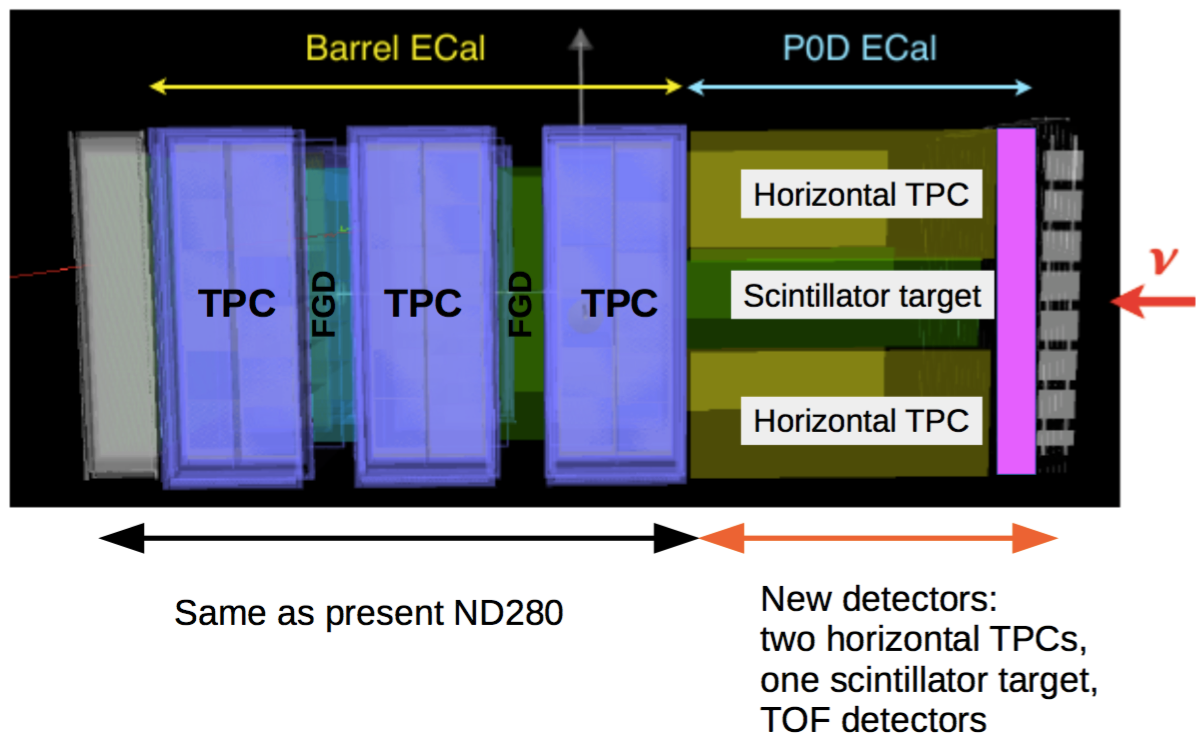} \\
    \end{center}
    \caption{The proposed configuration for the upgrade of the ND280 off-axis detector.  
    The neutrino beam goes from right to left. 
    The upgraded detector consists of the current tracker (three vertical TPCs and two FGDs)
    plus rotated TPCs and a neutrino target detector parallel to the beam.}
    \label{fig:bdl-ND280upgrade-configurations}
\end{figure}

The proposed ND280 upgrade is shown in Fig.~\ref{fig:bdl-ND280upgrade-configurations}.  The upgrade keeps the current
ND280 tracker, i.e. three vertical TPCs and two FGDs, and would replace the P0D detector with a longer horizontal target sandwiched between two 
horizontal TPCs. The upgraded detector also keeps the current Electromagnetic Calorimeter (ECal).
Furthermore, in order to veto the sand muons and $\gamma$'s produced in the magnet or
cosmic particles, both coming from the upstream part of the detector,
the Upstream ECal of the P0D detector (ECal-P0D) would be kept,
providing $\sim 5 X_{0}$.
In addition the Central ECal-P0D could also be used, providing in total a radiator of $\sim 10 X_{0}$.

In addition to the new tracker, a Time-of-Flight (ToF) detector would be built,
with the goal to improve the reconstruction of backward-going tracks, 
poorly detected by the current ND280 detector.
Studies are ongoing in order to understand the impact of the ToF  
on the particle identification, combined with the TPCs.
The ToF detector would be placed all around the new horizontal tracker.
Furthermore two vertical ToFs may be placed 
one between the Upstream ECal-P0D and the tracker  
and the other one between the Downstream ECal and the tracker.

\begin{figure}[tbh]
  \begin{minipage}[t]{0.48\hsize}
    \begin{center}
      \includegraphics[width=0.9\linewidth]{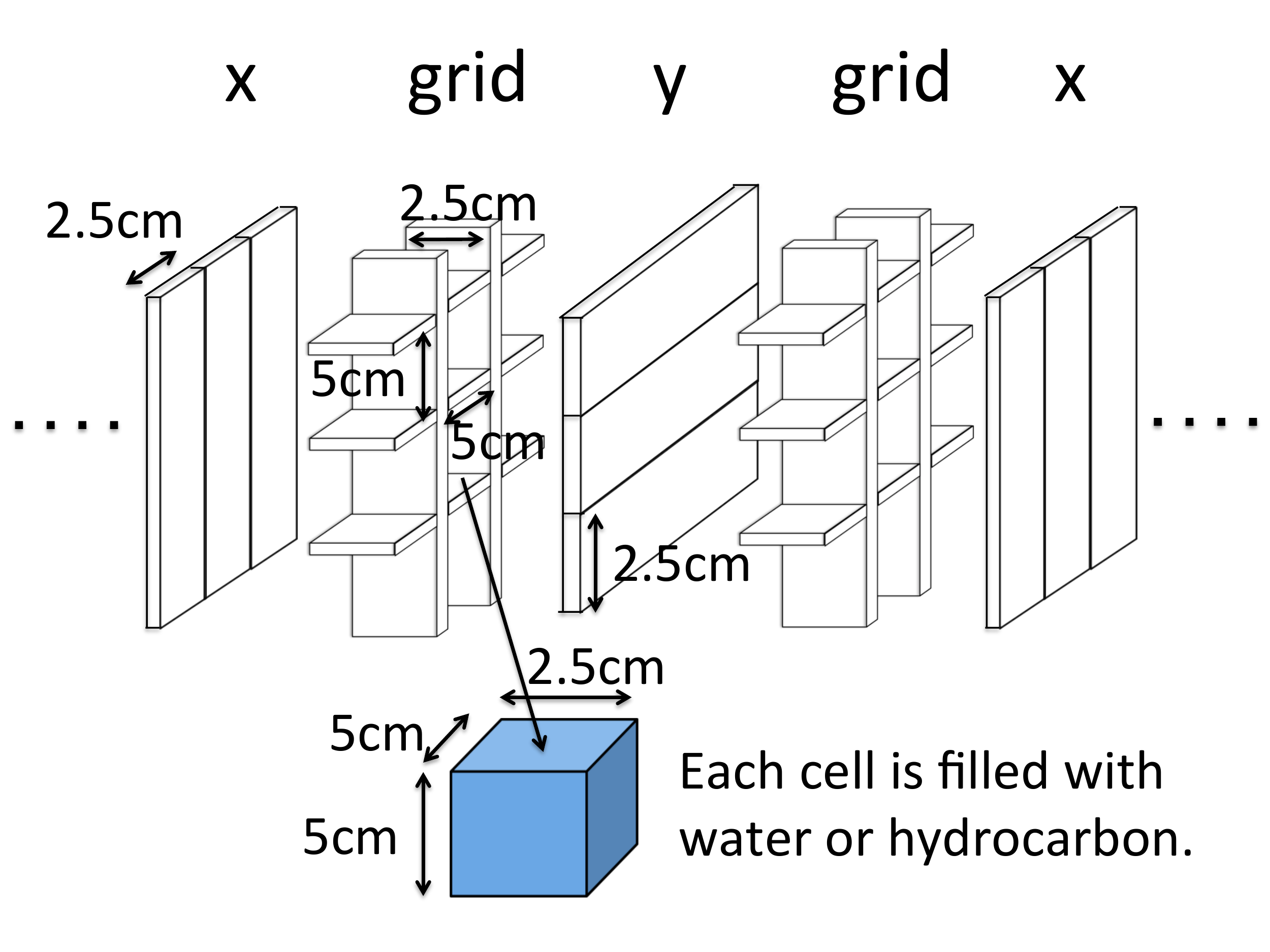}
    \end{center}
    \caption{Schematic view of 3D grid-like structure of plastic scintillator bars inside the WAGASCI detector.}
    \label{fig:3dgrid}
  \end{minipage}\hfill
  \begin{minipage}[t]{0.48\hsize}
    \begin{center}
      \includegraphics[width=0.9\linewidth]{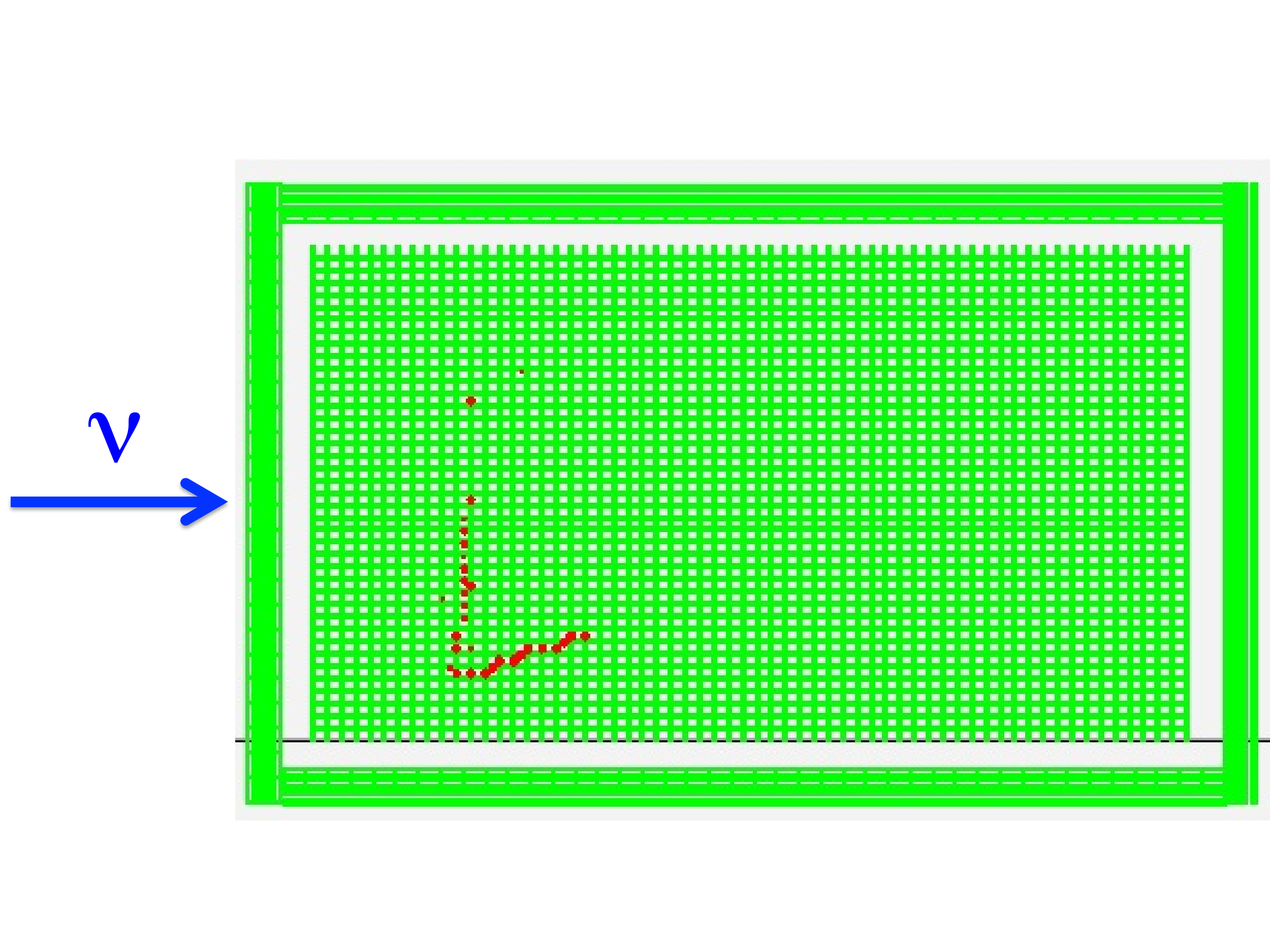}
    \end{center}
    \caption{MC event display of a charged current neutrino event in the WAGASCI detector.}
    \label{fig:wagasci_event_display}
  \end{minipage}
\end{figure}

Currently different neutrino target detectors are considered for the new horizontal tracker: 
a WAGASCI detector, empty or filled with water (described  in the next paragraph and shown in Fig.~\ref{fig:3dgrid}--\ref{fig:wagasci_event_display}),
an FGD-like detector rotated along the neutrino beam direction,
an FGD-3D, same as FGD but with additional plastic scintillator bars along the third axis,
a tracking fiber detector
and a SuperFGD, made of many plastic scintillator cubes, in order to detect tracks with a 4$\pi$ acceptance.

WAGASCI (Water Grid And SCIntillator detector), is a new neutrino detector with a water target under development to measure neutrino interactions with high precision and a large angular acceptance. The concept uses a 3D grid-like structure of scintillator bars to track charged particles across the full $4\pi$ solid angle (Fig. \ref{fig:3dgrid} and \ref{fig:wagasci_event_display}), which provides a larger angular acceptance and larger mass ratio of water to scintillator bars (80:20) than the current off-axis ND280 detector. 

\noindent
Such a detector could be installed within the ND280 magnet as part of the upgrade, but as a first step, WAGASCI modules with water and hydrocarbon targets will be installed on the B2 and SS floors of the near detector hall at J-PARC surrounded by massive muon range detectors to range out the muons~\cite{Asfandiyarov:2014haa}.  These detectors have been approved by the J-PARC PAC as test experiment T59 and will measure neutrino cross sections on water and hydrocarbon~\cite{WAGASCIProposal}. By comparing the observed interaction rate in the two targets, the inclusive water to hydrocarbon charged current cross section ratio can be measured with better than 3\% precision, an established technique that has been shown to work by the current INGRID detector \cite{Abe:2014nox}. 

 In Tab.~\ref{tab:nd280upgradethreshold} the momentum threshold for different particles
corresponding to a WAGASCI-like detector with $2.5 \text{cm}^3$ size is shown.
If WAGASCI is empty the momentum threshold for protons is about 300~MeV/c.

\begin{table}[tbh]
  \small
  \begin{center}
    \caption{\label{tab:nd280upgradethreshold} Momentum threshold of particles detected
    by a WAGASCI-like detector with a $2.5 \text{cm}^3$ cell size.}
    \begin{tabular}{|c|c|c|c|c|c|c|}
      \hline
      \hline
      Detector			& $p_{\mu}$ 	& $p_{\pi^\pm}$ thresholds 	& p$_{p}$ threshold \\ 
  	\hline
      Water-In 	 		&150~MeV/c  	& 150~MeV/c 				& 550~MeV/c\\
       Water-Out 		 	& 50 MeV/c 	& 50 MeV/c 				& 300 MeV/c  \\
      \hline
      \hline
    \end{tabular}
  \end{center}
\end{table}

Tab.~\ref{Table:nd280rateInclusive} shows the neutrino target mass for the current ND280 and proposed upgrade and
 the number of neutrino events 
obtained after applying the full event selection
for an exposure of $1\times 10^{21}$ POT.
A simplified MC study, without a full event reconstruction, has been used.  The ND280 upgrade increases the 
total target mass, giving twice the event rate for the same exposure.

\begin{table}[!tbp]
  \centering
  \caption{The neutrino target mass and predicted total number of selected neutrino events both in neutrino and antineutrino enhanced beam
  for the currrent ND280 detector and the proposed ugrade.
  The predictions corresponds to $1\times 10^{21}$ POT.  
  The out-of-FV and the wrong-sign backgrounds are not included.
  }\label{Table:nd280upgradeevents}
  \begin{tabular}{c|c|c|c|c}
   \hline
   \hline
                          &                   &  \multicolumn{3}{c}{Number of Selected Events} \\ \hline
   Detector Configuration & Target Mass (ton) & CC-$\nu_{\mu}$ ($\nu$ beam) & CC-$\bar{\nu}_{\mu}$ ($\bar{\nu}$ beam) & CC-$\nu_{\mu}$ ($\bar{\nu}$ beam) \\ \hline
   Current ND280          & 2.2               & 95,860                      & 27,443                                  & 14,862                             \\
   ND280 Updgrade         & 4.3               & 199,775                     & 54,249                                  & 28,370                             \\ \hline
  \end{tabular}
  \label{Table:nd280rateInclusive}
 \end{table}

One possibility is to replace the existing tracker region with a high pressure TPC, as was proposed for the LBNO design study~\cite{Stahl:2012exa}. This would allow full angular coverage and a very detailed view of the vertex as well as superb particle identification and sensitivity to low momentum protons. A strength of a gas TPC in the case of T2K and Hyper-K is that it will be possible to change the target gas, which allows testing the nuclear model components of the neutrino-nucleus interaction simulation.  Gas TPCs have been operated successfully with a wide range of gases, including He, CH$_2$, Ne, Ar, CF$_4$, and an N$_2$:CO$_2$ mixture that is 60\% oxygen by mass.  This would provide data samples unavailable to other experiments (or near detector designs), and that complementarity will help break degeneracies that arise from modelling neutrino-nucleus interactions. The ND280 tracker region could easily accommodate a TPC with 8~m$^3$ fiducial volume, which at 10~bar pressure would accumulate sizeable data sets of CC-inclusive $\nu_{\mu}$ interactions as shown in Table~\ref{tab:lbl-ndcross}. A gas TPC should also provide a relatively pure $\nu_e$ sample, because of the significantly reduced gamma backgrounds compared to the existing ND280, allowing $\nu_e$ data sets of hundreds of events per $1\times10^{21}$ POT on the heavier gas targets, as shown in Table~\ref{tab:lbl-ndcross}.

Another option to probe the hadronic final state and enhance our understanding of neutrino-nucleus interactions and to measure the low energy $\nu_e$ interaction cross-section is to add an emulsion detector to the ND280 suite. Emulsions provide 3D tracking with sensitivity to protons down to $\approx$20\,MeV, and $\sim4\pi$ sub micrometer position accuracy through offline scanning~\cite{Fukuda:2014vda}.

 \label{subsection:nd280}

    \subsubsection{Intermediate detector}
    
A water Cherenkov (WC) near detector can be used to measure the cross
section on H$_2$O directly, with the same solid angle acceptance as
the far detector with no need for a subtraction analysis.  This
approach was taken by K2K~\cite{Ahn:2006zza} and was proposed for
T2K~\cite{2km_proposal}.  The MiniBooNE experiment has also employed a
mineral oil Cherenkov detector at a short baseline to great
success~\cite{AguilarArevalo:2008qa}.  Additionally, WC detectors have
shown excellent particle identification capabilities, allowing for the
detection of pure $\nu_{\mu}$-CC, $\nu_{e}$-CC and NC$\pi^{0}$
samples.
The CC$\pi^{0}$ rate and kaon production in neutrino interactions,
which are backgrounds to nucleon decay searches, can also be measured.

These additional WC measurements are essential to achieve the low
systematic errors required by Hyper-K, but are complemented by the
ND280 magnetised tracking detector, which has the capabilities to
track particles below the threshold to produce Cherekov light in water
and to separate neutrino and antineutrino charged current interactions
via the lepton charge measurement. Hence a combination of a magnetized
tracking detector such as ND280 and the WC detector should have the
largest impact to reduce systematic uncertainties.

A WC near detector design should be large enough to contain muons up
to the momentum of interest for measurements at the far detector and
should be far enough from the neutrino production point that there is
minimal pile-up of interactions in the same beam timing bunch. These
requirements lead to designs for kiloton size detectors located at
intermediate distances, 0.7-2\,km from the target, for the J-PARC
neutrino beam.

Here we present the two main features that are being considered for the WC detector:
\begin{itemize}
\item A detector that spans an off-axis angle range of $1^{\circ}-4^{\circ}$ to measure the final state leptonic response over
a range of neutrino spectra peaked at different energies.  Measurements at multiple off-axis angles can be used to address the
limitation of different neutrino spectra at the near and far detectors.
\item Gd loading in the WC detector allows for the measurement of neutron multiplicities in neutrino and antineutrino 
interactions.  These measurements can be used for statistical separation of neutrino and antineutrino interactions as well
as different interaction modes and may be applied to the atmospheric and accelerator neutrino analyses in Hyper-K. The
measurements can also be used to predict the rejection of nucleon decay backgrounds in Hyper-K with neutron tagging.  
\end{itemize}

An international collaboration called E61 was formed in May 2017, combining groups working on the off-axis spanning detector and Gd loading features.  
The E61 collaboration is working closely with the Hyper-K near detector working group to develop a detector design that meets the goals of Hyper-K.
The E61 collaboration has proposed a staged approach to realize the final intermediate water Cherenkov detector (IWCD).  The initial phase would consist of 
a kiloton scale detector on the surface near the current ND280 detectors at an off-axis angle of 6--12$^{\circ}$.  The initial phase may provide
a test-bed for new detector technolgies to be used in the IWCD and Hyper-K, and will be used to prove that the kiloton scale water Cherenkov detector
can be calibrated with sufficient precision to make neutrino interaction measurements with percent level systematic errors.  The E61 collaboration
proposes the second phase of the intermediate water Cherenkov detector as an intermediate detector for Hyper-K.

The IWCD will also have its own physics program independent from
the measurements in service to Hyper-K.  At a 0.7-2\,km baseline, it can
search for neutrino oscillations through sterile neutrinos, on its own
and in combination with ND280.  The WC detector can also be used as an
independent supernova alarm, and with the addition of gadolinium,
would provide valued neutrino-type discrimination in the event of a
supernova in the local galaxy.
 
In the following text, the above main features of the intermediate
detector will be described in more detail, making use of the studies
from other detectors.

\subsubsection{Off-axis angle spanning configuration \label{sec:nuprism}}
\begin{figure}
\centering
\includegraphics[width=0.25\textwidth]{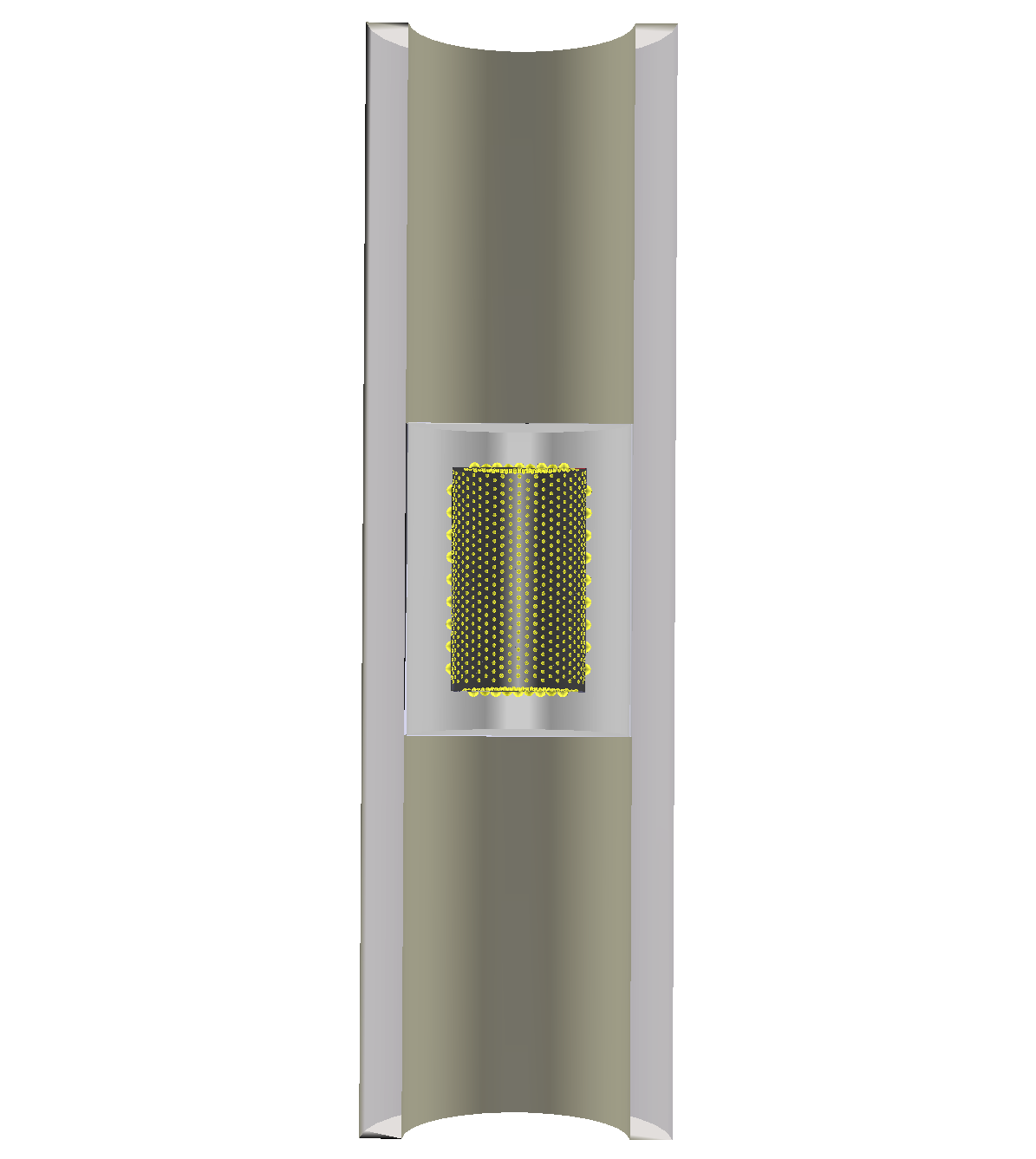}
\includegraphics[width=0.40\textwidth]{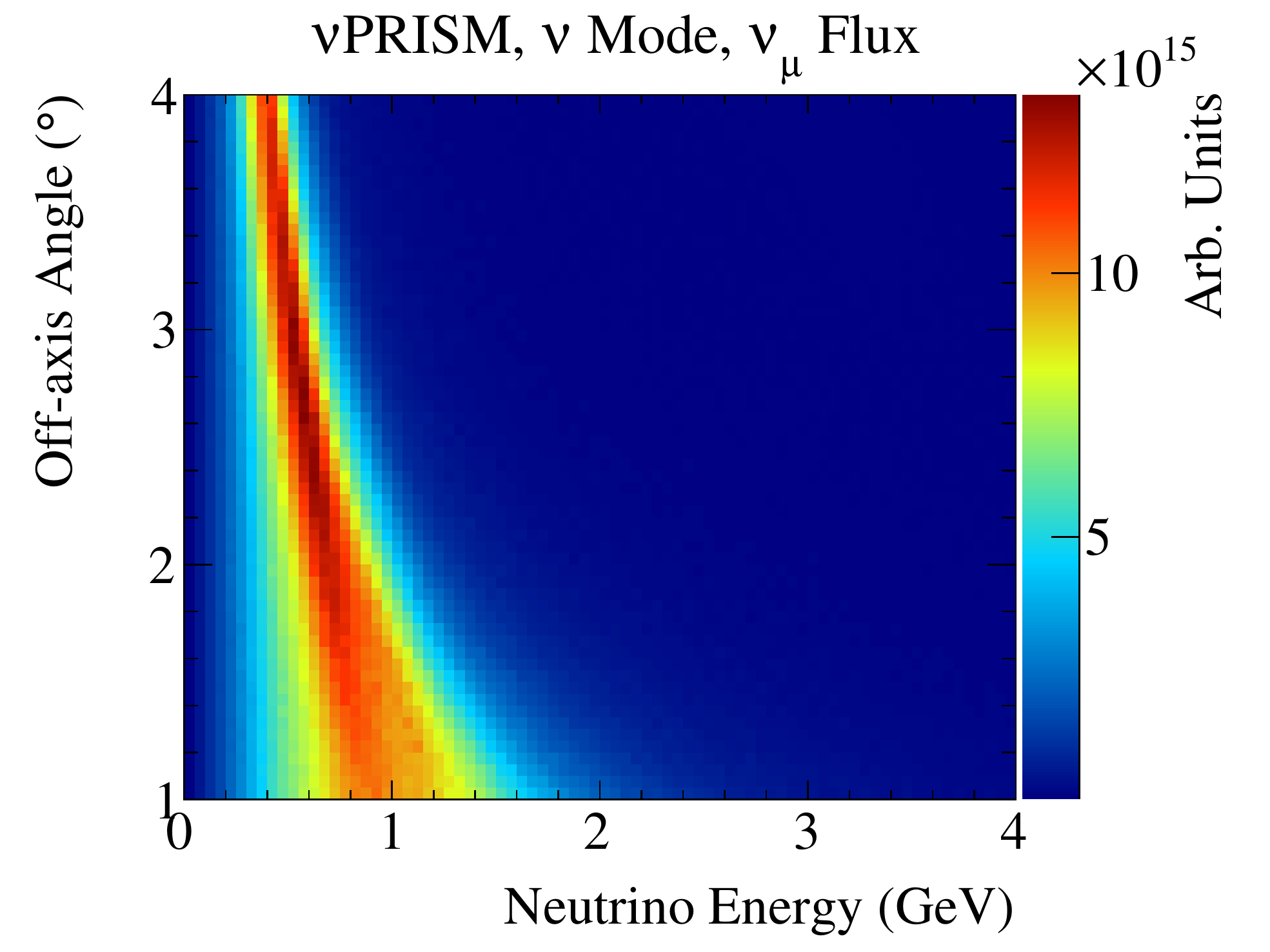}
\caption{Left: A conceptual drawing of the nuPRISM detector.  Right: the $\nu_{\mu}$ flux energy dependence for the
$1^{\circ}-4^{\circ}$ off-axis angle range.
\label{fig:nuprism_offaxis}}
\end{figure}

The intermediate WC detector can be oriented with the polar axis of
the cylinder in the vertical direction and the detector extending from
the ground level downward.  This configuration was originally 
proposed as the NuPRISM detector~\cite{Bhadra:2014oma}, located at a
baseline of 1\,km filling a 10\,m diameter, 50\,m deep pit.  
Fig.~\ref{fig:nuprism_offaxis} shows the conceptual drawing for the
NuPRISM detector and the $\nu_{\mu}$ spectra for the
$1^{\circ}-4^{\circ}$ off-axis angle range spanned by the detector.
The baseline design for the detector is an instrumented structure with
a 10\,m tall inner-detector containing 3215 8 inch inward facing
photomultiplier tubes to detect the Cherenkov light, giving 40\%
photo-coverage. A crane system will move the detector structure
vertically in the 50\,m pit to make measurements at different off-axis
angles. By orienting the long axis of the detector perpendicular to the beam
direction, the detector covers a range of angles relative to the beam
direction.  Hence, each vertical slice of the detector samples a
different neutrino spectrum due to the decay kinematics of the pions
and kaons producing the neutrinos, the so-called off-axis beam effect.

There are three primary motivations for making neutrino measurements
over a range of off-axis angles.  First, the change in the neutrino
spectrum with off-axis angle is well known from the flux model, so the
predicted off-axis spectra can be combined in a linear combination to
produce almost arbitrary neutrino spectra.  Measured distributions at
different off-axis angles can be combined in the linear combination to
produce the predicted measured quantity for the neutrino spectrum of
interest.  In this way, it is possible to measure the muon spectrum
for a nearly mono-chromatic neutrino spectrum, or a spectrum that
closely matches the oscillated spectrum that is expected at the far
detector.  This approach can nearly eliminate the main model dependent
uncertainty in near to far extrapolations, which arises from the
combination of two factors: the near and far detector do not see the
same flux due to oscillations, and the relationship between the true
neutrino energy and final state lepton kinematics strongly depends on
nuclear effects, which are not well modelled~\cite{Bhadra:2014oma}.

The second physics motivation is the measurement of the electron
neutrino cross section relative to the muon neutrino cross section.
At further off-axis positions, the fraction of intrinsic
$\nu_{e},\bar{\nu}_{e}$ in the beam becomes large, making the
selection of pure candidate samples possible.  By taking advantage of
the enhanced purity at large off-axis angles, a measurement of the
cross section ratio, $\sigma_{\nu_{e}}/\sigma_{\nu_{\mu}}$ with 3\%
precision or better may be possible.  A measurement of the
$\sigma_{\bar{\nu}_{e}}/\sigma_{\bar{\nu}_{\mu}}$ ratio is also
possible, although the precision is expected to be degraded due to the
larger neutral current background rate for electron antineutrino
candidates and the presence of a larger wrong-sign background for both
muon and electron antineutrino charge current interactions.

The third physics motivation is the search for sterile neutrino
induced oscillations that are consistent with the
LSND~\cite{Aguilar:2001ty} and
MiniBooNE~\cite{Aguilar-Arevalo:2013pmq} $\bar{\nu}_{e}$ and $\nu_{e}$
appearance anomalies.  At a 1\,km baseline, the $L/E$ of the neutrino
spectrum peak varies between 1.1\,km/GeV at $1^{\circ}$ off-axis to
2.5\,km/GeV at $4^{\circ}$ off-axis.  Since the neutrino spectrum
varies with off-axis angle, it is possible to search for the
oscillation pattern not only through the reconstructed energy of the
neutrino candidate events, but also through the reconstructed off-axis
angle.  This method provides a significant improvement in the electron
neutrino appearance search sensitivity, and preliminary studies with a
non-optimal detector configuration already show that much of the LSND
allowed region can be excluded at 5$\sigma$~\cite{NUPRISMProposal}.

\subsubsection{Gadolinium Loading}

Recent developments in the addition of gadolinium
(Gd)~\cite{Watanabe:2008ru} and Water-based Liquid Scintillator (WbLS)
compounds~\cite{Alonso:2014fwf} to water raise the possibility to
separate neutrino and antineutrino interactions by detecting the
presence of neutrons or protons in the final state.

Final state proton tagging has been studied intensively for an
application for LArTPC detectors~\cite{Acciarri:2014gev}, where final
state protons can be counted to further purify the sample to improve
the oscillation sensitivity~\cite{Mosel:2013fxa}.  An analogous
approach is possible for the larger WC detectors.
Namely, Gd-doped WC detectors possess neutron tagging
ability on top of the 4$\pi$ detector coverage~\cite{Abe:2014oxa},
which allows statistical separation of primary interaction modes,
otherwise impossible.  
In particular, the ability to tag neutrons provides charge separation
information due to the enhanced presence of neutrons in the final
state for $\bar{\nu}$ charged current interactions. This will allow
studies of neutrino:anti-neutrino cross-section ratios on water and constraints on wrong-sign backgrounds, thus
reducing a critical systematic uncertainty in both the beam
$\delta_{CP}$ and atmospheric neutrino oscillation analyses. Neutron
tagging also allows more detailed studies of the interaction modes,
and in particular final state interaction effects, for the main
backgrounds to proton decay. 

\begin{figure}[!tbp]\centering

\includegraphics[width=0.32\textwidth]{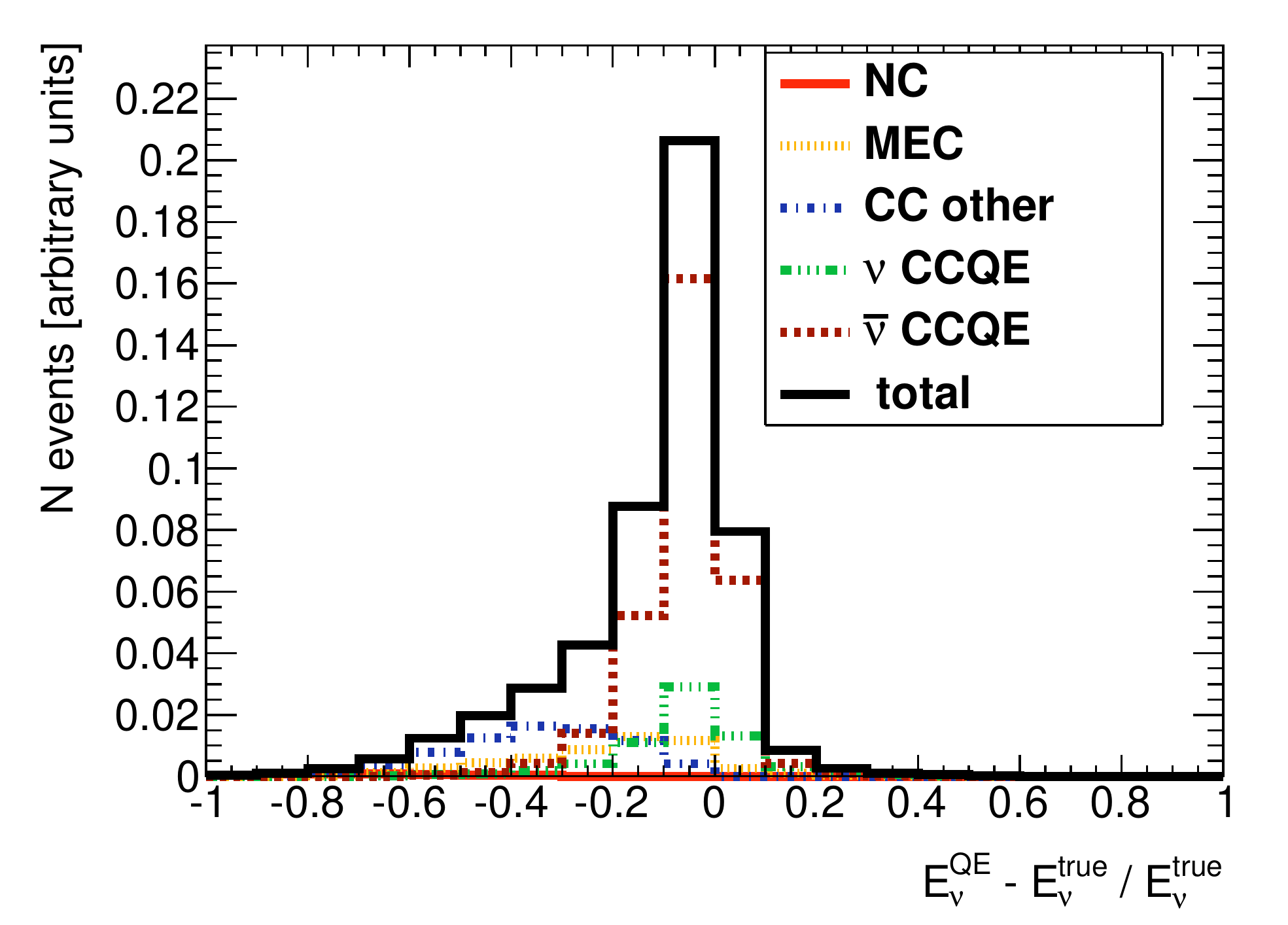}
\includegraphics[width=0.32\textwidth]{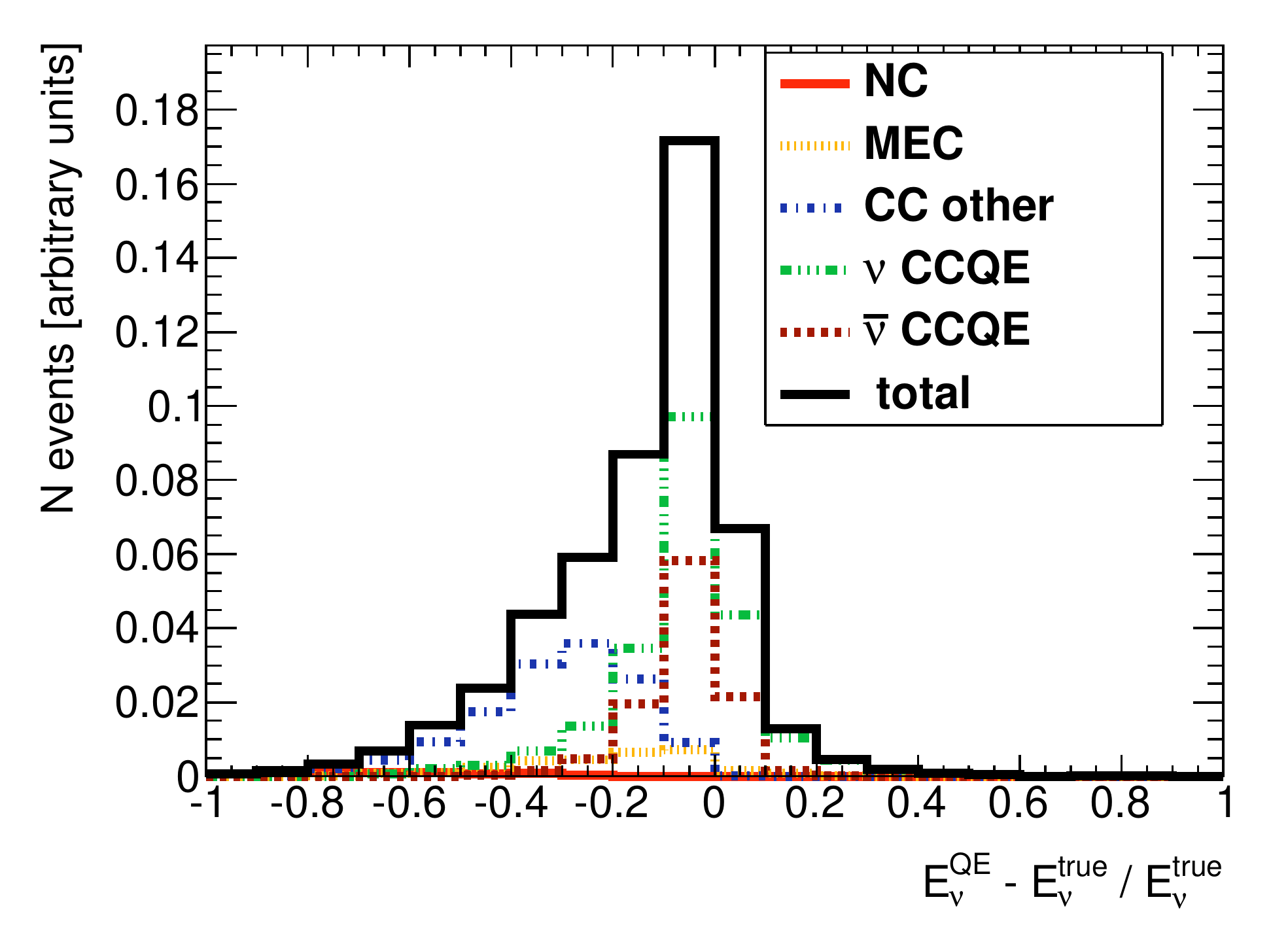}
\includegraphics[width=0.32\textwidth]{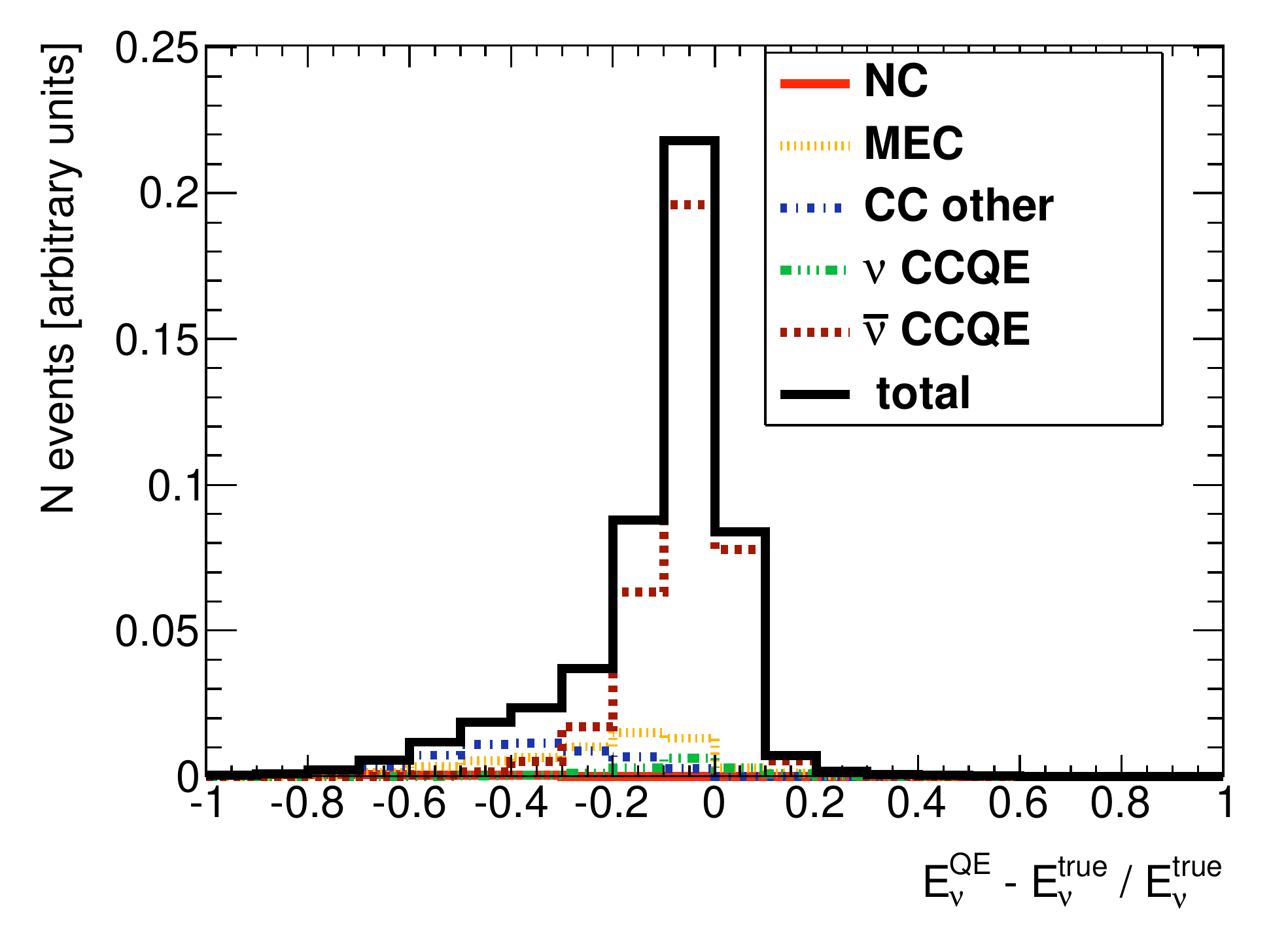}
\caption{The neutrino energy resolution due to the QE assumption in water Cherenkov near detector simulation 
for the TITUS detector~\cite{TITUSpreprint} during anti-neutrino mode running. 
The effect of different neutron selections is shown. From left to right, no neutron tagging, neutron number =0, 
and neutron number $>$ 0.\label{fig:neutrontaggingresolution}}
\end{figure}

The TITUS detector was originally proposed to provide an intermediate detector with neutron tagging capabilities for Hyper-K as described in Ref.~\cite{TITUSpreprint}. 
Figure~\ref{fig:neutrontaggingresolution} shows an example of a WC near detector simulation study for TITUS in which selecting ``neutron$\ge$1'' increases the selection purity for $\bar{\nu}$CCQE
interactions and hence improves the energy resolution. This technique
will be also applied to the ANNIE experiment~\cite{Anghel:2015xxt} in
the next years.

One aspect of the intermediate detector's design that needs to be
carefully considered with Gd-doped water is how to veto incoming
neutrons from beam-induced interactions in the material surrounding
the detector which will be the dominant contributor for the number of
particles entering the detector.  Vetoing most of these particles
requires at least 1\,m of water to reduce the low energy tail, plus a
fiducial cut on the reconstructed capture vertex. Preliminary studies
using spallation rates induced by muons~\cite{Galbiati:2005ft} and
interactions in the material surrounding the detector show this veto
can reduce the number of neutrons entering the detector's ID to just
10\% of all the events entering the tank and the fiducial region
further reduces this to approximately 7\% of the entering particles.

In principle, Gd loading is compatible with the off-axis spanning
detector configuration described in the previous section.  The
off-axis spanning detector should be as near as possible to the beam
origin to reduce the depth of the excavated volume, while the Gd
loaded detector should be far enough away to limit the beam induced
entering neutron background to the necessary level.  Preliminary
studies suggest that the entering neutron rate is sufficiently low for
the off-axis spanning detector located at 1\,km from the neutrino
production point.

\subsection{Summary}
This section has outlined the performance and requirements of the
accelerator complex, neutrino beamline and near detectors at J-PARC
that will be required for the Hyper-K physics program.  The J-PARC
accelerator chain has achieved 475 kW beam power extracted to the
neutrino beamline.  The accelerator upgrade plan, which includes the
upgrade of the MR magnet powers supplies and RF, is expected to achieve
1.3\,MW beam operation with $3.2\times10^{14}$ protons per pulse, as
early as 2026.

The neutrino beamline components require some upgrades to accept the
repetition rate, proton intensity and total beam power necessary to
achieve 1.3\,MW at $3.2\times10^{14}$ protons per pulse.  To achieve
the 1.16 Hz operation, each magnetic horn requires an individual power
supply utilizing an energy recovery scheme and low
inductance/resistance striplines.  
The treatment facilities for activated cooling water will be expanded to accept up to 2 MW operation as well.  
The current beam window and
target are rated to $3.3\times10^{14}$ ppp, however their lifetime at
$3.2\times10^{14}$ ppp and 1.16 Hz will be studied and upgrades may be
necessary.

The current T2K near detectors, including ND280 and INGRID, are used
to control neutrino flux and cross-section systematic errors at the
$\sim$5--6\% level.  Further upgrades to the ND280 data analyses with
water target measurements and large angle tracks will reduce the
systematic error, although the ultimate performance may be limited by
the relatively low water fraction and low efficiency for large angle
track reconstruction.  Upgrades to ND280 are being considered by T2K
and these include a high pressure TPC, the WAGASCI water/scintillator
3D grid detector and emulsion detectors.  A reconfiguration of the TPC
geometry is also being considered to give better reconstruction at
high angles.  It is expected that some of these upgrades may be
carried out during the T2K experiment and Hyper-K may benefit from
their continued use.  If they are not implemented during T2K, these
ND280 upgrades as well as the continued operation of ND280 are an
expected area for international contributions to Hyper-K.

An intermediate water Cherenkov detector provides a necessary
complement to the ND280 magnetized tracking detector in order to
constrain all the dominant systematics at the precision required.  The
WC detector requires a new facility off of the J-PARC site and the
excavation of a new pit to house the detector.

\clearpage
\color{black}
\section{Hyper-Kamiokande detector} \label{section:design}

%---------------------------------------------------------------------------
%%%%%%%%%%%%%%%%%%%%%%%%%%%%%%%%%%%%%
\subsection{Introduction of the Hyper-Kamiokande detector}
%%%%%%%%%%%%%%%%%%%%%%%%%%%%%%%%%%%%%
\graphicspath{{design-introduction/figures/}}

\begin{figure}[htbp]
  \begin{center}
  \includegraphics[width=0.69\textwidth]{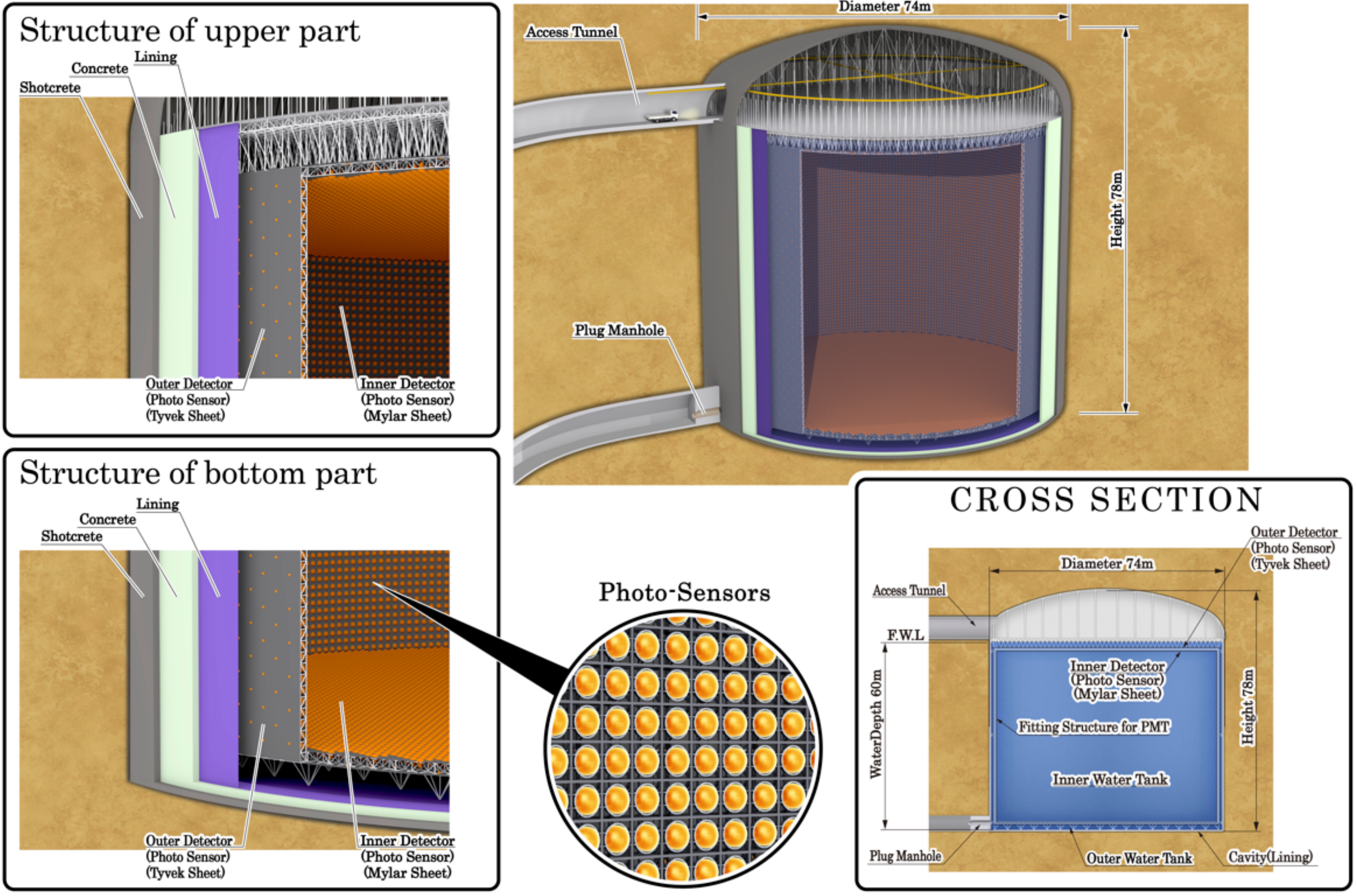}
  \caption{Schematic view for the configuration of single cylindrical tank instrumented with high density (40\% photocoverage) PMTs.
It is referred as \hksingletank{} in this report.}
  \label{fig:hk-perspective-1TankHD}
  \end{center}
\end{figure}

The \hksingletank{} design, i.e. one cylindrical vertical tank with
40\% phoocoverage as shown in Fig.\ref{fig:hk-perspective-1TankHD}, corresponds to the optimized configuration studied
by the proto-collaboration. It is referred as \hksingletank{} in this report.

The strategy considers also a staged second tank (2TankHK-staged)
currenlty being investigated (see appendix~\ref{sec:hakamagoshi}).

The Hyper-K experiment employs a ring-imaging water Cherenkov detector
technique to detect rare interactions of neutrinos and the possible
spontaneous decay of protons and bound neutrons.
Table~\ref{Table:detectorparameters} summarizes the key parameters of
the Hyper-K detector compared with other previous and currently
operating water Cherenkov detectors.  These types of detectors are
located deep underground in order to be shielded from cosmic rays and
their corresponding daughter particles and thereby to achieve a very
low background environment.

The detector mass -- or equivalently the underground detector cavern
size or water tank size -- is one of the key detector parameters that
determines the event statistics in neutrino observations and nucleon
(proton or bound neutron) decay searches.  The detector water plays
two roles: a target material for incoming neutrinos and source of
nucleons to decay.  We need a detector mass of at least $O(10^2)$
kton.  in order to accumulate $O(10^3)$ electron neutrino signal
events (as shown in Table~\ref{Tab:sens-selection-nue}) from the
J-PARC neutrino beam.  This is necessary to measure the $CP$ violation
effect with a few \% accuracy.  This mass of water contains
$O(10^{35})$ nucleons (protons and nucleons) which would give an
unprecedented sensitivity to nucleon lifetime at the level of
$10^{35}$ years.  The location and detailed designs of the Hyper-K
cavern and tank are presented in Section~\ref{section:location},
\ref{section:cavern}, and \ref{section:tank}.

The detector is filled with highly transparent purified water, as
shown in Section~\ref{section:water}. A light attenuation length above
100\,m can be achieved which allows us to detect a large fraction of
the emitted Cherenkov light around the periphery of the water volume.
Radon concentration in the supplied water is kept below 1\,mBq/m$^3$
to control the radioactive background event rate in solar neutrino and
other low energy observation.  An option being investigated is the
Gd-doping of the water.  This option, in addition to the nominal water
one, is presented in Section~\ref{section:water}.

\begin{table}[!tbp]
  \centering
  \caption{Parameters of past
    (KAM~\cite{Suzuki:1992as,Fukugita:1994wx}), running
    (SK~\cite{Fukuda:2002uc,Abe:2013gga}), and future
    HK-\hksingletank{}) water Cherenkov
    detectors.
  The KAM and SK have undergone several configuration changes
  and parameters for KAM-II and SK-IV are referred 
  in the table.
  The single-photon detection efficiencies are products of
  the quantum efficiency at peak ($\sim 400$\,nm), 
  photo-electron collection efficiency,
  and threshold efficiency.
  }\label{Table:detectorparameters}
  \begin{tabular}{lccc}
   \hline
   \hline
                        & KAM & SK & HK-\hksingletank{} \ \\
   \hline
   Depth   &    1,000 m    &  1,000 m  &    650 m  \\
   Dimensions of water tank & & &  \\
   ~~~~diameter &  15.6 m $\phi$ &  39 m $\phi$ & 74 m $\phi$ \\
   ~~~~height &  16 m &  42 m & 60 m \\
   Total volume          &  4.5 kton & 50 kton & 258 kton \\
   Fiducial volume        &    0.68 kton & 22.5 kton & 187 kton \\
   Outer detector thickness  &  $\sim$ 1.5 m & $\sim$ 2 m & $1 \sim 2$ m \\
   Number of PMTs    &  &  & \\
   ~~~~inner detector (ID)    &  ~~948 (50 cm $\phi$)~~ & ~~11,129 (50 cm $\phi$)~~ & ~~40,000 (50 cm $\phi$)~~ \\
   ~~~~outer detector (OD)   &  123 (50 cm $\phi$) & 1,885 (20 cm $\phi$) & 6,700 (20 cm $\phi$) \\
   Photo-sensitive coverage    &  20\% & 40\% & 40\% \\
   Single-photon detection & unknown  & 12\% & 24\%\\
   efficiency of ID PMT  &  &  &  \\
   Single-photon timing & $\sim 4$ nsec  & 2-3 nsec & 1 nsec \\
   resolution of ID PMT  &  & & \\
   \hline
   \hline
  \end{tabular}
  \label{table:example}
 \end{table}

The detector is instrumented with an array of sensors with
single-photon sensitivity in order to enable reconstruction of the
spatial and timing distributions of the Cherenkov photons which are
emitted by secondary particles from neutrino interactions and nucleon
decays.  The dimension of the photo-sensors and their density are
subject to an optimization that takes into account the required signal
identification efficiencies, background rejection power, and cost.  As
a reference, the Super-K detector shown in
Table~\ref{Table:detectorparameters} covers $40\%$ of the detector
wall with Hamamatsu R3600 50\,cm diameter hemispherical
photomultiplier tubes (PMTs) with the original goal to measure the
solar neutrino energy spectrum above $\sim$5\,MeV.

The Hyper-K detector is designed to employ newly developed
high-efficiency and high-resolution PMTs (Hamamatsu R12860) which
would amplify faint signatures such as neutron signatures associated
with neutrino interactions, nuclear de-excitation gammas and $\pi^+$
in proton decays into Kaons, and so on.  This increased sensitivity
greatly benefit the major goals of the Hyper-K experiment such as
clean proton decay searches via $p\rightarrow e^+ + \pi^0$ and
$p\rightarrow \bar{\nu} + K^+$ decay modes and observation of
supernova electron anti-neutrinos.  The characteristics of the R12860
tubes are shown in Section~\ref{section:photosensors}.  The
photo-sensors have vacuum glass bulbs and will be located as much as
60\,m underwater in the Hyper-K cavern.  At this depth, the applied
pressure is close to the manufacturers upper specification of the
Super-K R3600 PMT (0.65\,MPa).  Therefore, we need to develop a new
bulb design and a quality controlled production method to ensure that
the sensors can withstand this pressure.  Furthermore, PMT cases
will envelop each photo-sensors to avoid a potential chain reaction
accident due to the implosion of a glass bulb in the water.  The
designs of the bulb and case are also described in
Section~\ref{section:photosensors}.

The detector is instrumented with front-end electronics and a readout
network/computer system as shown in Section~\ref{section:electronics}
and \ref{section:daq}.  The system is capable of high-efficient data
acquisition for two successive events in which Michel electron events
follow muon events with a mean interval of 2\,$\mu$sec.  It is also
able to collect the vast amount of neutrinos, which would come from
nearby supernova in a nominal time period of 10\,sec.

Similar to Super-K, an outer detector (OD) is being envisaged that, in
addition to enabling additional physics, would help to constrain the
external background.  Sparser photo-coverage and smaller PMTs than
that for the ID is also planned.

\newpage
%-----------------------------------------------------------------------------
\def\makehako#1#2#3
  {\hbox{\vrule
       \vbox to #1{\hrule \vss
                   \hbox to #2{\hss#3\hss}\vss
                   \hrule}\vrule}}
\def\red#1{\textcolor{red}{#1}}
\def\blue#1{\textcolor{blue}{#1}}
%-----------------------------------------------------------------------------

\graphicspath{{design-location/figs/}}

%%%%%%%%%%%%%%%%%%%%%%%%%%%%%%%%%%%%%%%%%%%%%
\subsection{Detector site}\label{section:location}
%%%%%%%%%%%%%%%%%%%%%%%%%%%%%%%%%%%%%%%%%%%%%

\subsubsection{Detector location}
\begin{figure}[tb]
  \includegraphics[width=1.0\textwidth]{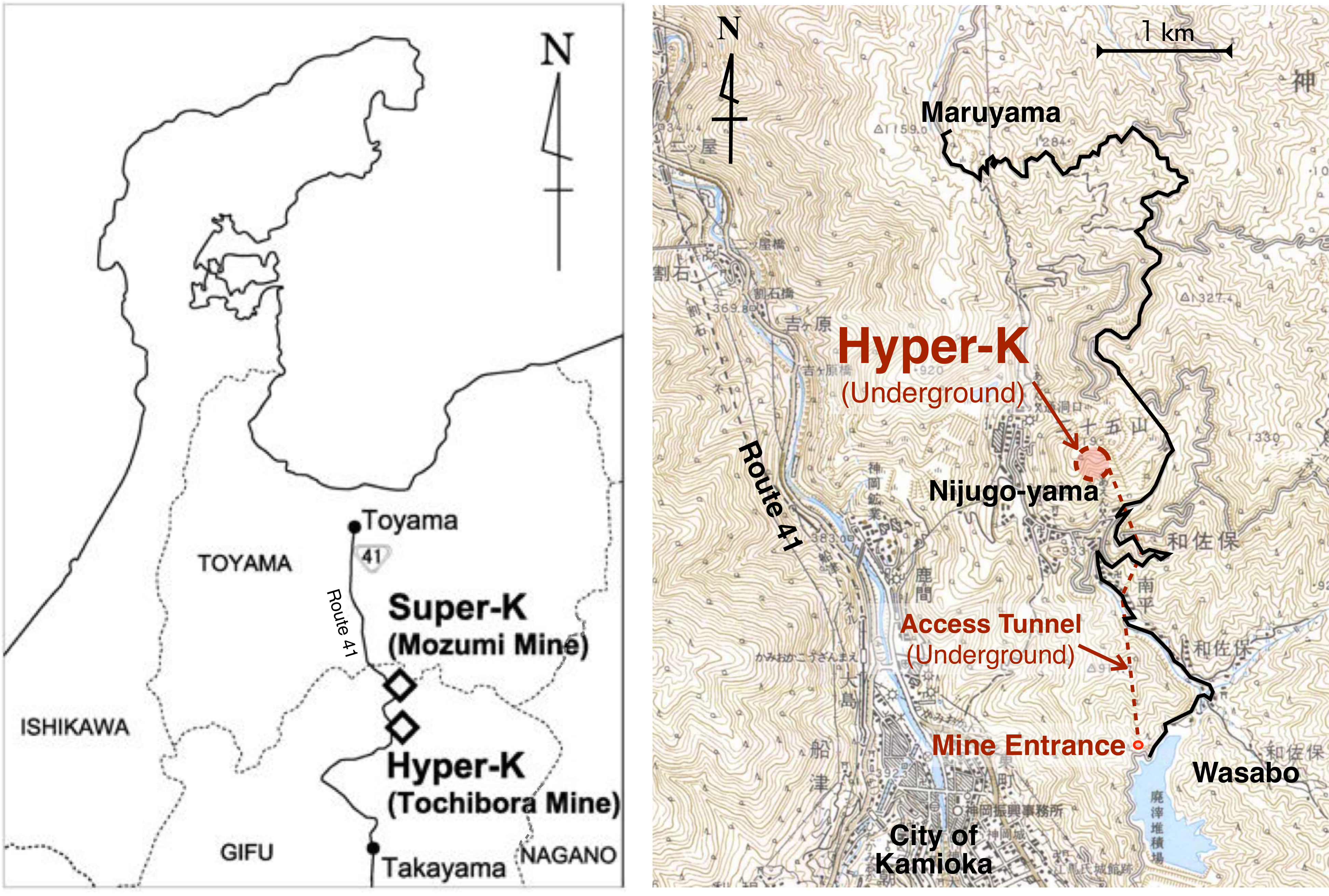} 
  \caption{The candidate site map. Broad area map (left) and detailed map
  (right).}  \label{fig:map}
\end{figure}

\begin{figure}[tb]
\centering
  \includegraphics[width=0.95\textwidth]{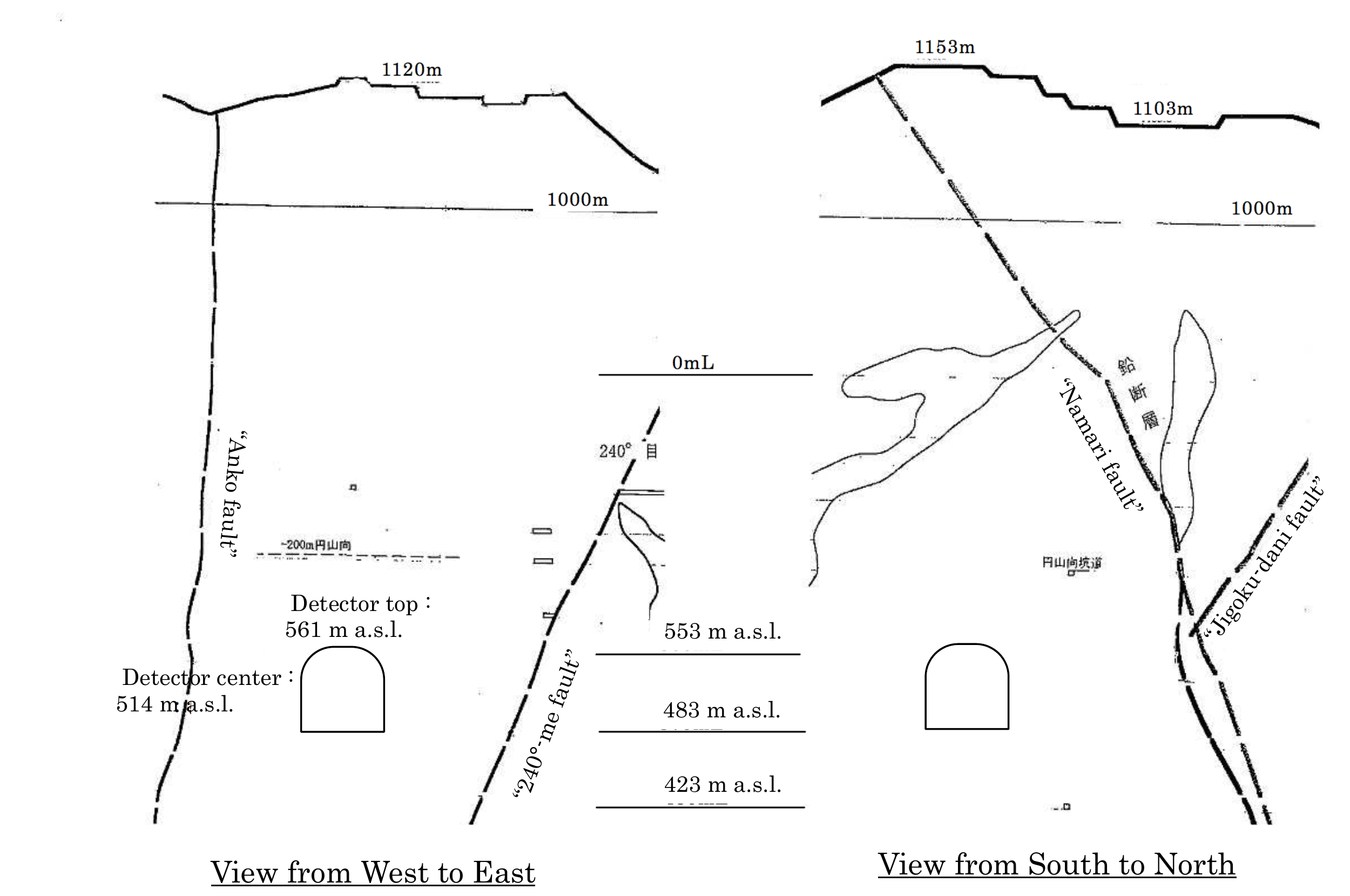} \caption{Location
  of faults and existing tunnels around the candidate site.  The
  existing tunnels are located at 423, 483, and 553\,m
  a.s.l.}  \label{fig:fault}
\end{figure}

The Hyper-K detector candidate site, located 8\,km south of Super-K,
is in the Tochibora mine of the Kamioka Mining and Smelting Company,
near Kamioka town in Gifu Prefecture, Japan, as shown in
Fig.~\ref{fig:map}.
The J-PARC neutrino beamline is designed so that the existing
Super-Kamiokande detector and the Hyper-K candidate site in Tochibora
mine have the same off-axis angle.  The experiment site is accessible
via a drive-in, $\sim$2.5\,km long, (nominally) horizontal mine tunnel.
The detector will lie under the peak of Nijuugo-yama, with an
overburden of 650\,meters of rock or 1,750\,meters-water-equivalent
(m.w.e.), at geographic coordinates Lat. 36$^{\circ}$21'20.105''N,
Long. 137$^{\circ}$18'49.137''E (world geographical coordinate
system), and
an altitude of 514\,m above sea level (a.s.l.).  The candidate site is
surrounded by several faults as shown in Fig.~\ref{fig:fault} and the
caverns and their support structure are placed to avoid a conflict
with the known faults.
The site has a neighboring mountain, Maruyama, just 2.3\,km away,
whose collapsed peak enables us to dispose of more than one million
m$^3$ of the excavated rock from the detector cavern excavation.

\subsubsection{Geological condition at the site vicinity}
\begin{figure}[tb]
\centering
  \includegraphics[width=1.0\textwidth]{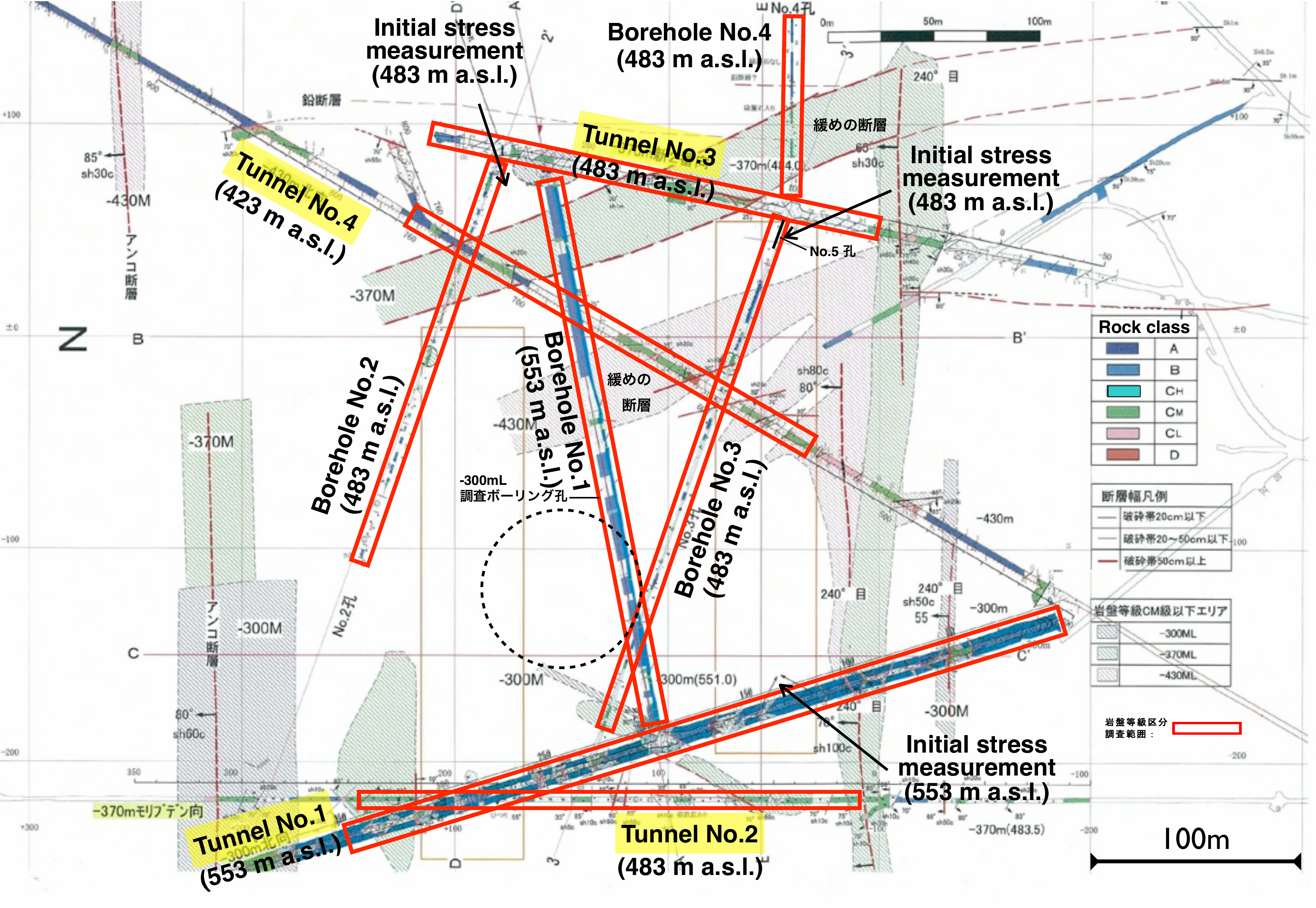} \caption{Location
  of rock quality measurements in existing tunnels and bore-hole cores
  at 423\,m, 483\,m, and 553\,m a.s.l. The red rectangulars show the
  surveyed regions in the measurements. 
  The black dashed circle indicates the Hyper-K cavern construction candidate site
  and size of the cavern.
  }  \label{fig:rock_qual_meas}
\end{figure}

Rock quality is investigated in the existing tunnels and in sampled
borehole cores near the candidate site.
Fig.~\ref{fig:rock_qual_meas} summarizes the geological surveys.
The rock wall in the existing tunnels and sampled borehole cores are
dominated by Hornblende Biotite Gneiss and Migmatite in the state of
sound, intact rock mass.  This is desirable for constructing such
unprecedented large underground cavities.  A rock mass classification
sytem developed by Central Research Institute of Electric Power
Industry (CRIEPI)~\cite{rock_class}, which is widely used for dams and
underground cavities construction for the electric power plants in
Japan,
is utilized to classify rock quality. The CRIEPI system categorizes
rock quality in six groups as A, B, CH, CM, CL, and D (in order of
good quality), among which the A, B, and CH classes are suitable for
cavern construction.  Fraction of rock quality at the measured sites
is summarized in Table~\ref{tab:rock_qual_dist}.
\begin{table}[htbp]
\caption{Summary of measured rock quality fraction. Sum of rock quality fraction
in some Bore-holes is not 100\% since a small fraction of sampled rock cores was broken
during the survey due to a sampling failure.
  \label{tab:rock_qual_dist}}
\begin{tabular}{l|c|c|c|c|c|c}
\hline\hline
Place           & \multicolumn{6}{c}{Rock quality fraction (\%)} \\ \cline{2-7}
                & A   & B    & CH          & CM  & CL  & D           \\\hline\hline
Tunnel No.1     & 0.0 & 51.6 & 43.6        & 3.0 & 1.8 & 0.0         \\\cline{2-7}
(553 m a.s.l.)  & \multicolumn{3}{c|}{95.2} & \multicolumn{3}{c}{4.8} \\\hline
Bore-hole No.1  & 0.0 & 67.9 & 27.7        & 4.0 & 0.4 & 0.0         \\\cline{2-7}
(553 m a.s.l.)  & \multicolumn{3}{c|}{95.6} & \multicolumn{3}{c}{4.4} \\\hline
Tunnel No.2     & 0.0 & 11.4 & 45.4        & 39.8 & 3.4 & 0.0        \\\cline{2-7}
(483 m a.s.l.)  & \multicolumn{3}{c|}{56.8} & \multicolumn{3}{c}{43.2}\\\hline
Tunnel No.3     & 0.0 & 4.9 & 55.7         & 25.0 & 14.4 & 0.0       \\\cline{2-7}
(483 m a.s.l.)  & \multicolumn{3}{c|}{60.6} & \multicolumn{3}{c}{39.4}\\\hline
Bore-hole No.2  & 2.4 & 10.5 & 49.2        & 29.7 & 5.7 & 0.2        \\\cline{2-7}
(483 m a.s.l.)  & \multicolumn{3}{c|}{62.1} & \multicolumn{3}{c}{35.6}\\\hline
Bore-hole No.3  & 0.0 & 19.2 & 59.2        & 16.5 & 3.8 & 0.3        \\\cline{2-7}
(483 m a.s.l.)  & \multicolumn{3}{c|}{78.4} & \multicolumn{3}{c}{20.6}\\\hline
Bore-hole No.4  & 6.6 & 20.5 & 36.4        & 22.6 & 7.1 & 3.1        \\\cline{2-7}
(483 m a.s.l.)  & \multicolumn{3}{c|}{63.5} & \multicolumn{3}{c}{32.8}\\\hline
Tunnel No.4     & 0.0 & 18.1 & 39.0        & 38.1 & 1.9 & 2.9        \\\cline{2-7}
(423 m a.s.l.)  & \multicolumn{3}{c|}{57.1} & \multicolumn{3}{c}{42.9}\\\hline\hline
\end{tabular}
\end{table}
The geological surveys are performed at three different altitudes
(423\,m, 483\,m and 553\,m a.s.l.)  and better fraction of B and CH
classes is observed at higher altitude.  The measured fraction of rock
quality is used for an assumption of rock quality distribution in
cavern stability analyses.
 
\begin{figure}[tb]
\centering
  \includegraphics[width=0.9\textwidth]{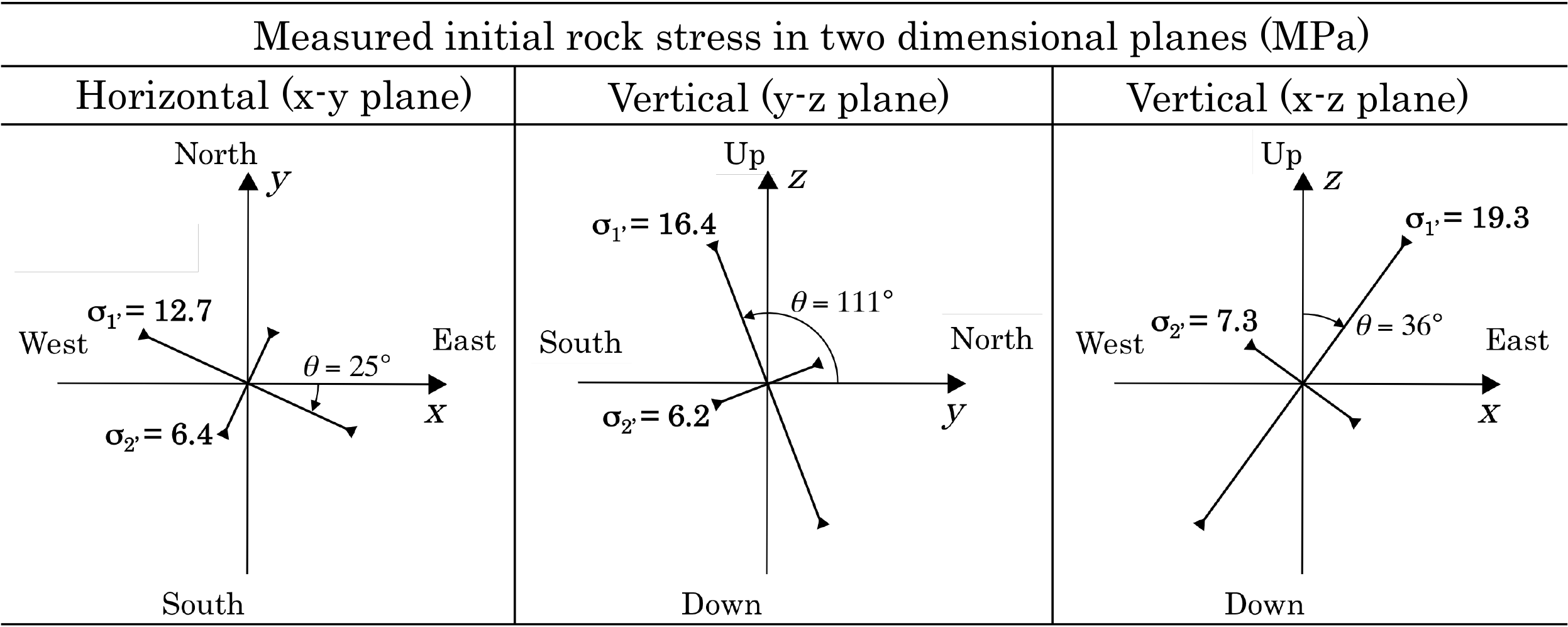}
  \caption{Results of initial rock stress measurement at 553\,m a.s.l.}
  \label{fig:ini-stress}
\end{figure}

The initial stress of the rock is also measured at three points, two
of which are located at the bottom of the detector cavern (483\,m
a.s.l.) and one at top (553\,m a.s.l.).
It was found that the two measurements at 483\,m a.s.l. are strongly
influenced by existing faults.
We aim to build our detector at a place where there is no interference
with any faults or fracture zone, and the inputs, like initial stress,
to the cavern stability analysis should not be influenced with faults.
Thus, the two measurements at 483\,m a.s.l. are eliminated and the one at
553\,m a.s.l. is used for a cavern stability analysis described later.
The measured rock stress at 553\,m a.s.l. is shown in
Fig~\ref{fig:ini-stress}.
Based on the {\it in-situ} measurements of the rock quality and the
rock stress, it is confirmed that the Hyper-K caverns can be
constructed with the existing excavation techniques (described in
section~\ref{section:cavern}).

\graphicspath{{design-location/figs_seismic/}}
\clearpage
%%%%%%%%%%%%%%%%%%%%%%%%%%%%%%%%%%%%%%%%%
\subsubsection{Refining the cavern construction candidate site\label{sec:seismic}}
%%%%%%%%%%%%%%%%%%%%%%%%%%%%%%%%%%%%%%%%%

As shown in previous section, the candidate site
for cavern construction has area of approximately 300\,m$\times$300\,m
(see Fig.~\ref{fig:rock_qual_meas}).
In order to further refine and narrow the candidate area where has the best geological condition
for the cavern construction,
a geological survey in wide-range, called ``seismic prospecting,'' has been carried out.

Seismic prospecting uses an artificially generated elastic wave that
transmits underground bedrock, and identifies physical properties
of bedrock and geological structure underground, based on the classical physics
principle of transmission, reflection, and refraction of the elastic wave.
For example, the speed of elastic wave transmission is varied
if the elastic wave propagates in a bedrock with different physical
properties, e.g. elastic modulus.

The target area of seismic prospecting is defined as
423$\sim$703\,m a.s.l., 400 meters from east to west and 400 meters
from south to north, that covers the entire Hyper-K candidate site
shown in Fig.~\ref{fig:rock_qual_meas}.
There are six existing tunnels around the target area and they locate
at different elevations between 423$\sim$723\,m a.s.l.
For the seismic prospecting, receivers or sensors, called `geophones,'
which detect the elastic wave, were installed in all the six tunnels at interval of 20\,m 
-- 111 locations in total and each location has three geophones to
capture triaxial components of the elastic wave.
A seismic source is set in the tunnels and generated the elastic wave
at all the six tunnels with interval of 2.5\,m in order -- 738 seismic source points in total.

\begin{figure}%[htbp]
  \begin{center}
  \includegraphics[width=0.88\textwidth]{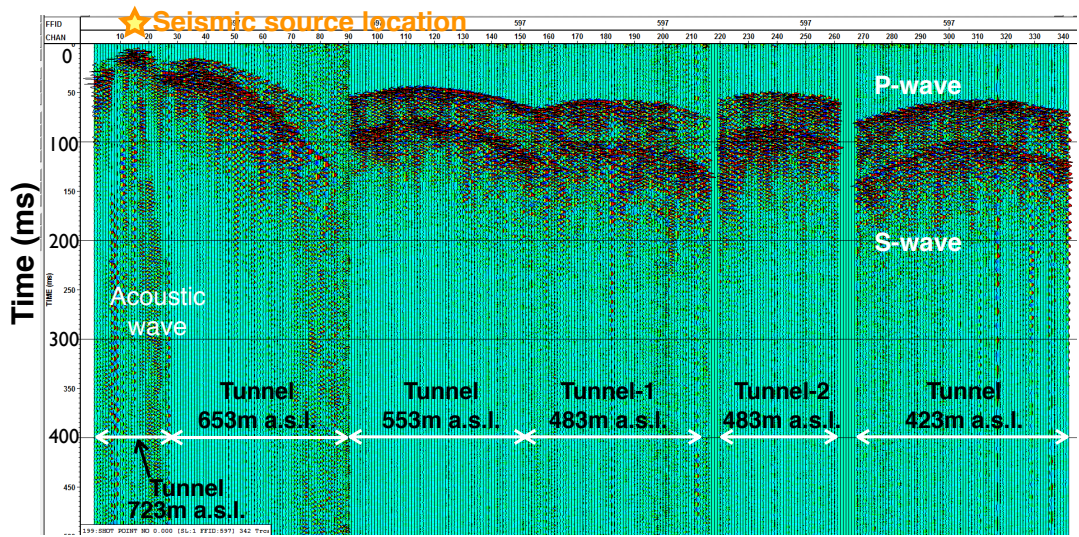}
  \caption{Seismic prospecting waveform data obtained with geophones
  when an elastic wave generated at a location in the tunnel at 723\,m a.s.l., as an example.
  Vertical axis is a time in millisecond and the origin of vertical axis is when
  an elastic wave is generated.
  Each line runs in vertical direction is a waveform data obtained with a geophone,
  and figure shows waveform data for all the 333 geophones, which are arranged in the
  lateral direction.
  In the figure, pulse heights of the waveform are shown with different colors,
  and the waveform in darker color corresponds to a time when geophones captured an elastic wave.}
  \label{fig:seismic_waveform}
  \end{center}
\end{figure}

Figure~\ref{fig:seismic_waveform} shows a waveform data obtained with all geophones
when an elastic wave generated at a location in the tunnel at 723\,m a.s.l.
One can find that an elastic wave transmitted from the seismic source point
through the tunnels in lower altitude.

\begin{figure}[p]%[htbp]
  \begin{center}
  \includegraphics[width=0.45\textwidth]{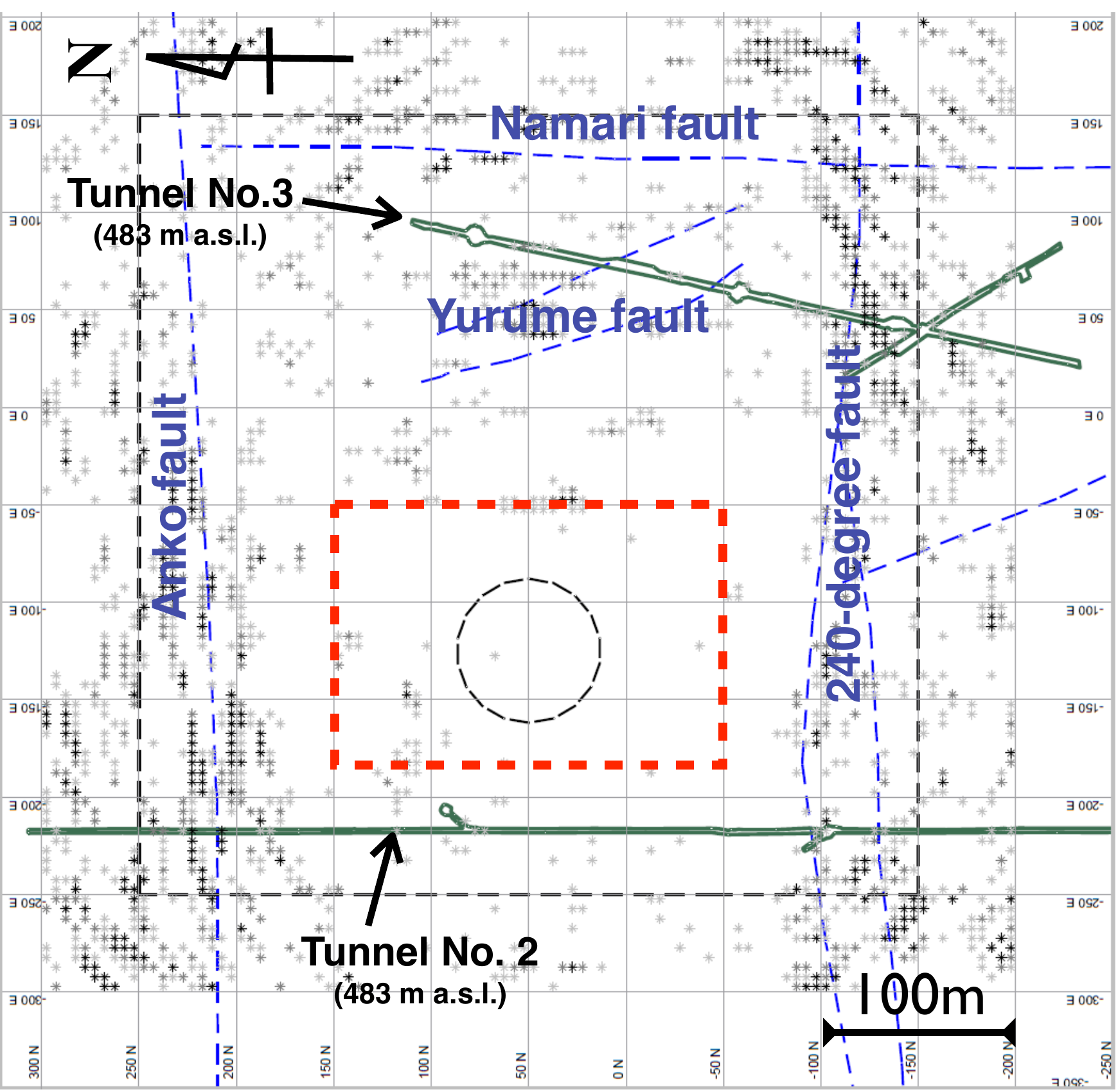}
  \includegraphics[width=0.45\textwidth]{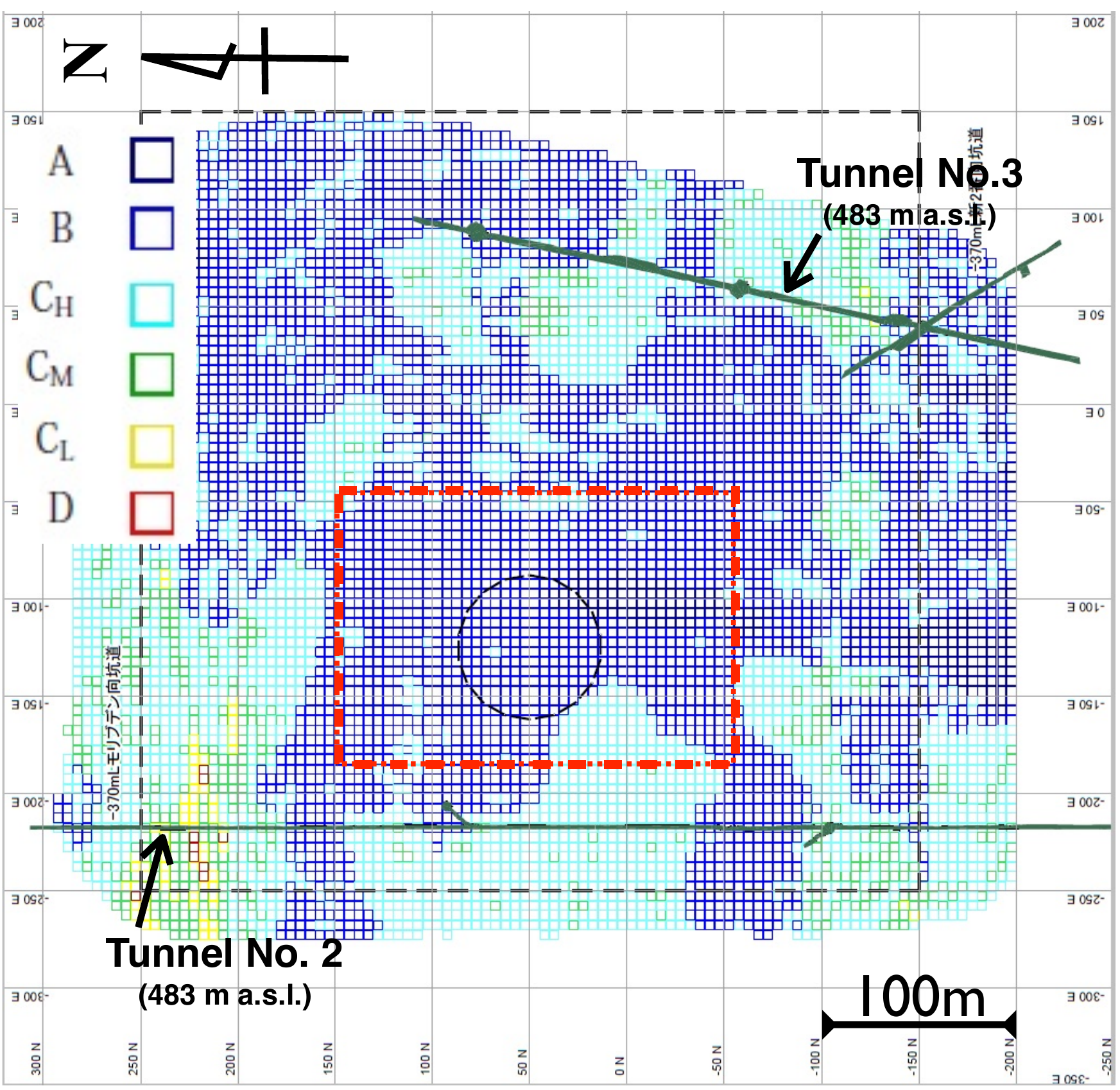}
  \includegraphics[width=0.45\textwidth]{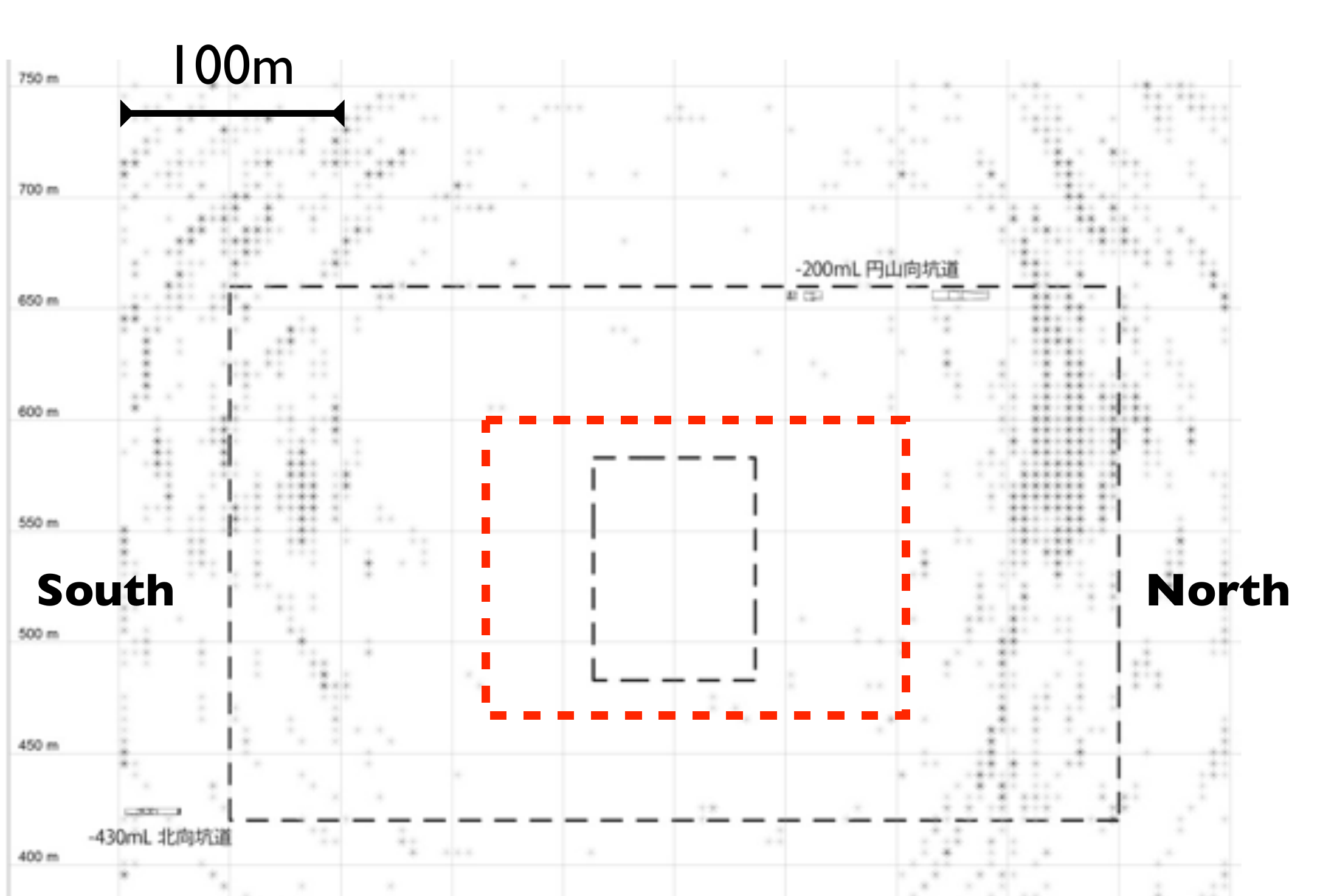}
  \includegraphics[width=0.47\textwidth]{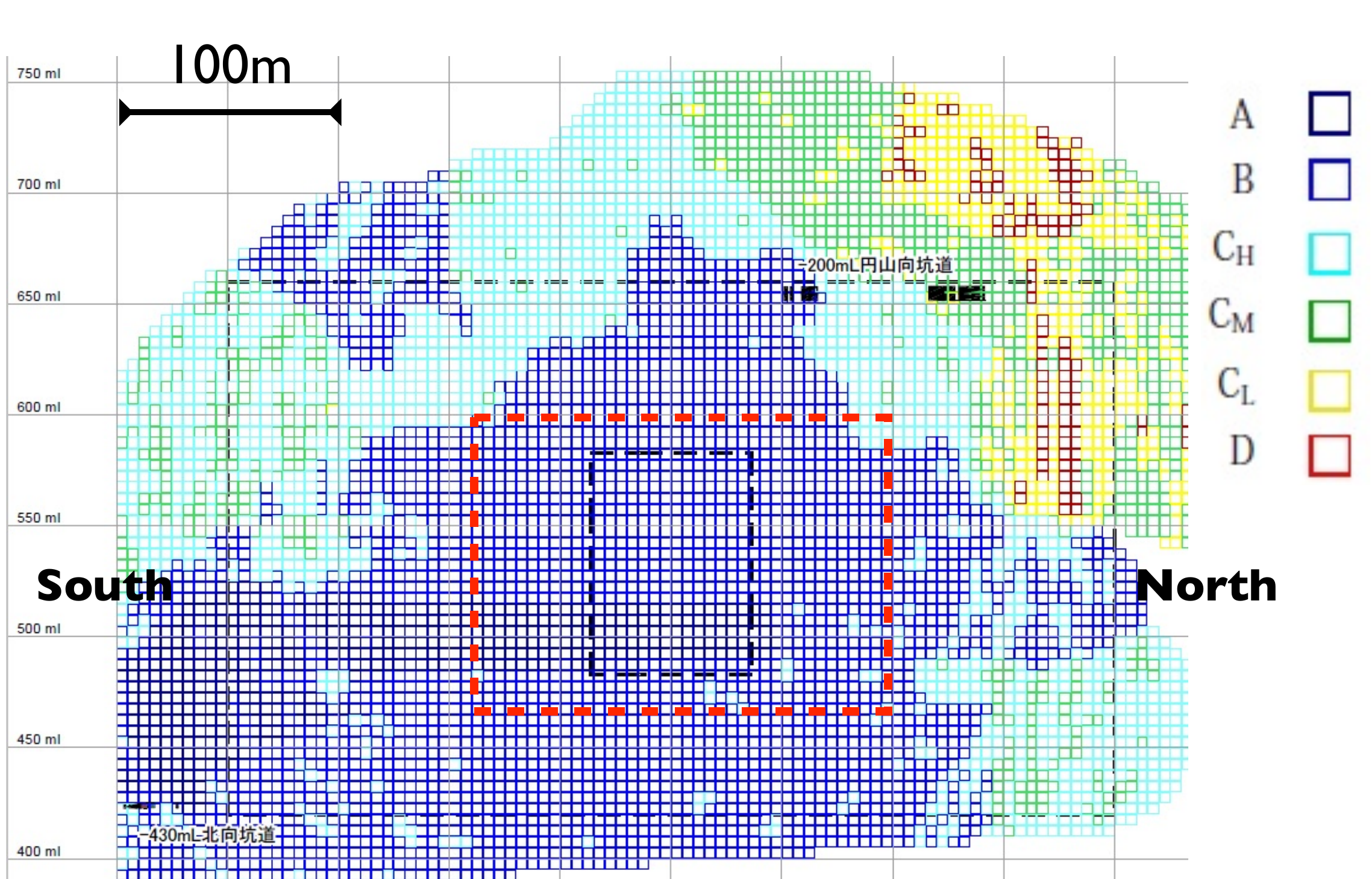}
  \caption{Results of seismic prospecting at Hyper-K candidate site.
  Top two figures are the results at altitude of 483\,m a.s.l.
  and lower two figures are results in vertical slice from south to north direction
  as an example.
  Top-left and lower-left plots show the results of reflection imaging,
  and asterisk (*) indicate the identified locations where have fault,
  fracture zone or open cracks in the bedrock.
  In the top-left figure, blue dashed lines indicates the location of known faults which
  are shown in Fig.~\ref{fig:rock_qual_meas}.
  Top-right and lower-right figures are rock class distribution obtained by combining the results of
  seismic tomography, reflection imaging and the geological survey results
  with borehole and the existing tunnels as shown in Fig.~\ref{fig:rock_qual_meas}.
  The red dashed rectangles in the figures denotes a region where has best rock
  quality and least uneven rock structure over the entire Hyper-K candidate site.
  Dashed circle indicates the size of Hyper-K cavern.}
  \label{fig:seismic_results}
  \end{center}
\end{figure}

The data obtained in the seismic survey were analyzed with two methods, seismic
tomography and reflection imaging.
Seismic tomography uses the speed of transmission of the elastic wave.
The speed of elastic wave transmission varies depending on the physical properties,
e.g. elastic modulus, density,
and seismic tomography identifies the physical properties of rock in the
target region, the entire Hyper-K candidate site.
Reflection imaging identifies fault, fracture zone and open cracks in rock
using a nature that an elastic
wave is reflected if there is a discontinuous or uneven structure, like a fault, in the bedrock.
Left figure in Fig.~\ref{fig:seismic_results} shows the results of reflection imaging.
In the figure, blue dashed lines indicate the known faults location.
As shown in the figure, the reflection imaging identified the known faults, and
confirmed that there is no major fault, fracture zone nor open cracks in the
Hyper-K cavern construction candidate site.
Right figure in Fig.~\ref{fig:seismic_results} is rock class distribution obtained by
combining the results of seismic tomography, reflection imaging and by comparing
those results with the geological survey results shown in Fig.~\ref{fig:rock_qual_meas}, 
which are obtained with borehole coring and investigation of the existing tunnels.
The red dashed rectangles in the figures denote a region where has the best rock
quality and the least uneven rock over the entire Hyper-K candidate site.
From the seismic prospecting results, the location for Hyper-K cavern
construction is narrowed down to approximately $200$\,m$\times150$\,m region
(red dashed rectangle in Fig.~\ref{fig:rock_qual_meas}).

In conclusion, the risk associated with insufficient geological information,
especially regarding geological discontinuities, e.g. faults, and low quality
rock mass with CM or lower classes, has been largely reduced by the seismic surveys.

\newpage
\def\makehako#1#2#3% vsize, hsize, inserted text
  {\hbox{\vrule
       \vbox to #1{\hrule \vss
                   \hbox to #2{\hss#3\hss}\vss
                   \hrule}\vrule}}
\def\red#1{\textcolor{red}{#1}}
\def\blue#1{\textcolor{blue}{#1}}
\setcounter{secnumdepth}{5}
\makeatletter
\newcommand{\subsubsubsection}{\@startsection{paragraph}{4}{\z@}%
{1.5\baselineskip \@plus.5\dp0 \@minus.2\dp0}%
{.5\baselineskip \@plus2.3\dp0}%
{\reset@font\normalsize\itshape}
}
\newcommand{\subsubsubsubsection}{\@startsection{subparagraph}{5}{\z@}%
{1.5\baselineskip \@plus.5\dp0 \@minus.2\dp0}%
{.5\baselineskip \@plus2.3\dp0}%
{\reset@font\normalsize\itshape}
}
\makeatother
\setcounter{tocdepth}{5}
\renewcommand{\theparagraph}{\thesubsubsection.\arabic{paragraph}}
\renewcommand{\thesubparagraph}{\theparagraph.\arabic{subparagraph}}
%---------------------------------------------------------------------------

\graphicspath{{design-cavern/figs/}}

\clearpage
%%%%%%%%%%%%%%%%%%%%%%%%%%%%%%%%%%%%%
\subsection{Cavern}\label{section:cavern}
%%%%%%%%%%%%%%%%%%%%%%%%%%%%%%%%%%%%%

%%%%%%%%%%%%%%%%%%%%%%%%%%%%%%%%%%%%%
\subsubsection{Cavern shape}
%%%%%%%%%%%%%%%%%%%%%%%%%%%%%%%%%%%%%

\begin{figure}%[htbp]
  \begin{center}
  \includegraphics[width=0.4\textwidth]{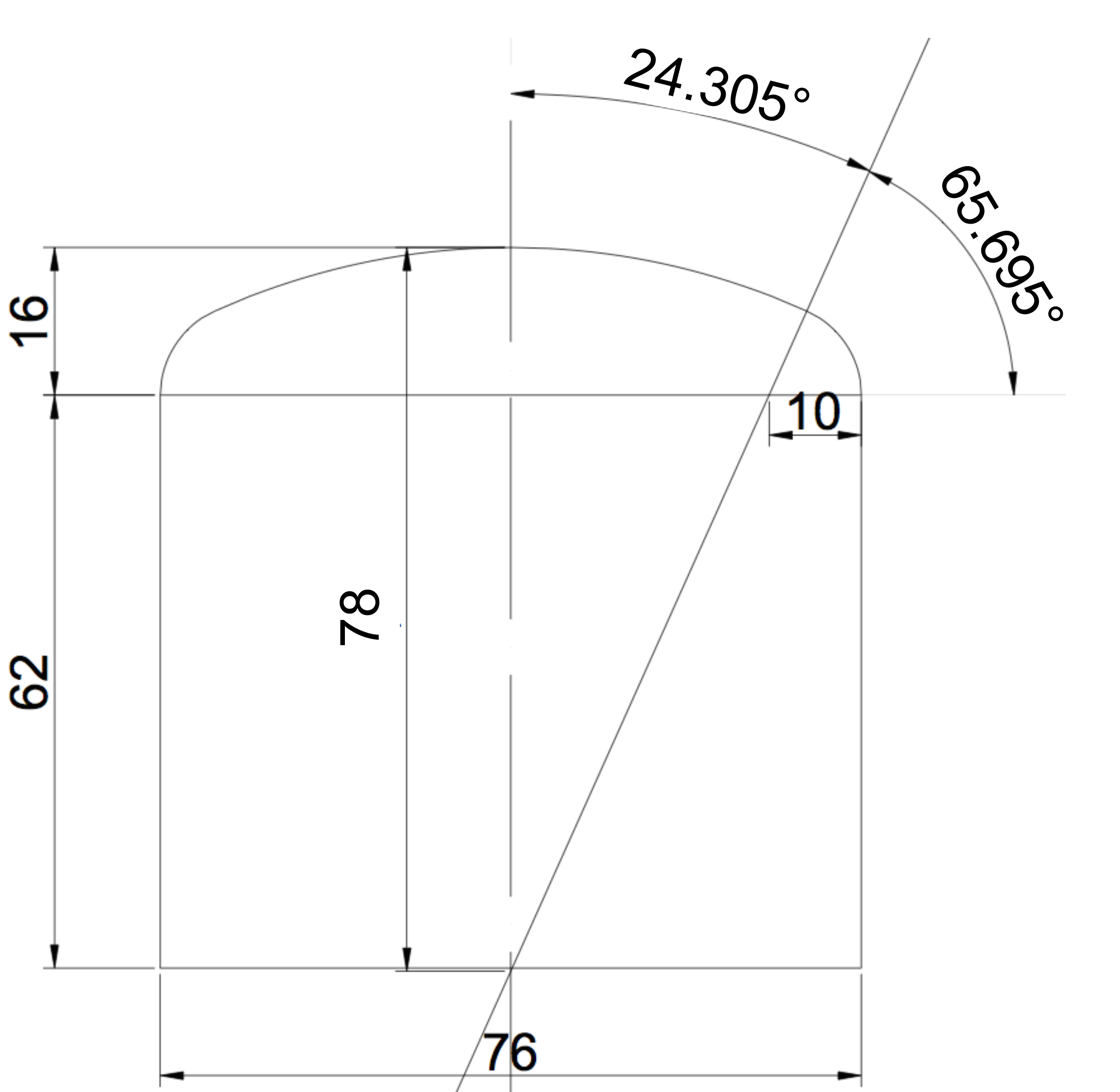}
  \caption{Cavern shape and dimension. The dimensions in the figure are in meter.
  The shape of the dome section (top portion of the cavern) is defined with 
  two different curvatures divided in 24.305\,degree and 65.695\,degree sections
  denoted in top-right of the figure.}
  \label{fig:cavern_dimension}
  \end{center}
\end{figure}

The Hyper-K cavern has 
a cylindrically shaped, barrel region 76\,meters in diameter and
62\,meters in height with a 16\,meter high dome above it.
Figure~\ref{fig:cavern_dimension} shows the cavern dimension.
The excavation volume of the cavern is approximately 0.34 Million m$^3$.
It should be noted that the dimension of the excavation volume will be
slightly larger than the detector dimension since the water
containment system, e.g.  a concrete lining, is constructed inside of
the excavated cavern surface.

%%%%%%%%%%%%%%%%%%%%%%%%%%%%%%%%%%%%%
\subsubsection{Cavern stability and support\label{sec:cavern_stability_support}}
%%%%%%%%%%%%%%%%%%%%%%%%%%%%%%%%%%%%%
The excavated rock wall is supported by rock-bolts, pre-stressed (PS)
anchors and shotcrete. A cavern structural stability analysis has been
carried out based on the geological condition obtained from the
geological surveys.
The vertical profile of rock quality is
assumed to have the uniform distribution of the CH-class, which is the
major component in the rock quality measurement.  The initial rock
stress for this analysis is based on the measured stress at 553\,m
a.s.l.  as shown in Fig.~\ref{fig:ini-stress} and the rock stress at
each depth is corrected by taking into account the depth, overburden.
The FLAC3D analysis software, which uses a finete difference method,
is adopted to perform a three-dimensional stability analysis.  The
Hoek-Brown model \cite{Hoek-1,Hoek-2,Hoek-3} is applied as a dynamic
model.  The Hoek-Brown model is the method to estimate physical
properties of rock by using results obtained from examinations of
sampled rock, and is widely used in the world.

\begin{figure}%[htbp]
  \begin{center}
  \includegraphics[width=0.38\textwidth]{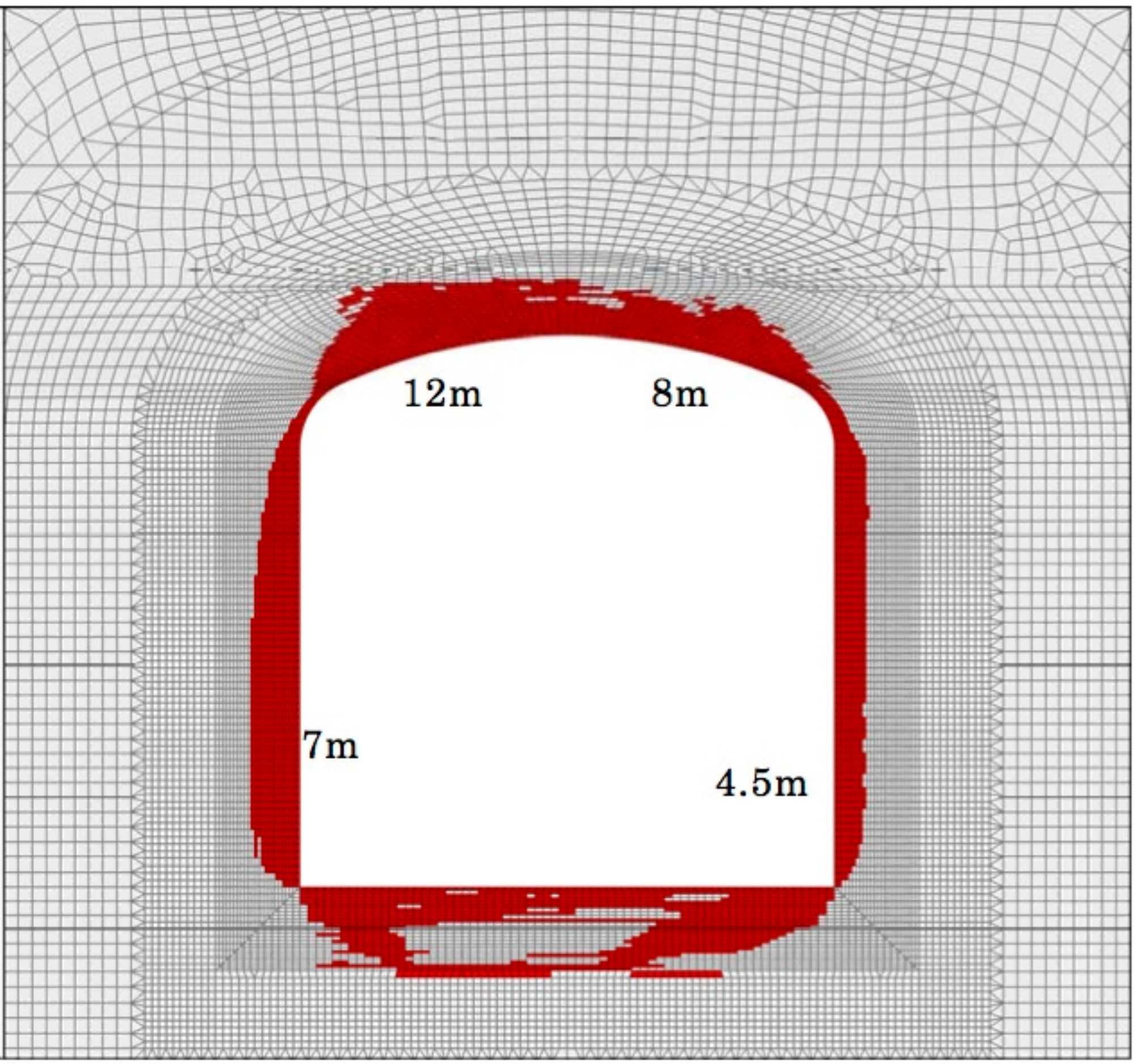}
  \includegraphics[width=0.38\textwidth]{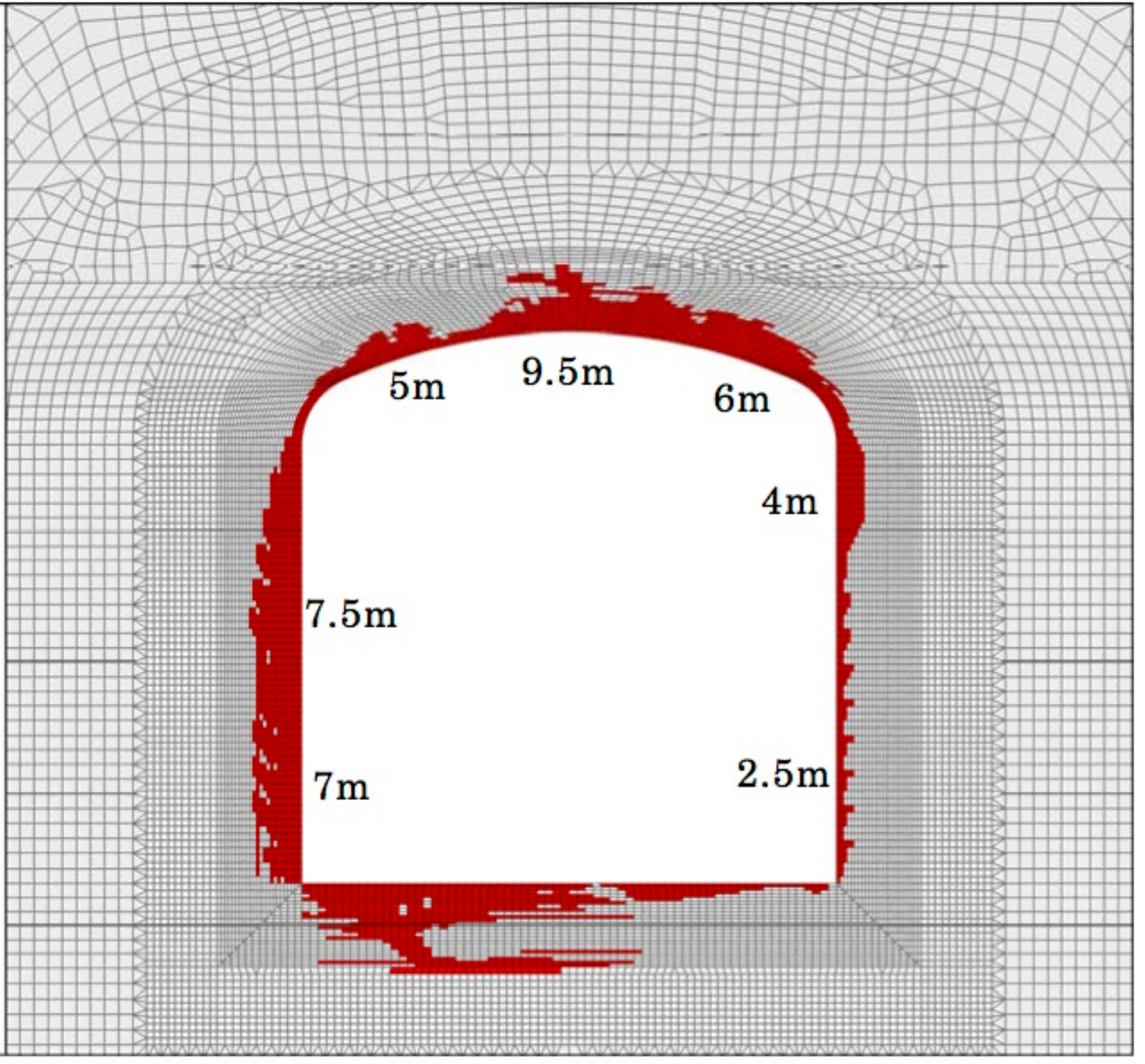}
  \includegraphics[width=0.18\textwidth]{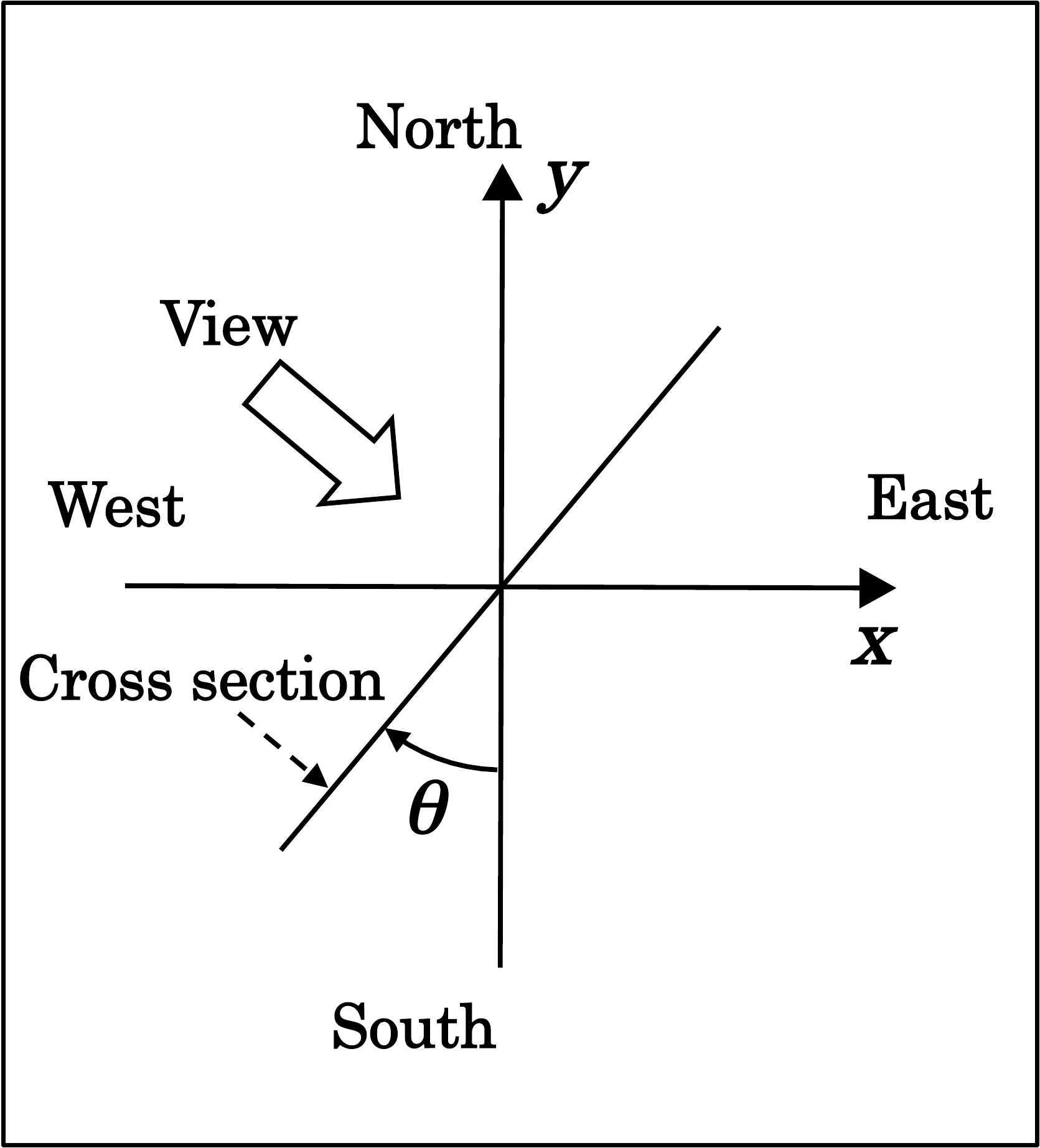}
  \caption{The plastic region at 45\,degree (left) and 105\,degree (middle)
  slices with assumption of uniform CH distribution. The right figure shows definition of the view angle.}
  \label{fig:cavern_plastic_region_CH}
  \end{center}
\end{figure}

\begin{figure}%[htbp]
  \begin{center}
  \includegraphics[width=0.48\textwidth]{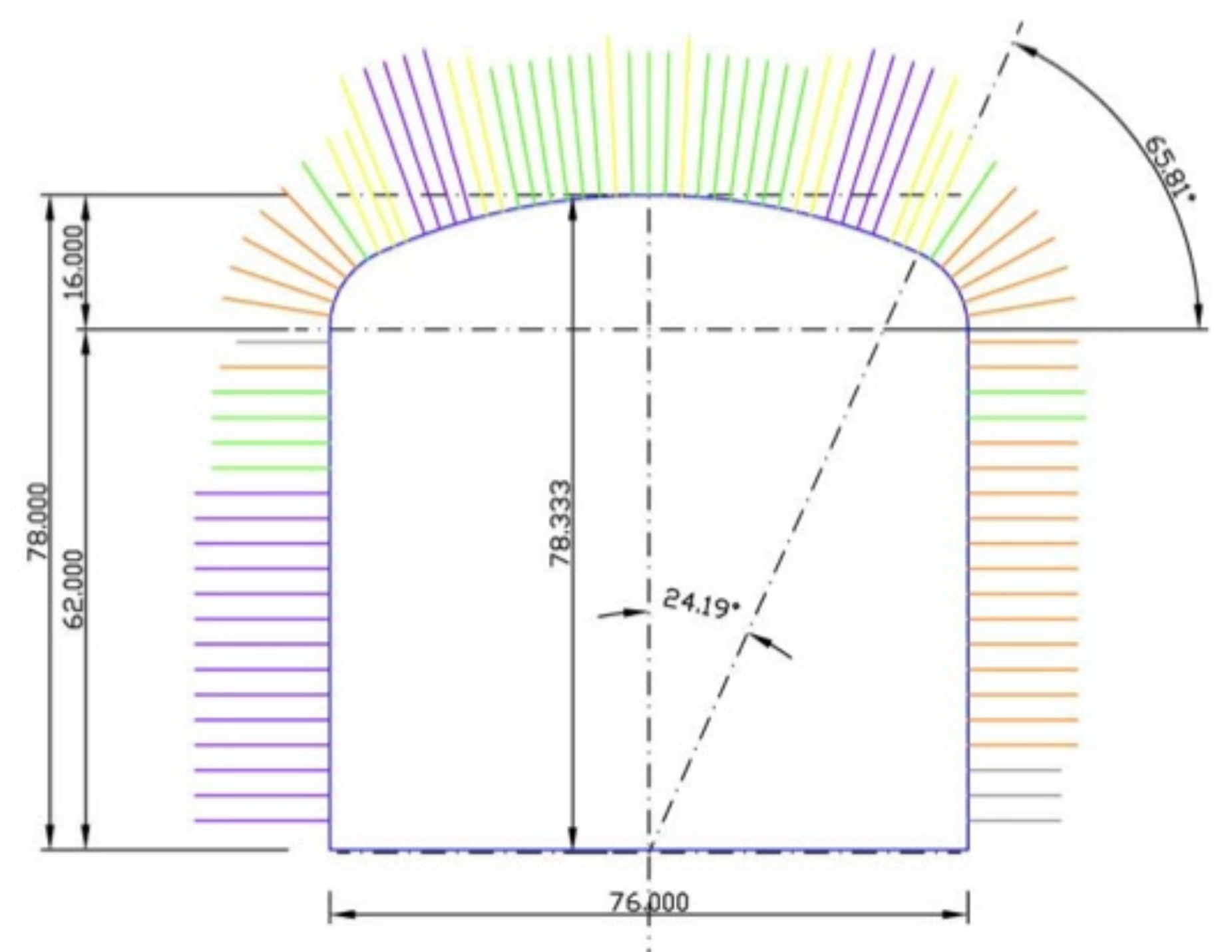}
  \includegraphics[width=0.48\textwidth]{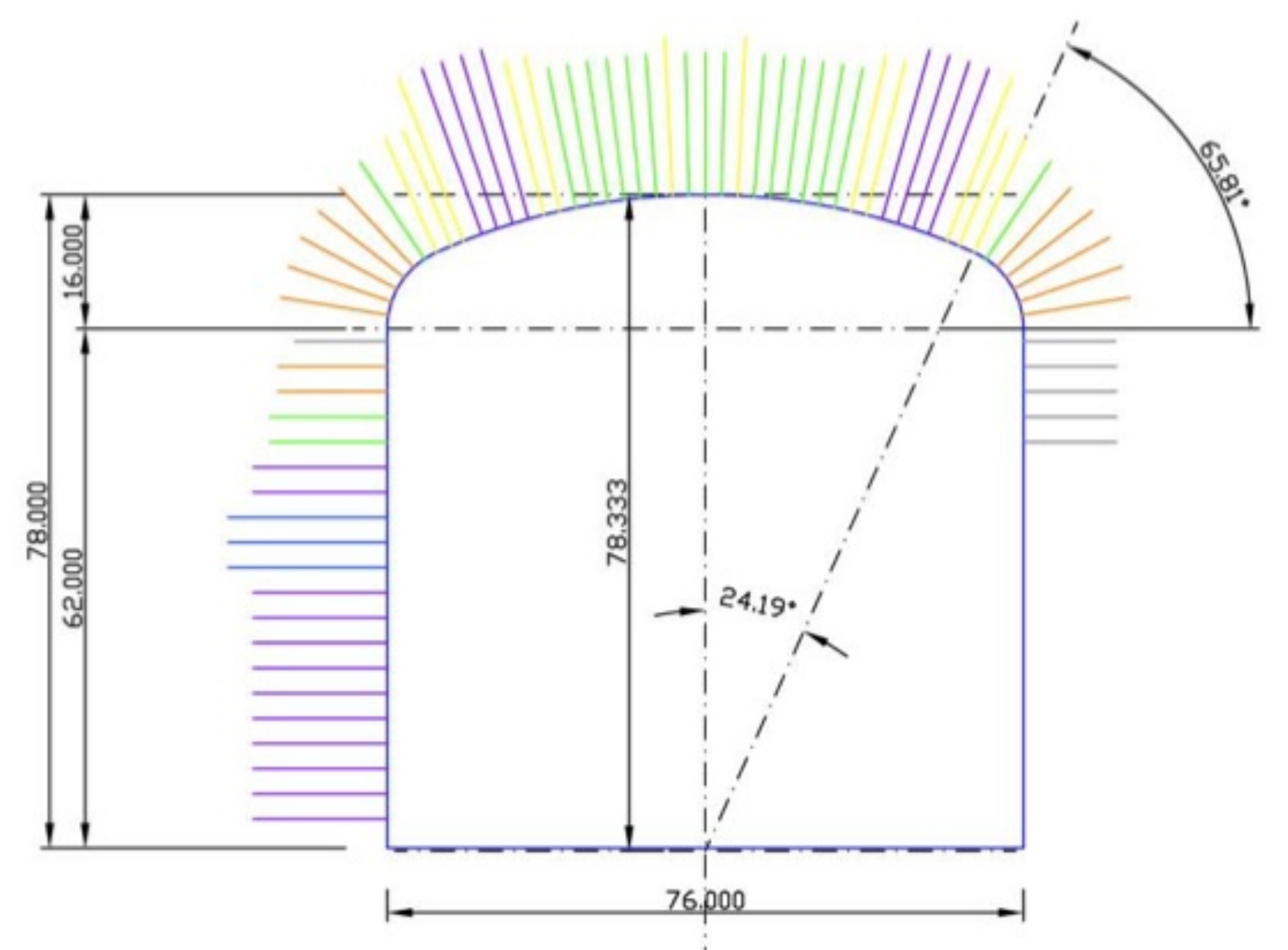}
  \includegraphics[width=0.27\textwidth]{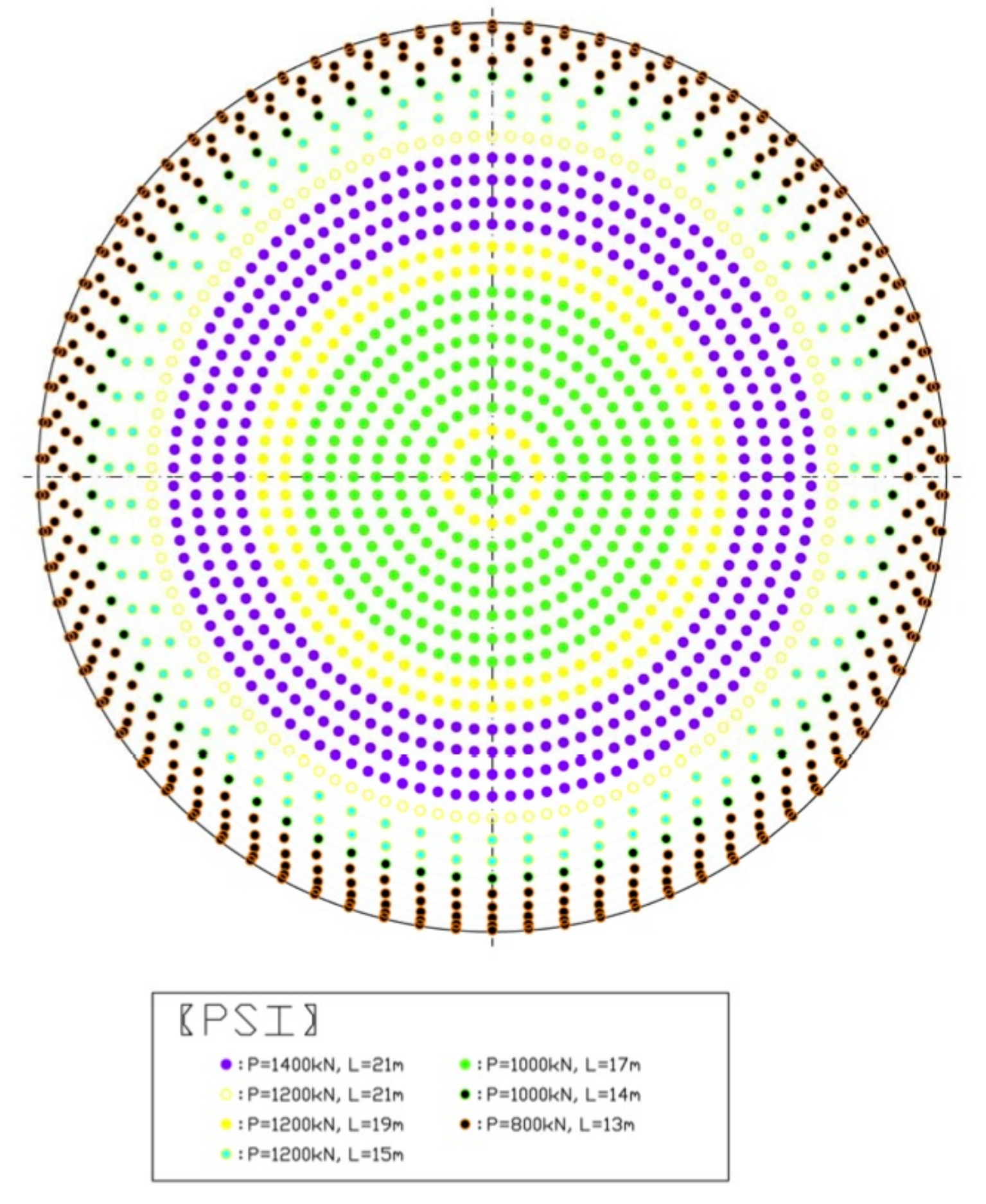}
  \includegraphics[width=0.68\textwidth]{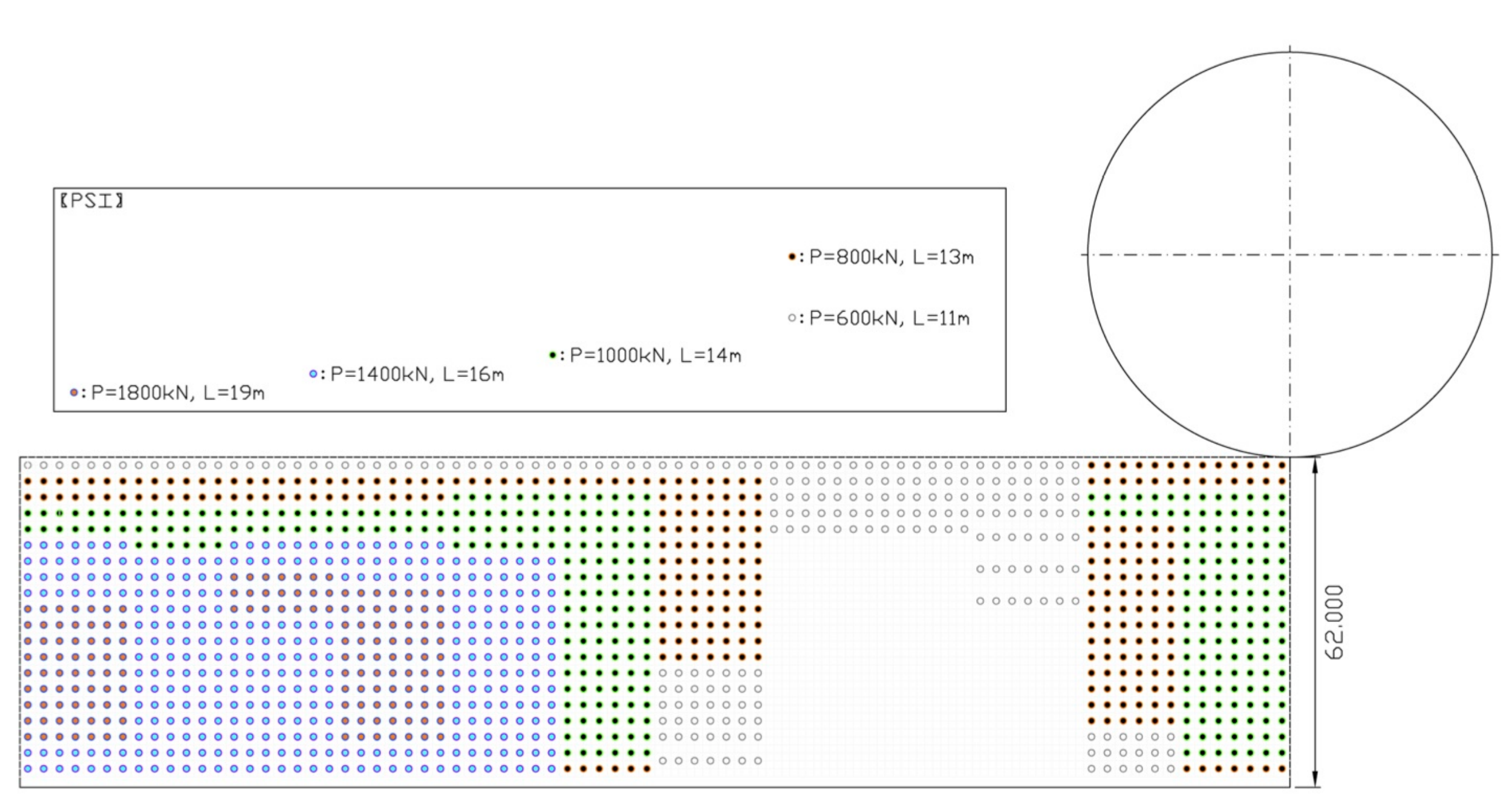}
  \caption{PS-anchors pattern at 45\,degree (top-left) and 105\,degree (top-right)
  slices with uniform CH distribution. 
  Colored lines indicate PS-anchors with different setting of initial force applied to
  PS-anchors. Bottom figures show developed figures of PS-anchors pattern for dome (bottom-left)
  and barrel (bottom-right) sections. Initial force and length of PS-anchors
  are indicated by colored circles.}
  \label{fig:cavern_anchor_pattern_CH}
  \end{center}
\end{figure}

Figure~\ref{fig:cavern_plastic_region_CH} shows the plastic region at
45\,degree and 105\,degree slices in the case of no support (i.e., no
rock-bolts, no PS-anchors, and no shotcrete).
The plastic region depth is estimated to be $\sim$2.5\,m to $12$\,m.
The variation of the plastic region depth is due to the geological
condition, e.g. initial stress direction.
Based on the plastic region obtained, the respective cavern support
(PS-anchors) patterns are considered as shown in
Fig.~\ref{fig:cavern_anchor_pattern_CH}.
The number of PS-anchors and the total length for the cavern construction are
summarized in Table~\ref{tab:PS_anchor_summary_CH}.
The total length of PS-anchors is estimated to be approximately
45\,km.

\begin{table}%[htbp]
\caption{Summary of total number of PS-anchors and total length for the cavern
excavation in case of uniform CH distribution.
\label{tab:PS_anchor_summary_CH}}
\begin{tabular}{c|r|r}
% ---------------------------------
\hline\hline
Section & \# of anchors & Total length (m) \\ \hline\hline
Dome    & 1,537 &  26,539 \\
Barrel  & 1,307 &  18,823  \\ \hline
Total   & 2,844 &  45,362 \\
\hline\hline
\end{tabular}
\end{table}

\begin{figure}%[htbp]
\centering
  \includegraphics[width=0.5\textwidth]{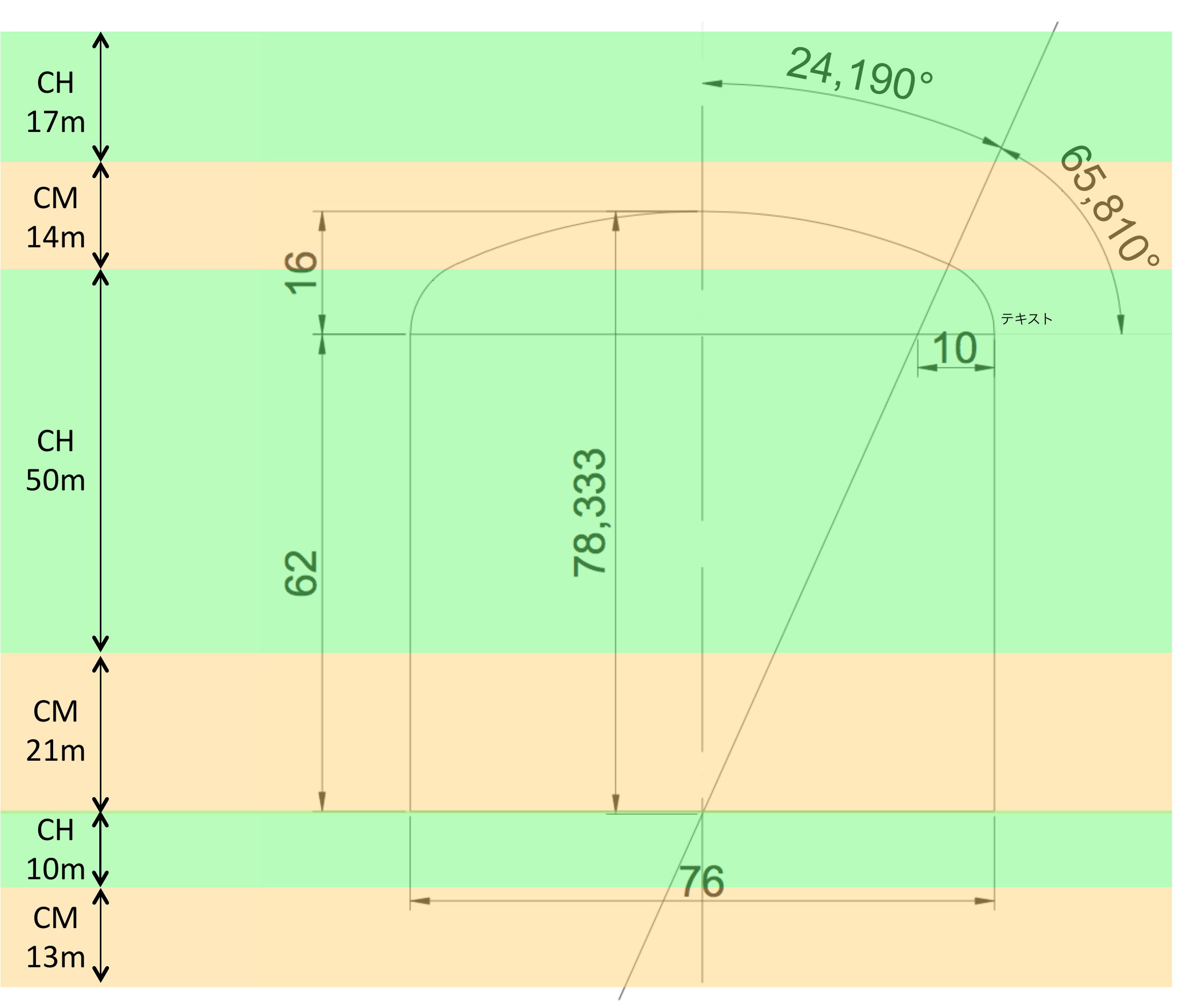}
  \caption{Assumed rock quality distribution in vertical direction.
  Rock quality in horizontal plane is assumed to be uniform.
  Two classes of the rock quality, CH and CM classes, are used for this analysis.}
  \label{fig:rock_qual_assum}
\end{figure}

\begin{figure}%[htbp]
  \begin{center}
  \includegraphics[width=0.38\textwidth]{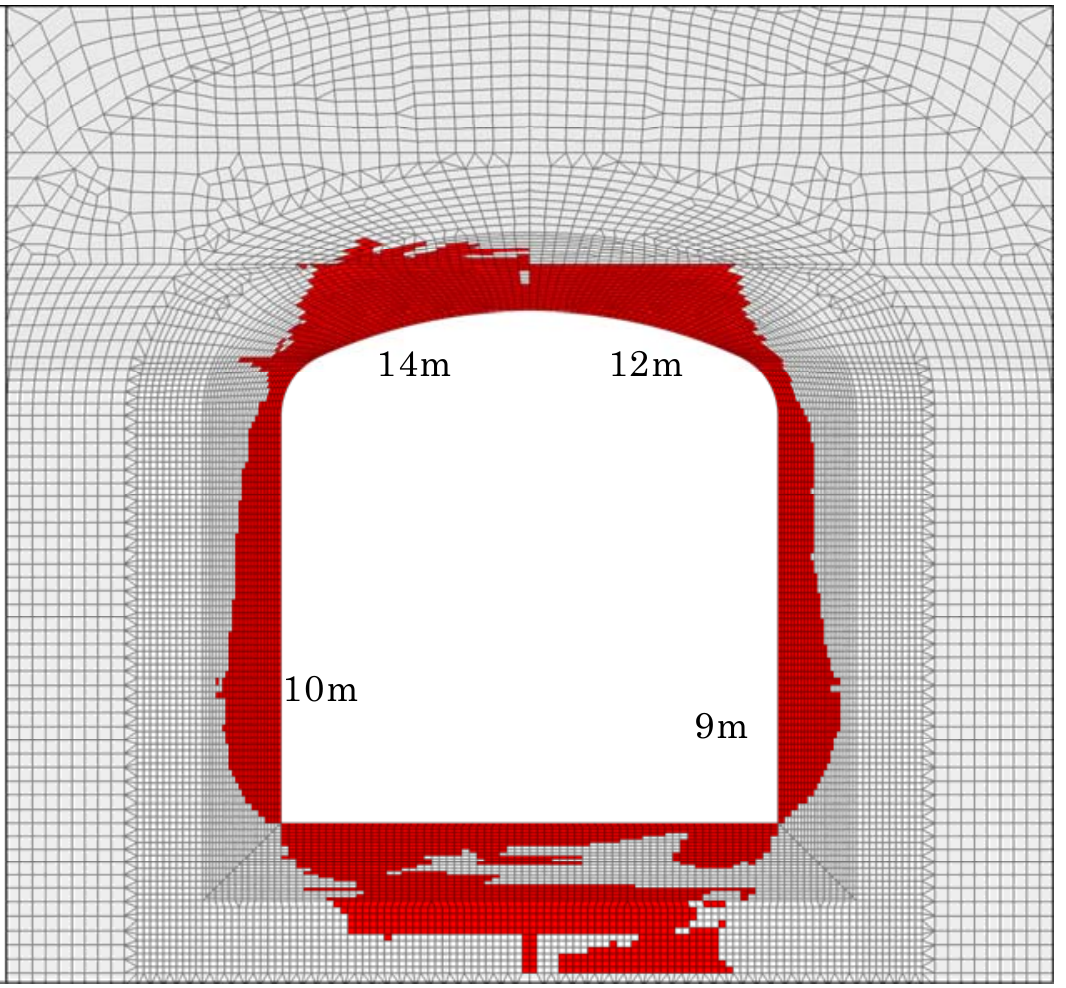}
  \includegraphics[width=0.38\textwidth]{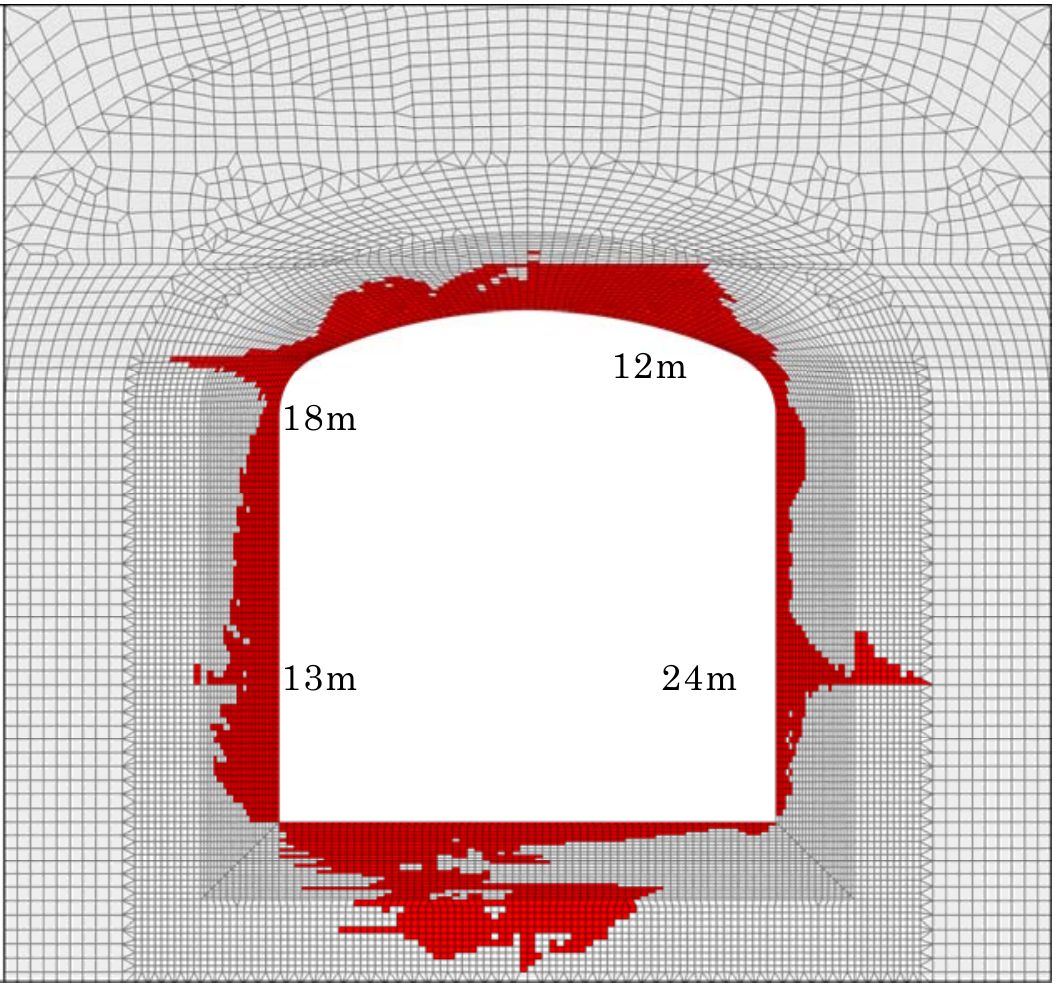}
  \caption{The plastic region at 45\,degree (left) and 105\,degree (right)
  slices with assumption of CH-CM mixed distribution.}
  \label{fig:cavern_plastic_region_CMCH}
  \end{center}
\end{figure}

\begin{figure}%[htbp]
  \begin{center}
  \includegraphics[width=0.48\textwidth]{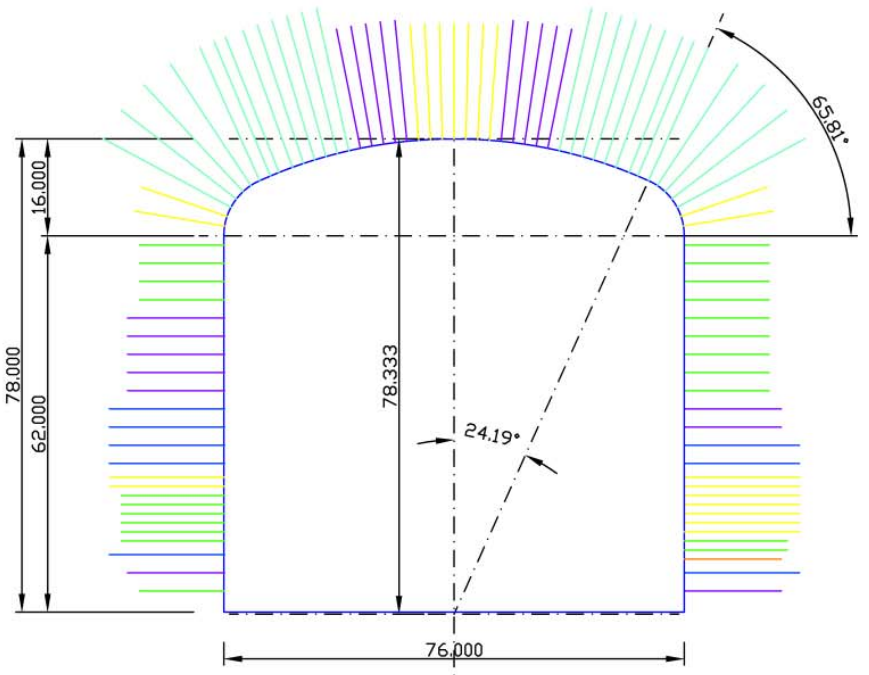}
  \includegraphics[width=0.48\textwidth]{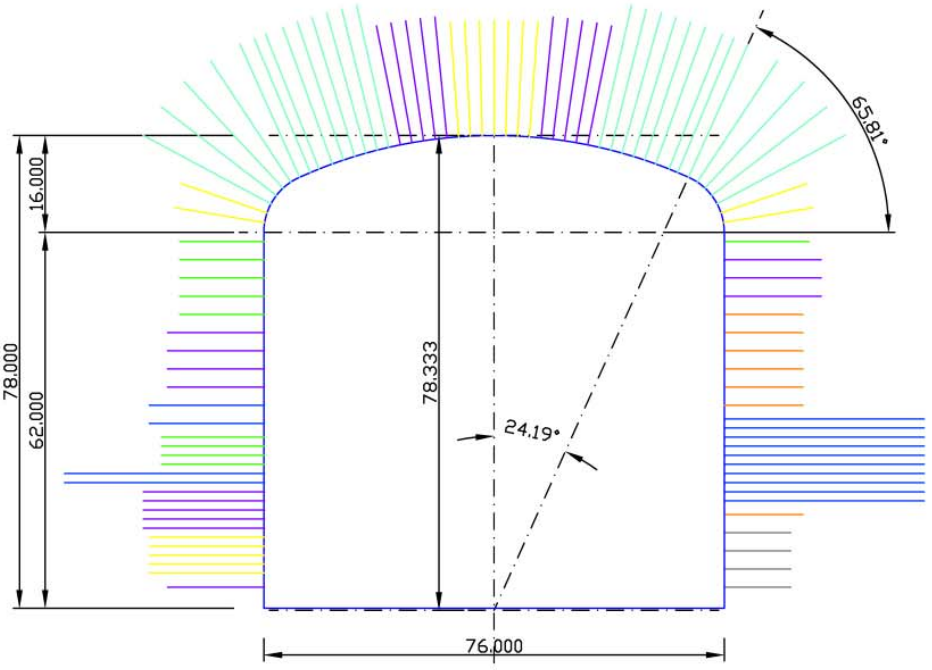}
  \includegraphics[width=0.27\textwidth]{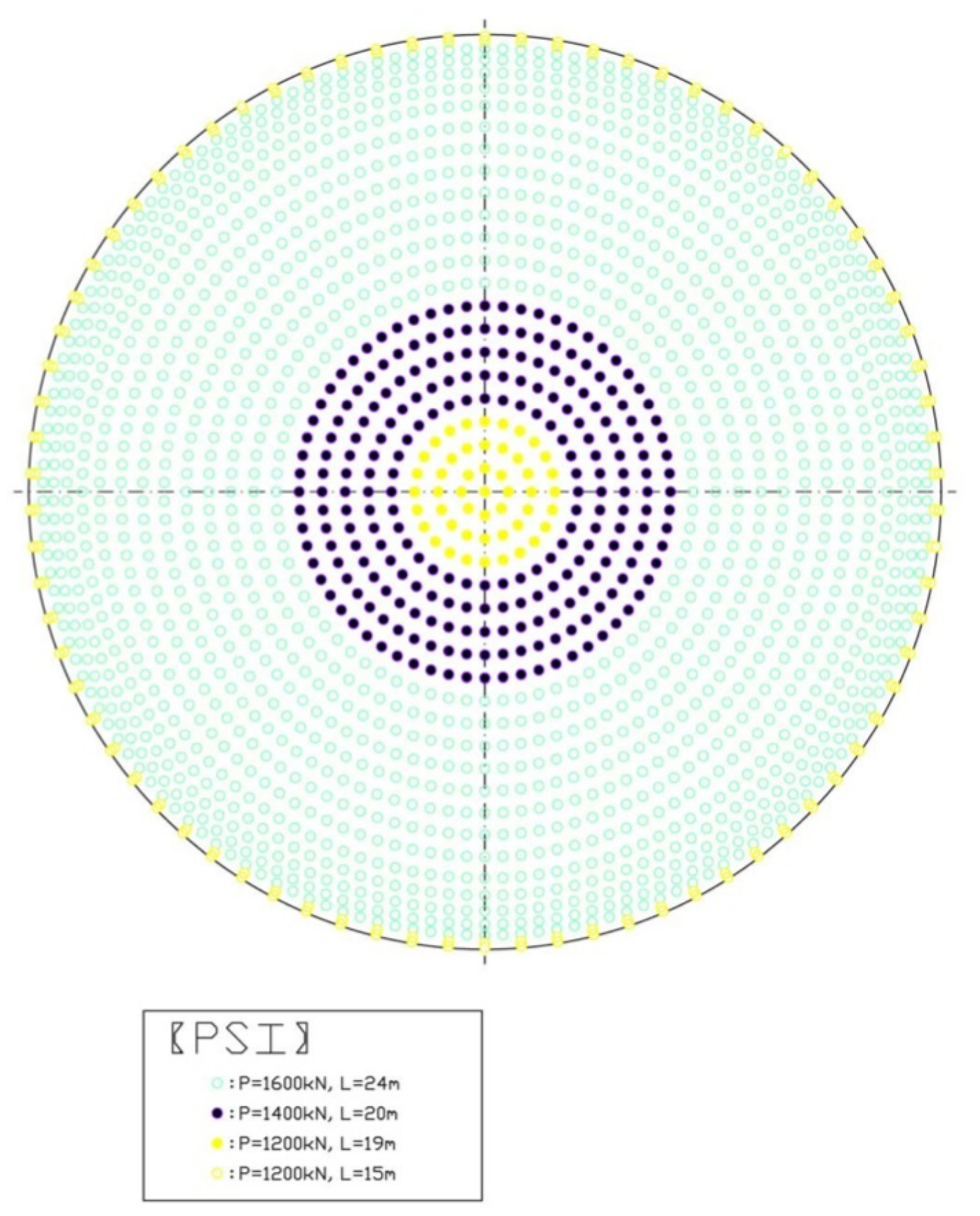}
  \includegraphics[width=0.68\textwidth]{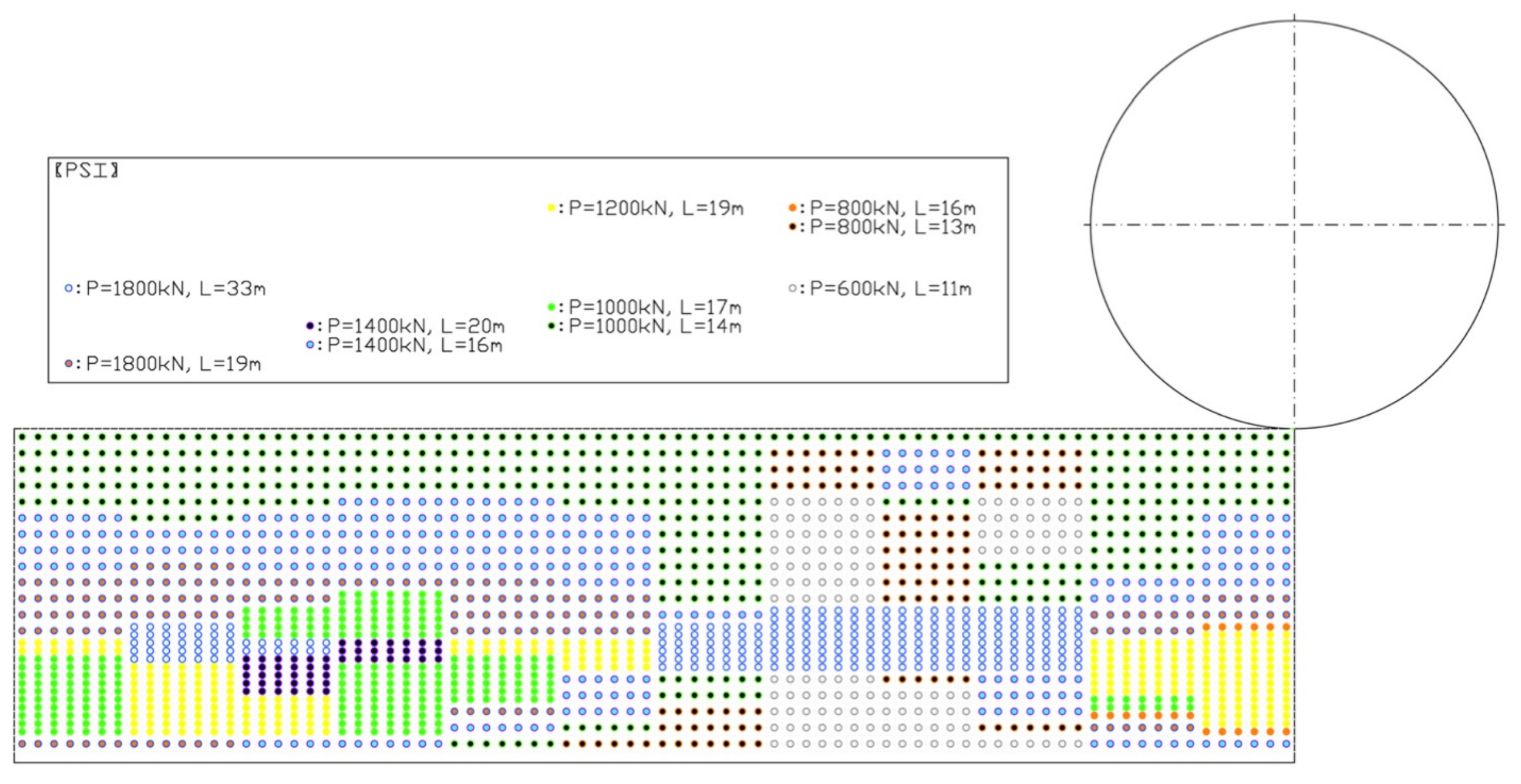}
  \caption{PS-anchors pattern at 45\,degree (top-left) and 105\,degree (top-right)
  slices with assumption of CH-CM mixed distribution.
  Colored lines indicate PS-anchors with different setting of initial force applied to
  PS-anchors. Bottom figures show developed figures of PS-anchors pattern for dome (bottom-left)
  and barrel (bottom-right) sections. Initial force and length of PS-anchors
  are indicated by colored circles.}
  \label{fig:cavern_anchor_pattern_CMCH}
  \end{center}
\end{figure}

Another analysis is performed with a different vertical profile of
rock quality, as shown in Fig.~\ref{fig:rock_qual_assum}.
In Fig.~\ref{fig:rock_qual_assum}, the fraction of rock quality is
based on the measurements of rock quality and CM-class is arranged to
the dome and bottom sections, which are structurally weaker due to its
shape, so as to perform an analysis with a severe condition.
Figure~\ref{fig:cavern_plastic_region_CMCH} shows the plastic region at
45\,degree and 105\,degree slices with the CH-CM mixed assumption in
the case of no support.
The plastic region depth is estimated to be $\sim$10\,m to $24$\,m.
In the plastic region at 105\,degree slice, two spikes in the plastic region
can be seen at the boundary between CH and CM classes. These sharp
spikes are due to the discontinuity of rock quality, which corresponds
to a discontinuous change in physical strength, and it is difficult to
correctly analyze the plastic region in such a discontinuous
condition.  PS-anchor pattern is also considered for this case, as
shown in Fig.~\ref{fig:cavern_anchor_pattern_CMCH}.
The number of PS-anchors and the total length for the cavern excavation are
summarized in Table~\ref{tab:PS_anchor_summary_CHCM}.
The total length of PS anchors is estimated to be approximately
81\,km 
in this assumption.  The difference in the total length
between two cases can be considered as an uncertainty on the
PS-anchors estimation.

\begin{table}%[htbp]
\caption{Summary of total number of PS-anchors and total length for the cavern excavation
in case of CH-CM mixed distribution.
\label{tab:PS_anchor_summary_CHCM}}
\begin{tabular}{c|r|r}
% ---------------------------------
\hline\hline
Section & \# of anchors & Total length (m) \\ \hline\hline
Dome    &  1,962 & 44,479 \\
Barrel  &  2,020 & 36,392 \\\hline
Total   &  3,982 & 80,871 \\
\hline\hline
\end{tabular}
\end{table}

While the geological surveys that have been completed already show the
feasibility of the required cavern construction, further detailed
surveys in the vicinity of the candidate site must be conducted for
the final determination of the cavern allocation and 
PS-anchors pattern before starting cavern excavation.
It should be stressed that structural stability of the detector cavern with the proposed
shape can be achieved by using existing cavern construction technologies.

A detailed plan for the additional geological surveys, which need to be done before actual
construction begins, has been established.
The surveys are divided into three steps:
{\it Step-1} begins with drilling boreholes in the vertical direction from
the existing tunnels at 653\,m a.s.l. to the cavern location where is defined
in Fig.\ref{fig:seismic_results},
{\it Step-2} carries out an `{\it in-situ} testing' to measure the physical properties
of the rock at around the cavern construction site using the existing tunnels,
{\it Step-3} is the final step of making the detailed cavern design (e.g. PS-anchor pattern)
for the actual cavern excavation, based on the geological survey results in {\it Step-1} and
{\it Step-2}.

%%%%%%%%%%%%%%%%%%%%%%%%%%%%%%%%%%%%%
\subsubsection{Cavern construction\label{sec:cavern_construction}}
%%%%%%%%%%%%%%%%%%%%%%%%%%%%%%%%%%%%%

This section describes the cavern construction
method and procedure.

The cavern excavation begins with construction of access tunnels and
approach tunnels.  The tunnels and cavern are excavated with a
blasting technique.  

\begin{figure}%[htbp]
  \begin{center}
  \includegraphics[width=0.98\textwidth]{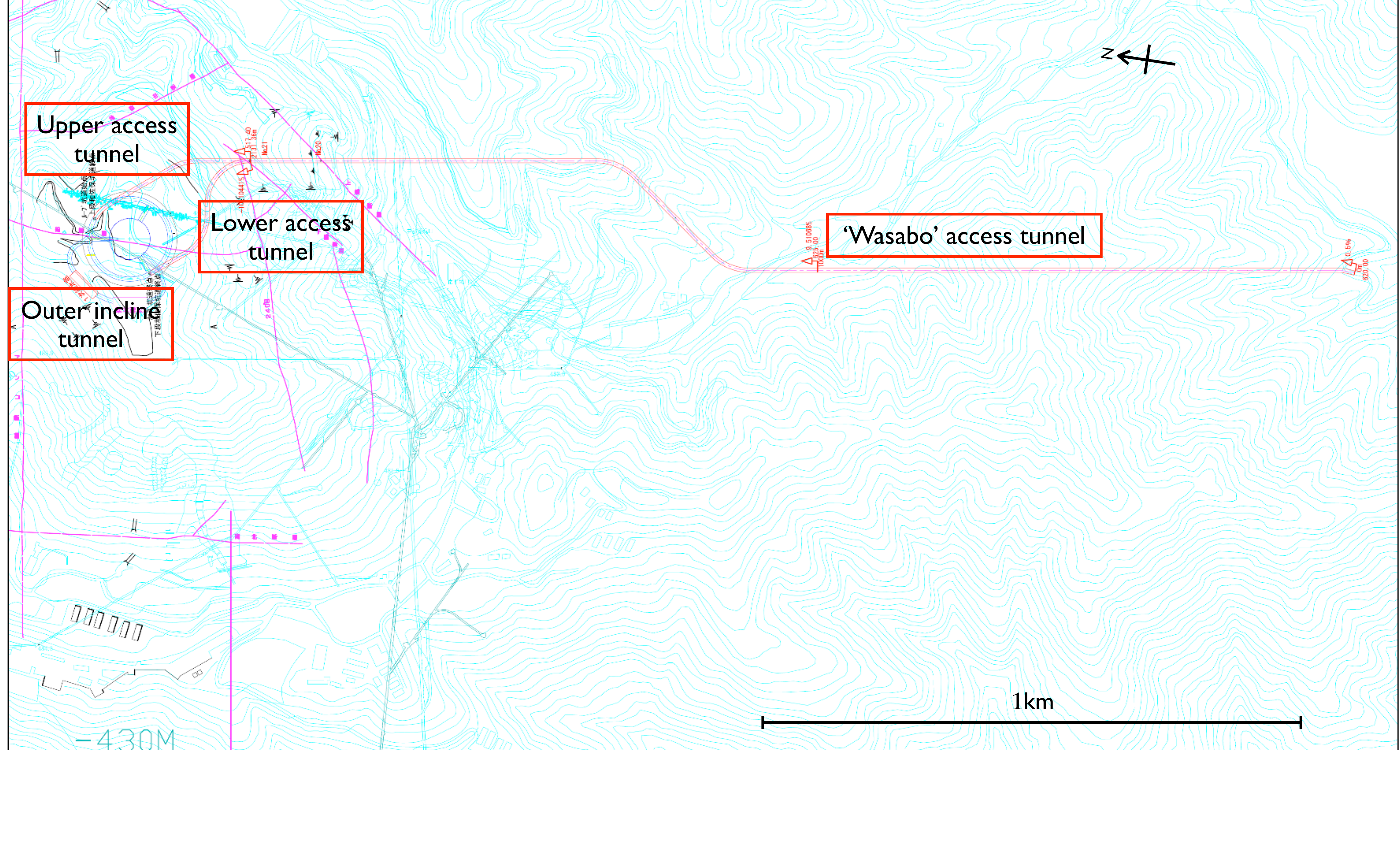}
  \caption{Overview of the access tunnels. For details of Hyper-K site, one can refer
  to Fig.~\ref{fig:approach_tunnels}.}
  \label{fig:tunnels_overview}
  \end{center}
\end{figure}

The tunnels leading from the mine entrance into the detector site
vicinity are called ``access tunnels,'' and the tunnels leading from
the access tunnels into the group of tunnels connected to the cavern
are collectively called ``approach tunnels.''
Figure~\ref{fig:tunnels_overview} shows overview of the access tunnels.
The access tunnel, named as `Wasabo' access tunnel, is also used to transport the excavated
rock to the Wasabo site where excavated rock is temporarily stored (described
in later section). 

\begin{figure}%[htbp]
  \begin{center}
  \includegraphics[width=0.9\textwidth]{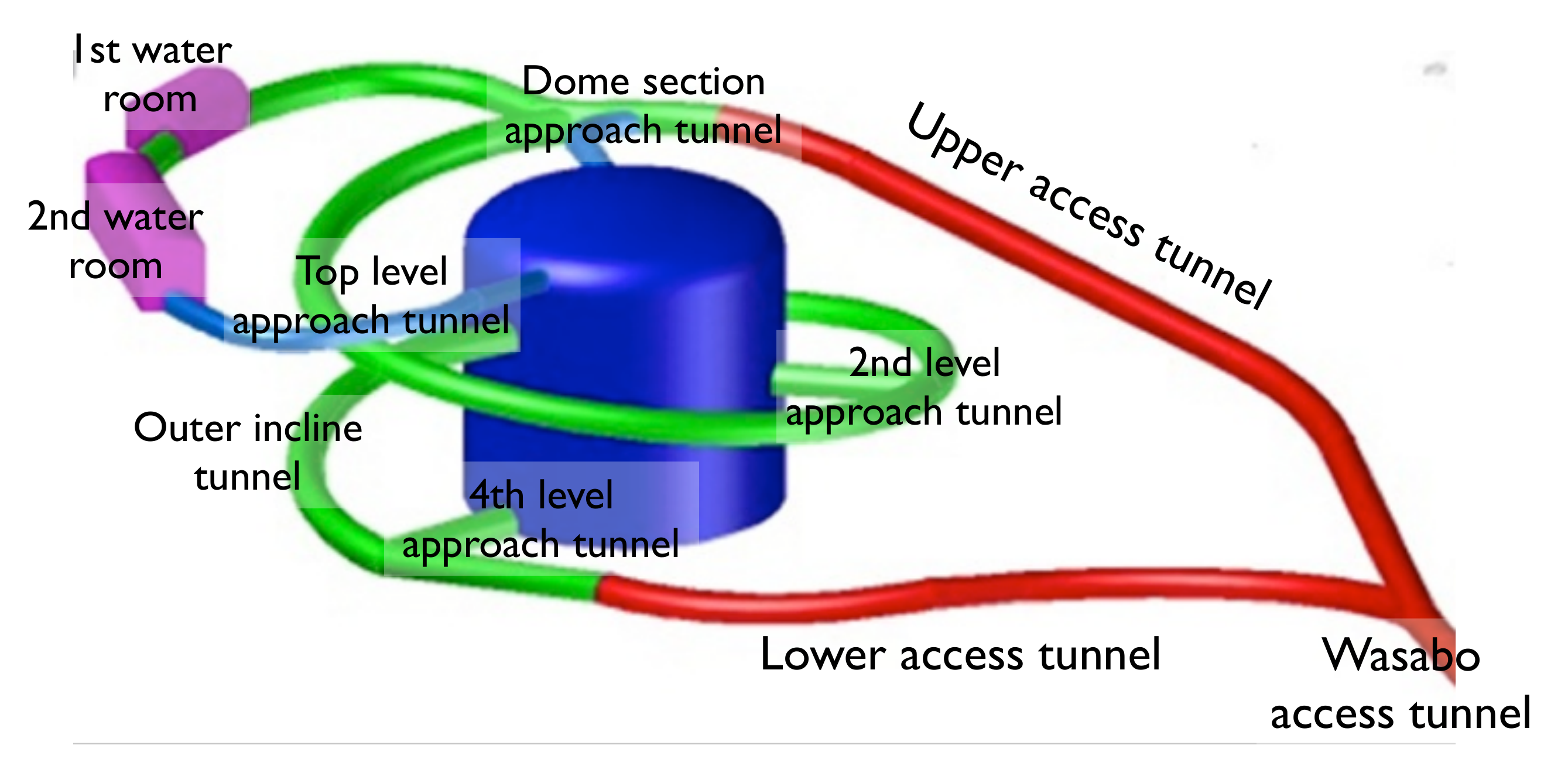}
  \includegraphics[width=0.9\textwidth]{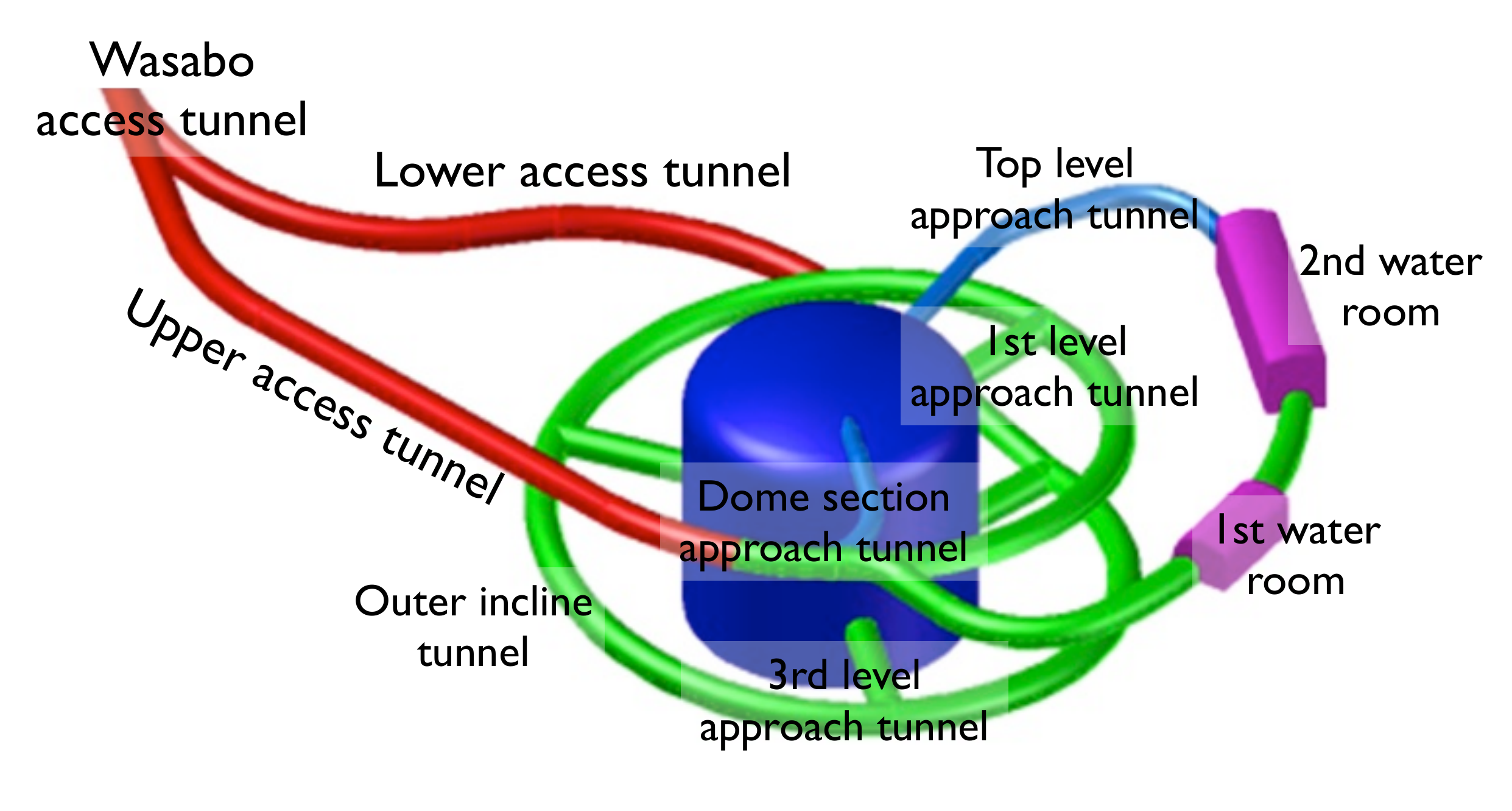}
  \caption{Layout of approach tunnels (see also Fig.~\ref{fig:approach_tunnels2}).
  The figure shows the `water rooms' as well,
  where the water purification systems are located.
  The ``electronics huts,'' (a.k.a. counting room) which stores the readout electronics
  and DAQ computers etc. (not shown in the figure), 
  will be built in the approach tunnel.}
  \label{fig:approach_tunnels}
  \end{center}
\end{figure}

\begin{figure}%[htbp]
  \begin{center}
  \includegraphics[width=0.9\textwidth]{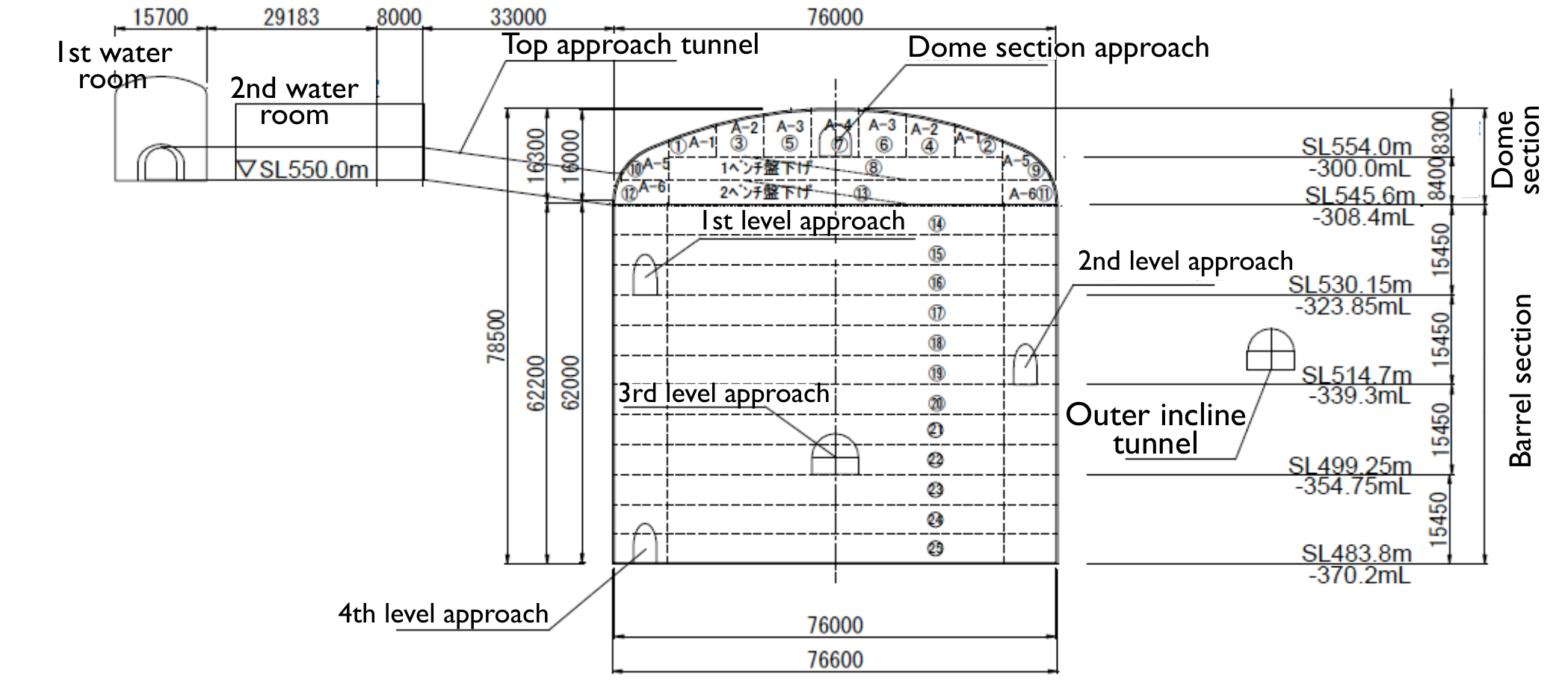}
  \caption{Layout of approach tunnels.} 
  \label{fig:approach_tunnels2}
  \end{center}
\end{figure}

The cavern is excavated from top to bottom, and there are two
general phases in the cavern construction -- excavation of ``dome''
and ``barrel'' sections.  The dome section is top portion of the
cavern, and the barrel section is vertical straight wall section of the cavern
(see Fig.~\ref{fig:cavern_dimension}).  The barrel section is further
divided into four stages.
Each stage has 15.5\,m height, and the barrel section is excavated in
stage by stage basis.  Top and bottom of each stage are connected to
approach tunnels, and excavation of each stage proceed from the top
approach tunnel to the bottom approach tunnel.
Figures~\ref{fig:approach_tunnels} and \ref{fig:approach_tunnels2}
illustrate the layout of the approach tunnels.
As shown in the figure, ``outer incline tunnel'' is helicoidally or spirally arranged around
the cavern and the outer incline tunnel works as an interface between access tunnels and
approach tunnels.
\begin{figure}%[htbp]
  \begin{center}
  \includegraphics[width=0.7\textwidth]{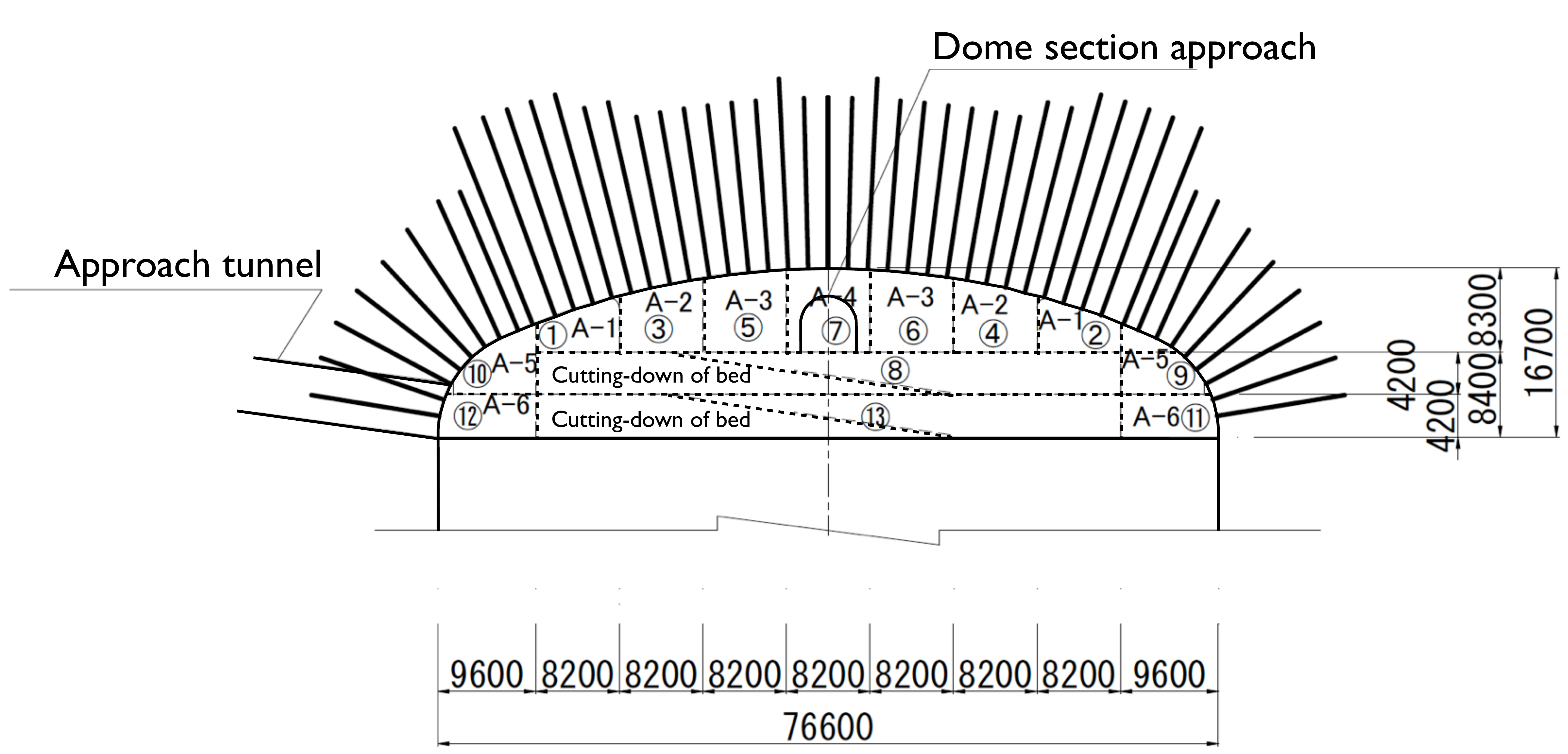}
  \includegraphics[width=0.7\textwidth]{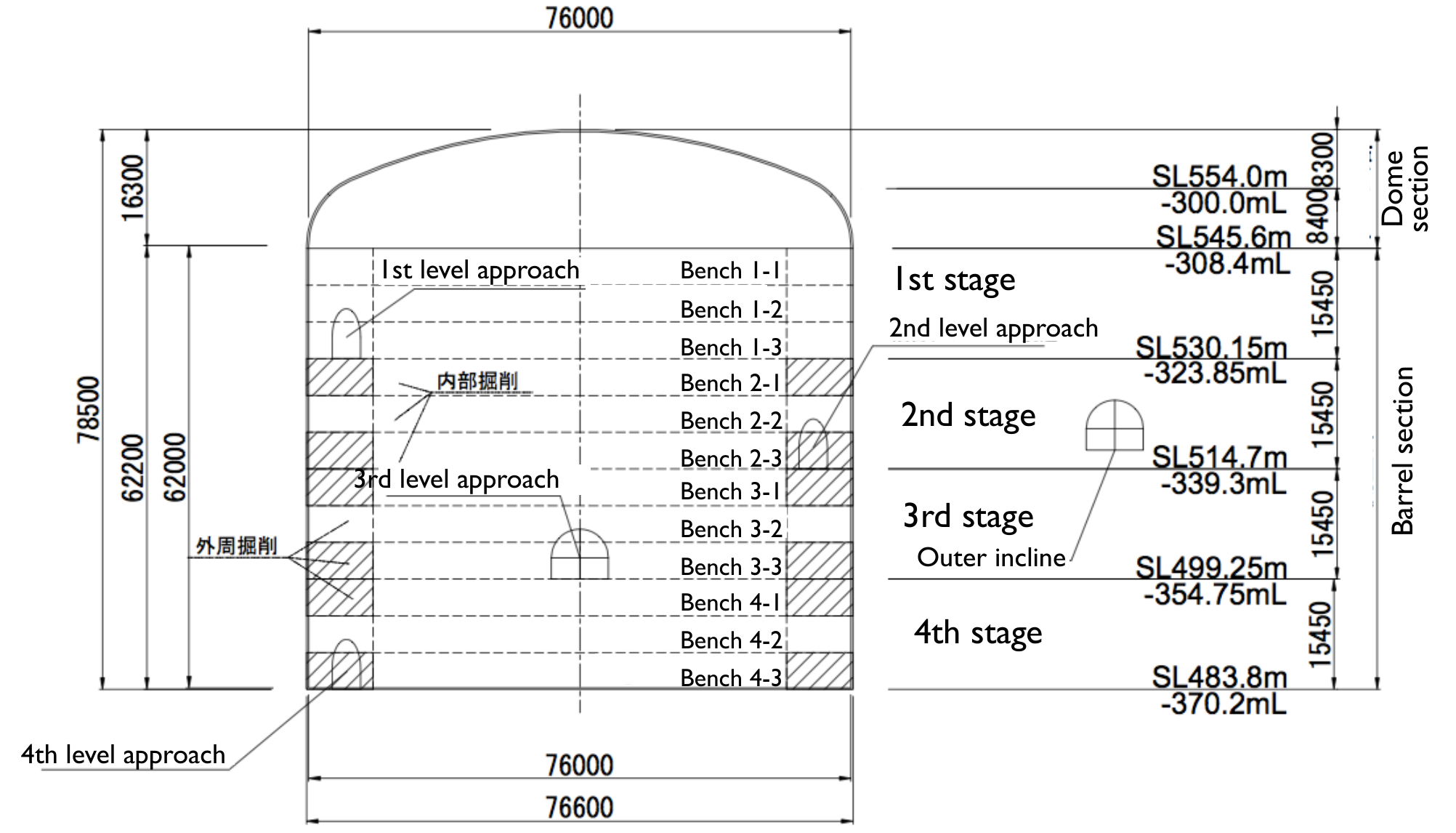}
  \caption{Illustration of the excavation steps for dome section (upper figure) and
  barrel section (lower figure).}
  \label{fig:excavation_step}
  \end{center}
\end{figure}
Figure~\ref{fig:excavation_step} shows schematic of the excavation steps
of the cavern construction.  The dome section is excavated with
thirteen steps (from section ``A-1'' through section ``A-6'').  The barrel
section is divided into four stages and each stage has three
``benches.''  The excavation of barrel section proceeds from ``bench
1-1'' through ``bench 4-3''.

%%%%%%%%%%%%%%%%%%%%%%%%%%%%%%%%%%%%%%%%%%%%%
\subsubsubsection{Excavated rock handling and disposal}
%%%%%%%%%%%%%%%%%%%%%%%%%%%%%%%%%%%%%%%%%%%%%

For the excavated rock handling and disposal, two sites are used for different purposes:
\begin{itemize}
\item{\bf Wasabo-site}\\
          Wasabo-site is an intermediate (temporary) excavated rock deposition site.
          All the excavated rock from cavern and tunnels excavation is transported and temporary
          stored at Wasabo-site.
\item{\bf Maruyama-site}\\
          Maruyama-site is the main rock disposal site for all the excavated rock
          from tunnels and cavern excavation.
          Capacity of the site is more than two Million-m$^3$.
          The total distance from Wasabo-site to Maruyama-site is about 14\,km.
\end{itemize}

\begin{figure}%[htbp]
  \begin{center}
  \includegraphics[width=0.8\textwidth]{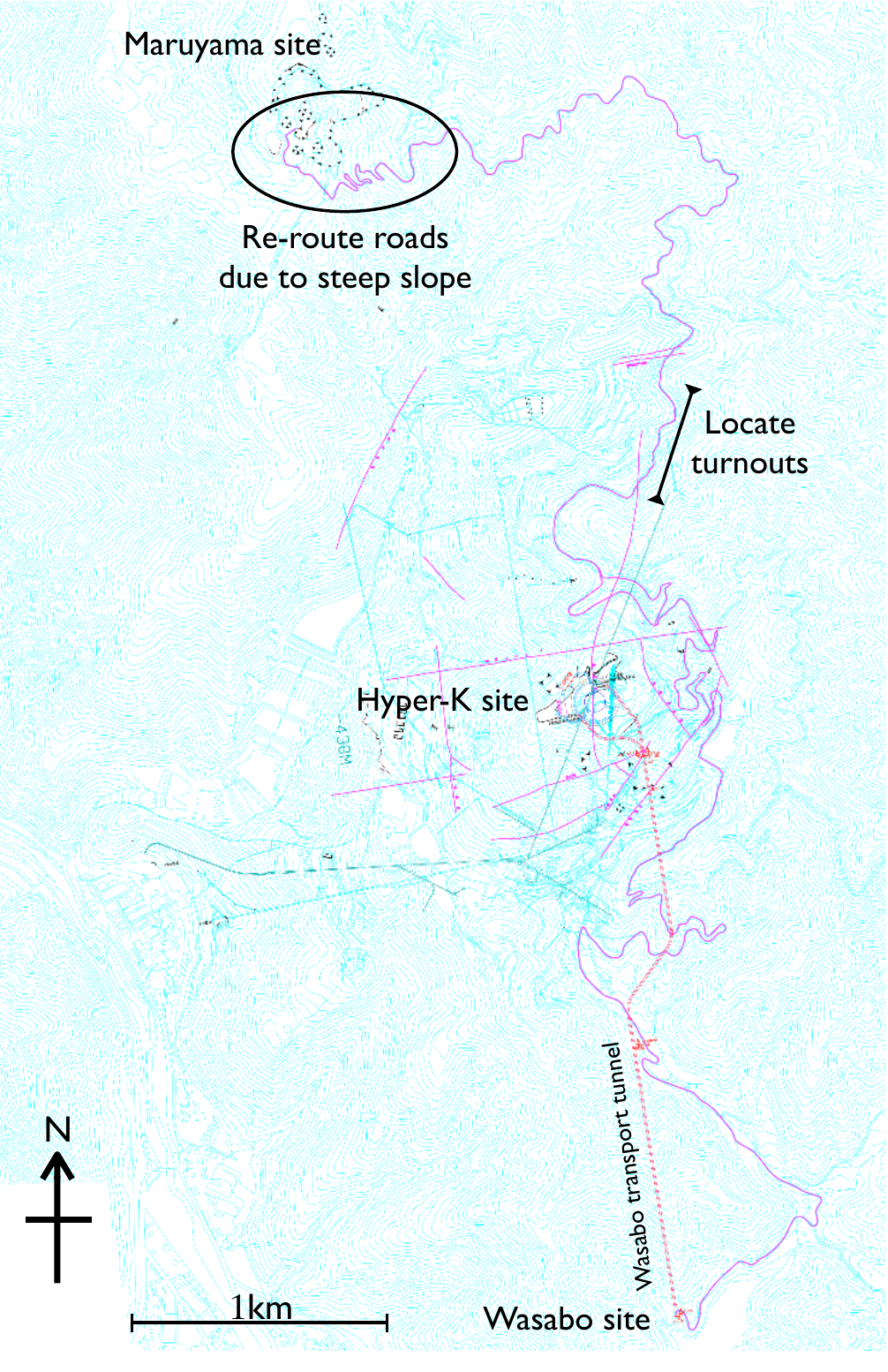}
  \caption{Overview of the excavated rock transportation routing from Hyper-K site to
  Maruyama-site.
  Magenta line denotes the transportation routing from Wasabo-site to Maruyama-site.
  }
  \label{fig:cavern_waste_rock_transportation}
  \end{center}
\end{figure}

The excavated rock will be transported with dump-trucks
from the detector site to Wasabo-site and from Wasabo-site to
the Maruyama-site.
Figure~\ref{fig:cavern_waste_rock_transportation} shows the route of
excavated rock transportation from the detector site vicinity through to
Maruyama-site.
There are the existing roads from Wasabo-site to Maruyama-site, that
can be used for the transportation of the excavated rock.  Some part of the
existing roads, however, need to be improved, e.g. widening the roads
or allocating turnouts (passing-places), in order to get a large
number of dump-trucks pass through.

\begin{figure}
    \centering 
    \includegraphics[width=1.0\textwidth]{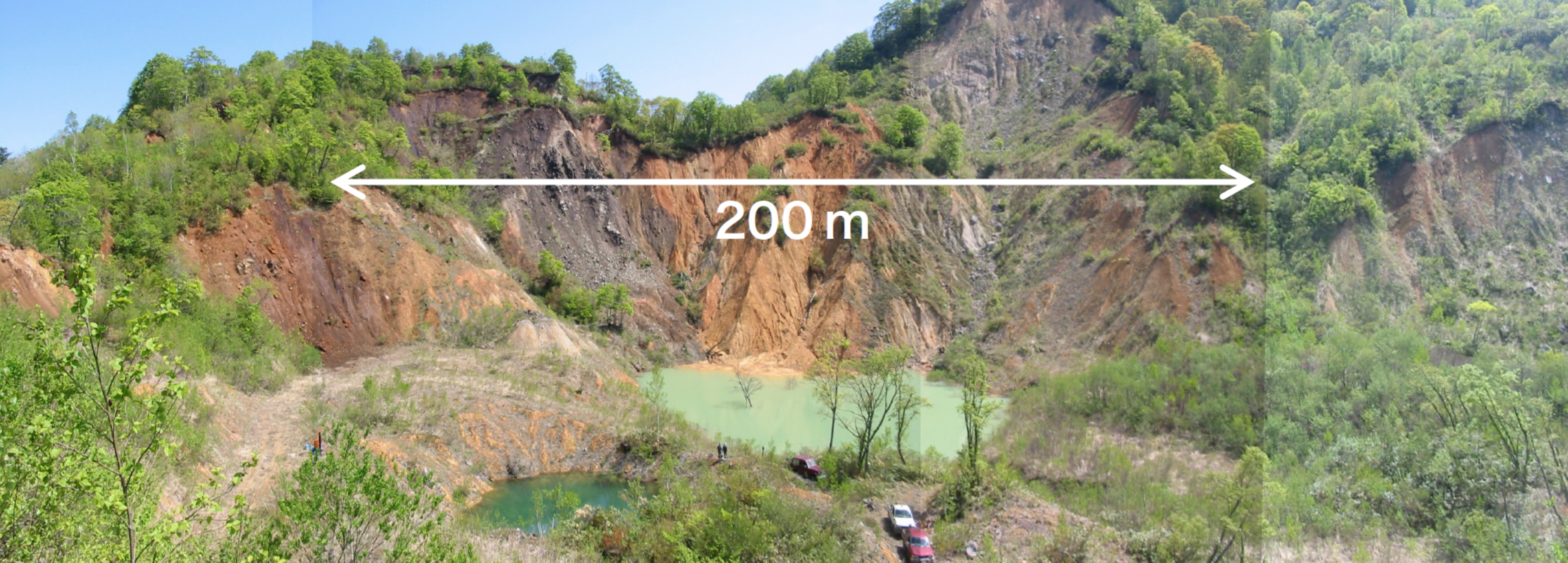}
    \caption{The rock disposal site at Maruyama.
    Its capacity is more than two Million-m$^3$.}
    \label{fig:maruyama_photo}
\end{figure}
The rock disposal site at Maruyama has a large sinkhole (see FIG.~\ref{fig:maruyama_photo})
induced by the past underground block caving.
The excavated rock with a soil volume of 570,000\,m$^3$ produced by the Hyper-K
cavern construction will be piled up on top of this sinkhole.
The base of such a sinkhole induced by block caving is mainly filled with caved waste.
To investigate the geological condition of the rock disposal place,
two vertical boring holes, No.1 and No.2, were excavated.

The No.1 borehole with a length of 24\,m was excavated near the edge of the sinkhole
to understand physical properties of the surface soil layer and its thickness.
First, the standard penetration test has shown that the thickness of
the surface layer (i.e. the depth to bedrock) was about 21\,m.
Then, core samples were subjected to various laboratory testings,
such as uniaxial compression test and consolidated-undrained triaxial compression test,
to measure their physical properties,
namely the wet bulk density, the uniaxial compression strength,
the deformation coefficient, the cohesion, and the internal friction angle.

The No.2 borehole with a length of 100\,m was excavated at the midpoint of the sinkhole
to understand the geological condition of the caved waste.
First, the core inspection has shown that
a gravel bed ($0.0\sim9.4$\,m) overlay a sandy clay layer ($9.4\sim100$\,m),
and no cavity was found in the surveyed range.
The laboratory testings of the core samples obtained by the No.2 borehole drilling
have indicated that the filling state and relative density of the sandy clay
were higher at a deeper position.
A suspension P-S velocity logging was also performed by using the No.2 borehole.
A several meters long probe, containing a source and two receivers
spaced 1\,m apart, was lowered into the borehole to a specific depth,
where the source generated seismic waves.
The elapsed time between arrivals of the waves at the receivers was used
to determine the average velocity of a 1\,m column of the ground around the borehole.
It was confirmed by the P-S logging that no cavity existed in the surveyed range
and the filling state of the ground were higher at a deeper position.

\begin{figure}
    \centering 
    \includegraphics[width=0.85\textwidth]{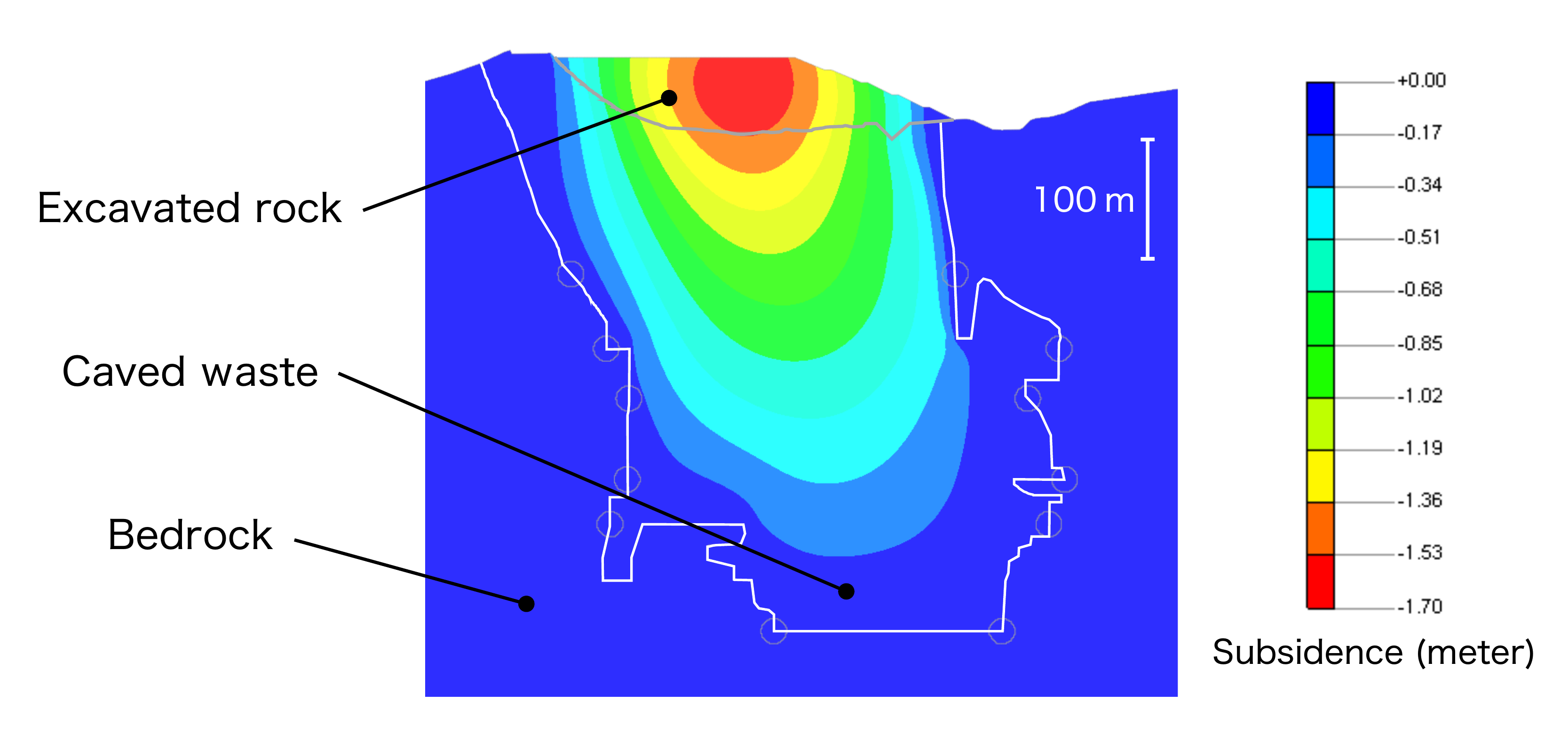}
    \caption{Ground subsidence which might arise by piling up the excavated rock
    on the Maruyama sinkhole. The figure shows a vertical section of the rock disposal place,
    and the colored region above the gray line represents the piled rock produced by
    the Hyper-K cavern excavation.
    The white line show the boundary between the region filled with the caved waste
    by past block caving and the surrounding bedrock.}
    \label{fig:maruyama_vertical_displacement2}
\end{figure}
By using geological information obtained by the boring survey,
we have performed two types of stability analyses.
One is a standard slope stability analysis considering circular slip surfaces.
In the analysis, safety factors during a normal period and those during earthquakes
were calculated both for the slope made of the piled excavated rock
and for the existing slope around the sinkhole.
The design horizontal seismic coefficient was set at 0.15
according to the technical guideline established by METI
(Ministry of Economy, Trade and Industry in Japan).
The safety factors were found to be above 1.20,
which is the reference value described in the guideline,
both for the slope of the piled rock and for the existing slope.

The other is an elasto-plastic analysis using the finite element method (FEM) of the sinkhole ground.
Distributions of stress, plastic region, and displacement were calculated
for both the situations before and after piling up the excavated rock
on top of the sinkhole, and were compared for estimating influences of the piling up.
In the analysis, physical properties of the excavated rock were set
according to results from the past boring survey at Tochibora
(i.e. Hyper-K tank construction site), and those of the caved waste and surrounding bedrock
were set based on the boring survey results at the Maruyama sinkhole.
As a result of the analysis,
FIG.~\ref{fig:maruyama_vertical_displacement2} shows the distribution of
the expected ground subsidence which might arise by piling up the Hyper-K excavated rock
on the Maruyama sinkhole.
The size of possible subsidence is expected to be at most about 1.7\,m,
which would not be difficult to deal with.
Monitoring subsidences during the rock piling work will be important for safety.

In the undergound of the sinkhole exist many mine tunnels, which were excavated
to extract ores in the past block caving.
Dead ends of such tunnels are located near the boundary between
the caved waste region and the surrounding bedrock.
Currently the caved waste including broken ores stay at the dead ends by an arch effect
and don't flow into the mine tunnels,
but it might be possible that they will start to move and flow in
due to the pressure from the excavated rock piled on the ground.
According to the FEM stability analysis mentioned above,
additional horizontal compression stresses at dead ends of existing mine tunnels,
which might cause such an inflow of rocks,
were calculated to be 0.8\,MN at most.
By constructing concrete plugs with a length of about 3\,m in front of dead ends,
such an inflow of rocks into existing tunnels can be prevented.

All the stability analyses of the Maruyama rock disposal site described in this section
were performed by assuming that
the volume of the piled excavated rock was 2,600,000\,m$^3$,
which was an estimation from the old 1 mega-ton Hyper-K design
and was much larger than 570,000\,m$^3$ from the current design.
Therefore, in the current rock disposal plan,
the safety factors of the slope stability during a normal period and during earthquakes
must be even higher,
the expected ground subsidence must be even smaller,
and the length of concrete plugs necessary to prevent possible inflow of caved wastes
to existing tunnels must be even shorter
than those described above.

%%%%%%%%%%%%%%%%%%%%%%%%%%%%%%%%%%%%%%%%%%%%%
\subsubsubsection{Cavern construction time}
%%%%%%%%%%%%%%%%%%%%%%%%%%%%%%%%%%%%%%%%%%%%%

The construction sequence has been established by making every effort
to minimize the total construction time and construction cost, for
example, 
the approach tunnels construction and cavern construction run in
parallel at different elevations, {\it etc}.

As described in the previous section, the cavern construction begins
with
Wasabo access tunnel constructions, and the
cavern constructions will follow.
The construction of the access
tunnels takes $\sim$17\,months, and the cavern
excavation takes $\sim$30\,months.
The total duration of cavern construction is estimated to be $\sim$4 years.

It should be noted that additional $\sim$10\,months will be required in the cavern
construction time if the cavern excavation volume
and/or PS-anchor supporting region is near a weak layer, such as a
fracture zone, that requires additional construction work.
Further detailed surveys in the vicinity of the candidate site is important
to minimize such slippage of the cavern construction,

\newpage
\graphicspath{{design-tank/figs/}}
\clearpage

\subsection{Water Tank}\label{section:tank}

\begin{figure}
  \begin{center}
  \includegraphics[width=0.9\textwidth]{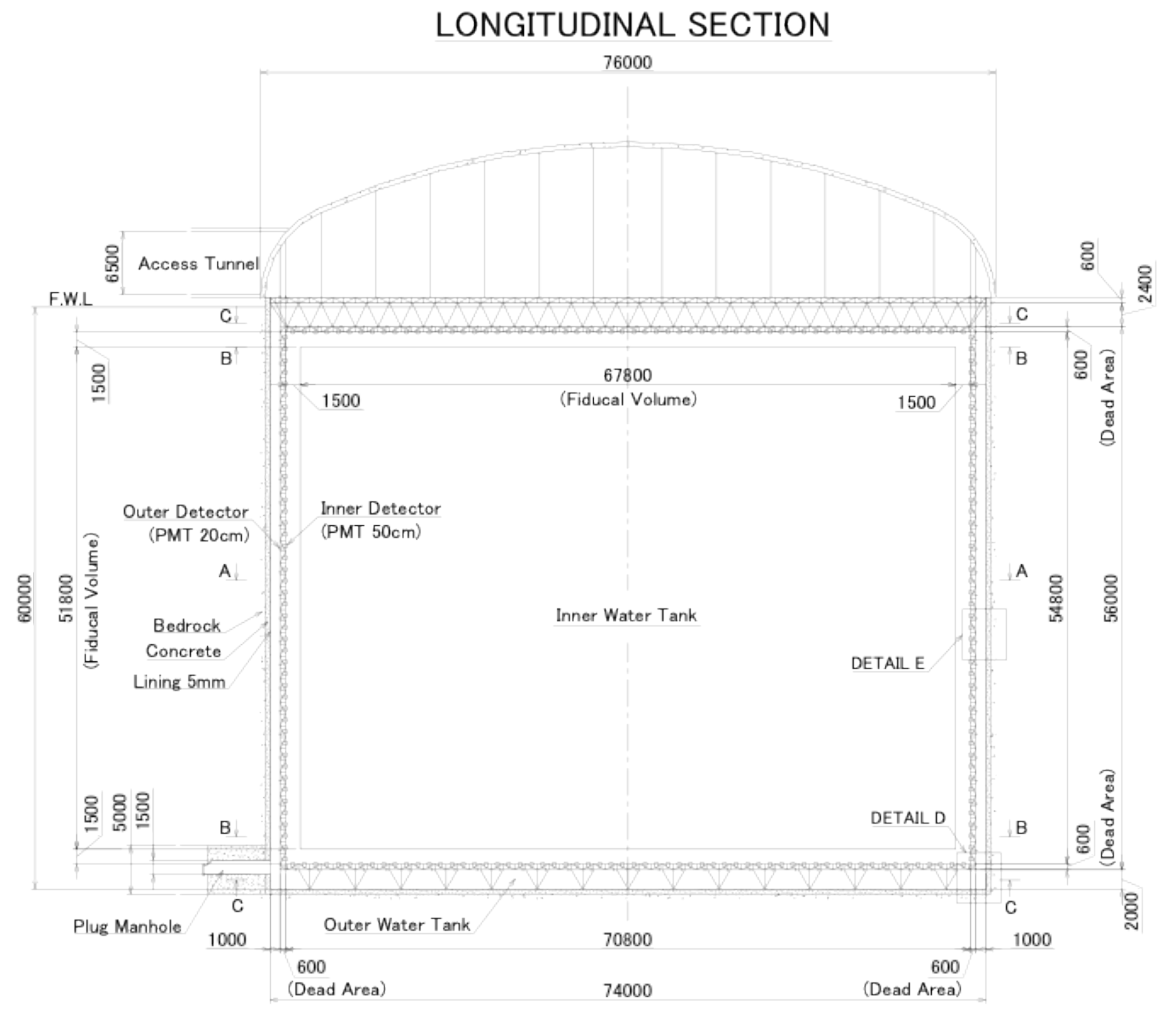}
  \caption{Schematic view of the Hyper-K water tank.}
  \label{fig:tank_schematic}
  \end{center}
\end{figure}

Figure~\ref{fig:tank_schematic} shows a schematic view of the Hyper-K water tank.
The excavated cavities for Hyper-K are lined with a watertight liner
to contain ultra-pure detector water. The liner surface bounding the
detector water is made of waterproof polyethylene sheets, which are
stuck on the backside concrete layer.
Dimensions of the cylindrical water volume is 74.0 m in diameter and
60.0 m in height, so the total water mass after the two Hyper-K tanks are
built is $0.258 \times 2 = 0.516$
million metric tons.
The water volume in each of the two tanks contains two photo-sensitive
segments optically separated by a 60 cm thick insensitive region.

The inner segment called the Inner Detector (ID)
has a cylindrical shape of 70.8 m in diameter and 54.8 m in height.
This main active volume for physics measurements is viewed by an array of
inward-facing $\sim40,000$ 50 cm $\phi$ photosensors per tank. The photocathode
coverage in the ID wall is 40\%, equivalent to that of SK-IV.
Since the new 50 cm $\phi$ PMTs (Hamamatsu R12860) developed for Hyper-K
have about twice higher single photon
detection efficiency than that of the Super-K PMTs (Hamamatsu R3600),
the overall photon detection efficiency in the ID
is almost double that of SK-IV.
A standard fiducial volume in each tank,
defined as the region inside a surface drawn 1.5 m from the ID wall,
is 0.187 million tons,
The Hyper-K total fiducial volume, $0.187 \times 2 = 0.374$ million tons,
is about 17 times the fiducial volume of Super-K. 

The outer segment monitored by outward-facing $\sim6,700$ 20 cm $\phi$
photosensors per tank is called the Outer Detector (OD), which acts mainly
as a veto for entering particles such as cosmic ray muons.
Another important task of the OD is to determine whether a particular event
occuring within the ID is fully contained in the ID or not.
The OD water thickness is 1\,m in the barrel region and 2\,m in the top
and bottom regions.
The number density of the OD photosensors is about
(1 photosensor)/(3 m$^2$), one sixth of that of the ID photosensors,
making the photocathode coverage of about 1\%.

\begin{figure}
  \begin{center}
  \includegraphics[width=0.5\textwidth]{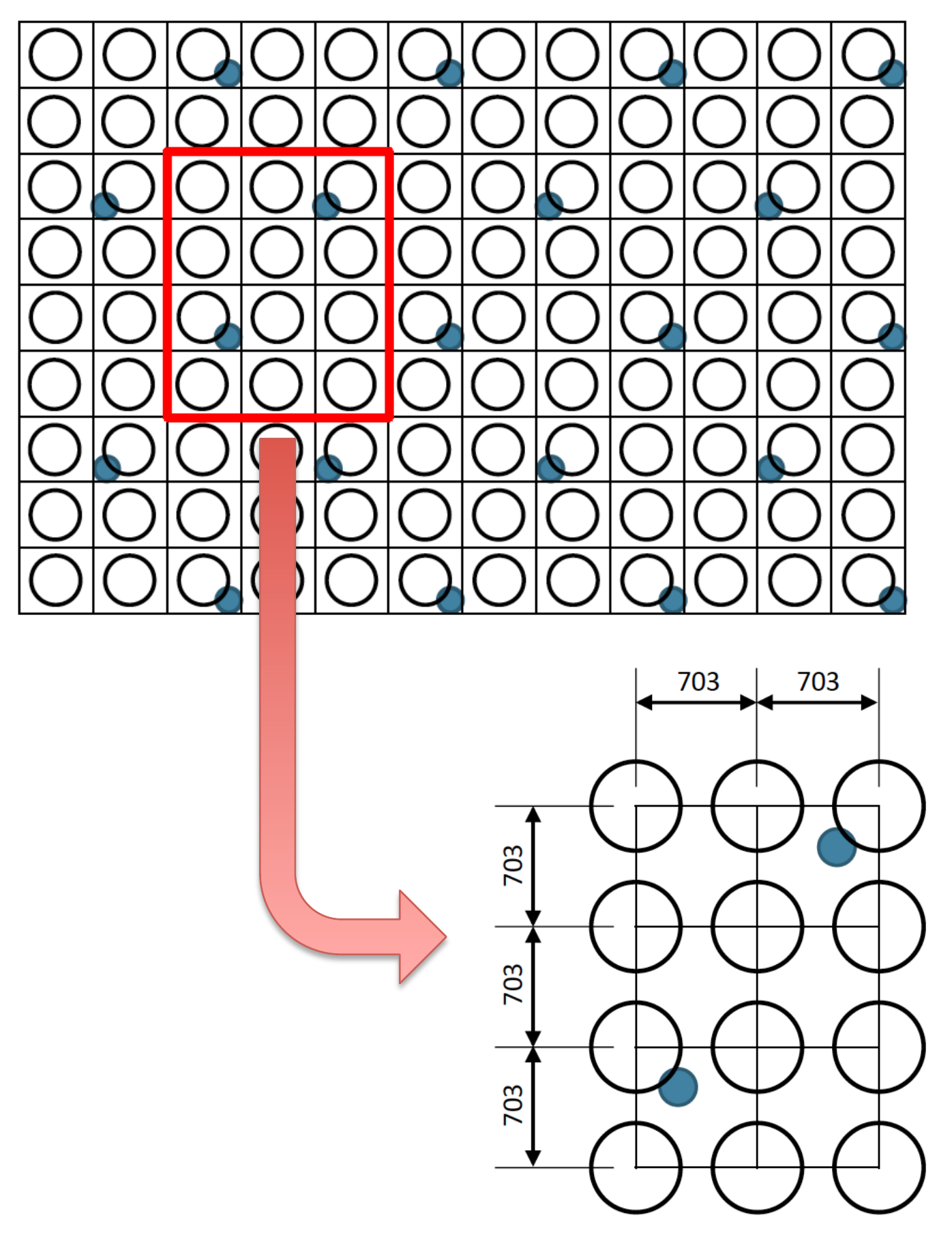}
  \caption{Arrangement of ID and OD photosensors.
The ID photosensors (open circles) are facing inward,
and the OD photosensors (blue circles) are facing outward.
See also Figure~\ref{fig:cavern_tank_boundary}.
}
  \label{fig:pmt_arrangement}
  \end{center}
\end{figure}

The photosensors for the ID and OD are mounted on stainless steel
supporting framework.
The arrangement of ID and OD photosensors is shown in
Figure~\ref{fig:pmt_arrangement}.
The grid size of the ID photosensor array is about 70 cm,
while the grid size of the OD photosensor array is roughly 2 m.
The space between the ID photosensors is lined
with opaque black sheets to prevent light leaks, while the gaps between the
OD photosensors are lined with reflective sheets to enhance light
collection in the OD. The stainless steel framework and the photosensor cables
are located in the 60 cm thick insensitive region between the ID and the OD.

Every component of the Hyper-K tank, such as the waterproof tank liner
and the photosensor supporting framework, was designed so that the tank
can be built with a construction cost as low as possible, while fulfilling
the requirements from the Hyper-K physics programs.
The design of each component and the tank construction procedure
will be described later in this section.

\subsubsection{Tank-Cavern Interface}  

After the cavity is excavated, shotcrete is sprayed onto the bedrock surface.
The tank liner, made of waterproof polyethylene sheets lying on a backfill concrete
layer, is constructed inside of the shotcrete surface.
The backfill concrete layer is reinforced by steel rods
to ensure  the integrity of the water tank.
In between the shotcrete layer and the backfill concrete liner,
waterproof sheets
are placed to prevent both the leakage of the detector water to the outside
and the penetration of external sump water into the tank.

\begin{figure}
  \begin{center}
  \includegraphics[width=0.65\textwidth]{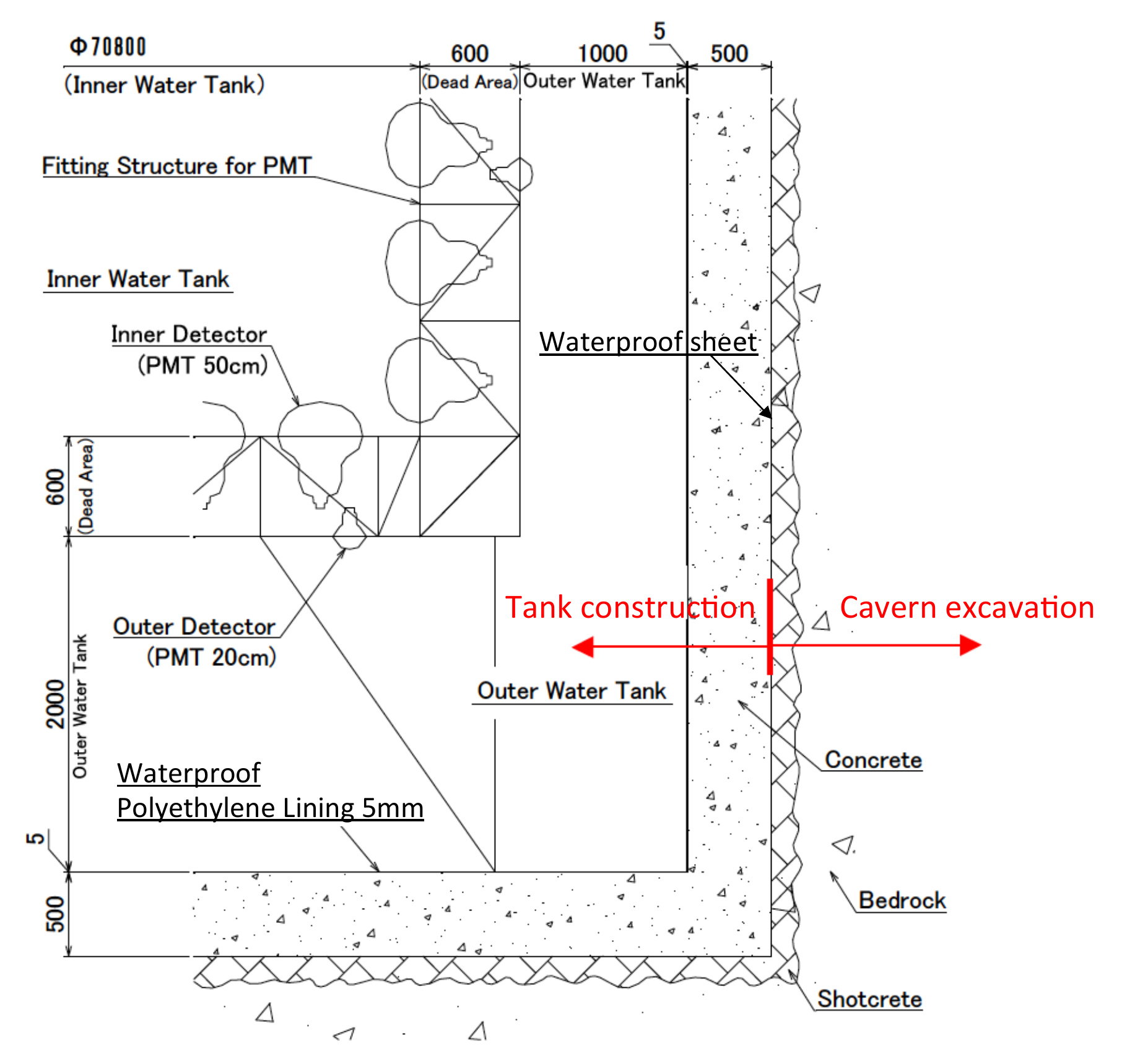}
  \caption{Boundary between the cavern excavation work and
  the water tank construction work.}
  \label{fig:cavern_tank_boundary}
  \end{center}
\end{figure}

The boundary between the cavern excavation
and the water tank construction is shown in Figure~\ref{fig:cavern_tank_boundary}.
As illustrated in the figure, the shotcrete spraying process is included in
the cavern excavation work, while the liner installation, including the middle
waterproof sheet layer, is included in the tank construction.

\subsubsection{Tank Liner}  
\label{section:tank-liner}
%
%======================%
%\subsubsection{Tank Liner}
%======================%
The lining covers inner surface of the Hyper-Kamiokande tank. It is to
contain ultra purified water (UPW) or gadolinium sulfate
(Gd$_2$(SO$_4$)$_3$) water solution inside of the tank ideally without
any leakage and without any dissolution of impurities into the
medium. Durability should be $\sim$30 years.
The lining structure is to be constructed inside of the cavern bedrock
coated with shotcrete. Between the shotcrete and the lining, a
backfill concrete is to be employed. As a former example, a
4\,mm-thick stainless steel membrane, backfilled with a reinforced
concrete, was adopted as the lining material for the Super-Kamiokande
tank~\cite{Fukuda:2002uc}.
In designing a similar lining structure for Hyper-K, we assume the
following conditions:
%---%
\begin{itemize}
%---%
\item Physical properties of the bedrock surrounding the tank are the same as those used for the Super-Kamiokande designing. For example, elastic modulus of the bedrock is 51 (20)\,GPa for non-damaged (damaged) region, respectively.
\item Physical properties of the backfill concrete are taken from {\it Standard Specification for Concrete Structure}\cite{liner:spec-concrete}.
\item The surrounding bedrock will not be displaced during/after tank construction.
\item The backwater is controlled so that there is no water pressure on the lining structure from the bedrock side. To satisfy this critical condition, location of the entire detector cavern(s) is to be chosen at the -370\,mL above the main water drainage level of the candidate mine site (-430\,mL).
%---%
\end{itemize} 
%---%

\begin {figure}[htbp]
  \begin{center}
    \includegraphics[width=0.8\textwidth]{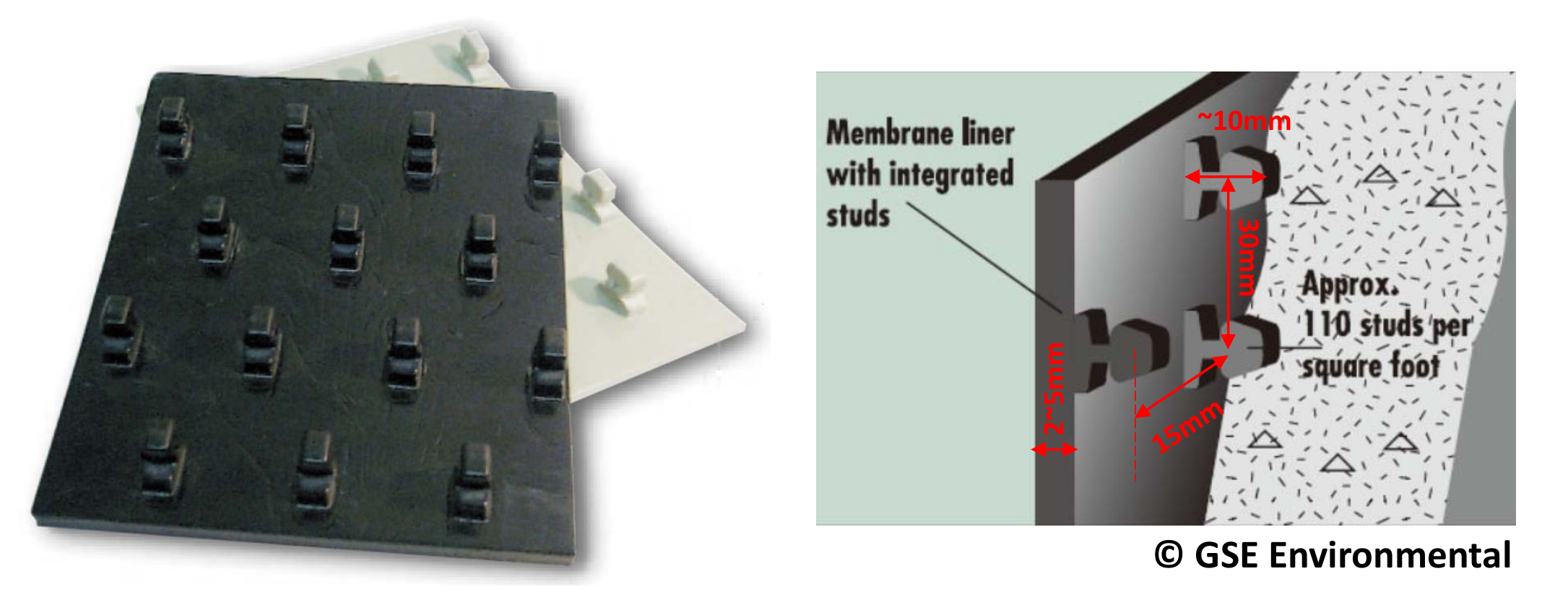}
\caption{Concrete Protective Liner made of High Density PolyEthylene, considered as the baseline tank lining material of Hyper-Kamiokande (Studliner, GSE Environmental).}
    \label{fig:liner-concept}
  \end{center}
\end {figure}

%-------------------------------------%
\paragraph{Liner sheet characteristics}  
%-------------------------------------%
As a lining material for the gigantic Hyper-K, firm adhesion to the
backfill concrete wall and enough elongation to follow possible
deficits and cracks of the concrete wall are both desirable
characteristics. To fulfill these functionalities, concrete embedment
liner, or the Concrete Protective Liner (CPL), made of High Density
PolyEthylene (HDPE), has been chosen as the baseline candidate lining
material. Figure~\ref{fig:liner-concept} shows schematic views of the
candidate CPL (Studliner, GSE Environmental). It has a
2.0$\sim$5.0\,mm thick section of HDPE with a number of studs
protruding from one side, that lock the liner into the surface of
concrete to prolong the service life of concrete structures.

\begin {figure}[htbp]
  \begin{center}
    \includegraphics[width=0.8\textwidth]{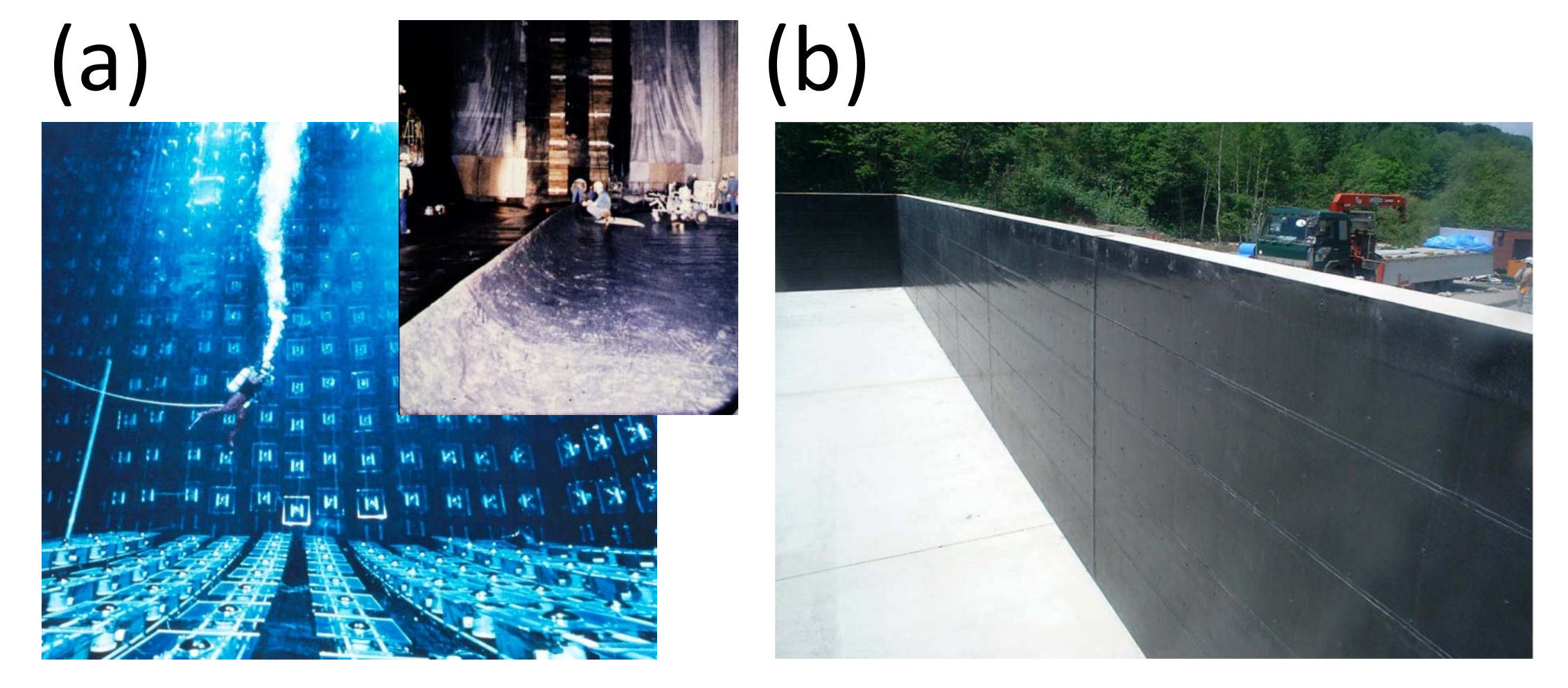}
\caption{Former examples of the HDPE lining. (a) IMB detector's 8\,kt water tank lined with HDPE geomembrane sheets (product of Schlegel Lining Technology, a precursor of GSE Environmental). (b) The CPL (GSE Studliner) applied to water-sealing walls of a large industrial waste processing trench.}
    \label{fig:liner-examples}
  \end{center}
\end {figure}

HDPE is a thermoplastic resin, a linear polymer prepared from ethylene
(C$_2$H$_4$) by a catalytic process. The absence of branching results in
a more closely packed structure with a higher density (greater than
0.94), and somewhat higher chemical resistance than Low Density
Polyethylene (LDPE). HDPE is also harder and more opaque, and it can
withstand higher temperatures (120$^\circ$\,Celsius for short periods,
110$^\circ$\,Celsius continuously). HDPE is known to maintain pure water
quality as shown later. Advantages of HDPE as the lining
material are: impact/wear resistance, flexibility (very high
elongation before breaking), good chemical resistance, very low water
permeability, good plasticity (particularly well to blow molding), and
low price. On the contrary, disadvantages of HDPE are: it may have
voids, bubbles or sink in the thick sections, poor dimensional
accuracy, and low mechanical and thermal properties.

A former example to apply HDPE liner to large water tank can be found
in the IMB detector\cite{BeckerSzendy:1992hr} as shown in
Fig.\ref{fig:liner-examples}(a). The 8\,kt water tank
(22.5$\times$17$\times$18\,m$^3$) utilized 2.5\,mm -thick double
layered non-reflective black HDPE liners, separated by a plastic
drainage grid allowing water to flow between the liners. They were
produced and installed by Schlegel Lining Technology, one of the
precursors of GSE Environmental. Figure~\ref{fig:liner-examples}(b)
shows an application of the CPL as the water-sealing walls of a large
industrial waste processing trench. Table~\ref{tab:liner-spec} shows
material parameters of the candidate CPL, GSE Studliner. It is also to
be noted that the original design for the far-site LBNE Water Cherenkov Detector (WCD) with
200\,kt volume adopted the water containment system option with
use of 1.5$\sim$2.5\,mm-thick Linear Low-Density PolyEthylene
(LLDPE) geomembrane.\cite{liner:LBNE-WCD-Golder}

\begin{table}[htbp]
\begin{center}
\caption{Material parameters, taken from specification of the candidate CPL (Studliner, GSE Environmental).} 
\label{tab:liner-spec}
\footnotesize
\begin{tabular}{lll}
\hline\hline
\multicolumn{2}{l}{Material property} & Nominal Value \\
\hline
Thickess	& (mm)		& 5.00 \\
Density	& (g/cm$^3$)	& 0.94 \\
Yield strength & (MPa)	& 15.2 \\
Elongation at break & (\%)	& 500 \\
Carbon black content & (\%) & 2$-$3 \\
Pigment content	& (\%) & 1.5$-$2.5 \\
Notched constant tensile load & (hours) & 400 \\
Thermal Expansion Coefficient & (C$^\circ$) & 1.20E-04 \\
Low temperature brittleness    & (C$^\circ$) & -77 \\
Dimensional stability in each direction & (\%) & $\pm$1.0 \\
Water vapor transmission  & (g/m$^2$/day) & $<$ 0.01 \\
Typical roll dimension & (m) & 2.44(W)$\times$59.73(L) \\
\hline\hline
\end{tabular}
\end{center}
\end{table}

\begin {figure}[htbp]
  \begin{center}
    \includegraphics[width=0.8\textwidth]{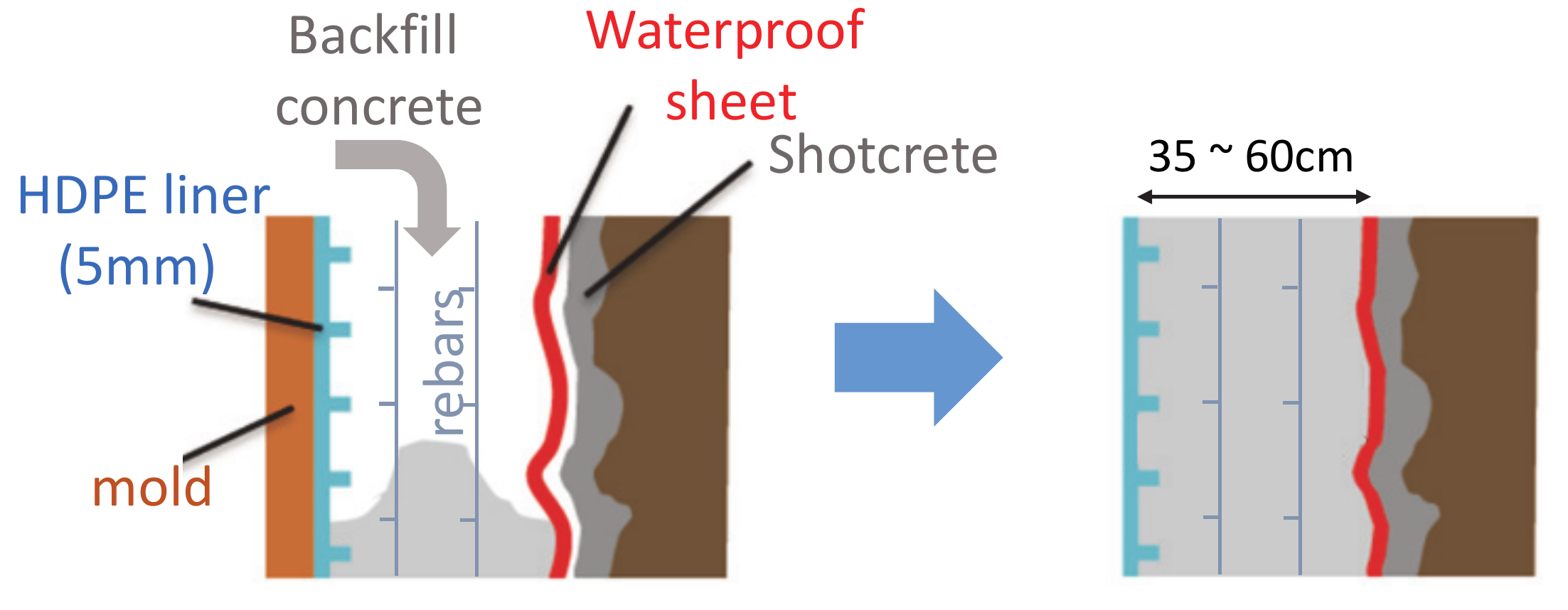}
\caption{Schematics of the planned procedures for CPL installation into the cavern 
         with shotcrete surface.}
    \label{fig:liner-install}
  \end{center}
\end {figure}
\begin {figure}[htbp]
  \begin{center}
    \includegraphics[width=0.8\textwidth]{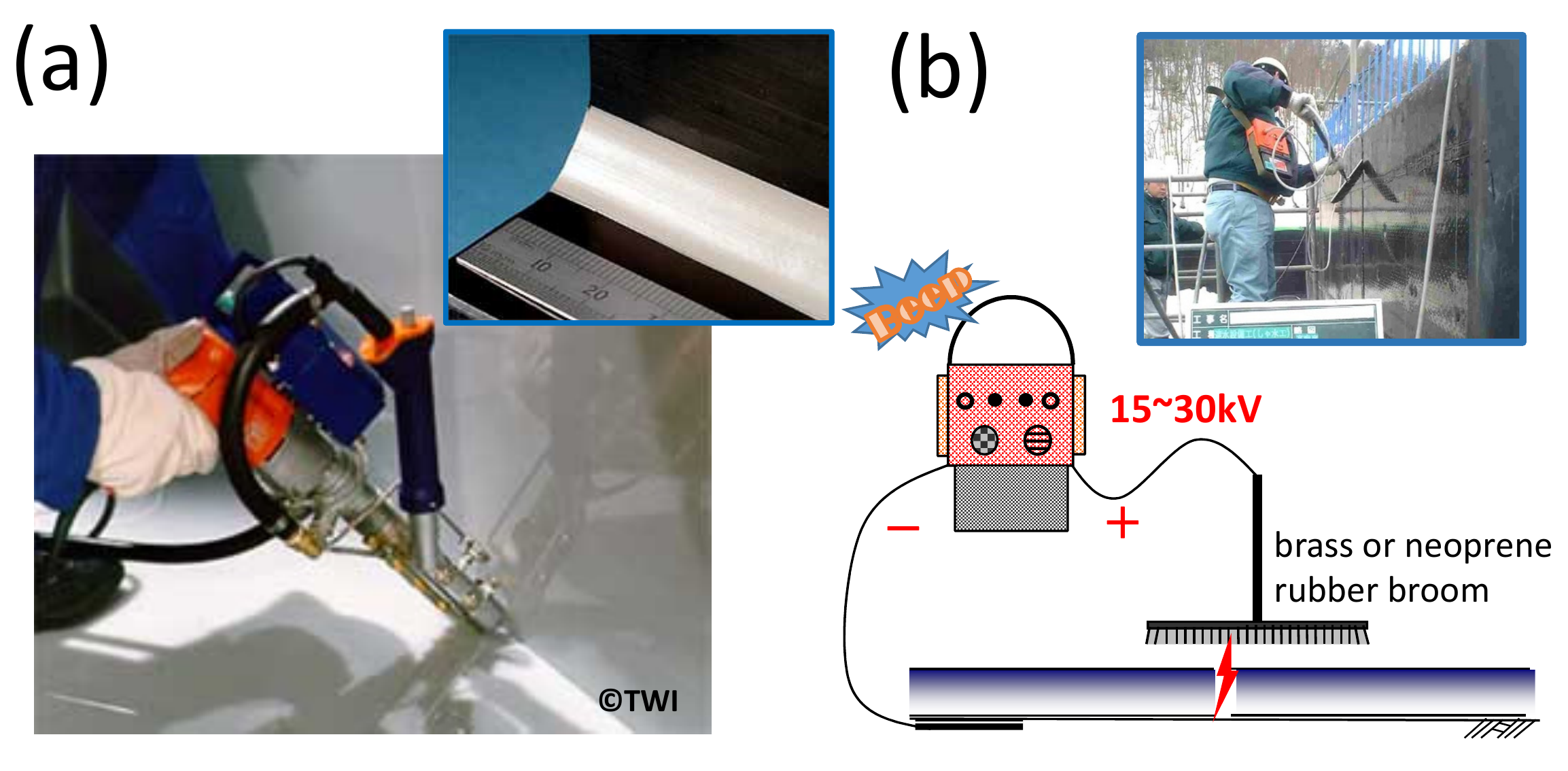}
\caption{(a) The extrusion welding and a welded seam. (b) The high-voltage pin-hole test.}
    \label{fig:liner-welding}
  \end{center}
\end {figure}

The planned procedures of the CPL installation into cavern are
illustrated in Fig.\ref{fig:liner-install}: At first CPL is fastened
to the inside of molds before concrete is poured to create a surface
lined with HDPE. The backfill concrete flows around the studs
anchoring CPL firmly in place, and fastens it securely to the surface
of the concrete. The waterproof sheet between bedrock-covering
shotcrete and backfill concrete aims conveyance of water, coming from
tiny leakage of water through the CPL and backfill concrete structure,
if any, and/or penetrating underground backwater from bedrock.

The adjacent CPLs are welded by the extrusion welding of
thermoplastics, which is used typically for assembly of large
fabrications (such as chemical storage vessels and tanks) with wall
thicknesses up to 50\,mm. Figure~\ref{fig:liner-welding}(a) shows
an extrusion welding work and close-up to the welded seam. In this
method, molten thermoplastic filler material is fed into the joint
preparation from the barrel of a mini hand-held extruder based on an
electric drill. For the CPL welding the same HDPE is used as the
filler material. The molten material emerges from a PTFE shoe shaped
to match the profile being welded. At the leading edge of the shoe a
stream of hot gas is used to pre-heat the substrate prior to the
molten material being deposited, ensuring sufficient heat is available
to form a weld.

For the quality control of the lining, the holes in the CPL sheets
with size of $>$0.5\,mm, including those on the welded seams, can
be identified by a high-voltage pin-hole testing
method~\cite{liner:industrialwaste}, as illustrated in
Fig.~\ref{fig:liner-welding}(b): It utilizes a charged metal or
neoprene-rubber broom above the liner. The power source is grounded to
the conductive deck and creates a high potential difference ($\sim$30
kV at maximum) with tiny current. When the metallic broom head is
swept over a breach or a hole in the insulating membrane surface,
current is detected by the test unit which turns off the power to the
broom and emits a beep sound to alert the test operator. The area is
then carefully swept again at $\sim$90 degrees to the original
sweeping direction to pinpoint the exact location of the
breach/hole. This process is continued until all areas of the CPL have
been tested. Occasionally negative pressure tests with a vacuum box
can be applied on the possible breaches and the welded seams. The
leakage water through the holes less than 0.5\,mm diameter, if
any, can be collected and controlled by a leakage detection and drain
system, as described later.
%

%----------------------------%
\paragraph{Liner sheet tests}
%----------------------------%
Various material tests were carried out for the candidate CPL, as are
described in Appendix~\ref{sec:linertests}.
To see the change of light absorbance and elusion of impurities,
specimens of the lining sheet were soaked both into ultra-purified
water and into 1\% gadolinium sulfate solution: increase of the light
absorbance were observed at the wavelength lower than 300\,nm, and
certain amount of material elution, i.e. total organic carbon, anions
and metals, were observed. The relation between material elusion and
change of light absorbance should be studied carefully. Meanwhile,
since PMT is sensitive for higher wavelength, the effect to the
experiment can be limited.

\begin{figure}[htbp]
   \begin{center}
      \includegraphics[width=0.8\textwidth]{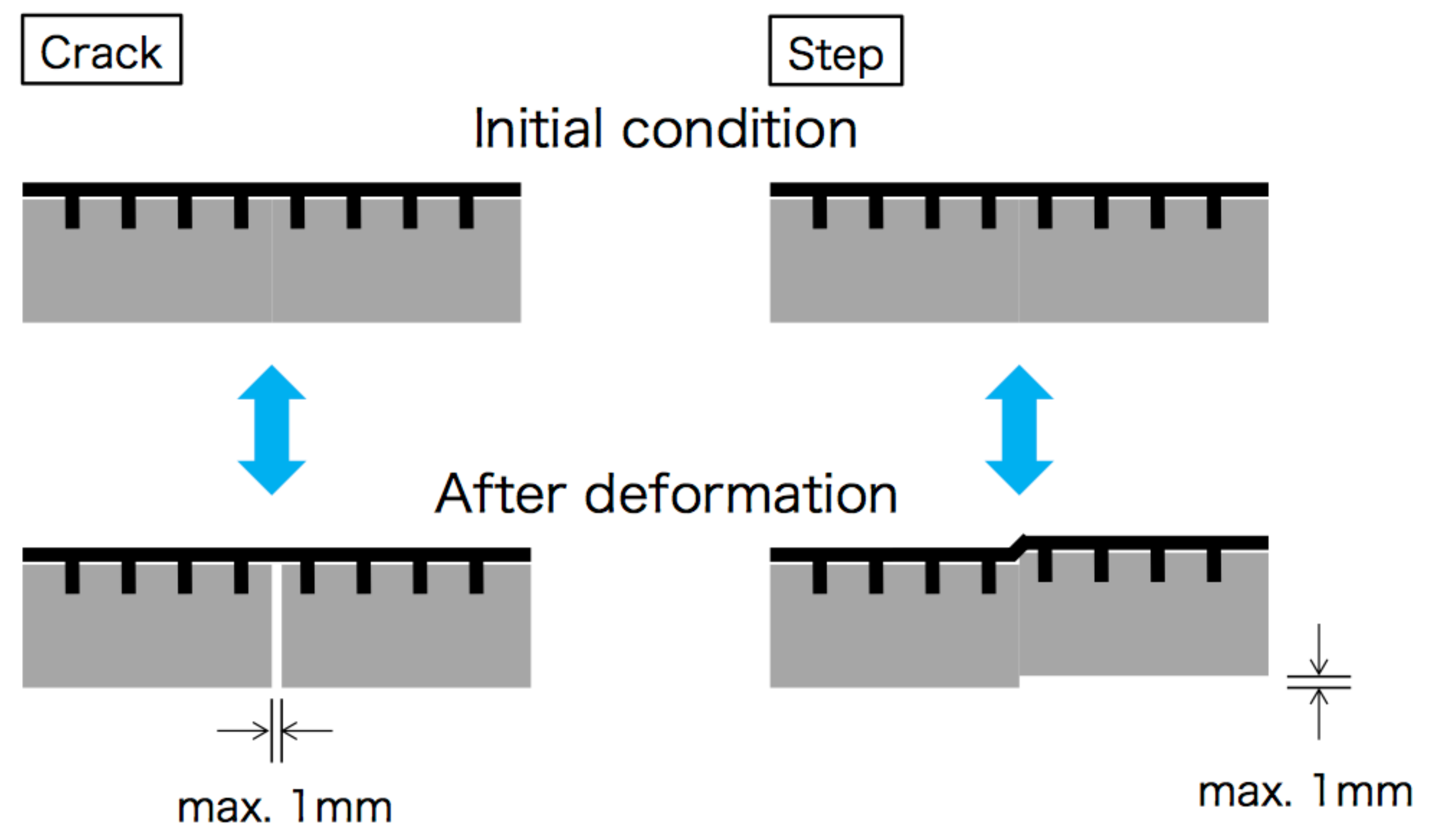}
      \caption{Two possible situations in deformation of backfill concrete, which
               can introduce tensile or shear stress in the liner sheet. Left:
               a crack is created. Right: a step is created.}
      \label{fig:liner_elongation_test_fig}
   \end{center}
\end{figure}

\begin{figure}[htbp]
   \begin{center}
     \includegraphics[width=0.8\textwidth]{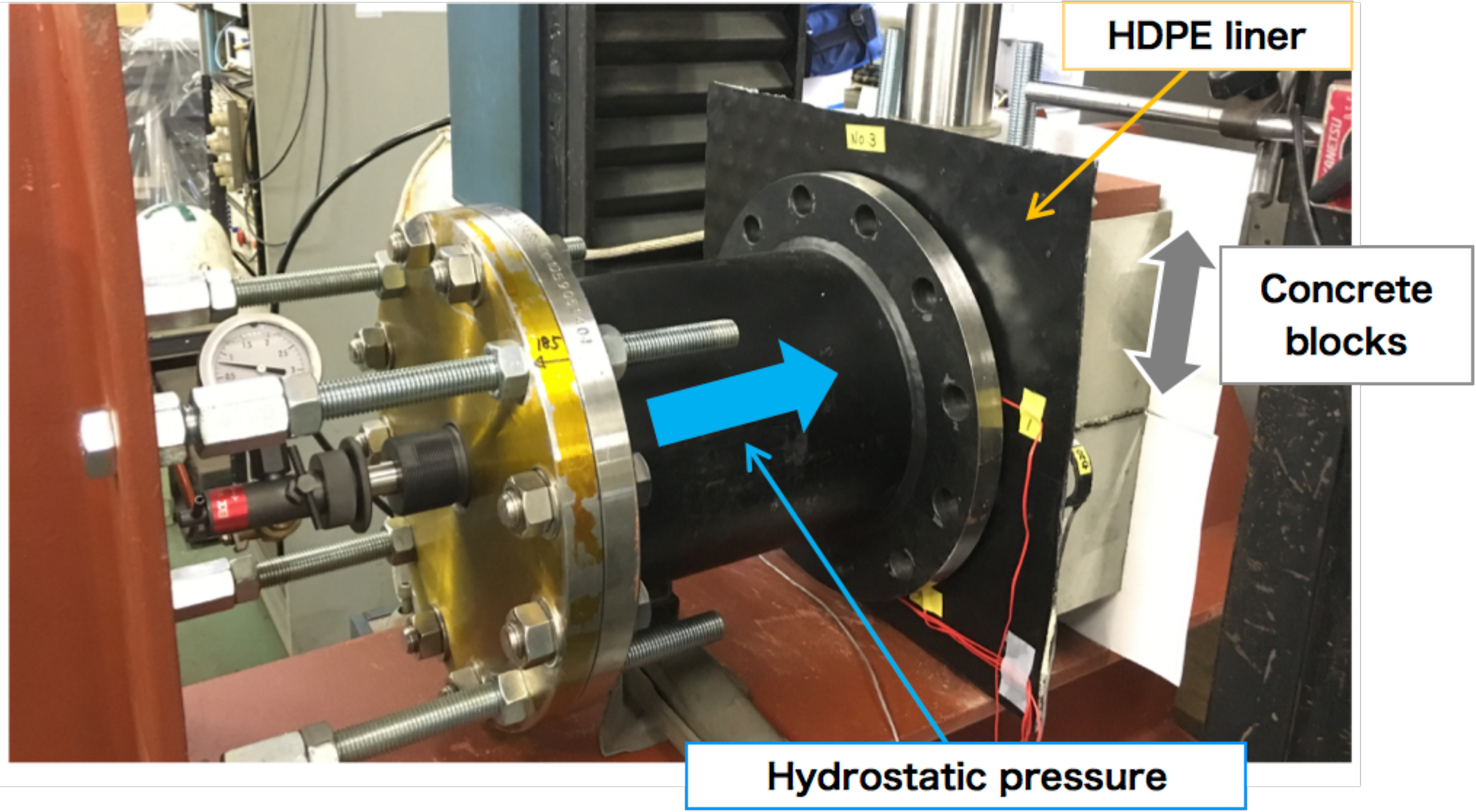}
     \caption{Picture of setup for crack elongation test. A HDPE sheet was
              attached to a concrete block and a crack or a step was intentionally
              produced in the concrete block. Strain on the liner sheet was then measured.}
     \label{fig:liner_elongation_test_pic}
   \end{center}
\end{figure}

Measurements on material strength, i.e. tension test and creep
test, were performed: the candidate CPL sheet has basically enough
strength. If cracks or rough holes happen in the backfill concrete,
the liner should locally stand for water pressure. To simulate the
situation, tests to apply localized water pressure on the lining were
performed with variety of slits and holes: For all cases, the liner
survived without breaking.
Another concern is that tensile or shear stress may be applied to the
liner sheet if deformation in the backfill concrete occurs after installation.
A crack elongation of the liner sheet was tested in the following way: 
A liner sheet was attached to a concrete block and a crack or a step
was created in the concrete as shown in Figure~\ref{fig:liner_elongation_test_fig},
then strain on surface of the liner sheet was measured. The surface of the liner sheet
was pressurized with 0.8~MPa during the measurement to simulate a water pressure.
Figure~\ref{fig:liner_elongation_test_pic} shows the actual setup of the crack
elongation test.
From the experience of Super-K operation\footnote{During Super-K operation, 
observed deformation was at most $O(0.01)$\%. Cracks in concrete tend to be created
in an even pitch. Assuming deformation of 0.1~\% (with a safety factor of 10) and
crack pitch of 600~mm, a gap of crack can be estimated to be 0.6~mm.},
a gap or a step of 1~mm was applied to the concrete block as shown in
Figure~\ref{fig:liner_elongation_test_fig}. 
In this test, the concrete was moved by 1~mm for 6 miniutes,
held for 30 minutes, and then moved back to the original position. This procedure was
repeated 10 times. 
The measured strain with 1~mm gap/step was at most 1.4\%/0.9\%, respectively.
These values were consistent with the simulated ones based on the test condition.
From the simulation results, it was found that strain of approximately 6\% at maximum could be
applied on the back (concrete) side of the liner sheet.
After this test, no damage was observed on the liner sheet by visual inspection.
Furthermore, water leak test with 0.8~MPa was also performed and no leakage was
observed. From the crack elongation test, it was concluded that the HDPE liner sheet
had enough strength against deformation and water tightness could be proved even in case that
deformation of the backfill concrete occured after installation.

The water leak can happen around components which penetrate the water
tank lining, such as anchors and water pipes. A possible design of the
penetration structure was developed. Its prototype was exposed to
series of pressure tests, which showed no leaks.
%

%----------------------------%
\paragraph{Long term stability}
%----------------------------%
Long term stability of the liner sheet is one of big concerns on a water tank.
Polyethylene materials have been used for sheeth of PMT cables and PMT endcap inside SK
for 20 years, and no obvious problem is observed. 
Possible sources of degradation of HDPE material are physical stress,
and some other effects (materials attached to the liner, oxidization, temperature, 
etc.)\footnote{Generally, a UV light is one of the big concern on long term stability of 
HDPE material, but this is not the case for underground area.}.
Effect of oxidization on lifetime of a HDPE material was reported in \cite{liner:HDPE_lifetime}.
Lifetime of HDPE liner can be strongly affected by temperature. At 20~$^{\circ}$C its lifetime
is predicted to be 446 years, however it reduces to 69 years at 40~$^{\circ}$C.
Since water temperature in HK tank will be controlled below 15~$^{\circ}$C,
lifetime can be expected to be more than 500 years. To estimate long term stability, 
an accelerating test to measure strength of a HDPE liner sheet after long-term soak 
in ultra pure water is being considered.

%----------------------------%
\paragraph{Leakage detection and drain system}
%----------------------------%

\begin {figure}[htbp]
  \begin{center}
    \includegraphics[width=0.8\textwidth]{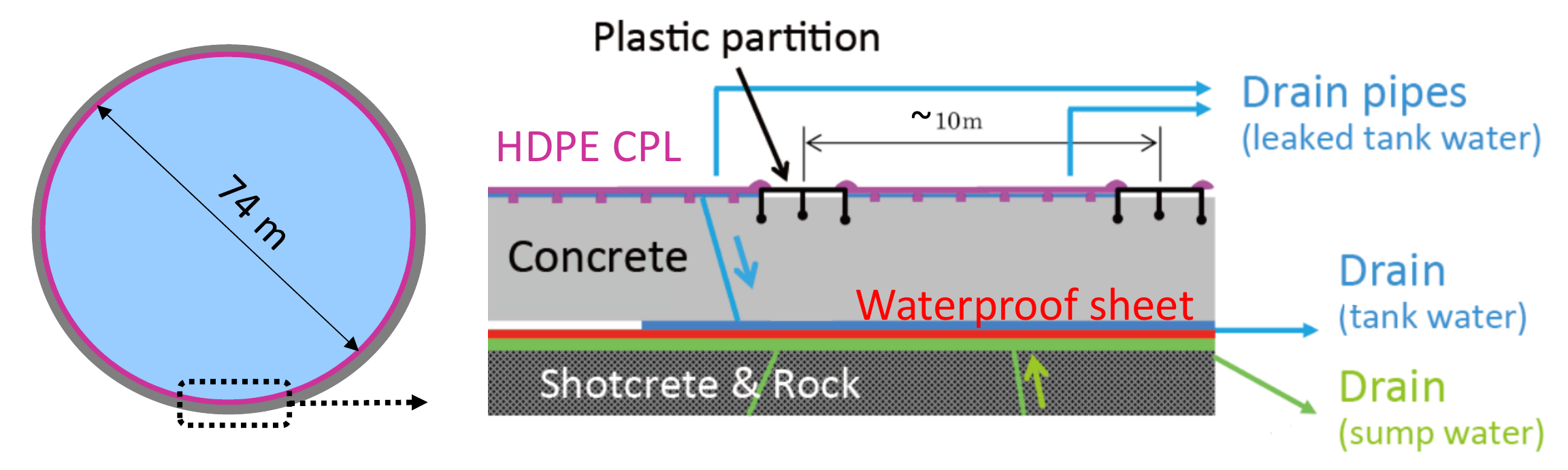}
\caption{Conceptual diagram of water leak detection/drain system.}
    \label{fig:liner-leakdrain}
  \end{center}
\end {figure}
The water leakage, if it happens, will be not through the sheets
themselves, but through small holes, which are undetectable by the
pin-hole/vacuum tests, or breaches, which are caused unavoidably by
works after the tests. To be prepared for these possible failures, a
leak detection/drainage system is to be
developed. Figure~\ref{fig:liner-leakdrain} shows preliminary concept
of the system. HDPE plastic moldings are embedded together with the
CPL in the backfill concrete, to work as partition at a pitch of about
10\,m in the direction of circumference of the tank. Water leaks
from the CPL(s) or seam(s) in each partition are to be collected
individually, so that leak detector installed at the bottom can
identify the partition with the problem.  Occasionally, water leak
through the CPL can flow into bedrock side through cracks of the
backfill concrete. Water-proof sheets (high panel signal sheet),
installed between bedrock and backfill concrete, can separate leakage
of inside (tank) water and sump water coming from outside
bedrock. These water will be drained separately, and tank water will
be treated with care especially for the case with gadolinium sulfate
solution is used in the tank.

\subsubsection{Photosensor Support Framework}  
The structural framework on which ID and OD photosensors are mounted
is basically made of commercially available SUS304 shaped steels, like
Super-K.  The stainless steel framework has been designed to support
the weight loads listed in Table~\ref{tbl:weight_loads}.

\begin{table}
\caption{List of the major weight loads taken into account for designing
the supporting framework.}
\label{tbl:weight_loads}
\begin{center}
\begin{tabular}{l|c}
\hline\hline
ID photosensor & ( / PMT ) \\
~~~~~~~50 cm $\phi$ PMT & 13 kg \\
~~~~~~~Protective cover & 39 kg \\
~~~~~~~Cable (for readout/power supply, 10 m) & 2 kg \\
\hline
OD photosensor & ( / PMT ) \\
~~~~~~~20 cm $\phi$ PMT &  2 kg \\
~~~~~~~Protective cover &  8 kg \\
~~~~~~~Wavelength shifting plate & 5 kg \\
~~~~~~~Cable (for readout/power supply, 10 m) & 2 kg \\
\hline
Underwater electronics (for readout/power supply) & 47\,kg / unit \\
Network cables connecting adjacent underwater electronics units & 2 kg/ unit \\ 
Water system pipes (65A PVC) & 1.4 kg/m \\ 
Calibration system (w/ 200A SUS pipe holes) & 1000 kg/m$^2 \times 4$ \\ 
& 100 kg/m$^2 \times 16$ \\ 
Other distributed load on the roof & 100 kg/m$^2$ \\
\hline\hline
\end{tabular}
\end{center}
\end{table}

\begin{figure}
  \begin{center}
  \includegraphics[width=0.55\textwidth]{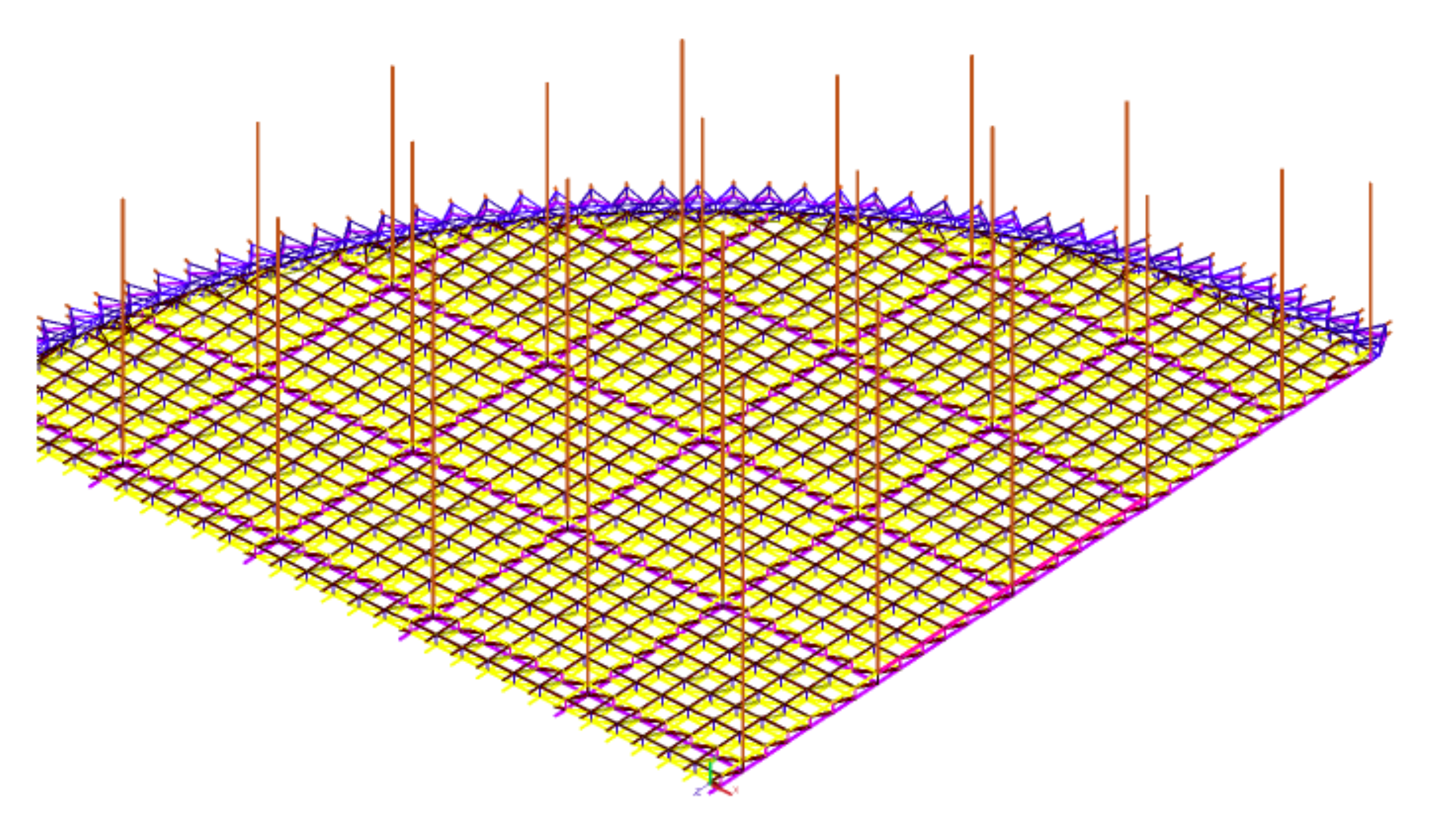}
  \includegraphics[width=0.43\textwidth]{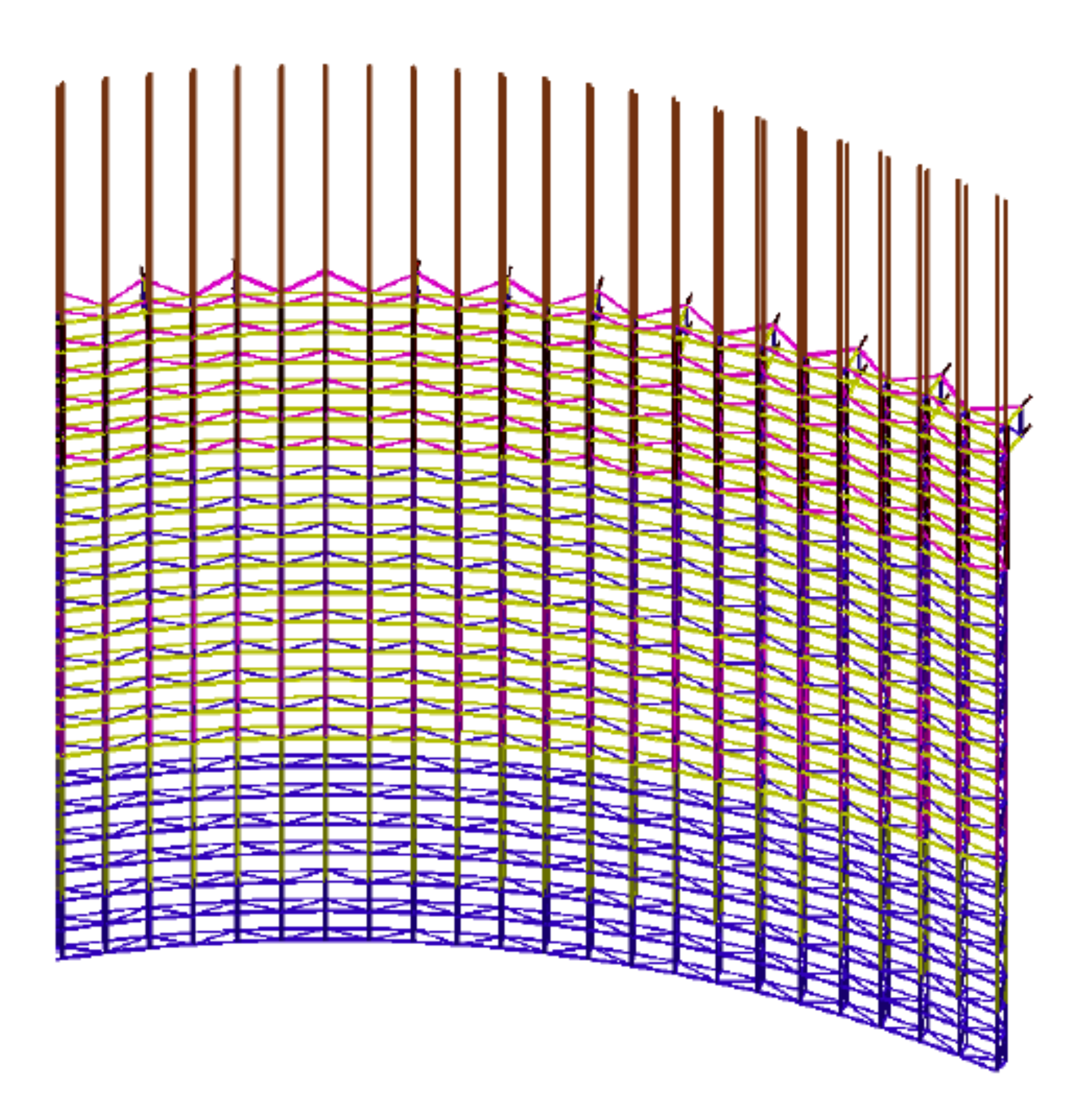}
  \caption{Schematic view of the top and barrel frameworks.
The lines extending upward show the vertical pipes which are anchored
to the dome ceiling for suspending the top/barrel frameworks.}
  \label{fig:framework_suspension}
  \end{center}
\end{figure}

The PMT supporting framework in the tank bottom part is constructed
on the floor, independently from the top/barrel frameworks.
The supporting frameworks in the top and barrel parts are truss
structures hung from the ceiling, unlike the Super-K's framework whose
barrel part is freestanding on the ground, while the top roof is
supported at the edge contacting the side concrete wall.  A suspension
structure can usually be built with relatively thinner and lighter
steel members, resulting in a lower construction cost, while a
freestanding structure usually needs thicker and heavier framework
members to avoid a buckling.  Figure~\ref{fig:framework_suspension}
shows part of the top and barrel frameworks along with their
suspension parts.  The weight load at each suspension point is about
110\,kN for the top structure including the photosensors and other
instruments on the top deck and about 70\,kN for the barrel structure
including the photosensors etc.  The anchor bolts for the suspension
points are embedded in the ceiling rock during the excavation of the
dome part of the cavern.

The top floor of the tank on which people can walk is made of the
stainless steel plates placed on the truss framework.  On this top
deck, the following penetrating components are built; the
``calibration holes'' through which various instruments for the
detector calibration are inserted into the water tank,
the ``water pipes'' for the tank water supply/drain,
and the ``cable holes'' for the photosensor/electronics cables.
As described in Section~\ref{section:electronics},
unlike the Super-K detector in which all the PMT cables are directly
connected to the readout electronics placed on the top deck,
the photosensor cables in Hyper-K are connected to nearby electronics
submerged to the water tank.
The underwater electronics modules are connected to each other
and only the top modules are connected to the readout computers,
thus the size of the cable holes can be reasonably small.
The top deck will be designed to also allow 
the penetration of the pipes to hang the barrel supporting structure.

When the tank is filled with water, the overall load to
the framework is reduced by the buoyancy of the various components
such as photosensors. Therefore, the design of the structural framework
has been made so that it has sufficient strength when the tank is empty
(i.e. no water inside).
As for the horizontal load in the case of an earthquake, the peak
horizontal acceleration is set as 0.15\,$g$ ($g$ = 9.8\,m$^2$/s), although
there is no official regulation for considering the effect of an
earthquake in ``the Law on Special Measures related to Public Use of
Deep Underground''.  The peak horizontal acceleration of 0.15\,$g$ is
equal to that used for designing the Super-K water tank, which is a
conservative assumption as it is derived based on ``the Seismic Design
Code for High-Pressure Gas Facilities of Japan'' , a standard for
facilities on the ground.

For the designed tank structure, a seismic response analysis has been
performed to estimate the maximum displacement of the tank structure
during an earthquake assuming various seismic waveforms.
The result shows that the framework displacement becomes maximum at
the bottom end of the suspended structure and its size is estimated to
be smaller than 50\,cm.
Since the Hyper-K water tank is built deep
underground, the actual displacement of the framework during an
earthquake is expected to be much smaller.
To ensure safety for people working inside the narrow outer detector layer
in time of an earthquake,
we are considering putting some temporary 'spacer' structure
for securing a safe space in the outer detector
only during tank construction and future detector maintenance periods.

Finally the tank will have an enough air space below the top deck so
that a water sloshing in the tank caused by an earthquake does not
damage detector components.

\subsubsection{Geomagnetic Field Compensation Coils}  
\label{section:tank-coil}
\paragraph{Introduction}

Photon collection efficiency (CE) decreases when a magnetic field is
applied on a PMT especially perpendicular to the PMT direction. For
example, the CE for box and line PMT decreases by about 1\%, 2\%, and
3\% at 100\,mG, 150\,mG, and 180\,mG, respectively. On the other hand,
decrease of the CE is negligible when a magnetic field is applied
parallel to the PMT direction even at 200\,mG. An initial goal of the
geomagnetic field compensation coil design is to keep the remaining
magnetic field perpendicular to the ID PMT ($B_{perp}$) smaller than
about 100\,mG.

\paragraph{Calculation of remaining geomagnetic field}

The remaining geomagnetic field at each ID PMT is calculated for a 60\,m
height and 74\,m diameter vertical cylindrical tank. The geomagnetic
field compensation coils are located along the tank inner wall. To
compensate for the geomagnetic field (x, y, z) = (-303, 0, -366)\,mG,
vertical rectangular coils along the x-axis and horizontal circular
coils along the z-axis (vertical direction) are assumed. For
simplicity after several initial studies, the interval between each
coil is set to be 2\,m for both vertical and horizontal coils, and the
following three currents are used to minimize $B_{perp}$: a common
constant current for all the vertical coils ($I_V$), a common constant
current for all the horizontal coils ($I_H$), and additional currents
only at the top and bottom horizontal coils ($I_V \times n$, where $n$
is a positive integer and corresponds to the number of winding of the coils). 
The last parameter $n$ is found to be
effective to reduce $B_{perp}$ at around both the top and bottom tank
corners. Distances between the tank inner wall and the ID surface are
(2+0.6)\,m at the top and bottom and (1+0.6)\,m at the barrel, where 2/1\,m
are for the OD regions and 0.6\,m is for a dead region. Note that the OD
region is reduced from 2\,m to 1\,m at the barrel (SK and the baseline
tank) to minimize cost. This is one of the HK detector design
parameters to be evaluated from this calculation of the remaining
geomagnetic field.

\begin{figure}
\centering
\includegraphics[width=0.49\textwidth]{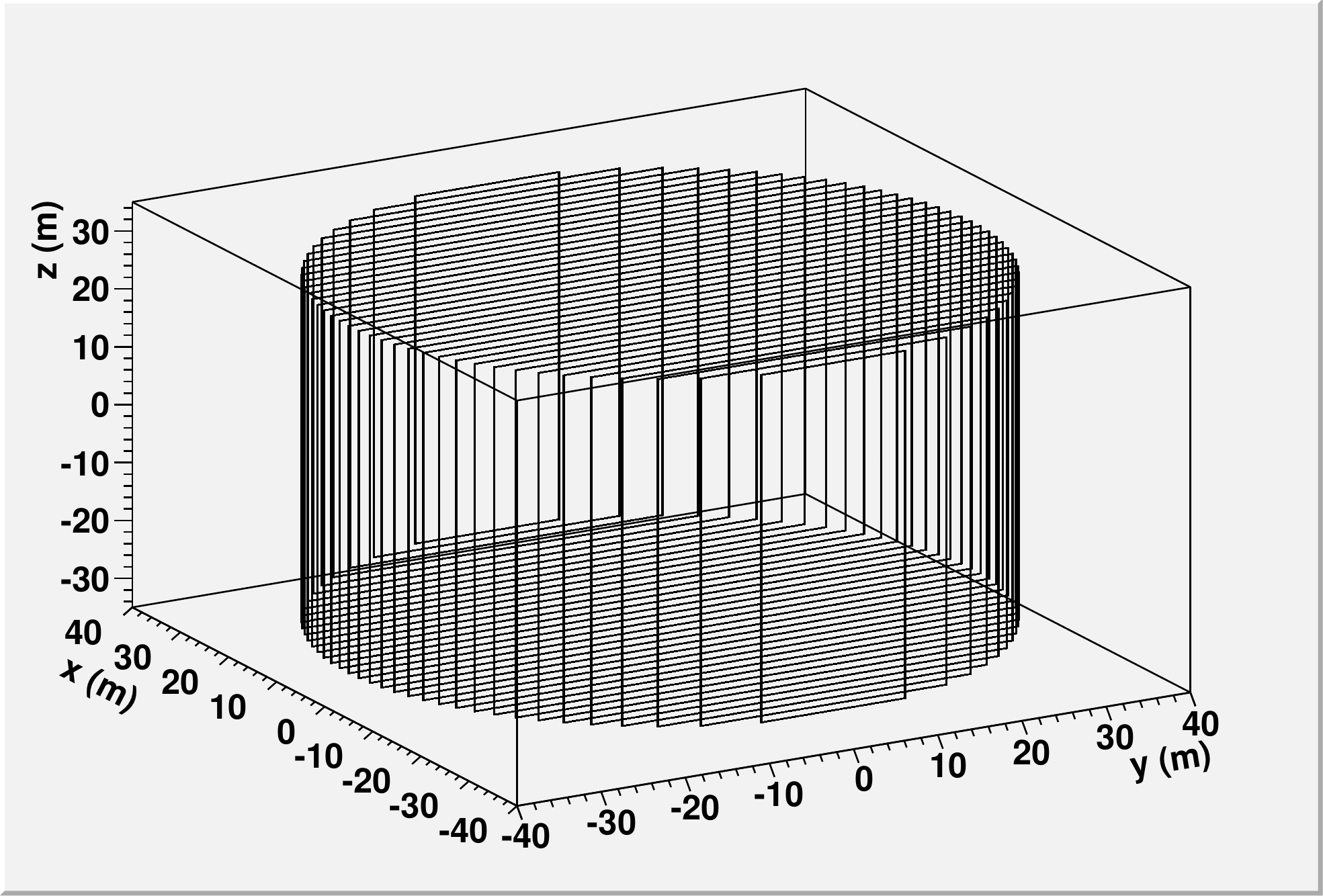}
\includegraphics[width=0.49\textwidth]{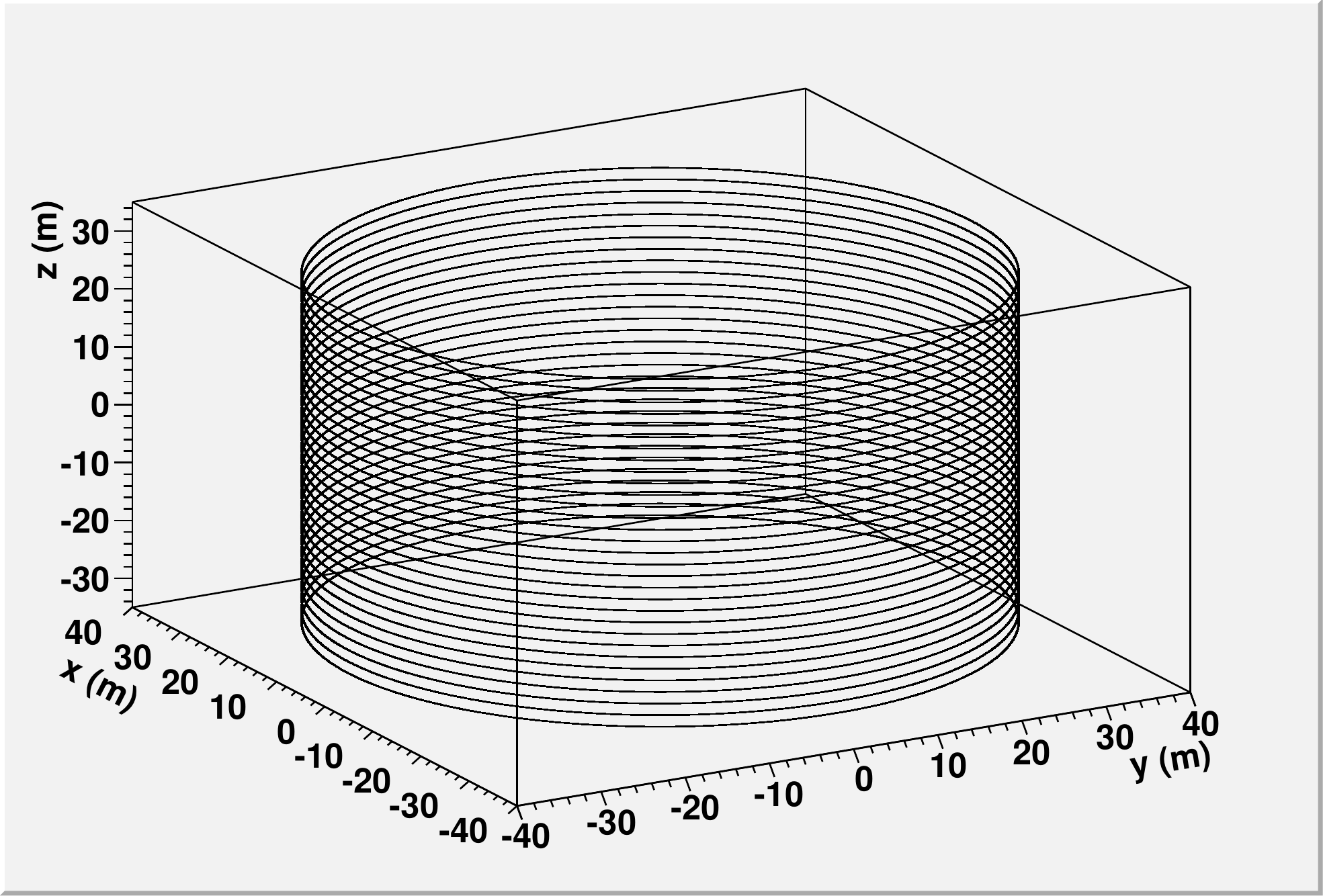}
\caption{Location of vertical rectangular coils (left) and horizontal circular coils (right).}
\label{fig:coil-location}
\end{figure}

The ID PMTs are mounted on the ID inner wall (54.8\,m height and 70.8\,m
diameter cylinder). The interval between each PMT is 70.7\,cm except in
the horizontal direction at the barrel (7080$\cdot \pi$/314 = 70.84\,cm,
where 314 is an appropriate integer value to make the interval close
to 70.7\,cm). There are 39,424 ID PMTs in total.
The coil locations are shown in Figure~\ref{fig:coil-location}.

\begin{figure}
\centering
\includegraphics[width=0.6\textwidth]{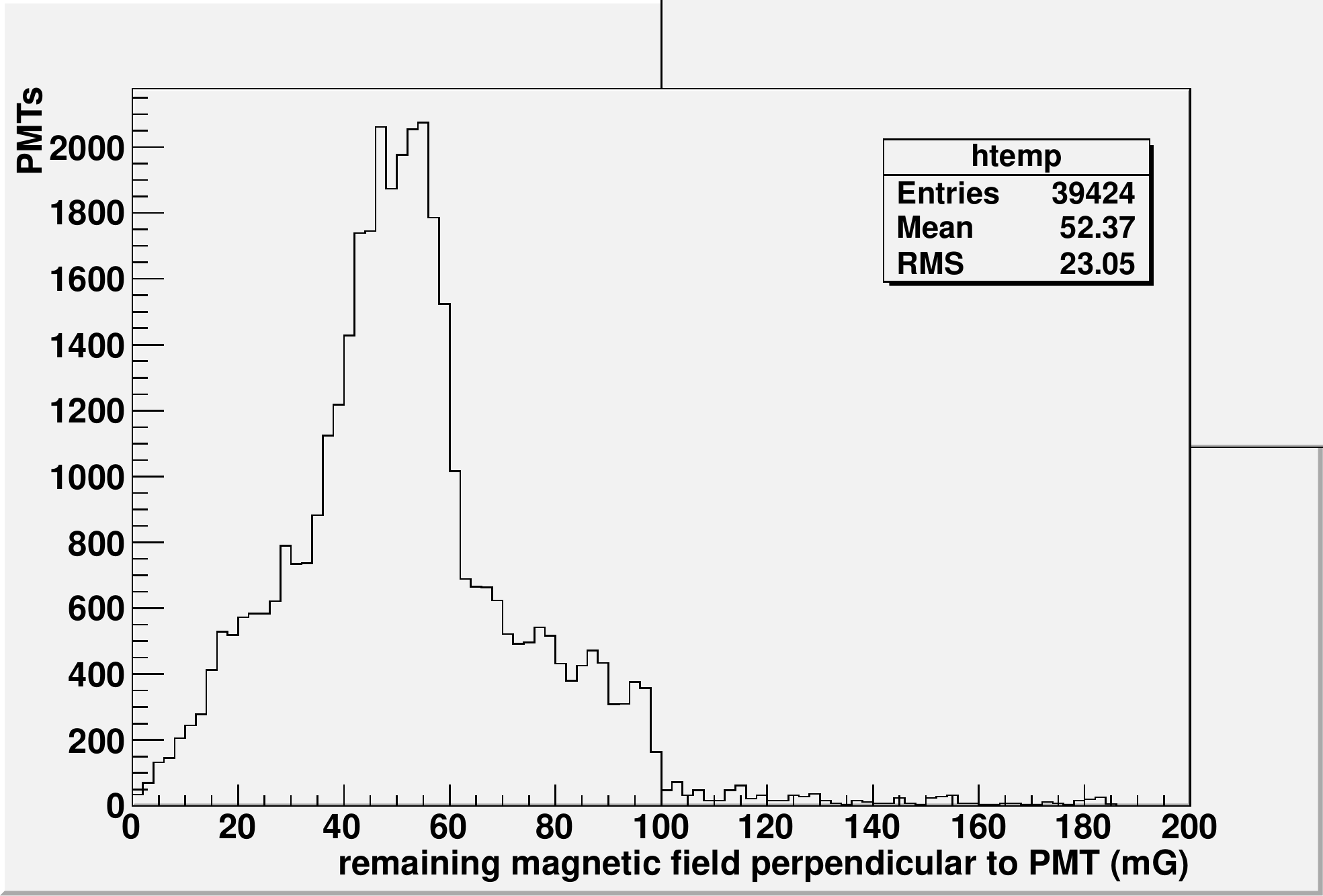}
\caption{\protect\small $B_{perp}$ distribution for all the ID PMTs at the best current set.}
\label{fig:residual_magnetic_field_perp}
\end{figure}

\begin{figure}
\centering
\includegraphics[width=0.49\textwidth]{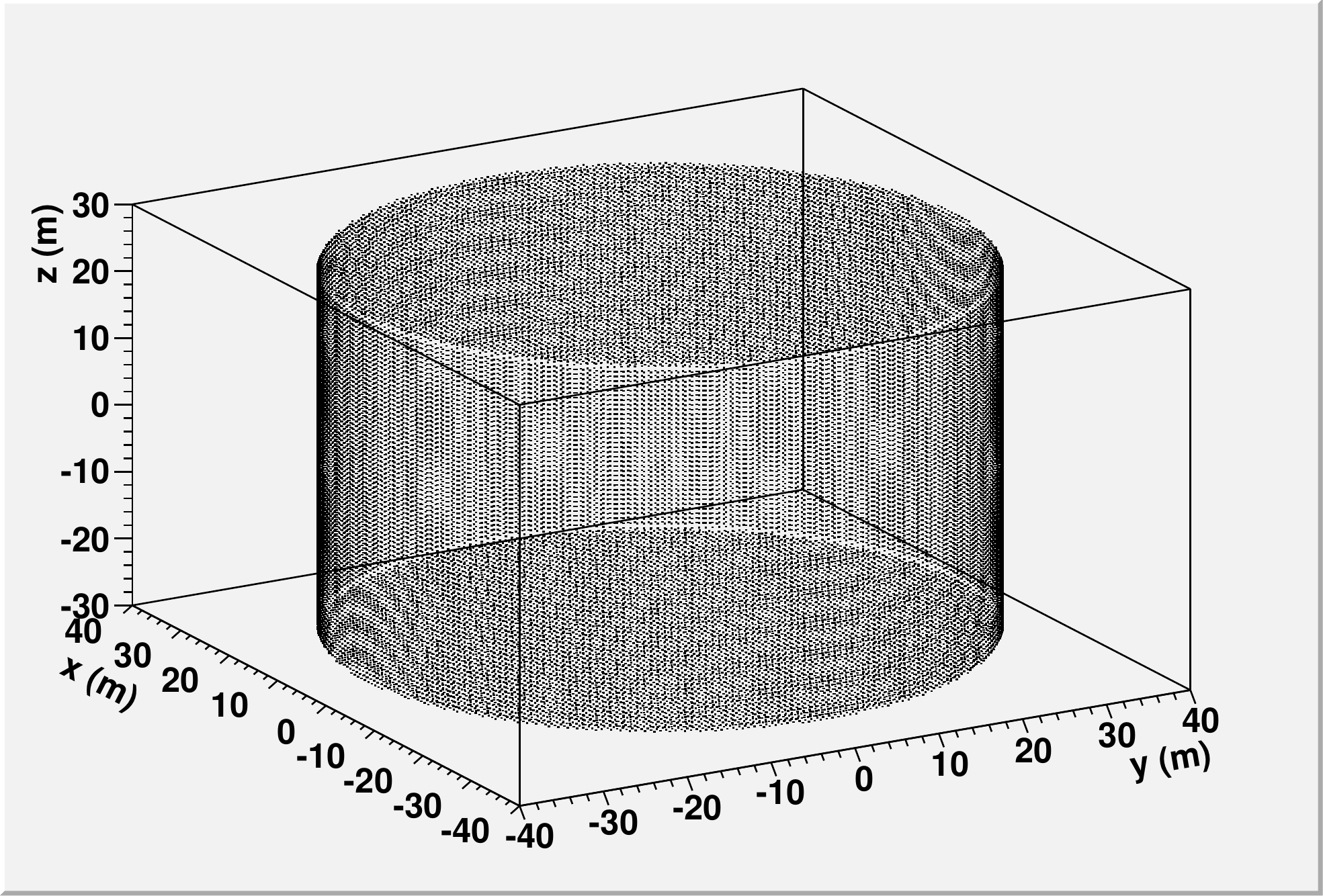}
\includegraphics[width=0.49\textwidth]{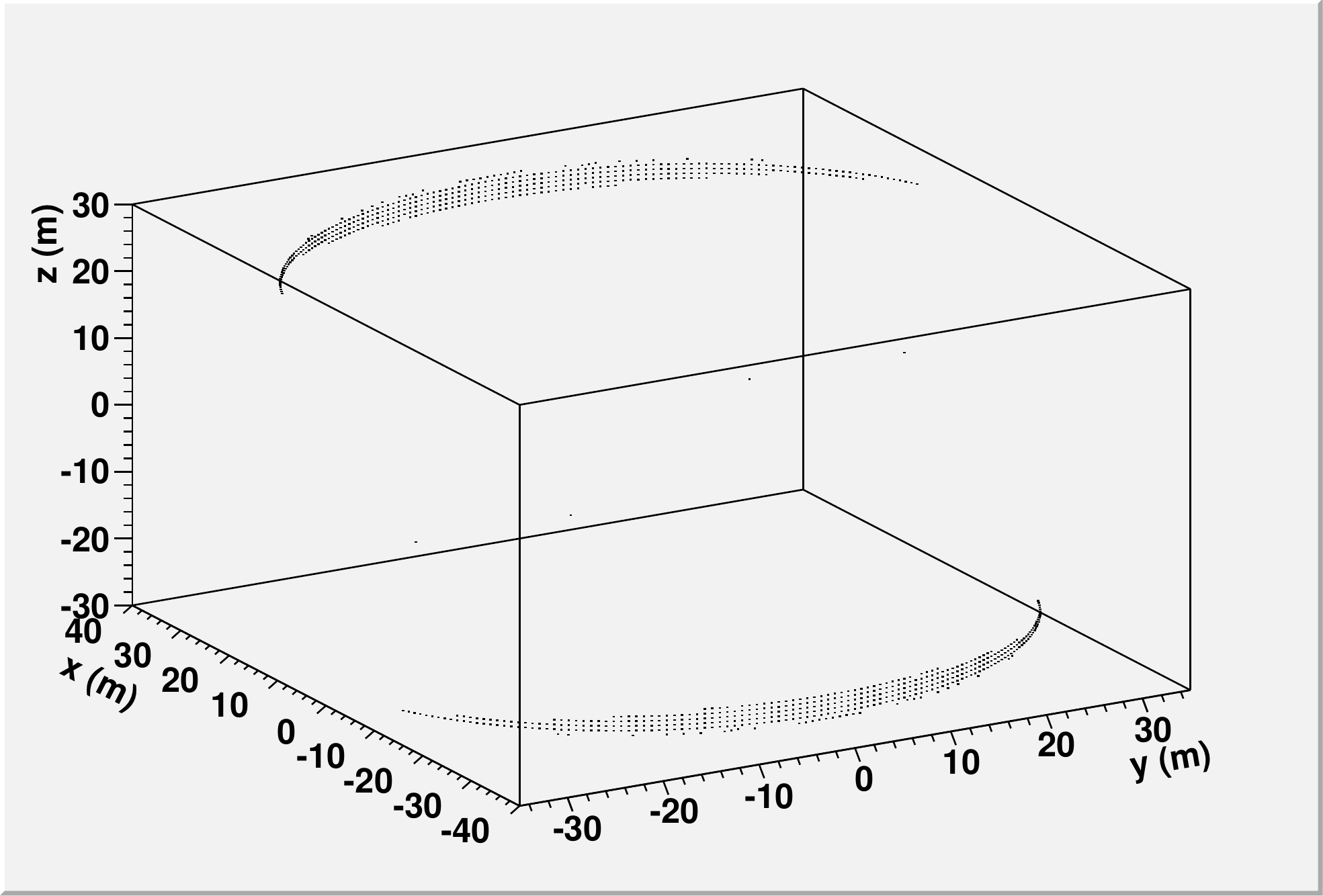}
\caption{\protect\small Location of all the ID PMTs (left) and ones with $B_{perp}>$100\,mG(right).}
\label{fig:pmtposition_over100mgauss}
\end{figure}

A current set to ($I_V$, $I_H$, $n$) = (60A, 67A, 7) is the best case
(the minimum fraction with
$B_{perp}>$100mG). Figure~\ref{fig:residual_magnetic_field_perp} shows
the $B_{perp}$ distribution for all the ID PMTs at the best
currents. The fraction of the number of the ID PMTs with
$B_{perp}<$100\,mG is 97.8\%. Even though the OD region at the barrel is
reduced from 2\,m to 1\,m, the fraction with $B_{perp}<$100\,mG is large
enough.
Figure~\ref{fig:pmtposition_over100mgauss} shows the location of the ID PMTs with $B_{perp} >$100\,mG at the best currents, along with positions of all ID PMTs.

From Figure~\ref{fig:residual_magnetic_field_perp}, decrease of CE is
at most 1\% on average. This satisfies the initial goal. If necessary,
$B_{perp}$ at the tank corners could be reduced by adding more coils
and so on.

\paragraph{Cables and power supplies (PSs)}

The cables for the geomagnetic field compensation coils 
are installed near the lining surface of the tank.
We are considering of placing the cable inside tubes
embedded in the backfill concrete layer.
In an alternative design, the cables are located in the tank water,
like those of the magnetic coils in Super-K,
supported by the fixtures attached to the lining sheets.
We are investigating which design is better for Hyper-K.

A 4-conductor cable (0.491\,$\Omega$/km) is assumed for the total power
consumption estimation.  Table~\ref{tbl:horips} and
Table~\ref{tbl:verps} show PSs used for the estimation and their
consumption for the horizontal and vertical coils, respectively.
\begin{table}
\caption{Power supply (PS) for horizontal circular coils per tank. The PS number is assigned along the z-axis of tank.}
\label{tbl:horips}
\begin{center}
\begin{tabular}{l|c|c|c|c}
\hline\hline
PS \# & Coil length [km] & Voltage [V] & $I_H$ [A] & consumption [kW] \\
\hline
1 (bottom) & 1.62 & 53.4 & 67 & 0.894 \\
2          & 1.62 & 53.4 & 67 & 0.894 \\
3          & 1.62 & 53.4 & 67 & 0.894 \\
4 (z=0)    & 0.23 &  7.6 & 67 & 0.127 \\
5          & 1.62 & 53.4 & 67 & 0.894 \\
6          & 1.62 & 53.4 & 67 & 0.894 \\
7 (top)    & 1.62 & 53.4 & 67 & 0.894 \\
\hline
Total      &      &      &    & 5.49  \\
\hline\hline
\end{tabular}
\end{center}
\end{table}
\begin{table}
\caption{Power supply (PS) for vertical rectangular coils per tank. The PS number is assigned along the x-axis of tank.}
\label{tbl:verps}
\begin{center}
\begin{tabular}{l|c|c|c|c}
\hline\hline
PS \# & Coil length [km] & Voltage [V] & $I_V$ [A] & consumption [kW] \\
\hline
1          & 1.18 & 34.8 & 60 & 0.522 \\
2          & 1.48 & 43.6 & 60 & 0.654 \\
3          & 1.59 & 46.8 & 60 & 0.702 \\
4 (x=0)    & 0.27 &  7.9 & 60 & 0.119 \\
5          & 1.59 & 46.8 & 60 & 0.702 \\
6          & 1.48 & 43.6 & 60 & 0.654 \\
7          & 1.18 & 34.8 & 60 & 0.522 \\
\hline
total      &      &      &    & 3.88  \\
\hline\hline
\end{tabular}
\end{center}
\end{table}
For the horizontal (vertical) coils, PSs are assigned for 7 circles (6
rectangular) along the z-axis (x-axis) except 1 PS for 1 circle
(rectangular) at z (x) = 0, respectively. The total power consumption
for the tank is 5.49\,kW (horizontal coils) + 3.88\,kW (vertical coils) =
9.37\,kW. This is higher than that for SK (about 6.5\,kW with 7 PSs in
total) but reasonable.

Note that the coils between PSs and both horizontal/vertical coils are
neglected (about 2/1\%, respectively).

\subsubsection{Construction}  
The water tank construction work includes laying the liner, building
the photosensor support framework, and the installation of the
underwater instruments (photosensors, electronics, light shielding
sheets, magnetic coils, etc.).  The planned construction procedure is
similar to that of the Super-K water tank.

\begin{figure}
  \begin{center}
  \includegraphics[width=0.6\textwidth]{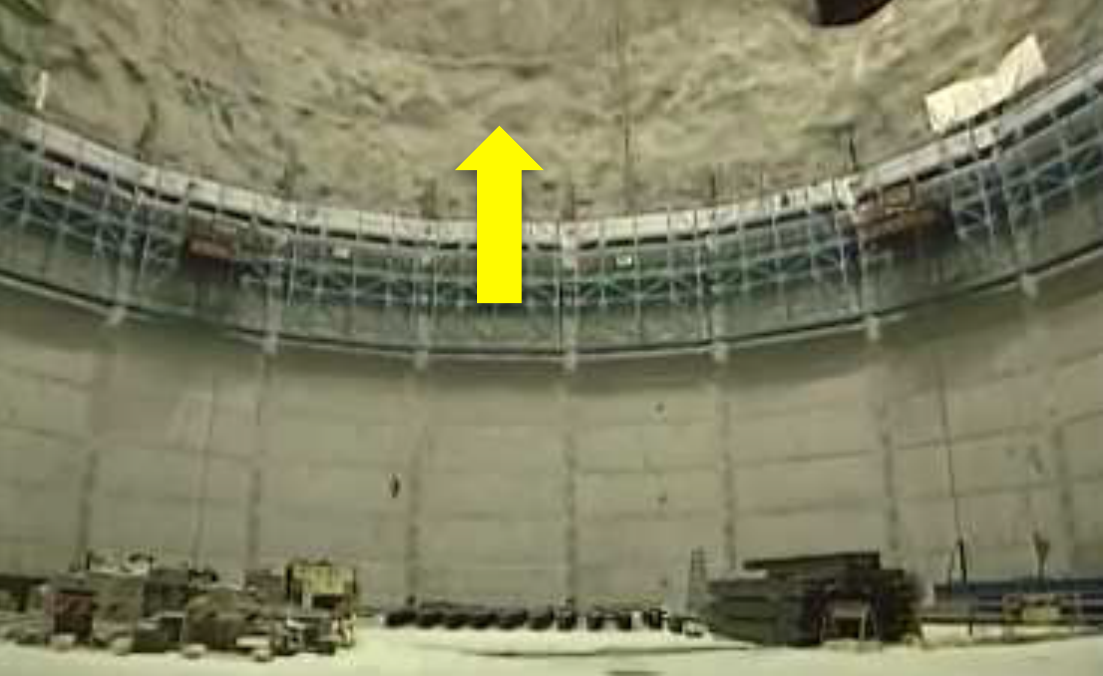}
  \caption{A photograph of the Super-K tank liner construction
using mobile scaffoldings.}
  \label{fig:sk_lining_scaffoldings}
  \end{center}
\end{figure}

First, the drainage structure and the leveling concrete layer are
constructed at the bottom of the cavern.  Then, the side-wall liner is
constructed ring-by-ring from bottom to top using mobile scaffoldings
built along the whole circumference.  Such a barrel liner laying using
elevating scaffoldings was successful in the Super-K water tank
construction (see Figure~\ref{fig:sk_lining_scaffoldings}).  To fasten
HDPE sheets on the concrete layer by their studs, the backfill
concrete is poured after setting a form with a HDPE sheet attached.
After the concrete sets hard, the form is disassembled.

\ 

After the barrel liner construction is completed, building the
photosensor support framework starts.

First, the roof structure and the support framework for the
photosensors in the top region are built on the bottom floor.  The
installation of the top photosensors is also done on the bottom floor.
Then the top framework with photosensors mounted is lifted up from the
bottom floor to its final destination by using the jacks.  The total
lifting load of the top structure with the photosensors mounted is
about 9,400\,kN. The number of hanging points is about 150, and the
weight load per jack is about 70\,kN.

The construction of the barrel framework and the photosensor installation
are performed layer-by-layer on the bottom floor.
Once the first stage of the barrel units with photosensors is completed,
the whole structure is lifted up by using jacks, and then the second stage of
barrel units with photosensor is hung just below the first stage one.
The total lifting load of the barrel structure with the photosensors mounted
is about 21,000\,kN. The number of hanging points is about 320, and the weight
load per jack is about 70\,kN.

Finally, after the barrel framework construction and the photosensor
installation is completed, the support framework in the bottom region
is built and the photosensors are mounted.

\ 

During the tank construction, most of the components such as the liner
sheets, the framework members, the photosensors are transported into
the tank through the bottom approach tunnel.  After the construction
of the support framework is completed, the tunnel is closed by using
the concrete and the waterproof liner sheets.  The plug manhole made
of the stainless steel is built there so that people can enter the
tank through this manhole for future maintenance works (see
Figure~\ref{fig:tank_schematic}).

\ 

As for the tank construction time, the liner laying work will take 14
months, then the support framework construction in parallel with the
photosensor installation will take 14 months.
In the current estimation, the time for the framework construction
and the photosensor installation is determined by the speed for
supplying the photosensor assemblies.
In total, about 2.5 years will be needed for the tank construction.

\newpage

\subsection{Water purification and circulation system }\label{section:water}

\subsubsection{Introduction}
Water is the target material and signal-sensitive medium of the
detector, and thus its quality directly affects the sensitivity. In
order to realize such a huge Cherenkov detector, achieving good water
transparency is the highest priority. In addition, as radon emanating
from the photosensors and detector structure materials is the main
background source for low energy neutrino studies, an efficient radon
removal system is indispensable.

In Super-Kamiokande the water purification system has been continually
modified and improved over the course of SK-I to SK-IV.  As a result,
the transparency is now kept above 100 m and is very stable, and the
radon concentration in the tank is held below 1\,mBq/m$^3$.  Following
this success, the Hyper-Kamiokande water system design will be based
on the current Super-Kamiokande water system.

Naturally, ever-faster water circulation is generally more effective
when trying to keep huge amounts of water clean and clear, but
increasing costs limit this straightforward approach so a compromise
between transparency and re-circulation rate must be found.  In
Super-Kamiokande, 50\,ktons of water is processed at the rate of
 60\,tons/hour in order to keep the water transparency (the attenuation
length for 400\,nm-500\,nm photons) above 100\,m, and 20\,Nm$^3$/hour of
radon free air is generated for use as a purge gas in degas modules,
and as gas blankets for both buffer tanks and the Super-Kamiokande
tank itself. For the 258\,ktons of water in
%%one tank of
the tank of Hyper-Kamiokande, these process speeds will need to be scaled-up to
310\,tons/hour for water circulation and 50\,Nm$^3$/hour for radon free
air generation.

\subsubsection{Source water}
The rate of initial water filling is restricted by the amount of available source water.
 In Mt.~Nijuugo-yama, the baseline location of Hyper-Kamiokande, the total
amount of the spring water is about 600\,tons/hour. (It varies
seasonally between 300\,tons/hour and 800\,tons/hour and it is above
600\,tons/hour except in Winter (December-March).) However, as the mine
company uses all the water for their smelting factory, the available
spring water for Hyper-Kamiokande is limited and cannot be allocated at this point. 
Therefore, the baseline plan is getting 105\,tons/hour of source water from the outside of the mine, making 78\,tons/hour of
ultra-pure water and filling the 258\,ktons for 180\,days.
The source water site is the well for snow-melting system in the Kamioka town at Oshima public hall which is
about 5\,km away from the tank position. 
Hida city is supportive in our use of the well and Gifu prefecture is also helping to
decide the route from the well to the entrance of the Tochibora mine.
Serious investigations and negotiations are ongoing with these local governments.

The water quality of  the snow melt water and Tochibora spring
water are compared with that of Mozumi spring water in Table~\ref{tab:source}.
In the Mozumi mine, the location of Super-Kamiokande, there is sufficient mine water and no mining/smelting activities.
\begin{table}[htb]
\begin{center}
\scalebox{0.8}[0.8]{
\begin{tabular}{lr|rrr}
\hline\hline
   &  & The well for Kamioka snow melt      &  Tochibora spring water  & Mozumi spring water  \\
   &  &  as of 21 Jun. 2016               &  as of 1 Mar. 2011       & as of 16 Mar. 2011   \\
\hline \hline 
Temperature(Typical) & $^{\circ}$C        &  11.9     &    11    &    12   \\
pH (25$^{\circ}$C) &                             &  7.1    &   7.8    &   7.8   \\
Conductivity     & $\mu$S/cm               & 101    &   170    &   221   \\
Turbidity        &degree(Kaolin)                & $<1$    &  $<1$   &   $<1$  \\
Acid consumption (pH 4.8) & mg CaCO$_3$/L &  27.9  &   40.0   &   75.8   \\
TOC      & mg/L                                   &  $<0.1$ &  $<1$   &   $<1$  \\
Phosphate & mg/L                               &  $<0.1$ &   $<0.1$   &  $<0.1$   \\
Nitrate   & mg/L                                  &  3.0    &   1.0    &  1.6  \\
Sulfate   & mg/L                                  &  4.4   &  36.4    &  30.2  \\
Fluoride  & mg/L                                  &   $<0.1$    &   0.3    &  0.4  \\ 
Chloride  & mg/L                                  &  8.6    &   1.6    &  1.8  \\
Sodium    & mg/L                                 &  4.6    &   4.9    &  6.2  \\
Potassium & mg/L                                &  0.8    &   0.5    &  0.5  \\   
Calcium   & mg/L                                  &  12.3   &  25.2    &  32.0 \\  
Magnesium & mg/L                                &  1.5    &   1.5    &  2.9  \\
Ammonium  & mg/L                              &  $<0.1$ &   $<0.1$   &  $<0.1$   \\
Ionic silicon dioxide  & mg/L                   &  12.8   &   17.1    &  11.8  \\
Iron      & mg/L                                    & $<0.01$ &  $<0.01$  &   $<0.01$  \\
Copper    & mg/L                                  & $<0.01$ &   $<0.01$  &   $<0.01$  \\
Zinc      & mg/L                                    &   -    &   0.09     &    $<0.01$  \\
Lead      & mg/L                                   & $<0.1$  &   $<0.1$  &   $<0.1$  \\
Aluminum & mg/L                                 & $<0.01$ &    $<0.01$  &   $<0.01$  \\
Boron     & mg/L                                   & $<0.01$ &    $<0.01$  &  0.2  \\
Strontium & mg/L                                  & -   &   0.18      &  0.52 \\
Barium    & mg/L                                  & $<0.01$ &    $<0.01$  &  0.03  \\
\hline\hline
\end{tabular}
}
\caption{Source water quality.}
\label{tab:source}
\end{center}
\end{table}

\subsubsection{Main system flows and layouts}
\begin{figure}[htb]
\includegraphics[width=0.95\textwidth]{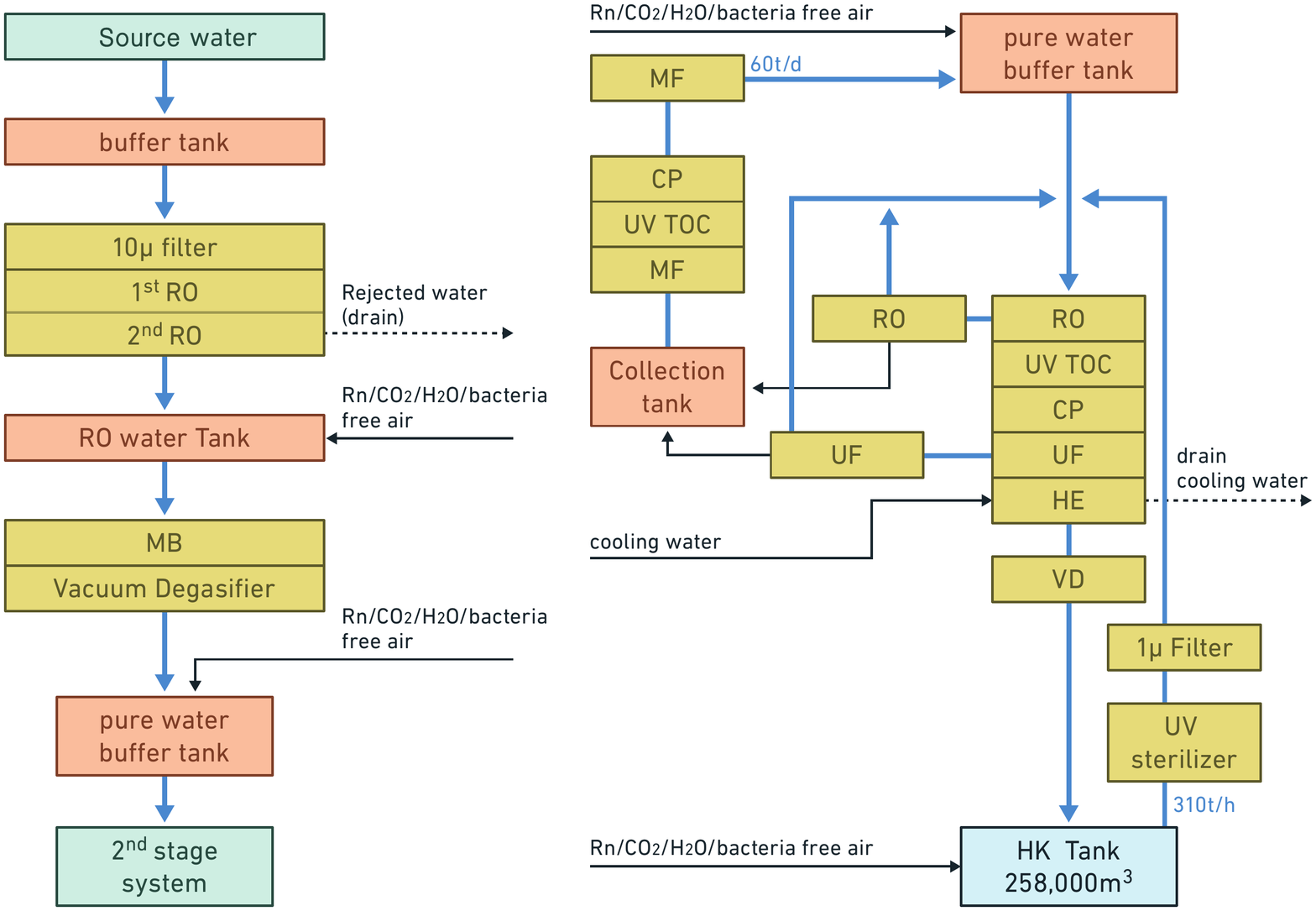}
\caption{1st stage and 2nd stage water systems.}
\label{fig:1st2ndsys}
\end{figure}

\begin{figure}[htb]
\includegraphics[width=1.0\textwidth]{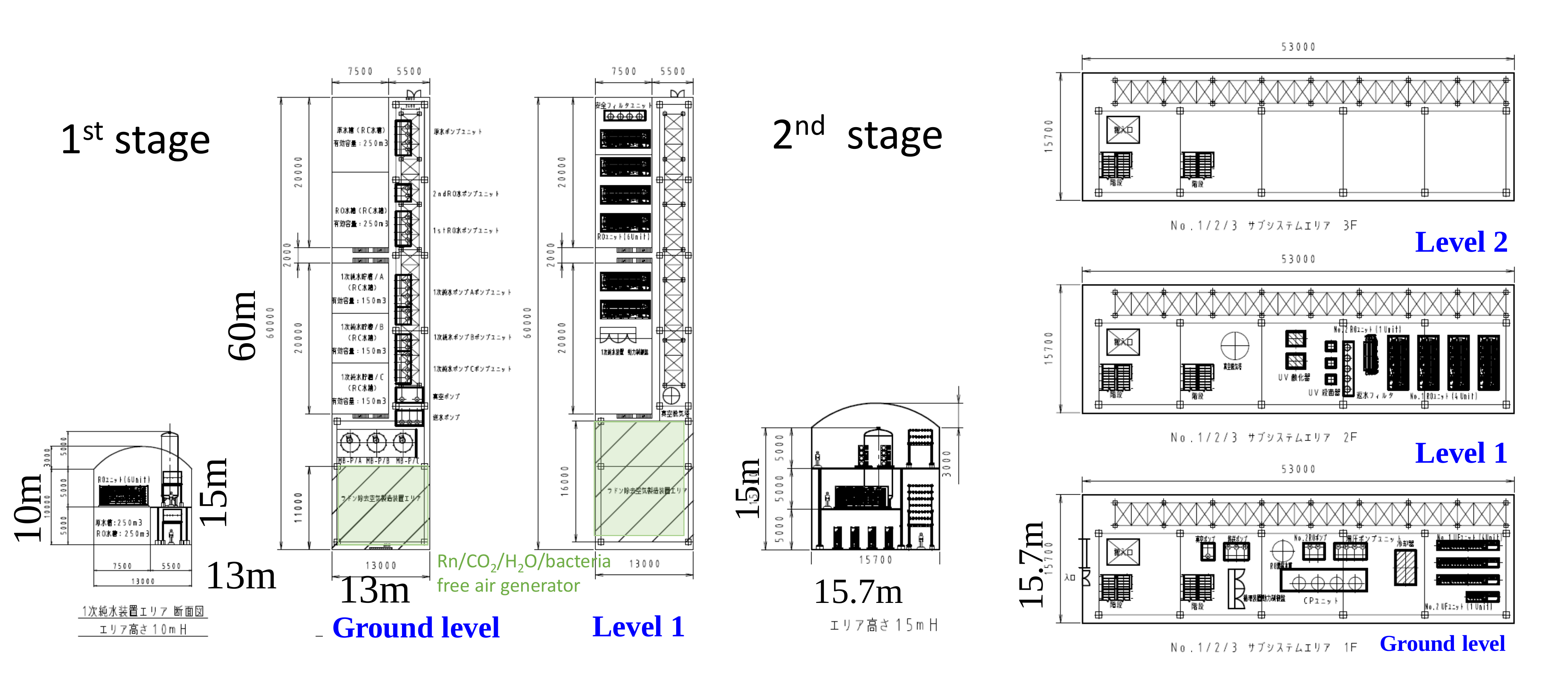}
\caption{Necessary space for the main systems.}
\label{fig:LO}
\end{figure}

The HK main water purification system consists of a 1st stage system
(filling) and a 2nd stage (re-circulation) system for the 258,000\,m$^3$ tank    
as shown in Figure~\ref{fig:1st2ndsys}.
Figure~\ref{fig:LO} shows their layouts.
The process power of the 1st stage system is 78\,m$^3$/h, and accordingly, it takes
138 days to fill the tank without consideration of any maintenance.
It may take about 180 days in the realistic case.
Preferably, an additional, same amount, of 11 $^{\circ}$C cooling water is required for the heat
exchangers.

The process power of the 2nd stage system for the recirculation is 
310\,m$^3$/h.

\subsubsection{Water flow simulation in the tank}
\begin{figure}[htb]
\begin{center}
\includegraphics[width=0.85\textwidth]{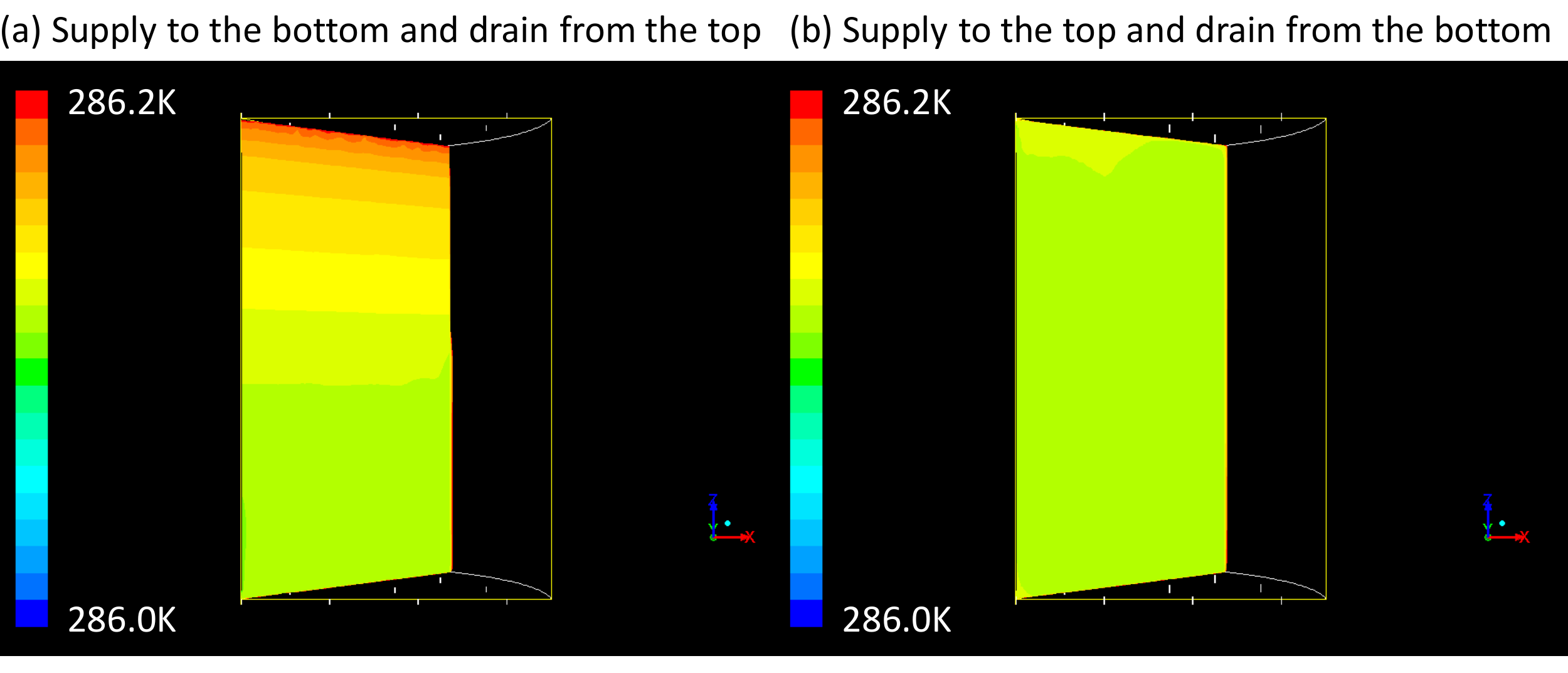}
\includegraphics[width=0.95\textwidth]{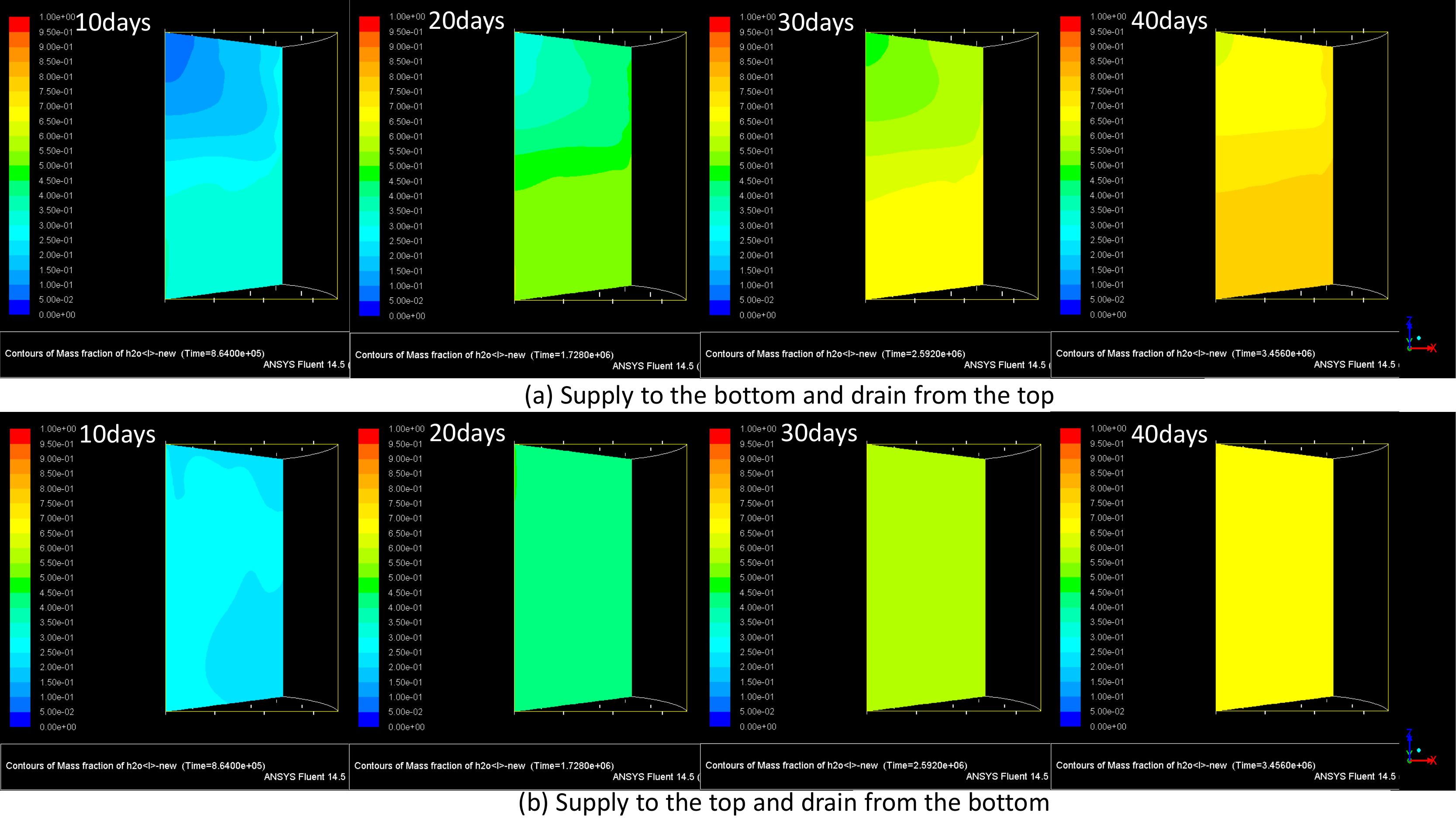}
\caption{Water temperature distributions (top 2 figures) 
  and water replacement efficiencies 
  as the result of water flow simulations. As the tank is in cylindrical shape and the 
 water inlets and outlets are distributed symmetrically,  
 only 1/6 of the tank was simulated with symmetric boundary condition and shown here.
  (a) The case for supplying water from the bottom of the
  tank and draining water from the top of the tank. (b) The case for
  supplying water from the top of the tank and draining water from the
  bottom of the tank. The elapsed days since the recirculation starts
  are indicated. In this simulation, at first the tank was filled with
  old water ($= 0$, blue), then new water ($= 1$, red) was supplied to the tank,
  therefore the color scale in the figures corresponds to the water
  replacement efficiency. After 40 days case (a) is more reddish, while case (b) is more uniform.}
\label{fig:flowsim}
\end{center}
\end{figure}

Water flow in the tank directly affects the water quality and the
physics results, therefore water flow simulations for the baseline
design tank were conducted. Water flow is determined not only by the
total water flow rate but also by detector geometry, the configuration
of water inlets and outlets, supply water temperature, heat sources in
the tank, surrounding rock temperature and so on. The input parameters
are summarized in Table~\ref{tab:simpara}, and the main results are
shown in Figure~\ref{fig:flowsim}. 
When cold water is supplied from the bottom of the tank, convection in the tank is suppressed
and the flow becomes laminar, resulting in effective water replacement. When cold
water is supplied from the top of the tank, large convection is evoked and the water quality in the
tank becomes uniform, spoiling effective water replacement. Actually this behavior was
confirrmed in Super-Kamiokande's 50 kton tank and seem to be common to cylindrical tanks; 
thus the water flow in Hyper-Kamiokande should be controlled as in Super-Kamiokande.

\begin{table}[htb]
\begin{center}
\scalebox{1}[1]{
\begin{tabular}{l|r}
\hline\hline
ID flow rate &    271.8 m$^3$/h      \\ 
OD flow rate &    37.9 m$^3$/h       \\  
Inlets/Outlets & 65A$\times$37/65A$\times$37 \\
ID boundary condition &  Inlet: 0.61 m/s, 286K Outlet: 0Pa \\
OD boundary condition &  Inlet: 0.67 m/s, 286K Outlet: 0Pa \\
Supply water temperature &  13.0 $^{\circ}$C   \\
Top level rock temperature &  16.7 $^{\circ}$C  \\
Bottom level rock temperature &  17.7 $^{\circ}$C \\
Heat flux from the PMT/electronics/coil & 3.2W/m$^2$\\
Total heat form ID top and bottom &  2100W and 2100W \\
Total heat from ID wall           & 6502W \\
Total heat from OD wall(rock)     & 5384W \\
Water density               & 999.4 kg/m$^2$ @286 K,  998.4 kg/m$^2$ @292 K \\
Water heat conductivity & 0.587 W/m/K  @286 K,  0.597 W/m/K  @292 K  \\
Water viscosity             & 0.0012 kg/m/s  @286 K,  0.0010 kg/m/s  @292 K \\
\hline\hline
\end{tabular}
}
\caption{Input parameters for the water flow simulations.}
\label{tab:simpara}
\end{center}
\end{table}

\subsubsection{Radon in the water}
\label{sec:radon-in-water}
The dominant low energy background is expected to be radon and 
the dominant radon source in the tank is expected to be the PMTs
themselves. The radon emanation from Hyper-Kamiokande photon sensors have 
not been measured yet, but each Super-Kamiokande ID PMT emanates about 10 mBq 
and the measured radon concentration in the Super-Kamiokande water is 2 mBq/$m^3$.
Super-Kamiokande has 11129 ID PMTs and 50\,ktons of water, therefore
the average radon concentration should be around
10 mBq/PMT $\times$ 11129 PMTs / 50000 tons = 2.2 mBq/$m^3$.
Accordingly, the radon concentration expected in one Hyper-Kamiokande HD tank is
10 mBq/PMT $\times$ 40000 PMTs / 258000 tons = 1.6 mBq/$m^3$.

Regarding radon suppression, the Hyper-Kamiokande water system includes Super-Kamiokande-based vacuum
degasifiers which reduce radon by about one order of magnitude as shown
in Figure~\ref{fig:1st2ndsys}. That being said, in the experience of Super-Kamiokande the best way to reduce
radon is by inducing a gentle laminar flow in the fiducial volume, 
allowing the radon to primarily decay close to the PMTs (i.e., not in
the fiducial volume) where it can do the least harm.

\subsubsection{Gd option}
In order to realize the many physics benefits provided by efficient
tagging of neutrons in water, it has been proposed (and recently approved) 
to add dissolved gadolinium sulfate to Super-Kamiokande.  As a result, 
over a period of
years much effort has gone into the design and demonstration of a
specialized water system capable of maintaining the exceptional water
transparency discussed above, while at the same time maintaining the
desired level of dissolved gadolinium in solution.  In other words,
somehow the water must be continuously recirculated and cleaned of
everything {\em except} gadolinium sulfate.

Built in 2007, a 0.2 tons/hour prototype selective filtration system at 
the University of California, Irvine, led in 2010 to the 3 tons/hour 
system at the heart of the Kamioka-based EGADS (Evaluating Gadolinium's 
Action on Detector Systems) project.  EGADS has now shown that this novel 
selective water filtration technology --- known as a "molecular band-pass 
filter" --- is both feasible and scalable.
It continuously improves and then maintains the transparency of water loaded with
\mbox{Gd$_2$(SO$_4$)$_3$} to SK ultrapure water levels, removing
unwanted impurities while simultaneously and indefinitely retaining
the desired levels of both the gadolinium and sulfate ions.

Since EGADS was built specifically to show that gadolinium loading
would be feasible in Super-K, scalability was always an important
design criterion. Therefore, from the beginning the EGADS band-pass
system was conceived of as a modularized design.  It uses
cost-effective, readily available components operating in parallel to
achieve the desired throughput and assure serviceability.

\begin{figure}
    \begin{center}
      \includegraphics[width=0.8\textwidth]{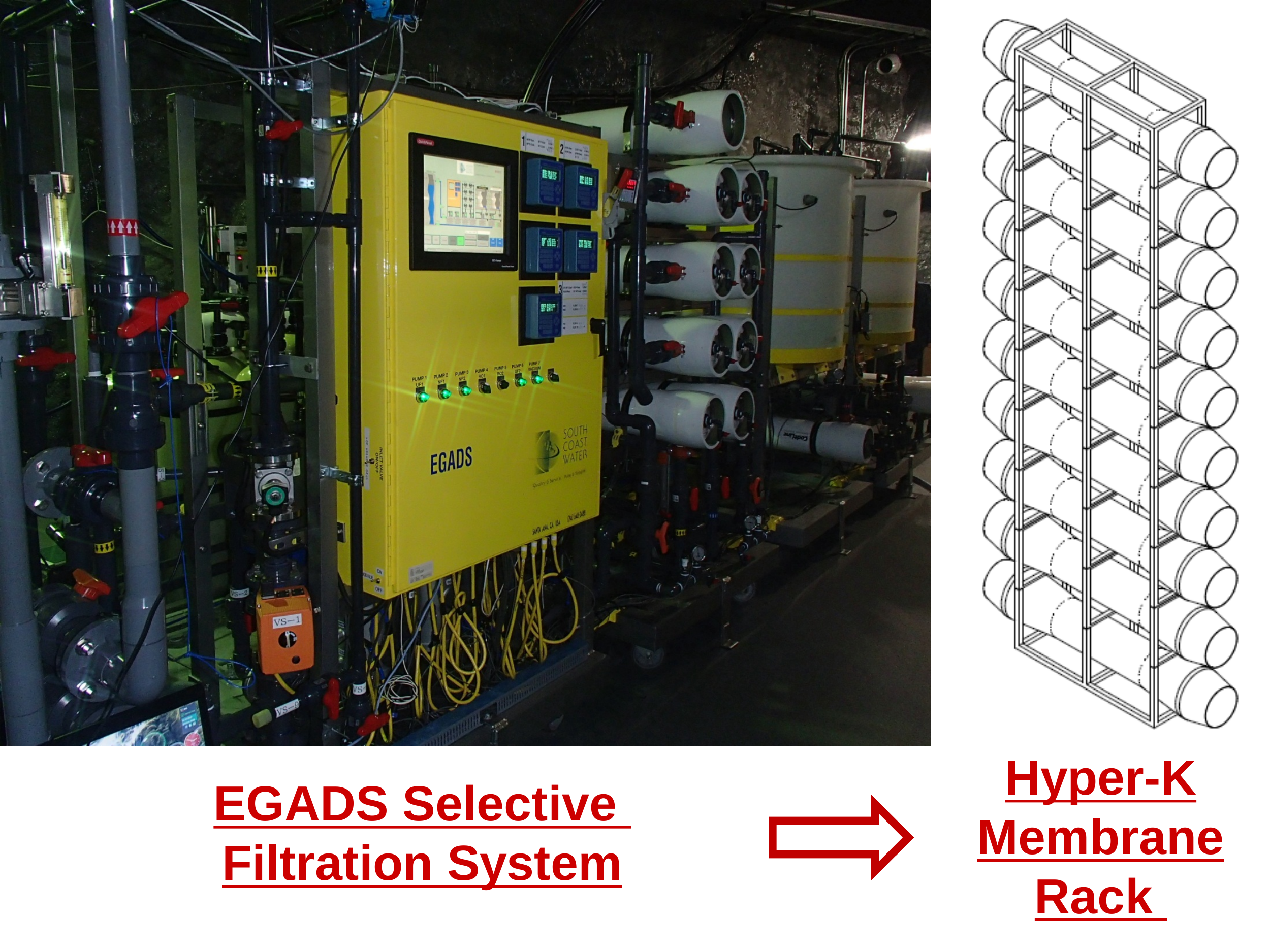}
      \caption[Scaling the EGADS Band-Pass to Hyper-Kamiokande]
              {Scaling the modular EGADS selective filtration
                band-pass for Hyper-Kamiokande. One rack of filtration
                membrane housings is shown here;
                Figure~\ref{fig:Gd_WS} shows many of them arranged
                into a functional selective filtration system.}
      \label{fig:rack}
    \end{center}
\end{figure}

As the band-pass design is modular and uses off-the-shelf equipment,
albeit in novel ways, scaling it up from the current 3~tons/hour to
60~tons/hour for Super-Kamiokande, or 310\,tons/hour for 
Hyper-Kamiokande, is straightforward.  Figure~\ref{fig:rack} indicates
how one rack of filtration membrane housings, the modular unit around
which the Hyper-K band-pass system is designed, is derived from the
operating EGADS selective filtration system.

\begin{figure}
    \begin{center}
      \includegraphics[width=0.8\textwidth]{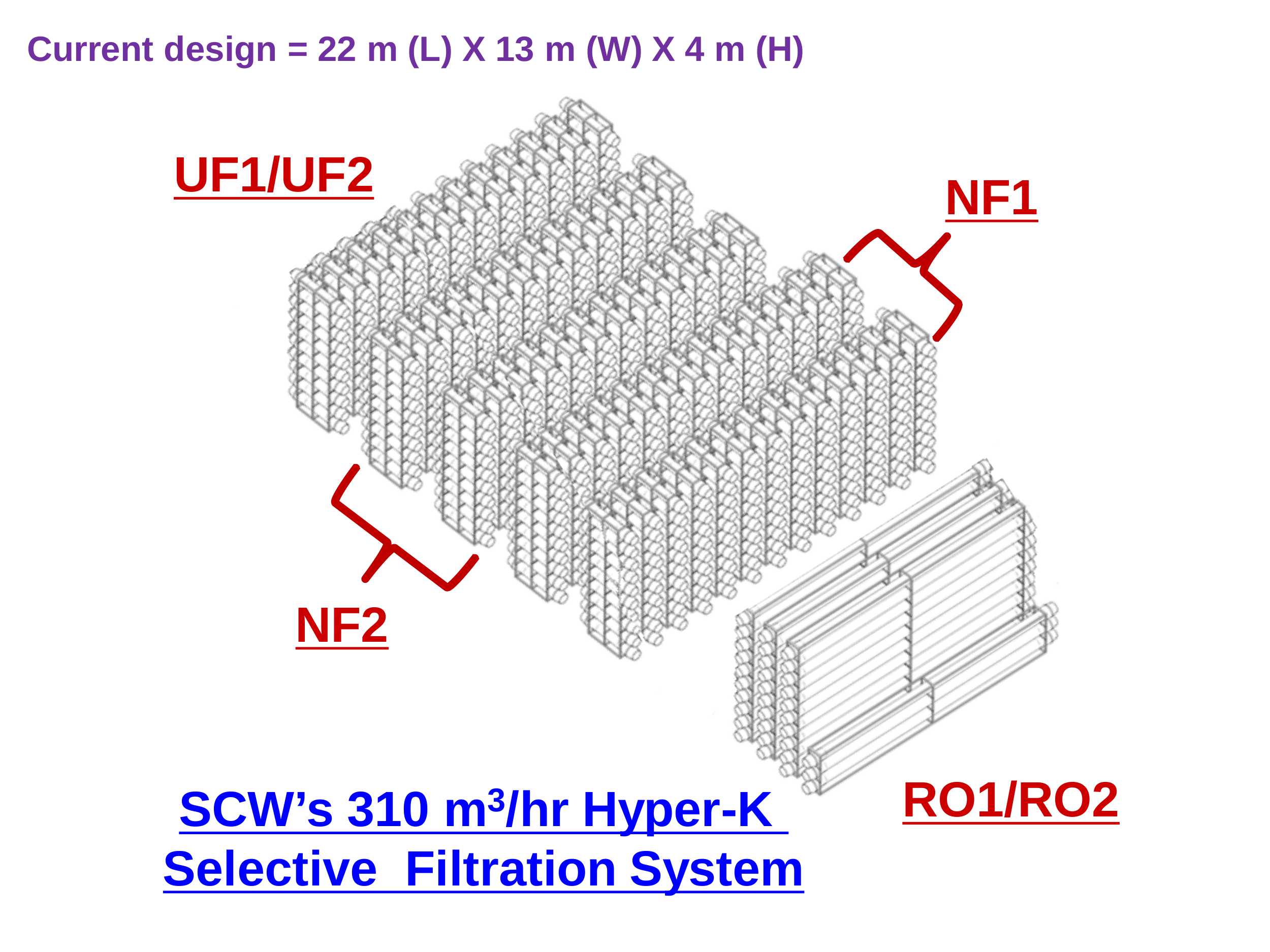}
      \caption[Gadolinium-capable water system for Hyper-Kamiokande.]
              {Gadolinium-capable water system for
                Hyper-Kamiokande. Two stages each of nanofilters (NF),
                ultrafilters (UF), and reverse osmosis (RO) membrane
                racks are shown, sufficient to provide selectively
                filtered water for Hyper-K.} 
      \label{fig:Gd_WS}
    \end{center}
\end{figure}

Figure~\ref{fig:Gd_WS} depicts how the modular rack from
Figure~\ref{fig:rack} may be duplicated and operated in parallel to
provide the needed throughput.  Further design simplification and cost
savings are achieved by using this standardized membrane housing array
and filling the housings with a variety of filter membranes, each of
which handles a different cleaning task.  These components include
nanofilters (NF), ultrafilters (UF), and reverse osmosis (RO)
membranes; in each case there are two stages.  Note that the layout
shown in Figure~\ref{fig:Gd_WS} is schematic in nature.  Due to space
constraints underground the illustrated system would likely be split
into two levels, one atop the other.

Using the baseline Hyper-K design, the system shown in
Figure~\ref{fig:Gd_WS} represents what is needed for selective
filtration following the addition of
gadolinium sulfate to the Hyper-K water. It is assumed that pure water
for filling the detector will be provided by the main, non-Gd-capable
water system described above.  The Gd-specific ``molecular band-pass''
system described here will be augmented with additional Gd-capable water  
handling equipment -- also demonstrated by and scaled up from a working 
EGADS version -- known as a ``fast recirculation'' system. The Hyper-K 
fast recirculation system will be used in conjunction with HK's 
band-pass to maintain the Gd-loaded water's quality.

Due to gadolinium sulfate's benign nature with regards to the usual
detector components (materials compatibility was another component of
the EGADS study), retaining the ability to add gadolinium to
Hyper-Kamiokande primarily means retaining the option of adding
gadolinium filtration capability to the Hyper-K water system.  Indeed,
if gadolinium works as well as expected in Super-Kamiokande over the
next few years, it is hard to imagine that a next-generation detector
like Hyper-K would not also want to enjoy the physics advantages a
gadolinium-loaded Super-K would already have.  Therefore, we have been
careful to keep the possibility of gadolinium loading in mind when
designing the overall Hyper-Kamiokande water system.

\clearpage
\graphicspath{{design-photosensor/figures/}}

%%%%%%%%%%%%%%%%%%%%%%%%%%%%%%%%%%%%%%%%%%%%%%%%%%%%%%%%%%%%%%%%%%%%%%%%%%%%%%%
%%%%%%%%%%%%%%%%%%%%%%%%%%%%%%%%%%%%%%%%%%%%%%%%%%%%%%%%%%%%%%%%%%%%%%%%%%%%%%%

    \subsection{Photosensors}\label{section:photosensors}

%%%%%%%%%%%%%%%%%%%%%%%%%%%%%%%%%%%%%%%%%%%%%%%%%%%%%%%%%%%%%%%%%%%%%%%%%%%%%%%
        \subsubsection{Introduction}\label{section:photosensors:intro} 

The Hyper-K photosensors detect the
Cherenkov ring pattern created in particle interactions. Photosensors
view the ID, as well as the OD where they are used to tag particles
entering or exiting the detector.   

\begin{figure}[htbp]
     \begin{center}
 \includegraphics[width=0.4\textwidth]{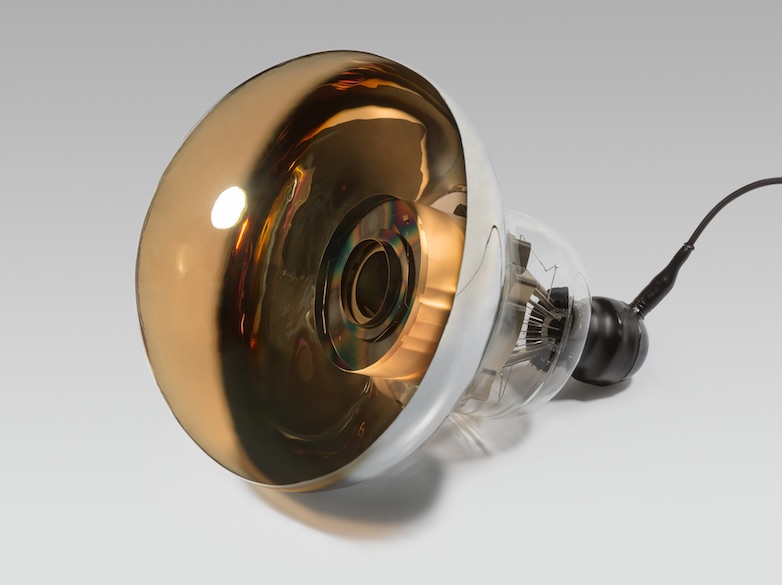}%
 \caption{Picture of the HQE 50\,cm box-and-line R12860 PMT.}
 \label{fig:R12860pic}
     \end{center}
\end{figure}

The ID photosensor was newly developed for Hyper-K to meet the
requirements listed in Table~\ref{photosensorrequirement}.  The new
photosensor is based on the well established and reliable design of
the 50\,cm R3600 PMT by Hamamatsu Photonics K.K. with a Venetian blind
dynode type, which is used for Super-K, and the 43\,cm PMT with a
box-and-line dynode (Hamamatsu R7250), which is used for the KamLAND
experiment.  Further improvements include a higher quantum
efficiency photocathode and an optimized box-and-line dynode, resulting in the new
photosensor, a Hamamatsu R12860-HQE PMT (Figure~\ref{fig:R12860pic}).

\begin{table}[htbp]
 \begin{center}
  \begin{tabular}{l|l|l|l}
   \hline \hline
     Requirements                     & Value                &      & Conditions \\                                    
   \hline
   Photon detection efficiency        & 26\%                 & Typ. & Quantum Efficiency $\times$ Collection Efficiency\\     
                                      &                      &      & ~around 400\,nm wavelength                        \\          %
                                      & (10\%)               &      & (including Photo-Coverage on the inner detection area)\\                
   Timing resolution                  & 5.2\,nsec            & FWHM, Typ. & Single Photoelectron (PE)              \\                          
   Charge resolution                  & 50\%                 & $\sigma$, Typ. & Single PE \\                          
   Signal window                      & 200\,nsec            & Max. & Time window covering more than 95\% \\           
                                      &                      &      & ~of total integrated charge\\
   Dynamic range                      & 2 photons/cm$^{2}$   & Min. & Per detection area on wall\\                     %&
   Gain                               & $10^{7} \sim 10^{8}$ & Typ. & \\                                              
   Afterpulse rate                    & 5\%                  & Max. & For single PE, relative to the primary pulse\\   
   Rate tolerance                     & 10\,MHz               & Min. & Single PE pulse, within 10\% change of gain \\   
   Magnetic field tolerance           & 100\,mG               & Min. & Within 10\% degradation \\                       
   Life time                          & 20\,years             & Min. & Less than 10\% dead rate \\                      
   Pressure rating                    & 0.8\,MPa              & Min. & Static, load in water \\                        
   \hline \hline
  \end{tabular}
  \caption{Minimum requirements of the Hyper-K ID photosensors.
           The dark rate is also an important parameter, but its
           minimum required value depends on the photosensor
           specification and is specific to each physics topic.
           It will be explored in the future using the Hyper-K simulation.
         }
  \label{photosensorrequirement}
 \end{center}
\end{table}

The OD photosensor design is based on the Super-K OD photosensor, the
20\,cm Hamamatsu R5912 PMT, with an improved high quantum efficiency photocathod and a hard waterproofing cover to stand the 60\,m deep water pressure in Hyper-K.

This section describes the characteristics of
these photosensors. In addition, we present prospects for alternative
options, which may be adopted in future.

%%%%%%%%%%%%%%%%%%%%%%%%%%%%%%%%%%%%%%%%%%%%%%%%%%%%%%%%%%%%%%%%%%%%%%%%%%%%%%%
        \subsubsection{Photosensor for Inner Detector}\label{section:photosensors:ID} 

%%%%%%%%%%%%%%%%%%%%%%%%%%%%%%%%%%%%%%%%%%%%%%%%%%%%%%%%%%%%%%%%%%%%%%%%%%%%%%%
        \subsubsubsection{Performance}\label{section:photosensors:IDperformance} 
A newly developed 50 \,cm R12860-HQE PMT (HQE, high quantum efficiency) for
Hyper-K by Hamamatsu (hereafter referred to as the HQE B\&L PMT), has a faster time response, better
charge resolution and a higher detection efficiency with a stable
mechanical structure, compared to the existing large aperture PMTs.
This section describes the specifications of the HQE B\&L PMT, as well
as the safety design that ensures longevity of operation.

\begin{figure}[htbp]
     \begin{center}
 \includegraphics[width=0.8\textwidth]{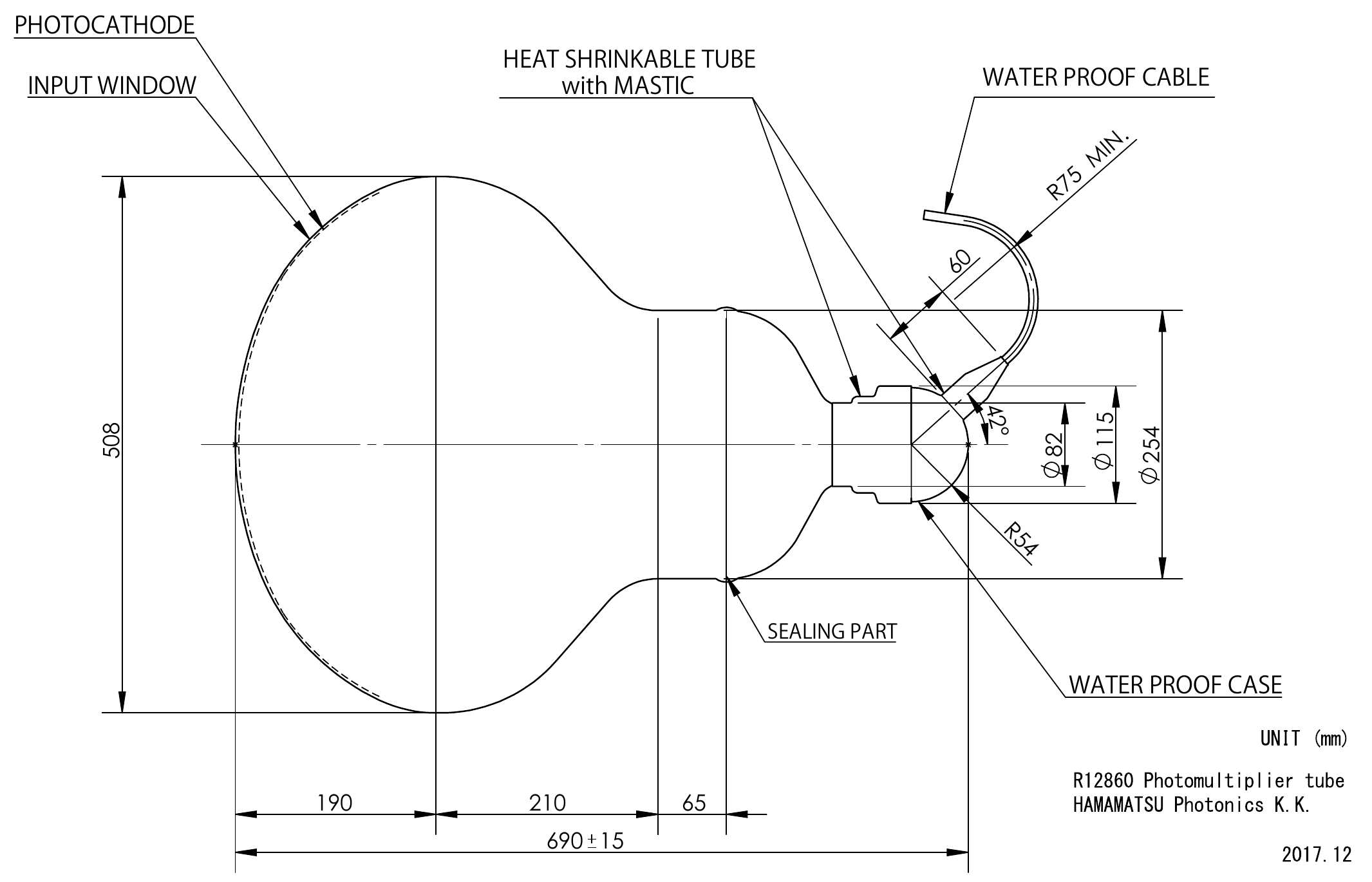}%
 \caption{Outline of the HQE 50\,cm box-and-line R12860 PMT.}
 \label{fig:R12860}
     \end{center}

\end{figure}

\begin{figure}[htbp]
     \begin{center}
 \includegraphics[width=0.8\textwidth]{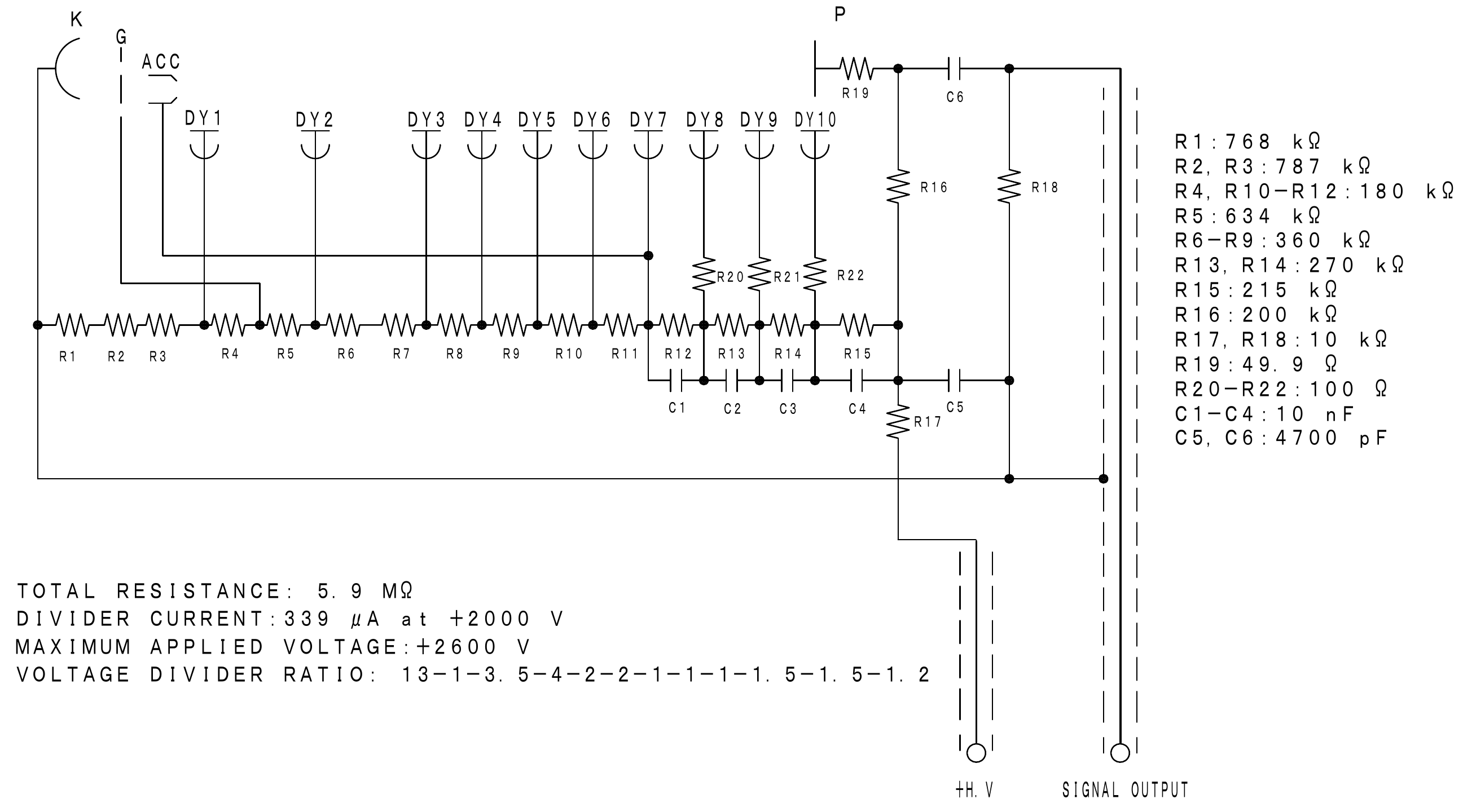}%
 \caption{PMT base circuit of the HQE box-and-line R12860 PMT.}
 \label{fig:R12860bleeder}
     \end{center}
\end{figure}

%%%%%%%%%%%%%%%%%%%%%%%%%%%%%%%%%%%%%%%%%%%%%%%%%%%%%%%%%%%%%%%%%%%%%%%%%%%%%%%
        \subsubsubsubsection{Design and
Specifications}\label{section:photosensors:IDperformance:design} 
Figure~\ref{fig:R12860} shows a side view of the HQE B\&L PMT, whose
shape is similar to the PMT used in Super-K. Hence, the support structure
developed in Super-K to attach the PMT is also appropriate for
the HQE B\&L PMT in Hyper-K. The dynode structure and the
surface curvature were improved.  A typical bias voltage of 2,000\,V is
divided to each dynode by a PMT base circuit such the one shown in
Figure~\ref{fig:R12860bleeder}.  The specifications for a typical HQE
B\&L PMT is listed in Table~\ref{pmtspec}.

\begin{table}[htbp]
 \begin{center}
  \begin{tabular}{l|l}
   \hline \hline
   Shape                  & Hemispherical \\
   Photocathode area      & 50\,cm diameter (20 inches)\\
   Bulb material          & Borosilicate glass ($\sim3$\,mm) \\
   Photocathode material  & Bialkali (Sb-K-Cs) \\
   Quantum efficiency     & 30\,\% typical at $\lambda=390$\,nm \\
   Collection efficiency  & 95\,\% at $10^7$ gain \\
   Dynodes                & 10\,stage box-and-line type\\
   Gain                   & 10$^7$ at $\sim2000$\,V \\
   Dark pulse rate        & $\sim8$\,kHz at $10^7$ gain (13 Celsius degrees, after stabilization for a long period) \\
   Weight                 & 9\,kg (without cable) \\ 
   Volume                 & 61,000\,cm$^3$ \\
   Pressure tolerance     & 1.25\,MPa water proof \\
   \hline \hline
  \end{tabular}
  \caption{Specifications of the 50\,cm R12860-HQE PMT by Hamamatsu.}
  \label{pmtspec}
 \end{center}
\end{table}

\begin{figure}[htbp]
 \includegraphics[width=0.4\textwidth]{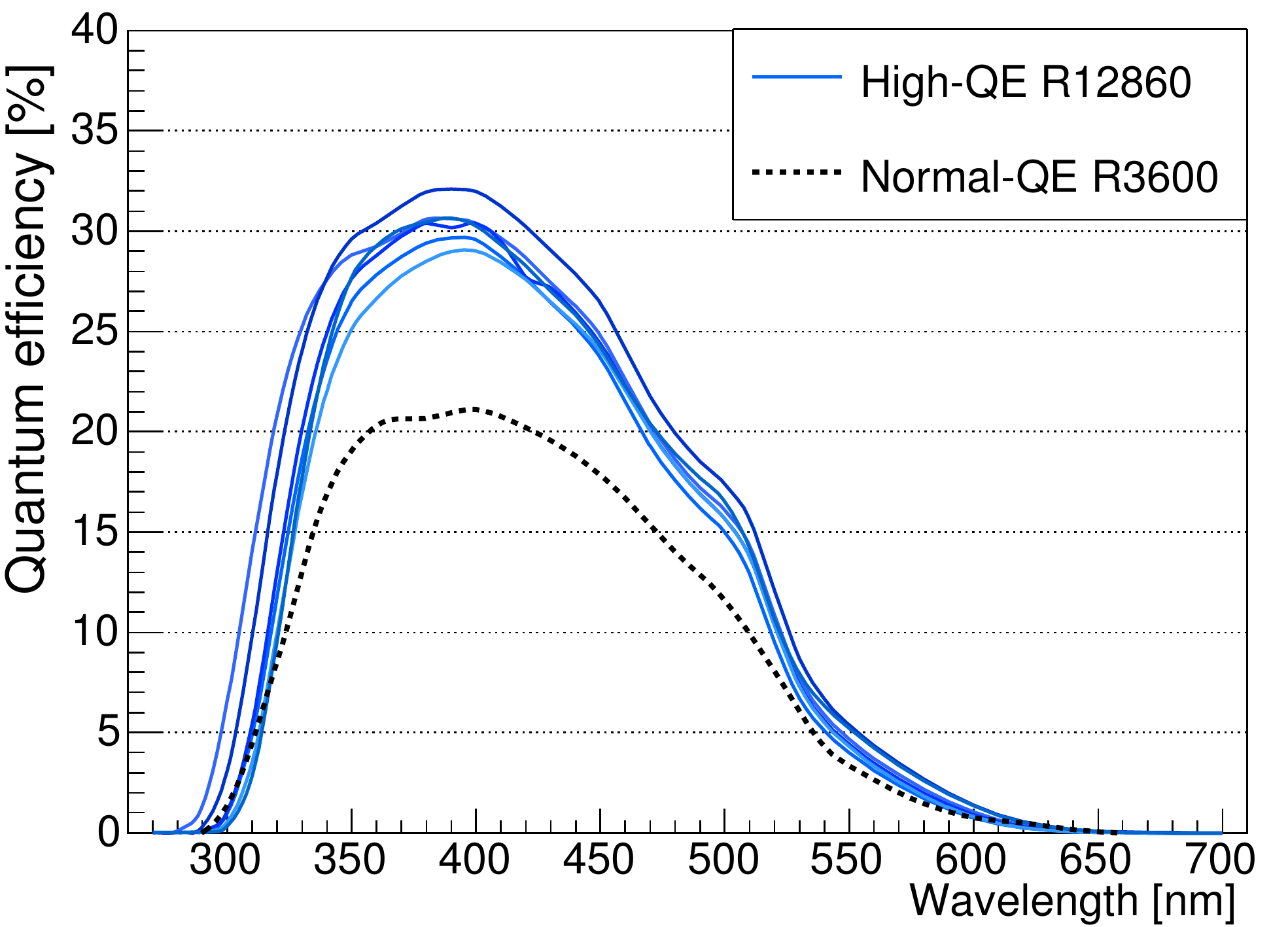}%
 \caption{Measured QE for six high-QE R12860 (solid lines) and a R3600 (dashed line).\label{fig:HQE}}
 \end{figure}

%%%%%%%%%%%%%%%%%%%%%%%%%%%%%%%%%%%%%%%%%%%%%%%%%%%%%%%%%%%%%%%%%%%%%%%%%%%%%%%
\subsubsubsubsection{Detection
  Efficiency}\label{section:photosensors:IDperformance:efficiency} 
The single photon detection efficiency of the HQE B\&L PMT is a factor of two
better than the conventional R3600 in Super-K (Super-K PMT).
Figure~\ref{fig:HQE} shows the measured quantum efficiency (QE) of
several HQE B\&L PMTs as a function of wavelength compared with a
typical QE curve of the Super-K PMT (dotted line).  
The QE of the R12860-HQE PMT is typically 30\% at peak wavelength around 390\,nm, while the peak QE of the Super-K PMT is about 22\%.

\begin{figure}[htbp]
 \includegraphics[width=0.7\textwidth]{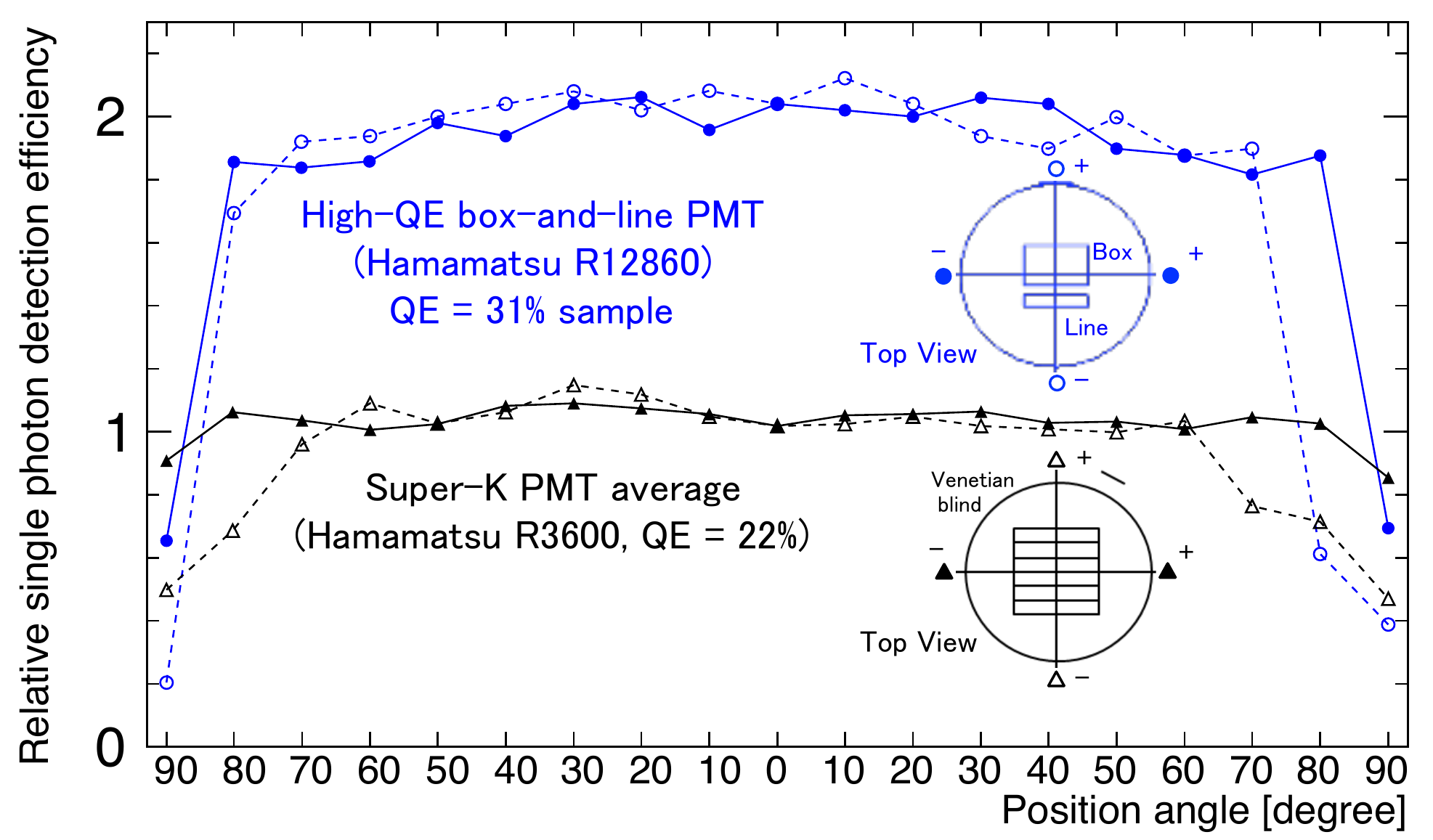}%
 \caption{Relative single photon detection efficiency as a function of the position in the photocathode, where a position angle is zero at the PMT center and $\pm 90^{\circ}$ at the edges.
          The dashed line is the scan along the symmetric line of the box-and-line dynode whereas the solid line is along the perpendicular direction of the symmetric line.
          The detection efficiency represents QE, CE and cut efficiency of the single photoelectron at 0.25 PE.
          A HQE B\&L PMT with a 31\% QE sample shows a high detection efficiency by a factor of two compared with normal QE Super-K PMTs (QE = 22\%, based on an average of four samples).\label{fig:EffUniformity}}
 \end{figure}

 The HQE B\&L PMT has a high collection efficiency and large sensitive photocathode area. The
 photocathode area with a collection efficiency (CE) of 50\% or better
 is 49.2\,cm for the HQE B\&L PMT, compared to 46\,cm in case of the
 Super-K PMT and 43.2\,cm in the KamLAND PMT.  Compared with 73\% CE
 of the Super-K PMT within the 46\,cm area, the HQE B\&L PMT reaches
 95\% in the same area and still keeps a high efficiency of 87\% even
 in the full 50\,cm area.  This high CE was achieved by optimizing the
 glass curvature and the focusing electrode, in addition to the use of a
 box-and-line dynode.  In the Super-K Venetian blind dynode, the
 photoelectron sometimes misses the first dynode while the wide first
 box dynode of the box-and-line accepts almost all the photoelectrons.
 This also helps improving the single photoelectron (PE) charge
 resolution, which then improves the hit efficiency at a single PE
 level.  By measuring the single PE level, we confirmed the CE
 improvement by a factor of 1.4 compared with the Super-K PMT, and 1.9
 in the total efficiency including HQE.
 Figure~\ref{fig:EffUniformity} shows that the CE response is quite
 uniform over the whole PMT surface despite the asymmetric dynode
 structure.

A relative CE loss in case of a 100\,mG residual Earth magnetic field is at most 2\% in the worst direction, or negligible when the PMT is aligned to avoid this direction on the tank wall.
The reduction of geomagnetism up to 100\,mG can be achieved by active shielding by coils.

\begin{figure}[htbp]
 \begin{tabular}{c}
   \begin{minipage}{0.45\hsize}
     \begin{center}
       \includegraphics[width=0.9\textwidth]{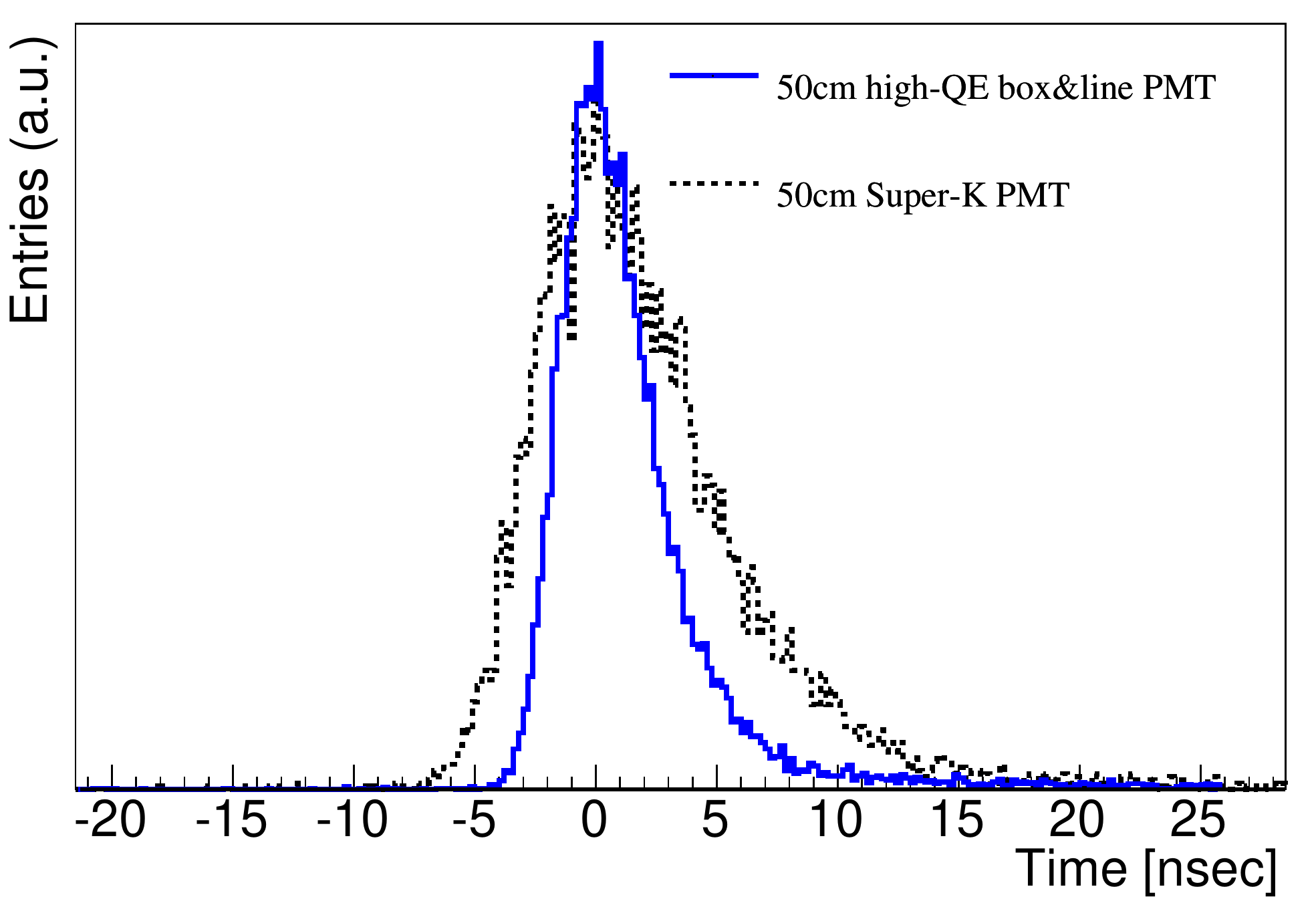}
       \caption{Transit time distribution at single photoelectron, compared with the Super-K PMT in dotted line. \label{fig:PMTTTS}}
     \end{center}
   \end{minipage}

\hspace{1cm}

   \begin{minipage}{0.45\hsize}
     \begin{center}
       \includegraphics[width=0.9\textwidth]{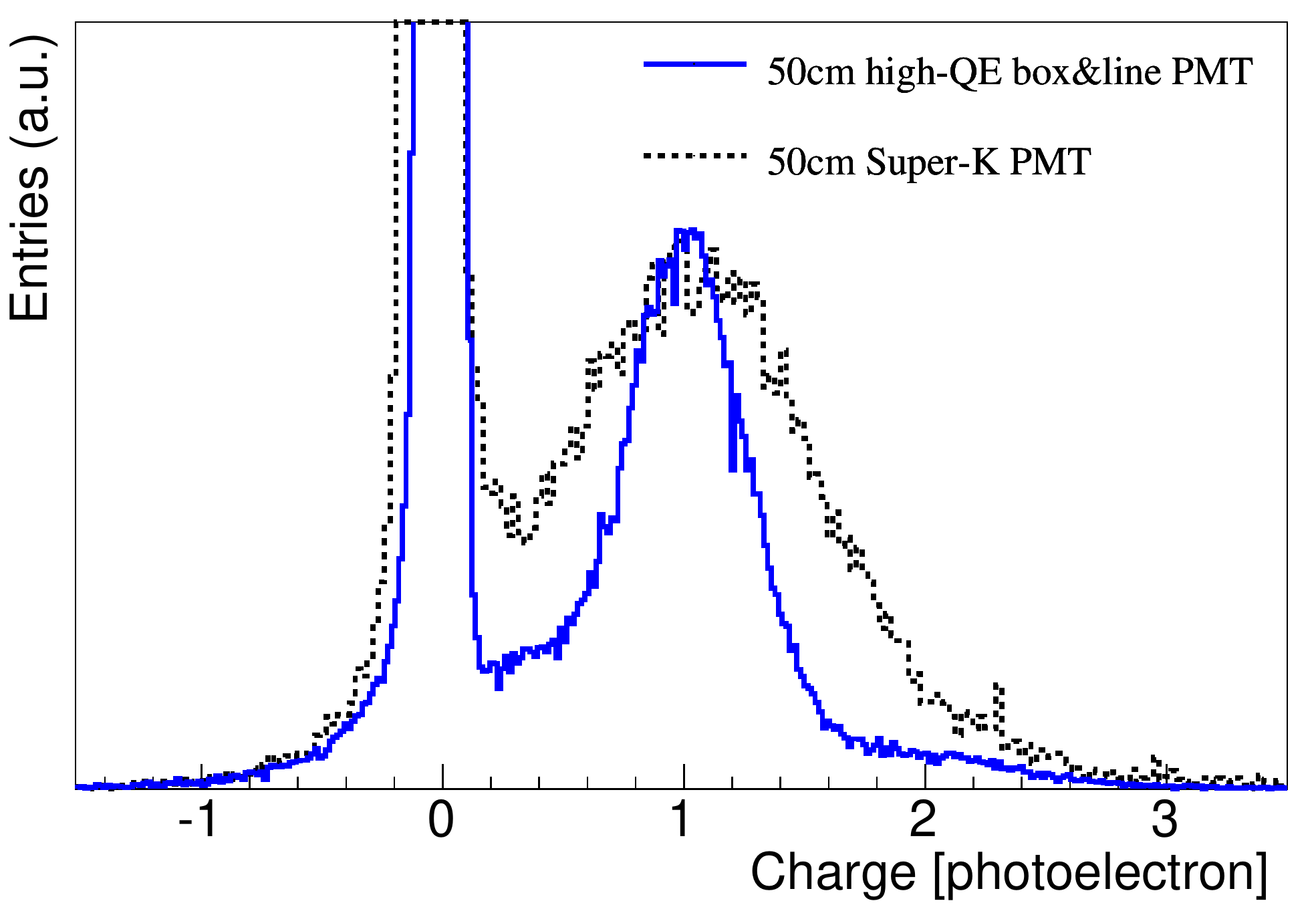}
       \caption{Single photoelectron distribution with pedestal, compared with the Super-K PMT in dotted line. \label{fig:PMT1PE}}
     \end{center}
   \end{minipage}
 \end{tabular}
\end{figure}

%%%%%%%%%%%%%%%%%%%%%%%%%%%%%%%%%%%%%%%%%%%%%%%%%%%%%%%%%%%%%%%%%%%%%%%%%%%%%%%
        \subsubsubsubsection{Performance of Single Photoelectron
Detection}\label{section:photosensors:IDperformance:1pe} 
The single photoelectron pulse in a HQE B\&L PMT has a 6.7\,nsec rise
time (10\% -- 90\%) and 13.0\,nsec FWHM without ringing, which is faster
than the 10.6\,nsec rise time and 18.5\,nsec FWHM in the Super-K PMT.  The
time resolution for single PE signals is 1.1\,nsec in $\sigma$ for the fast
left side of the transit time peak in Figure~\ref{fig:PMTTTS} and 4.1\,nsec
at FWHM, which is about half of the Super-K PMTs.  This would be an
important factor to improve the reconstruction performance of events
in Hyper-K.

The nominal gain is $10^7$ and can be adjusted for several factors in
a range between 1500\,V to 2200\,V.
Figure~\ref{fig:PMT1PE} shows the charge distribution, where the 35\%
resolution in $\sigma$ of the single PE is better for the HQE B\&L PMT compared to the 50\%
of the Super-K PMT.  The peak-to-valley ratio is about 4, defined by
the ratio of the height of the single PE peak to that of the valley
between peaks.

\begin{figure}
  \begin{center}
  \includegraphics[width=0.4\textwidth]{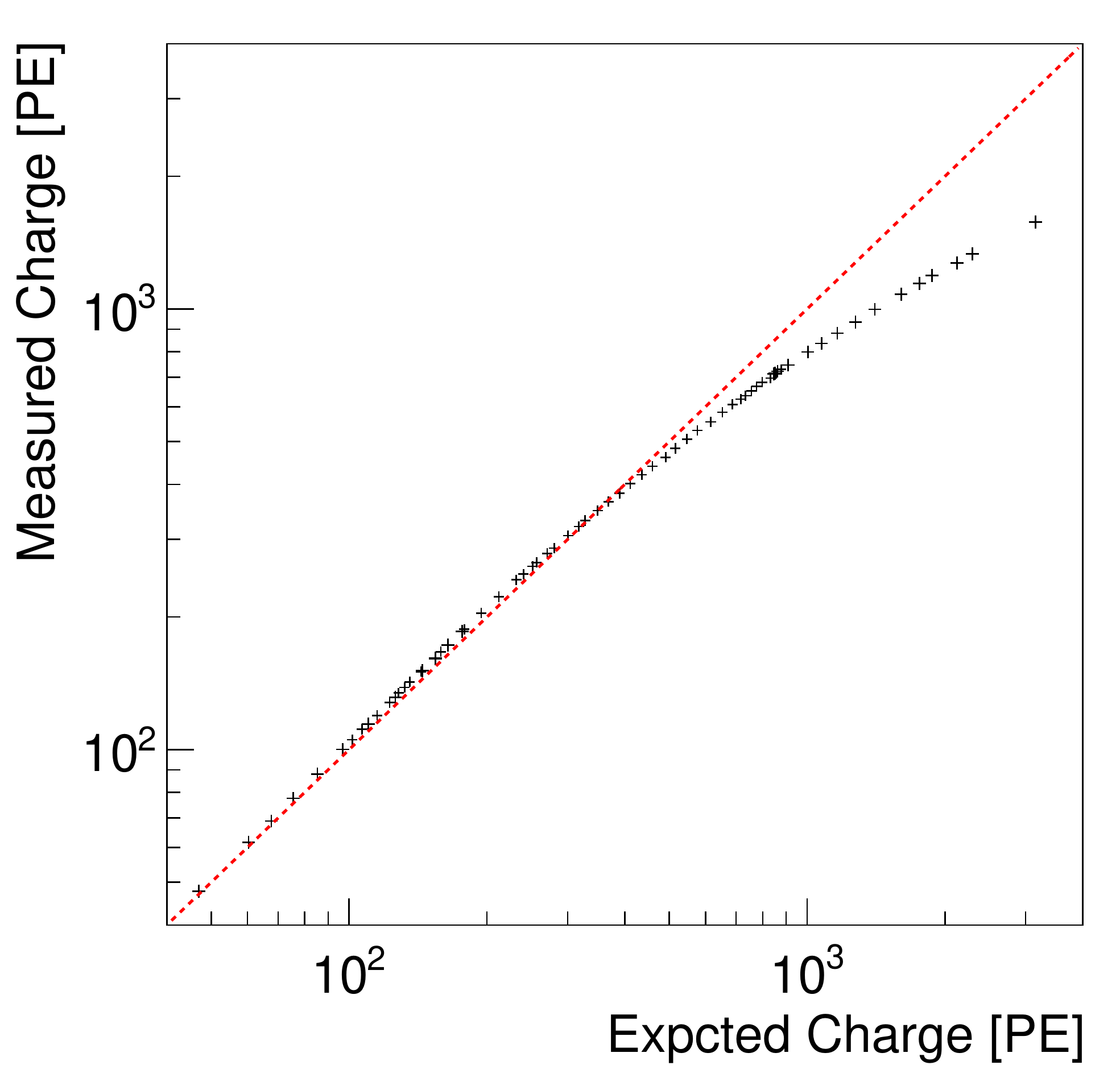}
        \caption{Output linearity of the HQE B\&L PMT in charge, where a dotted line shows an ideal linear response.
                 It is derived by measurements of a coincident emission by two light sources compared with an expectation by sum of individual detections. }
        \label{fig:pmtlinearity}
  \end{center}
\end{figure}

%%%%%%%%%%%%%%%%%%%%%%%%%%%%%%%%%%%%%%%%%%%%%%%%%%%%%%%%%%%%%%%%%%%%%%%%%%%%%%%
        \subsubsubsubsection{Gain Stability}\label{section:photosensors:IDperformance:gainstability} 
Because the Hyper-K detector aims for various physics subjects in a wide energy range, the PMT is required to have a wide dynamic range. 
The Super-K PMTs have an output linearity up to 250 PEs in
charge according to the specifications and up to about 700 PEs as measured in Super-K
(with up to 5\% distortion)~\cite{Abe:2013gga}, while the linearity of
the HQE B\&L PMT was measured to be within 5\% up to 470 PEs as seen
in Figure~\ref{fig:pmtlinearity}.  Even with more than 1,000 PEs, the
output is not saturated and the number of PEs can be calculated by
correcting the non-linear response.
The linearity range depends on the dynode current, and
can be optimized by changing the resistor values in the PMT base 
circuit.  This result demonstrates sufficient detection capabilities
in the wide MeV -- GeV region as in Super-K, as long as it is
corrected according to the response curve.

A fast recovery of gain for high signal rate is needed for supernova
observation, decay electrons from muons, and any accidental pileup
events. On the other hand, separation of individual signals in time is limited by the charge integration range,
that is 200\,nsec (equivalent to 5\,MHz) or more depending on the electronics.

\begin{figure}[htbp]
 \begin{tabular}{c}

   \begin{minipage}{0.46\hsize}
     \begin{center}
       \includegraphics[width=0.9\textwidth]{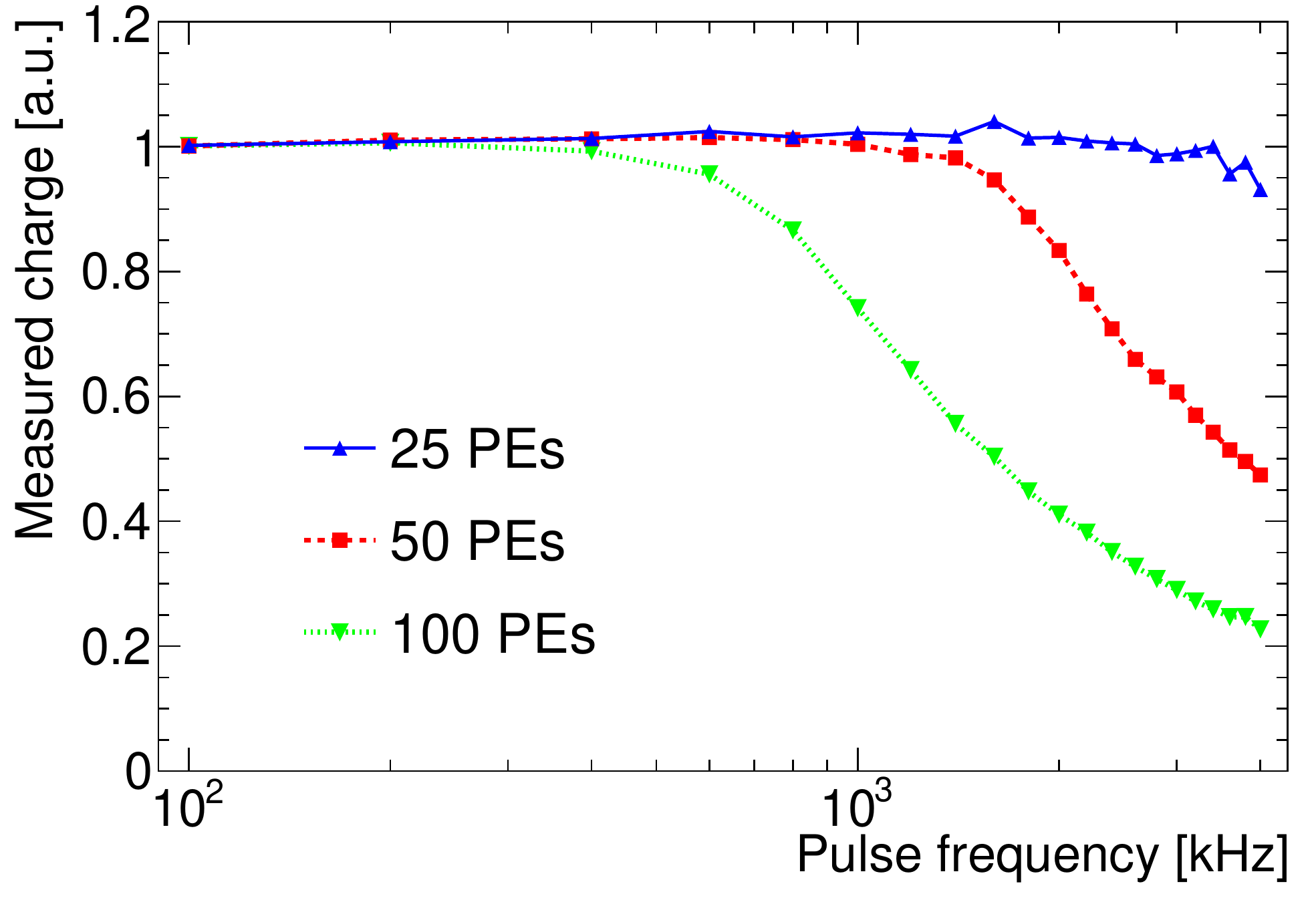}
       \caption{Measured charge as a function of the pulse rate in three light intensities of 25, 50 and 100 photoelectrons, relative to outputs at 100\,Hz.
                Each charge is calculated using the baseline just before the pulse. \label{fig:pmtratetolerance}}
     \end{center}
   \end{minipage}

\hspace{0.5cm}

   \begin{minipage}{0.46\hsize}
     \begin{center}
        \includegraphics[width=0.9\textwidth]{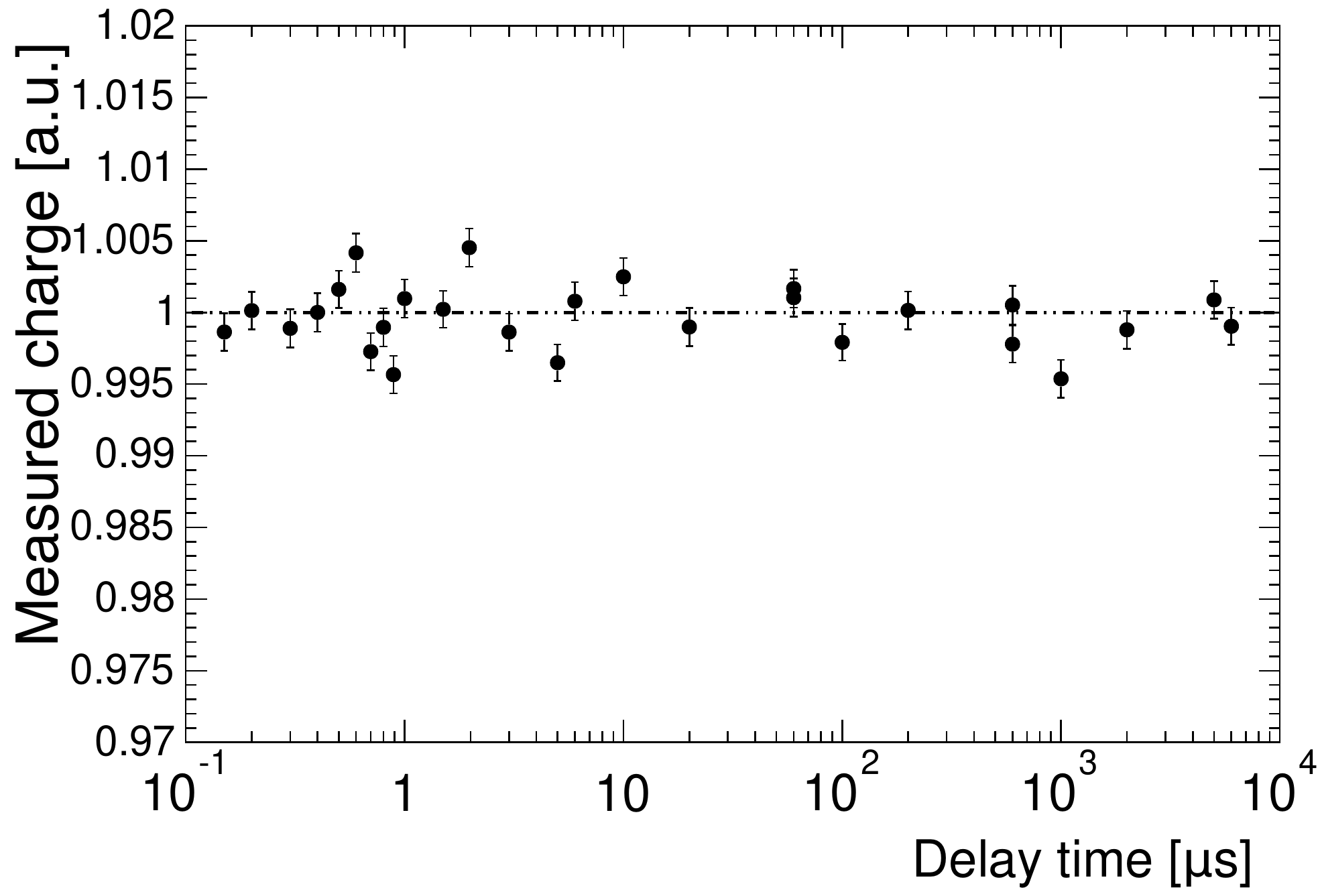}
        \caption{Output charge fidelity of a delayed pulse after a primary pulse, compared with no primary pulse.
                 The charge set is about 150 PEs at 10$^{7}$ gain for both primary and delayed pulses in various delayed time. }
        \label{fig:pmtgainrecovery}
     \end{center}
   \end{minipage}
 \end{tabular}
\end{figure}

The rate dependence of the output charge was measured at several light
intensities while varying the constant interval time of light pulses
(as shown in Figure~\ref{fig:pmtratetolerance}).  A 5\% drop is
observed at the output current of 170 $\mu$A.  It corresponds to 78
MHz in the single PE intensity or 1\,MHz in most of detected
intensities like at the level of few tens of PE.  This is sufficient
to detect possible burst physics events 
(Section~\ref{sec:supernova}).

Even when two near continuous events are detected, like in the case of
an event with a
decay particle, no loss of charge was observed for the
second delayed pulse.  By measuring two continuous pulses of about 150
PEs in both, the observed loss of gain is stable within 0.5\% as shown
in Figure~\ref{fig:pmtgainrecovery}. 
Therefore, the output charge fidelity for a delayed signal is sufficient with the HQE B\&L PMT.

\begin{figure}
  \begin{center}
  \includegraphics[width=0.5\textwidth]{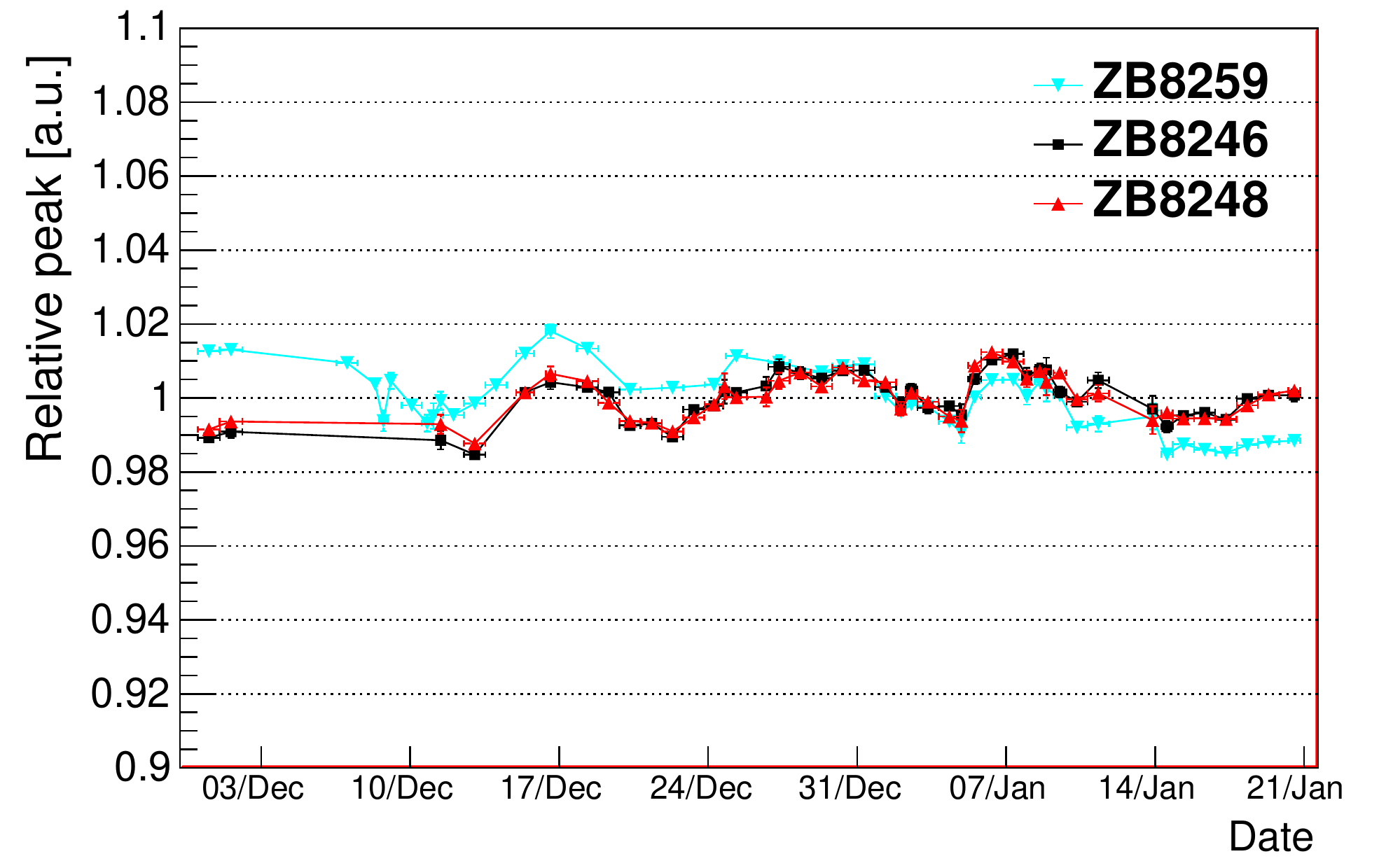}
  \caption{Relative charge of three HQE B\&L PMTs for two months in a 200\,ton water Cherenkov detector.
           Signals of several tens photoelectrons from a xenon light pulse were monitored. }
  \label{fig:pmtstability}
  \end{center}
\end{figure}

A long term stability test was performed on three HQE B\&L PMTs
in a 200\,ton water Cherenkov detector at the Kamioka mine, which was
constructed to evaluate the feasibility of anti-neutrino tagging via gadolinium doping
in water.  All the HQE B\&L PMTs functioned over two years, and
the gain measured using the charge peak resulting from a pulsed calibration source was stable
within 1\% RMS (Figure~\ref{fig:pmtstability}).

%%%%%%%%%%%%%%%%%%%%%%%%%%%%%%%%%%%%%%%%%%%%%%%%%%%%%%%%%%%%%%%%%%%%%%%%%%%%%%%
        \subsubsubsubsection{Backgrounds}\label{section:photosensors:IDperformance:BG} 
The dark hit rate originates from a thermionic emission on the
photocathode, and depends on the environmental temperature, the bias
high voltage and the accumulated operating time for stabilization.
For Hyper-K, the energy threshold for low energy physics studies depends largely on the PMT 
dark hit rate, because the sum of the PMT dark hits create fake event triggers.

The dark hit rates of several HQE B\&L PMTs were measured to be
8.3\,kHz at a temperature of 15\,$^\circ$C in air after a month-long
stabilization period. 
The adequacy of this dark hit rate on the physics sensitivities will be discussed in Section \ref{section:physics}.
Since the detection efficiency is doubled for the HQE B\&L PMTs, the obtained current dark hit rate is relatively high compared to the Super-K PMT (4.2\,kHz). 
A lower dark rate results in a better sensitivity to low energy events, thus
the HQE B\&L PMT production is being optimized to achieve a lower dark
hit rate. We expect further improvements within the next year.

\begin{figure}
  \begin{center}
  \includegraphics[width=0.85\textwidth]{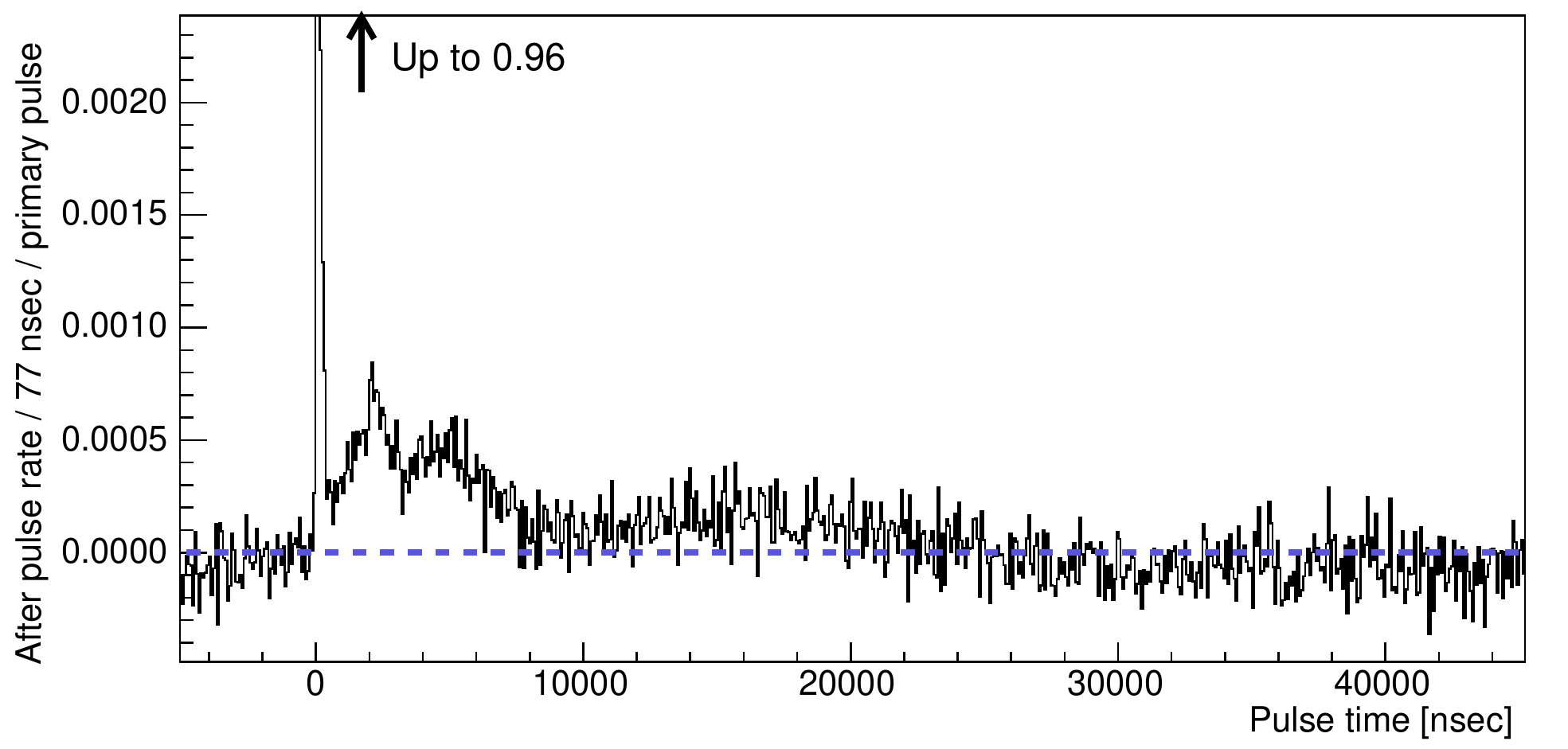}
  \caption{Time distribution of hits, where the primary single PE signal comes at zero and others are afterpulses.
           The dotted line represents the level of the dark hit which is set to zero.
           The expected value of the number of afterpulses is measured to be 0.05 for one primary pulse in this sample. }
  \label{fig:pmtafterpulse}
  \end{center}
\end{figure}

The afterpulse has a long delay of several microseconds order after
the primary PE, and can result in mis-reconstruction for tagging delayed
particles.  Afterpulsing is caused by a feedback of the ionized residual gas to
the photocathode, and several timing peaks appear due to the different gas
molecular masses.
As shown in Figure~\ref{fig:pmtafterpulse}, hit timing distribution of the HQE B\&L PMT has several peaks of afterpulses.
The afterpulse rate is, in total, less than 5\% relative to the main
pulse at single PE observation.  It is comparable to the Super-K PMT.

The radioactive contamination in the surface glass was measured by a
germanium semiconductor detector.  It is listed
in Table~\ref{pmtglassri} by Uranium series, Thorium series and
Potassium-40. The contamination of Potassium-40 was reduced by an
order of magnitude compared to the glass used in the Super-K PMT. 

\begin{table}[htbp]
 \begin{center}
  \begin{tabular}{l|lll}
   \hline \hline
           & U-chain & Th-chain & K$^{40}$ \\
   \hline
    Bq/kg  & 5.4 & 1.8 & 1.6 \\
    Bq/PMT & 34.5  & 11.3 & 10.5 \\
   \hline \hline
  \end{tabular}
  \caption{Radioactive contamination in glass for the HQE B\&L PMT (Hamamatsu R12860-HQE).}
  \label{pmtglassri}
 \end{center}
\end{table}

%%%%%%%%%%%%%%%%%%%%%%%%%%%%%%%%%%%%%%%%%%%%%%%%%%%%%%%%%%%%%%%%%%%%%%%%%%%%%%%
        \subsubsubsection{Mechanical Characteristics}\label{section:photosensors:Mechanical}
The HQE B\&L PMT bulb has been improved and proven to survive under
60\,meter water for Hyper-K as described in this section. It is
sufficiently better than the Super-K PMT, which is only specified for 40\,meter deep water. 

However, with the large number of photosensors in Hyper-K we expect
that even with a pre-selection (using a quick pressure test, etc.)
before installation, it is difficult to
ensure that there is no glass failure.
In 2001, a chain implosion of 6,779 PMTs out of 11,146 took place at Super-K.  
It was triggered by an accidental implosion which was transmitted to other PMTs as pressure pulse.  
In order to avert a similar accident, a protective cover made of a ultraviolet (UV) transparent acrylic cover for the detection area and a Fiber Reinforced Plastics (FRP) for the rear was introduced in Super-K.

Such a protective cover is needed to avoid any cascade implosion of the photosensors, making up for the difficult control of the glass quality in the production.
The cover was re-designed to further reduce the impact of pressure pulse, because a 60\,meter water depth boosts the peak pressure caused by the implosion by a factor of 1.6 from Super-K, corresponding to 6--7\,MPa.
The new design of the ID photosensor cover and its validation are explained in this section.

\begin{figure}[h]
    \centering
  \includegraphics[width=4.8cm]{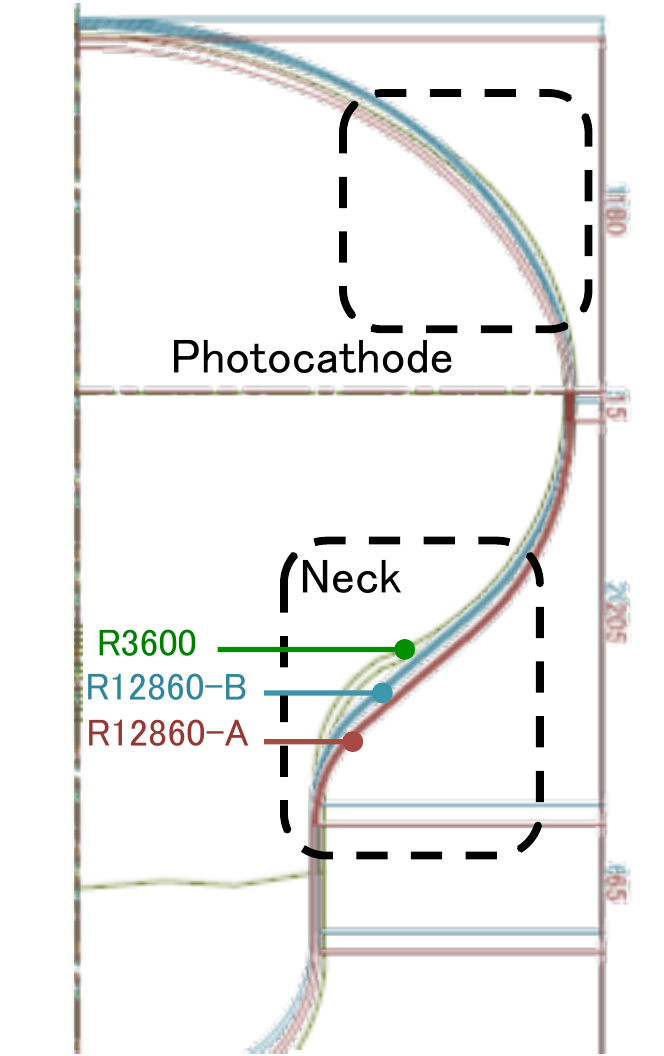}
    \caption{Comparison of the glass bulb curvature between R3600 and R12860 PMTs. }
    \label{fig:pmtcurvature}
\end{figure}

%%%%%%%%%%%%%%%%%%%%%%%%%%%%%%%%%%%%%%%%%%%%%%%%%%%%%%%%%%%%%%%%%%%%%%%%%%%%%%%
        \subsubsubsubsection{Design and Confirmation Test}\label{section:photosensors:Mechanical:Design} 
The weakest point of the Super-K PMT, which is around the largest reverse curvature
(Neck in Figure~\ref{fig:pmtcurvature}), was improved for the HQE B\&L
PMT.
Based on a stress analysis, the first bulb shape, R12860-A, was designed to reduce the stress concentration around the neck. 
Further improvement was achieved in R12860-B by optimizing the
curvature because there was a crack observed on the photocathode
surface of R12860-A (in Figure~\ref{fig:pmtcurvature}) after a high pressure water test. 

To validate the degree of improvement, a dedicated test of the PMT in
water under high pressure was performed.  In a high pressure vessel
filled with water, one PMT was tested by increasing the pressure in
steps of 0.1\,MPa above 0.5\,MPa and waiting for 7 minutes in each
step.

\begin{figure}[htbp]
 \begin{tabular}{c}
   \begin{minipage}{0.45\hsize}
     \begin{center}
 \includegraphics[width=0.9\textwidth]{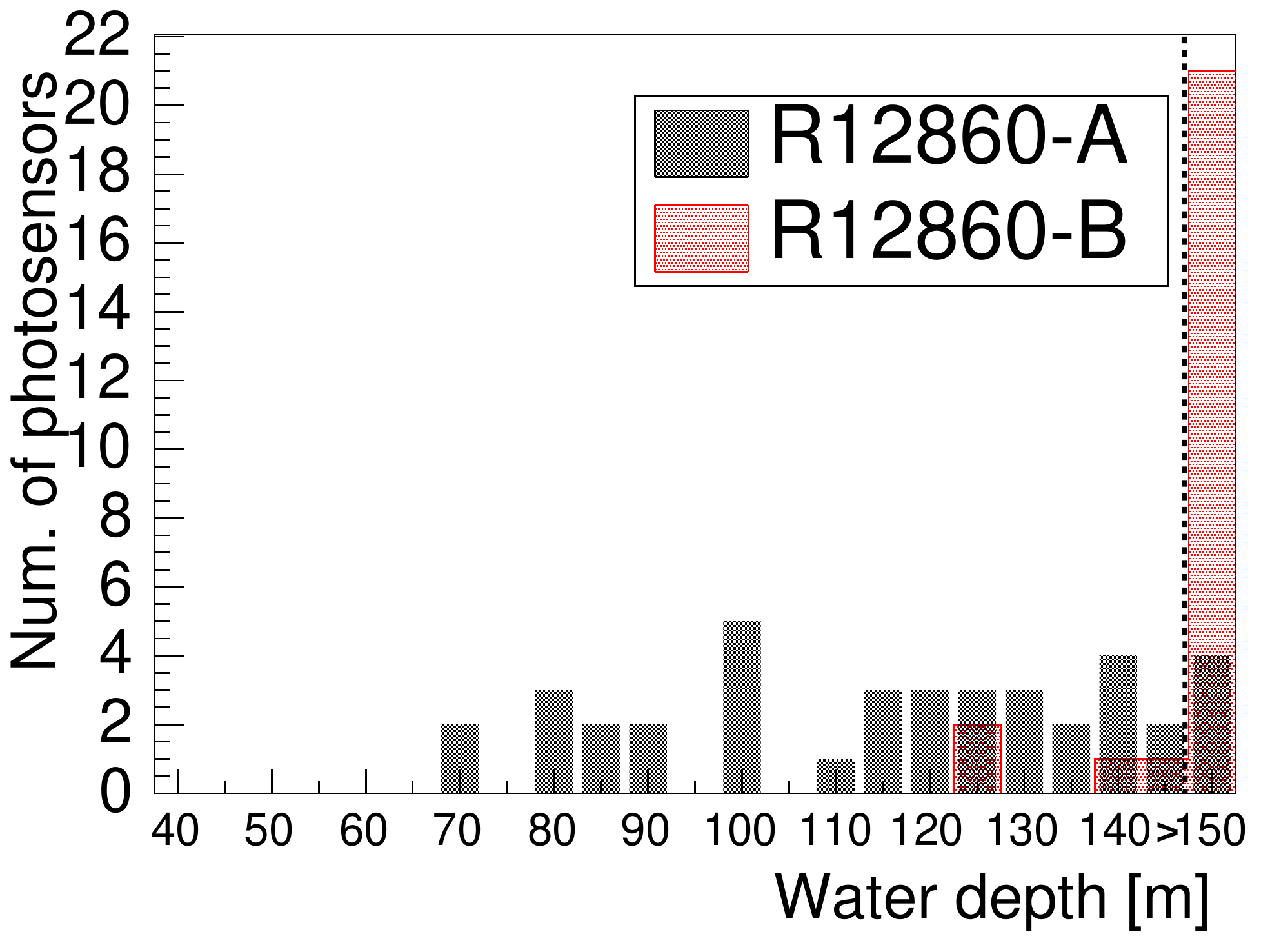}
 \caption{
Broken pressure in tested R12860-A and R12860-B samples up to 1.5\,MPa.
\label{fig:PMTdepth}}
     \end{center}
   \end{minipage}

\hspace{0.5cm}

   \begin{minipage}{0.5\hsize}
     \begin{center}
 \includegraphics[width=0.9\textwidth]{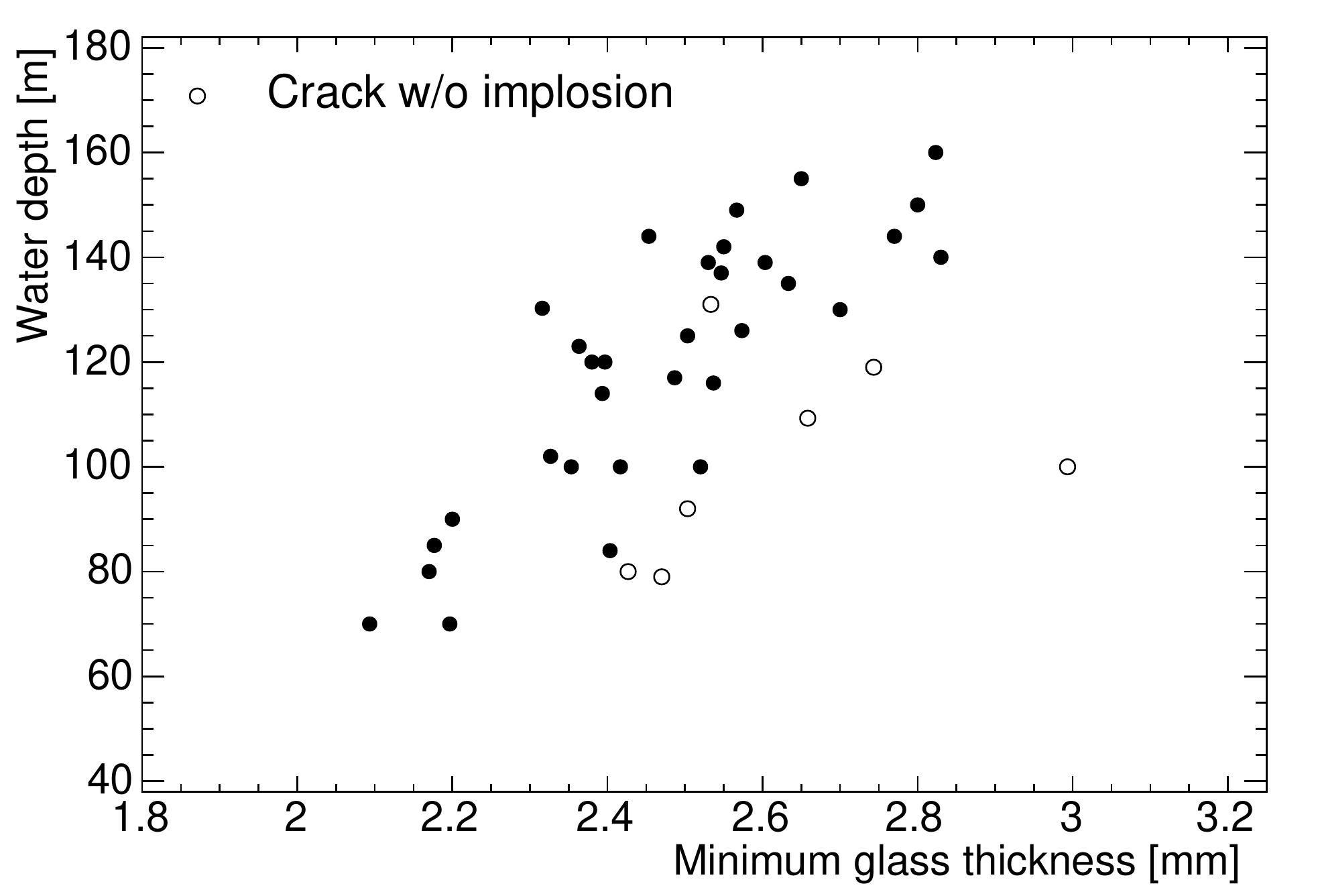}
 \caption{
Relation between a broken pressure in water depth and the minimum
glass thickness of the R12860-A tested samples. The minimum glass
thickness was estimated from several points around the neck of the
bulb and the photocathode glass.
Several PMTs that cracked, but did not implode, are shown in a blank
circle. \label{fig:PMTthickness}}
     \end{center}
   \end{minipage}
 \end{tabular}
\end{figure}

At first, we tested 35 samples of the initial prototypes
(R12860-A), and the depth range of implosion or crack was between 70
and 155\,meters pressure water (Figure~\ref{fig:PMTdepth}).
According to a survey of the glass thickness before the test, samples that
imploded around a shallow depth of 70--100\,meters were found to have a
relatively thin thickness of around 2.0--2.5\,mm at the thinnest point as in
Figure~\ref{fig:PMTthickness}. 
To mitigate this, a quality control step will be introduced whereby we
measure the glass thickness and reject bulbs with thin thicknesses.
The glass quality, in regards to bubbles, foreign matter, cracks and
thickness, is expected to be enhanced after improved training in bulb blowing
over the year prior to mass production.

The R12860-B PMT improved with the new shape was also tested. 21
R12860-B PMTs out of the total 25 did not implode up to 1.5\,MPa (150\,m equivalent) as shown in Figure~\ref{fig:PMTdepth}.
All tested PMTs had sufficient high pressure resistance
for the 60\,meter water depth of Hyper-K.
It should be noted that the test performed this time was in several
different conditions of PMT length or waterproofing, in order to find
the best design with high pressure bearing.

One of the two PMTs that imploded did so between 1.2 and 1.3\,MPa
(120\,m and 130\,m equivalent). This PMT had the smallest measured
glass thickness around the neck out of all of the 25 tested R12860-B
PMTs. Another PMT formed crack around the metal pins in the back,
which the stress analysis determined to be the weakest part in the
R12860.  This PMT had a waterproofed guard cover around the pins with
the same design as the R3600's, and its Polyethylene material is not
sufficiently hard to guard the glass against the high pressure water
that exists above 1\,MPa level.  Thus, the guard cover was improved
with a new hemisphere design made of PPS (Poly Phenylene Sulfide)
resin and adopted for a subsequent test of fifty PMTs.

%%%%%%%%%%%%%%%%%%%%%%%%%%%%%%%%%%%%%%%%%%%%%%%%%%%%%%%%%%%%%%%%%%%%%%%%%%%%%%%
        \subsubsubsubsection{Quality Control of the PMT Glass Bulb}
\label{section:photosensors:Mechanical:Test} 
The glass bulb is manufactured by hand; therefore it is difficult to expect uniform thickness throughout the mass production.  
In order to find an indication of a possible failure, the glass
thickness was checked by an ultrasonic thickness gauge at various measurement points. 
As indicated in Figure~\ref{fig:PMTthickness} there exists a relation
between the glass thickness and failure pressure, and therefore
screening PMTs based on glass thickness would be effective to minimize the failure. 
Also the bulb was inspected by eye to find unexpected cracks, foam and foreign matter.  

Eventually after the production, individual PMTs will be tested before installation.
It is planned to load a high pressure in water over a few minutes before the installation to the Hyper-K tank, in order to reject a bad bulb.
In case of Super-K, 4,727 PMTs were checked with 0.65\,MPa high pressure water to be safely used up to 40\,m water depth for the reconstruction after the accident.
Sixteen PMTs of 4,727 were rejected using the test, that is 0.3\% fraction.

Before the mass production of photosensors starts, we aim to establish
the quality control criteria.
We performed a screening test using fifty of the R12860 PMTs, in order
to set the best criteria that we will use for the mass production. 

\begin{itemize}
 \item Remarkable failures such as bubbles, foreign matter and striae
   were recorded and photographed during the quality check to measure
   the size and to count the number of occurances.
 \item The glass thickness was measured at 57 points.
\end{itemize}

It is noted that the photocathode and dynode are absent, but it does not matter because we only investigated the mechanical characteristics without measuring the detection performance.
Similar to the screening performed in Super-K, we tested all fifty PMTs in a high pressure water vessel. 
At this time, the load pressure is assumed to be 0.95\,MPa for the use in 60 meter water corresponding to 0.65\,MPa for 40 meter in Super-K.
So far, fifty PMTs were tested after the production and there was no damage at 0.95\,MPa.
All of the fifty PMTs were also tested up to 1.25\,MPa for further
investigation, but here we also found no damage in all PMTs.

According to this test, the selection criteria and condition is optimized. 
In the 2017 fiscal year, we will test 140 functional PMTs  
and will apply the criteria that will be used in Hyper-K. 
Although the bulb design of the R12860 PMT is same as the previous test, the production will be improved as below.
\begin{itemize}
 \item The metal mold is renewed. 
 \item The amount of glass is controlled by an automated machine to
   reduce variations in glass thickness.
 \item The automated measurement system of glass thickness is available for the precise and quick screening.
\end{itemize}

%%%%%%%%%%%%%%%%%%%%%%%%%%%%%%%%%%%%%%%%%%%%%%%%%%%%%%%%%%%%%%%%%%%%%%%%%%%%%%%
        \subsubsubsubsection{Degradation of the PMT Glass Bulb}
\label{section:photosensors:Mechanical:Degradation} 
As for a degradation of the glass material, the HQE B\&L PMT uses stable
glass made of borosilicate like Super-K, which is highly resistant to
an aqueous corrosion.
A general pressure cooker test for the glass plate, in 100\% humidity air at 121$^{\circ}$C and
0.2\,MPa for 17 hours, showed almost no optical degradation, where the
largest degradation is about 1\% only seen at around 350\,nm wavelength
and negligibly small.  A test to immerse a glass powder in
98$^{\circ}$C boiled pure water for an hour showed a small 0.03\%
dissolution in weight, which is very low compared to existing glass.

Mechanical characteristics were evaluated at
four different points on the PMT using PMT sampling glass taken from the bottom of the Super-K
tank after 5-years of running. A composition ratio of material and
bending strength are surveyed, and found to be comparable with other
Super-K PMTs stored in air at atmospheric pressure.  The high pressure
test was also performed on nine sample PMTs from Super-K -- three each
from the top, bottom and barrel sections after 5-years in water -- and
there was no implosion with a 0.65\,MPa load.

To examine the possibility of degradation by glass crystallization, the
glass surface was surveyed using X-ray diffraction at the four
different points as well.  There was no diffraction peak originating
from the crystallized glass, in either of the two samples;
one stored in air and the other from the bottom of
Super-K.

The number of dead channels of the ID photosensors is 0.4\% after
seven years of Super-K operation, including the ones with wrong
cable connections.  It might also include PMTs with an unknown crack
or implosion, and therefore we would expect a similar rate of glass
damage in Hyper-K, 1\% at maximum for twenty years.

It is concluded that the borosilicate glass used in the HQE B\&L PMT
is expected to have stable material characteristics over twenty years, and
several mechanical tests using the Super-K PMTs could
find no degradation in the long run.
The mass production and selection should be well managed such that the
physical damage rate is suppressed to the 1\% level. This level of 
failure is acceptable because the protective cover will avoid a chain implosion.

%%%%%%%%%%%%%%%%%%%%%%%%%%%%%%%%%%%%%%%%%%%%%%%%%%%%%%%%%%%%%%%%%%%%%%%%%%%%%%%
        \subsubsubsubsection{Shockwave Prevention Covers for PMTs}\label{section:photosensors:cover} 

\paragraph*{\underline{Necessity of PMT covers}}
As described in previous sections, every effort has been and will be
made to avoid a PMT implosion inside the Hyper-K water tank.  Based on
the knowledge of the mechanical characteristics of the Super-K PMT,
the new 50\,cm PMT has been designed to have enough strength for the
safe use at a water depth of 60\,m, and its performance has been
demonstrated by hydrostatic pressure tests.  The production of the
forty thousand PMTs for Hyper-K will be carried out under a strict
quality control, and the total inspection of the products including a
pressure test will get rid of any individual PMTs having a higher risk
of an implosion before their installation.

However, the possibility of a single PMT implosion cannot be zero.  To
prevent a chain reaction of imploding PMTs caused by the failure of a
single one, all PMTs in Hyper-K are housed in the shockwave prevention
covers, similar to those installed on the 50\,cm PMTs in Super-K after the catastrophic
accident.

\paragraph*{\underline{Basic design concept}}
The PMT cover for Hyper-K is designed on the same basic concept as
that for the Super-K PMT cover design.  In both detectors, the PMT
cover has several small holes, and the gap between the PMT surface and
the cover is filled with the tank water.  Since the covers themselves
are not usually exposed to the water pressure, there is no need to
care about any deformations caused by a long-term exposure to the high
water pressure.  On the other hand, the PMTs are constantly exposed to
the water pressure.  In the unlikely event of an imploding PMT, the
water pressure is immediately applied to the cover housing the broken PMT.  
The tank water slowly flows in through the small holes on the cover and fills up the vacuum region made by the PMT implosion.
The PMT cover is designed to have enough strength so that it can keep its shape even in such a case.  
Therefore, the peak amplitude of the pressure shockwave is significantly
reduced outside the PMT cover and thus cannot cause a chain reaction.

\paragraph*{\underline{Super-K PMT covers}}
In developing the Super-K PMT cover, there were strict restrictions on
its weight and shape, since the PMT supporting framework constructed
in the tank had originally been designed to support the bare PMTs.
Among three major candidate designs, the PMT cover formed by combining
an acrylic front window and a backside cover made of fiber-glass
reinforced plastic (FRP) was selected in Super-K, since it had been
demonstrated by a hydrostatic test and a PMT implosion test that the
cover would not be crushed even if the PMT inside would implode.
Thus, the Super-K group decided that the combination of the acrylic and FRP parts
with flange coupling bolts was suitable for a possible future PMT
replacement work and that the components can be readily produced.

A cover formed by combining two half bodies of a molded acrylic
product was another candidate. This full-acrylic cover also was found
to have enough mechanical strength for withstanding the water pressure
in the case of a PMT implosion, but it was not adopted due to its mass
production difficulty, a higher manufacturing cost and uncertainty of
the strength when combining two bodies.

A cover composed of an acrylic front cover and a stainless steel (SUS)
backside cover was also a candidate. 
The reproducibility of its shape and thickness is better than those of FRP cases. 
The SUS cover with a thickness of 2\,mm could not pass a PMT implosion test, 
while PMT covers with a thicker SUS component were not possible 
due to the restriction on the cover weight in Super-K, 

\paragraph*{\underline{Cover design selection for Hyper-K}}
Since the depth of the Hyper-K water tank (60\,m) is about 1.5 times
larger than that of the Super-K water tank (41.4\,m), the Hyper-K PMT
covers have to withstand a higher pressure in the case of an implosion
of the PMT inside.  The three cover designs which had been studied for
Super-K (i.e. acrylic+FRP, full acrylic, and acrylic+SUS covers) can
be also candidates for the Hyper-K PMT cover.

However, it is now known that FRP does not just contain much more
radioisotopes than those in the PMT glass but FRP itself also emits
light via chemiluminescence, as observed in Super-K.
Since these unwelcome things can produce more background events for
low energy physics, we have decided not to use any FRP-made covers in
Hyper-K.  As for the PMT cover formed by a molded acrylic product,
there are still the problems such as a higher cost for an initial prototype.
If these problems could be solved
within a reasonable time scale, the full-resin cover can be an
alternative design for Hyper-K.

The cover made of the stainless steel contains less radioisotopes and
is suitable for mass production.  Unlike the Super-K case, in which
SUS-made PMT covers were not adopted due to the weight limit coming
from the existing tank framework specification, the Hyper-K tank
framework can be designed so that it can support the PMT system
including the SUS covers of a sufficient strength.  Therefore, we have
adopted the PMT cover design of a combination of acrylic and SUS
components for Hyper-K.

\paragraph*{\underline{Hyper-K cover design}}
\begin{figure}
  \begin{center}
  \includegraphics[width=0.9\textwidth]{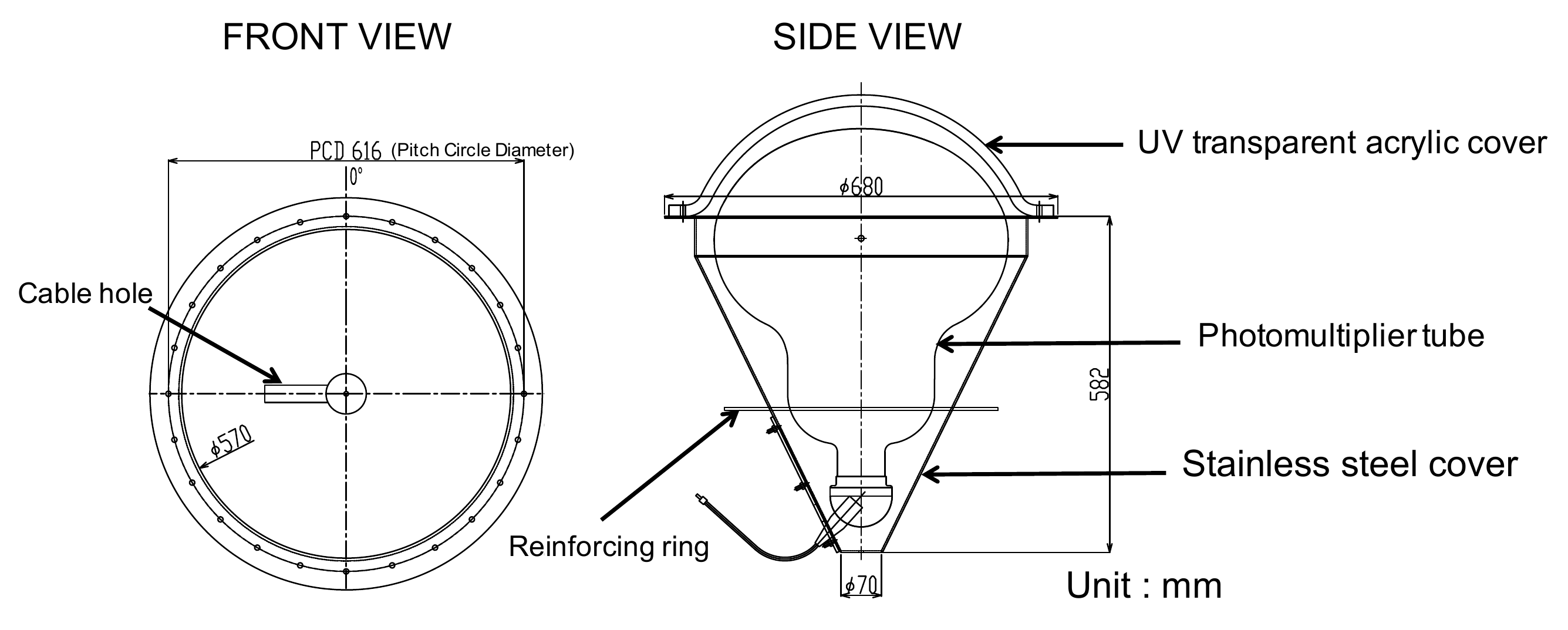}
  \caption{A schematic view of the shockwave prevention cover
for the Hyper-K ID PMTs.}
  \label{fig:protective_cover_shape}
  \end{center}
\end{figure}

\begin{figure}
  \begin{center}
  \includegraphics[width=0.4\textwidth]{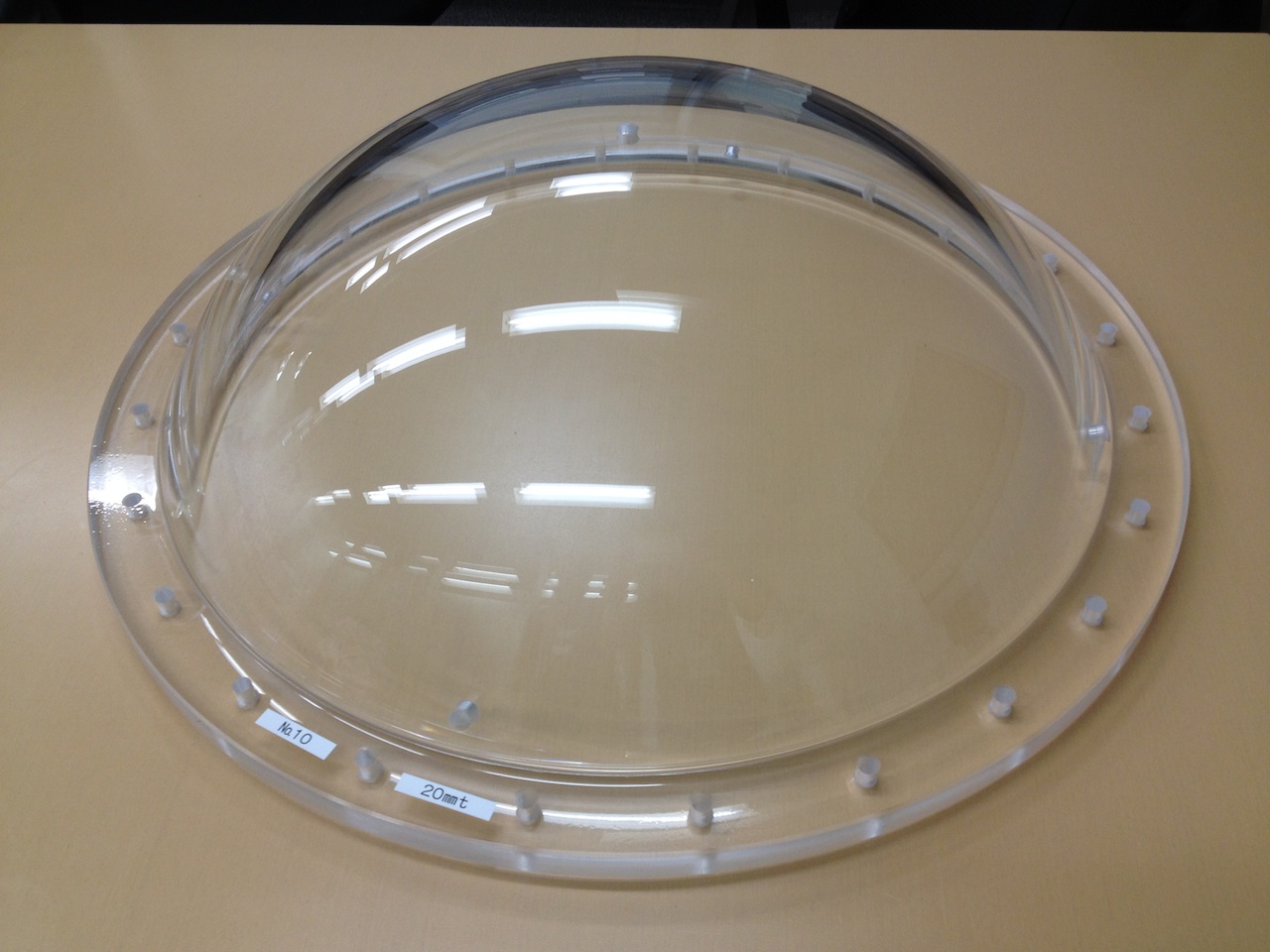}
  \caption{Acrylic window part of the PMT cover.}
  \label{fig:acrylic_cover_photo}
  \end{center}
\end{figure}

Figure~\ref{fig:protective_cover_shape} shows the shape of the Hyper-K
PMT cover.  The front-side cover with a partial spherical shape is
made of a UV transparent acrylic with thicknesses of 11\,mm at the
center position and 15\,mm at the flange part
(Figure~\ref{fig:acrylic_cover_photo}), which is about 1.2 times
thicker than the acrylic part of the Super-K PMT cover.  The light
transmittance of the acrylic cover measured in water is more than 95\%
for a wavelength longer than 350\,nm, which is reasonably good
considering the quantum efficiency of the 50\,cm Hyper-K PMT.  Since
the section modulus is proportional to the square of the thickness,
the Hyper-K acrylic cover could have about twice the strength of the
Super-K one, though this is just a crude estimation.  The backside
cover with a combination of ring and circular truncated cone shapes is
made of stainless steel with a thickness of 3\,mm.  The front acrylic
part and the backside SUS part are connected to each other by flange
coupling bolts.

The detailed design of the cover, such as a thickness of each part,
has been determined based on a dynamic behavior analysis, simulating
the situation after a PMT implosion.  The analysis has shown that the
PMT cover of the design mentioned above 
will not be crushed even if the PMT inside would implode at a
water depth of 100\,m.  The PMT implosion simulation should have some
uncertainties, but we think they cannot change the conclusion that the
cover will be functioning well at the depth of 60\,m.  This was
confirmed by the performance demonstration tests described later. 

On the acrylic cover, five holes with a diameter of 10\,mm are formed;
one at the center and four near the flange.  These number, diameter
and position of the holes on the acrylic cover are the same as those
for the Super-K PMT cover.  In developing the Super-K PMT cover, the
soundness of the holes on the acrylic cover against the water stream
caused by a PMT implosion had been checked. The test had shown that
the holes were not affected when exposed to a water stream with a
hydraulic pressure of 0.65\,MPa.  Therefore, the hole design should be
sufficient for the Hyper-K PMT cover.

It is measured that acrylic is degraded to 77\% by water absorption. 
Therefore, if it is confirmed by the PMT implosion test at 80\,m that the covers have enough performance
to prevent a chain implosion of PMTs, they are expected to be
functioning for the duration of the Hyper-K lifetime.

%%%%%%%%%%%%%%%%%%%%%%%%%%%%%%%%%%%%%%%%%%%%%%%%%%%%%%%%%%%%%%%%%%%%%%%%%%%%%%%
        \subsubsubsubsection{Demonstration Test of Shockwave Prevention Covers}
\label{section:photosensors:implosiontest}

The Hyper-K PMT cover has been designed so that it will not be crushed in the unlikely event of a PMT implosion and will prevent the occurrence of the shockwave causing a chain reaction of imploding PMTs.
To measure the performance of our PMT cover, we carried out two tests.
First, we carried out a hydrostatic pressure test of the PMT cover in Kamioka Mine to ensure the mechanical strength.
In this test, PMT cover was wrapped by a waterproof plastic bag and set in a high pressure vessel filled with water.
We pressurized the water surrounding the PMT cover, and the test showed that our PMT cover stood up to 1.1 to 1.5\,MPa as it was designed.

Next we carried out a prevention of the chain reaction of implosion with a mock-up of the PMT array of Hyper-K.
We utilized a test site formally used by Japan Microgravity Center (JAMIC), in Kami-Sunagawa, Hokkaido, Japan, for our chain implosion test.
The site has a vertical shaft depth of about 700\,m with about 4\,m diameter, filled with spring water.

\begin{figure}
  \begin{center}
   \includegraphics[width=0.4\textwidth]{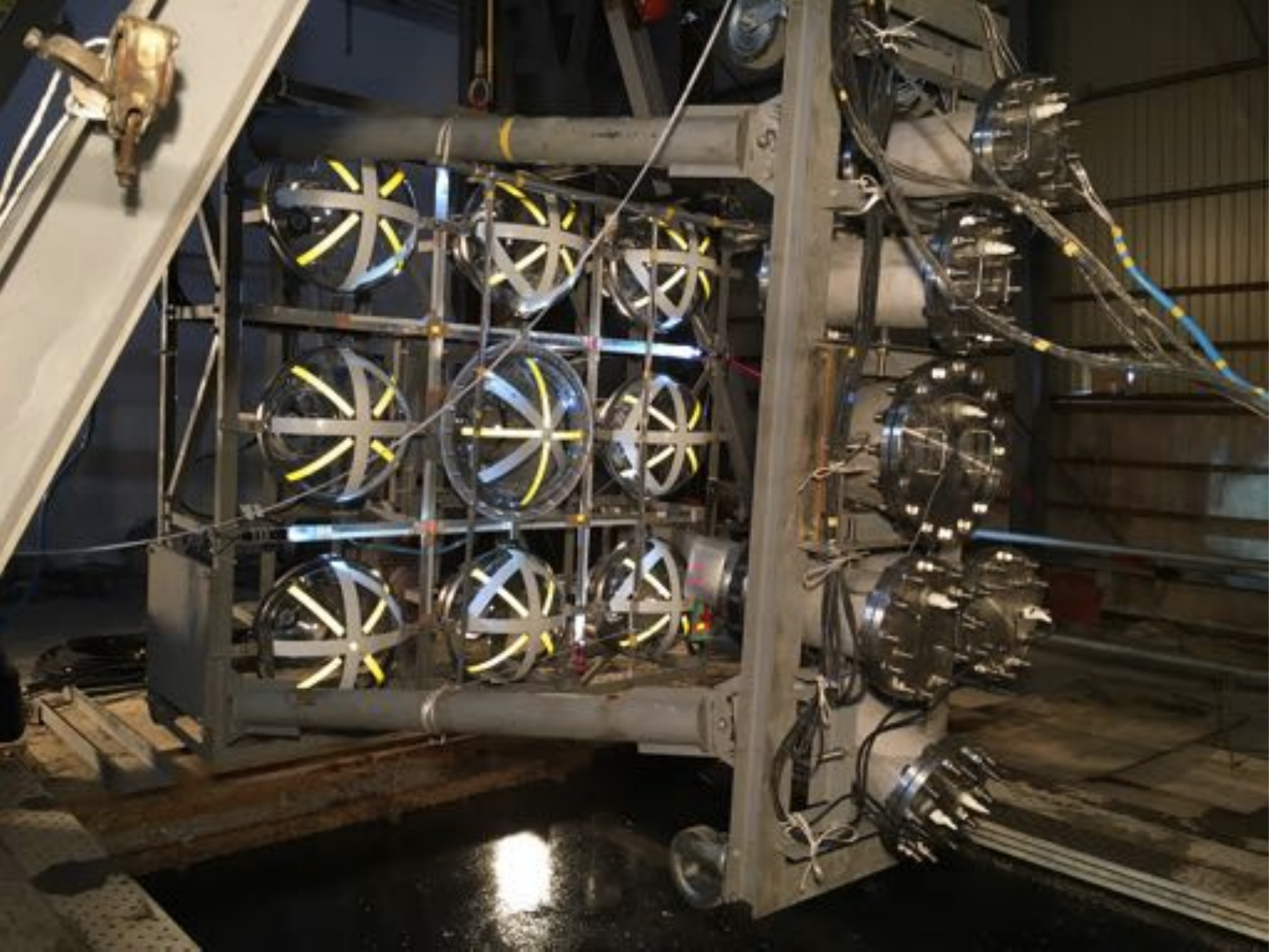}
   \caption{Mock of the chain implosion test. PMT encased in the cover was surrounded with bare PMTs to test the effect of the shockwave.
            Four pressure sensors were set in front of the PMT array to measure the pressure of the shockwaves.
            High-speed camera and lights are also set to investigate the crush visually.}
  \label{fig:mock_implosion_test}
  \end{center}
\end{figure}

In the test, nine PMTs were aligned 3 $\times$ 3 and mounted on a framework with same 70\,cm spacing in the Hyper-K inner detector.
Figure~\ref{fig:mock_implosion_test} shows the photograph of the mock-up just before sinking to deep water.
The central PMT was housed in the PMT cover, and implosion of the central PMT in the cover was induced by the apparatus which was designed to hit and crash the PMT remotely.
The others were bare to confirm there is no damage by a shockwave of the implosion at the center.

\begin{figure}
  \begin{center}
   \includegraphics[width=0.4\textwidth]{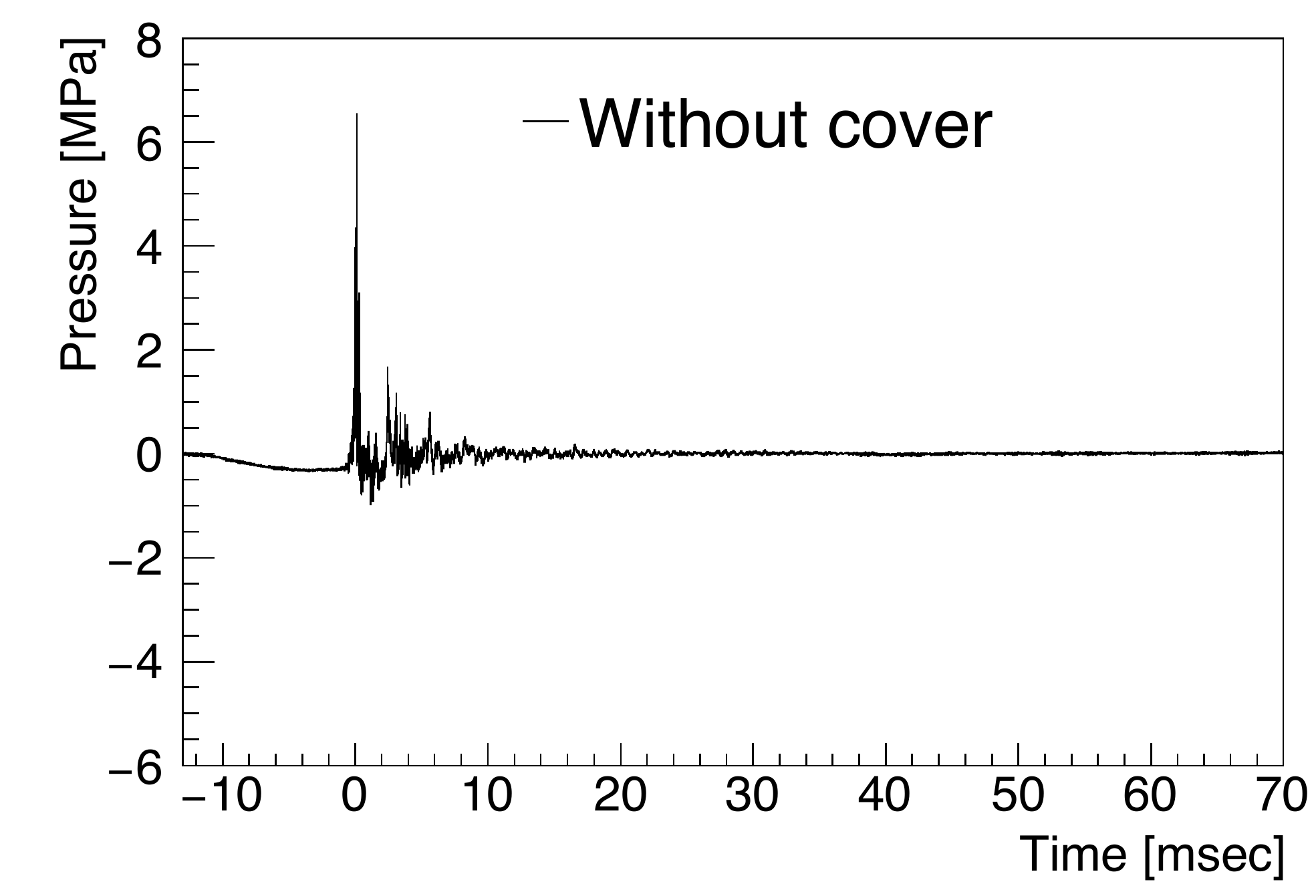}
   \includegraphics[width=0.4\textwidth]{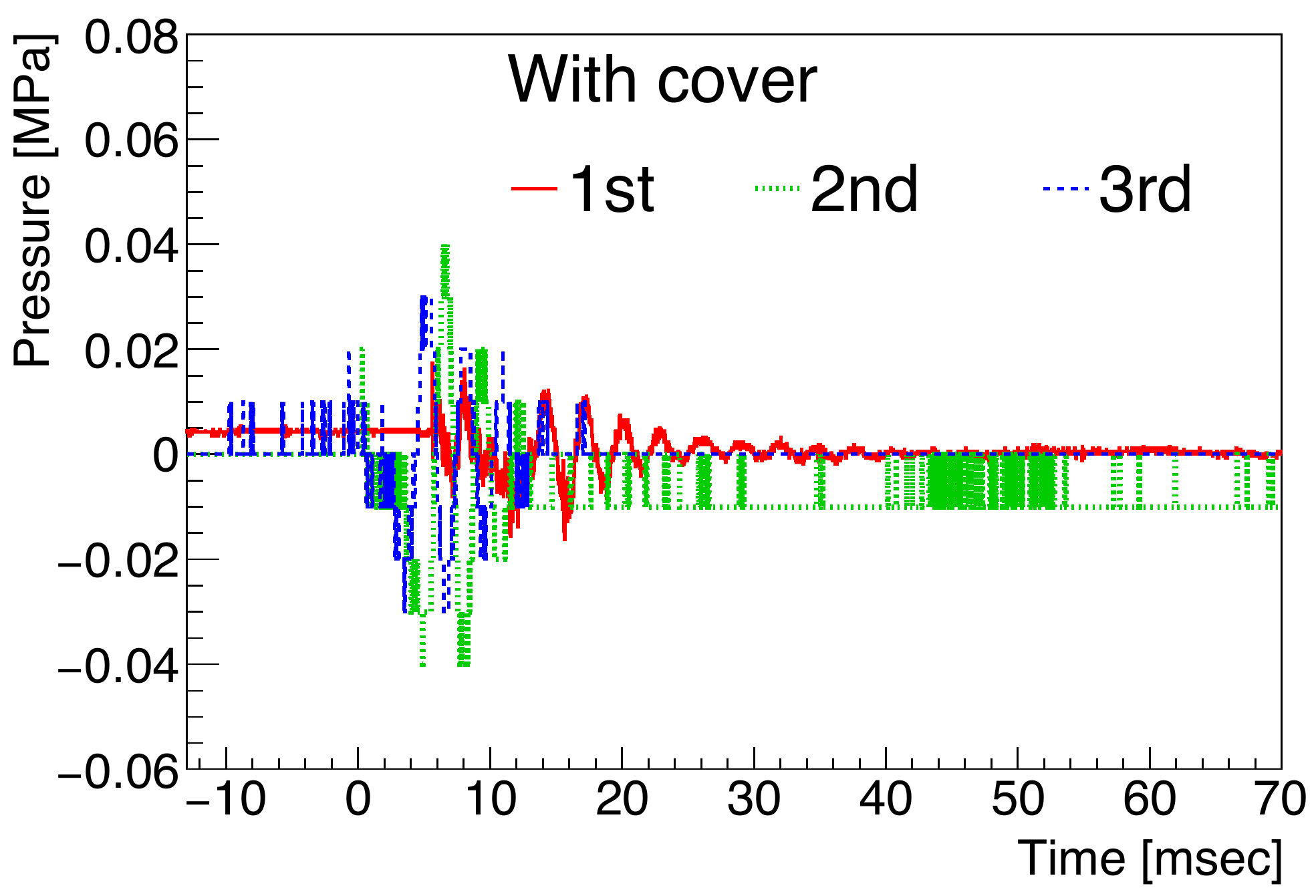}
   \caption{Measured pressure at 70\,cm ahead of the imploded PMT center at the 60\,m water depth.
            (Left panel) Shock wave without PMT cover. (Right panel) Shock wave with PMT cover. Three trials are shown in different colors.}
  \label{fig:shockwave_measurement}
  \end{center}
\end{figure}

This mockup was sunk into the deep water (60\,m or 80\,m) with pressure gauges and high-speed camera.
Three trials for each depth were done, and we found that the pressure was reduced to less than 1/100 and the cover was not crushed.
Figure~\ref{fig:shockwave_measurement} shows the measured pressure at the front of the central PMTs for each test at 60\,m water depth.
The PMTs surrounding the center one were used in the second and third tests, but no damage was incurred due to a sufficient suppression of the shockwave by the central cover.
We conclude that our PMT cover works to prevent the chain implosion in both 60\,m and 80\,m water depths.

We successfully established the cover design for Hyper-K using these tests.
In order to reduce cost and weight, improvement of the cover is on-going and several alternative ideas are under study.

%%%%%%%%%%%%%%%%%%%%%%%%%%%%%%%%%%%%%%%%%%%%%%%%%%%%%%%%%%%%%%%%%%%%%%%%%%%%%%%
        \subsubsubsubsection{Summary}\label{section:photosensors:prospect} 
\begin{figure}
  \begin{center}
  \includegraphics[width=1.\textwidth]{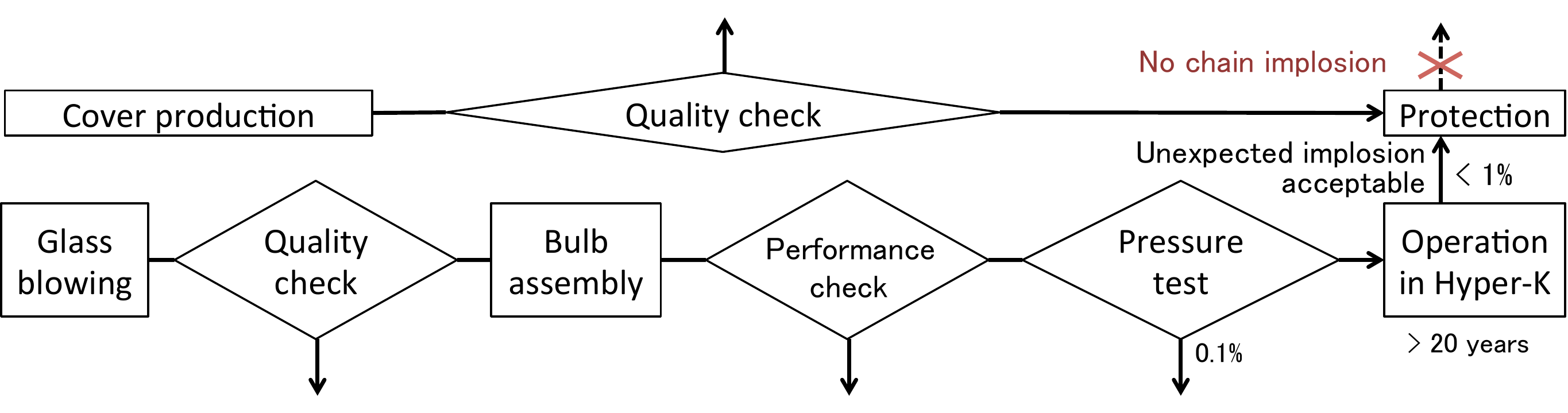}
  \caption{Flow from the bulb manufacture to the operation in Hyper-K. }
  \label{fig:pmtproduction}
  \end{center}
\end{figure}

The strength of the PMT bulb was enhanced by the improved bulb design,
quality check and pre-test in the high pressure vessel before the
installation.  A schematic flow diagram of various measures to avoid a
chain reaction of imploding PMTs is summarized in
Figure~\ref{fig:pmtproduction}.  In general, it is hard to expect
there will be no implosion in Hyper-K over the decades-long operation.
Thus, the protective cover is conservatively designed to avoid any
chain implosion by suppressing the shockwave, and this design will be
tested in advance of final production.

%%%%%%%%%%%%%%%%%%%%%%%%%%%%%%%%%%%%%%%%%%%%%%%%%%%%%%%%%%%%%%%%%%%%%%%%%%%%%%%
        \subsubsection{Photosensor for the Outer Detector}\label{section:photosensors:OD} 

        The primary function of the Outer Detector is to reject the
        incident cosmic ray muons that make up part of the background
        in the measurement of nucleon decays and neutrino interactions
        occurring in the Inner Detector.  The photosensor design for
        the Hyper-K Outer Detector will be similar to that of the
        successful Super-K Outer Detector using 20\,cm Hamamatsu R5912
        PMTs.  The OD PMT array is sparse relative to the ID PMT
        array, resulting in a 1\% photocathode coverage on the inner
        wall of the OD.  To improve the light collection efficiency by
        about a factor of 1.5, an acrylic wavelength shifting plate of
        a 60\,cm $\times$ 60\,cm square shape is placed around the
        glass bulb of each of the 20\,cm OD PMTs.

\begin{figure}[htbp]
 \begin{tabular}{c}
   \begin{minipage}{0.5\hsize}
     \begin{center}
  \includegraphics[width=1.0\textwidth]{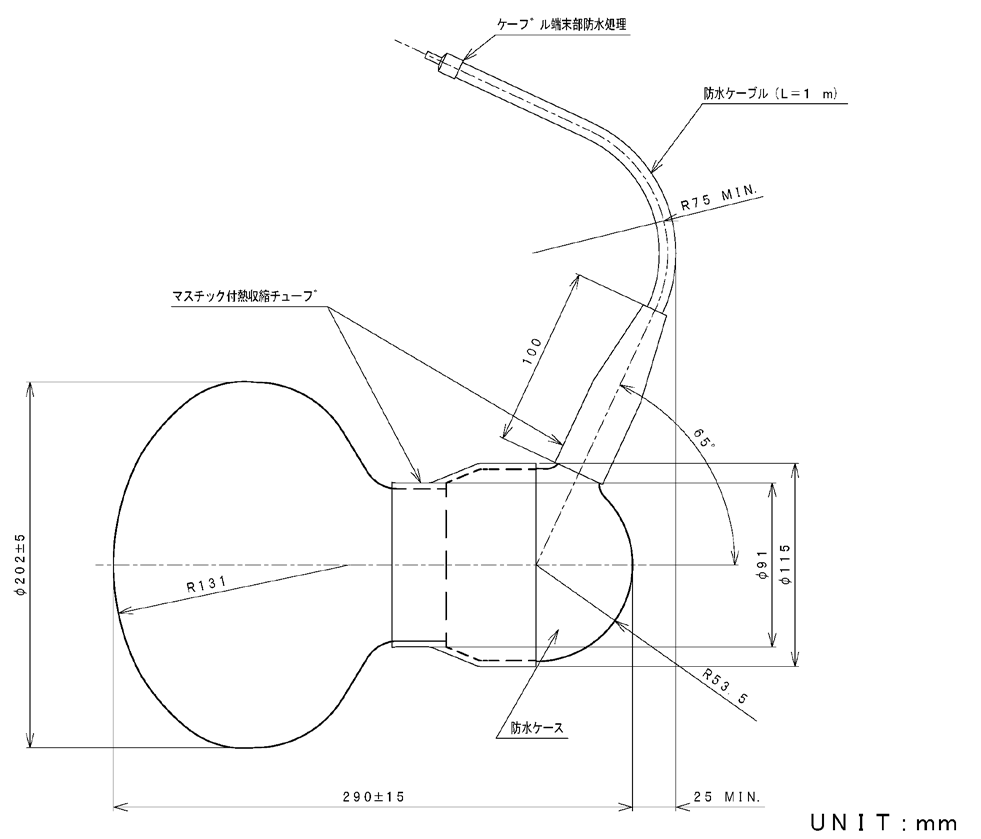}
  \caption{Design of the HQE 20\,cm box-and-line R5912 PMT.}
  \label{fig:R5912}
     \end{center}
   \end{minipage}
   \begin{minipage}{0.3\hsize}
     \begin{center}
 \includegraphics[width=1.0\textwidth]{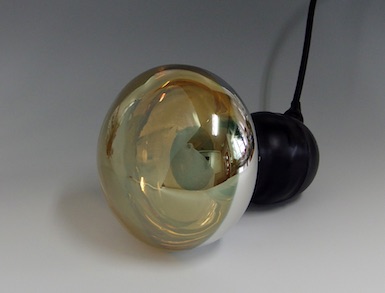}
  \caption{Picture of the HQE 20\,cm box-and-line R5912 PMT.}
 \label{fig:R5912pic}
     \end{center}
   \end{minipage}
 \end{tabular}
\end{figure}

The pressure tolerance tests have demonstrated that R5912 PMTs could withstand the water pressure at a depth of 60\,m.
For Hyper-K, the R5912 was improved using PPS resin for the guard cover like the R12860 PMT, as shown in Figures~\ref{fig:R5912} and \ref{fig:R5912pic}.
It also has improved high quantum efficiency of 30\%

In Super-K, there is no cover attached to the 20\,cm OD PMTs. 
It is noted that the volume of the 20\,cm PMT bulb is 6\% of the 50\,cm one, with the similar glass thickness and far distance among OD PMTs. 
Because of the 60\,m deep water level compared with Super-K, we would reconsider a safety use of the PMT and cover with appropriate evaluation and tests such as implosion. 

%%%%%%%%%%%%%%%%%%%%%%%%%%%%%%%%%%%%%%%%%%%%%%%%%%%%%%%%%%%%%%%%%%%%%%%%%%%%%%%
        \subsubsection{Alternative Designs}\label{section:photosensors:Alternative} 
There is still room to improve the Hyper-K performance with new possible
photosensors which are under development.  The key of the alternative
options is to show sufficient or superior physics sensitivities
while demonstrating safe use in the water tank over a long period of
time at a reasonable cost.
All the listed alternative candidates are expected to be ready before
the Hyper-K construction period, but are currently shown as options
because the product design is not finalized.

%%%%%%%%%%%%%%%%%%%%%%%%%%%%%%%%%%%%%%%%%%%%%%%%%%%%%%%%%%%%%%%%%%%%%%%%%%%%%%%
        \subsubsubsection{50\,cm High-QE Hybrid Photodetector}\label{section:photosensors:HPD} 
Another new 50\,cm photosensor with the better time and charge
resolution than the
existing 50\,cm photosensors is a combination semiconductor
device, called a hybrid photodetector (HPD), and is made by Hamamatsu
(R12850-HQE).

The HPD uses an avalanche diode (AD) instead of a metal dynode for the
multiplication of PEs emitted from a photocathode. 
A simple AD structure will have good quality
control in mass production, and a lower production cost than the
complex of metal dynodes.  In order to collect PEs in a small 20\,mm
diameter area of the AD, a high 8\,kV is applied.  Related items such
as the cable, connector and power supply were also developed.

\begin{figure}[htbp]
 \begin{tabular}{cc}
   \begin{minipage}{0.5\hsize}
     \begin{center}
 \includegraphics[width=1.\textwidth]{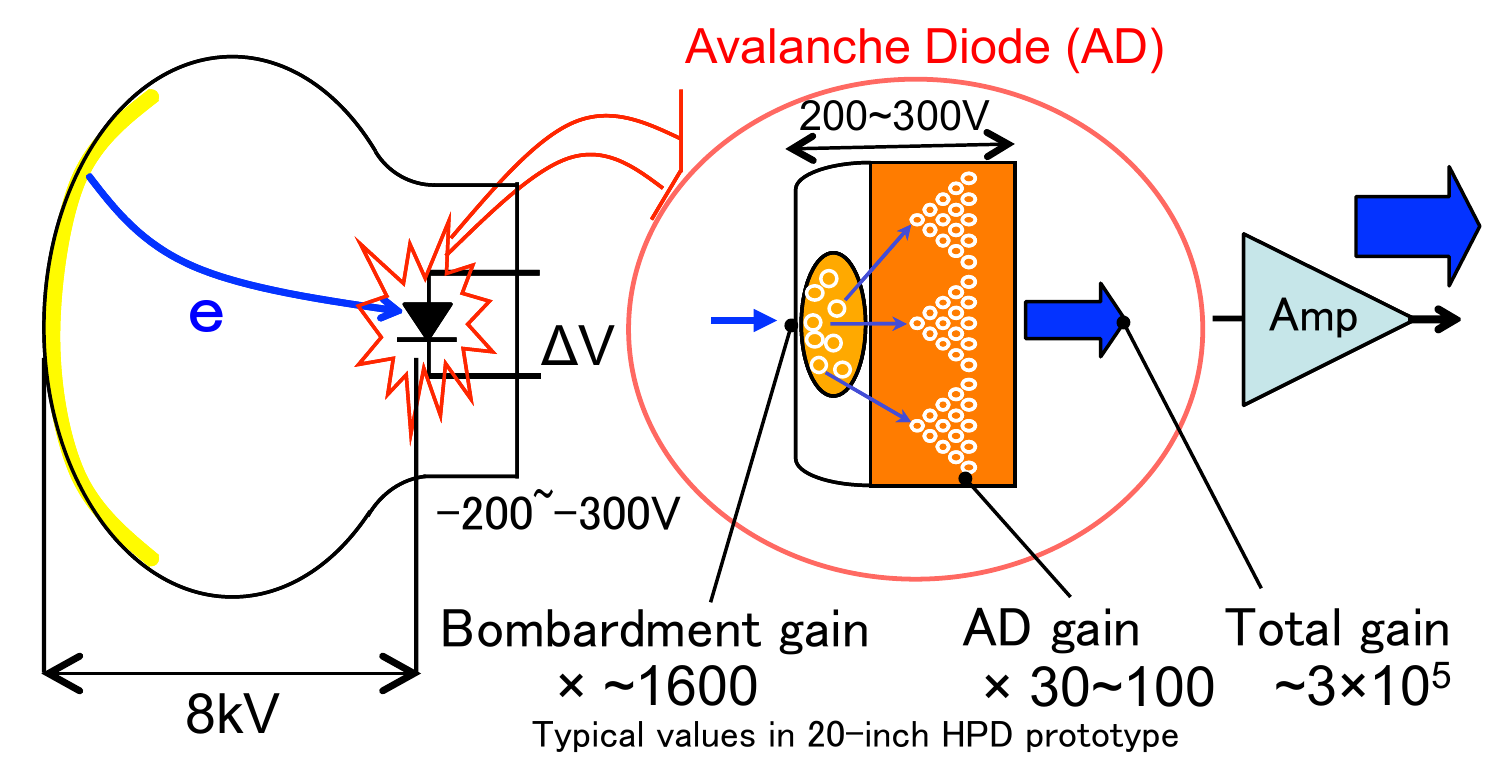}
 \caption{Schematic view of amplification system on the HQE 50\,cm HPD.\label{fig:HPDamplification}}
     \end{center}
   \end{minipage}

\hspace{0.5cm}

   \begin{minipage}{0.45\hsize}
     \begin{center}
 \includegraphics[width=0.6\textwidth]{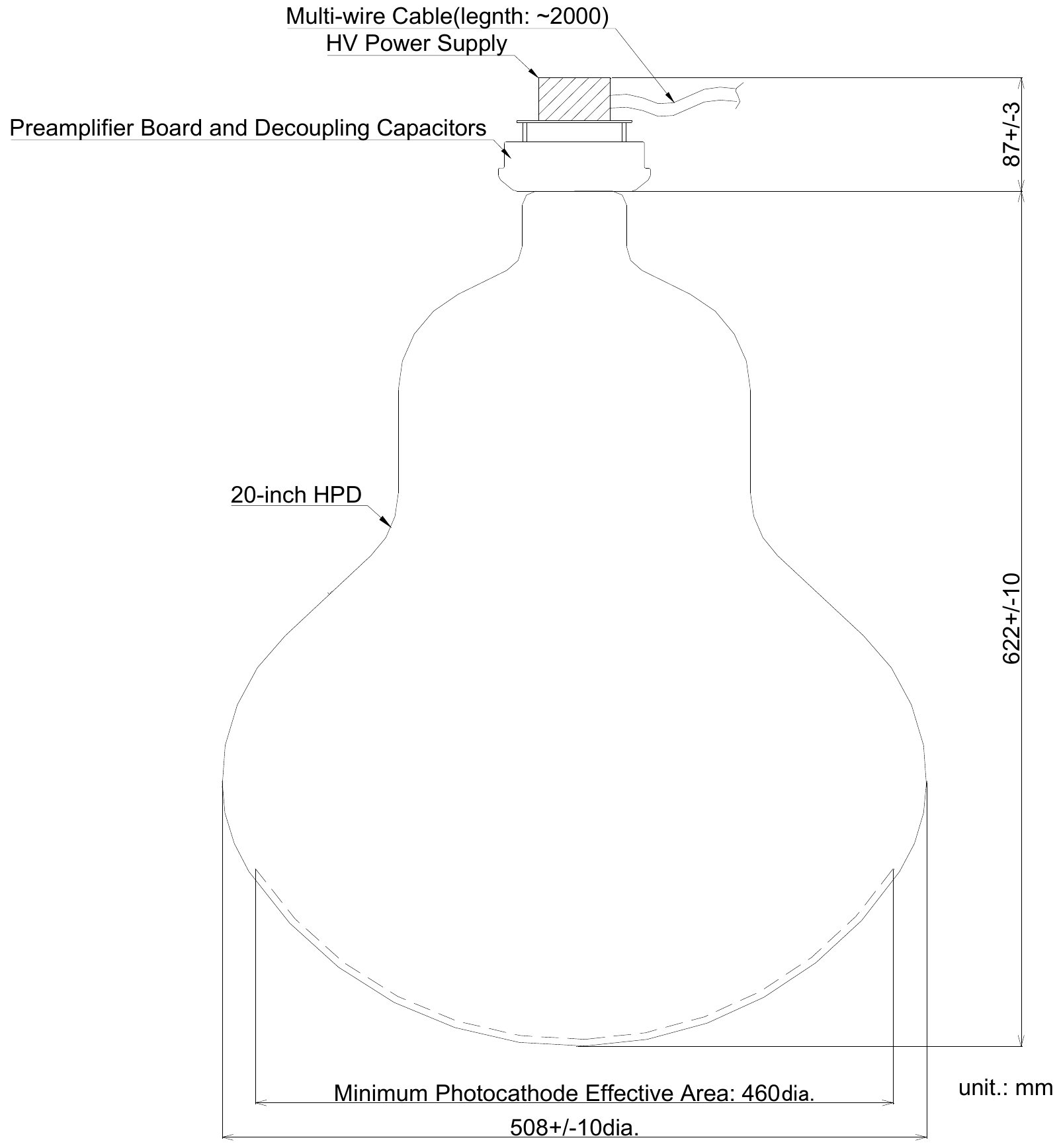}
 \caption{Design of the HQE 50\,cm HPD (before waterproofing).\label{fig:HPDdesign}}
     \end{center}
   \end{minipage}
 \end{tabular}
\end{figure}

Figure~\ref{fig:HPDamplification} shows that electrons are multiplied
by a factor of $10^{5}$ with a combination of a bombardment gain and
then avalanche gain.  The gain is adjusted by the bias voltage applied
on the AD, around a few hundred volt, while the 8\,kV is fixed.  The
HPD is equipped with a pre-amplifier, so the resulting gain is
equivalent to PMTs.
The size and surface material are almost the same as those of the 50\,cm
PMT as shown in Figure~\ref{fig:HPDdesign}, thus the same support
structure and protective cover can be used.

\begin{figure}[htbp]
 \begin{tabular}{cc}
   \begin{minipage}{0.45\hsize}
     \begin{center}
       \includegraphics[width=0.9\textwidth]{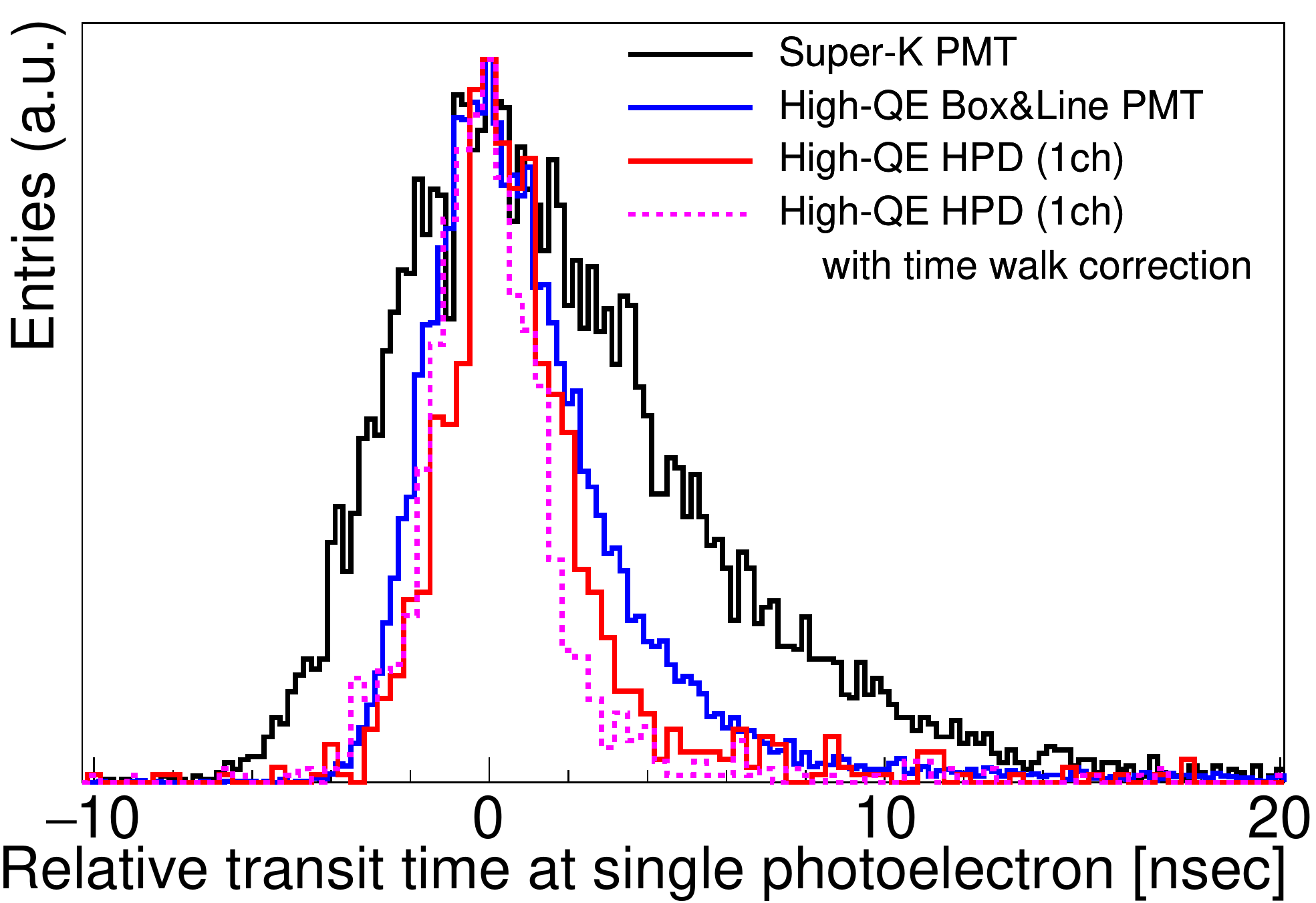}
       \caption{Transit time distribution at single photoelectron, compared with the Super-K PMT in dotted line. (HPD is added in Figure~\ref{fig:PMTTTS}.) \label{fig:HPDTTS}}
     \end{center}
   \end{minipage}

\hspace{1cm}

   \begin{minipage}{0.45\hsize}
     \begin{center}
       \includegraphics[width=0.9\textwidth]{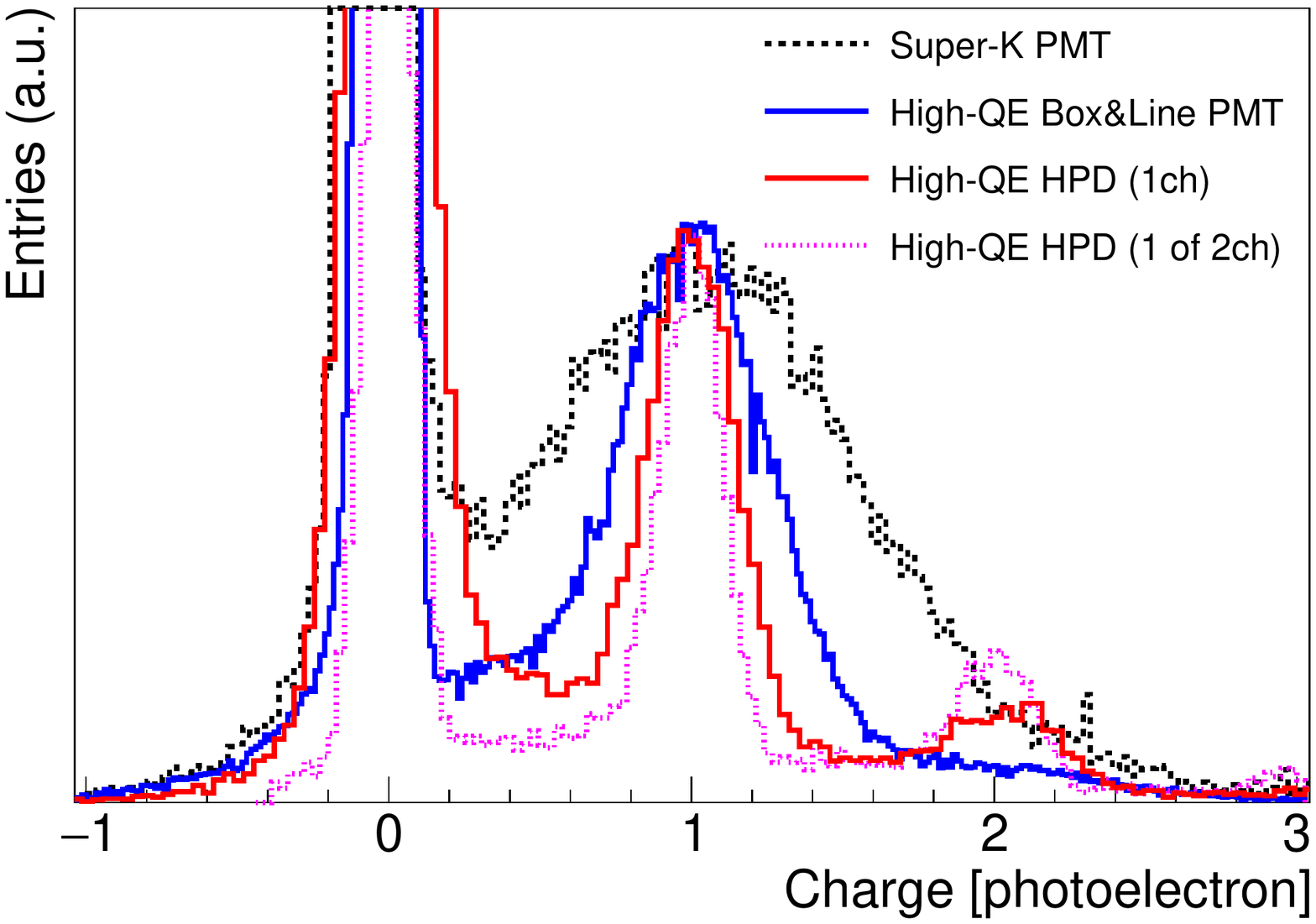}
       \caption{Single photoelectron distribution with pedestal, compared with the Super-K PMT in dotted line. (HPD is added in Figure~\ref{fig:PMT1PE}.) \label{fig:HPD1PE}}
     \end{center}
   \end{minipage}
 \end{tabular}
\end{figure}

The single PE detection is significantly better, though it is still limited by the pre-amplifier because of a large junction capacitance of the AD (400\,nF).
The transit time spread at single PE is 3.6\,nsec in FWHM measured at the fixed threshold, as drawn by the red line in Figure~\ref{fig:HPDTTS}.
It is superior to 7.3 and 4.1\,nsec for the Super-K PMT (black) and B\&L PMT (blue), respectively.
With the time walk correction using measured charge, the HPD resolution reached 3.2\,nsec in FWHM as shown by the dotted magenta line.

Figure~\ref{fig:HPD1PE} shows the 15\% charge resolution of the HPD as
the standard deviation at the single photoelectron peak as in the red line.
It is superior to both the 53\% and 35\% for the Super-K PMT (black) and B\&L PMT (blue).
If the AD is segmented into two channels as a test, individual readout of HPD showed better resolution of 10\% as shown by a dotted line due to a half junction capacitance.
Further five segmentation into a center channel and surrounding four channels brings a position sensitive detection with considering a hit probability in five channels
by a slight focusing shift by different arrival points of photons on the glass.
It might help a good event reconstruction and background rejection.

\begin{figure}[htbp]
   \begin{center}
     \includegraphics[width=0.4\textwidth]{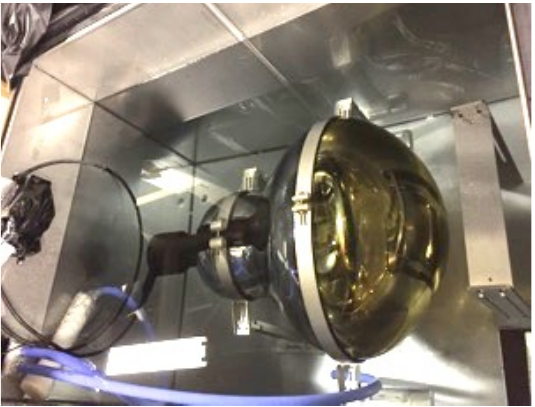}
     \caption{The HQE 50\,cm HPD (R12850) testing in water. \label{fig:HPDwater}}
   \end{center}
\end{figure}

The waterproof HPD was successfully operated for 20 days in a dark box filled with water as shown in Figure~\ref{fig:HPDwater}.
In very near future, the HPD would be a superior option to the PMTs, after successful long-term tests of all
performance and usability criteria.

%%%%%%%%%%%%%%%%%%%%%%%%%%%%%%%%%%%%%%%%%%%%%%%%%%%%%%%%%%%%%%%%%%%%%%%%%%%%%%%
        \subsubsubsection{Smaller Photosensors}\label{section:photosensors:SmallPD} 
Photosensors with a 20--30\,cm aperture
can also be an alternative option.
These are available with the HQE and by another manufacturer.
Using two small photosensors are comparable to a single 50\,cm photosensor as for the detection efficiency, or a little larger aperture size than the 20\,cm photosensor can be considered as the OD photosensor.
A smaller 7.7\,cm PMT with a large photo-collection plate is also a possible alternative option with a low cost for the OD photosensor system.

        \subsubsubsubsection{20--30\,cm High-QE PMTs}\label{section:photosensors:HQEODPMT} 
Based on the successful development of the 50\,cm HQE B\&L PMT, 20\,cm
or 30\,cm PMTs can obtain superior performance compared to the existing small PMTs.
By applying the same techniques, the 20\,cm and 30\,cm PMTs with a high
QE box-and-line dynode and improved performance will be available
easily by scaling the 50\,cm PMT down with a similar design.  The
performance is expected to be equivalent or better compared with the
50\,cm HQE B\&L PMT.

Prior to that, the HQE 30\,cm PMT, R11780-HQE by Hamamatsu, was
developed aimed at a large water Cherenkov detector planned in US for
the LBNE project.  It reached a QE of 30\% and pressure rating over 1
MPa.  Further improvement was tried, and a new bulb design of the HQE
30\,cm PMT based on the R12860-HQE and R11780-HQE was made.  In order
to validate the high pressure tolerance, a test in a high pressure
water was performed on three samples at Kamioka.  As a result, all the
samples got no implosion up to 150\,meter water equivalently. 

Until recently, photomultiplier tubes with an aperture over 25\,cm have
been almost exclusively supplied to the market by Hamamatsu.  It is
important that additional vendors come in the marketplace for price
competition and for additional supply capacity.

ET Enterprises Limited ADIT, now a US-based PMT manufacturer in Texas,
has been developing a large area PMT, financially supported by NSF.
Testing of the operational first generation 28\,cm HQE PMTs have been
performed at Pennsylvania, UC Davis, etc.  showing comparable
efficiency and charge measurement performance to those of
similarly-sized Hamamatsu HQE PMTs.

If successfully produced, the PMTs can be a cost-effective
alternative to Hamamatsu for Hyper-K OD PMTs.

        \subsubsubsubsection{7.7\,cm Photodetector Tube}
\label{section:photosensors:OD3inchPMT} 

\par
In this section we will focus on an alternative design of the
outer-detector using the 7.7\,cm photosensors. It is expected that the
costs will be lower than the nominal configuration as well as also
providing an extended market and production capacity.  At the moment
the nominal configuration of the outer-detector consists of an array
of 20\,cm hemispherical photosensors placed on a grid to reach a
photo-coverage of 1\%, as represented in
Fig.~\ref{fig:pmt_arrangement}.  These photosensors are mounted on
wavelength shifting plates, to improve the light collection.

By using 7.7\,cm photosensors instead of the 20\,cm, the number of
photosensors is multiplied by six to keep the same coverage.  The
photosensors are therefore closer to each other thus increasing the
number of hits collected, and improving the coincidence accuracy.
Moreover, the outer-detector water thickness is about 1\,m, and
particles will produce less light and in a narrower region ---
compared to the Super-Kamiokande detector where the outer-detector
water thickness is about 2\,m.  Thus, a setup with more photosensors
closer to each other should allow a better sensitivity, especially
with the configuration of Hyper-Kamiokande with a reduced water
thickness.

\begin{figure}[!htb]
  \centering \includegraphics[width=0.5\linewidth]{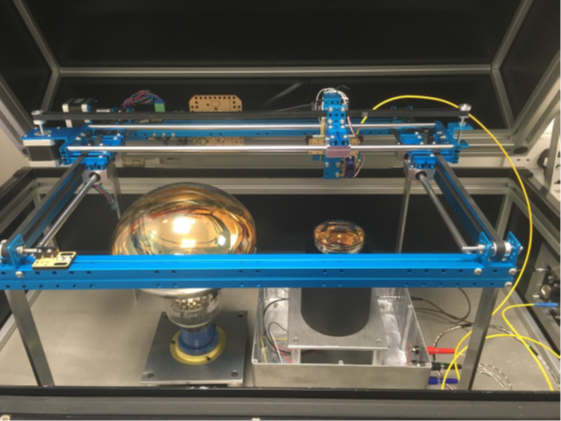} \caption{Setup
    for photosensors testing at Queen Mary University of London.  This
    picture was taken with a 20\,cm ET9354KB (on the left) and a
    7.7\,cm ET9302B (on the right) photosensors.  One can see the XY
    stage above the photosensors (blue) which moves the optical fibre
    (in yellow) along the X-Y axis. The optical fibre guides the light
    out from the LED driver to the blackbox.}  \label{fig:QMSetup}
\end{figure}

We are currently testing candidate 7.7\,cm photosensors to be used in
the outer-detector, see Fig.~\ref{fig:QMSetup} for the teststand, as
well as implementing the outer detector in that simulation and
reconstruction, which is needed for optimizing the configuration.

Our tests have been performed using the ET9302KB and other
photosensors.  The ET9302KB was extensively tested and many of its
parameters measured.  It has a QE of 30\% and a small dark current
rate which has been measured at 400\,Hz --- about ten times less than
typical rates for 20\,cm photosensors --- and a small after-pulse rate
with respect to the gain.

The ET9302KB would fit perfectly in a configuration with several
photosensors like the outer detector, since the number of accidental
coincidences inherently raised by the amount of photosensors used will
be reduced thanks to its low-noises properties.

\begin{figure}[!htb]
      \begin{minipage}[t]{1.\linewidth}
       \includegraphics[width=0.6\linewidth]{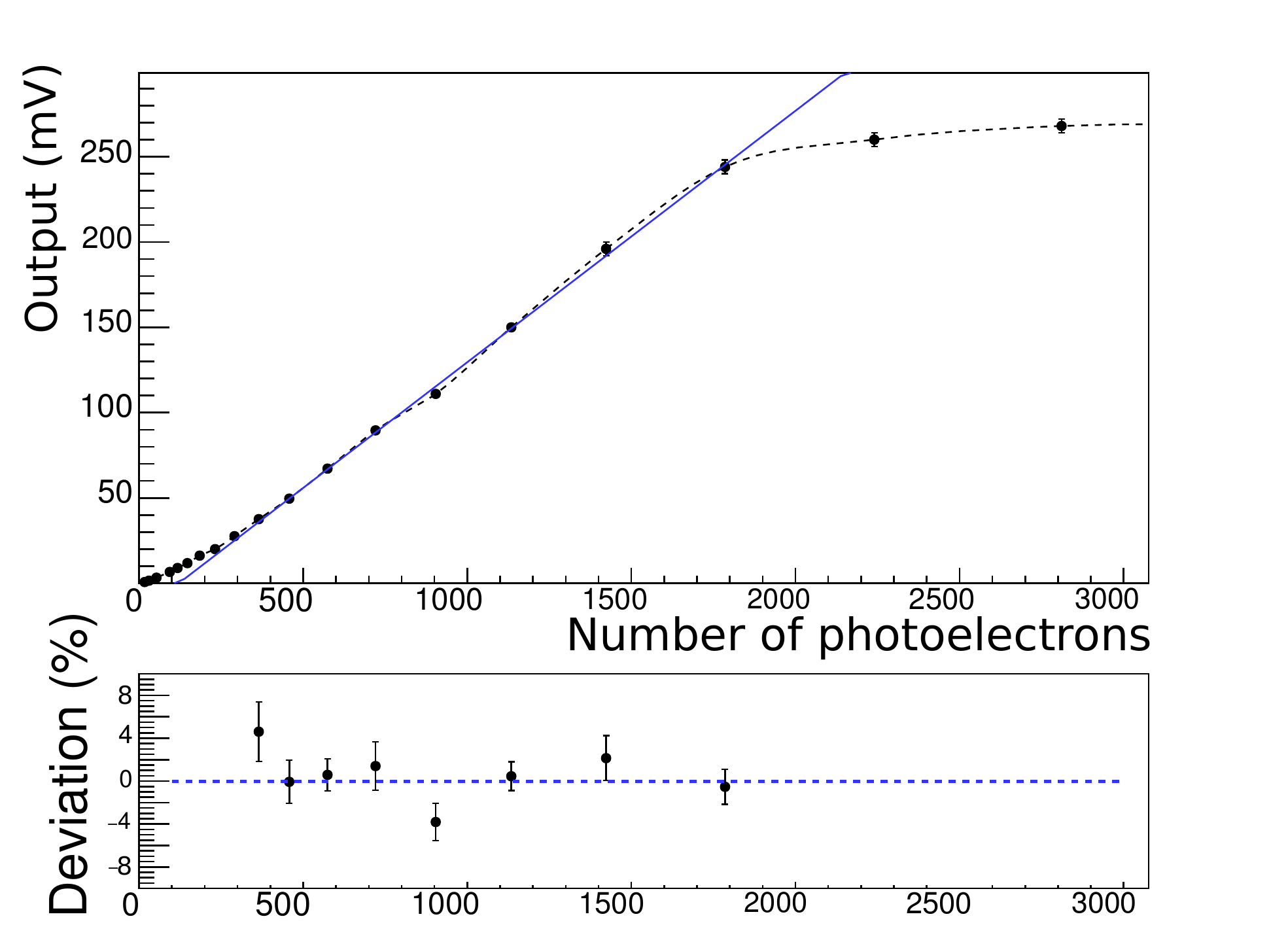}
       \caption{Linearity for the 7.7\,cm photosensor ET9302KB,
       measured with an LED driver ranged from few to several thousands
       photons.}
       \label{fig:lin:ET9302KB}
     \end{minipage}
  \end{figure}

Furthermore, these photosensors also showed excellent linearity and
resolution, allowing an accurate reconstruction of the energy of the
particles inside the outer-detector.
Figure~\ref{fig:lin:ET9302KB} shows the linearity of the photosensor,
where deviation was measured to be of the order of 1\%.
The resolution at 1000 photoelectrons collected
--- which is what is expected for a few-MeV gamma event for example ---
is measured to be 2\%.

The newly improved photosensor from ET Enterprises Ltd., D793KFLB, with a transit time spread of 1.6\,nsec, a nominal gain $6 \times 10^{6}$ at 850--1100\,V and nominal dark rate 1000\,Hz at 800\,V is currently under test. 
Preliminary results showed excellent linearity and resolution.

The photosensors are hemi-spherical and will also be mounted on
wavelength shifting plates, which allow an easy way to enhance the
light collection.  Wavelength shifting plates will reemit the absorbed
photons in the direction of the photo-cathode of the photosensors at a
wavelength around 400\,nm, where the quantum efficiency of the
ET9302KB is maximum.

Outer detector alternative configurations with 7.7\,cm photosensors
will be evaluated in simulation and the best setting will be selected
accordingly, using realistic cosmic flux files from the Super-K
collaboration to test the performance.

The chosen configuration should achieve at least the minimum detection
efficiency to have sufficient ability to veto cosmic rays or catch
through-going particles.

Finally, also another alternative of using 12.7\,cm photosensors,
i.e. an intermediate solution between 7.7\,cm and 20\,cm photosensors,
is planned to be addressed in both the optimization in simulation of
the configurations and in testing the photosensors.

%%%%%%%%%%%%%%%%%%%%%%%%%%%%%%%%%%%%%%%%%%%%%%%%%%%%%%%%%%%%%%%%%%%%%%%%%%%%%%%
        \subsubsubsection{Multi-PMT Optical Module }\label{section:photosensors:MultiPMT}
\par The concept of an optical module with a 20 to 40\,cm PMT housed in a glass pressure vessel
     has been developed the past decades for neutrino telescopes in
     water and ice (DUMAND~\cite{DUMAND}, Baikal~\cite{Baikal},
     NESTOR@~\cite{NESTOR,NESTOR-PMT}, ANTARES~\cite{Antares,
     Antares-PMT}, AMANDA~\cite{AMANDA} and IceCube~\cite{IceCube}).
     For KM3NeT~\cite{KM3NetDR}, the km$^3$ neutrino observatory under
     construction in the Mediterranean, a single large area phototube
     has been replaced by 31 7.7\,cm PMTs packed in the same glass
     pressure vessel. This new fly's eye photosensor concept, in which small PMTs replace a
     single PMT, has been dubbed the ``multi-PMT Optical Module (mPMT)''.
     The mPMT has several advantages as a photosensor module:
     \begin{itemize}
     \item Increased granularity\\
         The increased granularity of mPMT provides enhanced event reconstruction, in particular for
     multi-ring events, such as in proton decays and multi-GeV neutrinos which is important for mass hierarchy studies,
     and for events near the wall where fiducial volume is defined.
     Since each of the PMTs have different orientations with limited field of views,
     the mPMT carries information on the direction of each detected photon.
     This directional information effectively reduces the dark hit rate since the dark hits only applies to the field of view
     of the PMT. As a result, it improves signal-to-noise separation for low energy events such as solar/supernova
     neutrinos and the neutron tagging as discussed in Section~\ref{section:pdecay-coverage}.
     \item Mechanically safe pressure vessel with readout and calibration integrated\\
     The pressure vessel protect against implosion. Since the vessel is filled with
     small PMTs, electronics and the support structure without large void,
     shock waves would be suppressed even when implosion takes place.
     The pressure vessel contains digitization electronics and calibration sources,
     providing natural solutions for readout and calibrations.
    \item Cost, geomagnetic field tolerance, and fast timing for 7.7\,cm PMT's\\
     Economical mass-production of 7.7\,cm PMT's in particular for medical use
     provides almost the same cost per photocathode area for 7.7\,cm PMT compared
     to larger PMTs. For example, KM3NeT claims the cost per photocathode
     area is cheaper than for the 20 to 40\,cm PMTs.
     Additional advantages of small PMTs are their much lower sensitivity to the Earth's magnetic field,
     which makes magnetic shielding unnecessary~\cite{KM3Net:VLVnT15.1}, and potentially better
     timing resolution, which could further reduce the dark hit background and event reconstruction.
     The failure rate of small PMTs is of the order of 10$^{-4}$/year. Any
     loss of a single PMT would affect the detector performance
     minimally compared to the loss of one large area PMT.
     \item Complex structure, an opportunity for international contributions\\
     A potential draw back of mPMT compared to off-the-shelf 50\,cm PMT is its complicated structure
     with multiple components. This could be regarded as an advantage for international contributions
     where each partner can take on mPMT module parts and/or assembly and testing procedures,
     as is successfully done for the KM3NeT mPMT and IceCube mDOM (Digital Optical Module) constructions.
       \end{itemize}

\begin{figure}[htbp]
 \begin{tabular}{c}
   \begin{minipage}{0.55\hsize}  
     \begin{center}
 \includegraphics[width=0.8\textwidth]{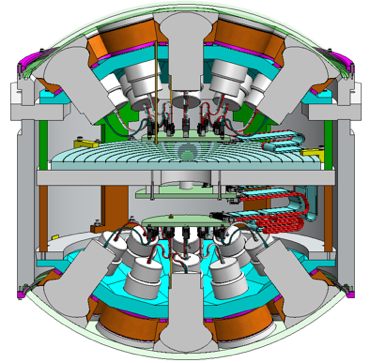}
 \caption{Multi-PMT conceptual drawing with 19 7.7\,cm PMTs as ID detectors and the OD detectors on the other half. Each small PMT has
 a reflector cone. An 50-cm acrylic covers on a cylindrical support is used as pressure vessel. Readout electronics and calibration
 sources are imbedded inside the vessel.
 \label{fig:mPMT}}
     \end{center}
   \end{minipage}
\hspace{0.5cm}
   \begin{minipage}{0.38\hsize}   
     \begin{center}
 \includegraphics[width=0.9\textwidth]{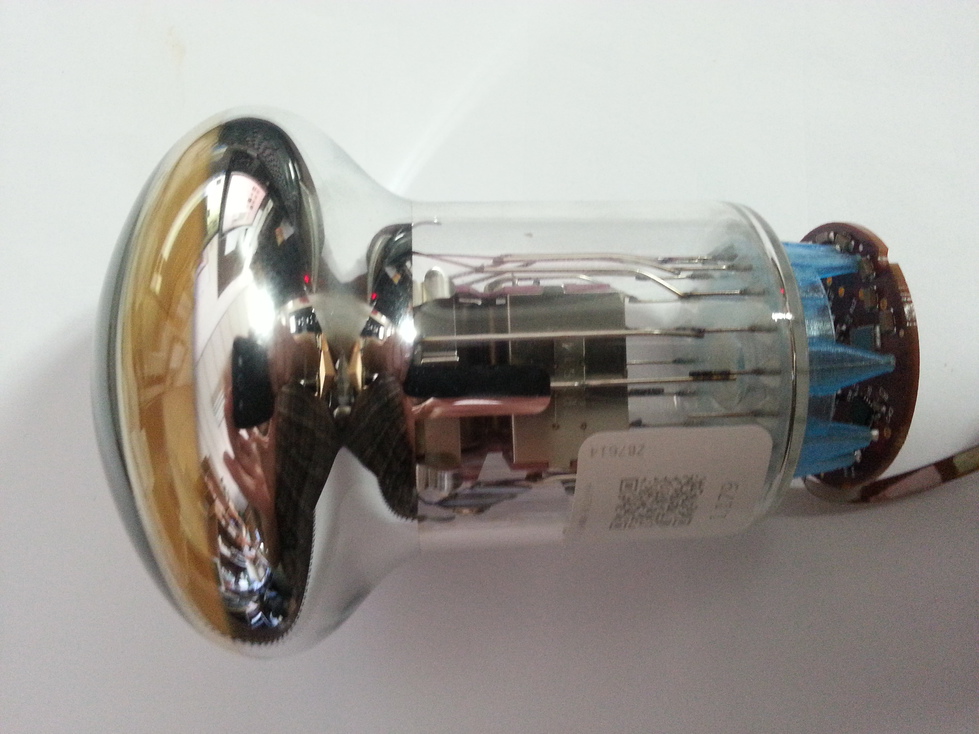}
 \caption{A Hamamatsu R12199-02 7.7\,cm PMT that is currently used in KM3NeT and considered for
 IceCube-Gen2 modules. As this passed the Hyper-K PMT requirements, it is also a good candidate
for a Hyper-K mPMT. \label{fig:3inchPMT}}      
     \end{center}
   \end{minipage}
 \end{tabular}
\end{figure}

\subsubsubsubsection{The Multi-PMT Reference Design} \label{multiPMT reference design}
As a reference design for Hyper-K, we consider 50\,cm size vessel, the same size as the baseline
50\,cm PMT so that the same mechanical support structure can be used. This would allow part
of the photocathode coverage to be replaced by an mPMT without major change in the support structure.
Fig.~\ref{fig:mPMT} shows the 50\,cm diameter mPMT design originally developed for the NuPRISM detector.
There are 19 7.7\,cm PMTs looking inner detector side and 6 7.7\,cm PMTs looking outer detector side.
The 7.7\,cm PMTs will be supported by a 3D printed foam structure and
optically and mechanically coupled by Silicon Gel to an acrylic pressure sphere.
When reflector cones are added to each 7.7\,cm PMT to increase the effective photocathode area
by about 30\%, we get about half of the effective photocathode area as a single 50\,cm PMT.
For the low energy events such as neutron tagging, the number of detected photons will be reduced
by a factor of two, which can be compensated by the reduction of the effective dark hit rate with the
limited field of view. A back-of-the-envelope calculation shows that the neutron tagging efficiency
would be similar or better than all 50\,cm PMT, although more detailed simulation studies are required.

\par Currently, there are two main candidate 7.7\,cm PMTs that have
     been developed specifically for KM3NeT, the Hamamatsu R12199-02
     (see Fig.~\ref{fig:3inchPMT}) and the ET Enterprises
     D792KFL/9320KFL. They have been measured in
     detail~\cite{VLVnT13.1,VLVnT13.2,VLVnT13.3} and would be adequate
     for Hyper-K. Both PMTs have a high peak QE of $\sim$ 27\% at
     404\,nm. Their collection efficiency is more than 90\%.  The
     transit time spread is about 4\,nsec at FWHM and the dark rate is
     200--300\,Hz. 
     
Recently, improved photosensors from both Hamamatsu and ET Enterprises have been made available, and are currently under test to fully characterize their performances.
Preliminary measurements of the newly developed Hamamatsu R14374 7.7\,cm PMT showed an improved TTS and dark rate respect to Hamamatsu R12199 PMT, finding a gain $\sim 5 \times 10^{6}$ at $\sim$ 1.2\,V with negative HV ($\sim$ 1.1\,V with positive HV) and a TTS of 1.35\,ns with negative HV ($\sim$ 1.58\,ns with positive HV) at gain $\sim 5 \times 10^{6}$. 
The dark hit rates were measured to be $\sim$ 0.3\,kHz at the 25\,$^\circ$C temperature in air. 
An improved photosensor from ET Enterprises Ltd., D793KFLB, which has been recently made available, and is currently under test. 
Preliminary results are given in Section \ref{section:photosensors:OD3inchPMT}.
In addition to Hamamatsu and ET Enterprises PMT's, HZC XP72B20 is currently reducing the TTS and dark rate and becomes a candidate.
HZC has a mass production capacity and potential to provide significantly lower cost. 
MELZ PMT, that is considered for the JUNO experiment, is another potential candidate.

\par The price for $\sim$19 7.7\,cm PMTs is comparable or cheaper than
one large area 50\,cm HQE B\&L PMT. In addition, the cost could be reduced
due to the competition between several companies like
Hamamatsu, ET Enterprises and HZC in the next couple of years.
The front-end electronics
will be situated in modules in the water near the PMTs which need to
be pressure tolerant, water-tight and use water-tight connectors (see
Section~\ref{section:electronics-general}). This cost may be reduced
by encasing the front-end electronics inside the same pressure vessel
as the ID and OD PMTs. The HV generation for each ID PMT can be done
on a board attached to the PMT base. Only one water-proof cable for
both communication, LV and signal can then be connected to the whole
module through a penetrator, as done in previous deep water neutrino
experiments.

\par
\begin{figure}[htbp]
     \begin{center}
\includegraphics[width=0.95\textwidth]{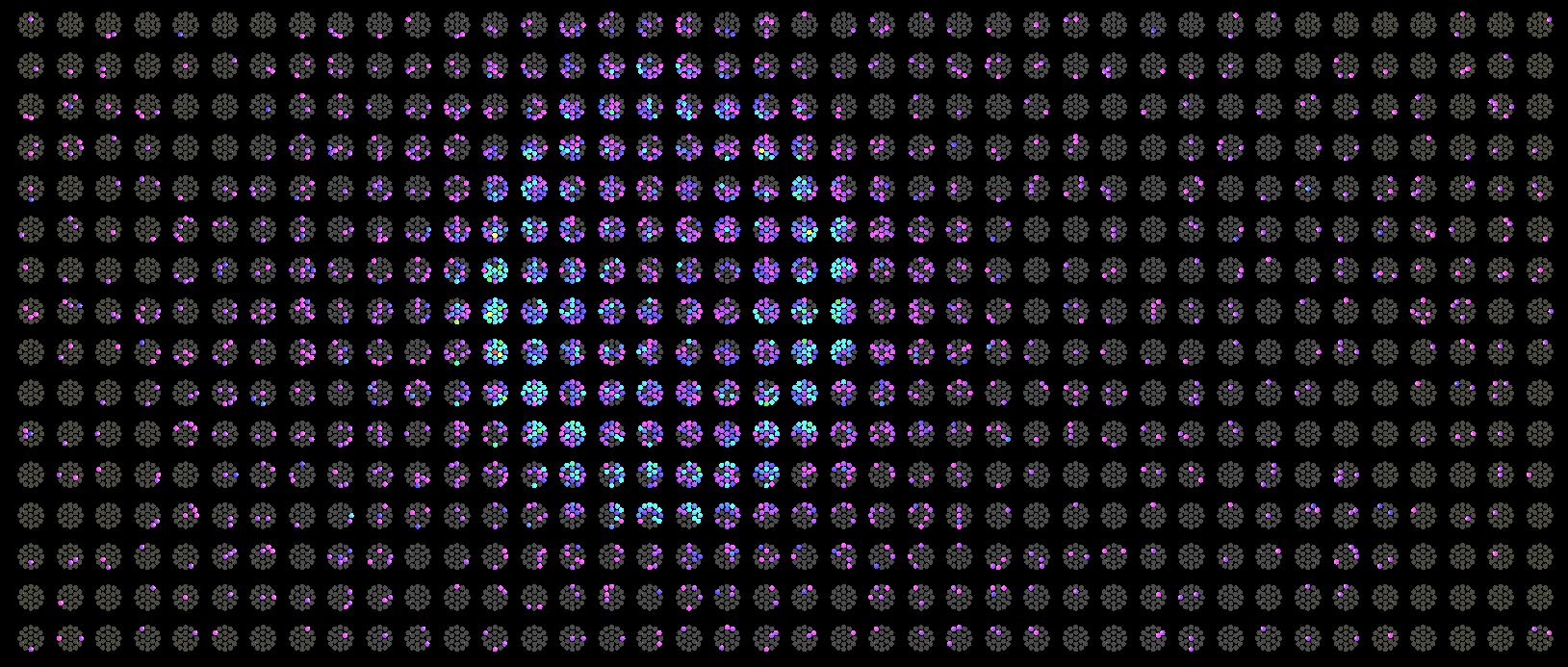}
\caption{An event display of mPMT for NuPRISM which assumes mPMT for all the photosensors. }
\label{fig:mPMT_display}
     \end{center}
\end{figure}

 A flexible implementation in the simulation software WCSim (see Section~\ref{software:WCSim}) makes
     optimization of the reference design possible and will be the
     main topic for further study together with further improvement of
     small photosensors. Based on the results of the optimization
     studies 
     facilities a prototype mPMT will be built and tested.
     Figure~\ref{fig:mPMT_display} shows an event display of mPMT for NuPRISM
     which assumes mPMT for all the photosensors. The event shows an improved
     ring pattern recognition by mPMT for an event.

%%%%%%%%%%%%%%%%%%%%%%%%%%%%%%%%%%%%%%%%%%%%%%%%%%%%%%%%%%%%%%%%%%%%%%%%%%%%%%%

\subsubsubsubsection{The mPMT Prototype}

\par    The design of a KM3NeT mPMT module is restricted by the size of
     commercially available transparent pressure vessels: borosilicate
     glass spheres with a diameter of 33.3\,cm and 43.5\,cm.
     The glass sphere also contains radioactive contaminants that emit Rn into the water,
     which is not a problem for KM3NeT where the radioactive background rate is limited
     by $^{40}$K in the sea water.
     Pure glass could be used for the vessel in Hyper-K, but the cost
     should increase, therefore a good alternative is given by
     acrylics.  A first prototype of an mPMT of the future Hyper-K
     experiment is under construction mainly to demonstrate the
     effectiveness of a vessel system based on acrylic and to study a
     better solution for the PMT Read-out system. Other prototypes
     will be built and tested when the final design of the mPMT will
     be defined on the basis of the optimization studies.  Several
     comparative tests are ongoing to identify the best acrylic for
     the experimental requirements, including optical tests, stress
     and compression mechanical tests and thermal tests. The water
     absorption tests are based on Nuclear Reaction Analysis (NRA),
     whereas radioactivity contamination measurements are based on
     gamma spectroscopy.  The UV transparency has
     been measured for several commercial acrylics.  For some sample,
     the light transmittance of the acrylic cover measured in water is
     greater than 95\% for a wavelength longer than 350\,nm.  The
     pressure vessel will be realized starting from two acrylic
     hemispheres.  The hemisphere fabrication processes starts by blowing the
     flat sheet onto a positive mold. A mold made from two parts
     (i.e., positive and negative) could be used to have
     a more uniform thickness. The final design of the vessel will be
     defined on the basis of the simulation studies for the mPMT
     optimization.

\par For the pressure vessel closure system, the two acrylic
hemispheres might be glued by using a specific glue for
acrylics. However, glue itself could emit light, producing more
background events for low-energy physics. A mechanical system has been
evaluated, both to avoid fluorescence emissions, and to guarantee a
longer endurance, and to simplify the anchorage to the tank frame and
the implementation of the cooling system of the mPMT. Thus a metallic
ring is being evaluated, modifying the spherical final shape by
extending the equatorial zone with a cylinder.
The acrylic vessel in the mPMT is similar in price to the protective cover required for the single large PMTs.

\begin{figure}[htbp]
     \begin{center}
 \includegraphics[width=0.7\textwidth]{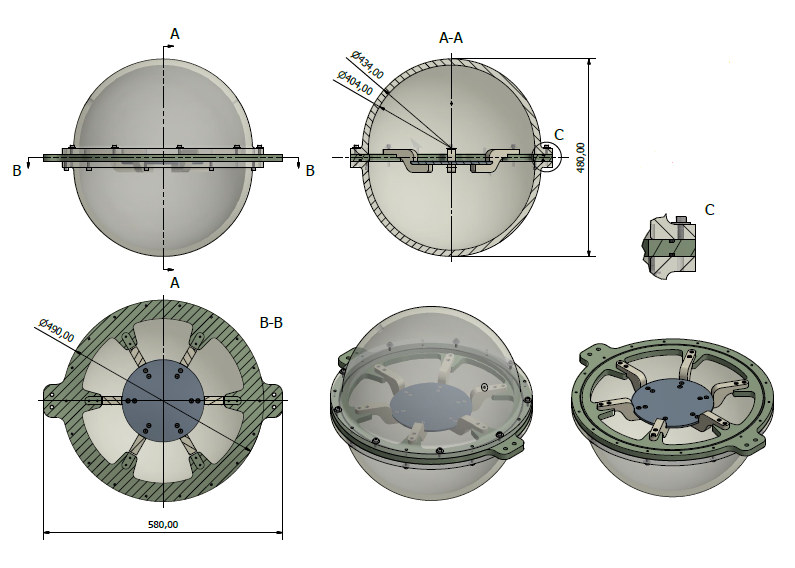}
 \caption{Preliminary
 design of the mPMT vessel with its cooling system in the equator of
 the sphere.}
 \label{fig:VesselDesign}
     \end{center}
\end{figure}

The first mPMT prototype will have an acrylic vessel with a diameter of 17 inches
(432\,mm, as in KM3NeT).  Several commercial acrylics have been studied
and tested. EVONIK UV transmitting PLEXIGLAS GS \footnote{\href{http://www.plexiglas.net}{http://www.plexiglas.net}} has been
chosen for the construction of the first prototype.
The thickness of the vessel has been studied on the basis of
simulation and three vessel prototype with thickness of 12, 15 and 18\,mm
will undergo to preliminary pressure tests.  The sphere will house
26 PMTs with a photocathode diameter of 7.7\,cm.

19 PMTs will view the inner detector side and 7 PMTs
will view the outer detector side.  The final number of PMTs in the
mPMT will be defined on the basis of simulation studies.
The PMTs will be placed into a 3D printed structure and will be
optically and mechanically coupled by Silicon Gel to an acrylic
pressure sphere.
The optical gel used for this prototype will be the same as in
KM3NeT. The compatibility between optical gel and acrylic has been
checked and the transparency of acrylic+optical gel has been measured.
For the final mPMT design, other options for the optical gel are under
study.

For the present prototype module, Hamamatsu R12199 PMTs are used. They
are arranged in 3 rings of PMTs in the hemisphere looking at the inner
detector with zenith angles of 33$^\circ$, 56$^\circ$, 72$^\circ$,
respectively. In each ring 6 PMTs are spaced at 60$^\circ$ in azimuth
and successive rings are staggered by 30$^\circ$. The central PMT in
the hemisphere point at a zenith angle of 0$^\circ$, looking at the
inner detector axis.  Seven PMTs are arranged in the hemisphere looking at
the outer detector. Six of them are arranged in one ring which opens a
half angle of 33$^\circ$ with respect to the nadir.

A basic Cockcroft-Walton voltage multiplier circuit design developed
by KM3NeT Collaboration~\cite{Timmer2010GF} is used to generate multiple
voltages to drive the dynodes of the photomultiplier tube. The system
draws less than 1.5\,mA of supply current at a voltage of 3.3\,V with
outputs up to -1400\,V$_{dc}$ cathode voltage.
A passive cooling system, based on the heat conduction mechanism,
aimed at keeping the temperature of the electronic components as low
as possible, thus maximizing their lifetime has been designed in order
to optimize the transfer of the heat generated by the electronics.
In KM3NeT the time over threshold (ToT) strategy is exploited; this is
not a good solution for Hyper-K project in which charge measurement is
important.
To fulfill the requirements of low consumption and charge and time
resolution, a solution based on a Sample\&Hold plus ADC has been
investigated. Several commercial low power and highly versatile ASIC
from Weeroc \footnote{\href{http://www.weeroc.com}{http://www.weeroc.com}} are under study.

For this prototype, the module was developed as a complete stand-alone
detector and it will fully test both in air and in water.

%%%%%%%%%%%%%%%%%%%%%%%%%%%%%%%%%%%%%%%%%%%%%%%%%%%%%%%%%%%%%%%%%%%%%%%%%%%%%%%
        \subsubsubsection{Alternative Cover Design}
\label{section:photosensors:coveralternatives}

\begin{figure}[h!]
 \begin{tabular}{cc}
   \begin{minipage}{0.35\hsize}
     \begin{center}
 \includegraphics[width=1.\textwidth]{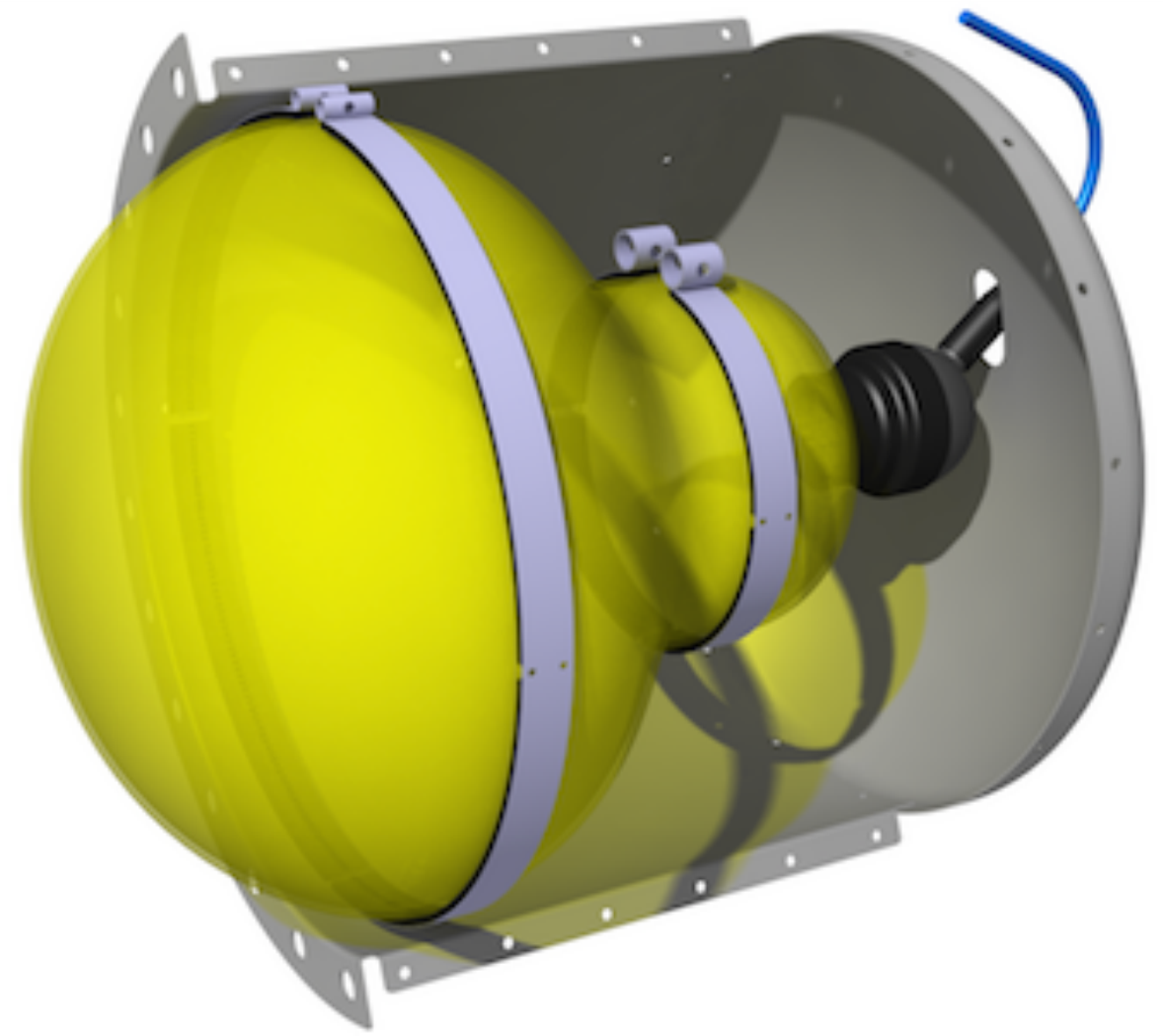}
 \caption{Alternative cover design comprised of a stainless steel tube and the acrylic window.\label{fig:AcrSUStubecover}}
     \end{center}
   \end{minipage}

\hspace{1.0cm}

   \begin{minipage}{0.4\hsize}
     \begin{center}
 \includegraphics[width=1.\textwidth]{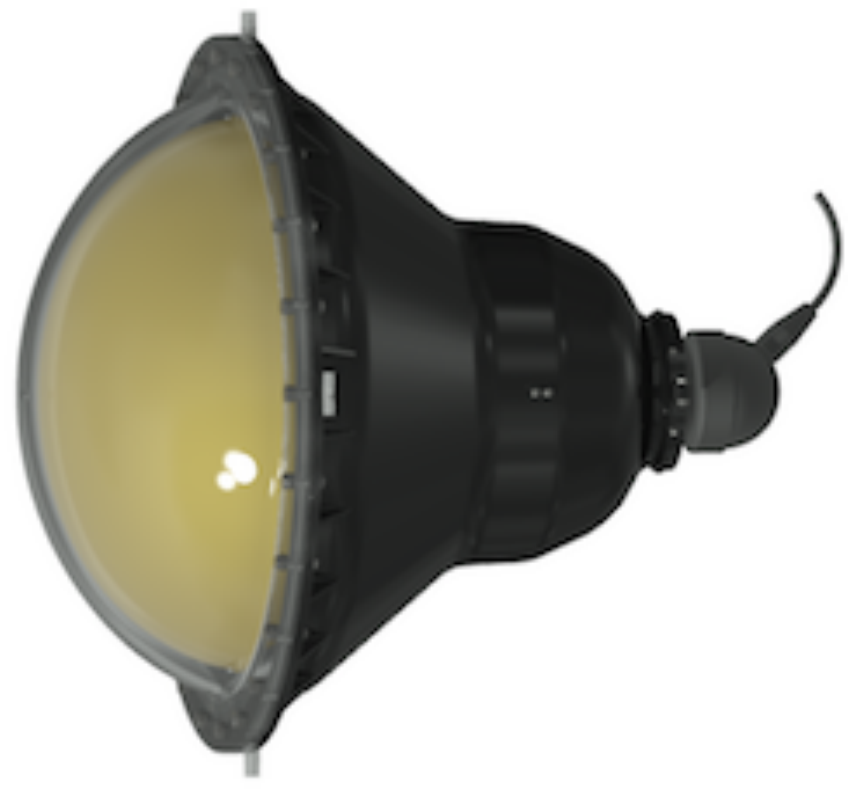}
 \caption{Alternative cover design comprised of a resin cover and the acrylic window.\label{fig:AcrPPScover}}
     \end{center}
   \end{minipage}
 \end{tabular}
\end{figure}

An alternative design of the shockwave prevention cover is a cheap stainless steel tube with the acrylic window, instead of the conical shape cover, as shown in Figure~\ref{fig:AcrSUStubecover}.
Benefit of the mass production and installation is expected due to the simple outer shape.

Figure~\ref{fig:AcrPPScover} shows another possible idea to use a resin instead of stainless steel.
It can give a cheap and light weight option.
One of the possible material is PPS resin mixed with a reinforced filler.

The light weight of the cover allows for a lower cost of the tank structure because the current cover weight is much heavier than the PMT and dominant in overall weight of the photosensor system.
Fast production and installation are also cost effective.
In both alternative cases, the design was evaluated by a simulation assuming a high pressure load outside, but it is still preliminary.
The hydrostatic test and demonstration test with implosion are necessary to ensure that the alternative covers are feasible for Hyper-K.

%%%%%%%%%%%%%%%%%%%%%%%%%%%%%%%%%%%%%%%%%%%%%%%%%%%%%%%%%%%%%%%%%%%%%%%%%%%%%%%
\subsubsection{Schedule}\label{section:photosensors:schedule} 

Before the photosensor mass production, it takes 0.5 years to complete
a design of a production line, and 1--1.5 years for the setup and
startup of the equipment in the factory.  The capacity of the factory
production is expected to be 11k 50\,cm and 4k 20\,cm photosensors per
year.

A test in Super-K using about a hundred of the B\&L PMTs starts
from 2018 to demonstrate the Hyper-K photosensor system.
Moreover, criteria for the quality control including a
selection with high pressure load will be established by the production.

Table~\ref{photosensoroption} summarizes the default design and
alternatives for photosensors described in this section.
Even without the alternatives, Hyper-K has sufficient
performance with a realistic time line using the default options.
Further improvements could be achieved and could be available in
time for Hyper-K tank construction.

\clearpage

\begin{table}[!h]
 \begin{center}
  \begin{tabular}{ll|ll|l}
   \hline \hline
      Items & & Type & & Remaining studies \\
   \hline
   \hline
     % ID B&L PMT
     \multicolumn{2}{l|}{{\bf ID photosensor}}  &
     \begin{minipage}{3.6cm}
     \begin{flushleft}
     {\bf HQE 50\,cm\\
          \ B\&L PMT}\\
     {\bf \small (Hamamatsu\\
                 \ R12860-HQE)} \\
     \end{flushleft}
     \end{minipage}
     &
     \begin{minipage}{3.cm}
     \includegraphics[width=\textwidth]{photosensor_R12860design.pdf}
     \end{minipage}
     &
     \begin{minipage}{7.cm}
     \begin{flushleft}
     Mass test using about 100 PMTs \\
     \ with reduced dark hit rate
     \end{flushleft}
     \end{minipage}
     \\
   \hline
     % ID HPD
     & (Alternative) &
     \begin{minipage}{3.6cm}
     \begin{flushleft}
     HQE 50\,cm HPD \\
     (Hamamatsu R12850-HQE)\\
     \end{flushleft}
     \end{minipage}
     &
     \begin{minipage}{2.4cm}
     \includegraphics[width=\textwidth]{photosensor_HPDwaterproof.pdf}
     \end{minipage}
     &
     \begin{minipage}{7.cm}
     \begin{flushleft}
     Electronics tuning,\\
     Confirmation of pressure resistance,\\
     Long-term proof test in water,\\
     Cost estimation
     \end{flushleft}
     \end{minipage}
     \\
   \hline
   \hline
     % OD
     \multicolumn{2}{l|}{ {\bf OD photosensor }} &
     \begin{minipage}{3.6cm}
     \begin{flushleft}
     {\bf HQE 20\,cm\\
          \ B\&L PMT}\\
     {\bf \small (Hamamatsu\\
                 \ R5912)}\\
     \end{flushleft}
     \end{minipage}
     &
     \begin{minipage}{2.5cm}
     \includegraphics[width=\textwidth]{photosensor_8pmtassy2.png}
     \end{minipage}
     &
     Test in high pressure water\\
   \hline
      & (Alternative)                      & HQE 20--30\,cm PMT                & &  HQE study using prototype,            \\
      &                                    &                                   & &  Trial manufacture and necessary test \\
   \hline
      & (Alternative) &
     \begin{minipage}{3.6cm}
     \begin{flushleft}
     7.7\,cm PMT
     \end{flushleft}
     \end{minipage}
     &
     \begin{minipage}{2.cm}
     \includegraphics[width=\textwidth]{3inchPMT.jpg}
     \end{minipage}
     &
     \begin{minipage}{7.cm}
     \begin{flushleft}
     Performance simulation in Hyper-K,\\
     Detection efficiency measured with prototype,\\
     Selection between design
     \end{flushleft}
     \end{minipage}
     \\
   \hline
   \hline
     \multicolumn{2}{l|}{
     \begin{minipage}{3.cm}
     \begin{flushleft}
     {\bf (ID and OD \\
     \ photosensor\\
     \ alternative) }
     \end{flushleft}
     \end{minipage}
     }
     &
     \begin{minipage}{3.6cm}
     \begin{flushleft}
     Multi-PMT \\ optical module\\
     \end{flushleft}
     \end{minipage}
     &
     \begin{minipage}{2.cm}
     \includegraphics[width=\textwidth]{mPMT_NuPRISM.png}
     \end{minipage}
     &
     \begin{minipage}{7.cm}
     \begin{flushleft}
     Performance simulation in Hyper-K,\\
     Cost estimation,\\
     Selection of 7.7\,cm photosensor,\\
     Trial manufacture and necessary test
     \end{flushleft}
     \end{minipage}
     \\
   \hline
   \hline
     \multicolumn{2}{l|}{ {\bf ID cover }}      &
     \begin{minipage}{3.6cm}
     \begin{flushleft}
     {\bf Acrylic and \\
      \ stainless steel\\
      \ cone}
     \end{flushleft}
     \end{minipage}
     &
     \begin{minipage}{2.8cm}
     \includegraphics[width=\textwidth]{protective_cover_shape.pdf}
     \end{minipage}
     &
     \begin{minipage}{7.cm}
     \begin{flushleft}
     Improved with light weight,\\
     Test of the improved design
     \end{flushleft}
     \end{minipage}
     \\
   \hline
      & (Alternative) &
     \begin{minipage}{3.6cm}
     \begin{flushleft}
     Acrylic and \\
     \ stainless steel\\
     \ tube
     \end{flushleft}
     \end{minipage}
     &
     \begin{minipage}{1.7cm}
     \includegraphics[width=\textwidth]{photosensor_AcrSUStubeCover_outside.pdf}
     \end{minipage}
     &
     \begin{minipage}{7.cm}
     \begin{flushleft}
     Design and simulation,\\
     Implosion test
     \end{flushleft}
     \end{minipage}
     \\
   \hline
      & (Alternative) &
     \begin{minipage}{3.6cm}
     \begin{flushleft}
     Acrylic and \\
     \ full resin
     \end{flushleft}
     \end{minipage}
     &
     \begin{minipage}{2.cm}
     \includegraphics[width=\textwidth]{photosensor_AcrPPSCover_outside.pdf}
     \end{minipage}
     &
     \begin{minipage}{7.cm}
     \begin{flushleft}
     Design and simulation,\\
     Prototype production,\\
     Implosion test
     \end{flushleft}
     \end{minipage}
     \\
   \hline \hline
  \end{tabular}
  \caption{ Summary of the current base design in bold type and alternative options related with photosensors.
            The remained studies of the baseline photosensors and cover will be finalized in fiscal year 2017.
           }
  \label{photosensoroption}
 \end{center}
\end{table}

%%%%%%%%%%%%%%%%%%%%%%%%%%%%%%%%%%%%%%%%%%%%%%%%%%%%%%%%%%%%%%%%%%%%%%%%%%%%%%%
%%%%%%%%%%%%%%%%%%%%%%%%%%%%%%%%%%%%%%%%%%%%%%%%%%%%%%%%%%%%%%%%%%%%%%%%%%%%%%%

\clearpage
\graphicspath{{design-electronics/figures}}
\subsection{Frontend electronics }\label{section:electronics}
\subsubsection{ General concept of the baseline design }\label{section:electronics-general}

 It is not possible to tell when and where a natural neutrino 
interacts in the detector. Therefore, the front-end electronics 
modules for the detectors, which are used to study neutrino from nature, 
are required to digitize all signals from photo-sensors that are above a certain
threshold -- i.e. the acquisition needs to be self-triggered. The digitized information 
is then either recorded or discarded, depending on the design of the detector-wide trigger system.

Current design of the HK detector is quite similar to the SK
detector, in terms of the required specifications and the number of
photo-sensors in one detector. Therefore, it is reasonable to 
start with the system used in the SK detector.

 The photo-sensor for the inner detector of HK is newly developed.
Based on the baseline option, around 40,000 20inch PMT R12860-HQE 
is used. The R12860-HQE PMT has better timing and charge resolution 
compared to the same diameter PMT (R3600), which has been used in SK.
The dark (noise) rate is required not to exceed 4~kHz, which is a 
similar requirement to the R3600 PMT. 
Based on these information, we have
estimated the total data rate and concluded that it is possible to
design the data acquisition system, which has similar to the concept 
of the SK-IV DAQ.

 As already realized in the SK-IV DAQ system, it is possible to 
read out all the hit information from the photo-sensors, including 
the dark noise hits. There is no technical problem in selecting 
the actual events to be recorded for the analyses by software.

 One difference is the size of the detector.  The total amount of
photo-sensors in one entire detector is expected to be up to $\sim$
47,000, including the sensors for OD.  If we locate the front-end 
electronics modules on the top of the detector, it is necessary 
to run the cable from the PMT to the roof and the detector structure 
has to support the weight of the cables, which is expected to be 800~tons. 
Thus, it would be possible
to simplify the detector structure if we can reduce the weight
of the cables. Also, the maximum length of the cable is $\sim$ 
30\% longer than in the SK case. This not only reduces the signal amplitude,
but also degrades the quality of the signal -- the leading edge is smoothed out due
to higher attenuation of the cable in the high frequency region.
Therefore, we plan to place the modules with the front-end 
electronics and power supplies for the photo-sensors
in the water, close to the photo-sensors. This configuration
makes it possible to have shorter signal cables from the photo-sensors 
and also allows for significant reduction of the weight that needs to be supported by the photo-sensor 
support structure.  Of course, it is necessary to place the front-end module and
power supply for the photo-sensors in a pressure tolerant enclosure, and also to 
use water-tight connectors. This kind of ``water-tight'' casing 
has been studied in other experiments and there are several 
possible options. One concern is the cost of the special cables
and connectors -- consequently, we are developing special cables and connectors
dedicated for HK.

 The other issue we have to keep in mind is an inability to do any repairs 
 to a broken module that will be submerged in water.  Furthermore, a failure of
one module could affect the data transmission from other modules, resulting in a much
larger region of the detector that could be lost.  Therefore, the system
must be redundant and care must be taken to avoid a single
point of failure. Also, careful design of the data
transport connections and the timing distribution system are
essential.
 
\begin{figure}[htbp]
\begin{center}
  \includegraphics[width=0.7\textwidth]{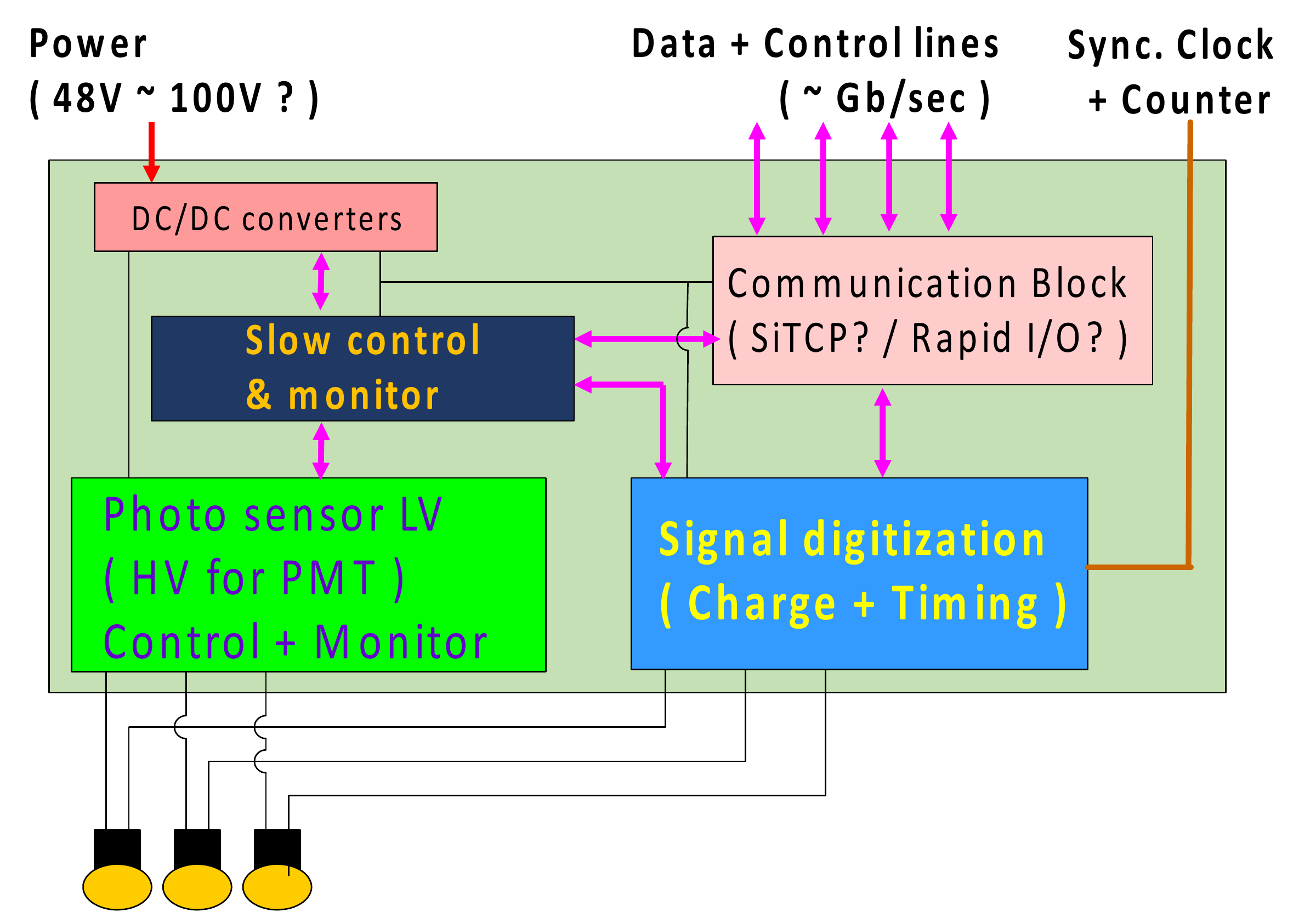}
  \caption{Schematic diagram of the front-end module.}
\label{schematic_frontend:daq}
\end{center}
\end{figure}

 The current baseline design of the front-end module is prepared
considering these requirements. The schematic diagram of the front-end
module is shown in Fig.~\ref{schematic_frontend:daq}.

 There are 4 main function blocks in the front-end board. The signal
digitization block, the photo-sensor power supply block, the slow
control block and the communication block. In the current baseline design, 
one module accepts signals from 24 photo-sensors, digitizes them and 
sends out the data.

 In the following sections, details of each component of the
front-end module are described, along with the data readout and processing parts.

\subsubsection{ Signal digitization block }

 The signal digitization block accepts the signals from the photo-sensors 
and converts them to the digital timing and charge data.  As
mentioned in the previous section, there is no way to tell when a
neutrino or a nucleon decay event happens. Therefore, the front-end
electronics module is required to have self-triggered analog to
digital conversion mechanism and to be dead-time free.  The actual
event rate is expected to be smaller than a tens of kHz, even with the
background events from gamma-rays from the surrounding wall or
from cosmic-rays. Also, the number of photo-sensors, which detect
sufficient charge in case of gamma-ray events is quite small, much 
less than 1\% of the photo-sensors in the detector.
Therefore, the time interval between photons hitting a single photo-sensor 
is rather long, much longer than the dark rate of the sensor.
However, muons decay into electrons in the detector and photons from
both of them may hit a single photo-sensor. Therefore, it is necessary 
to have the capability to detect both photons, generated by the parent 
muons and the decay electrons. The lifetime of muons are rather long, 
$\sim$ 2\,$\mu$ sec and thus, it it not necessary to be completely 
dead-time free but the dead-time should be as short as possible.

 One possible way to satisfy these requirements is to employ the 
charge-to-time conversion (QTC) chips. The QTC chip receives
the signal from the photo-sensor and produces the digital signal, 
whose width is linearly dependent on the amount of the input charge. Also, 
the leading edge of the output digital signal corresponds to the 
time when the input signal exceeded the pre-defined threshold to produce the
output digital signal. The digital output signal from the QTC chip
is read out by a TDC. Usually, the maximum width of the output 
signal may be slightly longer than the charge integration gate 
width. Therefore, there is a small dead-time after the first 
signal but it is no larger than several hundreds of ns and 
is acceptable for use in the water Cherenkov detector.

 The requirements of the charge and timing resolution are summarized
in Table~\ref{requirements1:daq}.

\begin{table}[htb]
\begin{tabular}{l|l}
\hline\hline
items & required values \\
\hline
Built-in discriminator threshold & 1/4 p.e ( $\sim$ 0.3 mV )\\
Processing speed & $\sim$ 1$\mu$sec. / hit \\
Charge resolution & $\sim$ 0.05 p.e. ( RMS ) for $<$ 5 p.e. \\
Charge dynamic range & 0.2 $\sim$ 2500 pC ( 0.1 $\sim$ 1250 pe. )\\
Timing response & 0.3 ns RMS ( 1 p.e. )\\
                & 0.3 ns RMS ( $\geq$ 5 p.e. )\\
Least time count & 0.52 ns \\
Time resolution & 0.25 ns \\
Dynamic range   & $\geq$ 15 bits \\
\hline\hline
\end{tabular}
\caption{ Specification of the signal digitization block. }
\label{requirements1:daq}
\end{table}

 The QTC chips ( CLC101 ) used in the front-end module of SK-IV, 
called the QBEE, are a good reference and satisfy all the requirements.  
The design rule of these chips is 0.35$\mu$m and it is possible to
produce them again. As for the TDC, the chips used in the QBEE, called AMT3,
have been discontinued. However, there are several implementations
of a `TDC in an FPGA' and some of them seem to have sufficient performance.
One candidate is the `wave union TDC' developed at FNAL. The performance 
of this TDC is expected to be better than that of the AMT3 and we are currently
evaluating this TDC design.

 Even though the current baseline design is to utilize the QTC-TDC 
approach, we are also investigating possibility of adopting Flash-ADC 
(FADC) type digitization. In this case, the FADC chip would run all 
the time and digitize the input signal. Afterwards, FPGA-based 
on-the-fly digital signal processing would be utilized to find the PMT 
pulse and determine its charge and time of arrival. An advantage of 
this approach is that it is completely dead-time free -- we would be 
able to detect photons both from prompt muons and from decay electrons, 
even if the latter happen only 100 ns after the initial interaction. 
We may also be able to distinguish photons from direct and reflected light.
 Another potential advantage is an ability to remove deterministic 
interference that may be present in the PMT signal. Example sources of 
such an interference are switching power supplies and high voltage 
supplies. The disadvantage is  potentially larger power consumption and
higher cost.

\begin{figure}[htb]
\centering
\includegraphics[width=0.49\textwidth,trim={12mm 74mm 22mm 74mm},clip]{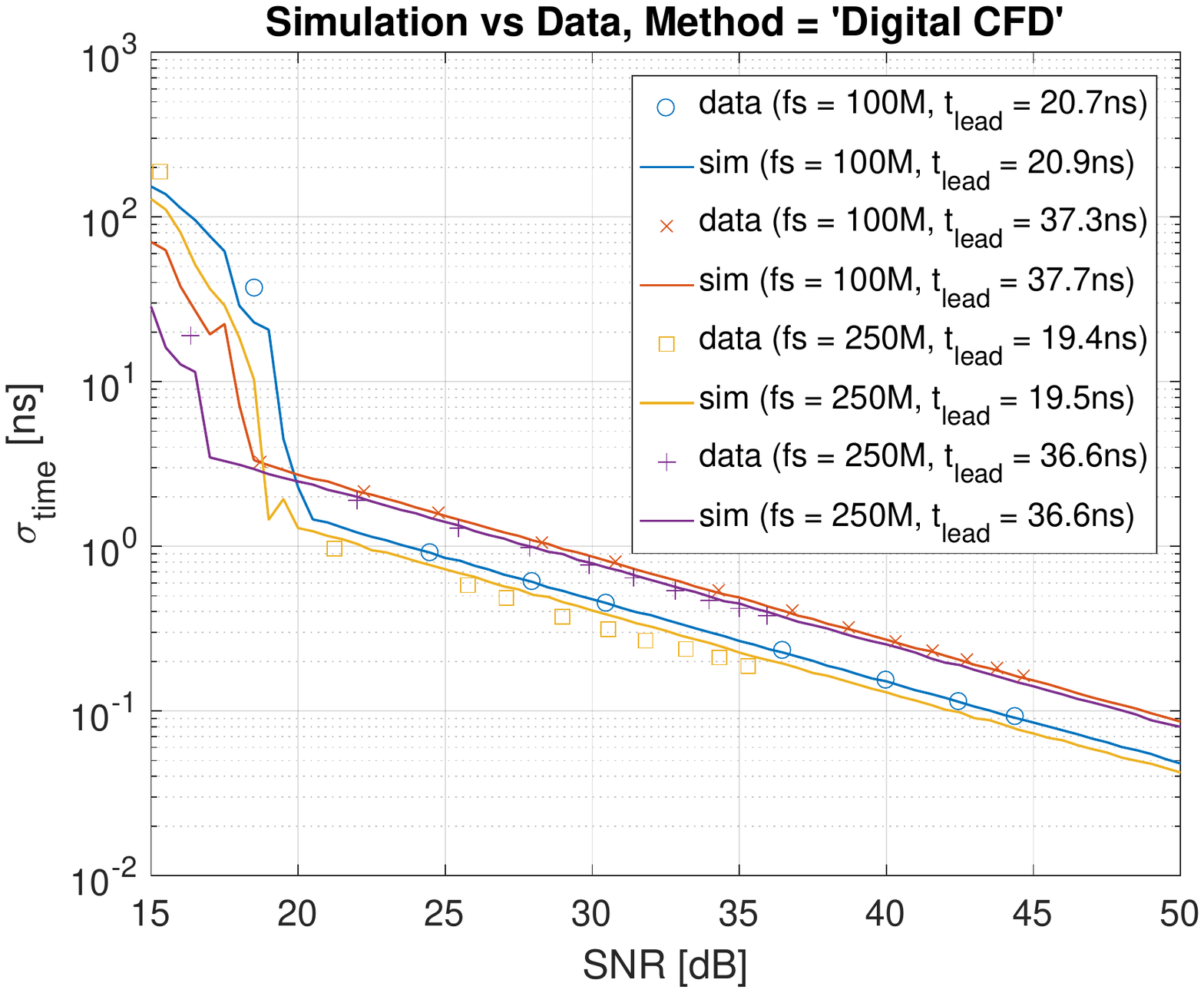}
\includegraphics[width=0.49\textwidth,trim={12mm 74mm 22mm 74mm},clip]{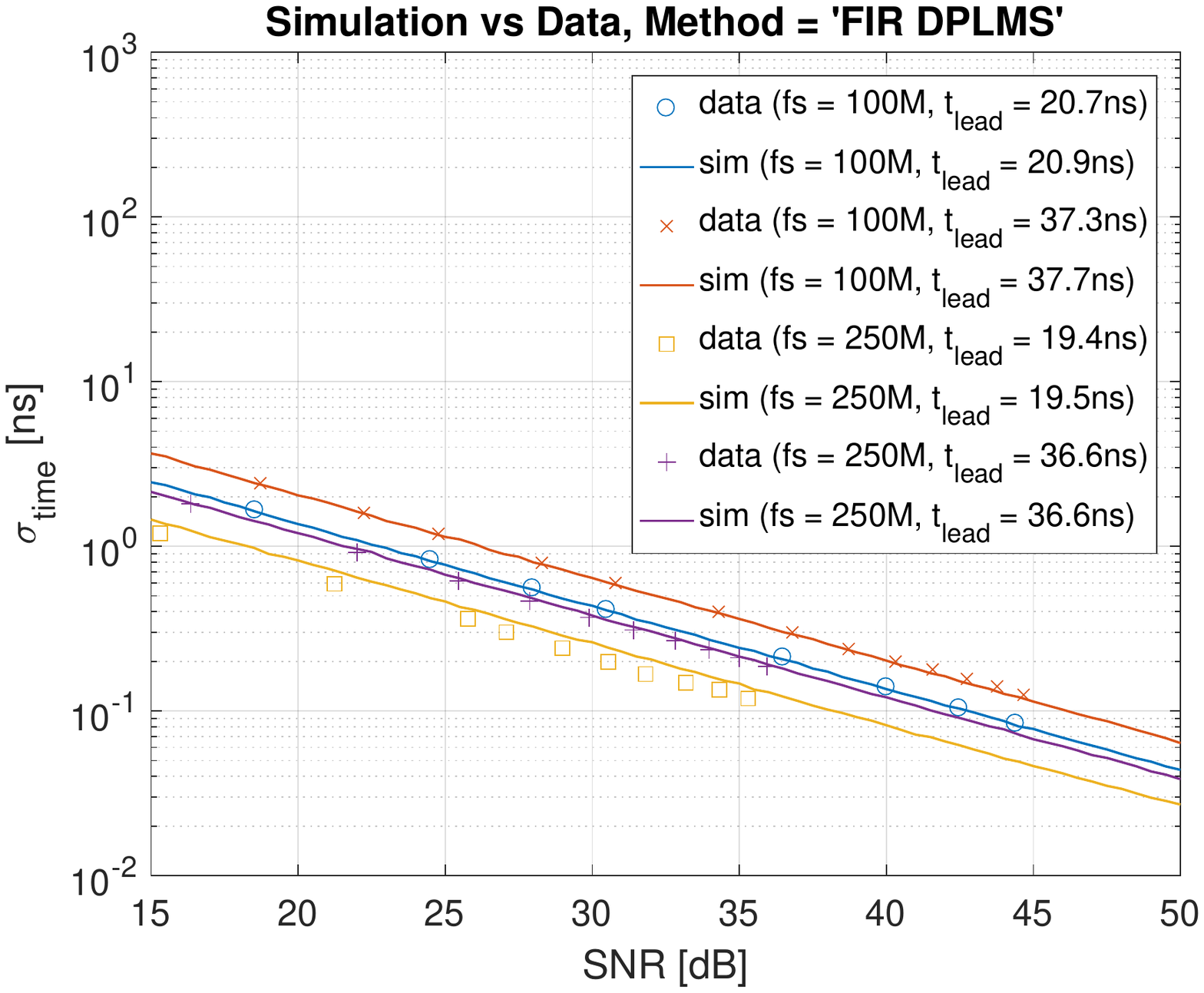}
\caption{Results of the study of the timing performance of FADC-type digitization, with pulse timing using digital constant fraction algorithm (left) and optimal filter (right). The precision of the ADC is expressed as signal-to-noise ratio -- $SNR = 6.02N + 1.76$~[dB], with $N$ being the effective number of bits.}
 \label{electronics:fadc:shaper_study}
\end{figure}

 Since both the power consumption and the cost are highly dependent on 
the speed and precision of the FADC ICs, it is advisable is to use the 
slowest possible configuration that will `do the job'. As such, a study 
has been performed in order to understand the performance of the system 
as a function of both the resolution and the sampling frequency of the 
FADC.  Furthermore, models of the system were developed and validated, 
so that further studies can be streamlined. Finally, various signal 
processing methods were tested for determining timing of the pulse -- 
among them the digital constant fraction discriminator \cite{Huber:2011dt}, 
optimal filters \cite{Gatti:2004lms,Abbiati:2006tim} and matched filters. 
The results of the study are presented in 
Fig.~\ref{electronics:fadc:shaper_study}. It is relatively easy to 
achieve the timing resolution that is below 10\% of the sampling period 
($T_S$). With sufficiently high SNR it is also possible to reach even 
better timing accuracy, well below 1\% of $T_S$. Based on these results,
we decided to use a 100 MSPS/14 bit FADC as the current baseline design for 
this digitization scheme. Two FADC channels will be used per single PMT (high-gain and low-gain channels), 
so that the dynamic range requirements are fulfilled. Current studies concentrate 
on optimizing an anti-aliasing filter, in order to achieve best possible 
timing resolution.
 
 Another solution, which allows for savings in power consumption and still 
maintains high sampling frequency, is to use switched capacitor 
arrays (SCA). The biggest advantage of this approach is that no 
bandwidth-limiting anti-aliasing filter is necessary. The price to pay 
is dead-time introduced due to `freezing' of the capacitor, which 
is necessary for its readout. One possibility of alleviating this problem is 
to utilize multiple SCA channels per photo-sensor. This way a dead-time free readout
can be provided up to a certain trigger frequency. A better option is to use SCAs
with segmented memory, so that one can avoid potentially expensive increase of the number
of SCA chips. Unfortunately, we are currently not aware of an existence of such an IC, with sufficiently 
large memory buffer. Therefore, we are considering possibility of a development of a new IC. The goal would be
to allow both a dead-time less readout, typical for pure FADC-type digitization,
 as well as high sampling speeds and low power consumption, provided by the SCAs.
 
 The basic assumption of the new design is that the chip would consist 
of both an SCA-type analog memory, a flash-ADC and a discriminator. The 
sampling speed of both the SCA and the flash-ADC would be equal. The
flash-ADC, which is the most `power hungry' part, would be kept off 
most of the time -- it would be activated only once a sufficiently high 
pulse is detected at the input. Since some time is required for resuming
the flash-ADC operation, the analog memory would be used to store the 
pulse -- sampled, but not yet quantized. Once the ADC is fully active, 
the analog memory would act as a first-in first-out buffer, without any freezing 
of its content. Thus, the system would be able to work as long as there 
is any useful signal, without  any dead-time.

 In either case, it is necessary to be prepared for a failure of the
digitization component. We are therefore considering to have a set of spare
digitizer channels and to insert analog switches between the signal 
inputs and the digitization block. With this configuration, we have 
some flexibility to change the assignment of input signals to 
digitizer channels. This also provides us with an additional way to
calibrate the digitization blocks. Of course, analog switch is known
to degrade the quality of the signal. Thus, we are going to study
carefully prior to implementing this solution. Furthermore, the photo-sensors, 
which are currently considered, produce faster pulses and, consequently, higher maximum
voltage, which is expected to exceed 6 volts. This is larger than
the maximum allowable voltage of the digitizer chips. Therefore, we need to
design the protection circuit, which will
not degrade the timing and charge resolutions. 

 Because the relative timing is used to reconstruct the event vertex 
in the detector, all the modules have to be synchronized.  Therefore, 
it is necessary to drive the TDC or FADC by a clock synchronized to 
the reference clock fed externally.  Also, the system-wide counter 
is attached to the data to combine the data from different modules 
at a later stage.

\subsubsection{ The timing synchronization block }

 Synchronization of the timing of each TDC or FADC is crucial for precise 
measurement of the timing of photon arrival. In Hyper-Kamiokande,
timing resolution of the photo-sensor is expected to be largely
improved.  Therefore, we have to be careful with the
synchronization of the modules -- the design should minimize the clock jitter, so that
the timing resolution of the whole system is as good as possible. We are planning to distribute the
common system clock and the reference counter to all the modules. We
have not yet started the actual design of this system, but there are
several existing examples. The first method is to send the clock and
serialized 32 bit counter information using special STP cable, called the
nano-skew cable, whose skew is less than a few nanoseconds for a 100\,m long
cable.  This system has been used in current SK DAQ system and the
skew is measured to be much smaller than 100\,ps. It requires
intermediate timing distributor and therefore needs to be modified for
use in Hyper-Kamiokande, but it is not difficult. The other
possibility is to use the idea of White Rabbit. The White Rabbit
system is designed for the synchronization in the accelerator complex
and corrects the timing differences with measured delay in each node.
It is not necessary to implement entire functionality of White Rabbit
for our case but employ the main part of the timing synchronization
and stabilization.

\subsubsection{ The photo-sensor power supply block }

 If HPDs are used as the photo-sensor, the high voltage supplies for the 
 acceleration voltage and the avalanche photodiode bias voltage will
 be put on the HPD base. In this case, the front-end module will control 
 both power supplies via control signals.

If the normal PMTs are used as the photo-sensors, we are considering to
put the high voltage module in the same enclosure as the front-end
module. In this case, the control signal would be fed internally between 
the two modules.

The control signals to the HPD base or to the internal high 
voltage modules will be controlled remotely through the communication 
block.

\subsubsection{ The slow control and monitor block }

 It is important to control and monitor the status of the power supply
for the photo-sensors. Also, the voltage, the current and the
temperature of the front-end module has to be monitored.  This slow
control and monitor block is prepared for this purpose.  It
accepts the commands from the communication block and also keeps the
current status, which is available for read-back. All communication with the
`external world' is done through the communication block.

\subsubsection{ The communication block }

\begin{figure}[htbp]
\begin{center}
  \includegraphics[width=0.4\textwidth]{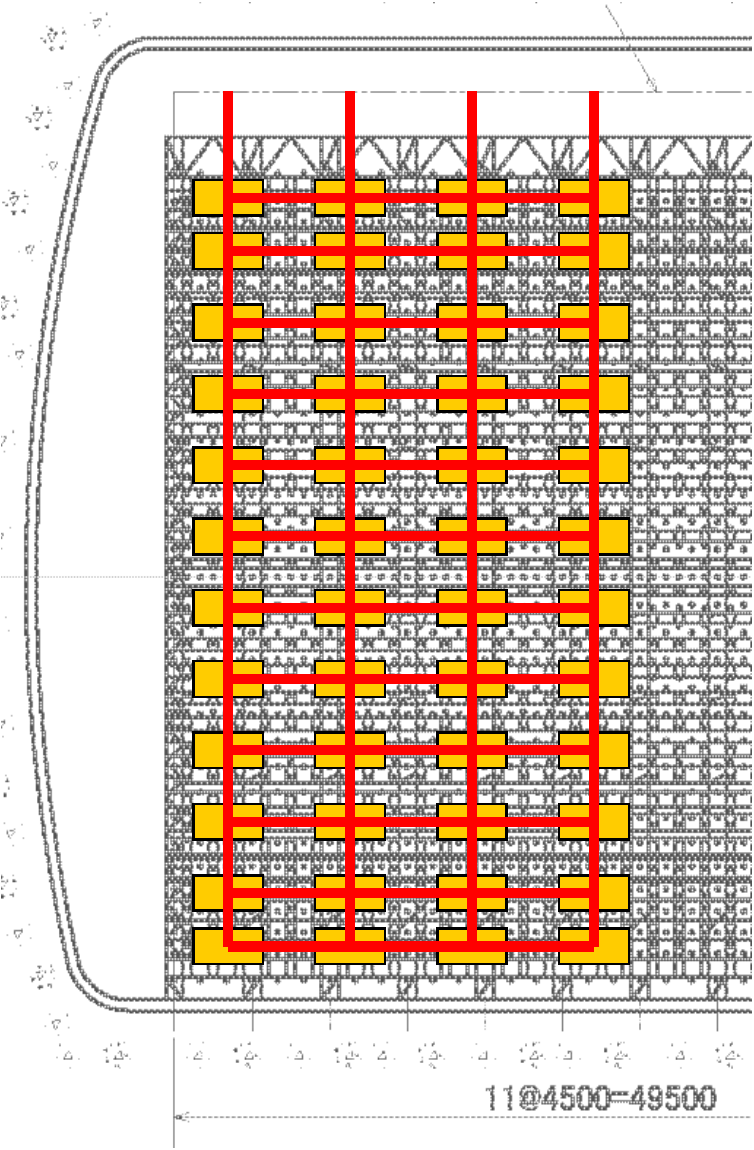}
  \caption{Schematic diagram of the connections between front-end modules.}
\label{connections:daq}
\end{center}
\end{figure}

In order
to reduce the amount of cables, we are planning to connect the modules
in a mesh topology, with each module connected to its neighbors -- Fig.~\ref{connections:daq}. 
Only the top modules would be connected to the readout computers. Each module will have
several communication ports, so that a single point of failure would be avoided. 
In case of failure of one of the modules, the data would simply be re-routed to one of the neighbors, 
thus ensuring that communication path will be secured.

 There are several possibilities for the connection, but one of
the promising ones is the SiTCP, an FPGA based TCP/IP stack. This TCP/IP
stack does not require a CPU core in the FPGA and is accessible like a simple
FIFO buffer. SiTCP acts as either a TCP/IP server or a client, so it is 
possible both to receive data from the other module and to add own data
 and later to send everything to the next module. Also, SiTCP has
registers, which can be accessed via UDP commands. With this 
functionality, it is possible to realize the slow control and
monitor system, such as setting the high voltage or monitoring
the status, for example read-back of voltages of the power supply.
Recently, CPU cores are embedded in the FPGA chip. With this kind
of chips, TCP/IP communication part is also possible to be handled
with the embedded CPU. We will investigate this possibility.
Apart from TCP/IP, there are several other industry standard
communication protocols available. One example is the Rocket-I/O.
Rocket I/O is the standard interface supported in Xilinx FPGA.
This allows us to transfer data at speeds exceeding gigabit per second.
We are also investigating this possibility for a faster 
communication between the modules.

\subsubsection{ Pressure tolerant cable and Water tight connectors }
 In order to connect the front-end electronics module with other modules, the photo-sensors
and the clock modules, we need to have water tight connectors and
pressure tolerant cables. It is known that normal Ethernet
cables are not capable of transmitting the data at full rate
under the pressure, because the characteristics of the cable
are changed when the cable is squeezed under the pressure.
Therefore, we have started the R\&D of the pressure-tolerant,
water tight Ethernet cable. This cable will use the similar sheath
material to the one used as the photo-sensor signal cable
not to affect the water quality of the detector.
We have also started designing the water tight connectors
for the PMT connection and the Ethernet connection. Both of the 
connectors are using screws and are easy to connect. This will 
reduce the time to connect cables during the construction.
The mock-up connectors have been designed and we are going
to produce samples and evaluate them in the coming years.

\subsubsection{ Timeline }
 Current plan from the finalization of the design to the completion
of the production and tests is shown in Table \ref{timeline1:daq}
\begin{table}[h]
\begin{tabular}{l l}
  Spring 2020 & Final design review of the system \\
  Autumn 2020 & Start the design of the system based on the design review \\
  Autumn 2021 & Start bidding procedure \\
  Autumn 2022 & Start mass production \\
  Autumn 2023 & Start final system test \\
  Autumn 2024 & Complete mass production \\
  Autumn 2025 & Complete system test and get ready for install
\end{tabular}
\caption{Timeline to complete the production for the installation.}
\label{timeline1:daq}
\end{table}

 In order to complete the design by Spring 2020, R\&D and evaluation
of each component have to be finished by then. Table \ref{timeline2:daq}
shows the deadlines for each component.
\begin{table}[h]
\begin{tabular}{l l}
  Digitizer & Autumn 2018 based on the decision of the photo sensors \\
  Timing and synchronization & Select technology by Autumn 2018 \\
  Communication block & Fix specification by Autumn 2018 \\
                      & Design by Spring 2019 \\
  High voltage system & Product selection and design by Autumn 2019 \\
  Water tight components & Technology choice by Spring 2019
\end{tabular}  
\caption{Deadlines for each components.}
\label{timeline2:daq}
\end{table}

 Considering the schedule, we need good coordination with the other
groups, including not only the photo-sensor groups but also the construction
groups. The allocated time for each item is not much but still achievable.

\newpage
\graphicspath{{design-daq/figures}}
\subsection{Data acquisition system }\label{section:daq}
\subsubsection{Data acquisition and triggering}

All PMT hits from the detector (above a threshold of $\sim
0.25$\,p.e.) will be delivered to the data readout and processing
system where they will be formed into events and recorded on disk for
further processing offline.  The overall rate of hits (mostly from
dark noise) from the inner detector will be about
460\,MHz, 
leading to a total input data rate of 5GB/s including additional data,
in the absense of waveform information.
The OD adds less than 10\% to this data load.  To reduce the data
recorded, trigger decisions will be made using real-time processing of
the hits in the detector.  Events will be formed from all hits within
a time-window surrounding the trigger and recorded to disk for offline
study.

The main trigger will be the same as that used in SK-IV; a trigger
will be generated when the total number of hits seen (NHITS) in a
sliding time-window exceeds a certain threshold (e.g. 27\,hits).  This
trigger will accept all the necessary data for studies of proton
decay, atmospheric neutrinos, beam neutrinos and cosmic ray muon
events.  It is important that there is no dead time in the triggering
or data collection so that delayed energy depositions following a
triggered event, such as from a Michel electron or neutron capture,
are recorded, either as part of the same event or separately.  More
sophisticated trigger algorithms, which can be added into the
architecture, are being studied to increase sensitivity to lower
energy events (by distinguishing events with fewer hits from random
combinations of dark-noise hits) and detection of supernova bursts (by
observing elevated trigger rates). As in Super-K, an additional
trigger input will be derived from the J-PARC beam-spill gate to
define readout windows around the beam spill time, independent of the
number of hits observed.  Triggers will be defined to receive
calibration events. Also, external trigger inputs are necessary to
take the calibration data with synchronized timing.  The estimated
rate of events is shown in Table~\ref{event_rates:daq} for readout
with the hit-only electronics, which require 12\,bytes per hit (the
waveform option needs a factor four higher bandwidth, $\sim 50$\,bytes
of information per hit).

\begin{table}[htbp]
\begin{center}
 \caption{Estimates of data rates.  The ``pre-trigger'' data rates for each physics process
    is given: (a) for events without dark noise and (b) for events including dark noise. 
    In these calculations, 12 Bytes per PMT hit is assumed. Event rates and event windows for background, muon, beam calibration and pedestal events are based on those from Super-Kamiokande.}
 \label{event_rates:daq}
 \begin{tabular}{l|l|r|r|r|r}
\hline\hline
Data source & Event rate & Hits/event         & Data rate  & Data rate & Data rate per\\
                     &                  & (event window) & pre-trigger(a)& pre-trigger(b)& event window\\
\hline
Dark noise                             & 10\,kHz each in 46,700\,PMT & 1          & 5.6\,GB/s & -- & --  \\
Very low energy background & 10\,kHz (1.5\,$\mu$s)             & 25       & 3\,MB/s & 84\,MB/s   & 200\,MB/s \\
Low energy background         & 35\,Hz (40\,$\mu$s)                & 50        &21\,kB/s & 7.8\,MB/s   & 15\,MB/s\\
Cosmic muons                       & 100\,Hz (40\,$\mu$s)             & 46,700 &56\,MB/s & 78\,MB/s & 15\,GB/s\\
Beam events                         & 1\,Hz (1\,ms)                          &      0      & 0\,MB/s & 5.6\,MB/s &5.6\,kB\\
Calibration                             & 2\,Hz                                      & 46,700  & 2\,MB/s & 2\,MB/s & -- \\
Pedestal                                & 1\,Hz                                      & 46,700  & 2\,MB/s & 2MB/s & -- \\

\hline
Total rate                &          &        &  5.6GB/s & 180MB/s &     \\
\hline\hline
 \end{tabular}
\end{center}
\end{table}

A requirement of the data acquisition system is to reliably trigger
and collect information from a supernova burst.  This is one of the
more challenging design aspects and benefits from the large memory
buffers available in modern commodity hardware, sufficient to retain
all raw hit information for around 100\,s of detector operation.  In 
the event of a supernova, the detector will self-trigger on a supernova 
burst by searching for an elevated rate of
individual triggers in a sliding time-window e.g.~0.1\,s.  If such a
trigger occurs, all raw hit information in a large time window around
the supernova trigger would be saved in the local hardware and then
slowly written to disk.  Note that even a close supernova burst
yielding 100,000~events per second of 50~PMT hits each (5~million
hits/s) will not overwhelm the readout which is designed to
continuously read 460\,MHz of dark noise hits.  A large storage buffer 
on hard disk drives is used to save
all the hit data, overwriting the oldest data.  An external
observation may reveal up to a few hours after the event that a
supernova signature was seen, perhaps in a neighbouring galaxy, in
which only a few events are expected to be observed in Hyper-K.

\begin{figure}[htbp]
\begin{center}
  \includegraphics[width=0.8\textwidth]{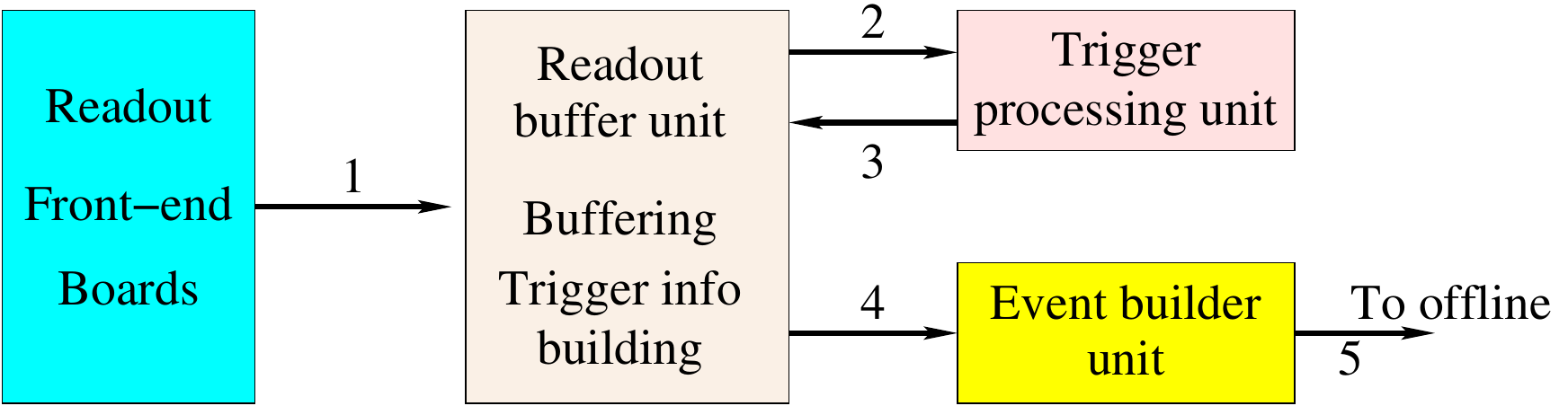}
  \caption{Simplified block diagram of the readout showing the
    sequence of data transfers.}
\label{online_schematics:daq}
\end{center}
\end{figure}
%%%%%%%%%%%%%%%%%%%%%%%%%%%%%%%%%%%%%%%%%%%%%%%%%%%%%%%%%%%%%%%%%%%%%%%%%%
%%%%%%%%%%%%%%%%%%%%%%%%%%%%%%%%%%%%%%%%%%%%%%%%%%%%%%%%%%%%%%%%%%%%%%%%%%
Figure~\ref{online_schematics:daq} is a schematic diagram of the data
readout and processing system showing the five steps in the sequence
of readout operations.  The main components are readout buffer units
(RBU) that receive PMT hit (and possibly waveform) data from the
front-end boards, trigger processing units (TPU) and event building
units (EBU).  The sequence of operation is indicated by the numbers
1-5 on Figure~\ref{online_schematics:daq}: (1) Data on hits in a group
of channels are streamed into buffers in the RBUs, where they are
retained for the duration of the trigger decision.  (2) A summary
block of information for triggering is sent to the TPUs. (3) Trigger
decisions are delivered back to the RBUs to extract event data from
the buffer.  (4) The event data are sent to the EBUs.  (5) Built
events are written to disk for later processing offline. A description
of each of the blocks shown is given in the following sections.  The
DAQ system will be designed to be homogeneous, flexible and scalable.

Parameters of the readout system, including data rates at each point
in the sequence of readout operation are shown in
Table~\ref{data_rates:daq}.  Since the architecture is flexible, the
additional data generated by the waveform option is easily
accommodated, although a larger number of Readout buffer units (RBUs) would be required.

\begin{table}[htbp]
\begin{center}
 \caption{Parameters of the readout design.}
 \label{data_rates:daq}
 \begin{tabular}{l|r|r}
\hline\hline
  Parameter &                              Hit-only option   & Waveform option \\
\hline
  Pre-trigger input data rate &                  5,600 MB/s  & 23,400 MB/s \\
  Number of RBUs &                                     38    & 122 \\
  Input rate to each RBU  &                         150 MB/s & 188 MB/s \\
  Latency provided by RBU (pre-trigger buffer length)& 109 s & 87 s \\
  Trigger info output rate per RBU &                50 MB/s  & 15 MB/s \\
  TPU data input rate (for 16 TPUs in detector) &  117 MB/s  & 117 MB/s \\
\hline\hline
 \end{tabular}
\end{center}
\end{table}
 
\subsubsection{Readout buffer unit (RBUs)}

The RBUs receive continuous streams of data from the front-end cards
and perform three main tasks.  First they store incoming data in
buffer storage.  A short-term store of the most recent data is
retained whilst trigger decisions are made and a long-term storage
area is reserved in case of a supernova event until it can be read
out.  Second, it forms a compressed block of data which is handed to
the TPU for trigger decision making.  Finally, the RBU handles trigger
requests to dispatch complete event data to the event builder unit,
and supernova trigger requests to move data to the long term storage
area.  Since the detector is large, and to allow scalability, there
are many RBUs working in parallel, each responsible for processing
data from a designated region of the detector.  In the current design
of 45,000~PMTs, between about 40~and 120~RBUs will be required
depending on the amount of data received per hit, and each will use
16\,GB of memory to provide the necessary storage.  The RBU design is
scalable and thus largely independent of the detector size.

The final design of the RBUs depends on the layout and interface of
the upstream electronics. Possible implementations use either
commodity computers with commodity network switches to direct the
data, or hardware receiver cards in a communication-crate such as 
Advanced Telecommunications Computing Architecture (ATCA)
with a commodity computer as a controller. The RBU may also include an
interface to the upstream link for monitoring information.

\subsubsection{Trigger processing unit}

The Trigger Processing Units (TPUs) will accept compressed trigger
data blocks from the RBUs and use these to form trigger decisions,
such as the simple, robust NHITS trigger.  Hooks will be provided to
allow for more sophisticated triggering.  The trigger will also search
for large collections of individual events, which may indicate a burst
of neutrinos from a supernova explosion in the galaxy.

The trigger will operate using windows of fixed time duration
(e.g.~60\,per second) in order to allow the trigger processing to be
parallelized.  The parallelisation allows for a scalable design that 
can be extended easily if workloads change due to increased darknoise, 
as well as handling any reprocessing due to errors or data corruption.  
One TPU is
allocated for a given time window and will process all the data in
that time window.  The TPUs are connected to the RBUs by a switched
Ethernet network to allow the data from the different RBUs (one for
each section of the detector) to be routed to the correct TPU.  The
data is transferred asynchronously and the TPU starts processing when
all data packets have arrived.

The trigger information can be compressed into a 32-bit word per hit
to identify the channel number within the RBU (10\,bits) and the time
within the window (21\,bits) to 10\,ns accuracy (better time
resolution is not needed in the trigger) and one spare status bit.
The trigger information can be truncated if it is clear that the NHITS
trigger will be satisfied from the data in that one RBU alone, as in
this case, the event is guaranteed to be collected without further
transfer or processing of trigger data.  For this level of packing,
the output rate per RBU will be about 30\,MB/s and if a farm of
16\,TPUs is used, the input rate to each will be around 100\,MB/s.
The final design of the TPUs is largely independent of the choice of 
front-end electronics, because the RBUs provide the trigger data in the same 
format regardless. The TPU design depends on the type of processing required 
for the sophisticated triggers. The base line design is to use commodity 
computer componets, where data packets would be received by the computer 
and complex trigger processing would occur in FPGAs or GPUs that read the 
data over PCIe links and deliver the trigger verdict for a given time 
block back to the main memory. 

\subsubsection{Event Building Unit}

Once the trigger decision has been made for a time-window, the
decision is reported back to a central trigger control process, from
where it is delivered to the RBUs.  The request contains an event
number, and the definition of the position and width of the trigger
window.  The central trigger control process also looks for an
increased rate of positive triggers, which would be indicative of a
supernova burst, and in such case sends the RBUs an instruction to
save the relevant data in the long-term part of the buffer memory.  

On receipt of a normal trigger, the RBUs send all the hits in the
trigger window to a designated event-building node, which puts the
event together and writes it to the output file.  Once the file
reaches a certain size, it is closed and released for offline
processing, and new events are recorded in a new output file without
interruption.  The event builders also allow events to be read by
monitoring and event display programs.  Once a supernova trigger has
occurred, a separate event-building stream is used to gather the data
in the long-term part of the RBU memory and output it to a separate
file.  There will be one file per supernova trigger.  A
straightforward implementation of the event builders is to use
commodity computers.

\subsubsection{Triggering}
A trigger will be issued if any event exceeds a pre-determined threshold 
of hits (NHITS) in a sliding time window.  This trigger will accept with 
high efficiency, all events for studies of proton decay, atmospheric 
neutrinos, beam neutrinos and cosmic ray interactions.  Aside from the 
NHITS requirement, the trigger must have no dead time as this could lead 
to the loss of information from associated delayed energy deposition events 
such as Michel electrons or neutron captures.

The low energy threshold limited by the dark noise rate of the PMTs, which 
prevents the use of such a simple NHITs trigger for low energy physics events 
such as solar neutrinos, supernova and neutron capture.  To allow sensitivity 
to low energy physics, the Hyper-K DAQ system must incorporate intelligent, 
fast trigger algorithms that can reject noise events but retain the low energy 
physics events of interest.  Any event that fails the NHITS threshold will be 
passed to the low energy trigger.  

The base line low energy trigger design uses a grid of test vertex positions 
with 5 m spacing inside the detector.  The distance between each test-vertex 
and PMT position is used to produce a look up table of time of flight 
information for each test-vertex PMT pair.  For an event, the ID and TDC 
information for each tube are recorded and 
passed to the test-vertices algortihm.  The algorithm loops over the test-vertices 
and corrects the measured TDC by the time of flight for each test-vertex PMT pair.  
This corrected value represents the time at which the photon was emitted by the 
test vertex to produce the observed PMT hit. A one dimensional histogram of photon 
emission times is produced with a carefully chosen bin width corresponding to half 
of the time resolution.  The best vertex is found by selecting the histogram bin 
with the highest number of entries and if this number of entries is greater than 
a pre-determined threshold, the trigger is accepted. Due to the computational 
intensity of this process, it will be executed using high performance Graphical 
Processing Units (GPUs).  Results from Monte Carlo simulations show high noise 
rejection and good low energy acceptance.

A second trigger that uses convolution neural nets and GPUs to perform real-time 
image recognition for triggering is also being investigated.  Initial studies have proven 
very promising, with good performance even at very low energies.  Other specialist 
triggers will be developed for calibration sources and supernova detection. 

Adaptations will be made to trigger algorithms for the use in the intermediate detector.  
As the low energy physics reach of the intermediate detector is less than that of 
the far detector, the triggering scheme can be simplified.  However, low energy 
triggers will be required to accept gadolinium gamma cascades.  Work in this area 
is continuing.

\newpage

\graphicspath{{design-calibration/figs/}}

\subsection{Detector calibrations \label{sec:calibration}}

The Hyper-K detector consists of an Inner Detector and an Outer
Detector and both detectors need to be calibrated.
The Super-K detector has been successfully operated for about two
decades and established several techniques to calibrate a large water
Cherenkov detector~\cite{Abe:2013gga}.  Hyper-K calibrations are
designed based on and extension of the Super-K calibrations.
This section describes the calibration strategy and methods for
Hyper-K detector.

%%%%%%%%%%%%%%%%%%%%%%%%%%%%%%%%%%%%%%%%%
\subsubsection{Inner Detector Calibration}
%%%%%%%%%%%%%%%%%%%%%%%%%%%%%%%%%%%%%%%%%

Hyper-K detector calibrations consist of two major phases:
calibrations of the detector system, e.g. PMTs, and calibrations
dedicated for physics analyses.
The calibrations for the detector system is to characterize the PMT
responses, including readout electronics, and the optical properties
of detector material, e.g. water, PMT (glass bulb and housing
material), black sheet, and tyvek sheet. 
The calibrations for physics analyses are designed to satisfy the
requirements from Hyper-K physics program which consists of variety of
physics topics covering wide range of energy region, from a few MeV to
several hundred GeV.
For these calibrations, various calibration techniques have been
developed and established by Super-K and they are applicable to
Hyper-K calibrations without major technical difficulties.

Table~\ref{tab:calib_items} summarizes the calibration items and the
calibration sources used in Super-K calibrations for reference.
These calibration items are important to characterize the Hyper-K
detector responses, and the calibration results need to be implemented
in the detector simulation.
\begin{table}
\begin{center}
\caption{Calibration items for the detector systems and calibration sources used in Super-K.}
\label{tab:calib_items}
\begin{tabular}{l|l}
\hline\hline
Calibration items & Calibration sources used in SK \\
\hline
% ---------------------------
Photosensor \& electronics calibrations & \\
\quad High-voltage tuning                         & Xe flash lamp \\
\quad Single photo-electron charge (gain)         & Nickel source \\
\quad Electronics threshold effect                & Nickel source \\
\quad Photo-detection efficiency                  & Nickel source \\
\quad Non-linearity (photosensor and electronics) & Nitrogen-dye laser \\
\quad `Time-walk' correction (TQ map)             & Nitrogen-dye laser \\
\quad Timing resolution                           & Nitrogen-dye laser \\
\quad Dark noise                                  & Off-timing hits \\
% ---------------------------
\hline
Optical properties of detector material & \\
\quad Light transparency of water (absorption, scattering) & Nitrogen laser, laser diodes, Xe flash lamp\\
\quad Optical properties of PMT glass \& housing material & Nitrogen laser, laser diodes, Xe flash lamp\\
% ---------------------------
\hline
Calibrations dedicated for physics analyses & \\
\quad Solar and supernova $\nu$ etc.: energy scale and vertex & LINAC, DT generator, Nickel source\\
\quad Beam and atmospheric $\nu$ etc.: energy/momentum scale & Cosmic-ray muons, decay-e's, $\pi^0$ etc.\\
\hline\hline
\end{tabular}
\end{center}
\end{table}
As shown in the Table~\ref{tab:calib_items}, the detector calibrations
require various calibration sources: light sources, radioactive
sources and physics events such as cosmic rays.
Note that ``Nickel source'' listed in Table~\ref{tab:calib_items} is a
calibration source to generate single photo-electron (p.e.)  level of
light, which isotropically emits $\sim$9~MeV gamma rays from thermal
neutron capture on nickel with the reaction of
$^{58}$Ni($n,\gamma$)$^{59}$Ni, where $^{252}$Cf is used as
a neutron source.
With the use of a Nickel source, for example, ``relative''
photo-detection efficiency,
(quantum efficiency times collection efficiency) of each single PMT is
calibrated by evaluating relative differences of the hit rates between
PMTs for single photon level of light.
The ``low-energy physics'' events like solar neutrinos mostly consist of
1~p.e.\ hits and thus the relative photo-detection efficiency
calibration plays an important role for the low energy physics
analyses.
On the other hand, PMT/electronics linearity calibration for the full
dynamic range is important for ``high-energy physics'' events like
atmospheric neutrinos, which can involve 100 GeV-scale muons.
For the linearity calibration, a nitrogen-dye laser is used in
Super-K, which is utilized with a variable light attenuation filtering
system to obtain a variety of light intensities covering the full
dynamic range of PMT and readout electronics ($\sim$0.25\,p.e. to
1000's\,p.e.).
In addition to calibrations of the PMTs and electronics, understanding of the
optical properties of the water plays a key role in all physics analyses
of Super-K and Hyper-K.
The water optical properties, absorption and scattering of light, are
calibrated as a function of time and position in the detector volume
using a series of laser diodes with various wavelengths.
Since the detector condition can change from time to time, the
continuous monitoring and periodical calibrations of these detector
components are also indispensable over the lifetime of the experiment.
There are many other calibrations to characterize/monitoring the
detector system, as explained in Ref.~\cite{Abe:2013gga} for further
details.

The Hyper-K physics program consist of various physics topics which
cover a wide range of energy from a few MeV to several hundred GeV.
For example, low energy events like solar neutrinos and supernova
neutrinos are in a few MeV up to tens of MeV, neutrinos from the J-PARC beam and nucleon decay events are
around 1\,GeV, and atmospheric neutrinos are in a range between
$\sim$1\,GeV and several hundred GeV.
For physics analyses in each energy regime, there are a series of
dedicated calibrations, e.g. energy scale, which are described in the
later sections.

For all necessary detector calibrations in Hyper-K, the calibration
methods and techniques established in Super-K are sufficiently good
enough, and we do not see any technical difficulty to apply them to
Hyper-K.
We have, however, been developing more sophisticated calibration sources/systems 
for Hyper-K in order to minimize the detector downtime and manpower
required for the detector calibrations since Hyper-K detector has a
larger detector volume than Super-K.  The following sections discuss
the status of research and development for Hyper-K calibrations.

\subsubsubsection {Detector Calibration Systems}
In order to achieve the level of calibration required for the
Hyper-K physics program we require an extensive calibration
infrastructure.  Data from in situ sources such as cosmic muons or
Michel electrons will be used to supplement the information from these
systems, but they cannot replace them.  In Hyper-K we propose
two systems, a source deployment system to allow the deployment of
calibration sources across the detector inner region and an integrated
light injection system for the inner and outer detector regions.
These two calibration systems provide the data that will be required
to characterise the detector and reduce the systematic uncertainties
to the required level.  The integrated calibration system allows for
calibration and monitoring of the detector without the deployment of
specialised manpower while calibration sources can be used for more
extensive calibrations during the time when the neutrino beam is off.

\subsubsubsubsection {Calibration sources and deployment system} 
Similar to other water Cherenkov detectors the ability to deploy
calibration sources is essential to understand the Hyper-K
detector.  The Hyper-K source deployment system will consist
of a number of source deployment points above the detector down which
calibration sources may be lowered down to any required detector level
with high precision.  In addition to the $z$-axis option, that these
points provide optionally a full 3D manipulator system may be
developed allowing sources to be deployed over a wider range of the
detector volume.

\begin{figure}[htbp]
  \begin{center}
  \includegraphics[width=0.3\textwidth]{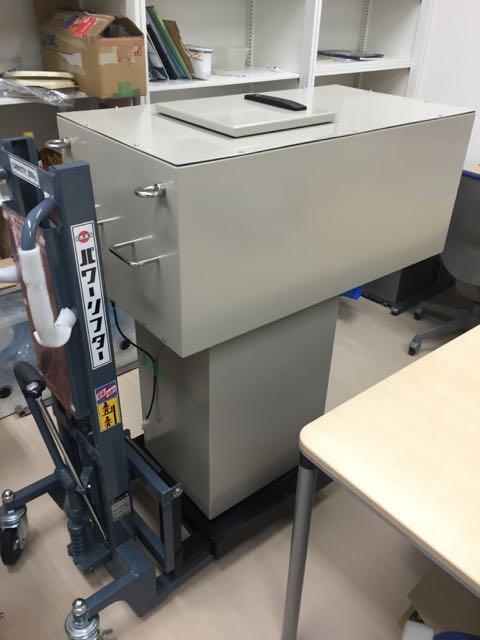}
  \includegraphics[width=0.4\textwidth]{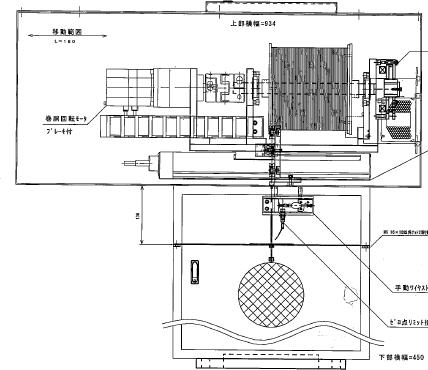}
  \caption{Photograph (left) and sketch (right) of the first prototype of the
  calibration sources deployment system.}
  \label{fig:calib_skcalib_system}
  \end{center}
\end{figure}

In order to calibrate the detector efficiently, a
computer-controlled source deployment system is desirable.  The first
prototype of the source deployment system has been developed
(Fig.~\ref{fig:calib_skcalib_system}).  The system will be installed
in Super-K in 2018, and we plan to further develop the source
deployment system for Hyper-K based on the Super-K operational
experience.
% -------------------------------
%

A variety of sources can be deployed through this system including
optical sources for PMT and measurements of detector optical
properties, and various radioactive sources.  Based on previous
experience at Super-K this would include a deuterium-tritium base
reaction pulsed neutron generator (DT generator) to create $^{16}$N
for calibration in the solar neutrino energy regime.  A number of
sources can be developed to provide neutrons for calibration.  These
include $^{252}$Cf and AmBe sources that were previously used to
calibrate the neutron response of the SNO detector.  These sources
produce neutrons at a known rate; measuring the neutron detection
efficiency and its possible variations for capture on H. If the
decision is made to deploy Gd in Hyper-K, these measurements will also
need to be made for neutron capture on Gd.  In addition to the
conventional neutron sources, we currently explore ideas for compact
and pulsed neutron generator~\cite{NeutronGenerator:6275497} for
Hyper-K detector calibrations, that uses a similar technology to DT
generator but is a palm-size portable neutron generator and can
manoeuvre more effectively in detector calibrations. The neutron
producion of these sources will ideally be tunable to provide single
neutron calibration and also to provide the copious neutron required
for $^{16}N$ calibration.
Development of additional calibration sources to exploit
this system can be expected over the lifetime of Hyper-K.

\subsubsubsubsection {Integrated light injection system}
Hyper-K will include an integrated light injection system for
optical (absorbtion and scattering) and PMT (timing, gain and
multi-photon) calibrations.  This system consists of a number of light
injection points connected via optical fibres to light pulsers in the
electronics.  Light pulses of 1-2 ns can be produced using LEDs, laser
diodes (LDs) or similar solid state optical devices can be produced
relatively inexpensively.  This allows multiple optical sources to be
deployed around the edge of the detector that can then be used for
calibration.  This system consists of an LED (or similar) coupled to
an optical fibre, which is then connected to an optical diffuser on
the PMT support structure.  The optical diffuser is used to shape the
light inside the detector and designs can provide different
calibration pulses for different needs. 

\begin{figure}[htbp]
  \begin{center}
  \includegraphics[width=0.5\textwidth]{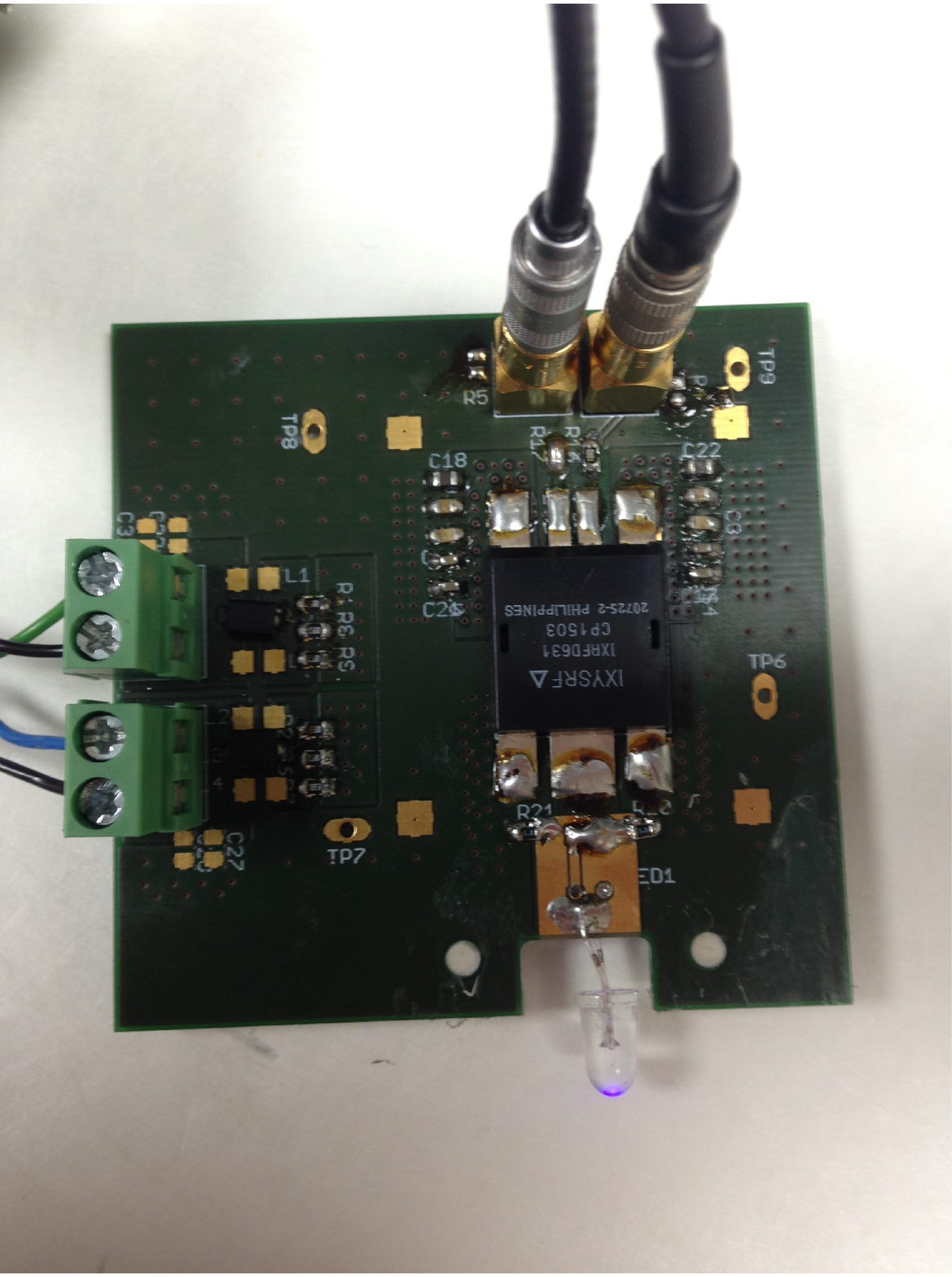}
  \includegraphics[width=0.41\textwidth]{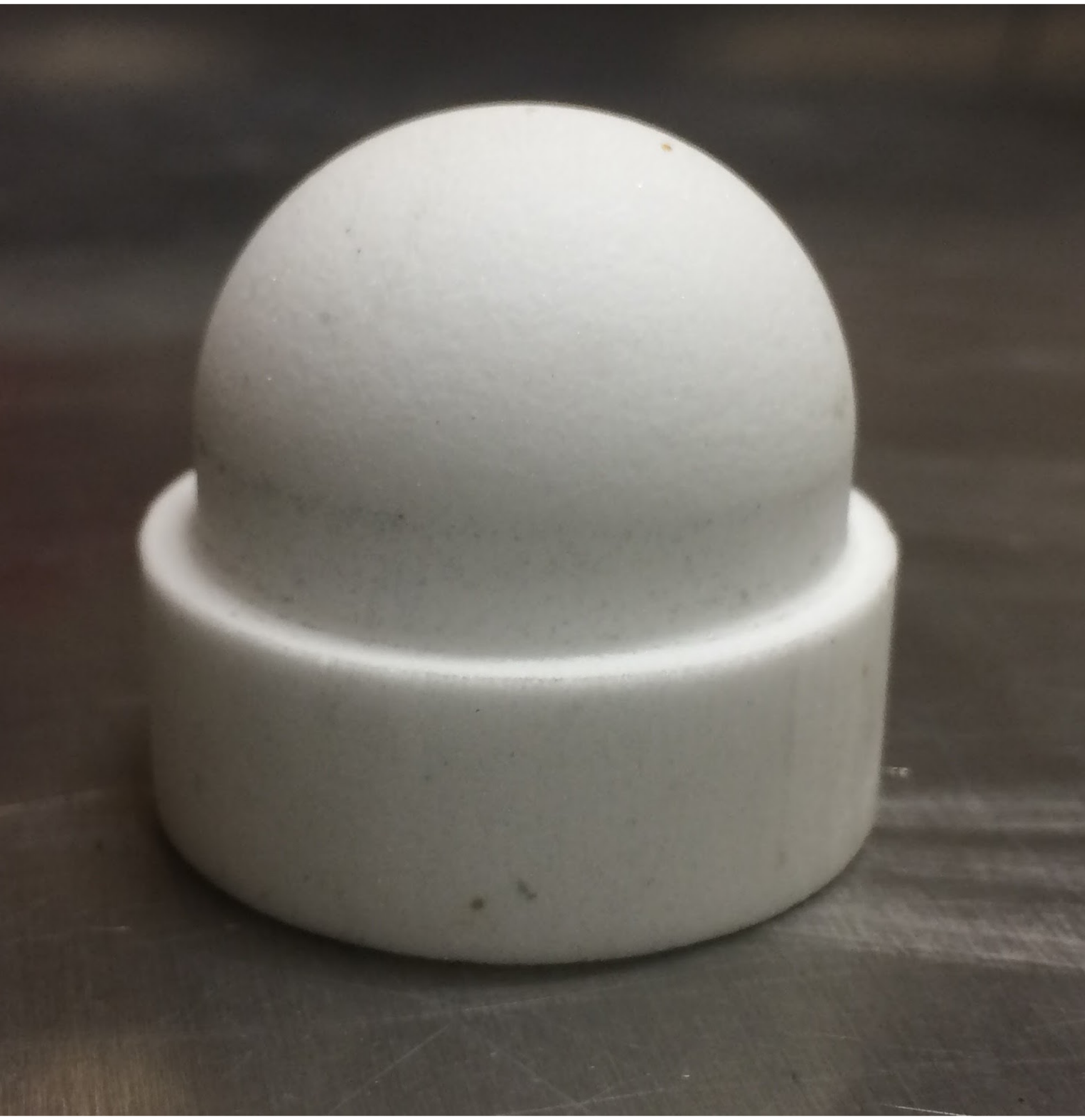}
  \caption{Photograph of the prototype of LED driver board (left) and
  the wide angle diffuser (right).}
  \label{fig:calib_LEDLD}
  \end{center}
\end{figure}

In order to maintain fast light pulses over the order 100~m distance
of optical fibres required for Hyper-K graded index fibre is
required rather than the step index fibre used in for example SNO+.
Graded index fibre has a small active core complicating the challenge
of light collection.  The key challenges of this system are the
coupling of the LEDs to the optical fibre, minimising dispersion in
the fibre to maintain short optical pulses and achieving the required
dynamic range without compromising the fast optical pulses.  Research
and development is currently underway in the UK to solve these
problems.
Figure~\ref{fig:calib_LEDLD} shows the current prototypes of the LED driver unit and the wide angle optical diffuser..
% ----------------------------------

For some calibrations, it will be essential to monitor the light
injected into the detector and this will be part of the light
injection system. There are multiple options for to do this. One
option is to use light that is not collected into the fibre,
alternatively fibre splitters are commercially available with 99:1
split ratios allowing a small fraction of the light to be collected
for monitoring. The monitoring light can be measured via optical
sensors in the calibration equipment such as a monitoring PMT. The
monitoring PMT signal will itself be calibrated and monitored by a
dedicated monitoring channel where the main pulse is not injected into
the detector, but is instead delivered to a well calibrated optical
power meter. The cross calibration will allow the absolute optical
signal of the monitoring system to be determined allowing comparison
of light injected pulse to pulse and across the running time of
Hyper-K.

This system allows PMT and optical calibration data to be taken
without manpower intensive calibration source deployment that has been
previously used in water Cherenkov detectors.  These data either can
be collected in either dedicated high rate calibrations or
interspersed during data taking.  This system also allows calibration
to be conducted during extended periods of beam running where
deployment of calibration sources would otherwise result in detector
downtime.  Given the systematic error budget of Hyper-K this
system will mean we do not have to compromise between efficiency and
the collection of sufficient calibration data.

The calibration of the PMT timing requires a short duration light
pulse of known origin and time.  The integrated light injection
system, from any given fibre, provides this but clearly cannot
illuminate the entire PMT array at once.  To minimise the number of
fibres required the optical diffuser for the PMT calibration is
required to provide a wide opening angle, to illuminate of order 1000
PMTs on the far side of the detector.  The diffuser must be carefully
designed to ensure that there is no time dependence as a function of
angle.  To achieve the overall calibration of global time offset of
the array PMTs must be illuminated by at least two fibres to allow the
fibre times to be cross calibrated.  We target a six-fold degeneracy
of the PMT calibration fibre points to allow to for improved cross
calibration and to provide redundancy against single point failures in
the fibres.  This system will allow for the calibration of PMT timing,
the dependence of time on charge and the PMT time response.

The integrated calibration system can also be used to measure optical
scattering, extinction and the PMT response.  While the basic elements
of the system are the same as that used for PMT calibration, a number
of changes are required meaning that fibres and diffusers used for
these calibrations are different.  These properties are required as a
function of wavelength, thus several LED types will be used to provide
light at six different wavelengths between 320\,nm and 500\,nm.  To
measure scattering a narrow beam is required from the optical
diffuser; the scattering length is measured by monitoring the light
level of PMTs outside the narrow beam as a function of the path length
of the beam through the detector.  Optical absorbtion is measured by
monitoring the light levels on given PMTs inside the optical beams;
unlike scattering wide angle beams are important in this calibration
to provide a variety of path lengths.  The optical calibration system
must be constructed at multiple detector levels to allow for any
variation of optical properties with detector height.  The pulse by
pulse monitoring of the calibration system is essential for this
calibration as the light level at given PMTs is the key measurement of
the system.  The measured light level at the PMTs is a combination of
absorbtion and PMT response as a function of angle and thus several
light paths and angles are required for these to be decoupled in
analysis requiring a variety of diffuser points and diffuser
directions to be deployed. A prototype of these systems is planned for
deployment in SK-Gd. The calibration procedures and performance for
the full system planned for Hyper-K will be determined using this
system.
%

%%%%%%%%%%%%%%%%%%%%%%%%%%%%%%%%%%%%%%%%%%%%%%%%%%%%%%%%%%%%%%%%%%%%%%%%%%%%%%
\subsubsubsection{Photosensor calibration ex-situ}
%%%%%%%%%%%%%%%%%%%%%%%%%%%%%%%%%%%%%%%%%%%%%%%%%%%%%%%%%%%%%%%%%%%%%%%%%%%%%%

As discussed in the previous sections, many of PMT properties can be
calibrated {\it in-situ} by deploying several calibration sources in
the detector.
We will have additional {\it ex-situ} measurements to understand
further details of PMT properties that are difficult to measure
in-situ.

A number of PMTs that are installed into Hyper-K must first be
precalibrated to allow the gain of the detector to be tuned for
uniformity. The charge recorded per photoelectron is a function of the
high voltage applied to the PMT and of the PMT itself. It is important
to characterise the response of a number of PMTs that are installed
uniformly into the detector. Using these known PMTs once the detector
is running, the high voltage applied to each PMT will be tuned such
that each PMT has the same gain. In Super-K 420 PMTs were
precalibrated using a Xe lamp connected via an optical fibre to a
scintillator ball in a shielded dark box in which the PMT to be
calibrated was mounted.

Large photo-cathode area PMTs have non-uniform charge and time
responses.  The photo-detection efficiency for example can vary
depending on the photon incident angle and position on the
photo-cathode of the PMT (e.g. \cite{Goon:2012if}).
Such non-uniformity of PMT responses need to be understood and are
required to build a better model of PMT responses, which is then
adopted in the detector simulation.
There are some difficulties to measure such non-uniformity of PMT
responses after they are installed in the detector since, for example,
a small non-uniformity of water transparency can make an apparent
variation of PMT responses.
Thus, we need to establish a special test stand for the measurements.

\begin{figure}[htbp]
  \begin{center}
  \includegraphics[width=0.55\textwidth]{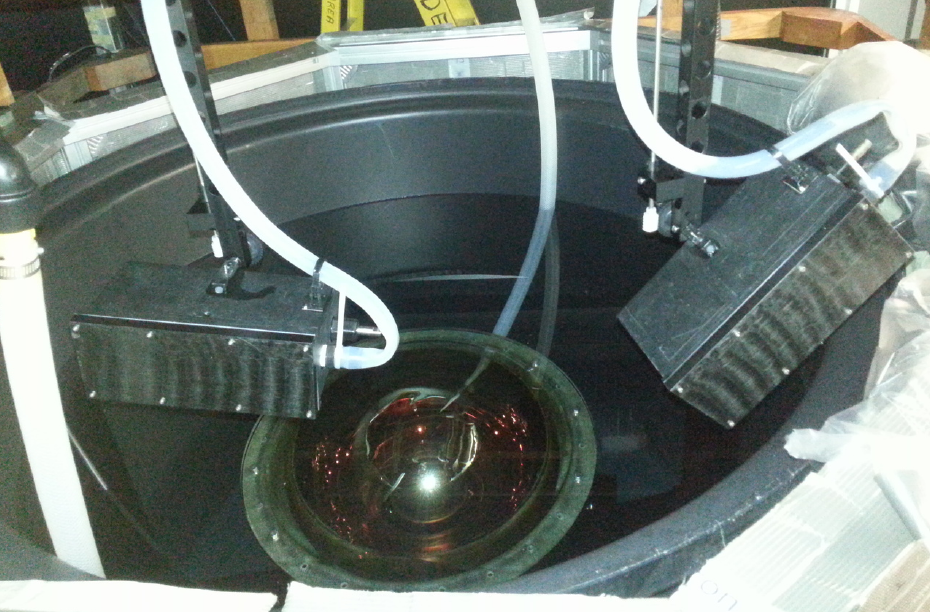}
  \includegraphics[width=0.35\textwidth]{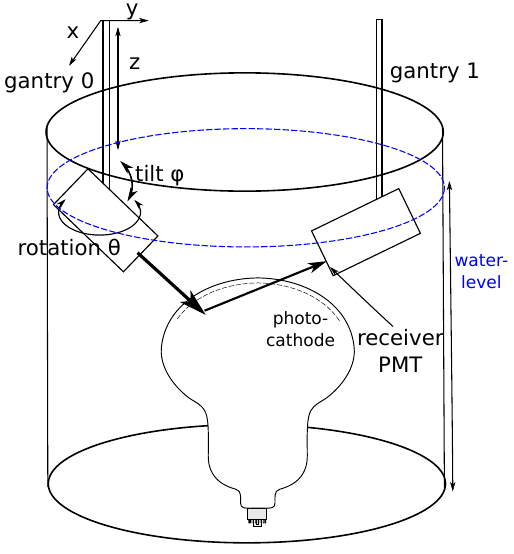}
  \caption{Photograph (left) and schematic diagram (right) of the photosensor test facility
    at TRIUMF.}
  \label{fig:calib_PTF}
  \end{center}
\end{figure}

\begin{figure}[htbp]
  \begin{center}
  \begin{tabular}{c}
  \includegraphics[width=0.65\textwidth]{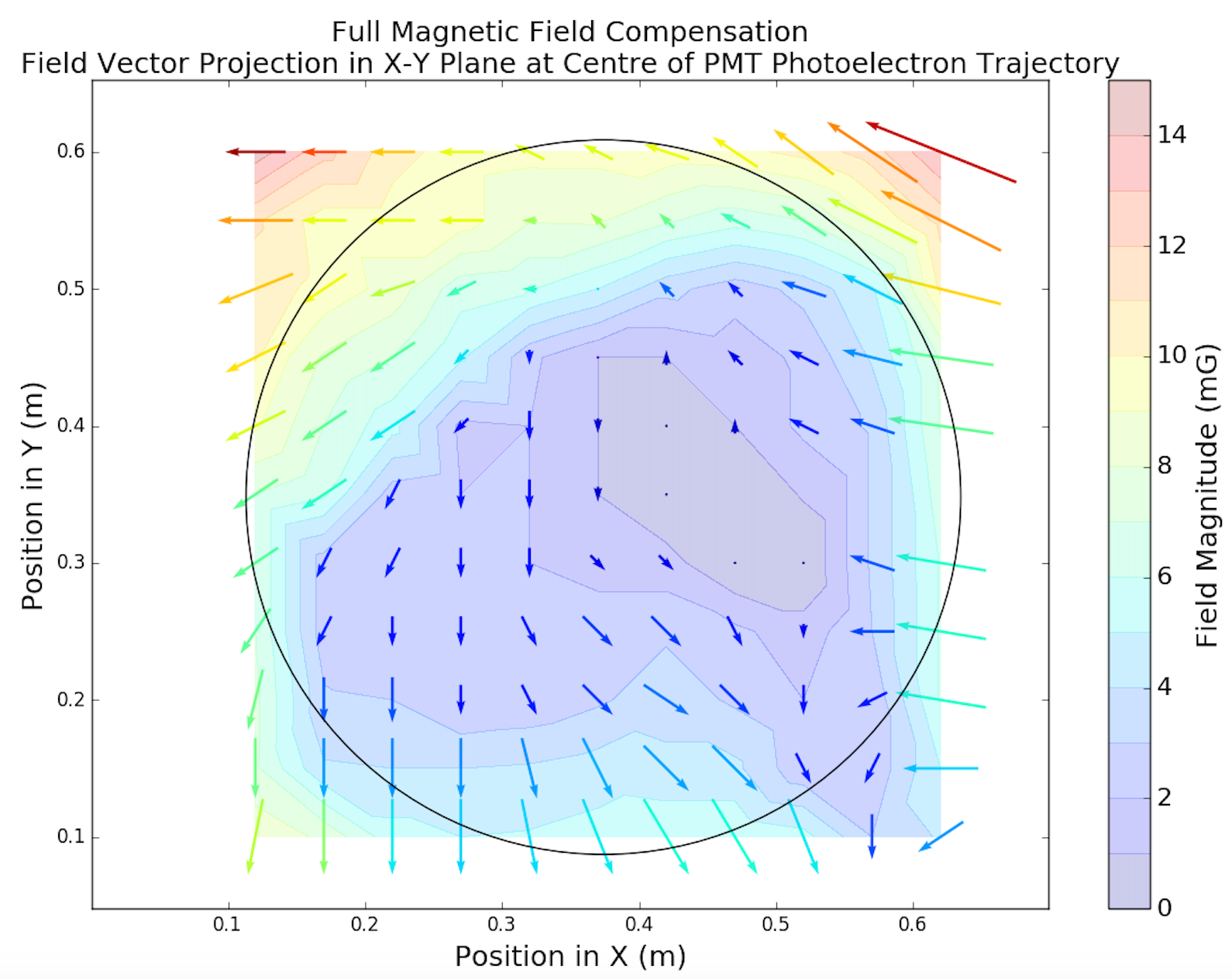} \\
  \includegraphics[width=0.65\textwidth]{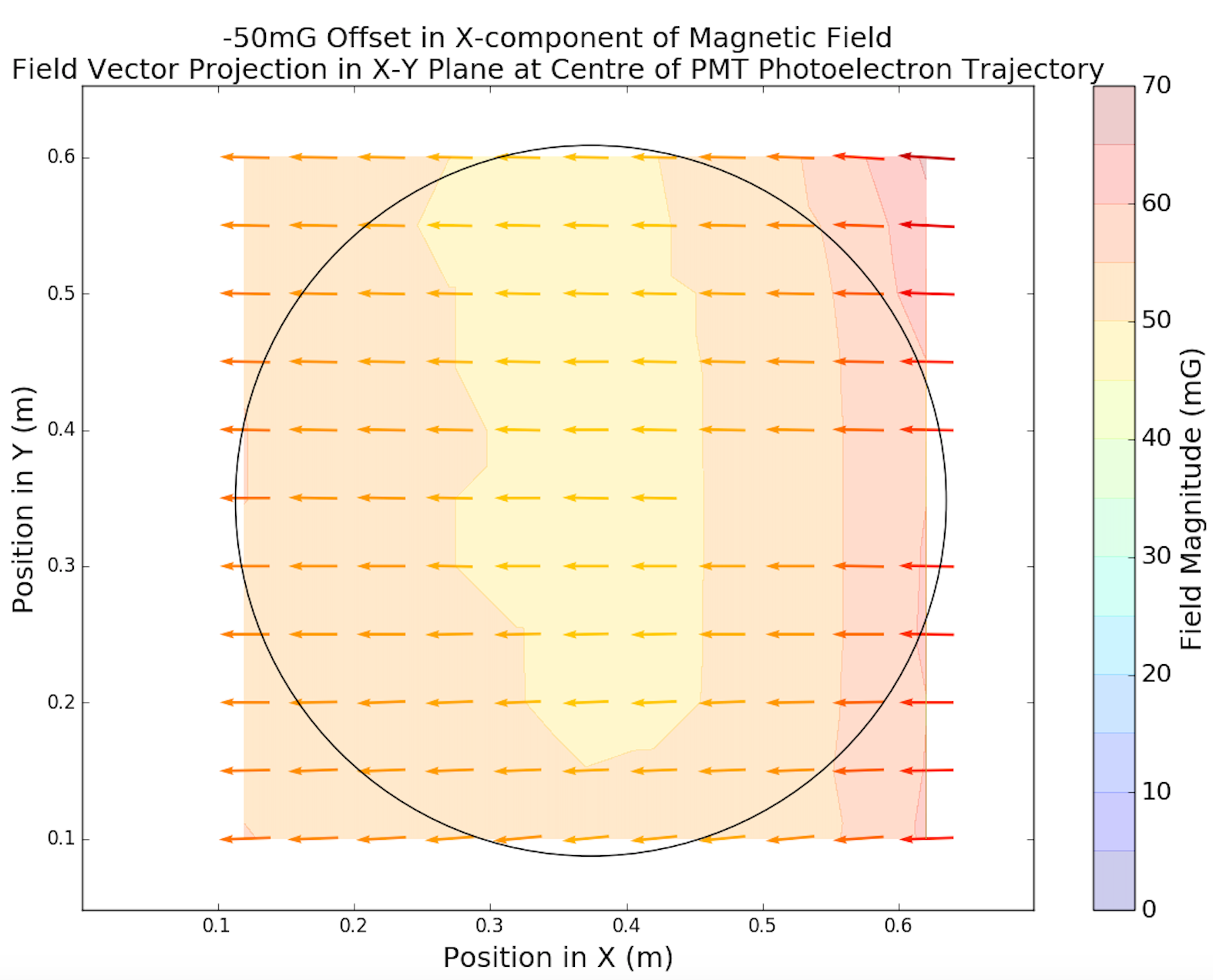} \\
  \end{tabular}
  \caption{Map of the compensated magnetic field at PTF (top) and 
 with 50mG field in the -x direction  (bottom).}
  \label{fig:PTF_field}
  \end{center}
\end{figure}

\begin{figure}[htbp]
  \begin{center}
  \includegraphics[width=0.6\textwidth]{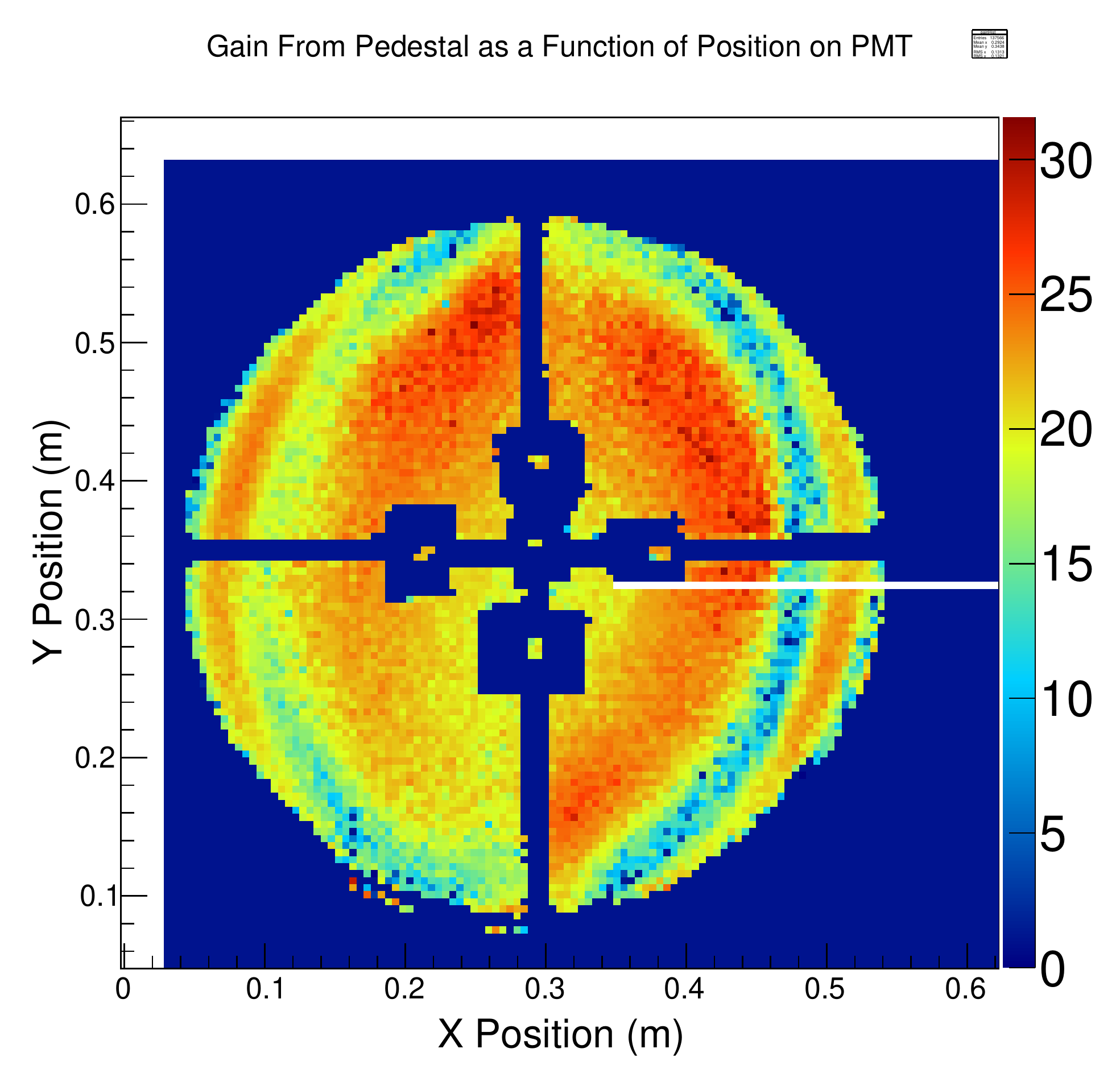}
  \caption{SK-PMT gain distribution under fully compensated field. The low gain valley
  comes from the photoelectrons escaping the first stage of the venetian blind dynode. 
  Also visible in the figure is the affect of adhesive tape applied to the PMT to allow 
  precise determination of position and orientation of the system.}
  \label{fig:SKPMT_PTF}
  \end{center}
\end{figure}

A test facility, called `photosensor test facility (PTF),' has been
built at TRIUMF.  Figure~\ref{fig:calib_PTF} shows a photograph and
schematic diagram of PTF.
The PTF has two manipulator arms (gantries) which are motorized and
move independently in the $x$-, $y$-, $z$-direction, rotation, and
tilt.  Each gantry is equipped with an optical box that contain a
light source with a chosen wavelength, a (monitor) PMT to measure the
intensity of the injected light and a (receiver) PMT which is used for
measurement of reflectivity.  The PTF is equipped with a water
purification system, which generates ultra-pure water, and can measure
PMT responses under water.
As discussed in the photosensor section, Hyper-K PMT will be
completely encased in a pressure housing.  The optical properties of
the PMT housing in ultra-pure water will also be measured by the PTF.
Currently, characterization of the 20-inch SuperK PMT is under way at PTF.
The ambient magnetic field is compensated down to 5mG level by a three sets
of Helmholtz coils and the desired magnetic field can be applied to study the
impact of the ambient magnetic field (Figure~\ref{fig:PTF_field}). Five milli-meter grid 
scan on the SK PMT photocathode revealed significant gain variation due to the 
venetian blind dynode structure. Significant gain variation due to the magnetic field 
is also observed (Figure~\ref{fig:SKPMT_PTF}). 

Calibration of the SK PMT along with the establishment of the PMT characterization procedure
at PTF will continue till the end of 2017. From 2018, PTF will focus on the characterization of
mPMT prototype for NuPRISM and Hyper-K and the final version of the 20-inch B\&L PMT for 
HyperK. From 2019 PTF will focus on the design study of mPMT for NuPRISM, and then test
the sampled modules for NuPRISM production. From 2022, PTF will test sampled 
photosensors for HyperK production till the end of 2024 when the HyperK photosensor installation 
starts.

In addition to PMT studies at the PTF facility, measurements of
Super-K PMT properties continue at Kamioka. Initial measurements from
these studies are complete and have been implemented into
SKDETSIM. The impact of these measurements on physics analyses are
ongoing.

%%%%%%%%%%%%%%%%%%%%%%%%%%%%%%%%%%%%%%%%%
%%%%%%%%%%%%%%%%%%%%%%%%%%%%%%%%%%%%%%%%%
\subsubsection {Detector Monitoring}
To ensure optimal detector performance, minimise systematic
uncertainties and facilitate timely responses to detector problems,
robust detector monitoring is essential. This will be enabled by the
integrated light injection system. The integrated light injection
system is being designed to allow calibration of PMT response and
detector optical properties. A dedicated calibration can be performed
using the system at high rate (1-10kHz) collecting sufficient
statistics for each calibration channel in a short period of
time. However, the system can also operate at low rate, (0.1-10 Hz)
cycling through calibration channels allowing lower statistics
calibration samples to be collected over the period of a few
hours. Using these low statistics samples the detector calibration can
be monitored for changes that might require a dedicated calibration
from the inbuilt system to update the detector calibration. These
monitoring pulses will also allow the health of other detector systems
such as the PMT efficiency to be monitored, tracking any changes to
the system as a function of time. Data from this system can be used
both by experts, automated systems and detector shifters to track
detector performance and understand the health of the detector.

%%%%%%%%%%%%%%%%%%%%%%%%%%%%%%%%%%%%%%%%%
\subsubsection{Calibrations dedicated for physics analyses}
%%%%%%%%%%%%%%%%%%%%%%%%%%%%%%%%%%%%%%%%%

\subsubsubsection{Calibrations for low energy physics}
In this section, detector calibration methods for so-called `low
energy' physics, which is a range from a few MeV to a few tens' MeV
energy region, are described.  The physics targets are solar
neutrinos, supernova neutrinos and geo neutrinos \textit{etc.}

\subsubsubsubsection{Review of calibration in Super-K}
In the low energy physics, electrons generated by neutrino-electron
elastic scattering and positrons by inverse beta decay are the
detection particles.  It is essential to reconstruct their vertex,
direction and energy accurately for a precise observation.  For this
purpose, Super-K employed multiple calibration techniques.

\begin{figure}[!h]
\centering
\includegraphics[width=0.6\textwidth]{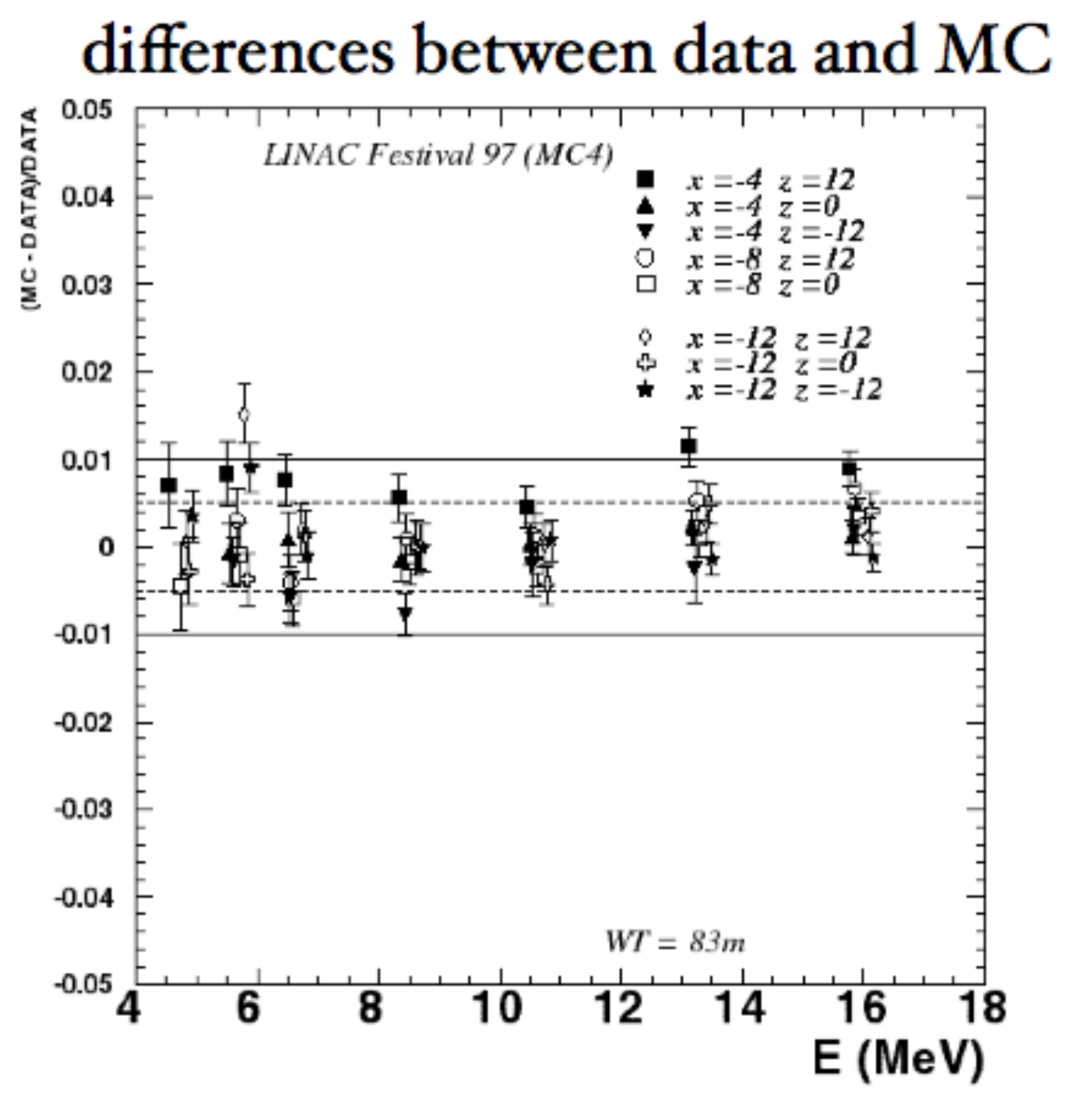}
\caption{Comparison of absolute energy scale between LINAC calibration data and Monte Carlo simulation after tuning.}
\label{fig:linac}
\end{figure}

First, an electron LINAC is used to calibrate the absolute energy
scale, energy resolution, vertex and direction resolution.  Single
electrons with mono-chromatic energy, from 5 to 16.3~MeV, are injected
from the LINAC into the detector downward at nine different detector
heights.  A comparison
between the observed LINAC data and the Monte Carlo simulation is
shown in Fig.~\ref{fig:linac}.  As shown in the figure, the Monte Carlo
prediction and data are in good agreement after tuning. Since
the LINAC calibrations require a long detector downtime and
significant manpower,
LINAC data are taken once per year or two years.  Detail of the LINAC
system and its performance can be found in
reference~\cite{Nakahata:1998pz}.

The decay of $^{16}$N is used to independantly check the energy
calibration across the detector. Decay of  $^{16}$N is dominated by an electron with a
4.5~MeV maximum energy, coincident with a 6.1~MeV $\gamma$ ray.  To
create $^{16}$N, a DT generator deployed within the detector to generate the (n,p)
reaction on $^{16}$O in the water, which decays with a 7.13 s half-life. After a neutron pulse the DT generator is retracted from the activated water to reduce shadowing from the calibration hardware. The DT calibrations are carried
out more frequently than LINAC, that is three or four times per year.
The DT calibration results are good agreement with LINAC calibration.
Since $^{16}$N decay generates electrons and gammas isotropically unlike the 
LINAC, the DT calibration technique allows to estimation of 
systematic uncertainty not measured by the LINAC calibration; both the
position and angular dependence of the energy scale uncertainty, for example.
Detail of the $^{16}$N calibration system can be found in
reference~\cite{Fukuda:2002uc}. From these results, the
energy scale uncertainty is estimated to be around 0.5\%.

The `Nickel source' is used for the calibration of the event vertex and
detector uniformity.  This is a $\sim$9~MeV $\gamma$ source generated
by thermal neutron capture on nickel.
Neutrons are produced by the spontaneous fission of $^{252}$Cf, and a 20~cm diameter
`nickel ball', which consists of 6.5~kg of NiO and 3.5~kg of Polyethylene, is used.
%%%%%
Since nickel source is easy to handle and can be deployed in the detector with
good position accuracy (less than 1~cm uncertainty),
nickel source calibration is carried out about every month by deploying the source at
multiple locations in the detector.
From this calibration, the vertex reconstruction uncertainty is estimated to be less than 5~cm.
The monthly nickel source calibration also measures the variation of the detector uniformity which is
caused by the water quality change.

\subsubsubsubsection{Toward Hyper-K}
For the solar and supernova neutrino detection, the similar level of
precise calibration to Super-K is needed for Hyper-K.  Many of
calibration sources used in Super-K, like nickel source, can also be
used in Hyper-K, however, employing an electron LINAC system requires
further considerations since Hyper-K will have a larger volume than Super-K, that causes a much larger
burden to operate LINAC calibration system at Hyper-K.
Thus, alternative approaches to the LINAC calibration are being developed.
One possible approach can be the following.  Super-K detector
simulation is tuned with LINAC calibration data, and the LINAC-tuned
simulation reproduces DT calibration data very well, as described
above.  By adopting the tuning parameters of the LINAC-tuned Super-K
simulation in Hyper-K, DT calibration data of Hyper-K can provide a
good calibration of the absolute energy scale without employing LINAC
calibrations in Hyper-K.
This way does not requires LINAC calibrations at Hyper-K but requires
an intensive DT calibrations.  For the purpose, a compact DT generator
or an alternative neutron generator would be desirable.  In order to
calibrate the energy regime, up to a few tens MeV, which is higher
energy than DT calibration, we need to develop new calibration
sources.
These new systems are under development.

The decision for the inclusion of the LINAC within HK will require resources such as the purchase of a new LINAC and the production of an extra cavity to house it. As such this decision is a significant milestone for the HK calibration design requiring further study before it can be made. We aim to make this decision prior to the end of 2018.

\subsubsubsection{Calibrations for high energy physics}

In this subsection, the energy scale calibration of the inner detector for the higher-energy
physics analyses (atmospheric neutrino oscillation, nucleon decay search, T2K, \textit{etc.})
is described.

\subsubsubsubsection{Review of calibration in Super-K}

The charge scale of the SK detector simulator (SKDETSIM) is initially
calibrated by measuring each PMT quantum efficiency, single
photo-electron charge distribution, \textit{etc.} by using various
control samples such as the nickel source.
A global correction factor which scales the total photo-electron yield is then tuned in SKDETSIM using cosmic-ray through-going muons.

After this correction, particle momentum is reconstructed as follows: the total integrated
corrected charge from all the hit PMTs within a 70~degree cone from
the reconstructed vertex with respect to the reconstructed ring
direction is measured for each ring.  The correction of the charge is
done for each hit PMT by taking into account the photon acceptance as
a function of the incident angle to the PMT and the light attenuation
length in water.  The corrected charge is converted to momentum by
using a conversion table made with MC for each particle type
assumption.  For multi-ring events, the momentum for each ring is
determined by separating charge for each hit PMT to maximize agreement
of observed and expected ring charge pattern according to the particle
type assumption determined by PID.  For the expected charge
calculation, photon scattering in water and reflection at the inner
detector wall are taken into account.  Time variation of the
attenuation length and PMT gain is taken into account and the momentum
is corrected in the real data.

\begin{table}
\begin{center}
\caption{The uncertainty in the energy scale in Super-K I-IV}
\label{tab:skescalephase}
\begin{tabular}{l|c|c|c|c}
\hline\hline
 & SK-I & SK-II & SK-III & SK-IV \\
\hline
Absolute Energy Scale $(\%)$ & 0.74 & 1.60 & 2.08 & 2.4 \\
Time Variation  $(\%)$ & 0.88 &  0.55 & 1.79 & 0.41 \\
\hline
Total Energy Scale  $(\%)$ & 1.1 & 1.7 & 2.7 & 2.4 \\
\hline
Directional Dependance  $(\%)$ & 0.6 & 0.6 & 1.3 & 0.8 \\
\hline\hline
\end{tabular}
\end{center}
\end{table}

\begin{figure}[!h]
\centering
\includegraphics[width=0.6\textwidth]{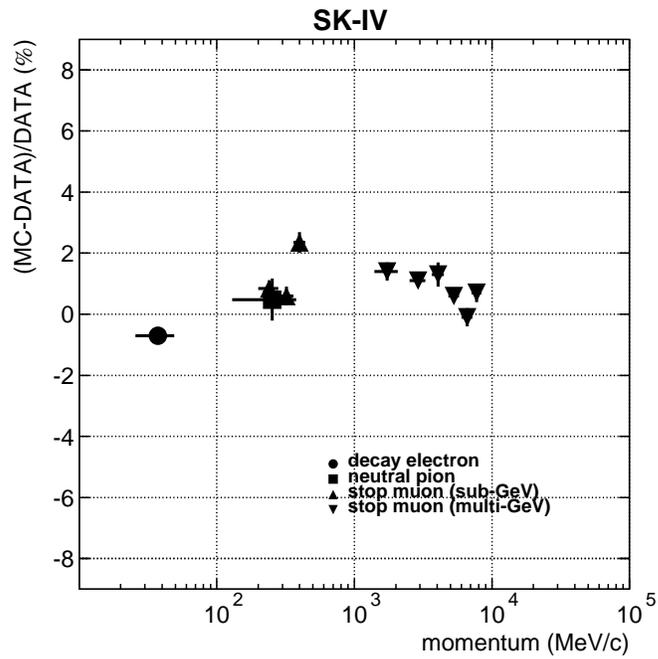}
\caption{The absolute energy scale check in SK-IV.}
\label{absescale}
\end{figure}

\begin{figure}[!h]
\centering
\includegraphics[width=0.6\textwidth]{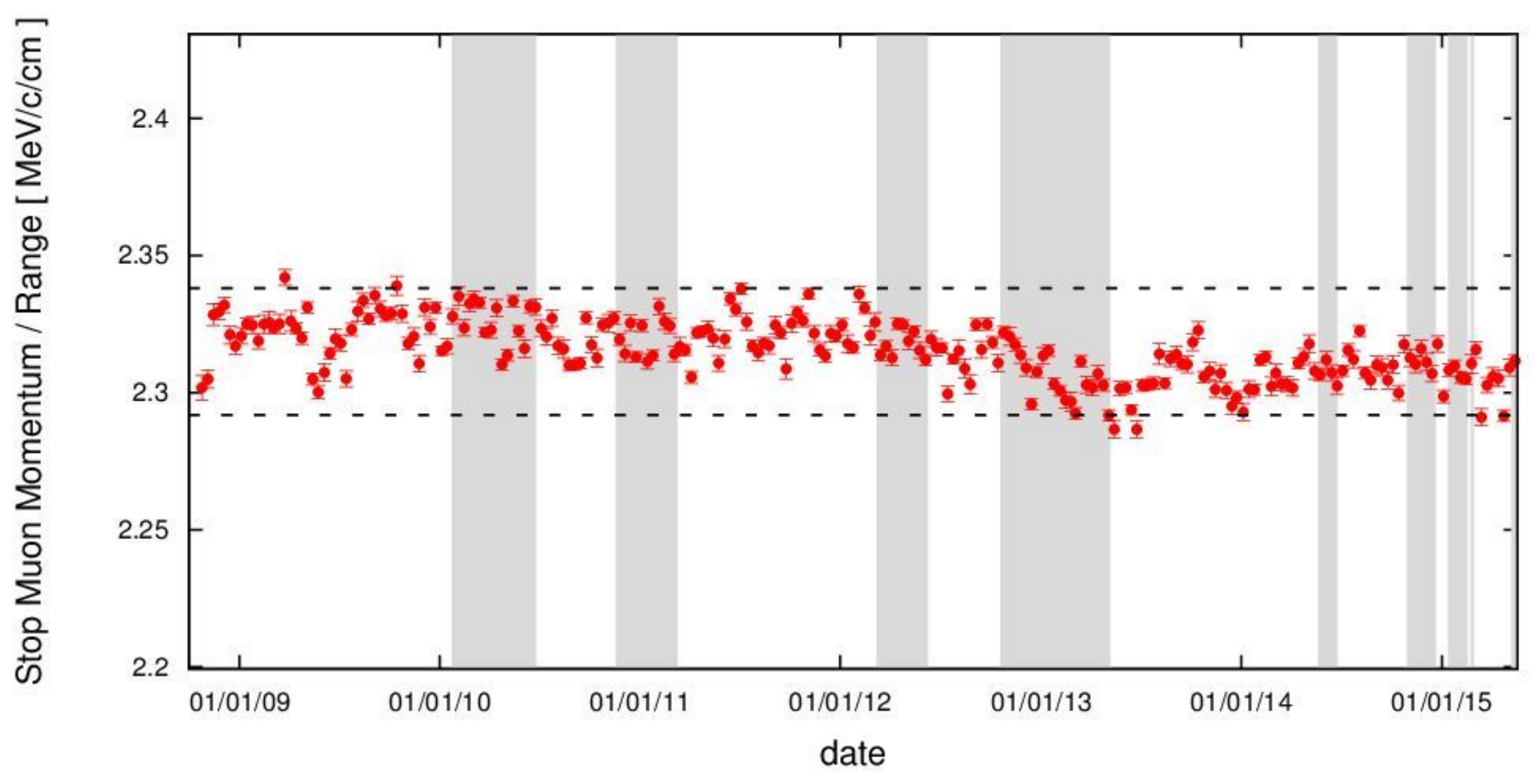}
\caption{The momentum time variation for cosmic-ray stopping muon in SK-IV.
The horizontal broken lines correspond to $\pm$ 1\% with respect to
the average.  The shaded regions correspond to T2K run periods.}
\label{escaletvari4mu}
\end{figure}

\begin{figure}[!h]
\centering
\includegraphics[width=0.6\textwidth]{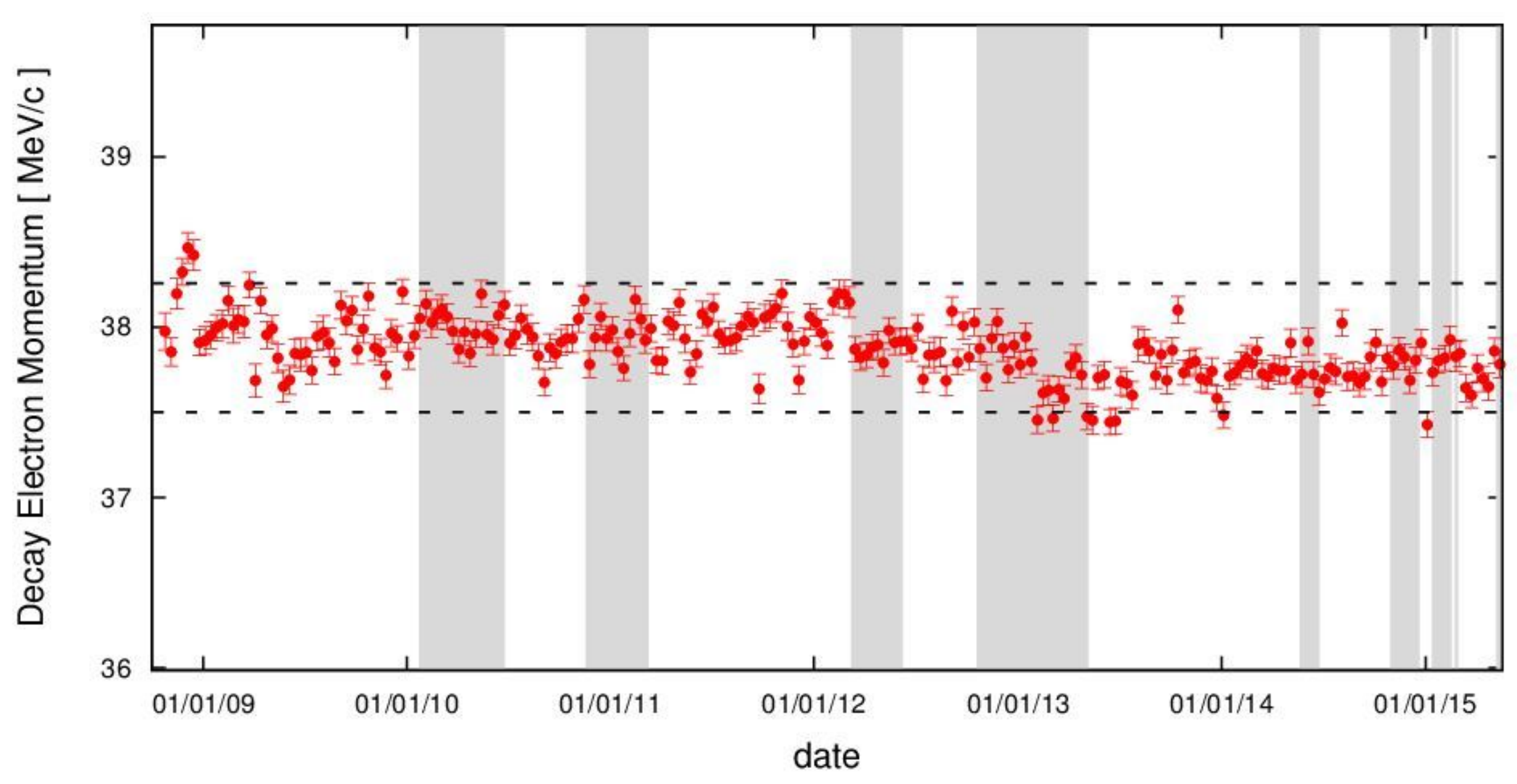}
\caption{The momentum time variation for decay electron in SK-IV.
The horizontal broken lines correspond to $\pm$ 1\% with respect to the average.
The shaded regions correspond to T2K run periods.}
\label{escaletvari4e}
\end{figure}

Tabel~\ref{tab:skescalephase} summarises the energy scale uncertainties in
Super-K. Figure~\ref{absescale} shows the result of a comparison of
the energy scale between data and MC from around 10~MeV/c to 10~GeV/c,
using cosmic-ray stopping muon and associated decay electron data, and
$\pi^0$ data produced in atmospheric neutrino interactions in SK-IV.
The absolute scale error is estimated to be 2.4\% from the largest
difference between data and MC which occurs in the sub-GeV stopping
muon sample.  Note that the global momentum scale for data was
adjusted typically in the end of each SK detector period to minimize
the absolute scale error but has not been done yet in SK-IV.
The time variation of reconstructed momentum is shown in
Fig.~\ref{escaletvari4mu} and Fig.~\ref{escaletvari4e}.
For the entire SK-IV time period, the momentum has been stable to within
1\% and the time variation (RMS/mean) is estimated to be about 0.4\%,
taken as the largest value between these two calibration sources.
By taking the quadratic sum of the absolute scale error and the time
variation, the energy scale error is estimated to be 2.4\% in SK-IV.
Further, the directional dependence of decay electron momenta is used
to estimate the asymmetry of the energy scale. The asymmetry was
estimated to be 0.8\% in SK-IV.

\begin{figure}[!h]
\centering
\begin{tabular}{c}
\includegraphics[trim={0 210 0 210},clip,width=0.6\textwidth]{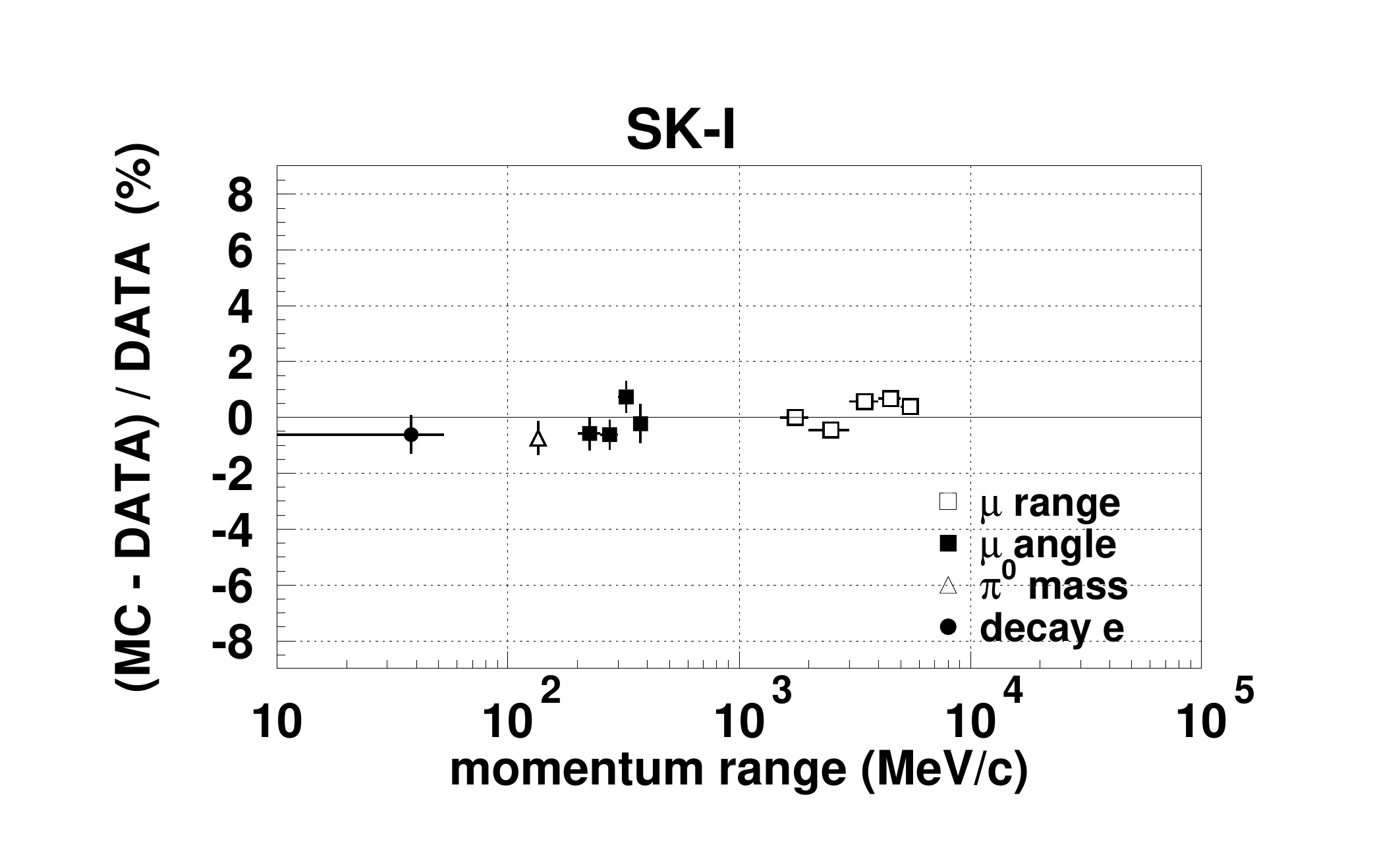} \\
\includegraphics[trim={0 210 0 210},clip,width=0.6\textwidth]{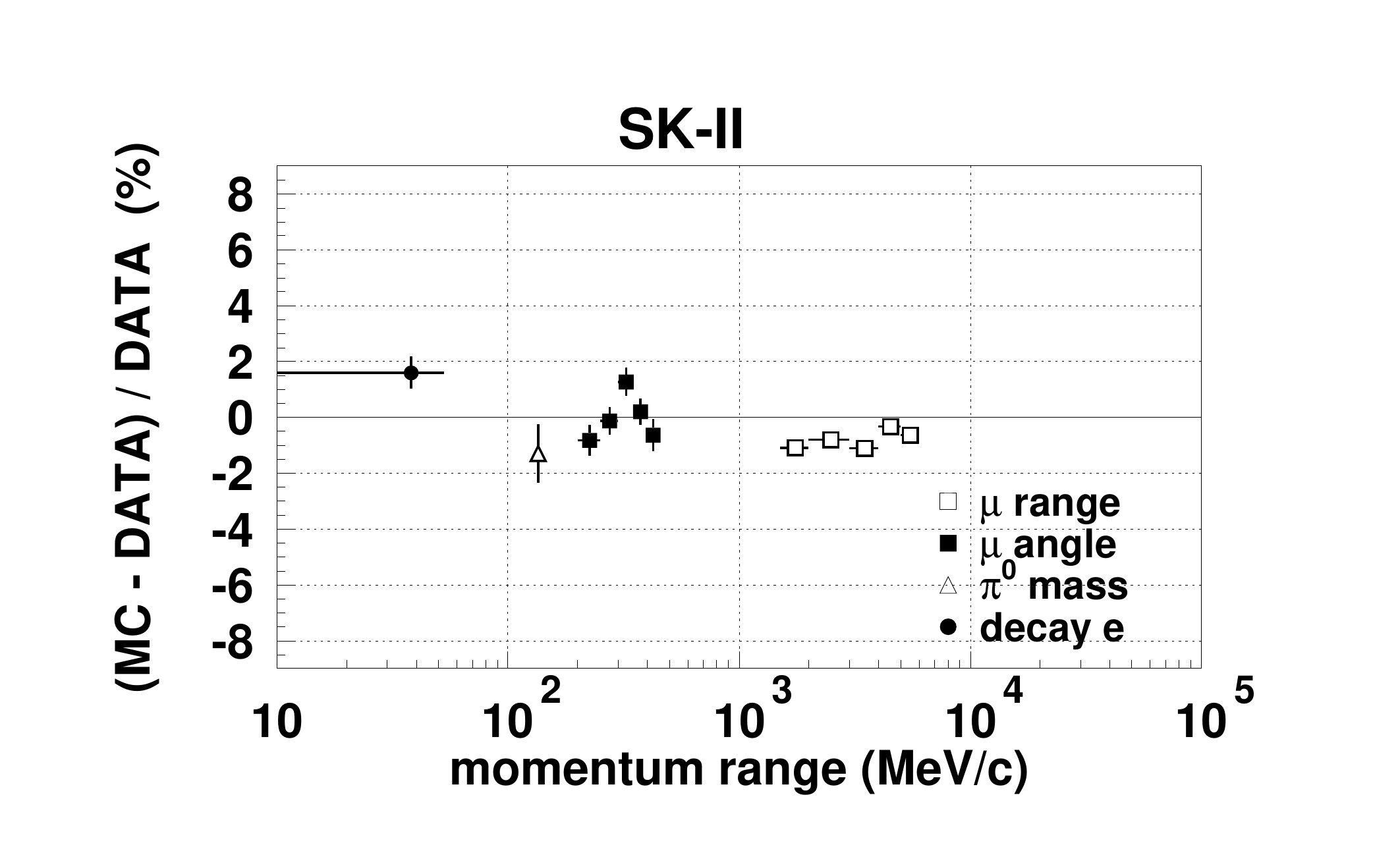} \\
\includegraphics[trim={0 210 0 210},clip,width=0.6\textwidth]{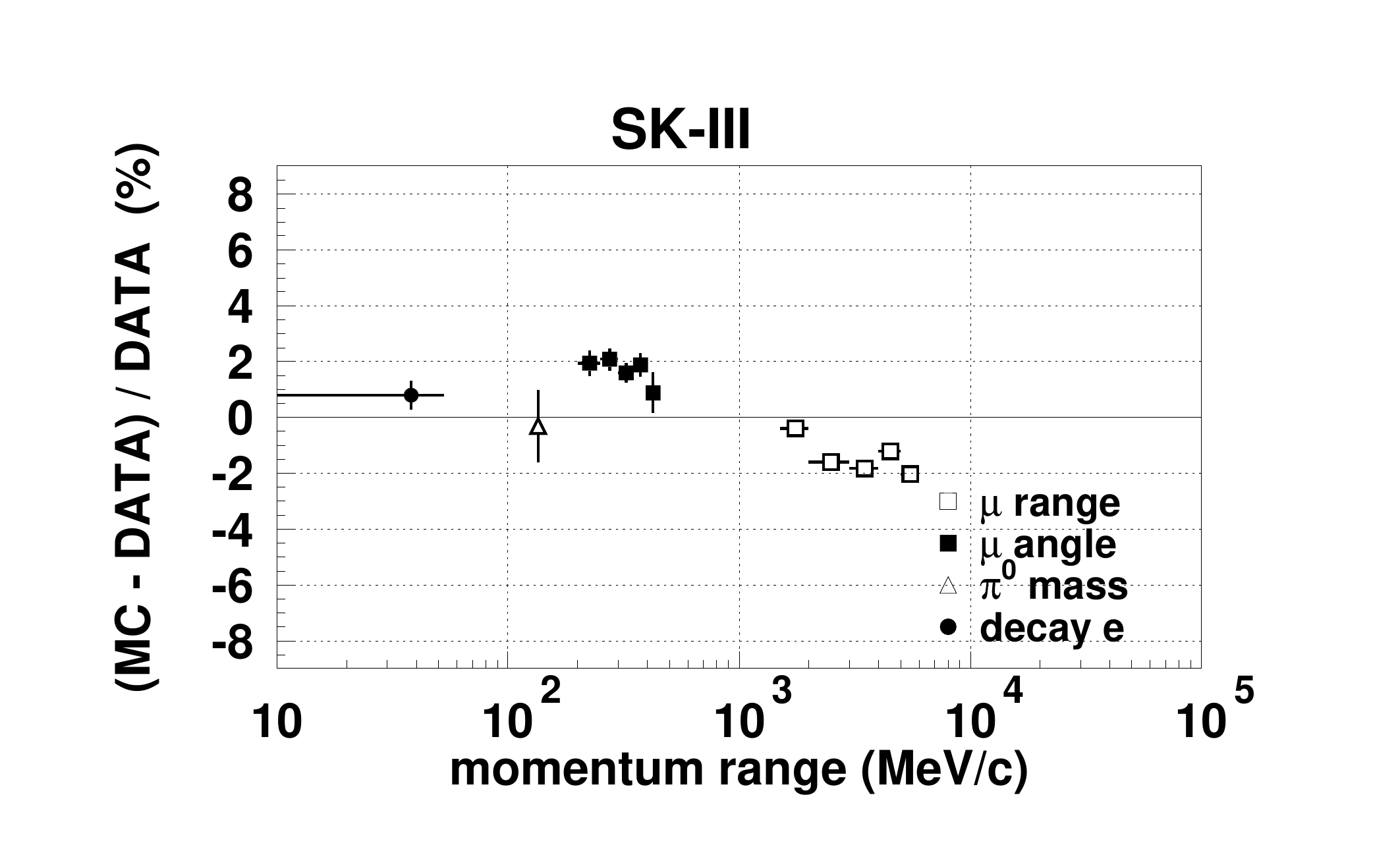}
\end{tabular}
\caption{Comparison of absolute energy scale checks from SK-I to SK-III.}
\label{compabsescale}
\end{figure}

Fig.~\ref{compabsescale} shows the absolute energy scale checks from SK-I to SK-III.

There have been a number of changes to SK over its running period. In SK-II the photo-coverage was reduced to 20\% from 40\%, and PMT acrylic covers were installed from this point on. From SK-III photo-coverage was restored to 40\% and new readout electronic modules were installed from SK-IV.
Despite these changes, the absolute scale error remains at about 1-2\% from
SK-I through SK-IV. The absolute energy error is estimated to be 0.74\%, 1.60\%,
and 2.08\% for SK-I, SK-II, and SK-III, respectively.

\begin{figure}[!h]
\centering
\includegraphics[width=0.6\textwidth]{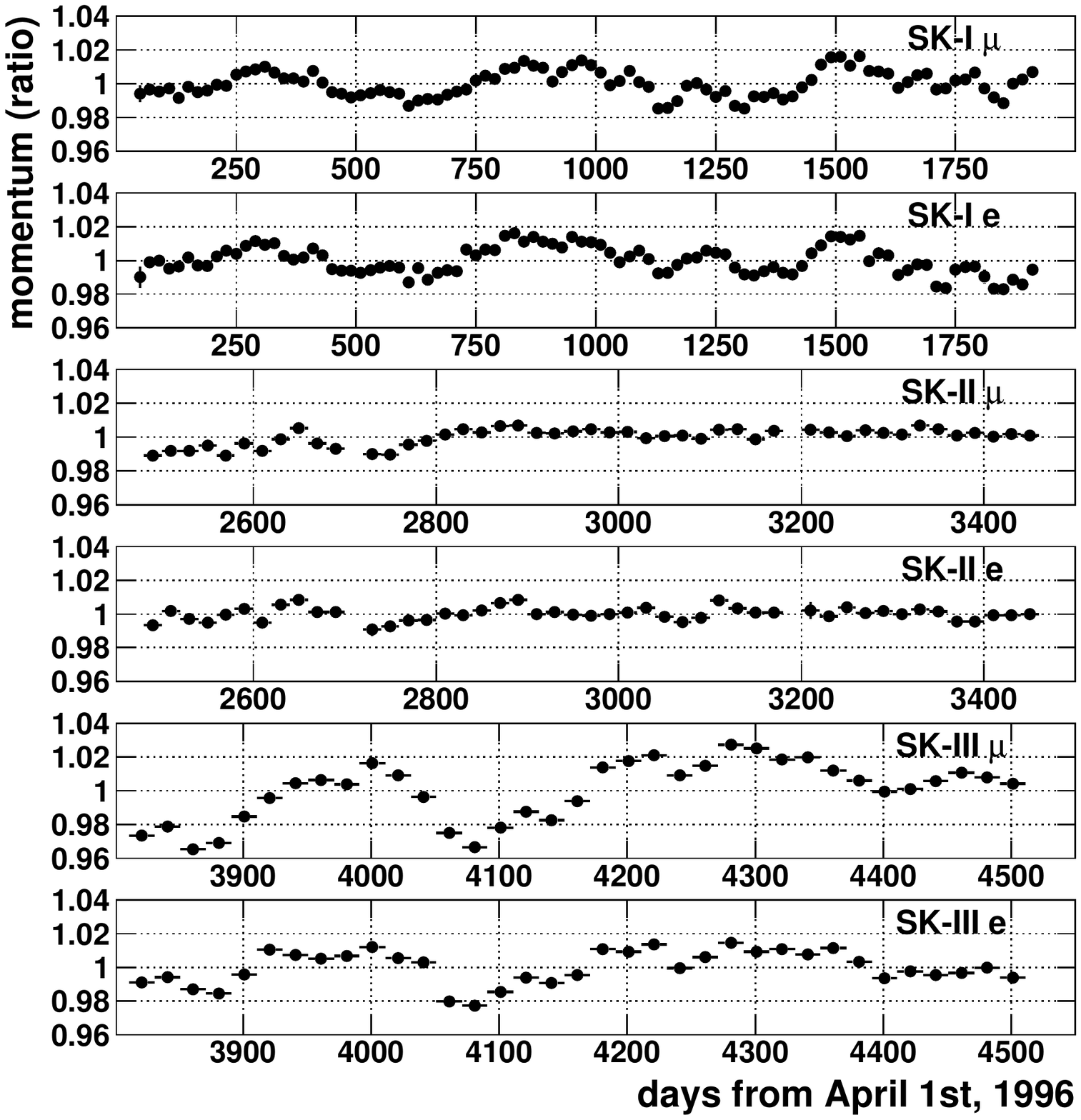}
\caption{Comparison of the momentum time variation from SK-I to SK-III.}
\label{escaletvari1-3}
\end{figure}

Figure~\ref{escaletvari1-3} shows the result of the momentum time
variation (RMS/mean) from SK-I to SK-III.
The time variation during SK-I, SK-II, and SK-III was 0.88\%, 0.55\%, and 1.79\%, respectively.
The time variation in SK-III was relatively larger due to worse water quality and it is at
a minimum in SK-IV due to improved momentum time variation correction for SK-IV.
During SK-I, SK-II, and SK-III, the directional dependance of the
momentum was estimated to be 0.6\%, 0.6\%, and 1.3\% respectively.

\subsubsubsubsection{Toward Hyper-K}

To achieve the physics goals of Hyper-K we must achieve an uncertainty
in the momentum scale as has been achieved in Super-K. In current
analyses of Super-K data involving, atmospheric neutrinos, neutrinos
from the T2K beam and nucleon decay, systematic uncertainties are
dominated by uncertainties in the physics model, such as cross section
uncertainties. We aim to reduce these for Hyper-K, but they are still
likely to be larger than the current Super-K momentum scale
uncertainty. We aim, in the first instance, to match the current
Super-K uncertainties, and to over the duration of the experiment make
further improvements to reduce these uncertainties further.

Thanks to the similar vertical cylindrical tank, there is no major concern which makes the
energy scale error significantly larger in Hyper-K.

The current energy scale error in SK-IV comes from the unknown
momentum dependence of the absolute scale, especially as seen in the
sub-GeV stopping muon sample.  The muon momentum of the sub-GeV sample
is below about 500~MeV/c where the Cerenkov angle significantly
depends on the momentum.  Each sub-sample of the sub-GeV sample is
defined by using the reconstructed Cerenkov angle.  The Cerenkov angle
reconstruction depends on the charge profile shape which is especially
affected by the tuning of the scattering parameters in SKDETSIM.  As
Hyper-K is significantly larger than Super-K the variation of water
properties across the detector, especially the vertical dependence,
becomes more of a concern. To address this analysis of light from
horizontal injectors at different depths is being developed in Super-K
and will form an important part of the Hyper-K calibration program
where better understanding of the water quality is important.  

As the sophistication of reconstruction and calibration continues to
increase, addressing the various contributions to the momentum scale
uncertainty and improving our understanding of their effects, we can
expect the overall systematic uncertainty to continue to improve. A
recent example of this includes the improvement of the PMT gain time
variation correction in to SK-IV analysis, reducing the time
variation with respect to that from SK-I-III. Further improvements can
be expected over time, through Super-K and into the Hyper-K era.

%%%%%%%%%%%%%%%%%%%%%%%%%%%%%%%%%%%%%%%%%
%%%%%%%%%%%%%%%%%%%%%%%%%%%%%%%%%%%%%%%%%

\subsubsection{OD calibration system}
\label{sec:hk_od_calibration}

The major task of the OD part of the Hyper-K detector is not to
obtain the exact energy deposited but to identify
the neutrino events out of the cosmic-ray muons.
For example, ``fully contained'' events are identified by requiring no
energy deposition in OD, and ``partially contained'' events and
``upward-going muon'' events, which are important sub-samples in
atmospheric neutrino analyses, are identified with OD hits
coinciding with ID hits.
For these physics analyses, an `inter-calibrations' between OD and ID,
e.g. timing calibrations between OD and ID, is also important in
addition to the calibrations of OD itself.

Compared the ID part, the OD has various disadvantages in having a
calibration system.  They are: 1) many light injection points are
necessary to illuminate all the light sensors in the OD area to an
intensity level of a few 100 PE's, 2) there are sensor support
structures which can hinder the delivery of calibration light, and 3)
there is no easy way to deploy additional light injection points to
replace non-functional ones once the detector is filled with
water. The latter 2 points can be mitigated by having redundant light
injectors, but this will certainly increase the total cost of the
system.

In the case of Super-Kamiokande experiment, the OD calibration system
consists of a $N_2$ and a dye laser, monitoring PMT's, a variable
attenuation wheel, optical switches, and 52 fibers.  Each fiber is
equipped with a light diffusing tip at the end. Of these fibers, 24
are placed in wall section and 14 each are placed in top and bottom
sections. They are 72\,m long, except for those placed in the bottom
section which is 110\,m long. In average, each wall fiber covers
160\,m$^2$ of OD sensor area and about 2.5\,m away from the OD PMT
plane. Top and bottom fibers cover 64\,m$^2$ per fiber and about 1.6~m
away. These fibers are reasonably redundant and a little over a half
of them are actually used to calibrate all the OD PMT's.

For the SK OD calibration system to be adopted to the Hyper-K detector, 79
fibers are required to achieve the same fiber density as the SK for
the Hyper-K wall section. For top and bottom, 61 each is necessary. In
total, 201 fibers are needed for the entire Hyper-K detector.  In terms of
length, 200~m fibers for bottom and 120~m ones for other sections are
necessary to compensate the linear dimension difference between the SK
and Hyper-K detectors.

%%%%%%%%%%%%%%%%%%%%%%%%%%%%%%%%%%%%%%%%%
%%%%%%%%%%%%%%%%%%%%%%%%%%%%%%%%%%%%%%%%%

\newpage
\graphicspath{{design-computation/figures}}

    \subsection{Computing}
\begin{figure}[htbp]
  \begin{center}
    \includegraphics[scale=0.4]{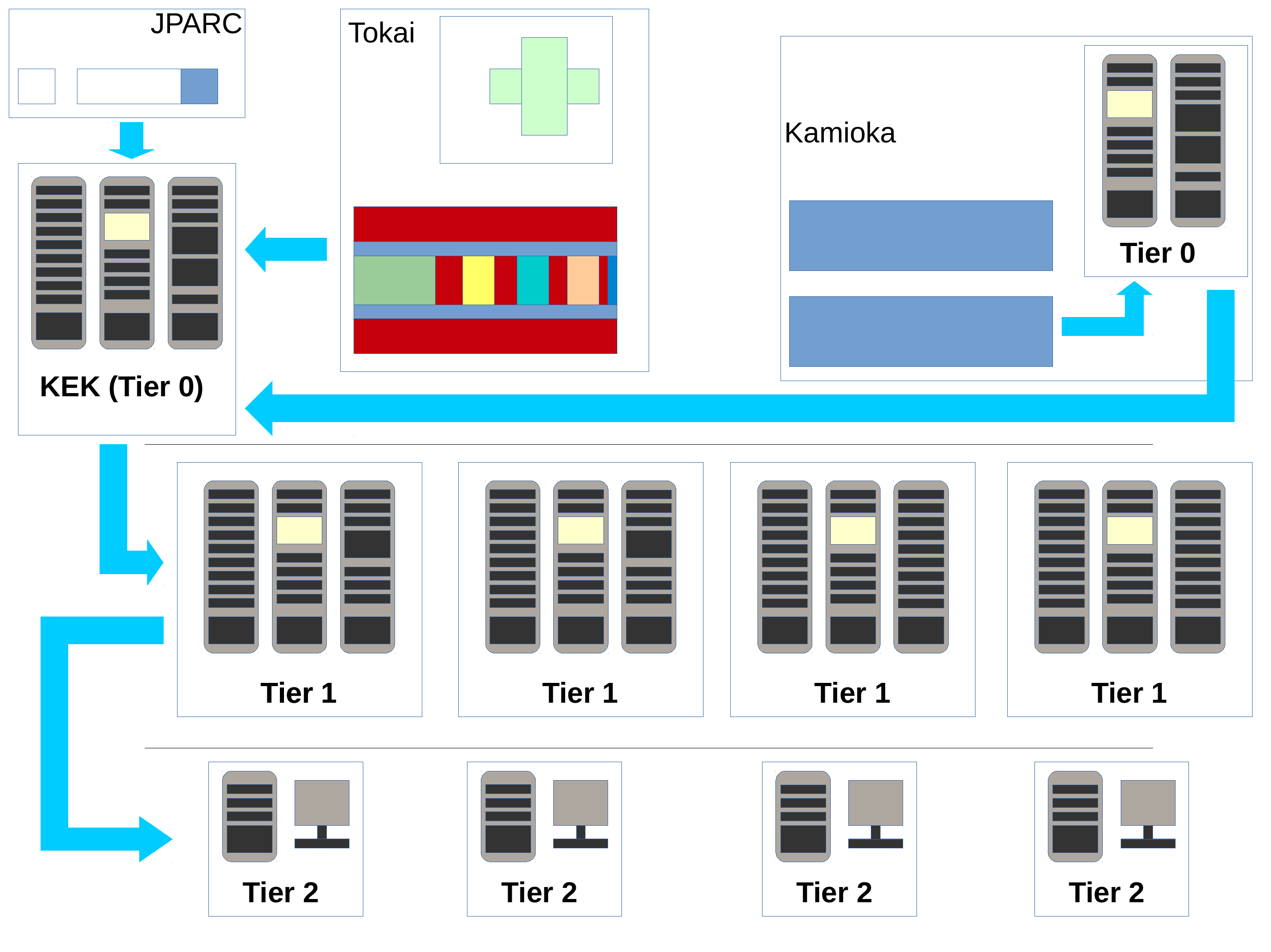}
  \end{center}
\caption {General overview of a possible Hyper-K tiered system.}
  \label{fig:computing_tiers}
\end{figure}

Hyper-K adopts a Tiered computing model where Kamioka and KEK sites form the
Tier-0 due to the distributed nature of the experiment. A general overview of a future 
Hyper-K tiered system is shown in Fig.~\ref{fig:computing_tiers}.

The Tier0 sites will hold the raw experiment data as well as the processed data. The KEK
Tier0 will also contain a  
The Tier1 centres (such as RAL, TRIUMF, ccin2p3) would hold portions of the raw, processed
and simulated data and provide computational resources for the
simulation, processing and reprocessing. The Tier2 sites which
typically consist of universities will provide computational and
storage resources (the storage is usually used to hold specific
subsets of the data or simulation). The model makes efficient use the
available computing resources that exist at collaborating sites. The
model will be regularly reviewed as changes to the computing landscape
take place.

A current estimate of the rate of raw data to be stored is ~20\,TB/day. 
About 80\,PB of the disk space
is necessary to store the raw data for the 10 years operation.  

Reduction and reconstruction software will be
applied to all the data in the Kamioka Tier0  as soon as the data are
taken to provide different samples for different energy regions or
different analysis groups, i.e. low energy region mainly for the study
of solar neutrinos, higher energy for the study of nucleon decay and
atmospheric neutrinos, downward going muons for the background study
of solar neutrinos, the data during the beam timing from the
accelerator for the beam neutrino analyses. These data sets also
provide timely feedback to the experiment and beam operations groups
on the quality of the beam and performance of the detector.  Also,
early detection of a supernova burst is crucial and dedicated realtime
analysis has to be performed in the independent system.  The required
computing power for data reduction and reconstruction at this level,
together with the supernova detection system, is not so huge and 1000
cores of the current Intel IA64 CPUs will be sufficient.

\subsubsection{Simulation production}

Mass production of the simulation data sets and their analyses are
expected to be performed in the Tier-1 centres because the required
CPU resource for the huge amount of simulation data is expected to be
at least a few tens of times larger than the ones necessary for the
real time data processing. On the other hand, the simulated data set
is not extremely large and the cost of the storage could be less than
10\% of the storage for the real data sets.  All the data sets after
reduction and the processed simulation data sets are shared among the
geographically distributed analysis working groups. The Tiered model
ensures results in a more scalable architecture capable of meeting the
computational and storage demands of the experiment.

The Monte Carlo simulation production currently makes use of existing
HEP computational Grid resources to produce sufficient quantities of
physics events necessary to optimise the detector design for maximum
efficiency. The data are managed by the iRODS data management system
(\url{http://irods.org/}) that enables distributed storage to be
managed and accessed in an uniform manner. Collaborators access the
stored data using the intuitive and simple iRODS client API.

\clearpage
\graphicspath{{simulation/figures}}

\section{Hyper-Kamiokande software} \label{section:software}

The Hyper-K software system is designed around the following
principles:
\begin{itemize} \itemsep0em
\item Adaptable. The Hyper-K experiment is expected to run for more
  than a decade.  This period typically spans more than one
generation of software and infrastructure. The Hyper-K offline system
is being designed to be flexible enough to accommodate changes in
tools or infrastructures.
\item Reliable. Each component needs to demonstrate it's reliability
by exhibiting well defined behaviour on control samples.
\item Understandable. Documentation on what the component does, what
it's dependencies are as well as test samples and outputs are
essential in being able to use it successfully.
\item Low overhead. The management and maintenance should be as
automated as possible to free collaborators to focus on the challenge
of extracting the high-quality physics measurements.
\end{itemize}

The software consists of a collection of loosely-coupled packages,
some of which are open-source and some of which are specific to
Hyper-K. The distributed code management system Git \cite{git} is used to manage
the software. Each package is hosted on a third-party central
repository (\url{https://github.com/}) that provides distributed
access to the packages. The distributed nature of the code management
allows researchers the possibility to develop the software independently without
impacting other researchers. The loose-coupling between packages
allows those that reach their end of life to be replaced by better
alternatives with minimal impact on the rest of the system.  Where
possible standard particle physics software libraries are used to
reduce the burden of support of experiment-specific code. The working
language for the Hyper-K software packages is C++, with the output
files being written in ROOT \cite{Brun97} format.

\begin{figure}[htbp]
  \begin{center}
    \includegraphics[scale=0.4]{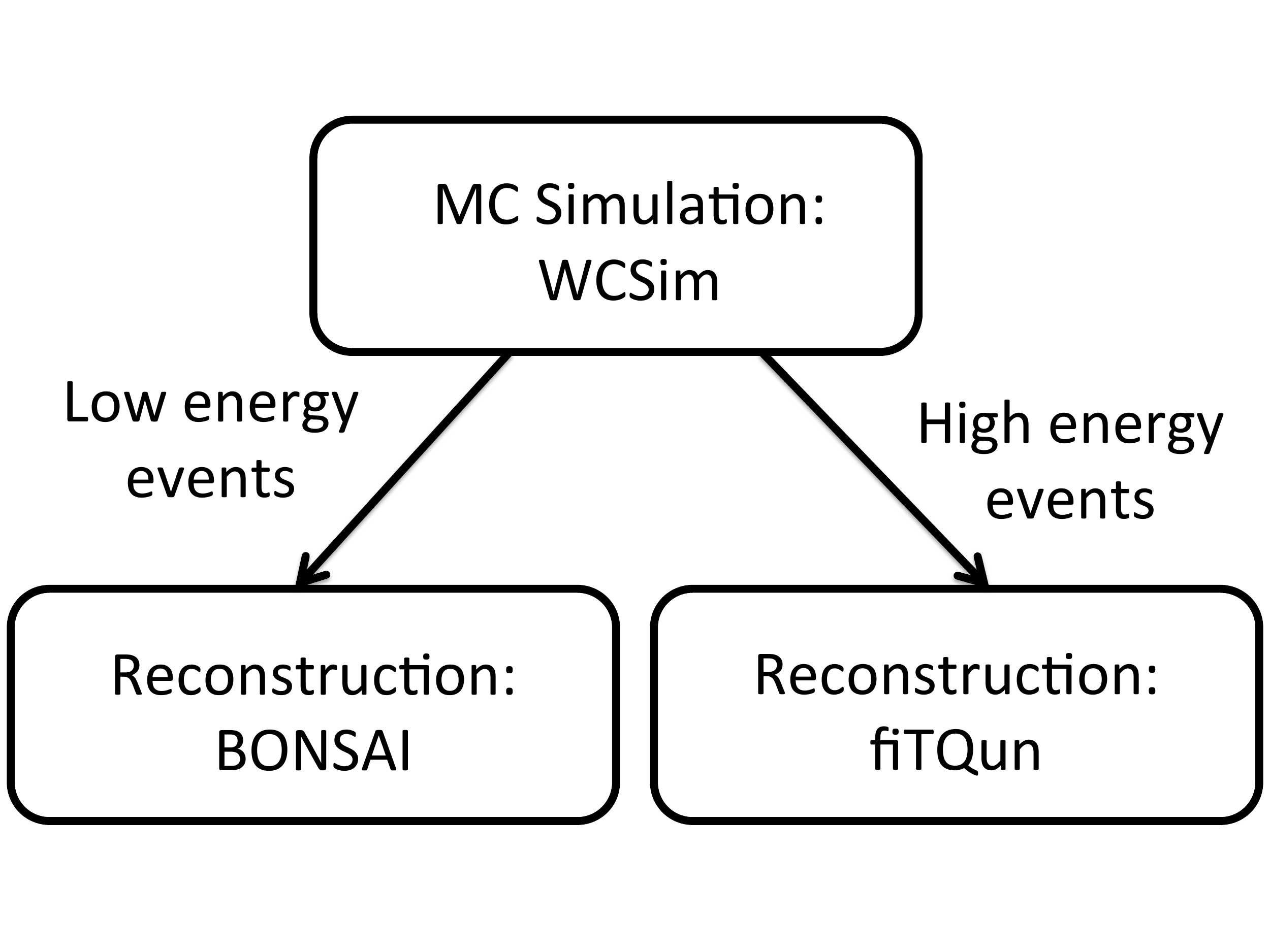}
  \end{center}
\caption {Flowchart of the simulation process.}
  \label{fig:software_flow}
\end{figure}

The flow for the simulation is as follows: The event topologies are
generated by a neutrino interaction package(GENIE
\cite{Andreopoulos:2009rq} and NEUT \cite{Hayato:2009zz}, for
example), and modeled by a Monte Carlo detector response code called
WCSim \cite{WCSim}. The event information is reconstructed using
either BONSAI \cite{icrc0213smy} (for low energy events) or fiTQun
(for high energy events) \cite{Tobayama_2016}. This is shown
schematically in Figure~\ref{fig:software_flow}. These packages will
be described in more detail in the next Sections.

An online workbook is also maintained to provide higher-level
documentation on overall procedures and information for new users of
the software and developers.  An overall software control package
allows for the fully automated download, compilation and running of
the software, based on user requests.

\subsection{WCSim}\label{software:WCSim}

The Water Cherenkov Simulation (WCSim) package is a flexible,
Geant4-based code that is designed to simulate the geometry and
physics response of user-defined water Cherenkov detector
configurations. WCSim is an open-source code and is available for
download at \url{https://github.com/WCSim/WCSim}.

The final performance of the Hyper-Kamiokande detector depends on the
detector geometry, the type of photodetectors, and the photocoverage
that will be used. WCSim takes these variables as inputs and simulates
the detector response, which can then be used to determine the physics
potential. WCSim users specify the type of photodetectors, the number
of photodetectors, the detector diameter and radius, and whether the
water should be doped with gadolinium. The outer detector volume is
currently not implemented in WCSim, though it is actively being
developed for a future release.

For this report, the relevant photodetectors in WCSim are the R3600
20'' diameter PMTs, as well as the R12850 20'' and 12'' diameter box
and line photodetectors. Photodetector parameters in the simulation
include the timing resolution, dark noise rate, and the overall
efficiency for a photon to register a charge (including the quantum
efficiency, collection efficiency, and hit efficiency as described in
Section \ref{section:photosensors}). For the R3600 PMTs, the
parameters were taken from the Super-Kamiokande simulation code
SKDETSIM. The parameters for the R12850 are taken from measurements as
described in Section \ref{section:photosensors}. Some higher-level photodetector effects
such as after-pulsing are not currently simulated in WCSim, though
this is a planned upgrade for a future releases.

Geant4 \cite{Agostinelli:2002hh} is used to track
the particles as they pass through the detector and compute the
final deposited energy. Particles that reach the photodetector glass
and pass the quantum efficiency and collection efficiency cuts are
registered as a hit. The hits are then digitized based on the SK-I
electronics scheme, though the code has the flexibility for users to
include their own custom electronics configurations. 

\begin{figure}[tbp]
 % \begin{center}
    \includegraphics[scale=0.4]{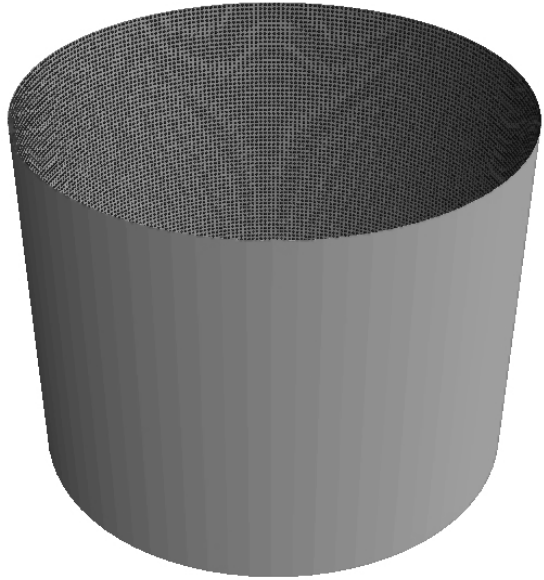}
  %\end{center}
\caption {Geant4 visualization of the Hyper-Kamiokande detector
configuration. The top cap of the detector has
been removed for visualization purposes. Phototubes are shown in
black, while the walls of the detector are shown in grey.}
  \label{fig:HK-WCSim}
\end{figure}

\begin{figure}[tbp]
   \centering
   \begin{tabular}{cc}
    \centering
    \includegraphics[scale=0.4]{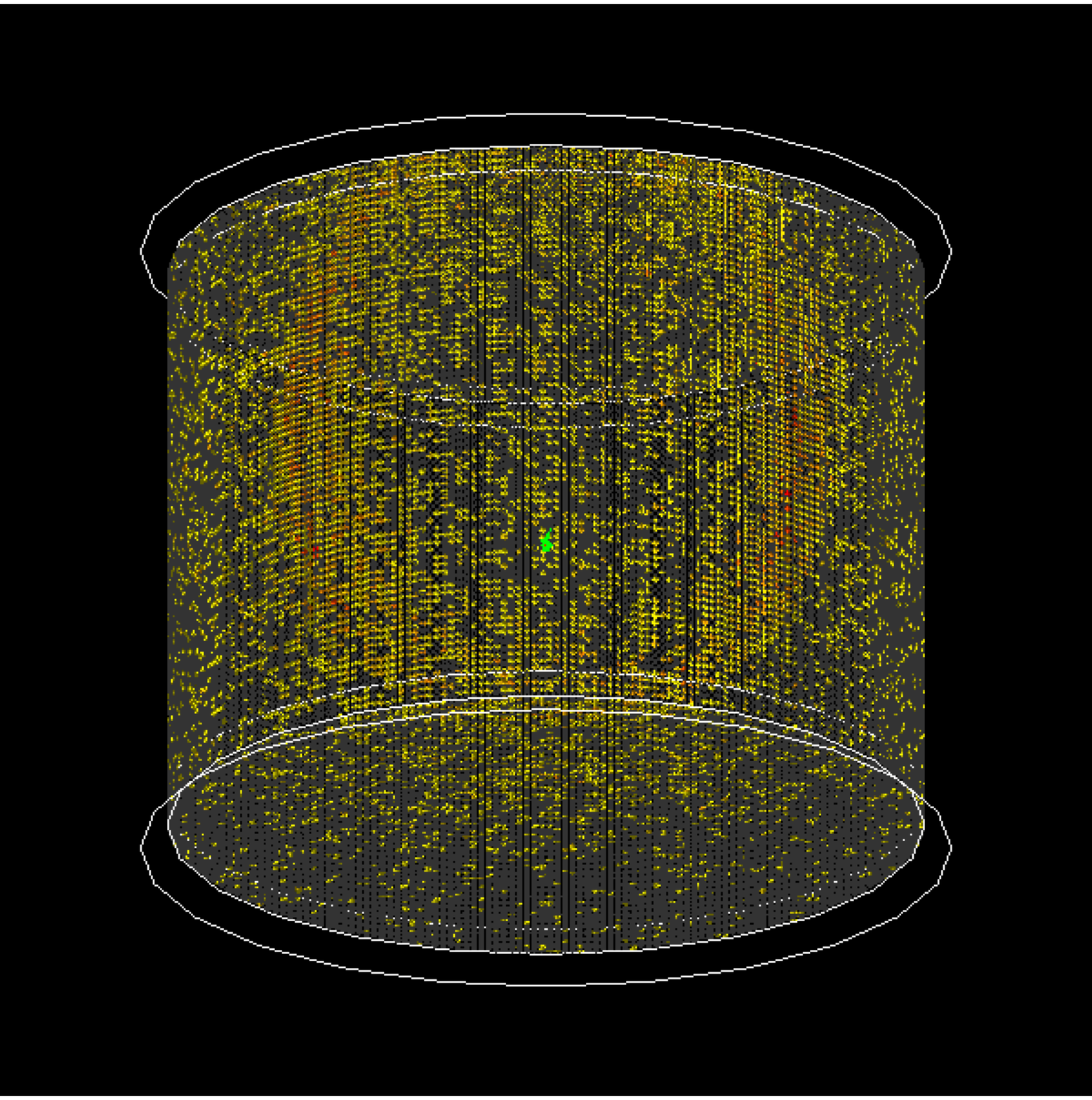}
   &
    \centering
    \includegraphics[scale=0.4]{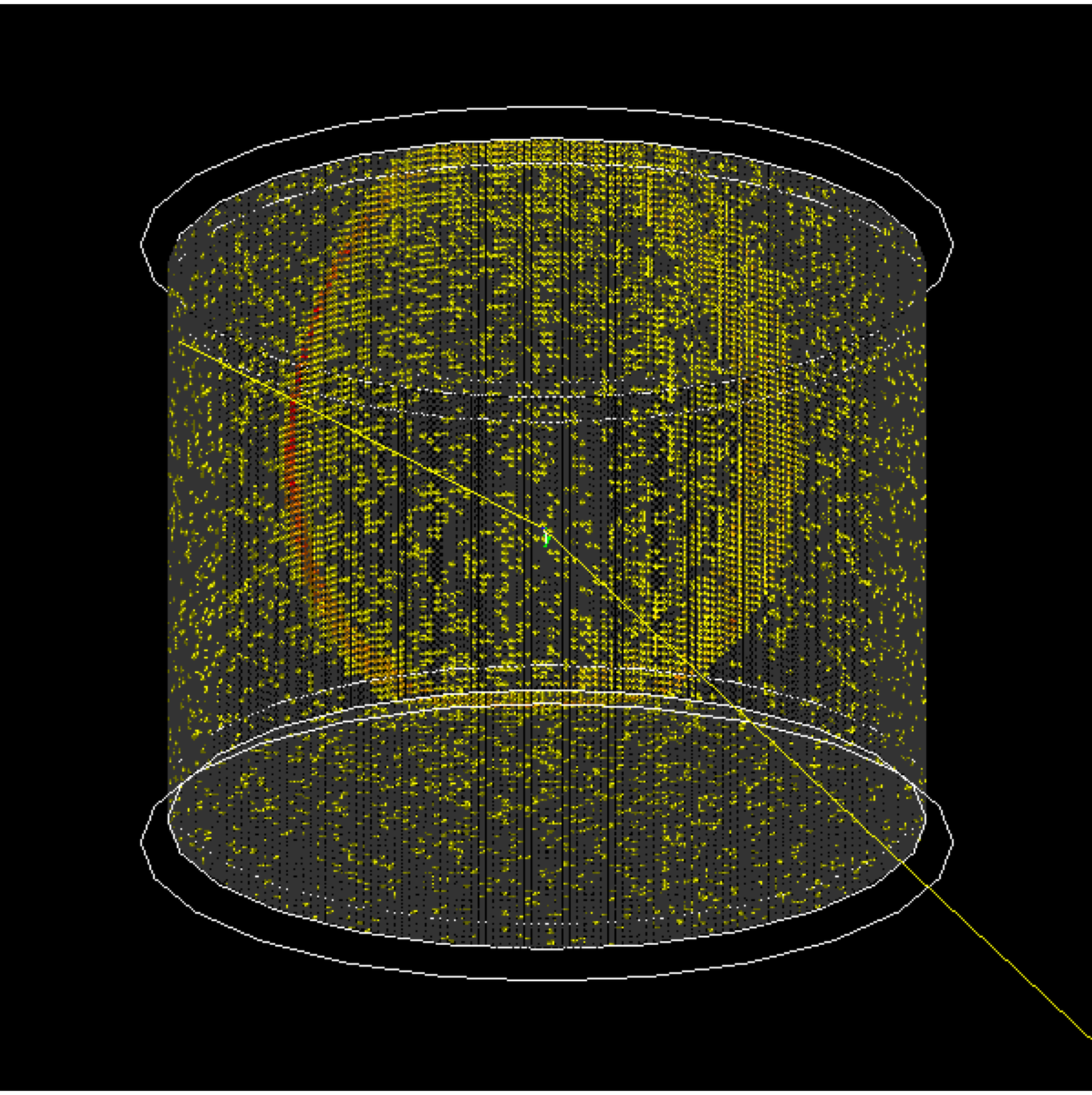}
  \end{tabular}
  \caption {Event displays in the HK detector for a 1 GeV electron
    (left) and a 1 GeV muon (right).}
  \label{fig:HK-WCSim-eventdisplay}
 \end{figure}

The output for the WCSim code includes both the raw hit and the
digitized information. The raw hit information includes which tubes
were hit and how many times each tube was hit. The digitized
information includes the number of hits in a trigger window, as well
as the charge and time of the hit tubes. WCSim output files can be
used for event reconstruction by fiTQun or BONSAI, which are described in the
following subsections. Geant4 visualization tools can be used to display the
detector geometry and particle tracks. Figure~\ref{fig:HK-WCSim} is a
rendering of one of the proposed Hyper-K
tanks. Figure~\ref{fig:HK-WCSim-eventdisplay} shows an example of an
event display for an electron and for a muon, each with 1 GeV kinetic energy. 

\begin{figure}[tbp]
   \begin{tabular}{cc}
  % \begin{minipage}[t]{0.55\hsize}
    \centering
    \includegraphics[trim=0cm 7.0cm 2.5cm 7.0cm, clip=true, scale=0.4]{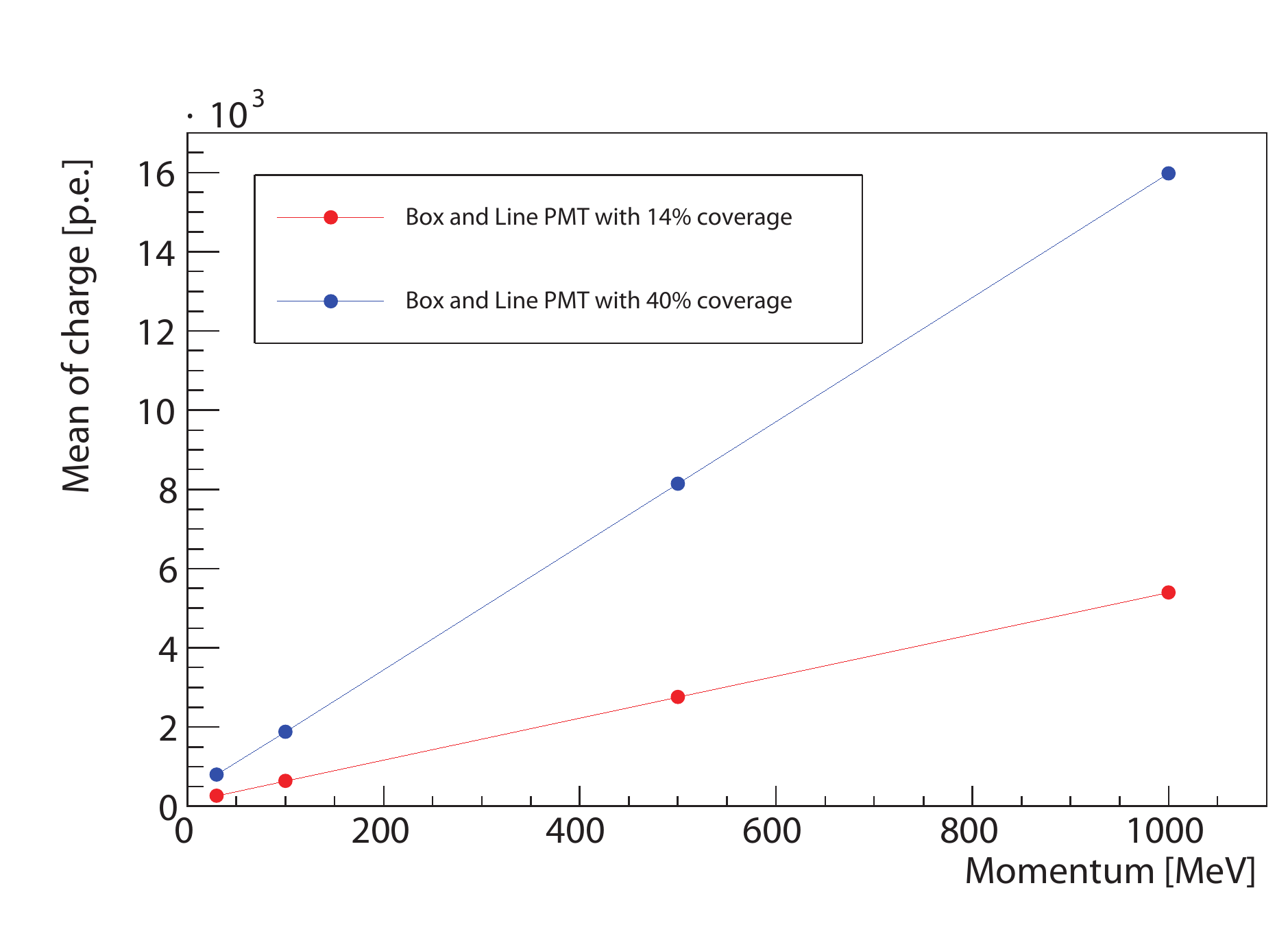}
   %\end{minipage} 
&
%   \begin{minipage}[t]{0.55\hsize}
    \centering
    \includegraphics[trim=0cm 7.0cm 2.5cm 7.0cm, clip=true, scale=0.4]{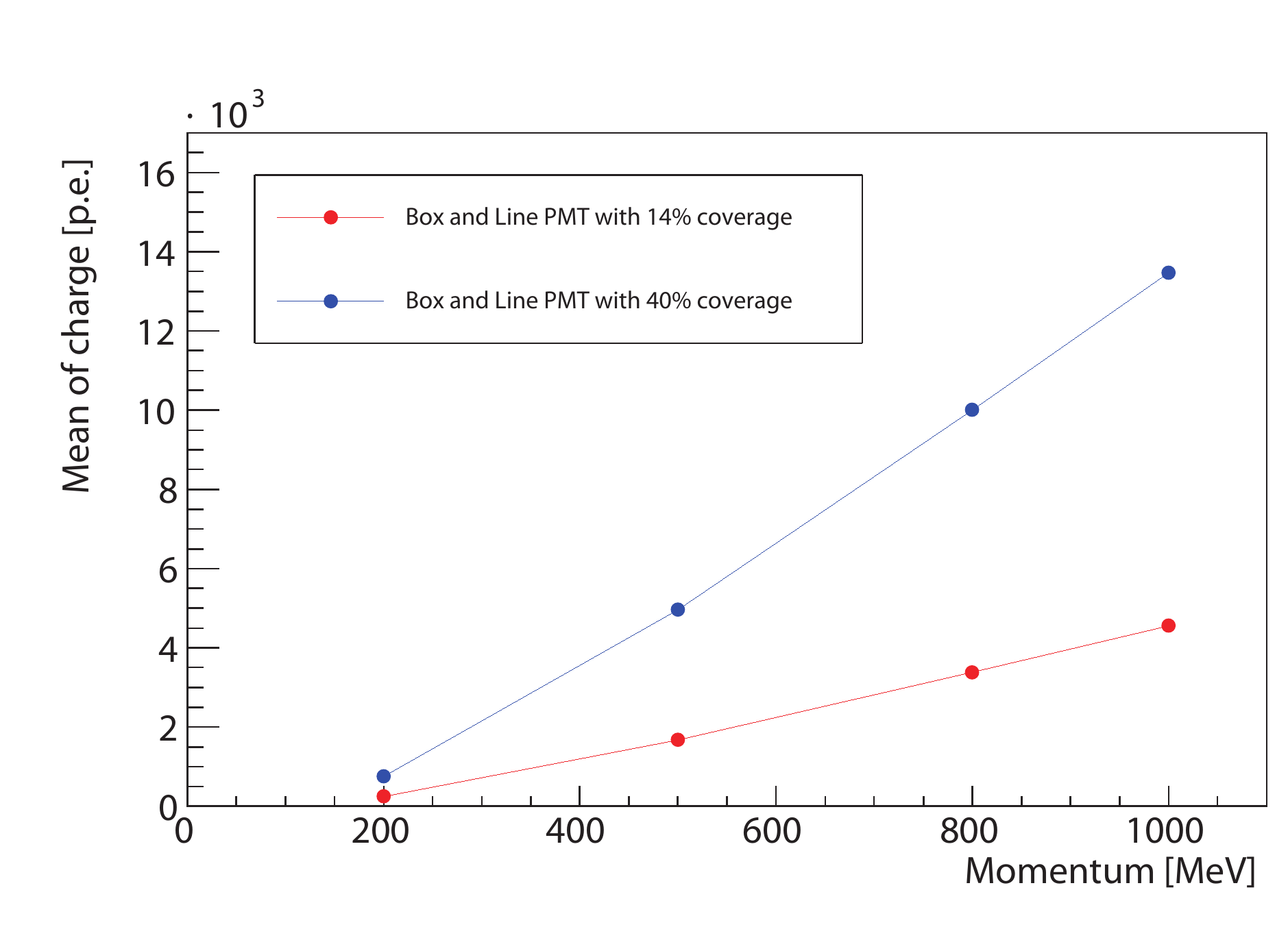}
 %  \end{minipage}
  \end{tabular}
  \caption {Total charge distributions for electrons (left) and muons (right)
  with several momenta in the Hyper-K detector. The red line
  corresponds to 14\% photocoverage, while the blue line corresponds
  to 40 \% photocoverage.}
  \label{fig:total-charge}
 \end{figure}

\begin{figure}[htbp]
  \begin{tabular}{cc}
 %  \begin{minipage}[t]{0.55\hsize}
    \centering
    \includegraphics[trim=0cm 7.0cm 2.5cm 7.0cm, clip=true, scale=0.4]{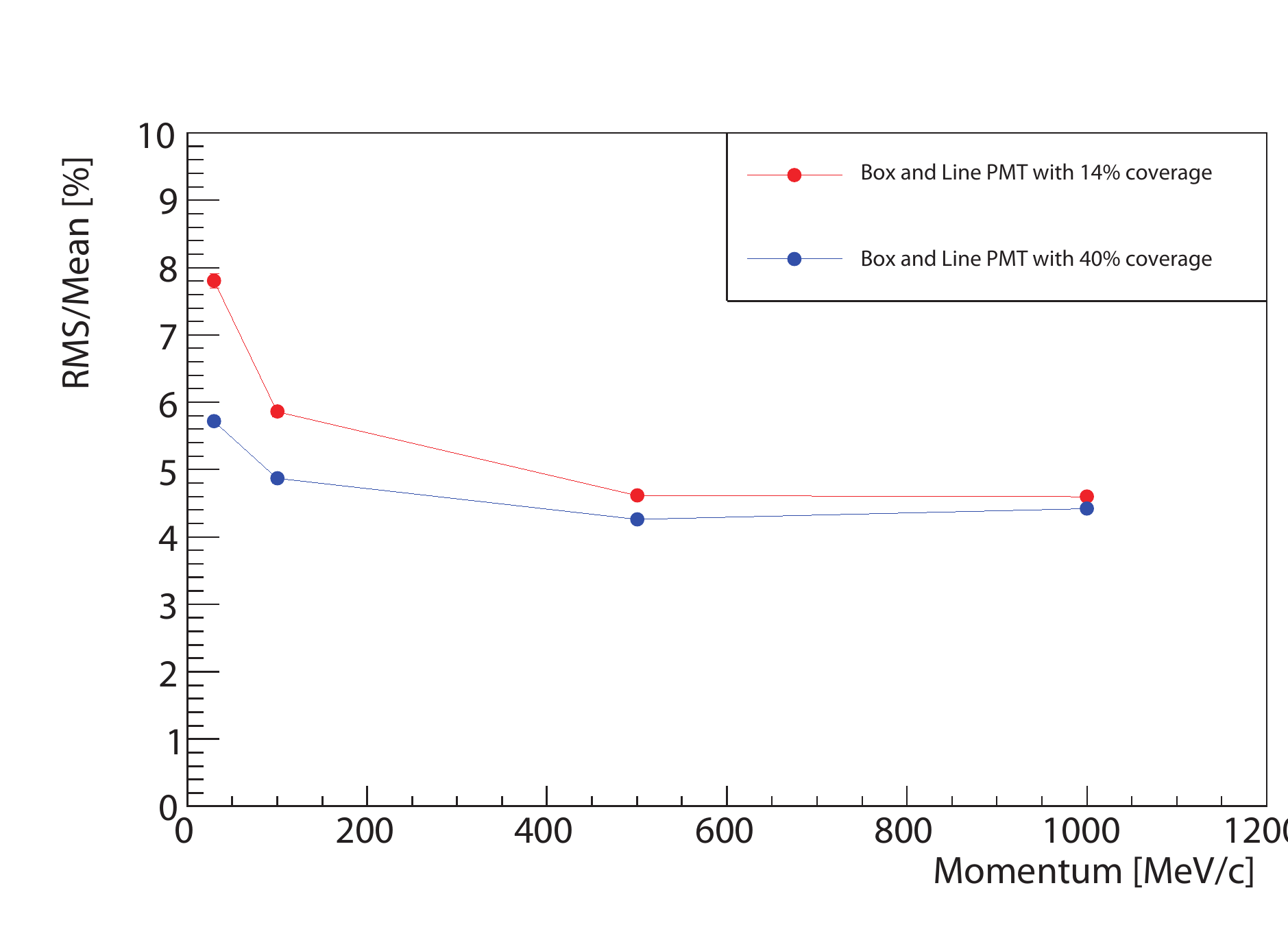}
 %  \end{minipage} 
&   
%\begin{minipage}[t]{0.55\hsize}
    \centering
    \includegraphics[trim=0cm 7.0cm 2.5cm 7.0cm, clip=true, scale=0.4]{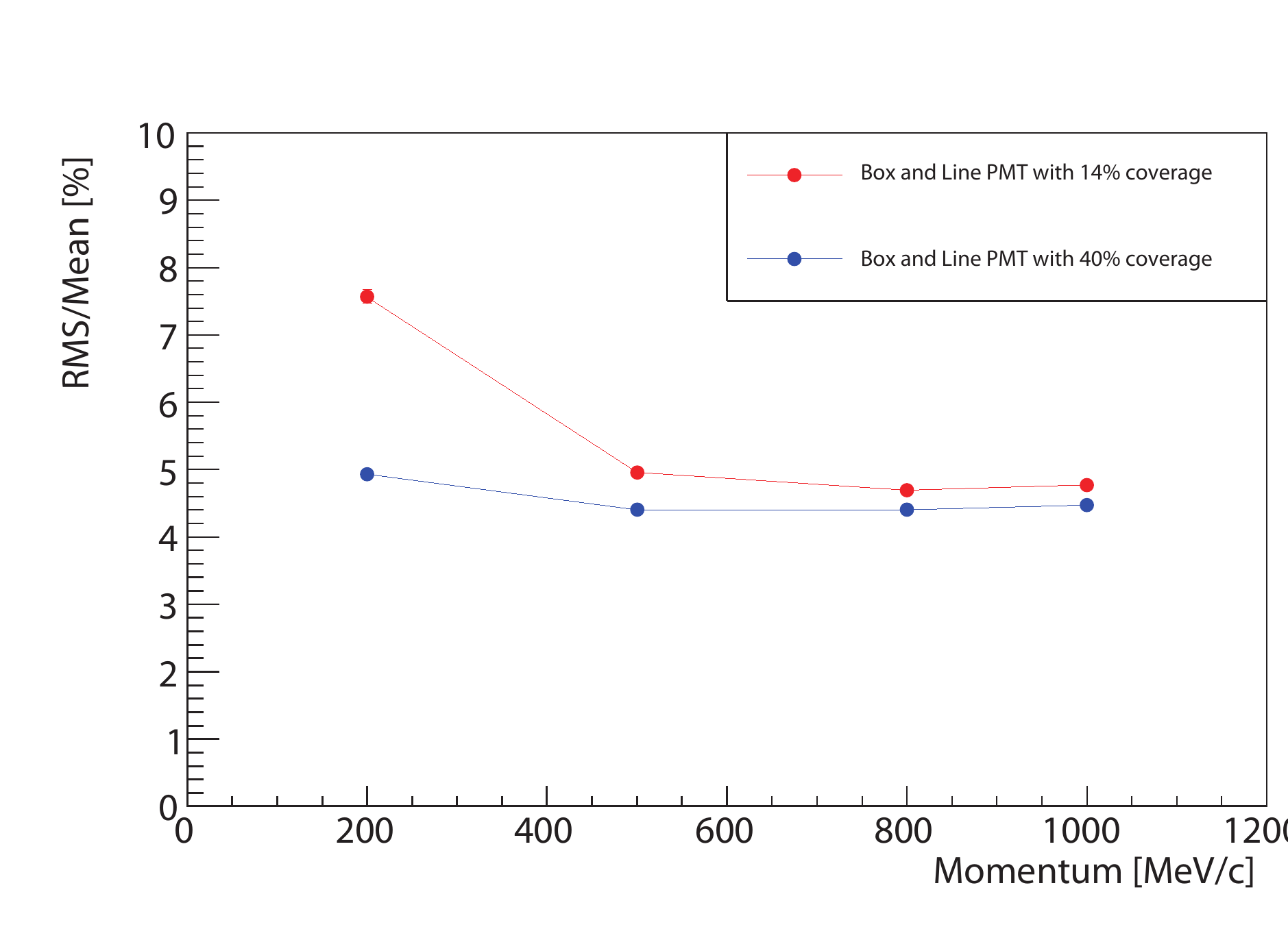}
 %  \end{minipage}
  \end{tabular}
  \caption {RMS/Total charge distributions for electrons (left) and muons (right)
  with several momenta. The red line corresponds to the configuration
  with 14\% photocoverage, while the blue line corresponds to the
  configuration with 40\% photocoverage.}
  \label{fig:rms-mom}
 \end{figure}

\begin{figure}[htbp]
   \begin{tabular}{cc}
   %\begin{minipage}[t]{0.55\hsize}
    \centering
    \includegraphics[scale=0.47]{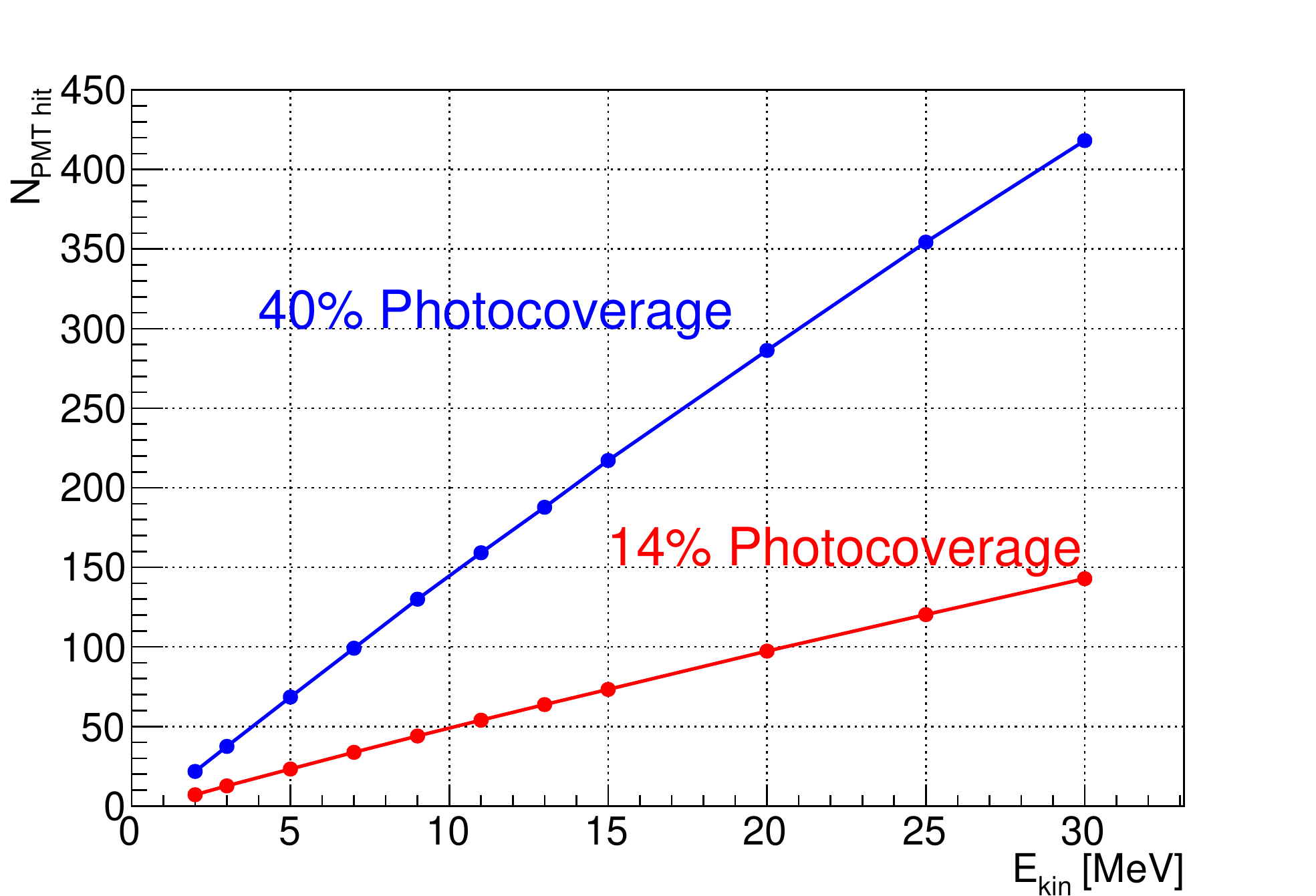}
   %\end{minipage} 
&
  % \begin{minipage}[t]{0.55\hsize}
    \centering
    \includegraphics[scale=0.47]{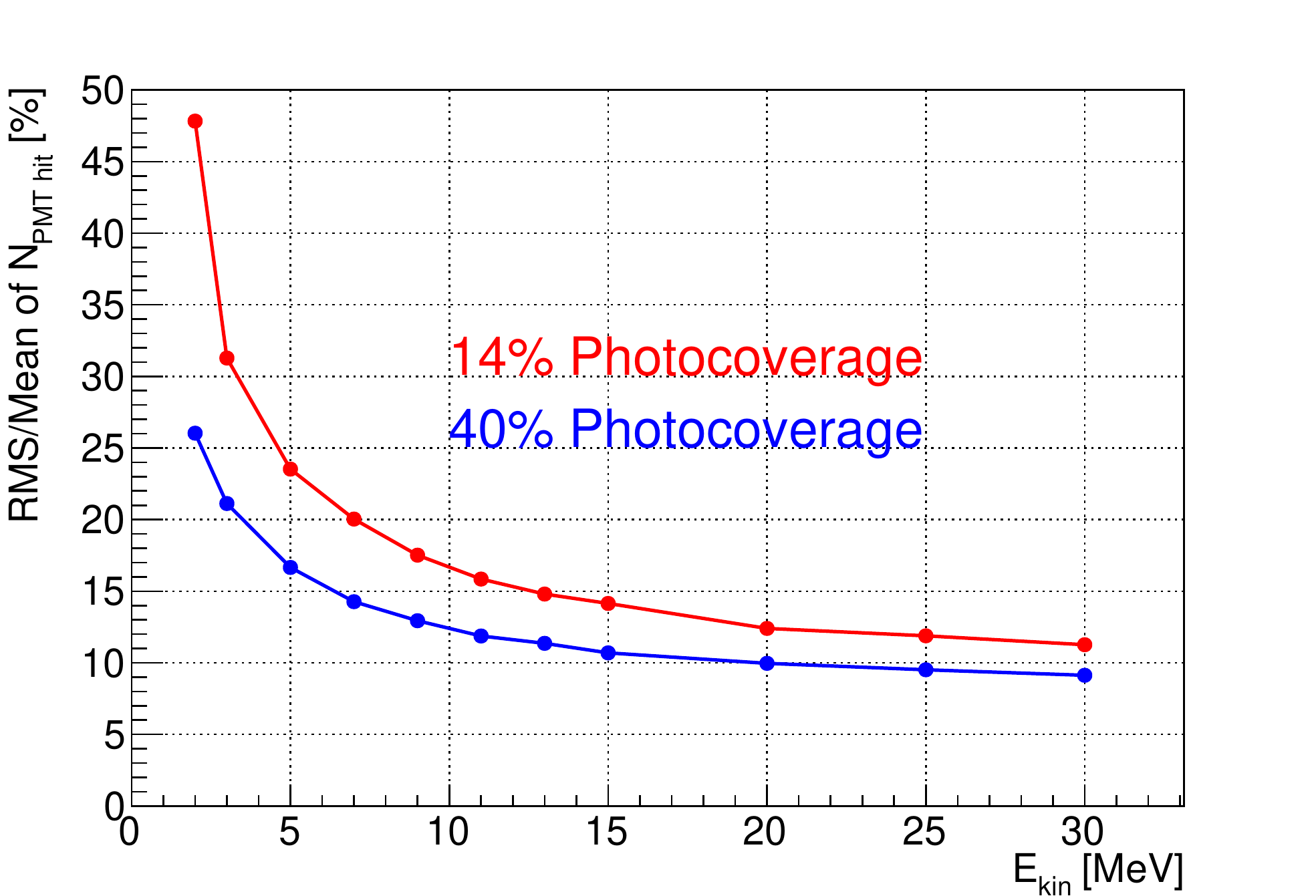}
   %\end{minipage}
  \end{tabular}
  \caption{ Expected number of PMT hits (N$_{PMT hits}$) and the RMS of
    the N$_{PMT hit}$ distributions. 
WCsim is used for simulating the injection of electrons with several
values of kinetic energy (E$_{kin}$).
The initial position is uniformly distributed inside the fiducial
volume ($>$2~m from inner detector wall). The red line corresponds to
the 14\% photocoverage configuration, while the blue line corresponds to
the configuration with 40\% photocoverage.}
  \label{fig:nhits}
 \end{figure}

Figure~\ref{fig:total-charge} shows how the flexibility of WCSim can
be used to explore different detector configurations. Here, the total
charge distribution for electrons and muons at several momenta in
the Hyper-Kamiokande detector with two different photocoverage options
are shown. RMS divided by mean charge
is plotted in Figure~\ref{fig:rms-mom} indicating better resolution
with 40\% photocoverage than with
14\% photocoverage. For lower energy particles, the resolution can be
approximated using nhits (the number of phototubes that register a
hit). The nhit distribution for both 14\% photocoverage and
40 \% photocoverage are shown in Figure~\ref{fig:nhits}. 

%\begin{center}
%\end{center}
%
%\begin{center}
%\end{center}
%
%\begin{center}
%\end{center}

\subsection{FiTQun} FiTQun is an event reconstruction
package initially developed for the Super-K detector based on the
formalism used by the MiniBooNE experiment
\cite{Patterson:2009ki}. The reconstruction algorithm allows for
single- and multiple-ring event hypotheses to be tested against
observed data. For a given event hypothesis, a prediction is made for
the complete set of observables at each PMT. This includes whether or
 not the PMT was hit, and for hit PMTs, the hit time and integrated 
charge. The hypothesis and associated kinematic parameters that best 
describe a given event are found by maximizing a likelihood function 
of the prediction with respect to the observed data.

\begin{figure}[htbp]
  \begin{center}
    \includegraphics[width=0.6\textwidth]{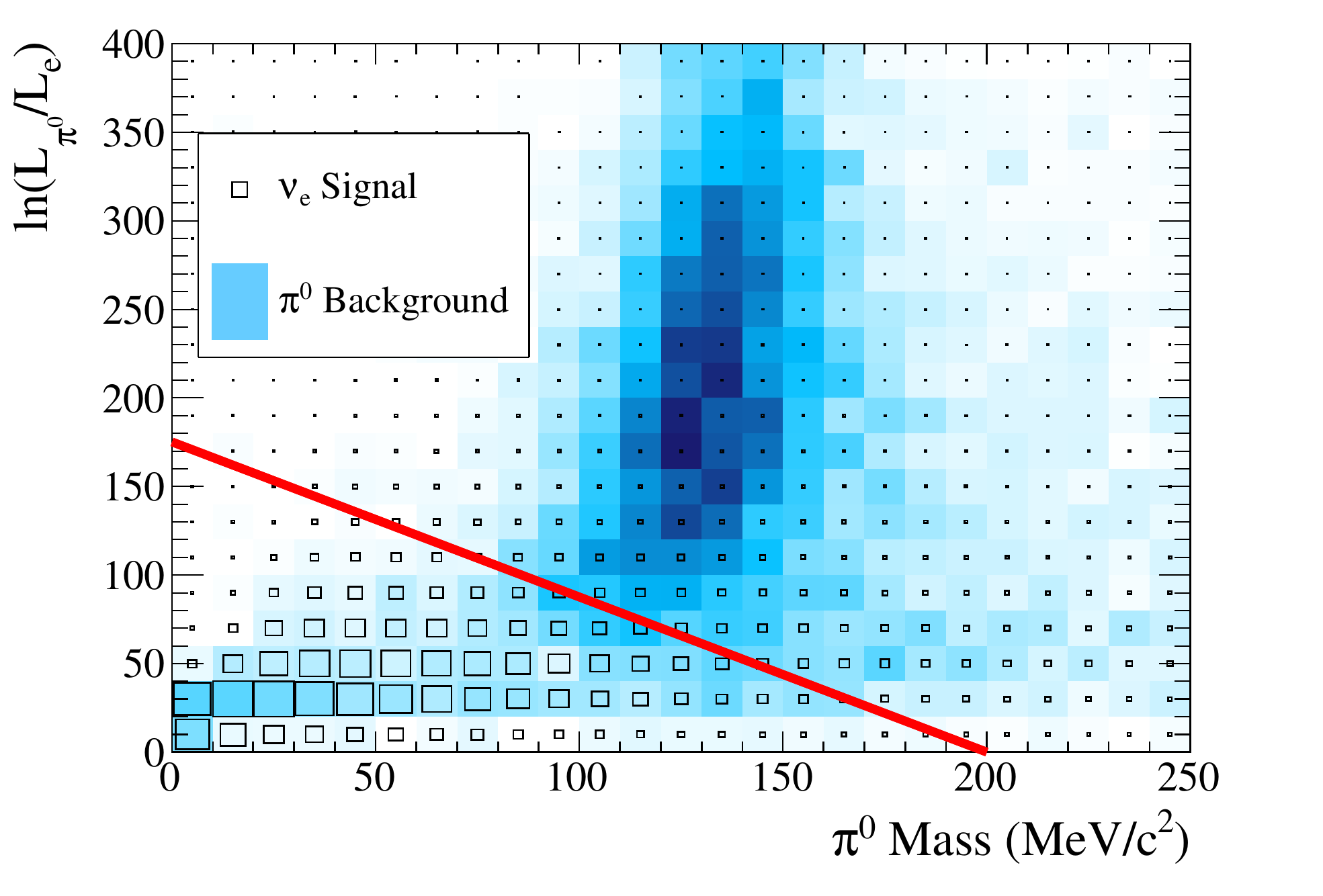}
  \end{center}
  \caption {FiTQun used for $\pi^0$ rejection in the $\nu_{e}$
    selection for the T2K $\nu_{e}$ appearance
    analysis\cite{Abe:2013hdq}. The red line represents the cut, with
    events above the line being rejected as $\pi^0$ background.}
  \label{fig:pi0plot}
\end{figure}

FiTQun has been shown to perform well on Super-K data, with
significant improvements on vertex, angle and momentum resolutions, as
well as particle identification when compared to previous
reconstruction algorithms. In particular, fiTQun was successfully
deployed to reject $\pi^0$ events from the Super-K $\nu_{e}$ sample in
the T2K $\nu_{e}$ appearance analysis (Figure~\ref{fig:pi0plot}) \cite{Abe:2013hdq,Abe:2015awa}.

\subsubsection{Reconstruction algorithm} 
At the core of fiTQun lies a likelihood function, which is evaluated
over all the PMTs in the detector:
\begin{equation} 
L \left( {\bf x} \right)= \prod_j^{unhit} P_j\left(unhit | {\bf x} \right) \prod_i^{ ihit} P_i \left(hit | {\bf x}\right) f_q\left(q_i|{\bf x}\right) f_t \left(t_i | {\bf x} \right) \,.
\end{equation} 

Event hypotheses are characterized by ${\bf x}$, which
includes the time and position of the interaction vertex, momentum and
direction of the charged particle tracks, and any other relevant
kinematic parameters such as the distance or time interval between
tracks, or the energy lost between track segments. For a given ${\bf
x}$, a prediction of the amount of charge at each PMT, $\mu_i$, is
made and the time at which the light is expected to arrive each PMT is
calculated.

The detector response is folded into these predictions to give the
probabilities $P$ of a PMT being hit and the read-out time and charge
probability distribution functions $f_t$ and $f_q$, respectively. The
negative log-likelihood ($-log\left( L \right)$) is maximised to
obtain the ${\bf x}$ that best describe the event according to some
event topology (\emph{e.g.}, single electron-like ring).

Once the best-fit parameters have been obtained for several
topologies, the ratio between their likelihoods is used to determine
which topology gives the best match to the event. This can be used as
a particle identification tool if simple one-particle hypotheses are
used, or as a more complex selection criterion if multiple final-state
particles are included in the hypothesis (\emph{e.g.}, a nuclear
de-excitation photon followed by a $K^+$ decay muon for the selection
of $p\rightarrow K^+ \nu$ events).

\subsubsection{Integration with WCSim and tuning} FiTQun has been
adapted to reconstruct events simulated with WCSim, in the various
detector configurations implemented in the simulation software. A
C++ class was written (WCSimWrapper) that reads in both detector 
geometry (positions and radius of PMTs) and event data from the WCSim
ROOT output files. A preprocessor flag allows fiTQun to be compiled 
against WCSim libraries, removing its dependence on Super-K software.

In the context of WCSim, events can be generated in an arbitrary
number of detector configurations. For fiTQun to adequately
reconstruct events in any given configuration, some of its components
have to be re-evaluated. For example, the charge and time response of
PMTs must be accurately known in order to obtain unbiased estimates of
particle momentum and vertex position. The tuning procedure developed
for Super-K and SKDETSIM was adapted to be used with WCSim and with
generalized cylindrical geometries.

The tunes produced for each simulated detector consist of ROOT files
and run-time parameters. A configuration file (that contains the 
information needed by fiTQun to load the appropriate files and 
parameters) is given for each tune. These configuration files are 
packaged with fiTQun, such that any tune available can be selected
in a single step.

\subsection{BONSAI} 
For event reconstruction at low
energy, i.e. few MeV - few tens MeV, a reconstruction algorithm BONSAI
(Branch Optimization Navigating Successive Annealing Iterations) is
supplied for Hyper-Kamiokande.  BONSAI was originally developed for
Super-Kamiokande \cite{icrc0213smy} and written in C++. It has been used for the low
energy physics analysis of SK-I to SK-IV. In the low energy region,
most of the photosensor signals are single photon hits.  BONSAI uses
this relative hit time information to reconstruct the position of the
Cherenkov light source, i.e. the position of low energy event.
For Hyper-K analysis, a wrapper library (libWCsimBonsai) is supplied for ROOT environment.

\subsubsection{Vertex reconstruction} 
For the vertex reconstruction, BONSAI performs a maximum likelihood fit using the
photosensor hit timing residuals.  This likelihood fit is done for the
Cherenkov signal as well as the dark noise background for each vertex
hypothesis.  The likelihood of the selected hypothesis is compared
to the likelihood of a hypothesis in an area nearby.  Highly ranked
hypotheses and new points in the likelihood will survive this step.
Finally, after several iterations, the hypothesis with the largest
likelihood is chosen as the reconstructed vertex.

\begin{figure}[htbp]
  \begin{center}
    \includegraphics[width=0.6\textwidth]{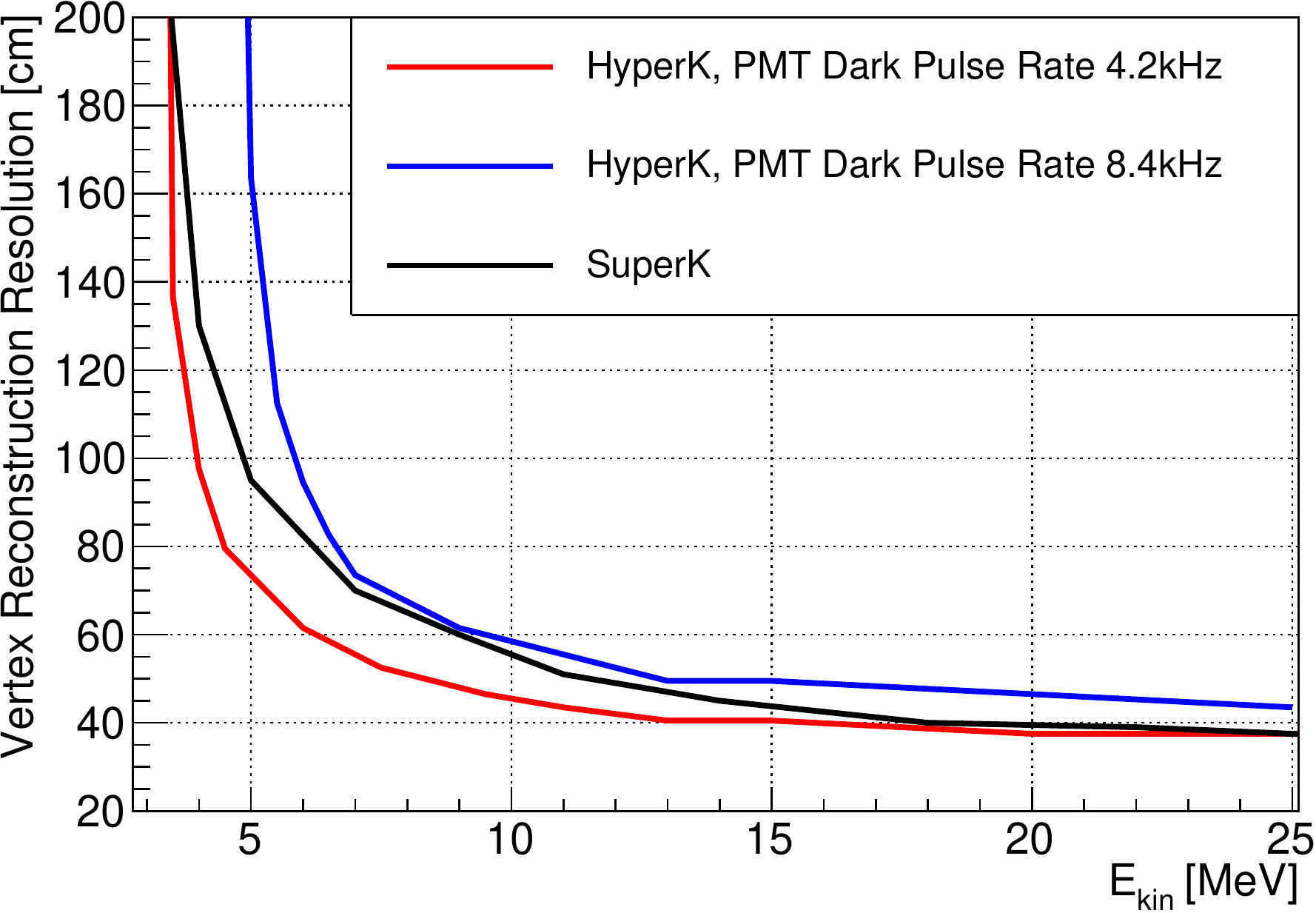}
  \end{center}
  \caption {
	  Vertex reconstruction resolution for electrons with BONSAI for Hyper-K and Super-K detectors.
	  Here, WCSim is used for Hyper-K detector simulation.
	  Red line shows the resolution with the PMT dark pulse rate of 8.4\,kHz, as seen in \ref{section:photosensors}.
	  Blue is for the case of PMT dark pulse rate of 4.2\,kHz, which is same as the rate of Super-K photosensors.
	  Black line shows the performance with Super-K detector, simulated with SKDETSIM.
  \label{fig:bonsaivtx}
    }
\end{figure}

The vertex goodness criterion testing the time residual distribution
is defined as follows:\\
\begin{equation} g
	(\vec{v})=\sum_{i=1}^{N}w_{i}\exp{-0.5(t_i-|\vec{x_i}-\vec{v}|/c_{wat})/\sigma)^2}
\end{equation} where $t_i$ are the measured PMT hit times, $\vec{x_i}$
the photosensor locations, $\vec{v}$ is reconstructed event vertex,
$\sigma$ is the effective timing resolution expected for Cherenkov events (total of photosensor and DAQ resolution).
$c_{wat}$ is the group speed of light in the water, i.e. $c/n$ with the speed of light in vacuum $c$ and refractive index $n$.
$\omega_i$ are Gaussian hit weights also based on the hit time residuals, but with a much wider effective time resolution.
The weights reduces the dark noise contamination of the Cherenkov light.
A result of vertex reconstruction performance study with BONSAI and WCSim can be found in the figure \ref{fig:bonsaivtx}.
More Cherenkov photons could be detected with new photosensors for Hyper-K and it improves the reconstruction results, comparing to those of Super-K
Though, at same time, the random photosensor signals caused by their dark pulse can spoil the merit.
So reducing dark pulse is a crucial factor to improve the low energy event detection.
Many efforts for the dark pulse reduction (\ref{section:photosensors:IDperformance:BG}) and improvements of the softwares are being continued.

\subsubsection{Energy and direction reconstruction}
BONSAI and its related subroutines also determine the energy and the
event direction reconstruction.
Because most of the photosensor signals consist of single photon hits at low energy below few tens MeV,
the total number of photosensor hits is the leading parameter for
reconstructing the energy of events.  First, time-of-flight values are
subtracted from each of the hit timing values based on the position of
each photosensor and the result of the BONSAI vertex reconstruction.
Next, the number of photosensor hits around the expected event timing
is calculated, considering its cross-section and the local
photocoverage with neighboring photosensors.  Finally, the number of
hits are scaled to energy using the information from detector
simulations and calibrations.\\
The direction reconstruction is also performed on the photosensor hit
patterns using a circular KS test that checks the azimuthal symmetry
around the Cherenkov cone.
As the result, the vertex position, direction and energy of low-energy
events are available after BONSAI reconstruction.

Several likelihoods to test mis-reconstruction are also available
during the reconstruction.  Likelihoods calculated using photosensor
hit patterns are also used in particle identification, e.g. to
differentiate between electron and gamma events.

\clearpage
\graphicspath{{background/figures}}

\section{Background rate estimation} \label{section:background}

\subsection{Background rate estimation for low energy neutrino study \label{sec:lowe_bg}}

In this subsection, we will show the background rate estimation used
for the study of low energy neutrinos, such as solar neutrinos,
supernova neutrinos, and relic-supernova neutrinos. The most important
background sources are the radioactive isotope of $^{222}$Rn contained
in water, and radioactive spallation products created by cosmic-ray
muons. Other important radioactive isotopes
are U and Th in the water, and $^{40}$K in the PMT glass; we need to
suppress the concentration of those isotopes to the similar level in
Super-K as well. On the other hand, the major background sources for
the anti-neutrino measurement is anti-neutrino backgrounds from
nuclear power reactors, which limit the energy threshold at around
10\,MeV; the smaller contribution comes from the radioactive
background of $(\alpha, n)$ reactions. In the background calculation,
the most complicated task is the estimation of the muon spallation
productions and its background reduction by the analyses, because it
depends on the detector location and the detector performance. In the
following paragraphs, after discussing radon backgrounds,
we focus on the discussion of the muon spallation backgrounds.

\subsubsection{Radon background}

Radon ($^{222}$Rn) is a radioactive noble gas, with a half-life of 3.8 days.  
$^{222}$Rn occurs as a daughter nuclide in the $^{238}$U decay scheme,
via the decay of $^{226}$Ra ($\tau_{1/2} = 1599$ years).
Small but finite quantities of $^{226}$Ra exist in all materials and
therefore, every material can produce $^{222}$Rn.
As a gaseous isotope, $^{222}$Rn can easily escape from materials used
in the construction of Hyper-K and the radioactivity content
of construction materials must be carefully screened.
The decay of $^{222}$Rn produces several daughter isotopes,
most of which are not sufficiently energetic to produce Cherenkov
light in the Hyper-K detector.  
The most serious background for solar neutrino measurements
is the radon daughter bismuth-214 ($^{214}$Bi)
which decays via beta emission with a Q-value of 3.27 MeV.
This limits the energy threshold of the solar neutrino
measurements in which a neutrino-electron elastic scattering reaction
is used.

In the same energy region, $^{208}$Tl in thorium series could become
another serious source of the background.
However, from the results of radon assay with special radon
detectors~\cite{70Lradon,700Lradon} in Super-K, 
the contamination of the radon in thorium series ($^{220}$Rn)
looks much smaller than that of $^{222}$Rn in Super-K water.
So, we discuss only $^{222}$Rn in this section. 

\begin {figure}[htbp]
  \begin{center}
   \includegraphics[width=1.00\textwidth]{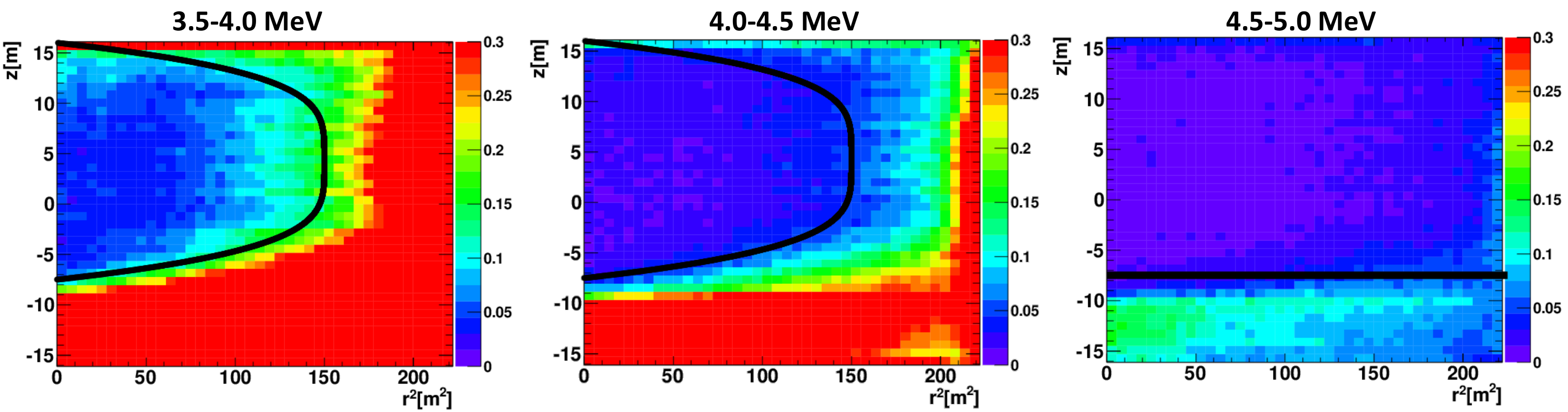}
   \caption{
   Vertex distribution of the SK-IV 2645-day final data sample in
   solar neutrino analysis in different electron kinetic energy regions.
   The black lines indicate fiducial volume region in each energy range.
   Whole area corresponds to 22.5 kton volume,
   and the fiducial volume above 5.0 MeV is 22.5 kton.
   The color shows the event rate (/day/bin). The R and Z correspond to
   the detector horizontal and vertical axis, respectively.}
   \label{fig:sk4-vertex}
  \end{center}
\end {figure}

Typical radon concentrations in Mozumi mine and Super-K are
summarized in Table~\ref{tab:radon_concentrations}.
\begin{table}[htbp]
 \caption{\label{tab:radon_concentrations}
 Typical radon concentrations in Mozumi mine and in
 Super-K.}
\begin{ruledtabular}
\begin{tabular}{lc}
Detector site & Radon concentration \\
\hline
Mine air in Mozumi~\cite{Takeuchi:1999}                      & $\sim1200$ [Bq/m$^3$]\\
SK-I inner detector water (upper half)~\cite{Takeuchi:1999}  & $<1.4$ [mBq/m$^3$]\\
SK-I inner detector water (bottom)~\cite{Takeuchi:1999}      & $3 \sim 5$ [mBq/m$^3$]\\
SK-IV supply water            & $< \sim 1$ [mBq/m$^3$]\\
SK-IV return water            & $8 \sim 10$ [mBq/m$^3$]\\
\hline
\end{tabular}
\end{ruledtabular}
\end{table}
The SK-IV return water is taken at $\sim 18$ ton/hour from inner detector
and at $\sim 42$ ton/hour from outer detector~\cite{Abe:2013gga}.
So, we think the radon concentration in the SK-IV outer detector
is close to (or larger than) that in the SK-IV return water.
From the event rate comparison between SK-IV and SK-I, the radon concentration
in SK-IV inner detector is similar or less than that of SK-I.
Therefore, the radon concentration would be different by several factors
between inner and outer detectors in SK-IV.
However, in Super-K detector, water flow is controlled well by temperature and flow
rate balances~\cite{Abe:2013gga}, and then the water condition in inner
detector has been stable.
Figure~\ref{fig:sk4-vertex} shows a vertex distribution of the final
data sample of the solar neutrino analysis in SK-IV.
The remaining event rate in the central detector is kept low
while the barrel and bottom of the detector are higher than that.
This shows the water flow control in Super-K detector works well.
For the solar neutrino measurements, a threshold of 4.5 MeV (electron
kinetic energy) in 22.5 kton fiducial volume was achieved in Super-K I.

In this design report, we estimate the radon concentration in
the Hyper-K tank water would be about 1.6\,${\rm mBq}/{\rm m}^{3}$
in average in whole detector, as described in section~\ref{sec:radon-in-water}. 
When applying the same water flow control technique as Super-K,
the radon concentration in the central Hyper-K detector will be reduced.

\begin {figure}[htbp]
  \begin{center}
   \includegraphics[width=0.70\textwidth]{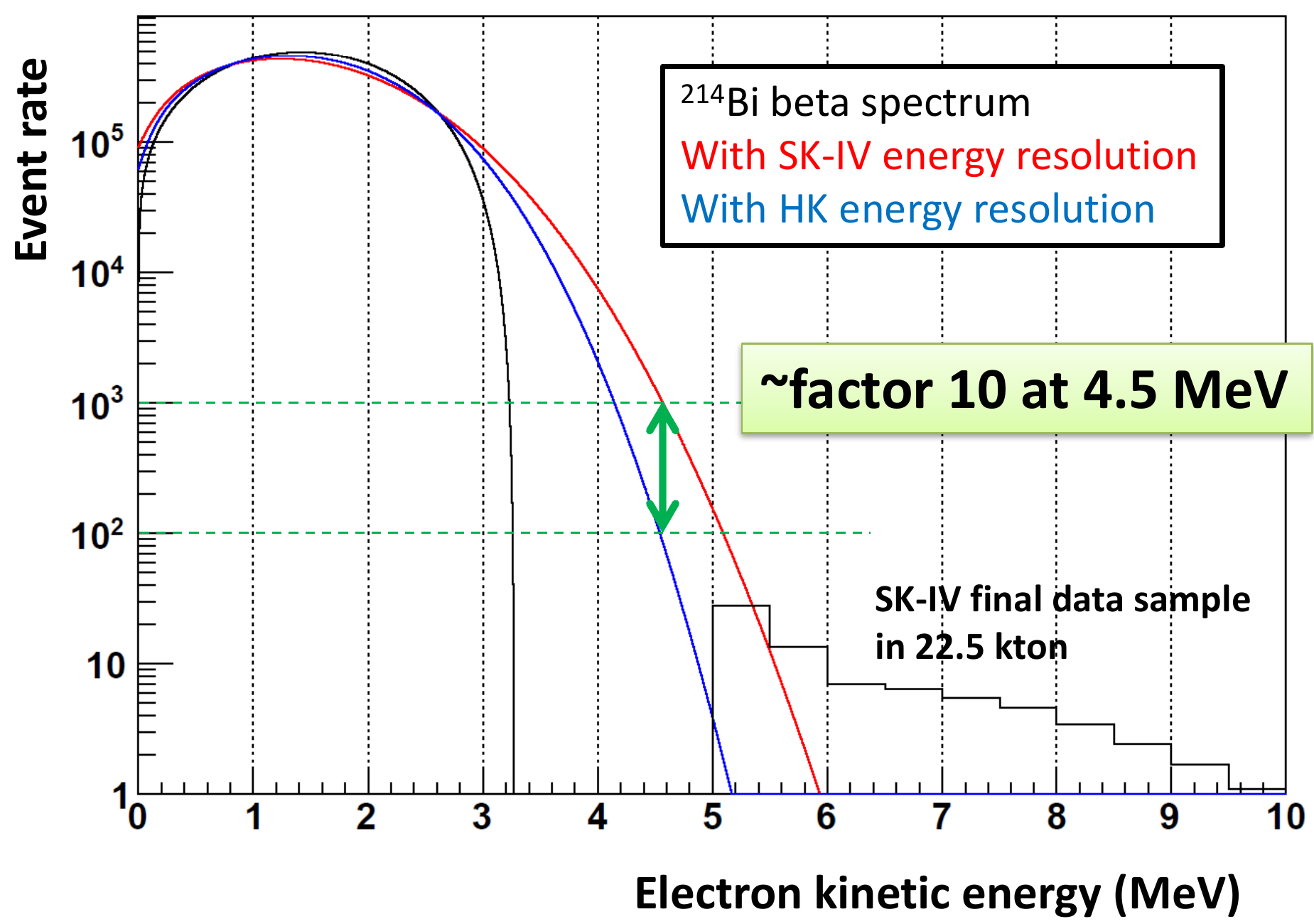}
   \caption{An estimation of expected Bi-214 energy spectrum.
   The black, red and blue lines are the original beta decay spectrum,
   the expected observed spectrum in SK-IV,
   and expected observed spectrum in Hyper-K, respectively.
   The black histogram is observed event rate of the SK-IV final data
   sample in 22.5 kton fiducial volume.
   The vertical axis is event rate in arbitrary unit.}
   \label{fig:bi214}
  \end{center}
\end {figure}

Moreover, we could suppress $^{214}$Bi events thanks to better
photon yield in Hyper-K. 
Figure~\ref{fig:bi214} shows expected observed energy spectrum of
beta decay of $^{214}$Bi in Super-K and Hyper-K.
At 4.5 MeV (electron kinetic energy), a suppression about factor 10 is expected
in Hyper-K detector comparing to Super-K detector. 
Actually, we have observed an expected energy resolution difference
due to photon yield change between SK-I and SK-II.

Another possible difference among Super-K and Hyper-K detectors related
to radon background would be the lining material of the detector.
As discussed in section~\ref{section:tank-liner}, 5 mm thickness HDPE
will be used to line the Hyper-K detector, and radon could permeate a HDPE sheet.
Typical radon permeability through a HDPE sheet 
is reported by various groups as $O(10^{-8}) \sim O(10^{-7})$
cm$^2$/s~\cite{radon1,radon2,radon3,radon4,radon5}. 

As shown in Table~\ref{tab:radon_concentrations}, the typical radon
concentrations are different by about 5 order of magnitude
between mine air and Super-K OD water.
For Hyper-K, radon concentration in OD water through the
radon permeation from outside detector into Hyper-K detector
could be estimated as $O(10)$ mBq/m$^3$ in Hyper-K OD water
under several assumptions, like
(1) a radon permeability of the Hyper-K HDPE sheet is $10^{-8}$ cm$^2$/s,
(2) radon concentration in mine water (spring water in the mine) is $10^{3}$ Bq/m$^3$,
(3) there is no water flow between Hyper-K ID and Hyper-K OD, and
(4) the volumes of mine water and Hyper-K OD water contributing to this effect are similar.
This estimation gives a similar radon concentration in Super-K OD detector,
though the uncertainty is large.

\begin {figure}[htbp]
  \begin{center}
   \includegraphics[width=0.70\textwidth]{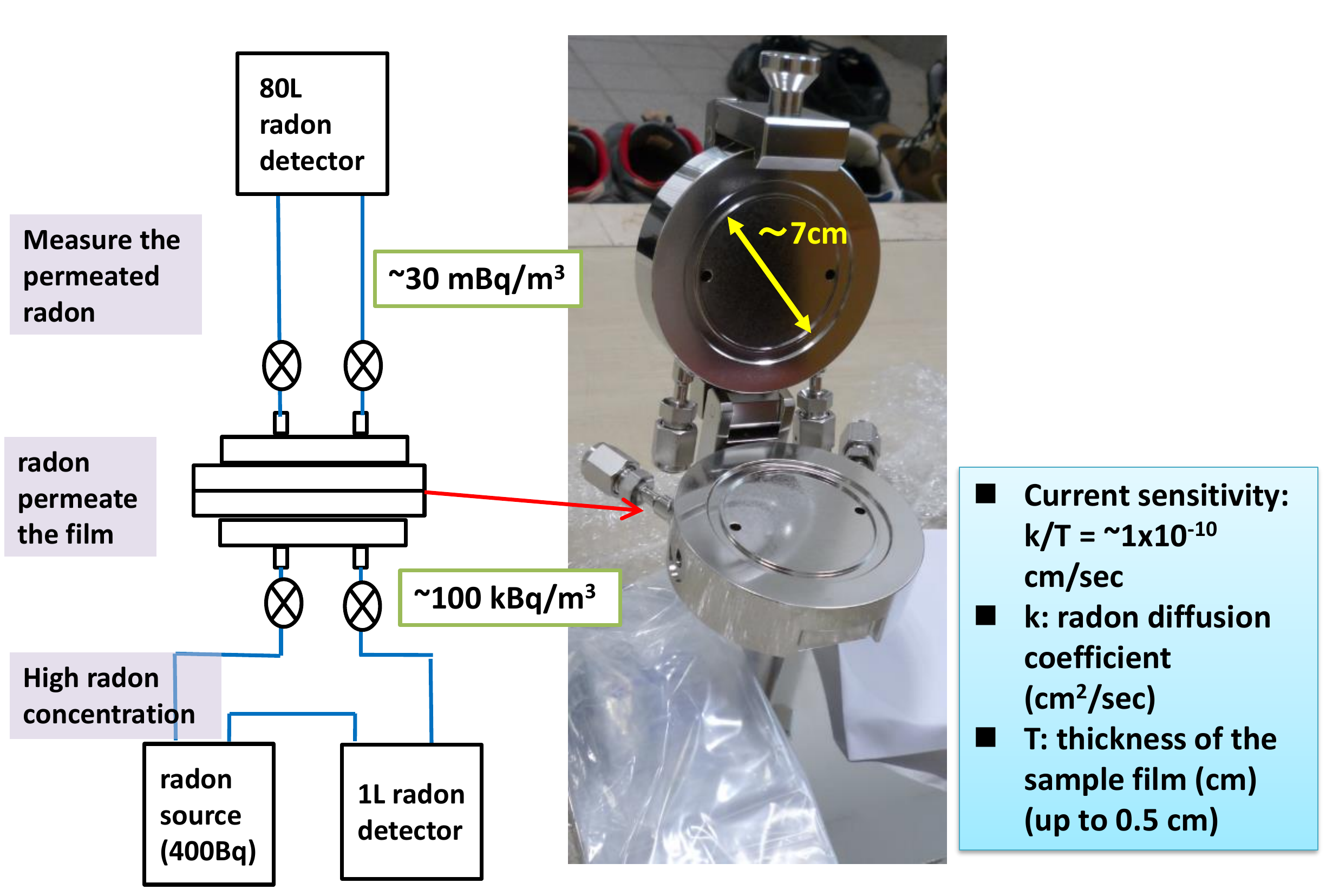}
   \caption{A system to measure radon permeation of a sheet in Kamioka.}
   \label{fig:radon-permeation}
  \end{center}
\end {figure}

In order to reduce the uncertainty of (1) in this estimation, we are planning to
measure radon permeability of a Hyper-K HDPE sheet.
Figure~\ref{fig:radon-permeation} shows a device to assay radon
permeability through a sheet in Kamioka.
The performance of the device is still under tuning, but
the current sensitivity on radon permeation assay is about $O(10^{-9})$ cm$^2$/s
for a 1 mm thickness sheet. Therefore, this device has enough
sensitivity to test the Hyper-K HDPE sheet.

In order to achieve (3) in Hyper-K, we are designing more hermetic ID
detector to reduce water flow between ID and OD.
We also apply the same water flow control techniques realized in Super-K
to keep radon around detector wall in Hyper-K ID.

As a summary, applying the same radon reduction techniques developed in
Super-K, a similar radon concentration in Hyper-K inner detector water is expected.
The radon permeation through lining material might increase, but an
initial estimation shows a similar level of the radon concentration in
outer detector water in Hyper-K. In order to avoid increase of radon in inner
detector, we are planning more hermetic inner detector.
We are going to measure radon permeation of our liner candidates, too.
Moreover, about factor 10 tolerance is expected at 4.5 MeV
thanks to improvement of the energy resolution.
Therefore, we think 4.5 MeV (electron kinetic energy)
threshold for solar neutrino measurements
would be feasible in the Hyper-K detector.

\subsubsection{Muon spallation}
	        
Radioactive isotopes produced by cosmic-ray muon-induced spallation
are potential backgrounds for low energy neutrinos. Generally, the
production rate depends strongly on the muon flux and the average
energy, and the delayed radioactive decays cause the backgrounds in
the energy region below about 20\,MeV. If the lifetime of radioactive
isotope is relatively short on the order of a few seconds or less, the
spallation backgrounds can be mitigated by time/volume cuts based on
the reconstructed muon track. Therefore, the detailed estimation of
the cosmic-ray muon intensity and the spallation production rate are
of great importance in demonstrating the sensitivity of Hyper-K to low
energy neutrinos.

\begin {figure}[htbp]
  \begin{center}
    \includegraphics[width=0.70\textwidth]{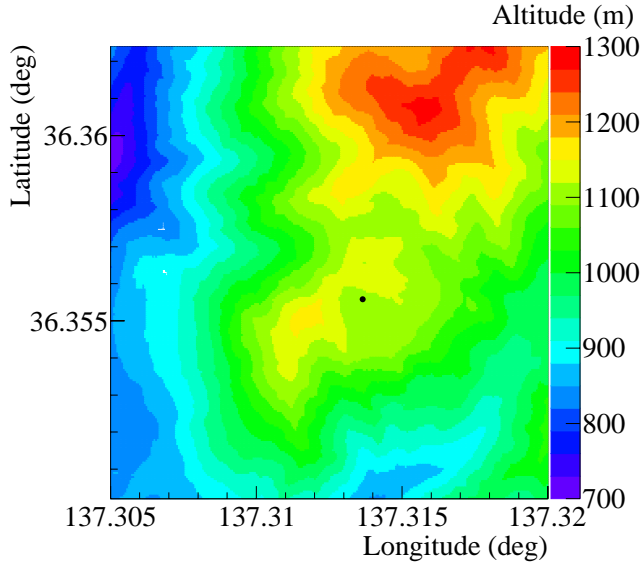}
    \caption{Topological profile of Nijuugo-yama~\cite{GeographicalSurvey:2010}. The black point is the basing point for the Hyper-K site.}
    \label{fig:profile_nijuugoyama}
  \end{center}
\end {figure}

The muon intensity at the planned site can be estimated using the
calculated surface muon flux and energy, the mountain profile, the
rock density and compositions. The muon flux ($J_{\mu}$) and average
energy ($\overline{E}_{\mu}$) at underground sites are estimated by
the muon simulation code (\texttt{MUSIC})~\cite{Antonioli:1997}, a
three-dimensional MC tool dedicated to muon transportation in
matter. In this MC, surface muons are generated according to
the \textit{Modified Gaisser Parameterization}~\cite{Tang:2006}
sea-level muon flux distribution. A digital map of the topological
profile of Nijuugo-yama with a 5\,m mesh
resolution~\cite{GeographicalSurvey:2010} is shown in
Fig.~\ref{fig:profile_nijuugoyama}. The Hyper-K detector will be
located around the basing point at the altitude of 508\,m, referenced in
Section~\ref{section:location} corresponds to a position under the old mountain peak
before the surface mining. Based on this elevation data, we calculate
slant depths as a function of zenith and azimuth angle at an arbitrary
point of Hyper-K candidate sites, and estimate the survival probability of muons
after the muon transportation through the rock for each angle using
the \texttt{MUSIC} simulation.

\begin {figure}[htbp]
  \begin{center}
    \includegraphics[width=0.48\textwidth]{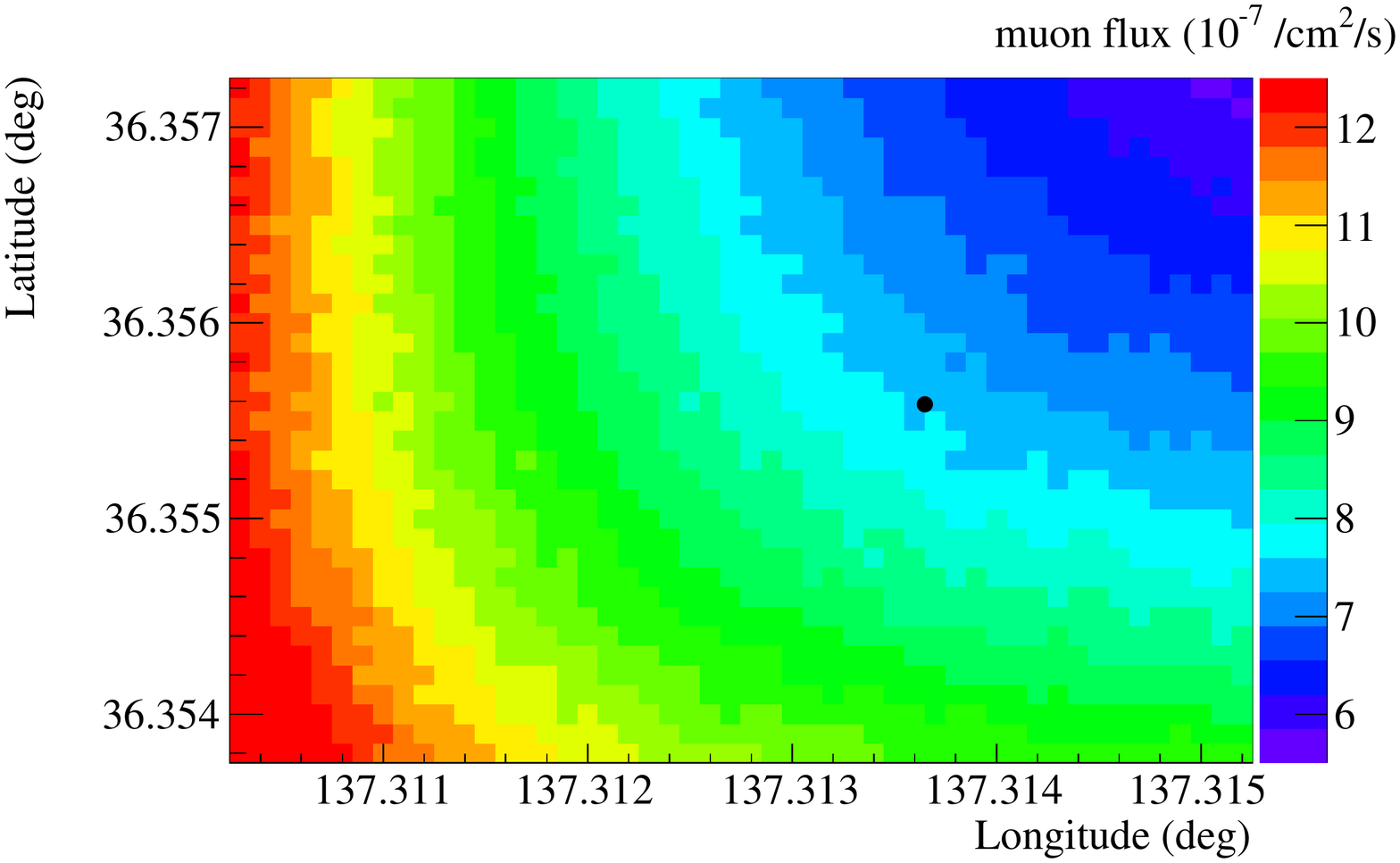}
    \includegraphics[width=0.48\textwidth]{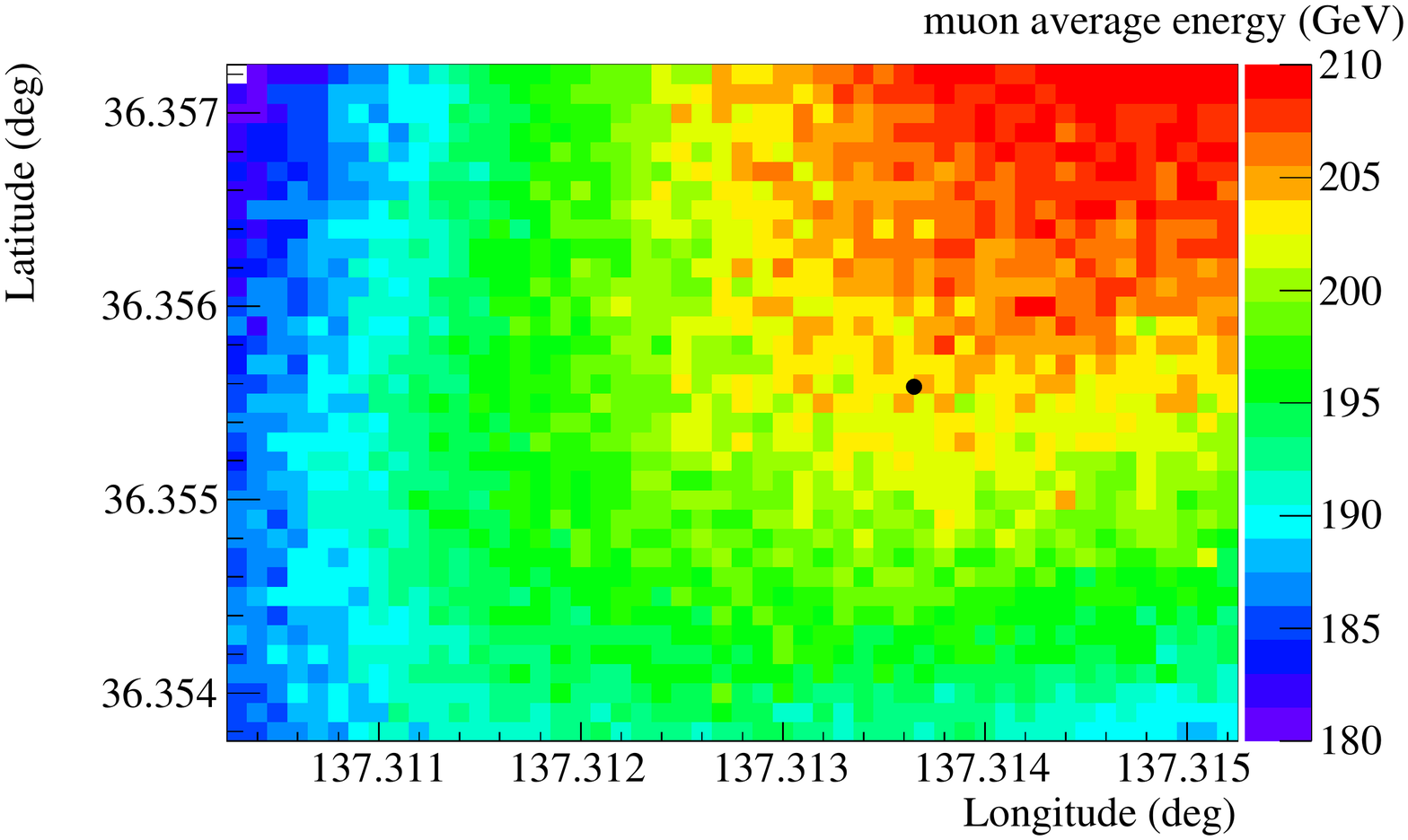}
    \caption{Calculated muon flux ($J_{\mu}$) and average energy ($\overline{E}_{\mu}$) at the altitude of the Hyper-K (508\,m) detector. The black point is the basing point for the Hyper-K site.}
    \label{fig:muon_flux_and_energy}
  \end{center}
\end {figure}

Previously, the value of $J_{\mu}$ and $\overline{E}_{\mu}$ at KamLAND
in Ikeno-yama are evaluated based on the \texttt{MUSIC} simulation for
various rock types~\cite{Tang:2006,Abe:2010}. The value of $J_{\mu}$
is dependent on the type of rock. Varying the specific gravity of rock
from 2.65 to 2.75\,${\rm g}/{\rm cm}^{3}$ in the \texttt{MUSIC}
simulation yields values of $J_{\mu}$ that agree with the KamLAND muon
flux measurement~\cite{Abe:2010}. As both Nijuugo-yama and Ikeno-yama
are skarn deposit, which are common characteristic in the Kamioka
mine, the simulation for Hyper-K assumes the same rock type used in
Ref.~\cite{Abe:2010}. Figure~\ref{fig:muon_flux_and_energy} shows the
calculated $J_{\mu}$ and $\overline{E}_{\mu}$ for Hyper-K at the
altitude of 508\,m for 2.70\,${\rm g}/{\rm cm}^{3}$ specific gravity
Ikeno-yama rock. The values of $J_{\mu}$ and $\overline{E}_{\mu}$ vary
greatly depending on the shallowest rock thickness on the west or
south side, as indicated in Fig.~\ref{fig:profile_nijuugoyama}.

\begin {figure}[htbp]
  \begin{center}
    \includegraphics[width=0.60\textwidth]{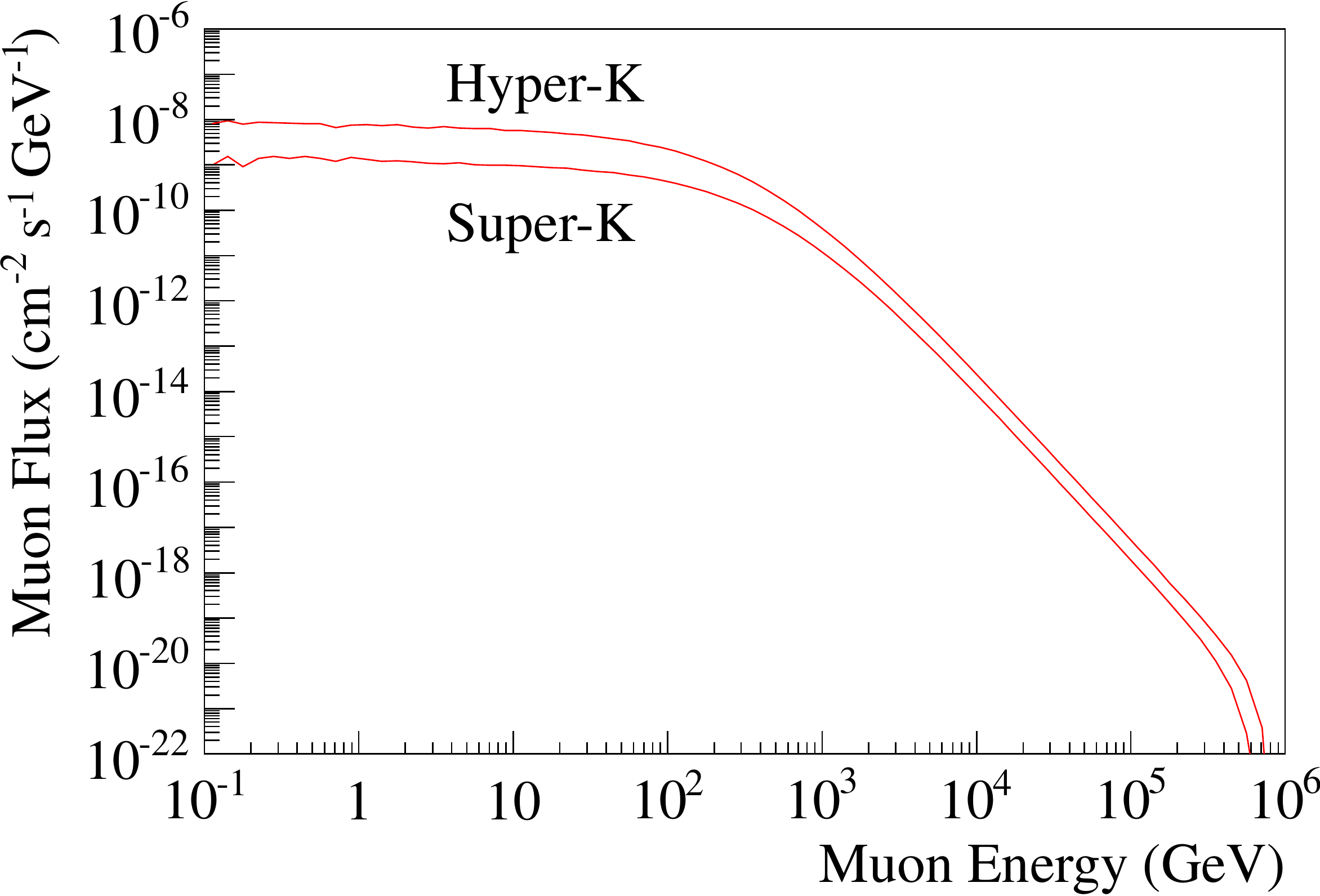}
    \caption{Calculated muon energy spectra for Super-K and Hyper-K at the basing point based on the \texttt{MUSIC} simulation.}
    \label{fig:muon_flux_energy}
  \end{center}
\end {figure}

\begin {figure}[htbp]
  \begin{center}
    \includegraphics[angle=270,width=0.60\textwidth]{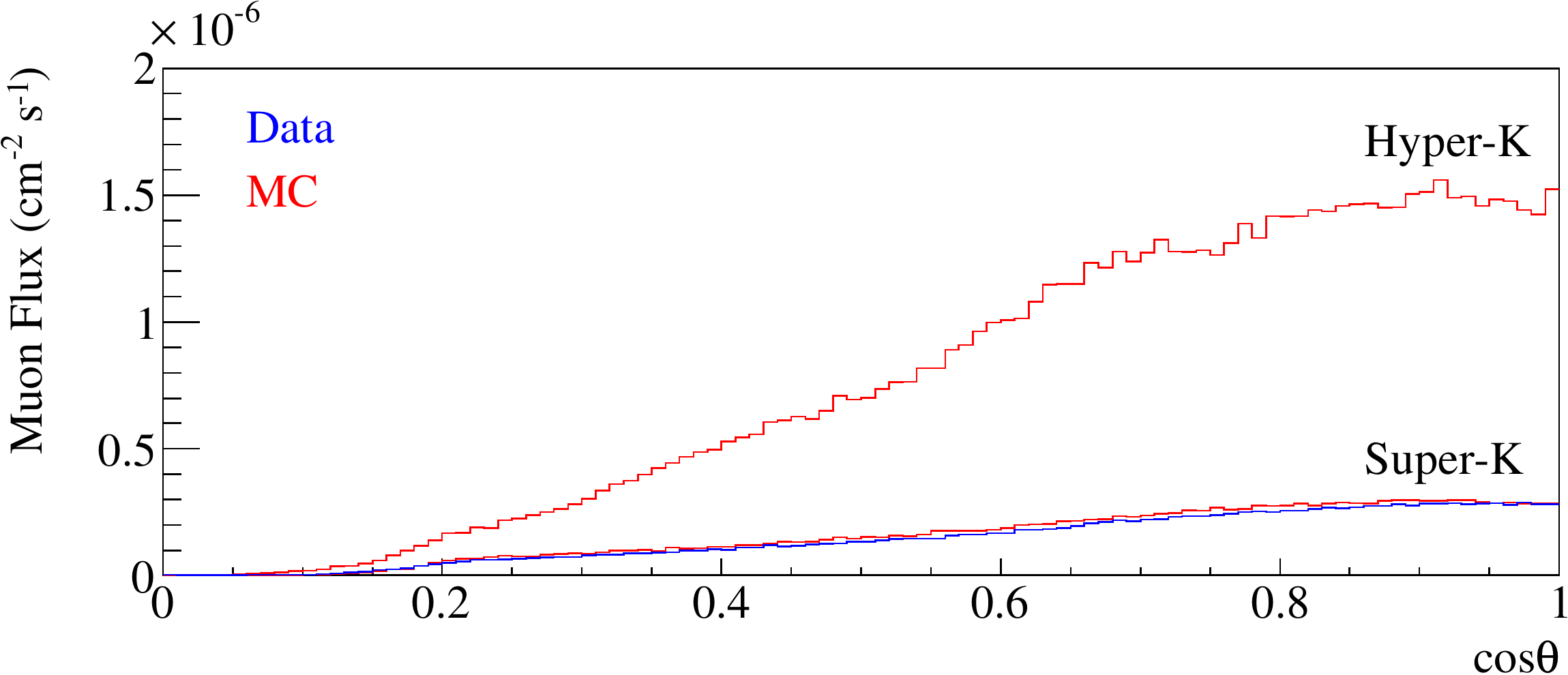}
    \includegraphics[angle=270,width=0.60\textwidth]{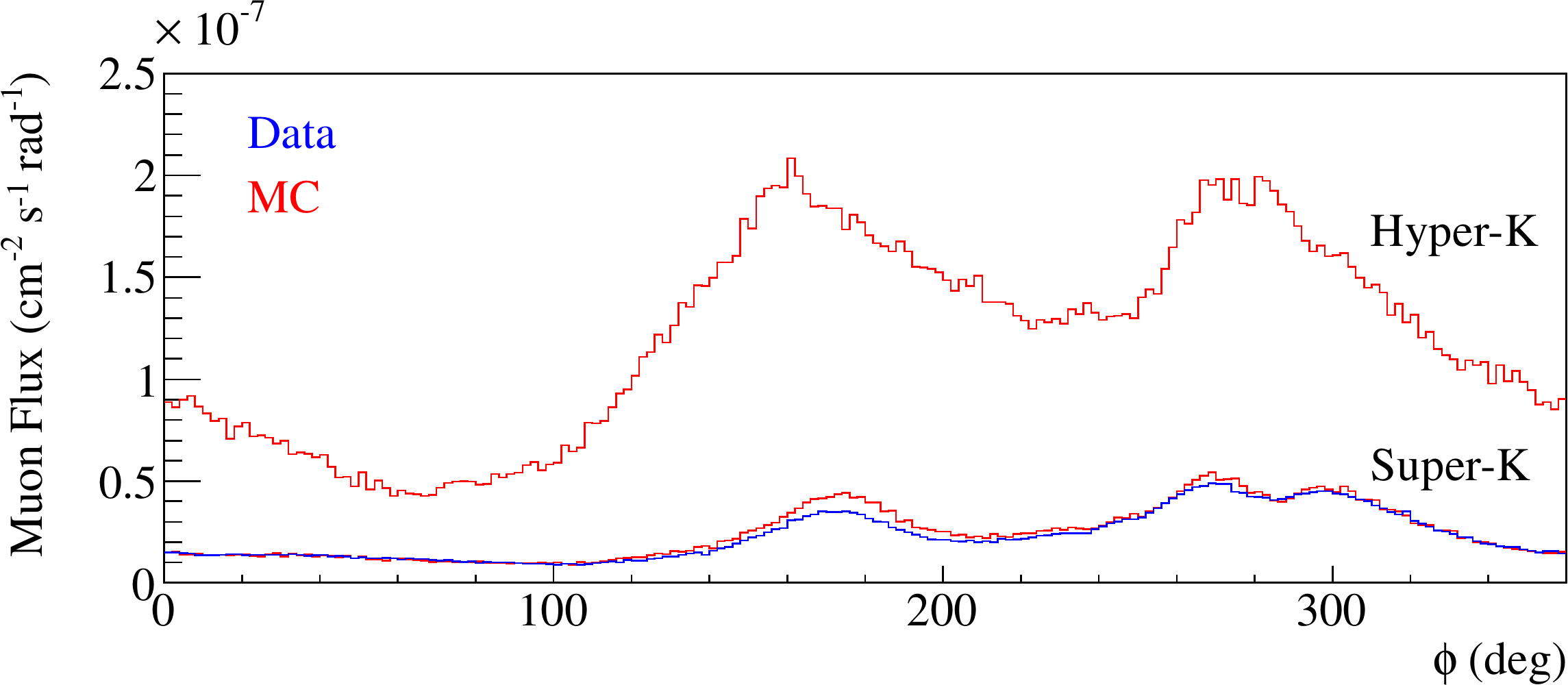}
    \caption{Muon flux as a function of zenith angle $\theta$ (upper) and azimuth angle $\phi$ (lower) for Super-K and Hyper-K at the basing point. The east corresponds to the azimuth angle of zero degree. The blue lines show the data for Super-K, and the red lines show the MC predictions for Super-K and Hyper-K based on the \texttt{MUSIC} simulation. The absolute flux and the shape of the Super-K data, which are determined by slant depths for each angle, are well reproduced by MC.}
    \label{fig:muon_flux_angle}
  \end{center}
\end {figure}

Table~\ref{tab:muon_simulation} summarizes the calculated $J_{\mu}$
and $\overline{E}_{\mu}$ in Hyper-K and Super-K for 2.70\,${\rm
g}/{\rm cm}^{3}$ specific gravity. Considering the variation of
$J_{\mu}$ for different rock types, we assume uncertainties of
$\pm$20\% for $J_{\mu}$. Because the Super-K site is deeper, the value
of $J_{\mu}$ for Hyper-K is higher than Super-K by a factor of 4.9. On
the other hand, the value of $\overline{E}_{\mu}$ for Hyper-K is
smaller than Super-K as indicated in Fig.~\ref{fig:muon_flux_energy},
because the relative contribution of lower energy muons becomes larger
at a shallower site. Figure~\ref{fig:muon_flux_angle} shows the muon
flux as a function of zenith angle $\theta$ (upper) and azimuth angle
$\phi$ (lower) for Super-K and Hyper-K at the basing point. We
confirmed that the \texttt{MUSIC} Monte Carlo simulation has been shown to be in good agreement with the Super-K data, as shown in Figure~\ref{fig:muon_flux_angle}. In Hyper-K, the major contribution of muon flux is introduced by
the flux in the west and the south.

\begin{table}[t]
\caption{\label{tab:muon_simulation}Calculated muon flux ($J_{\mu}$) and average energy ($\overline{E}_{\mu}$) in Hyper-K and Super-K for 2.70\,${\rm g}/{\rm cm}^{3}$ specific gravity Ikeno-yama rock based on the simulation method~\cite{Tang:2006}. The basing point in Hyper-K is illustrated in Fig.~\ref{fig:profile_nijuugoyama}.}
\begin{ruledtabular}
\begin{tabular}{lccc}
Detector site & Vertical depth (m) & $J_{\mu}$ ($10^{-7}\,{\rm cm}^{-2} {\rm s}^{-1}$) & $\overline{E}_{\mu}$ (GeV) \\
\hline
Hyper-K (basing point) & 600 & 7.55 & 203 \\
Super-K & 1,000 & 1.54 & 258 \\
\end{tabular}
\end{ruledtabular}
\end{table}

Based on the muon flux and energy spectrum calculated by
the \texttt{MUSIC} simulation, we can estimate the isotope production
rates by muon spallation in a planned detector. \texttt{FLUKA} can be 
used to reliably model nuclear and particle physics processes
involved in muon spallation. Previously, the measured isotope
production rates in a underground detector were compared with
the \texttt{FLUKA} simulation~\cite{Abe:2010}. However, owing to
large uncertainties on the isotope production cross sections by muons
or their secondaries, the production rate between data and MC differ
by up to a factor of two, as shown in Table V of
Ref.~\cite{Abe:2010}. In order to minimize the uncertainties, we use
the isotope production rates observed in Super-K as a basis, and the
values in Hyper-K are estimated based on the muon flux ratio
calculated by \texttt{MUSIC} and the isotope yield ratio
by \texttt{FLUKA},
\begin{equation}
R_{i} ({\rm Hyper\mathchar`-K}) = R_{i} ({\rm Super\mathchar`-K}) \times \frac{J_{\mu} ({\rm Hyper\mathchar`-K})}{J_{\mu} ({\rm Super\mathchar`-K})} \times \frac{Y_{i}({\rm Hyper\mathchar`-K})}{Y_{i} ({\rm Super\mathchar`-K})}
\end{equation}
where $R_{i}$ is the production rate per unit volume for isotope $i$,
$J_{\mu}$ the muon flux (${\rm cm}^{-2} {\rm s}^{-1}$), $Y_{i}$ the
yield per muon track length ($/\mu/{\rm m}$) for isotope
$i$. We use \texttt{FLUKA} version 2011.2b to estimate the isotope
yields in Hyper-K and Super-K. A water-filled volume of 40-m square
and 40-m length is used in the simulation, and the analysis of isotope
productions is limited within the inner volume of 40-m square and 20-m
length in order to avoid a boundary effect. To include the muon charge
and energy dependence in isotope production yields, beams of both
$\mu^{+}$ and $\mu^{-}$ with a calculated energy spectrum produced
by \texttt{MUSIC} were simulated, and the isotope yields by $\mu^{+}$
and $\mu^{-}$ are combined based on their weighted average assuming
that a relative intensity of $\mu^{+}$ to $\mu^{-}$ is
1.3. Table~\ref{tab:isotope_yield_estimation} shows the estimation of
isotope production yields for Hyper-K and Super-K, the ratio of
isotope yields, $Y_{i} ({\rm Hyper\mathchar`-K}) / Y_{i} ({\rm
Super\mathchar`-K})$. The ratio of the production rates, $R_{i} ({\rm
Hyper\mathchar`-K}) / R_{i} ({\rm Super\mathchar`-K})$, are also
calculated by multiplying the isotope yield ratio by the muon flux
ratio, $J_{\mu} ({\rm Hyper\mathchar`-K}) / J_{\mu} ({\rm
Super\mathchar`-K}) = 4.9 \pm 1.0$, which was evaluated from
the \texttt{MUSIC} simulation. We assume uncertainties of $\pm$20\%
for the muon flux ratio considering the possibility of different rock
types for the Hyper-K and Super-K sites. The resulting increase in
isotope production rate per unit volume from Super-K is approximately
a factor of $4 \pm 1$ in Hyper-K, which is used for studies of the
Hyper-K physics potential in the following sections.

\begin{table}[t]
\caption{\label{tab:isotope_yield_estimation}Estimation of isotope production yields for Hyper-K and Super-K by muon spallation with \texttt{FLUKA}. The ratio of the production yields for Hyper-K compared with Super-K are also listed. The ratio of the production rates are calculated by multiplying the isotope yield ratio by the muon flux ratio of $4.9 \pm 1.0$, evaluated by the \texttt{MUSIC} simulation.}
\begin{ruledtabular}
\begin{tabular}{lcccc}
 & \multicolumn{2}{c}{Isotope yield by \texttt{FLUKA} ($\mu$/m)} &Ratio of isotope yield & Ratio of production rate \\
\raisebox{1.5ex}[1.5ex][0.75ex]{Isotope} & Hyper-K & Super-K & (Hyper-K / Super-K) & (Hyper-K / Super-K) \\
\hline
$^{12}$B & $8.05 \times 10^{-5}$ & $9.93 \times 10^{-5}$ & $0.811 \pm 0.078$ & $3.98 \pm 0.88$ \\
$^{12}$N & $8.70 \times 10^{-6}$ & $1.11 \times 10^{-5}$ & $0.785 \pm 0.075$ & $3.84 \pm 0.85$ \\
$^{9}$Li & $1.23 \times 10^{-5}$ & $1.68 \times 10^{-5}$ & $0.732 \pm 0.070$ & $3.59 \pm 0.80$ \\
$^{8}$Li & $8.67 \times 10^{-5}$ & $1.08 \times 10^{-4}$ & $0.805 \pm 0.077$ & $3.95 \pm 0.87$ \\
$^{15}$C & $5.12 \times 10^{-6}$ & $6.68 \times 10^{-6}$ & $0.768 \pm 0.073$ & $3.76 \pm 0.83$ \\
$^{16}$N & $2.74 \times 10^{-4}$ & $3.41 \times 10^{-4}$ & $0.804 \pm 0.077$ & $3.94 \pm 0.87$ \\
$^{11}$Be & $5.32 \times 10^{-6}$ & $7.76 \times 10^{-6}$ & $0.685 \pm 0.065$ & $3.36 \pm 0.74$ \\
\end{tabular}
\end{ruledtabular}
\end{table}

\subsubsection{Muon spallation background reduction}

The spallation products as backgrounds have their origin in the
spallation reaction of the cosmic muons.  Therefore, we can identify
and remove these spallation products if we compare their
four-dimensional correlation with corresponding muons.  In this
section, we will discuss the muon spallation backgrounds using
these correlations.  The spallation reduction method is being used in
supernova relic neutrino searches at Super-K.

\paragraph{Method of muon spallation background reduction\\}

We apply a likelihood ratio test on low-energy events to reduce
spallation backgrounds.  The detail is discussed below.\\ With water
Cherenkov detectors, such as Super-K or Hyper-K, we
can measure the times, positions and energies of the spallation
products and their preceding muon tracks.  It is also possible to
measure the energy deposit per unit length $dE/dx$ along muon tracks,
by deconvoluting Cherenkov ring hits.  The peak position of $dE/dx$
distribution can be assumed as the position where muon spallation
occurs.\\

We use following valiables for spallation background reduction. 
\begin{itemize}
	\item Time difference $\delta t$, between the low-energy event and the preceding muon.
	\item Transverse distance $l_{trans}$, which is defined as the perpendicular distance from the muon track to the low-energy event.
	\item Longitudinal distance $l_{long}$, which is defined as the horizontal distance from the peak position of $dE/dx$ on reconstructed muon track and the low-energy event.
	\item Residual charge $Q_{peak}$, which is the amount of light seen in $the$ dE/dx distribution in a width of 4.5 m centered on the peak.
\end{itemize}

\begin {figure}[htbp]
\begin{center}
	\includegraphics[width=0.35\textwidth]{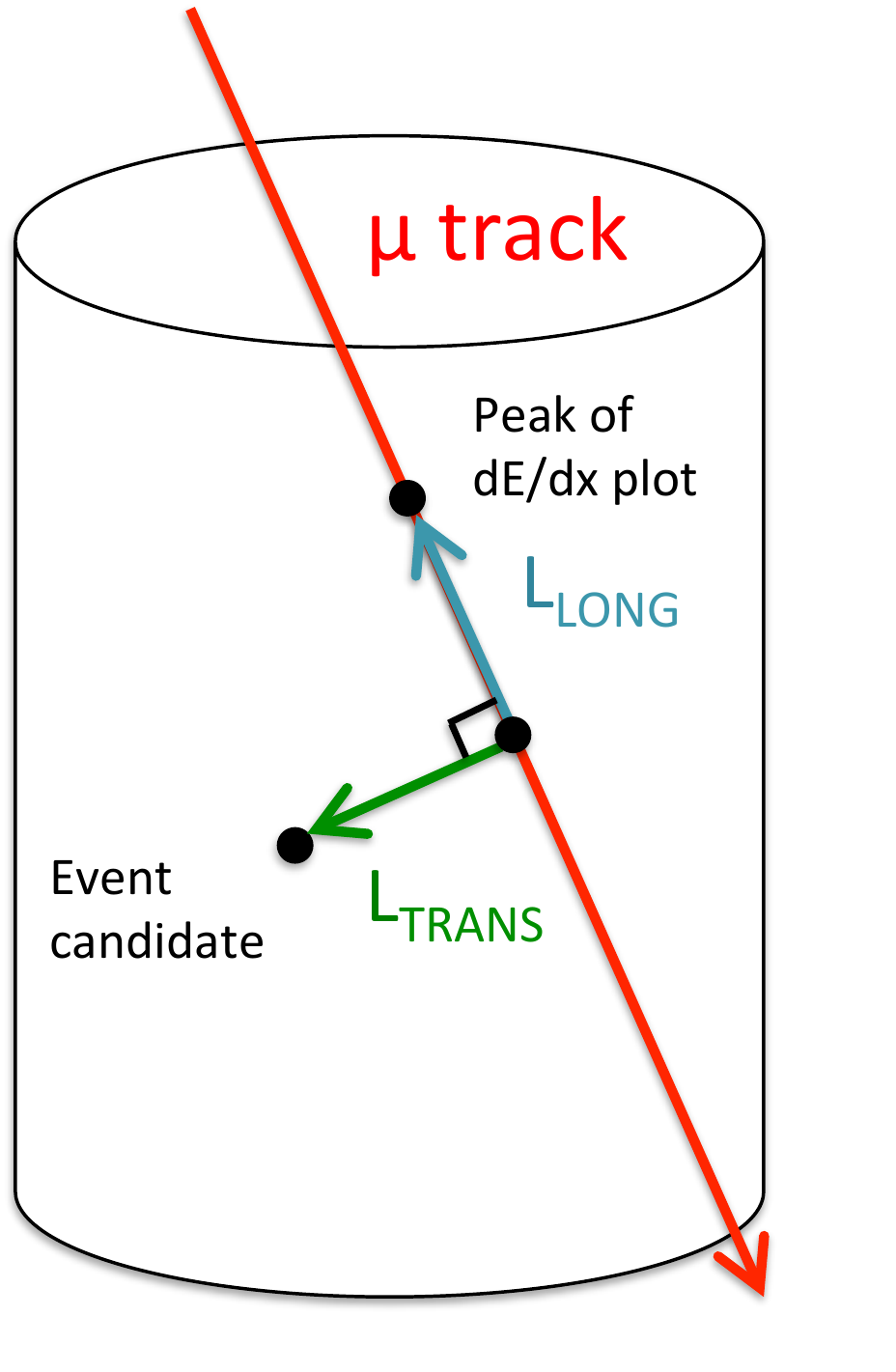}
	\caption{
		Schematic figure for showing spallation distance variables\cite{KirkRyanDrThesis}.
	}
	\label{fig:relic_spacut}
\end{center}
\end {figure}

A schematic figure of spallation distance variables are shown in Figure~\ref{fig:relic_spacut}.
Their actual distributions can be found in a reference \cite{KirkRyanDrThesis}.
The probability density functions $PDF_{spa}$ and $PDF_{rand}$ of each valuables are given from the fitting results of actual spallation candidates and non-spallation (random) event samples, respectively.
The likelihood ratio $\Lambda$ is defined using PDFs as follows:
\begin{equation}
\begin{split}
	\Lambda = -2\log{\frac{PDF_{spa}(Q_{peak}) \times PDF_{spa}(\delta t) \times PDF_{spa}(L_{trans}) \times PDF_{spa}(L_{long})}
	{PDF_{rand}(Q_{peak}) \times PDF_{rand}(\delta t) \times PDF_{rand}(L_{trans}) \times PDF_{rand}(L_{long})}}
\end{split}.
\end{equation}
Because any preceding cosmic muon can be a cause of spallation backgrounds, we calculate the likelihood ratio $\Lambda$ with all muons within 30\,seconds before the low energy event.
The largest likelihood ratio $\Lambda$ is adopted as the $\Lambda$ value for the low energy event.
We can arbitrarily choose the cut value for the likelihood ratio, which defines the reduction efficiency and the signal efficiency.  

\paragraph{Estimated muon spallation background after reduction\\}\label{par:spallation_reduction}
When we have more cosmic muon flux, the number of preceding muons
that are randomly paired with a low-energy event will be increased.  As a result, we
will have more chance to have ``more spallation like'' $\Lambda$ value
for a low-energy event even if it is a non-spallation event.  On the
other hand the likelihood ratio is not changed for real spallation
events, because they will be paired to their mother muons regardless
of the number of preceding muons.  Consequently we will have worse
separation between the likelihood distribution of spallation
backgrounds and that of non-spallation events with more cosmic muon flux.
Here we studied the spallation reduction efficiency and the signal efficiency in Hyper-K, based on the data of SK-II.
The spallation reduction efficiency is defined as the rate of spallation events that survived the likelihood cut.
The likelihood distribution of
non-spallation event sample is also made from the data, pairing the
low energy events with muons which were detected 300$\sim$330\,s before these events.
The effect of shallower Hyper-K location on the spallation reduction method is studied by increasing the proceding muons with random muon data.
\\
In following, we will show two cases of spallation cuts.
One is defined to keep the signal efficiency of 80\%, which is applied for usual analysis, e.g. solar neutrino analysis.
To estimate the performances of this cut, those are signal efficiency and background reduction efficiency,
we use the real events of SK-II between 17.5 and 20\,MeV.
We assume the same signal efficiency and the same reduction efficiency below 17.5\,MeV.

As the result, the spallation reduction efficiency ($\epsilon_{reduction}$) will be 1.2\% and 3.9\% for the same and 5 times larger amount of the
cosmic muons for this criteria, respectively.
Finally, the ratio of remaining spallation events in the solar neutrino analysis is calculated as follows:
\begin{equation}
	R_{spallation}({\rm Hyper\mathchar`-K/}{\rm Super\mathchar`-K}) =  \frac{R_{production}({\rm Hyper\mathchar`-K})}
	{R_{production}({\rm Super\mathchar`-K})} \times
	\frac{\epsilon_{reduction}({\rm Hyper\mathchar`-K})}
	{\epsilon_{reduction}({\rm Super\mathchar`-K})}.
	\label{eq:spallation_production_rate}
\end{equation}
Here, $\epsilon_{reduction}({\rm Hyper\mathchar`-K})$ is found to be 3.9\% for Hyper-K at Tochibora-site as discussed above.
${\epsilon_{reduction}({\rm Super\mathchar`-K})}$ of $\sim$6\% is taken form SK-II solar neutrino analysis.
$R_{production}({\rm Hyper\mathchar`-K})$ and
    $R_{production}({\rm Super\mathchar`-K})$ are the rate of spallation isotope production per unit volume in Hyper-K and Super-K respectively.
	Referring to the result of the former section, ${R_{production}({\rm Hyper\mathchar`-K})}/{R_{production}({\rm Super\mathchar`-K})}$ is assumed to be $4\pm1$.
	Since, we conclude the ratio of remaining spallation events of Hyper-K to Super-K is $R_{spallation}({\rm Hyper\mathchar`-K/}{\rm Super\mathchar`-K})=2.7$.\par
More strict spallation cut is applied for very low background analysis, e.g. supernova relic neutrino search.
In this case, the cut value is defined to remove the spallation backgrounds to the level of less than 1 event left between 17.5 and 20\,MeV or between 20 and 26 MeV.
So, the signal efficiency will be affected by the increased amount of muons.
The signal efficiency will be 79\% (29\%) for the energy range of 17.5$\sim$20~MeV and 90\% (54\%) for 20$\sim$26~MeV, for the same (5 times larger) amount of the cosmic muons.
Because the amount of spallation backgrounds decreases exponentially at the higher energies, no spallation background is expected above 26\,MeV.
The results are shown in Table~\ref{tab:spacut_solar} and Table~\ref{tab:spacut_relic}.

\begin{table}
\begin{center}
	\caption{The expected spallation background reduction efficiency for solar neutrino analysis.
	The signal efficiency of 80\% is kept for the event selection.}
\label{tab:spacut_solar}
\begin{tabular}{lcccc}
\hline \hline
Cosmic muons rate, comparing to Super-K   &&  $\times 1$ & $\times 5$ (Tochibora site) \\
\hline 
Signal Efficiency                                         &&  80\%        &     80\%                    \\
Spallation Reduction Efficiency                           &&  1.2\%       &     3.9\%                   \\
\hline \hline 
\end{tabular}
\end{center}
\end{table}

\begin{table}
\begin{center}
	\caption{The expected spallation background reduction efficiency for supernova relic neutrino searches.
	The spallation background is reduced to the level of less than 1 event for the each energy range.}
\label{tab:spacut_relic}
\begin{tabular}{lcccc}
\hline \hline
Cosmic muons rate, comparing to Super-K   &&  $\times 1$ & $\times 5$ (Tochibora site) \\
\hline 
Signal Efficiency ( - 20\,MeV)                         &&  79\%            &     29\%                    \\
Signal Efficiency (20   - 26\,MeV)                         &&  90\%            &     54\%                    \\
\hline \hline 
\end{tabular}
\end{center}
\end{table}

\subsection{Neutron background estimation for atmospheric neutrino/proton decay study  \label{sec:neutron_bg}}

This subsection will discuss the possible cosmic-ray
backgrounds for the atmospheric neutrino and proton decay analyses,
which visible energy is greater than 30~MeV.  In this energy range the
spallation background caused by cosmic muons, which is described in
Section~\ref{sec:lowe_bg}, can be neglected.

In the Super-K detector case, thanks to the double structure of the
inner and outer detector, cosmic muons entering the detector can be
easily rejected by looking at hit clusters around the entering and
exiting points of muons.  According to Super-K's experience, the
estimated background of the cosmic muons are negligible ($\sim$0.1\%
in the final atmospheric neutrino fully-contained sample).
Considering that Hyper-K design is basically same structure, similar
level of the background rejection performance for cosmic muons by the
outer detector is expected even if the cosmic muon rate is increased
by several factor due to the shallower site of Hyper-K.

\begin{figure}[htb]
\includegraphics[width=0.6\textwidth]{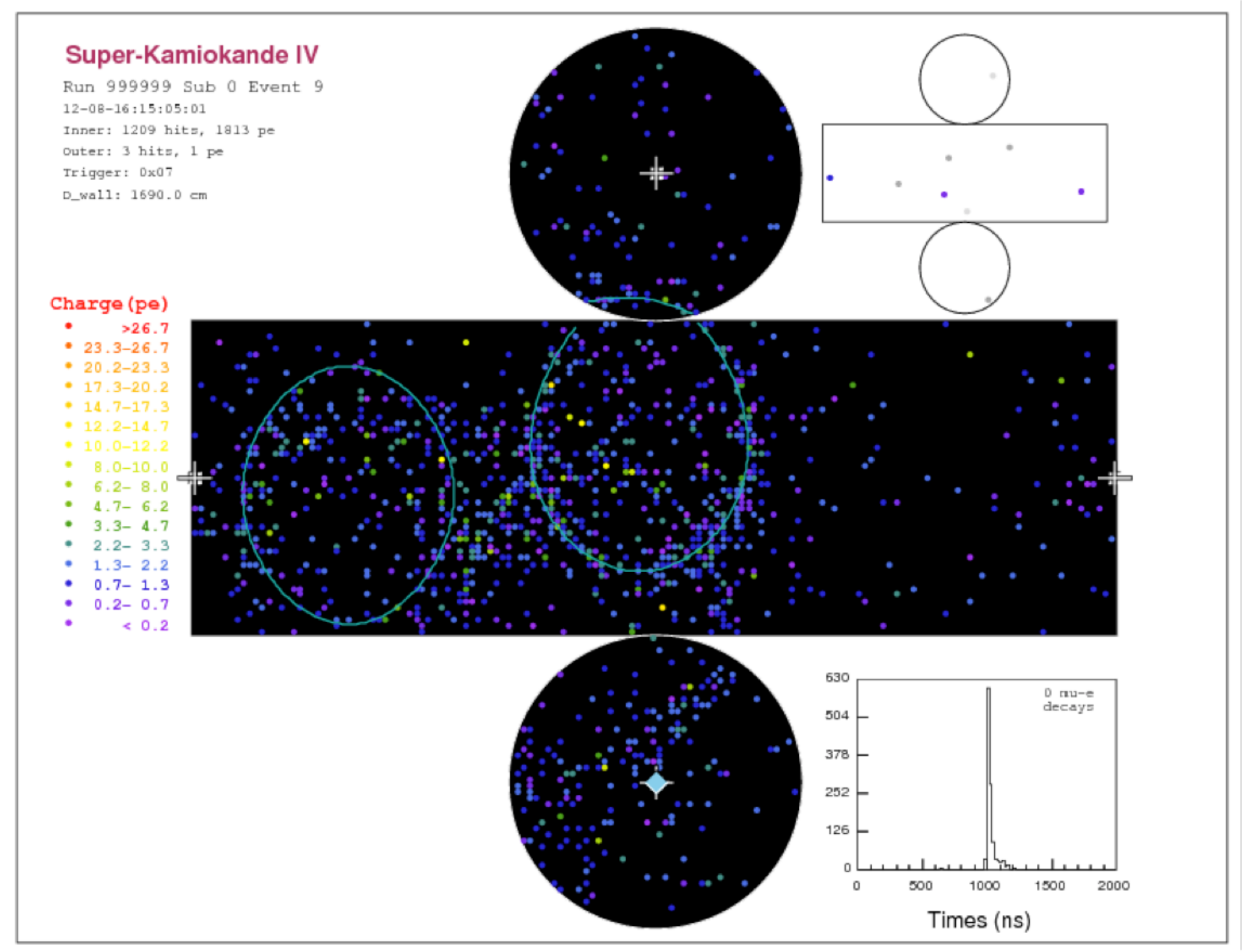}
\caption{Event display of a neutron background simulation. $\pi^0$ is produced by the interaction of $n + X \to X' + \pi^0$. A neutron is simulated with an energy of 1~GeV of the center of the detector.}
\label{fig:neutron_pi0_event}
\end{figure}

One possible concern about the background due to neutral particle,
such as neutrons and neutral kaons, which are produced by hadronic
interaction of cosmic muons near the detector, and enter the detector
without being detected by the outer detector.  Such particles may
penetrate deep into the detector and produce hadrons, such as $\pi^0$,
by interacting with water, which could become electron-like
backgrounds.  Figure~\ref{fig:neutron_pi0_event} shows a Super-K event
display of the simulated neutron background events which produced
$\pi^0$ particle in the detector.

For the study of neutron backgrounds, the flux of cosmic neutron at the detector site are estimated 
based on \cite{PhysRevD.73.053004} and shown in Table~\ref{tab:neutron_flux}. 
According to this table, the neutron flux of $E>100$~MeV at Hyper-K site will increase by a factor of $\sim$8 
than that of Super-K site.

\begin{table}
\begin{center}
\caption{Comparison of various parameters related to neutron background estimation between Super-K and Hyper-K.
         The estimation of these values are based on \cite{PhysRevD.73.053004}. }
\label{tab:neutron_flux}
\begin{tabular}{lccc}
\hline \hline
                                          &&  Super-K site &  Hyper-K site \\
\hline 
Site depth (m.w.e.)                       &&  2700        &    1750  \\
Cosmic muon rate (10$^{-6}$/cm$^2$/sec)   &&  0.13$\sim$0.14 &  1.0$\sim$2.3      \\
Effective depth (m.w.e.)                  &&  2050        &     1170  \\
$<E_\mu>$  (GeV)                          &&  219        &      146  \\
$\Phi_n$  (10$^{-9}$/cm$^2$/sec)          &&  12.3        &     101   \\
~~~~~~~  ($>$100~MeV)                        &&  0.81        &      6.7  \\
$<E_n>$  (MeV)                            &&  76        &        53   \\
\hline \hline 
\end{tabular}
\end{center}
\end{table}

Though detecting neutron is difficult, their backgrounds can be reduced by two ways; 

\begin{figure}[htb]
\includegraphics[width=0.6\textwidth]{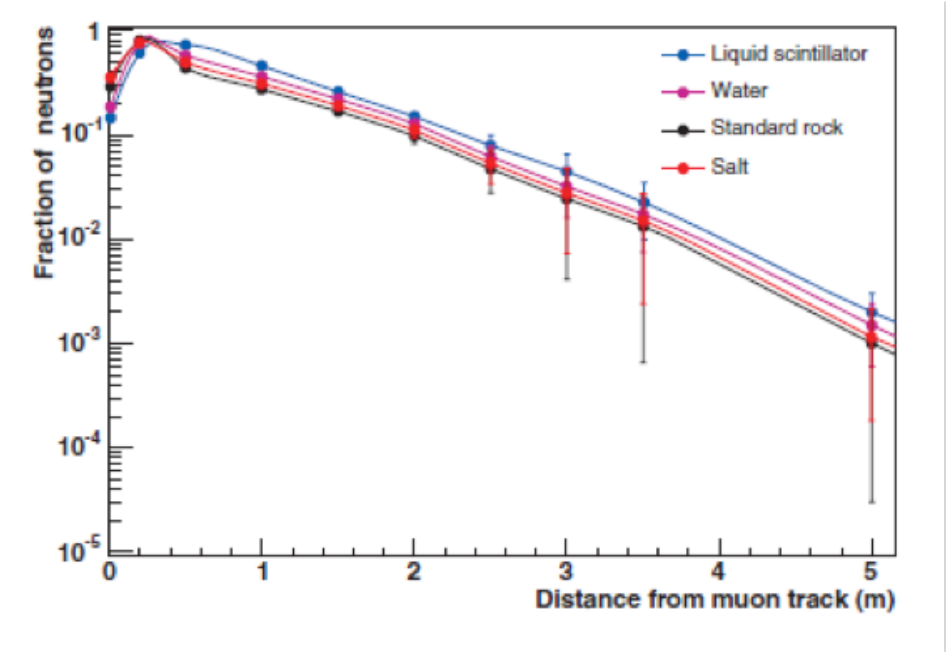}
\caption{Lateral distribution of cosmic neutron from parent muon track. Figure was taken from \cite{PhysRevD.73.053004}  }
\label{fig:neutron_lateral}
\end{figure}

\begin{itemize}
\item Self-shielding effects due to surrounding water around fiducial volume. In Super-K case, 
a water volume of $\sim$ 4.6~m thick (2.0~m in the inner detector and
2.6$\sim$2.8~m in the outer detector) is surrounded around fiducial
volume.  Since the neutron is reduced by hadronic interactions in
water in a scale of several 10~cm, neutrons is expected to be reduced
significantly before reaching the fiducial volume.
\item By detection of the accompanying cosmic muon. As seen in Fig~\ref{fig:neutron_lateral},
cosmic neutron and its parent muon are correlated spatially. This means that  neutrons are reduced after traveling in several meter from muon track in the rock. Considering the detector size of the Hyper-K, when neutrons comes into the detector, 
it is supposed that accompanying muons go through the detector also in most case, and rejected by the signal in
the outer detector.
\end{itemize}

In order to estimate the neutron background in Hyper-K,
neutron background simulations that took into account 
the effect of the accompanying muons were performed.
The detector simulation of Super-K was used.  
Since most of the neutrons are expected to be rejected by
taking the coincidence with muon signal, as described above, simple
toy Monte Carlo simulation considering the detector geometry are
performed, and then events in which only neutron is entering are
simulated with the Super-K simulator.  The detail procedure of the
simulation is described as follows:

\begin{enumerate}
\item Determine muon track. The starting position of muon track is in the plane about 20~meter above the top of Super-K detector 
and the vertex is randomly determined within the region of 200~meter from the detector center. 
\item Determine the point at which neutron enters the detector according to the neutron lateral distribution from muon track. 
      If there is no neutron which track does not hit the detector, this event is not counted. 
\item Rejection by muon track. If muon track goes through the detector region, this event is discarded. 
\item For the events which pass the previous step, neutron vector information, such as vertex, energy, direction, are fed into Super-K detector simulator and simulate neutron interactions.
\item Apply simple fully-contained (FC) reduction cut to simulated neutron events. Criterion that the number of hits in the outer detector ($nhitac$) is less than 16 and the visible energy in the inner detector ($E_{vis}$) is greater than 30~MeV are required. 
\end{enumerate}

The energy and directional angle distributions of neutrons are determined based on \cite{PhysRevD.73.053004}. 
According to the toy simulation,  97\% of events are rejected by the criteria of muon coincidence with neutron in step 3. 

\begin{figure}[htb]
\includegraphics[width=0.8\textwidth]{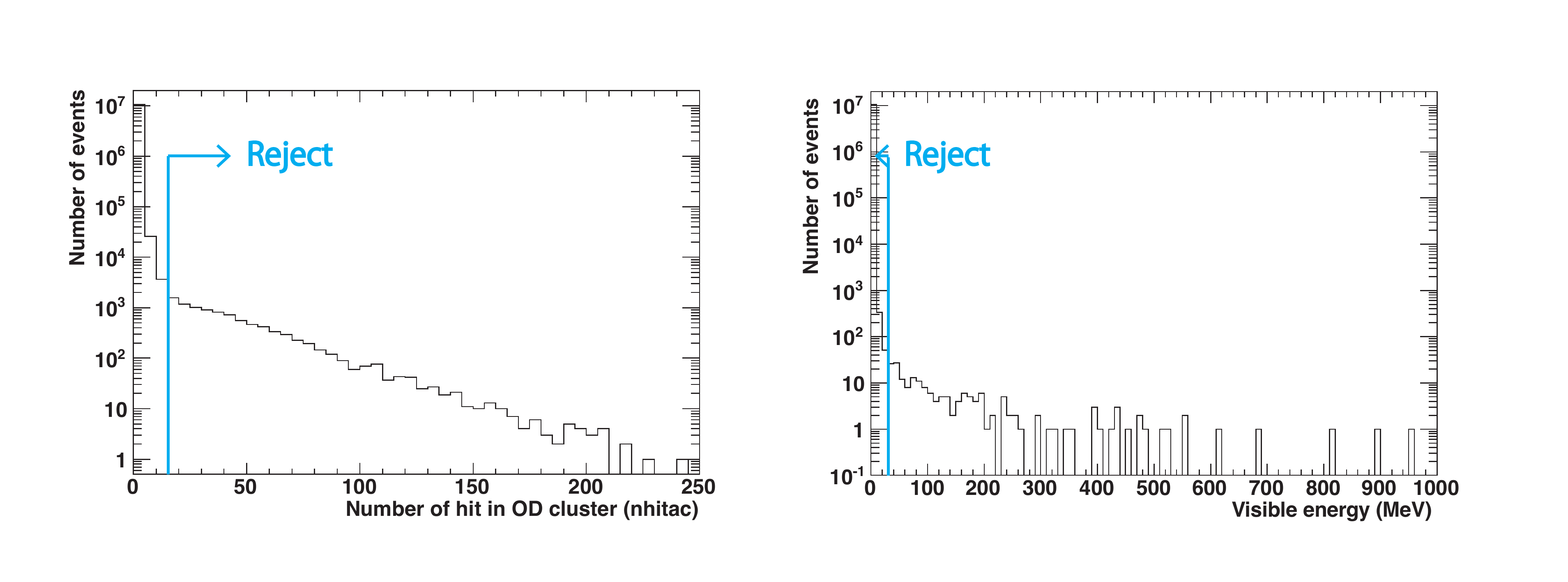}
\caption{Distributions of number of hits in the outer detector ($nhitac$) (left) and visible energy in the inner detector ($E_{vis}$) (right) for simulated neutron background events. }
\label{fig:neutron_reduction}
\end{figure}

\begin{figure}[htb]
\includegraphics[width=0.6\textwidth]{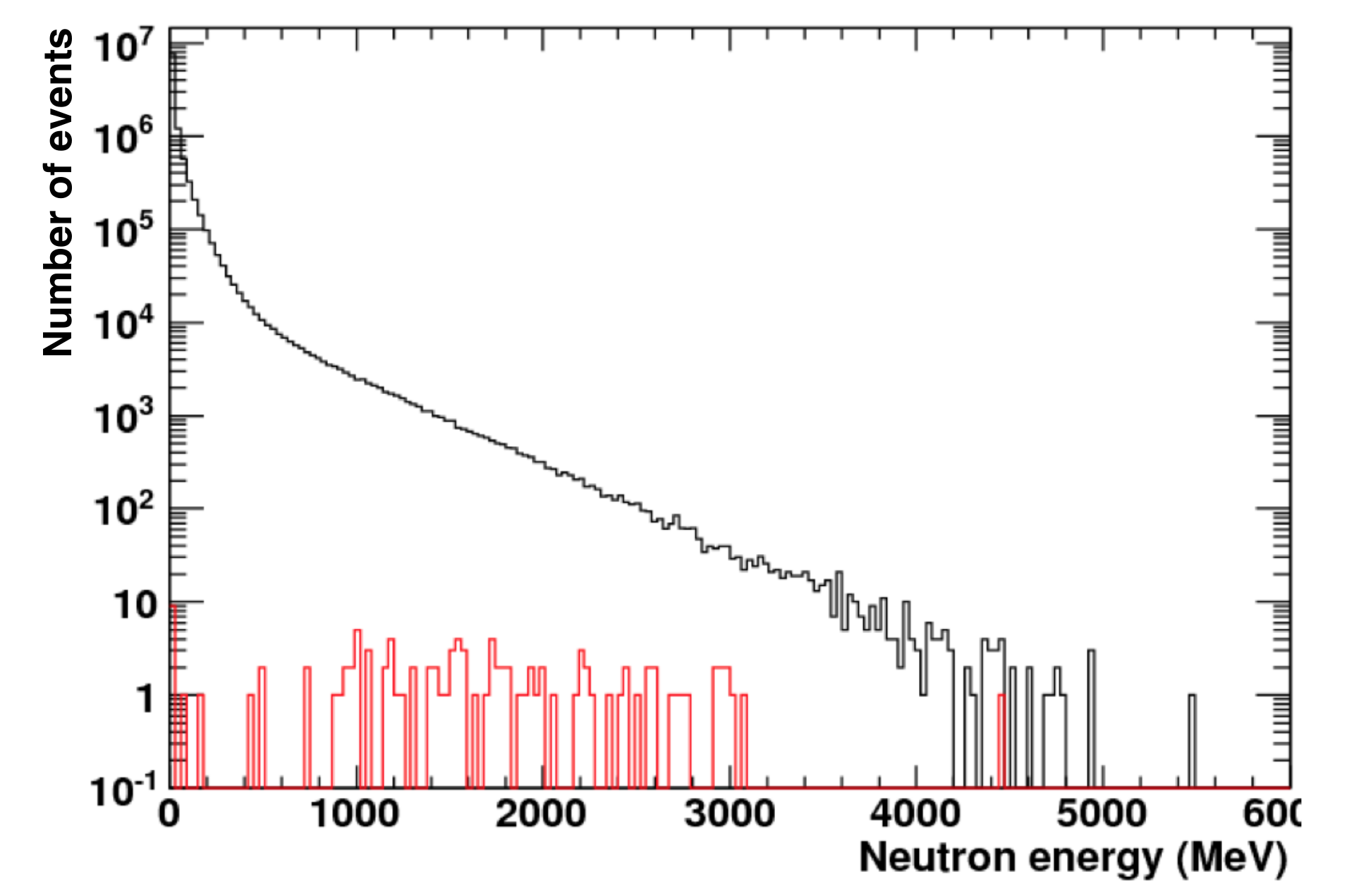}
\caption{Distributions of true neutron kinetic energy for all simulated events (black) and remaining events after FCFV cut (red).}
\label{fig:neutron_energy}
\end{figure}

\begin{figure}[htb]
\includegraphics[width=0.8\textwidth]{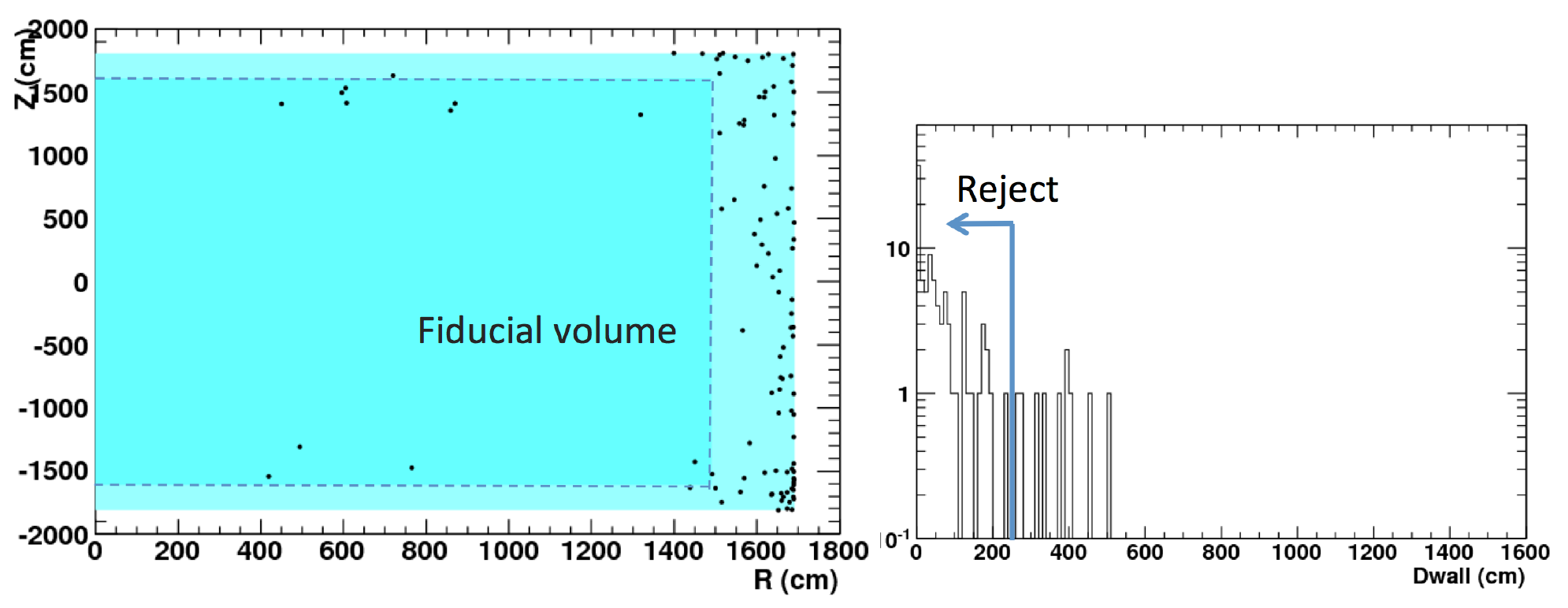}
\caption{Reconstructed vertex distributions of neutron background events which pass after fully-contained reduction (left),
and $D_{wall}$ distribution ,which corresponds to the distance between reconstructed vertex to the detector wall, 
for same event (right).}
\label{fig:neutron_vertex}
\end{figure}

Fig~\ref{fig:neutron_reduction} shows the distributions of the reduction parameters, $nhitac$ and $E_{vis}$. 
Fig~\ref{fig:neutron_energy} shows the distribution of neutron kinetic energy for all simulated events and the 
remaining events after FC and fiducial volume (FCFV) cut.
Fig~\ref{fig:neutron_vertex} shows the vertex distribution in Z (height)  vs R (radius) of the detector, $D_{wall}$ distribution, which 
is the distance to the wall. 
In the vertex distribution events are gathering around the side wall and fewer events around top, 
suggesting the vertical muons passing nearby the detector produced neutrons entering the detector.  

Table~\ref{tab:neutron_MC_summary} shows the summary of the neutron
background MC events in each reduction step.  Normalizing to the
number of events per one year at Super-K detector condition, 2.1 and
0.2 events are expected for fully-contained (FC) and fully-contained
fiducial volume (FCFV) event, respectively.  Considering the event
rate of $\sim$3000 atmospheric neutrinos, this corresponds to
$0.2/3000=7\times10^{-3}$\% background rate in FCFV sample.  As for
the case of Hyper-K site, neutron flux is increased by about factor of
eight according to Table~\ref{tab:neutron_flux} due to the shallower
overburden of detector site condition.
When the background rate is simply scaled by the factor of eight according to the increase of the muon flux,
$7\times10^{-2}$~$\times$~8~$=$~$5\times10^{-2}$\% 
of the neutron background rate is estimated for Hyper-K case, which
seems to be negligible level for physics study.

\begin{table}
\begin{center}
\caption{ Summary of the number of events in neutron background using Super-K detector simulation.}
\label{tab:neutron_MC_summary}
\begin{tabular}{lccc}
\hline \hline
                                        &&  All simulation    &  Event/1year at Super-K \\
\hline 
entering neutrons                       &&  4.5$\times$10$^8$   & 8.9$\times$10$^6$     \\
w/o muon coincidence                    &&  1.1$\times$10$^7$   & 2.1$\times$10$^5$      \\
passed fully-contained cut (FC)         &&   105       &    2.1   \\
vertex is in FV (FCFV)                 &&   11       &     0.2   \\
\hline \hline
\end{tabular}
\end{center}
\end{table}

Another possible neutral particle which could be the background is
neutral kaon.  According to the calculation of neutral kaon flux in
underground~\cite{JHEP04.041}, neutral flux is estimated to be
significantly smaller; 0.3\% of neutron flux at 3~km m.w.e..  

The neutral kaon background is also estimated by the same simulation 
method as in the neutron case with the estimated flux and energy spectrum 
described in \cite{JHEP04.041}. The simulation data corresponding to 
50~years livetime in Hyper-K are produced. After applying
FCFV selection, no background events are remained in the fiducial volume,
concluding that the background from neutral kaon is negligible for the 
atmospheric neutrino analysis.

It would plausible to consider that the impact of the neutron and 
kaon backgrounds on the proton decay analysis
is negligible as in the atmospheric neutrino case since 
the there is no reason that those backgrounds have the same event topologies
as proton decay. These backgrounds will be also reduced similarly as 
atmospheric neutrino background by the proton decay selection cuts.

\clearpage
\part{Physics Potential}
\label{section:physics}

\section{Neutrino Oscillation}

\graphicspath{{physics-lbl/figures/}}

\subsection{Accelerator based neutrinos \label{sec:cp}}

A long baseline neutrino oscillation experiment with the J-PARC
neutrino beam is one of the key elements of Hyper-K physics program.
Especially, a precise study of $CP$ asymmetry in the lepton sector is
one of the major goals of Hyper-K.  The existence of $CP$ violation is
one of necessary conditions to explain the matter-antimatter asymmetry
of the Universe.

In the Standard Model, extended to accommodate non-zero neutrino
masses, the source of $CP$ violation is (apart from the QCD phase)
attributed to irreducible complex phases in the flavor mixing
matrices.  In the quark sector, with more than fifty years of
extensive study after the initial discovery of $CP$ violation in kaon
decays, all measurements related to $CP$ violation are consistently
explained by the Kobayashi-Maskawa phase~\cite{Amhis:2014hma}.  On the
other hand, in the lepton sector, experimental study of $CP$ asymmetry
has just begun.  Recently, the T2K collaboration reported the first
constraint on $\deltacp$~\cite{Abe:2015awa, Abe:2017uxa} with $\theta_{13}$
measured by reactor experiments assuming the standard mixing
framework.  It suggests that the $CP$ violation in the lepton sector
may be large, although the statistical significance is not yet sufficient.
This is the first step in an experimental program towards
the test of leptonic $CP$ symmetry following the discovery of the relatively large
value of $\theta_{13}$.  Now, it is necessary to build a next
generation experiment with a definitive sensitivity to study the $CP$
asymmetry.

For a direct and model-independent measurement of the $CP$ asymmetry,
a comparison of oscillation probabilities between neutrino and
anti-neutrino is necessary.  Measurements of $\numu \to \nue$ and
$\numubar \to \nuebar$ oscillations are practically the only possible
way to study the lepton $CP$ asymmetry.  Furthermore, it will be
possible to check the consistency of the mixing framework by comparing
the accelerator ($\nu_\mu$ to $\nu_e$ appearance of GeV neutrino over
295~km) and reactor ($\nuebar$ disappearance of MeV neutrino over
$\sim$1~km) measurements, which are related to the same parameters in
the standard framework but may receive different contribution from new
physics.

The observation of $CP$ violation in the lepton sector will open a new
field of research.  A quest will start to understand its origin with
precision measurements, as it has been done in the quark sector.  A
measurement of $CP$ violating phase, together with precision
measurements of mixing angles and mass differences, will provide
crucial information to discriminate the fundamental physics behind
mass and mixing generation.  There are various models proposed based
on flavor symmetries or other methods, and many of them give testable predictions for 
relations among those parameters~\cite{Mohapatra:2006gs,Altarelli:2010gt,King:2013eh,King:2015aea}.
Precision measurements of oscillation parameters require both large statistics
and well controlled systematics.  Combining an intense ($>$MW) and
high quality neutrino beam from J-PARC, the huge mass and high
performance of Hyper-K detector, a highly capable near/intermediate
detector complex, and the full expertise obtained from ongoing T2K
experiment, Hyper-K will be the best project to probe neutrino $CP$
violation and new physics with neutrino oscillation.

\subsubsection{J-PARC to Hyper-Kamiokande long baseline experiment}
%\subsubsection{Experimental configuration}

The neutrino energy spectrum of J-PARC neutrino beam is tuned to the
first oscillation maximum with the off-axis technique, which enhances the
flux at the peak energy while reducing the higher energy component
that produces background events.  The peak energy, around 600~MeV, is
well matched to the water Cherenkov detector technology, which has an excellent
$e$/$\mu$ separation capability, high background rejection efficiency
and high signal efficiency for sub-GeV neutrino events.
Due to the relatively short baseline of 295~km and thus lower neutrino
energy at the oscillation maximum, the contribution of the matter
effect is smaller for the J-PARC to Hyper-Kamiokande experiment
compared to other proposed experiments like DUNE 
in the United States~\cite{Acciarri:2015uup}.
Thus the $CP$ asymmetry measurement with the J-PARC to Hyper-K long
baseline experiment has less uncertainty related to the matter effect,
while other experiments with $>1000$~km baseline have much better
sensitivity to the mass hierarchy (the sign of $\Delta m^2_{32}$) 
with accelerator neutrino beams.
Nevertheless, Hyper-K can determine the mass hierarchy using atmospheric 
neutrinos as described in Section~\ref{section:atmnu}.  The sensitivities for
$CP$ violation and mass hierarchy can be further enhanced by combining
accelerator and atmospheric neutrino measurements.

The focus of the J-PARC to Hyper-K experiment is the measurements of $|\Delta
m^2_{32}|$, $\sin^2\theta_{23}$, $\sin^2\theta_{13}$ and $\deltacp$.
The standard flavor mixing scenario is assumed in the following as a baseline study, although it is possible that new physics is involved in neutrino oscillation and will be revealed by Hyper-K.
The analysis presented in this report is based on \cite{Abe:2015zbg} but with an updated treatment of systematic uncertainties.

\subsubsection{Oscillation probabilities and measurement channels}
In what follows, the oscillation probabilities and sensitivities to
oscillation parameters with $\nue$ appearance and $\numu$
disappearance measurements are discussed.  The analysis will be
performed by a combination of these two channels.

\paragraph{$\numu \to \nue$ appearance channel}
The oscillation probability from $\nu_\mu$ to $\nu_e$ %in accelerator experiments
is expressed, to the first order of the matter effect, as follows~\cite{Arafune:1997hd}:
\begin{eqnarray}
P(\numu \to \nue) & = & 4 c_{13}^2s_{13}^2s_{23}^2 \cdot \sin^2\Delta_{31}  \nonumber \\
& & +8 c_{13}^2s_{12}s_{13}s_{23} (c_{12}c_{23}\cos\deltacp - s_{12}s_{13}s_{23})\cdot \cos\Delta_{32} \cdot \sin\Delta_{31}\cdot \sin\Delta_{21} \nonumber \\
& & -8 c_{13}^2c_{12}c_{23}s_{12}s_{13}s_{23}\sin\deltacp \cdot \sin\Delta_{32} \cdot \sin\Delta_{31}\cdot \sin\Delta_{21} \nonumber \\
& & +4s_{12}^2c_{13}^2(c_{12}^2c_{23}^2 + s_{12}^2s_{23}^2s_{13}^2-2c_{12}c_{23}s_{12}s_{23}s_{13}\cos\deltacp)\cdot \sin^2\Delta_{21} \nonumber \\
& & -8c_{13}^2s_{13}^2s_{23}^2\cdot \frac{aL}{4E_\nu} (1-2s_{13}^2)\cdot \cos\Delta_{32}\cdot \sin\Delta_{31} \nonumber \\
& & +8 c_{13}^2s_{13}^2s_{23}^2 \frac{a}{\Delta m^2_{31}}(1-2s_{13}^2)\cdot\sin^2\Delta_{31}, \label{Eq:cpv-oscprob}
\end{eqnarray}
\noindent where $s_{ij} = \sin\theta_{ij}$, $c_{ij}=\cos\theta_{ij}$, $\Delta_{ij} = \Delta m^2_{ij}\, L/4E_\nu$, 
and $a =2\sqrt{2}G_Fn_eE_\nu= 7.56\times 10^{-5}\mathrm{[eV^2]} \times \rho \mathrm{[g/cm^3]} \times E_\nu[\mathrm{GeV}].$
$L$, $E_\nu$, $G_F$ and $n_e$ are the baseline, the neutrino energy, the Fermi coupling constant and the electron density, respectively. 
The corresponding probability for a $\numubar \to \nuebar$ transition is obtained by replacing $\deltacp \rightarrow -\deltacp$
and $a \rightarrow -a$.
The third term, containing $\sin\deltacp$, is the $CP$ violating term
which flips sign between $\nu$ and $\bar{\nu}$ and thus introduces
$CP$ asymmetry if $\sin\deltacp$ is non-zero.  The last two terms are
due to the matter effect.  Those terms which contain $a$ change their
sign depending on the mass hierarchy.  As seen from the definition of
$a$, the amount of asymmetry due to the matter effect is proportional
to the neutrino energy at a fixed value of $L/E_\nu$.

\begin{figure}[tbp]
\centering
\includegraphics[width=0.48\textwidth]{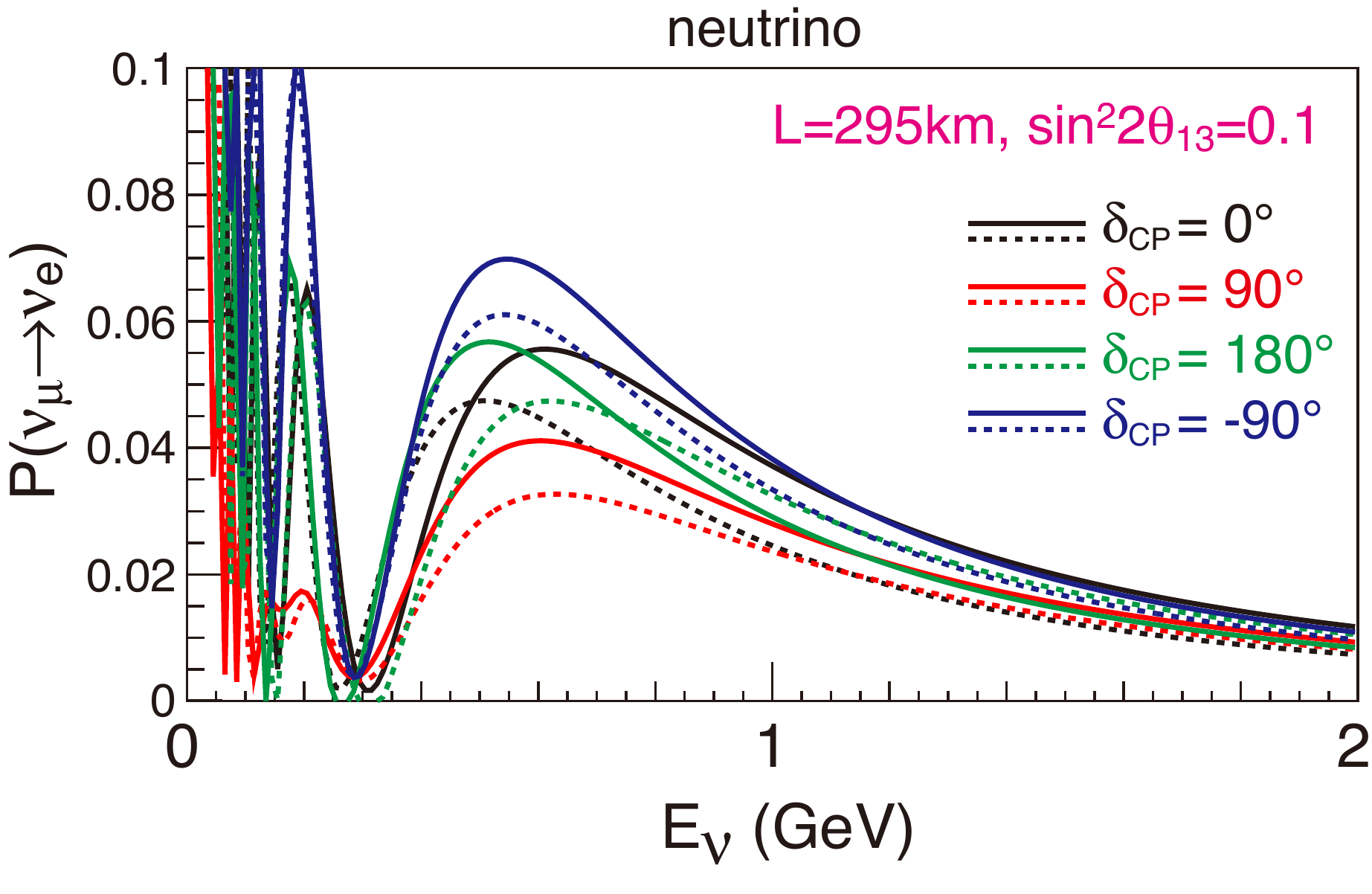}
\includegraphics[width=0.48\textwidth]{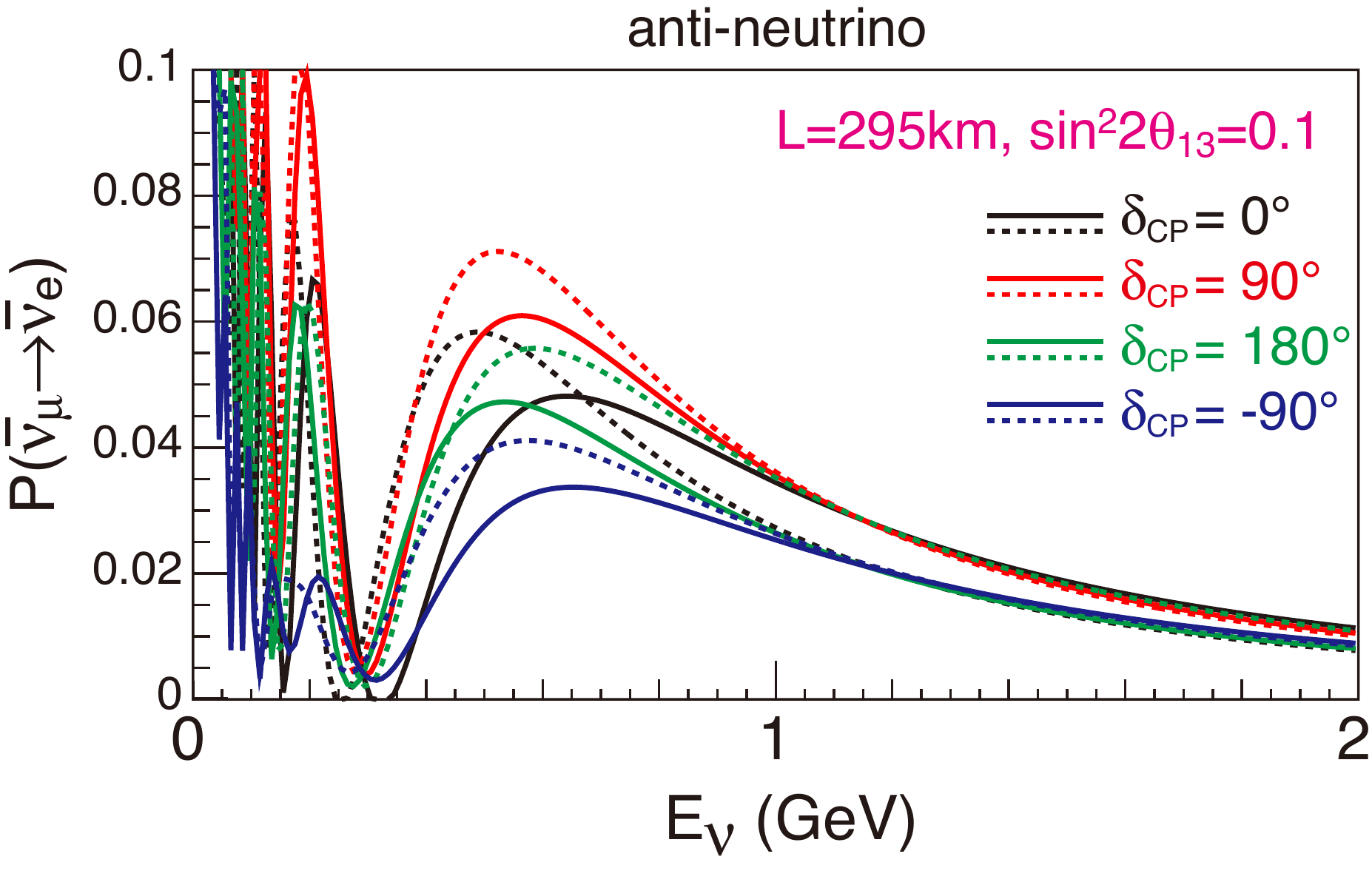}
\caption{Oscillation probabilities as a function of the neutrino energy for $\numu \to \nue$ (left) and $\numubar \to \nuebar$ (right) transitions with L=295~km and $\sin^22\theta_{13}=0.1$. 
Black, red, green, and blue lines correspond to $\deltacp = 0^\circ$,
90$^\circ$, 180$^\circ$ and $-90^\circ$, respectively.  Solid (dashed)
line represents the case for a normal (inverted) mass hierarchy.
\label{fig:cp-oscpob}}
\end{figure}

Figure~\ref{fig:cp-oscpob} shows the $\numu \to \nue$ and
$\numubar \to \nuebar$ oscillation probabilities as a function of the
true neutrino energy for a baseline of 295~km.  The Earth matter
density is assumed to be 2.6\,$g$/cm$^3$.  The cases for $0^\circ$,
90$^\circ$, 180$^\circ$ and $-90^\circ$, are shown together.  One can
see the effect of different $\deltacp$ values on the oscillation
probabilities.  For example, if $\deltacp = -90^\circ$, the appearance
probability will be enhanced for neutrino but suppressed for
anti-neutrino.  By comparing the oscillation probabilities of
neutrinos and anti-neutrinos, one can measure the $CP$ asymmetry. 
The information on the $CP$ phase can be derived from not only the total
number of events but also the energy spectrum of the oscillated events. 
For example, for both $\deltacp = 0^\circ$ and $180^\circ$, $CP$ is conserved
($\sin\deltacp=0$) and the oscillation probabilities in vacuum are the
same for neutrino and anti-neutrino, however those two cases can be
distinguished using spectrum information as seen in
Fig.~\ref{fig:cp-oscpob}.

Also shown in Fig.~\ref{fig:cp-oscpob} are the case of normal mass
hierarchy ($\Delta m^2_{32}>0$) with solid lines and inverted mass
hierarchy ($\Delta m^2_{32}<0$) with dashed lines.
There are sets of different mass hierarchy and values of $\deltacp$
which give similar oscillation probabilities, resulting in a potential
degeneracy if the mass hierarchy is unknown.  By combining information
from experiments currently ongoing~\cite{Abe:2011ks,Ayres:2004js,
Guo:2007ug,Ahn:2010vy,Ardellier:2006mn} and/or planned in the near
future~\cite{Aartsen:2014oha,Katz:2014tta,Ahmed:2015jtv,An:2015jdp,Kim:2014rfa},
it is expected that the mass hierarchy will be determined by the time
Hyper-K starts to take data.  If not, Hyper-K itself has a sensitivity
to the mass hierarchy by the atmospheric neutrino measurements as
described in the next section.  Thus, the mass hierarchy is assumed to
be known in this analysis, unless otherwise stated.

\begin{figure}[tbp]
\centering
\includegraphics[width=0.45\textwidth]{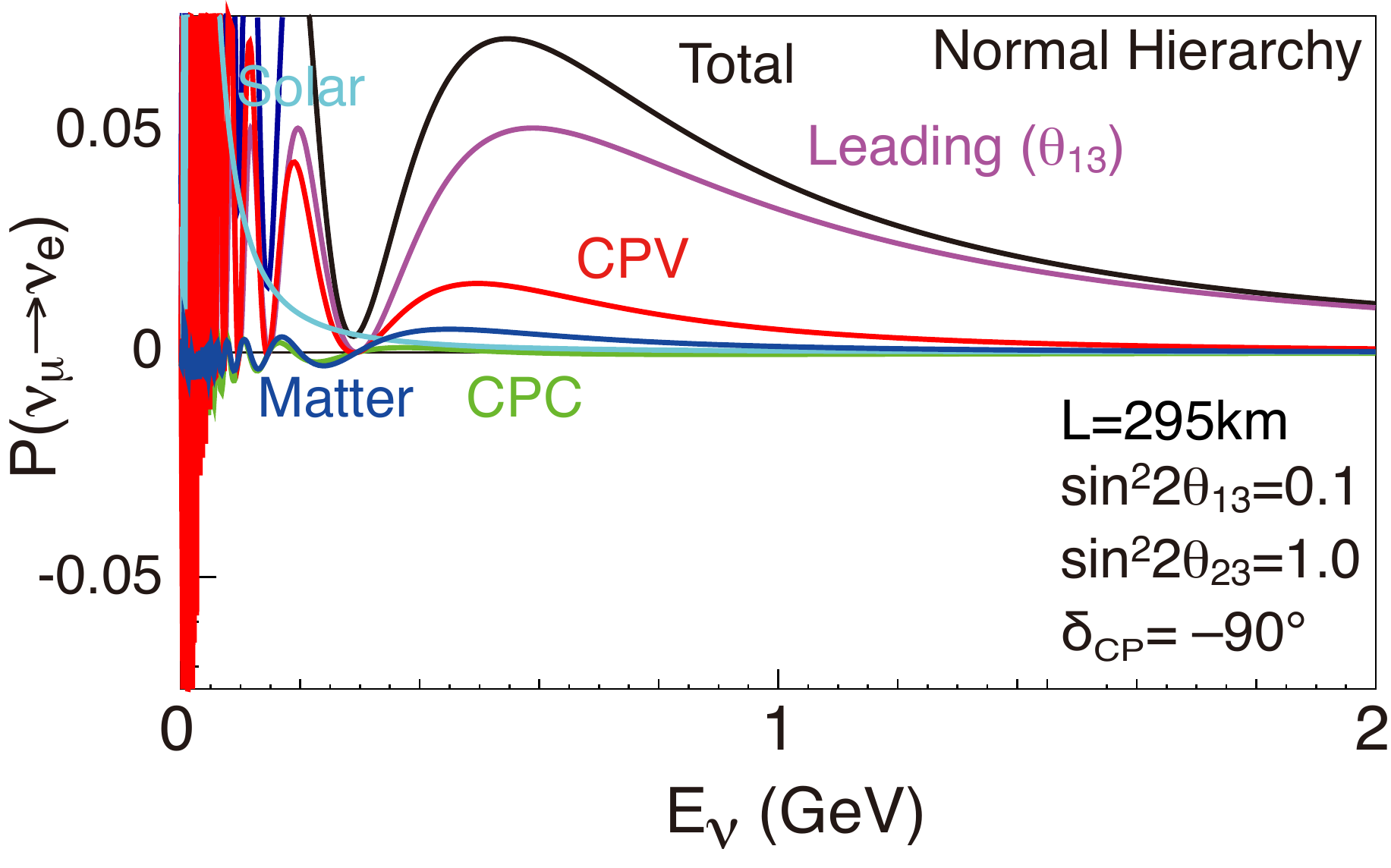}
\includegraphics[width=0.45\textwidth]{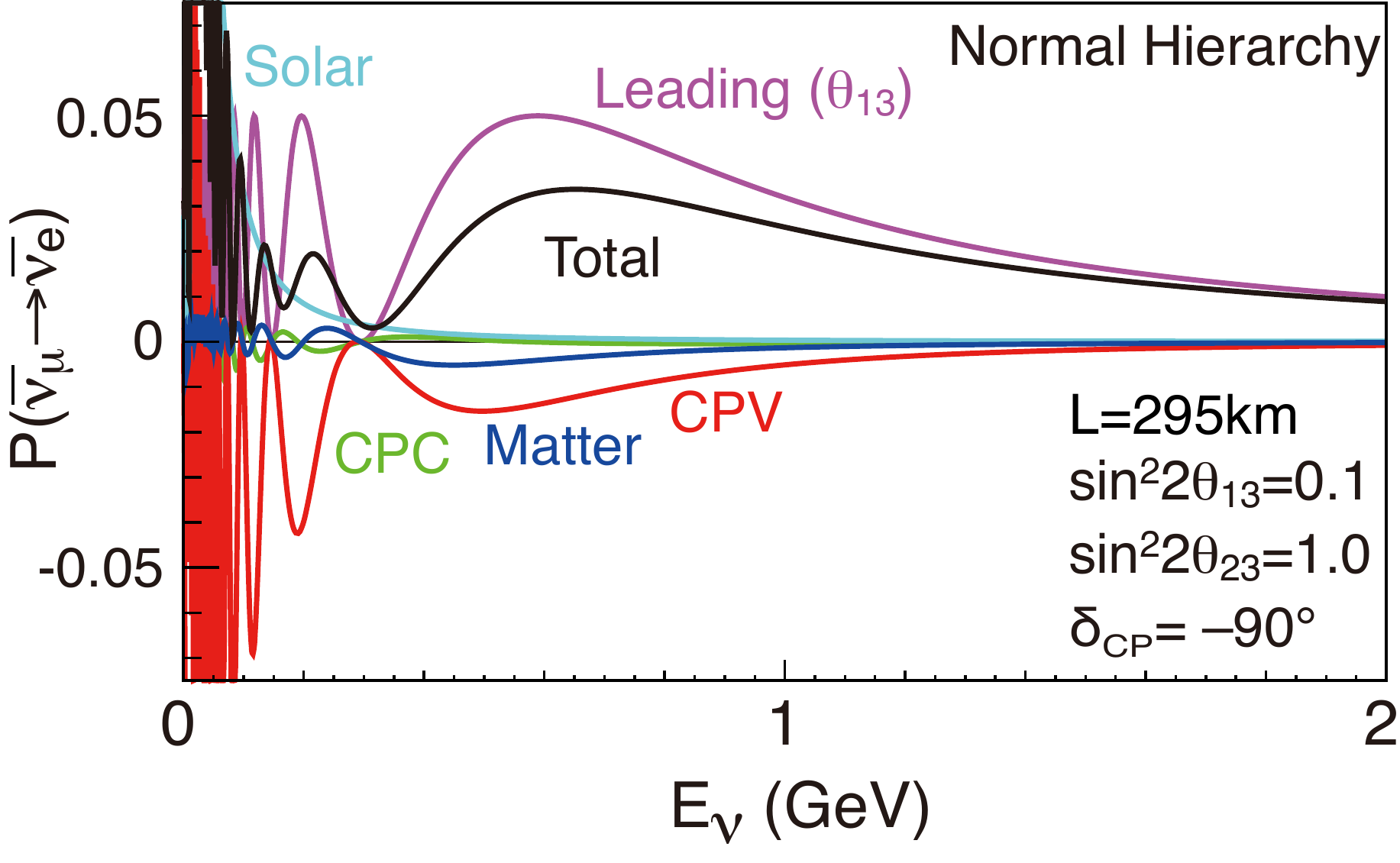}
\caption{Oscillation probabilities of $\numu \to \nue$ (left) and  $\numubar \to \nuebar$ (right) as a function of the neutrino energy with a baseline of 295~km.  $\sin^22\theta_{13}=0.1$,
$\deltacp = -90^\circ$, and normal hierarchy are assumed.
Contribution from each term of the oscillation probability formula is shown separately.
\label{fig:cp-oscpob-bd}}
\end{figure}

Figure~\ref{fig:cp-oscpob-bd} shows the contribution from each term of the $\numu \to \nue$ and $\numubar \to \nuebar$ oscillation probability formula, Eq.(\ref{Eq:cpv-oscprob}),
for $L=295$\,km, $\sin^22\theta_{13}=0.1$, $\sin^22\theta_{23}=1.0$, $\deltacp = -90^\circ$, and normal mass hierarchy.
For $E_\nu\simeq 0.6$\,GeV which gives $\sin\Delta_{32} \simeq \sin\Delta_{31} \simeq 1$,
\begin{eqnarray}
\frac{P(\nu_\mu \to  \nu_e) - P(\bar\nu_\mu \to  \bar\nu_e)}{P(\nu_\mu \to  \nu_e) + P(\bar\nu_\mu \to  \bar\nu_e)} &\simeq& 
\frac{-16J_{CP}\sin \Delta_{21} + 16 c_{13}^2s_{13}^2s_{23}^2 \frac{a}{\Delta m^2_{31}}(1-2s_{13}^2) }{8c_{13}^2s_{13}^2s_{23}^2}\\
&\simeq& -0.28 \sin\delta + 0.09,
\end{eqnarray}
where $J_{CP} = c_{12}c_{13}^{2}c_{23}s_{12}s_{13}s_{23}\sin\delta$ is called Jarlskog invariant.
The effect of $CP$ violating term can be as large as 28\%, while that of the matter effect is 9\%.
The first term will be $-0.31 \sin\delta$ with $\sin^{2}2\theta_{13}=0.082$~\cite{Olive:2016xmw}.

The uncertainty of the Earth's density between Tokai and Kamioka is estimated to be at most 6\%~\cite{Hagiwara:2011kw}. Because the matter effect contribution to the total $\numu \to \nue$ appearance probability is less than 10\% for 295km baseline, the uncertainty from the matter density is estimated to be less than 0.6\% and neglected in the following analysis.

\paragraph{$\numu$ disappearance channel}
The currently measured value of $\theta_{23}$ is consistent with maximal mixing, $\theta_{23} \approx \pi/4$~\cite{Abe:2014ugx, Adamson:2014vgd, Himmel:2013jva},
while NOvA collaboration recently reported a possible hint of non-maximal mixing~\cite{Adamson:2017qqn}.
It is of great interest to determine if $\sin^22\theta_{23}$ is maximal or not, and if not, whether $\theta_{23}$ is less or greater than $\pi/4$, as 
it could constrain models of neutrino mass generation and quark-lepton unification~\cite{King:2013eh,Albright:2010ap,Altarelli:2010gt,Ishimori:2010au,Albright:2006cw,Mohapatra:2006gs}.
When we measure $\theta_{23}$ with the survival probability $P(\numu \to \numu)$ which is proportional to $\sin^22\theta_{23}$ to first order, 
\begin{eqnarray}
P(\nu_\mu \rightarrow \nu_\mu) &\simeq& 1-4c^2_{13}s^2_{23} [1-c^2_{13}s^2_{23}]\sin^2(\Delta m^2_{32}\, L/4E_\nu) \\
&\simeq & 1-\sin^22\theta_{23}\sin^2(\Delta m^2_{32}\, L/4E_\nu), \hspace{2cm} \textrm{(for $c_{13}\simeq1$)}
\end{eqnarray}
there is an octant ambiguity, as for each value of $\theta_{23} \le
45^\circ $ (in the first octant), there is a value in the second
octant ($\theta_{23} > 45^\circ$) that gives rise to the same
oscillation probability.  As seen from Eq.~\ref{Eq:cpv-oscprob},
$\nu_e$ appearance measurement can determine
$\sin^2\theta_{23}\sin^22\theta_{13}$.  In addition, the reactor
experiments provide an almost pure measurement of $\sin^22\theta_{13}$.
Thus, the combination of those complementary measurements will be able
to resolve this degeneracy if $\theta_{23}$ is sufficiently away from
$\frac{\pi}{4}$~\cite{Fogli:1996pv,Minakata:2002jv,Hiraide:2006vh}.

Measurement of $\nuebar$ disappearance by reactor neutrino experiments
provides a constraint on the following combination of mass-squared
differences,
\begin{equation}
\Delta m^2_{ee} = \cos^2\theta_{12}\Delta m^2_{31}+\sin^2\theta_{12}\Delta m^2_{32}.
\end{equation}
while $\numu$ disappearance measurement with Hyper-K provides a different combination~\cite{Nunokawa:2005nx, deGouvea:2005hk}
\begin{equation}
\Delta m^2_{\mu\mu} = \sin^2\theta_{12}\Delta m^2_{31}+\cos^2\theta_{12} \Delta m^2_{32}
+ \cos\deltacp \sin\theta_{13} \sin2\theta_{12}\tan\theta_{23} \Delta m^2_{21}.
\end{equation}
Because the mass squared difference measurements by Hyper-K and by
reactor experiments give independent information, by comparing them
one can check the consistency of the mixing matrix framework, and
obtain information on the neutrino mass hierarchy.  In order to have
sensitivity to the mass hierarchy, uncertainties of both measurements
must be smaller than 1\%.  Future medium baseline reactor experiments,
JUNO~\cite{An:2015jdp} and RENO-50~\cite{Kim:2014rfa}, plan to measure
$\Delta m^2_{ee}$ with precision better than 1\%.  Thus, precision
measurement of $\Delta m^2$ by Hyper-K will provide important
information on the consistency of three generation mixing framework
and mass hierarchy.

\subsubsection{Analysis overview}
The analysis used in this report is based on a framework developed for
the sensitivity study by T2K presented in~\cite{Abe:2014tzr}.  A
binned likelihood analysis based on the reconstructed neutrino energy
distribution is performed using both \nue\ (\nuebar) appearance
and \numu\ (\numubar) disappearance samples simultaneously.  A full
oscillation probability formula, not the approximation shown in
Eq.~\ref{Eq:cpv-oscprob}, is used in the analysis.
Table~\ref{Tab:oscparam} shows the nominal oscillation parameters used
in the study presented in this report, and the treatment during the
fitting.  Parameters to be determined with the fit are
$\sin^2\theta_{13}$, $\sin^2\theta_{23}$, $\Delta m^2_{32}$ and
$\deltacp$.

An integrated beam power of 13~MW$\times$10$^7$~sec is assumed in
this study, corresponding to $2.7\times10^{22}$ protons on target
with 30\,GeV J-PARC beam.
It corresponds to about ten Snowmass years with 1.3~MW.
We have studied the sensitivity to $CP$
violation with various assumptions of neutrino mode and anti-neutrino
mode beam running time ratio for both normal and inverted mass
hierarchy cases.  The dependence of the sensitivity on the
$\nu$:$\overline{\nu}$ ratio is found not to be significant between
$\nu$:$\overline{\nu}$=1:1 to 1:5.  In this report,
$\nu$:$\overline{\nu}$ ratio is set to be 1:3 so that the expected
number of events are approximately the same for neutrino and
anti-neutrino modes.

\begin{table}[htbp]
\caption{Oscillation parameters used for the sensitivity analysis and treatment in the fitting. The \textit{nominal} values are used for figures and numbers in this section, unless otherwise stated.}
\centering
\begin{tabular}{cccccccc} \hline \hline
Parameter & $\sin^22\theta_{13}$ & $\deltacp$ & $\sin^2\theta_{23}$ &
$\Delta m^2_{32}$ & mass hierarchy & $\sin^22\theta_{12}$ & $\Delta
m^2_{21}$ \\ \hline Nominal & 0.10 & 0 & 0.50 &
$2.4\times10^{-3}~\mathrm{eV}^2$ & Normal & $0.8704$ &
$7.6\times10^{-5}~\mathrm{eV}^2$ \\ Treatment & Fitted & Fitted &
Fitted & Fitted & Fixed & Fixed & Fixed \\ \hline \hline
\end{tabular}
\label{Tab:oscparam}
\end{table}%

Interactions of neutrinos in the Hyper-K detector are simulated with
the NEUT program
library~\cite{hayato:neut,Mitsuka:2007zz,Mitsuka:2008zz}, which is
used in both Super-K and T2K.  The response of the detector is
simulated using the Super-K full Monte Carlo simulation based on the
GEANT3 package~\cite{Brun:1994zzo}, although some improvements are expected
with new photo-sensors with higher photon detection efficiency and timing
resolution (see Section~\ref{section:photosensors}). 
The simulation is based on the SK-IV configuration, which has an upgraded electronics and DAQ system compared to SK-III.
Events are reconstructed with the Super-K reconstruction software, which
gives a realistic estimate of the Hyper-K performance.

Based on the experience with the SK-II period when the number of PMT
was about half compared to other periods (corresponding to 20\%
photocoverage with the Super-K PMT R3600), the reconstruction
performance for beam neutrino events with around 1~GeV energy is known
not to degrade significantly with reduced photocathode coverage down to 20\% (with
R3600).  
Thus, the performance for the beam neutrino interactions is comparable within the range of 
the photocathode coverage considered for Hyper-K.
There will be additional capabilities such as neutron tagging with higher coverage, 
but they are not yet taken into account in the current study.

In what follows, results are presented assuming ten years of running with a single tank detector with 187\,kton fiducial volume unless otherwise noticed.
Also shown for comparison are results from the staging approach with ten years of running, with a single tank for the first six years, and two tanks starting in the seventh year.

\subsubsection{Expected observables at the far detector}
The criteria to select \nue\ and \numu\ candidate events are based on
those developed for and established with the Super-K and T2K
experiments.  Fully contained (FC) events with a reconstructed vertex
inside the fiducial volume (FV), which is defined as the region more than 1.5\,m away from inner detector wall, and visible energy ($E_\mathrm{vis}$)
greater than 30\,MeV are selected as FCFV neutrino event candidates.
In order to enhance charged current quasielastic (CCQE, $\nu_l +
n \rightarrow l^- + p$ or $\overline{\nu}_l + p \rightarrow l^+ + n$)
interaction, a single Cherenkov ring is required.

Assuming a CCQE interaction, the neutrino energy ($E_\nu ^{\rm rec}$)
is reconstructed from the energy of the final state charged lepton
($E_\ell$) and the angle between the neutrino beam and the charged
lepton directions ($\theta_\ell$) as
\begin{eqnarray}
E_\nu ^{\rm rec}=\frac {2(m_n-V) E_\ell +m_p^2 - (m_n-V)^2 - m_\ell^2} {2(m_n-V-E_\ell+p_\ell\cos\theta_\ell)},
\label{eq:Enurec}
\end{eqnarray}
where $m_n, m_p, m_\ell$ are the mass of neutron, proton, and charged
lepton, respectively, $p_\ell$ is the charged lepton momentum, and $V$
is the mean nuclear potential energy (27\,MeV).
It was shown in T2K analysis that the sensitivity can be slightly improved
by using two-dimentional information of $(p_\ell, \theta)$ in oscillation fit.

Then, to select \nue/\nuebar\ candidate events the following criteria are applied;
the reconstructed ring is identified as electron-like ($e$-like),
$E_\mathrm{vis}$ is greater than 100 MeV, there is no decay electron
associated to the event, and $E_\nu^\mathrm{rec}$ is less than
1.25~GeV.  Finally, in order to reduce the background from
mis-reconstructed $\pi^0$ events, additional criteria using the
reconstructed $\pi^0$ mass and the ratio of the best-fit likelihoods
of the $\pi^0$ and electron fits~\cite{Abe:2013hdq} are applied.

\begin{figure}[tbp]%
\includegraphics[width=0.48\textwidth]{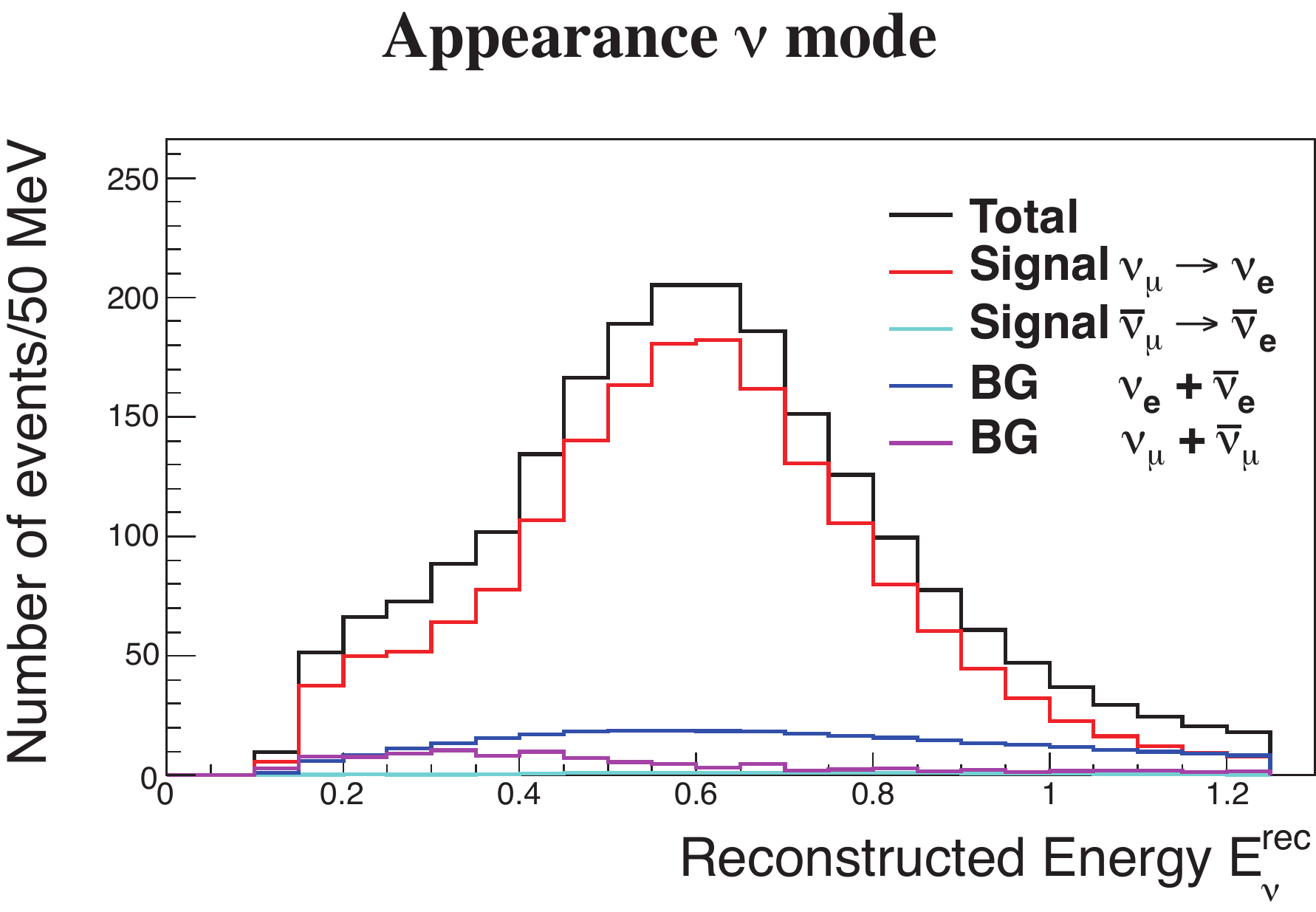}
\includegraphics[width=0.48\textwidth]{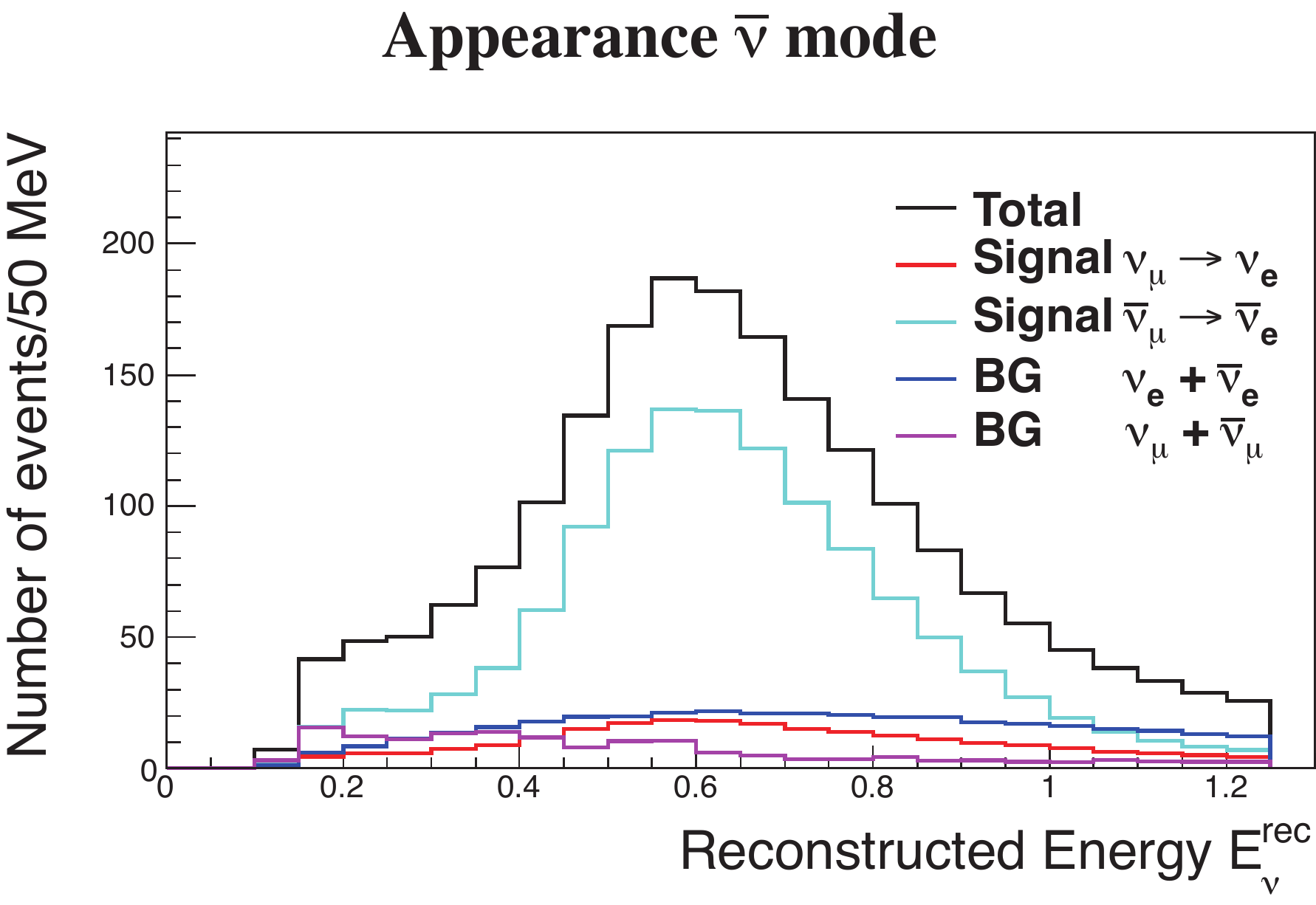}\\
\caption{
Reconstructed neutrino energy distribution of the $\nue$ candidate events.
Left: neutrino beam mode, right: anti-neutrino beam mode. Normal mass
hierarchy with $\sin^22\theta_{13}=0.1$ and $\deltacp=0^\circ$ is
assumed.
Compositions of appearance signal, $\numu \to \nue$ and $\numubar \to \nuebar$,
and background events originating from ($\numu + \numubar$) and ($\nue + \nuebar$) are shown separately.
\label{Fig:sens-enurec-nue}
}
\end{figure}

\begin{table}[tbp]%
\caption{\label{Tab:sens-selection-nue}%
The expected number of $\nue/\nuebar$ candidate events and
efficiencies with respect to FCFV events.
Normal mass hierarchy with
$\sin^22\theta_{13}=0.1$ and $\deltacp=0$ are assumed.  Background is
categorized by the flavor before oscillation.}
\begin{center}%
\begin{tabular}{cc|cc|ccccc|c|c} \hline \hline
&	& \multicolumn{2}{c|}{signal} & \multicolumn{6}{c|}{BG} & \multirow{2}{*}{Total} \\ 
&	&~$\numu \to \nue$~	& ~$\numubar \to \nuebar$~ 	&~$\numu$ CC~	&~$\numubar$ CC~	&~$\nue$  CC~& ~$\nuebar$ CC~ & ~NC~ & ~BG Total~	&  \\ \hline  
\multirow{2}{*}{$\nu$ mode~~} & Events	& 1643	&	15& 7 	& 0	 & 248	&11	& 134	&	400 & 2058 \\ 
 & Eff.(\%)  & 63.6 & 47.3 & 0.1 & 0.0 & 24.5 & 12.6 & 1.4 & 1.6 & --- \\
 \hline
\multirow{2}{*}{$\bar{\nu}$ mode~~} & Events	& 206	&	1183& 2	& 2	& 101	& 216	& 196&	517 & 1906 \\ 
 & Eff. (\%) & 45.0 & 70.8 & 0.03 & 0.02 & 13.5 & 30.8 & 1.6 & 1.6 & --- \\
\hline \hline
\end{tabular}%
\end{center}
\end{table}%

Figure~\ref{Fig:sens-enurec-nue} shows the reconstructed neutrino
energy distributions of $\nue/\nuebar$ events after all the
selections. 
The expected number of
$\nue/\nuebar$ candidate events is shown in
Table~\ref{Tab:sens-selection-nue} for each signal and background
component.  The efficiencies of selection with respect to FCFV events
are also shown in Table~\ref{Tab:sens-selection-nue}.  In the neutrino
mode, the dominant background component is intrinsic $\nue$
contamination in the beam.  The mis-identified neutral current $\pi^0$
production events are suppressed thanks to the improved $\pi^0$
reconstruction.  The total rejection factor, including FCFV selection,
for NC $\pi^0$ interactions is $>99.5$\%.  In the anti-neutrino mode,
in addition to $\nuebar$ and $\numubar$, $\nue$ and $\numu$ components
have non-negligible contributions due to larger fluxes and
cross-sections compared to their counterparts in the neutrino mode.

\begin{figure}[tbp]%
\includegraphics[width=0.48\textwidth]{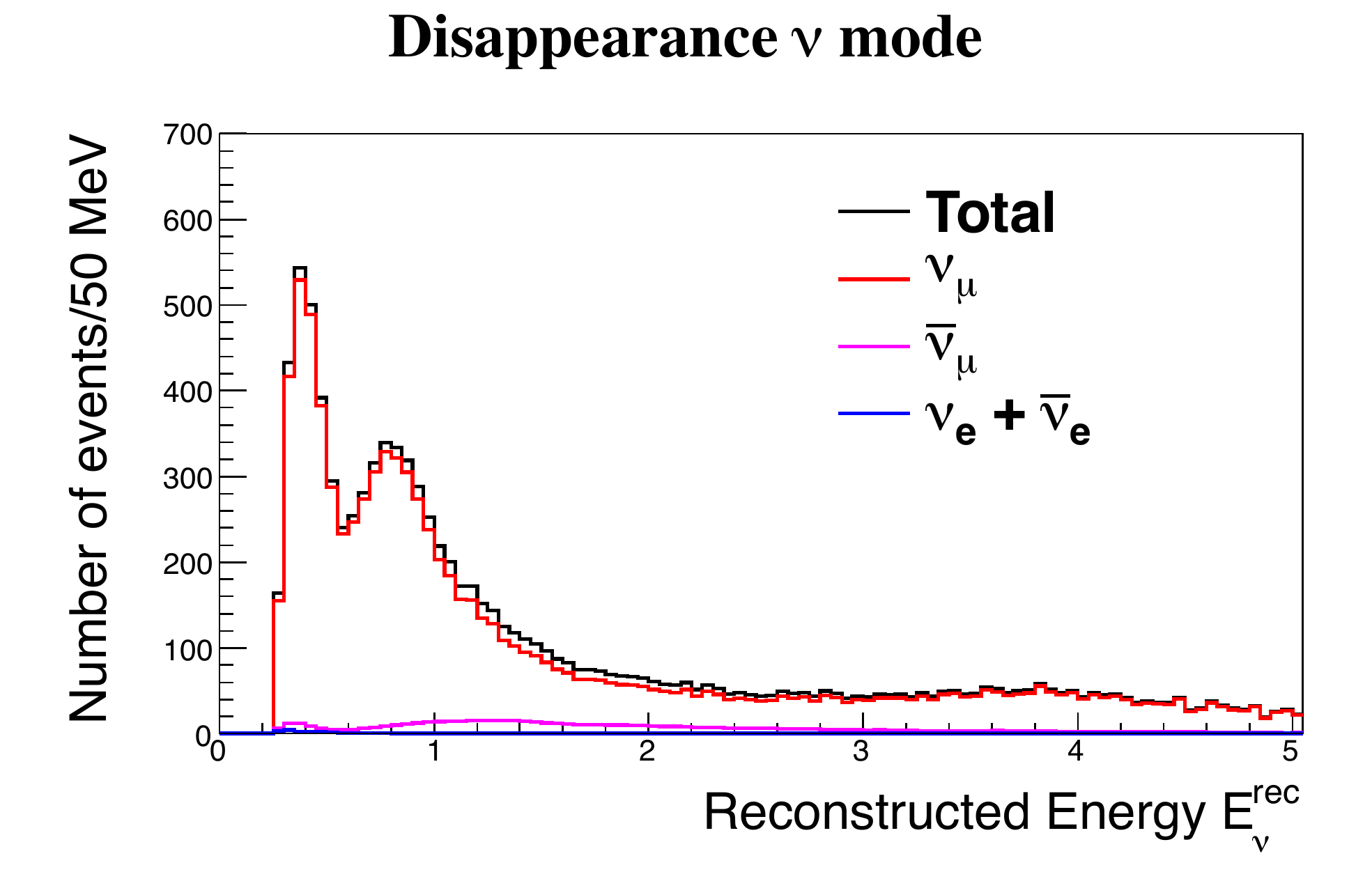}
\includegraphics[width=0.48\textwidth]{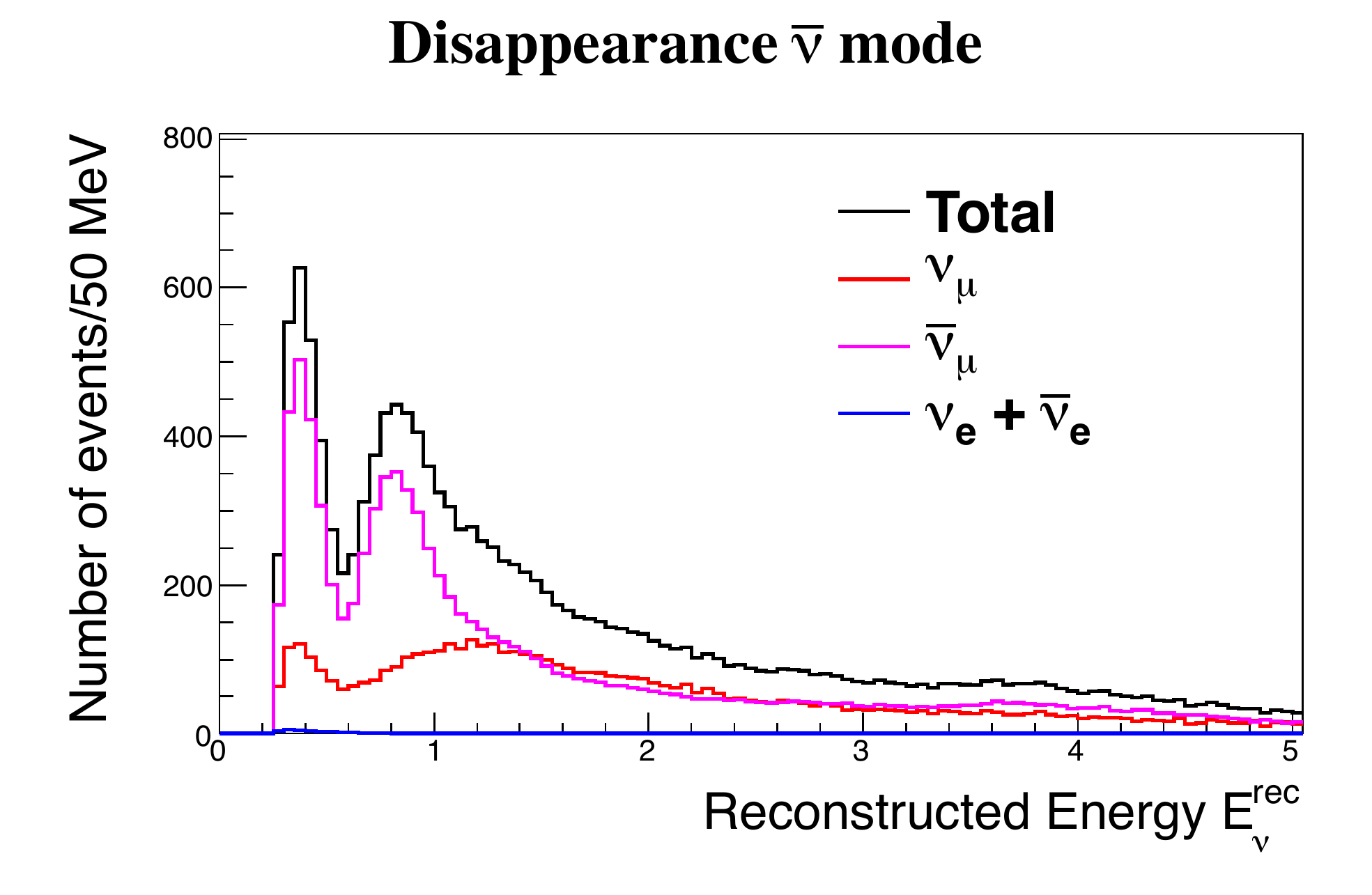}
\caption{%
Reconstructed neutrino energy distribution of the $\numu/\numubar$ candidate events after oscillation.
Left: neutrino beam mode, right: anti-neutrino beam mode.
\label{Fig:sens-enurec-numu}
}
\end{figure}

\begin{table}[tbp]%
\begin{center}%
\caption{\label{Tab:sens-selection-numu}%
The expected number of $\numu/\numubar$ candidate events and efficiencies (with respect to FCFV events) for each flavor and interaction type.
}
\begin{tabular}{lcccccccccc} \hline \hline
			&	&~$\numu$CCQE	& ~$\numu$CC non-QE & ~$\numubar$CCQE	& ~$\numubar$CC non-QE &~$\nue+\nuebar$ CC 	&~NC~ 	& ~$\numu \to \nue$		& ~total~ 		\\ \hline 
\multirow{2}{*}{$\nu$ mode}	& Events	& 6043 & 2981	 &	348 & 		194	& 6				& 480 		& 29			& 10080		 \\ 
 & Eff. (\%) & 91.0 & 20.7 & 95.6 &  53.5 &  0.5 & 8.8 & 1.1& --- \\ \hline
\multirow{2}{*}{$\bar{\nu}$ mode} & Events	& 2699 & 	2354	&	6099 &	 1961 & 7		& 603		& 4  			& 13726		 \\ 
 & Eff. (\%) & 88.0 & 20.1 & 95.4 & 54.8 & 0.4 & 8.8 & 0.7 & ---\\
\hline \hline
\end{tabular}%
\end{center}
\end{table}%

For the \numu/\numubar\ candidate events the following criteria are applied;
the reconstructed ring is identified as muon-like ($\mu$-like),
the reconstructed muon momentum is greater than 200 MeV/$c$, and
there is at most one decay electron associated to the event.

Figure~\ref{Fig:sens-enurec-numu} shows the reconstructed neutrino
energy distributions of the selected $\numu$/$\numubar$ events.
Table~\ref{Tab:sens-selection-numu} shows the number of
$\numu/\numubar$ candidate events for each signal and background
component.  In the neutrino beam mode, the purity of $\numu$ CC
events, after oscillation and for $E_{rec}<1.5$~GeV, is 89\%.  For the
anti-neutrino mode data, the contribution of wrong-sign $\numu$ CC
events is significant because the cross section for neutrino interactions is about
three times larger than anti-neutrino interactions in this energy range.  The
fractions of $\numubar$ and $\numu$ CC events in anti-neutrino beam
mode data after selection, for $E_{rec}<1.5$~GeV, are 66\% and 26\%,
respectively.

\begin{figure}[tbp]
\centering
\includegraphics[width=0.48\textwidth]{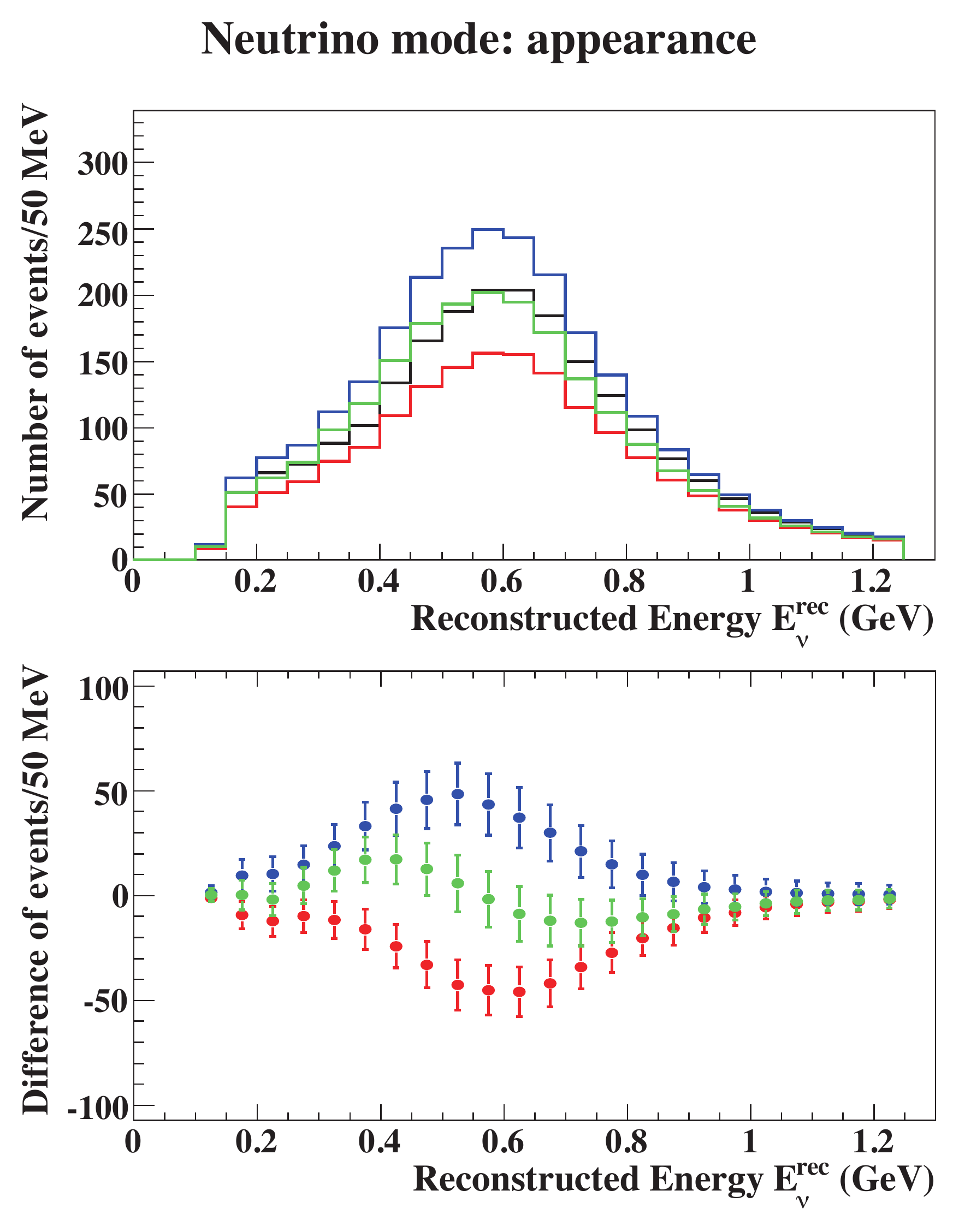}
\includegraphics[width=0.48\textwidth]{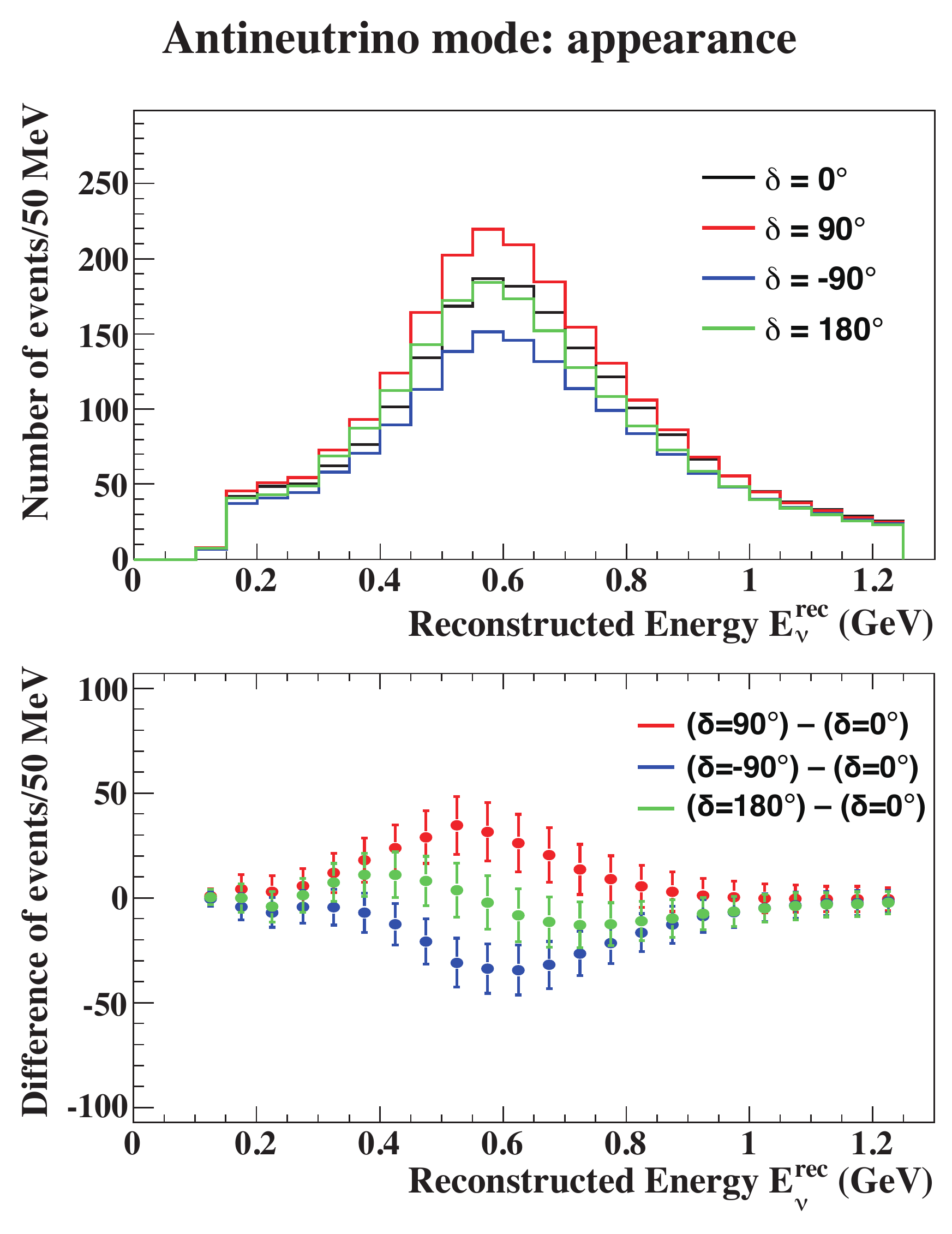} \\
\caption{
Top: Reconstructed neutrino energy distribution for several values of
$\deltacp$.  $\sin^22\theta_{13}=0.1$ and normal hierarchy is assumed.
Bottom: Difference of the reconstructed neutrino energy distribution
from the case with $\deltacp=0^\circ$.  The error bars represent the
statistical uncertainties of each bin.
}
\label{enurecdiff-nue}
\end{figure}

\begin{figure}[tbp]
\centering
\includegraphics[width=0.48\textwidth]{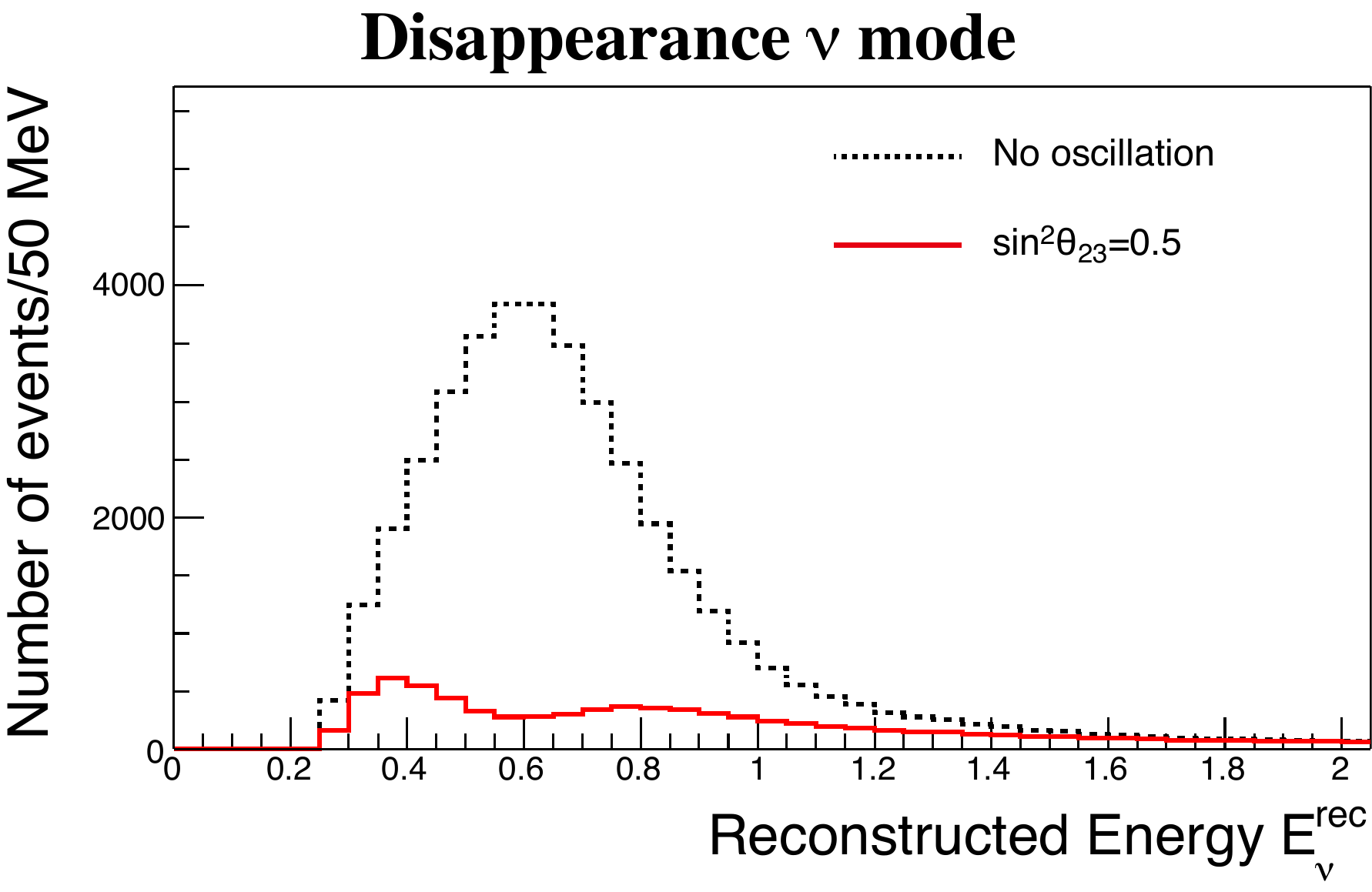}
\includegraphics[width=0.48\textwidth]{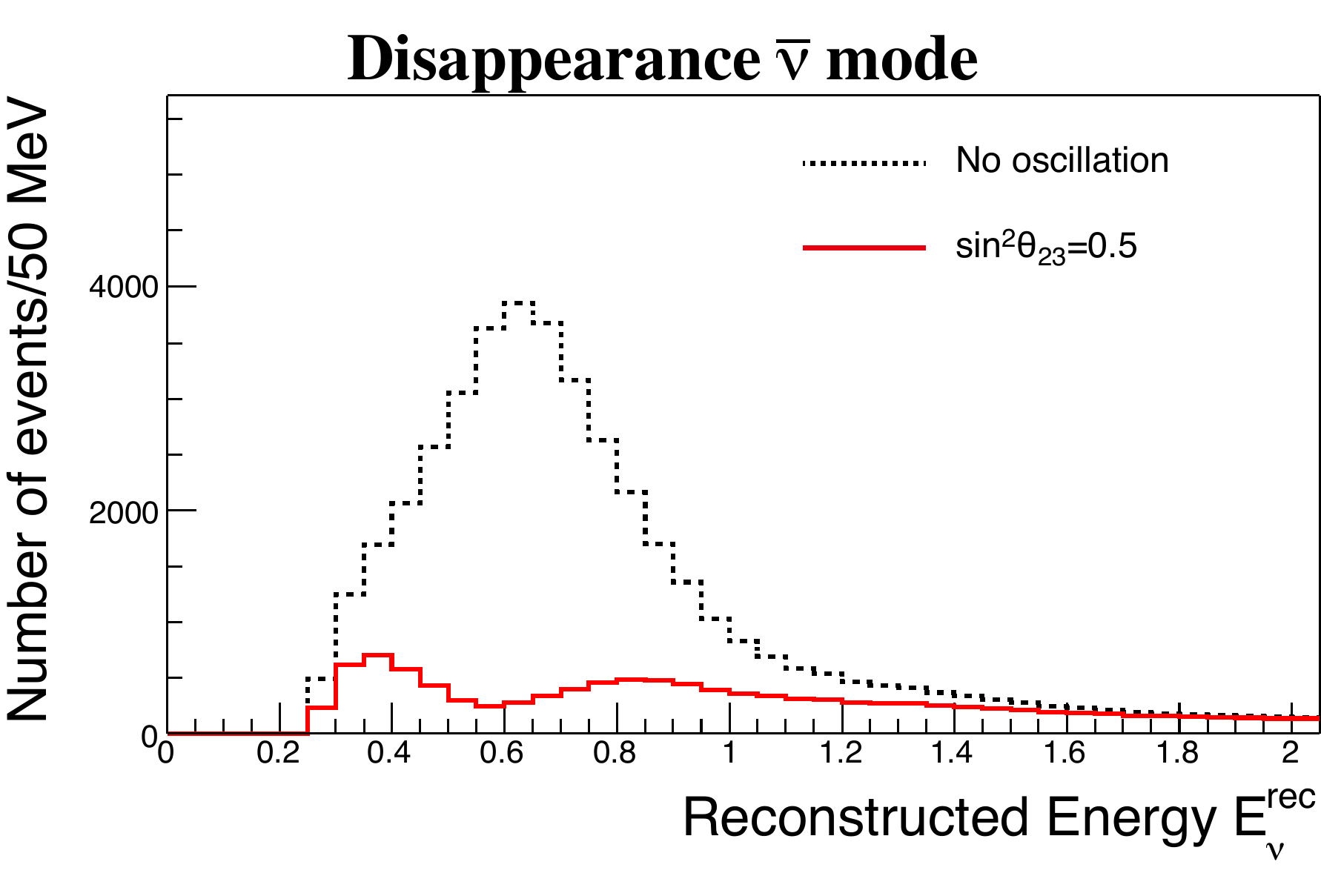}
\caption{
Reconstructed neutrino energy distributions of $\numu$ candidates.
Dotted black lines are for no oscillation case,
while solid red lines represent prediction with oscillation.
Left: for neutrino beam mode. Right: for anti-neutrino beam mode.
}
\label{enurecdiff-numu}
\end{figure}

The reconstructed neutrino energy distributions of $\nue$ events for
several values of $\deltacp$ are shown in the top plots of
Fig.~\ref{enurecdiff-nue}.  The effect of $\deltacp$ is clearly seen
using the reconstructed neutrino energy.  The bottom plots show the
difference of reconstructed energy spectrum from $\deltacp=0^\circ$
for the cases $\deltacp = 90^\circ, -90^\circ$ and $180^\circ$.  The
error bars correspond to the statistical uncertainty.
By using not only the total number of events but also the reconstructed energy
distribution, the sensitivity to $\deltacp$ can be improved and one
can discriminate all the values of $\deltacp$, including the
difference between $\deltacp = 0^\circ$ and $180^\circ$ for which CP
symmetry is conserved.

Figure~\ref{enurecdiff-numu} shows the reconstructed neutrino energy
distributions of the $\numu$ sample, for the cases with
$\sin^2\theta_{23}=0.5$ and without oscillation.  Thanks to the narrow
energy spectrum tuned to the oscillation maximum with off-axis beam,
the effect of oscillation is clearly visible.

\subsubsection{Analysis method}
As described earlier, a binned likelihood analysis based on the
reconstructed neutrino energy distribution is performed to extract the oscillation parameters. 
Both \nue\ appearance and \numu\ disappearance samples, in both neutrino and
antineutrino mode data, are simultaneously fitted.

The $\chi^2$ used in this study is defined as 
\begin{equation} \label{eq:sens:chi2}
\chi^2 =  -2 \ln \mathcal{L}  + P,
\end{equation}
where $\ln \mathcal{L}$ is the log likelihood for a Poisson distribution,
\begin{equation}
-2\ln \mathcal{L} = \sum_k \left\{ -{N_k^\mathrm{test}(1+f_i)} + N_k^\mathrm{hyp} \ln \left[ N_k^\mathrm{test}(1+f_i) \right] \right\}.
\end{equation}
Here, $N_k^\mathrm{hyp}$ and $N_k^\mathrm{test}$ are the number of
events in $k$-th reconstructed energy bin for the hypothesis and test oscillation parameters, respectively.
The index $k$ runs over all reconstructed energy bins for muon and electron neutrino samples and for neutrino and anti-neutrino mode data.  
The parameters $f_i$ represent fractional variations of the bin entries due to systematic uncertainties.

The penalty term $P$ in Eq.~\ref{eq:sens:chi2} constrains the systematic parameters $f_i$ with the normalized covariance matrix $C$,
\begin{equation}
P = \sum_{i,j} f_i (C^{-1})_{i,j} f_j.
\end{equation}
In order to reduce the number of the systematic parameters, several
reconstructed energy bins that have similar covariance values are
merged for $f_i$.

A robust estimate of the uncertainties is possible based on the T2K experience.
For each of three main categories of systematic uncertainties, we have made the following assumptions taking into account improvements expected with future T2K data and analysis improvements.
\begin{description}
\item[i) Flux and cross section uncertainties constrained by the fit to near detector data] 
Data from near detectors will be used in conjunction with models for
the neutrino beam, neutrino interactions, and the detector performance
to improve our predictions of the flux at SK and some cross-section
parameters.  The understanding of the neutrino beam, interaction, and
detector is expected to improve in the future, which will result in
reduction of uncertainties in this category.
On the other hand, the near detector analysis is expected to include
more samples to reduce the uncertainty for category ii), which will
result in migration of some errors into this category.  This category
of uncertainties is assumed to stay at the same level as currently
estimated by T2K.
\item[ii) Cross section uncertainties not constrained by the fit to near detector data ]
This category of error stems from the cross-section parameters which are independent between 
the near and far detectors because of their different elemental composition and the cross-section 
parameters for which the near detector is insensitive.
In T2K, an intensive effort has been made to include more samples into analysis, such as data 
from FGD2 containing a water target~\cite{Abe:2017uxa} and large scattering angle events, to 
provide more constraints on the cross section models.
Further improvement is expected in future analysis as T2K accumulates and analyze more data.
In addition, the intermediate detector will significantly reduce the uncertainty due to 
the neutrino interaction models.
\item[iii) Uncertainties on the far detector efficiency and reconstruction modeling]
Because most of the uncertainties related to far detector performance
are estimated by using atmospheric neutrinos as a control sample and
the current error is limited by statistics, errors in this category
are expected to decrease with much larger statistics available with 
Hyper-K than currently used for T2K.
Uncertainties arising from the energy scale are kept the same because
they are not estimated by the atmospheric neutrino sample, although it
could be also reduced with a larger statistics control sample and
better calibration of the detector.
\end{description}
Compared to the systematic uncertainty used for the past 
publication~\cite{Abe:2015zbg}, the uncertainties for anti-neutrino beam 
mode have been reduced to a similar level as those for neutrino beam mode, 
based on the experience with T2K anti-neutrino oscillation analysis.
The flux and cross section uncertainties are assumed to be
uncorrelated between the neutrino and anti-neutrino data, except for
the uncertainty of \nue/\numu\ cross section ratio which is treated to
be anti-correlated considering the theoretical uncertainties studied
in~\cite{Day:2012gb}.  Because some of the uncertainties, such as
those from the cross section modeling or near detector systematics,
are expected to be correlated and give more of a constraint, this is a
conservative assumption.  The far detector uncertainty is treated to
be fully correlated between the neutrino and anti-neutrino data.

\begin{figure}[tbp]
\centering
\includegraphics[width=0.48\textwidth]{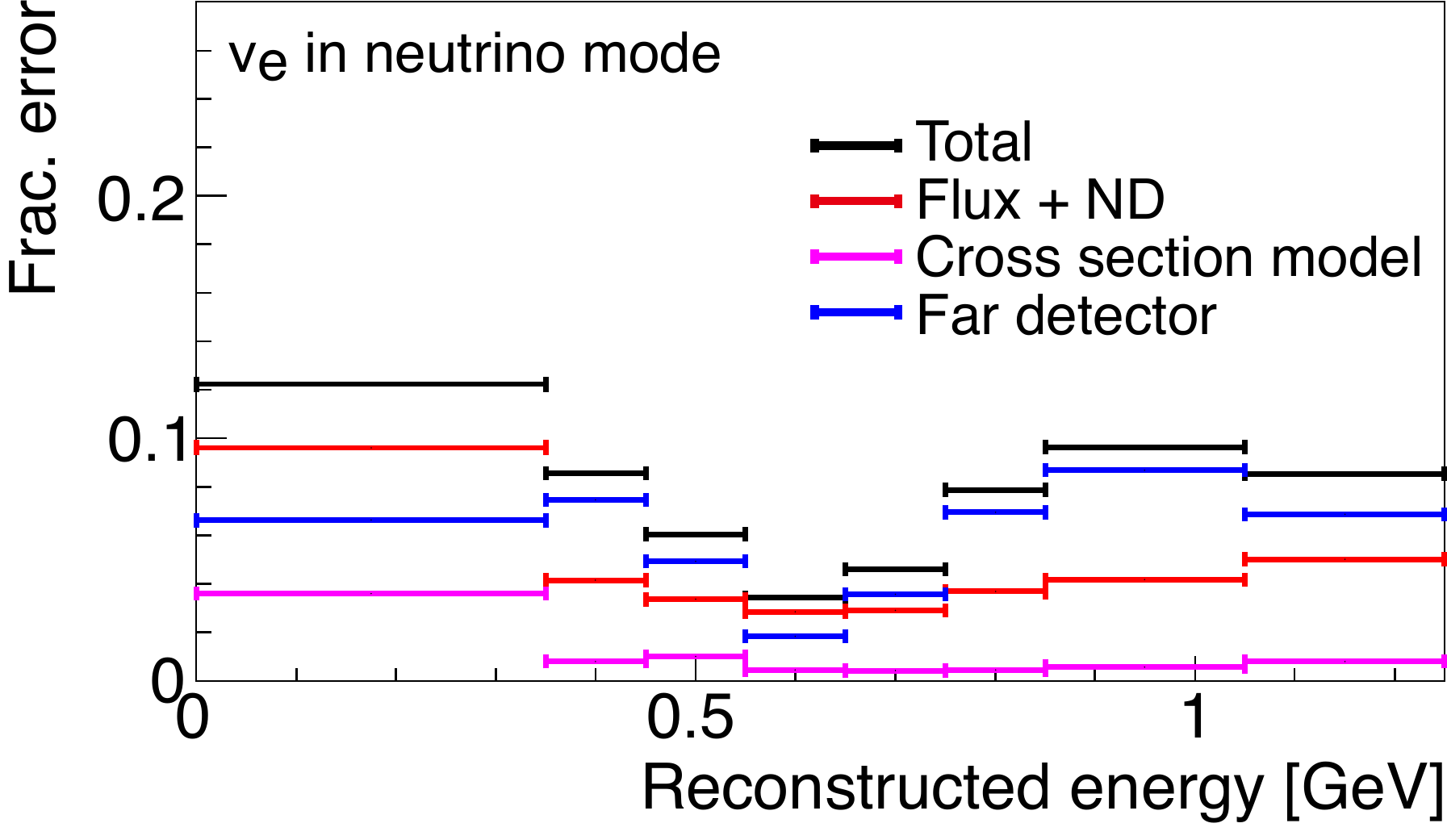}
\includegraphics[width=0.48\textwidth]{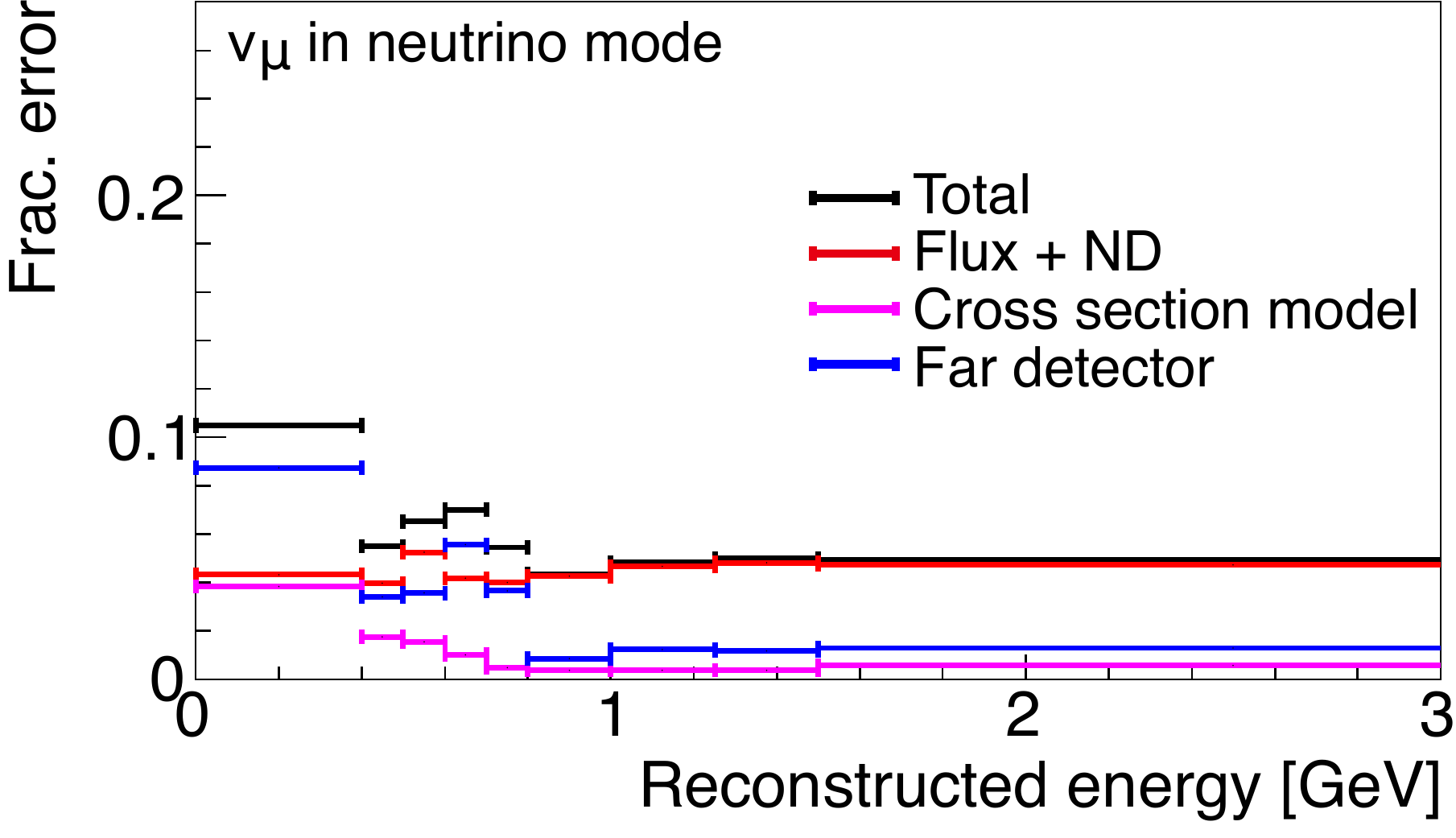}
\caption{
Fractional and total systematic error size for the appearance (left) and the disappearance
(right) samples in the neutrino mode.  Black: total uncertainty, red:
the flux and cross-section constrained by the near detector, magenta:
the near detector non-constrained cross section, blue: the far
detector error.
\label{Fig:systerror}
}
\end{figure}

\begin{figure}[tbp]
\centering
\includegraphics[width=0.48\textwidth]{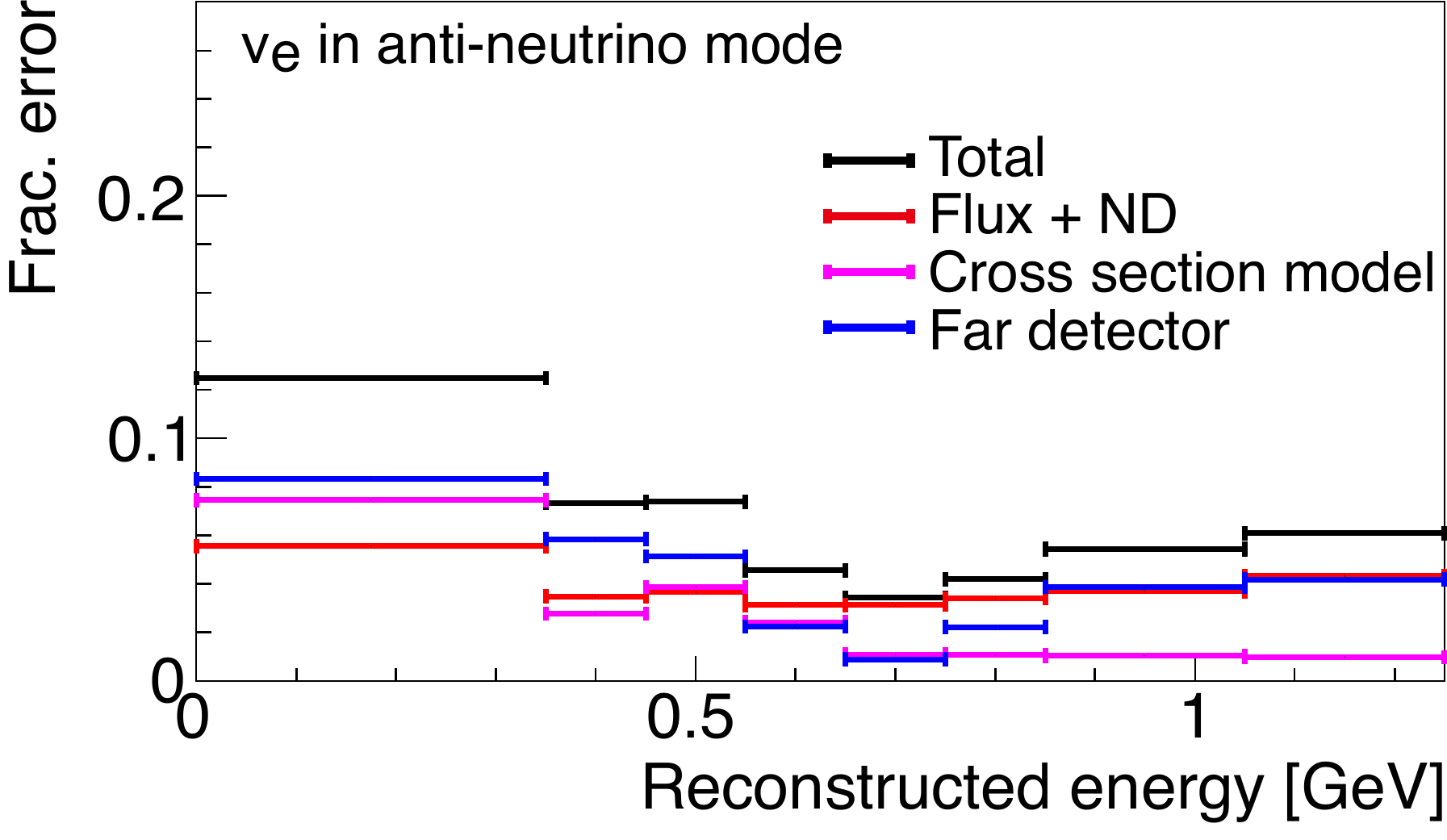}
\includegraphics[width=0.48\textwidth]{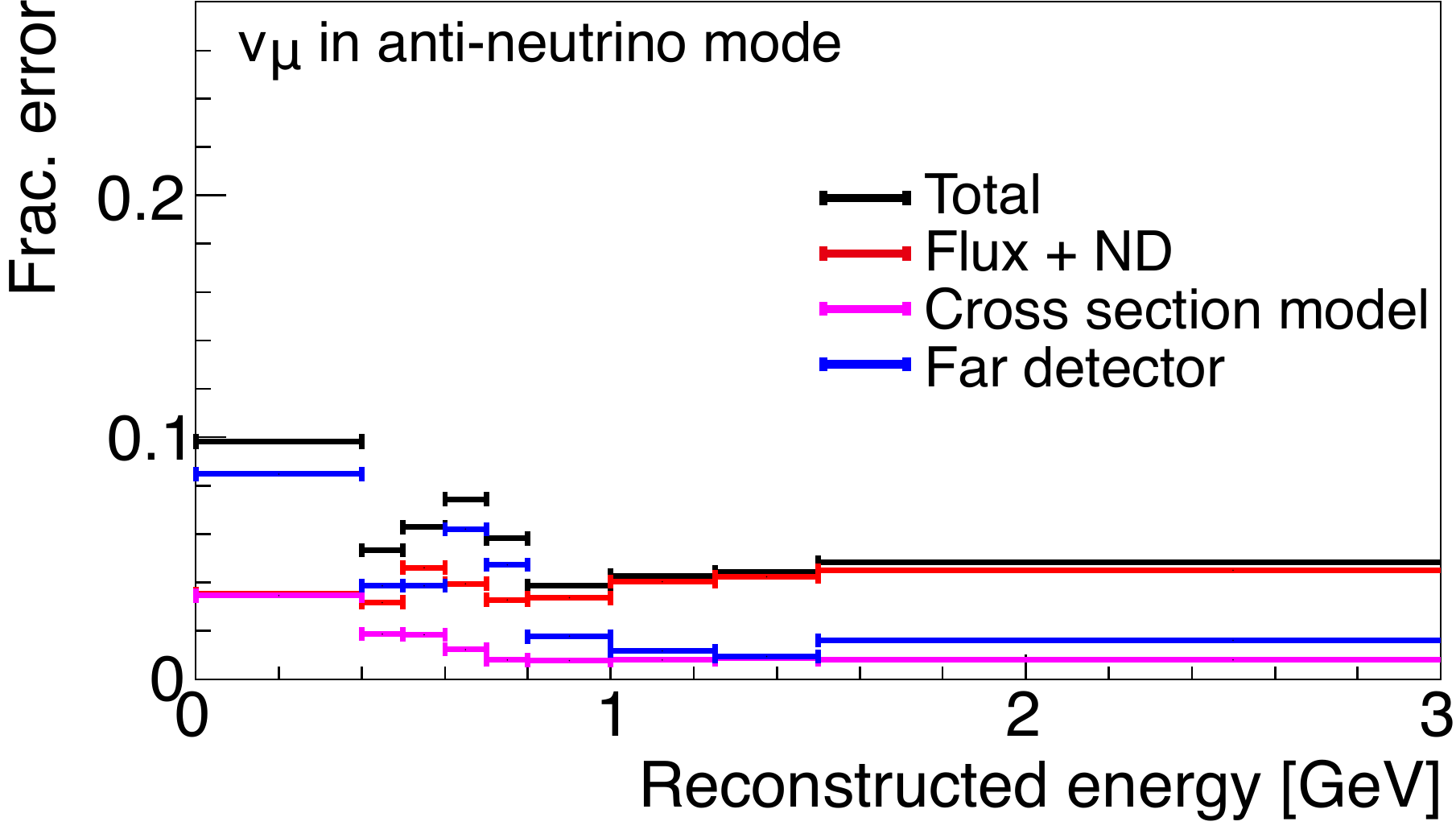}
\caption{
Fractional and total systematic error size for the appearance (left) and the disappearance
(right) samples in the anti-neutrino mode.  Black: total uncertainty,
red: the flux and cross-section constrained by the near detector,
magenta: the near detector non-constrained cross section, blue: the
far detector error.
\label{Fig:systerror-anti}
}
\end{figure}

\begin{figure}[tbp]
\centering
\includegraphics[width=0.6\textwidth]{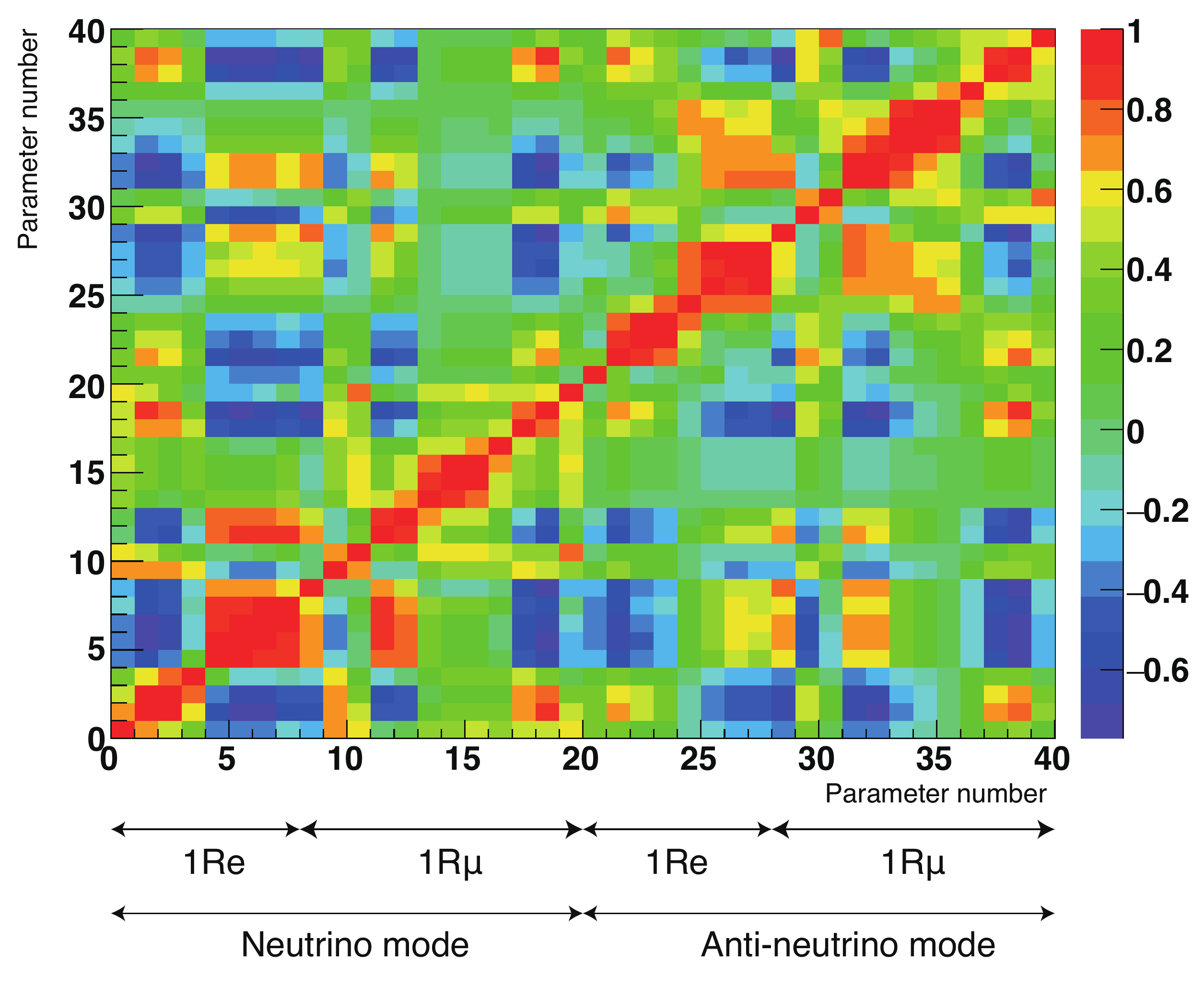}
\caption{
Correlation matrix between reconstructed energy bins of the four samples due to the systematic uncertainties.
Bins 1--8, 9--20, 21--28, and 29--40 correspond to 
the neutrino mode single ring $e$-like,
the neutrino mode single ring $\mu$-like,
the anti-neutrino mode single ring $e$-like, and
the anti-neutrino mode single ring $\mu$-like samples, respectively.
\label{Fig:correlationmatrix}
}
\end{figure}

Figures~\ref{Fig:systerror} and \ref{Fig:systerror-anti} show the
fractional and total systematic uncertainties for the appearance and
disappearance reconstructed energy spectra in neutrino and
anti-neutrino mode, respectively.  Black lines represent the prior
uncertainties and bin widths of the systematic parameters $f_i$, while
colored lines show the contribution from each uncertainty source.  It
should be noted that because some uncertainties are correlated between
bins, the uncertainty on the total number of events is not a simple
flux-weighted sum of these errors.  For example, the energy scale
uncertainty of the far detector has a large contribution around the
flux peak, but it does not change the total number of events.
Figure~\ref{Fig:correlationmatrix} shows the correlation matrix of the
systematic uncertainties between the reconstructed neutrino energy
bins of the four samples.  The systematic uncertainties of the number
of expected events at the far detector are summarized in
Table~\ref{tab:sens:systsummary}.

\begin{table}[htbp]
\caption{Uncertainties for the expected number of events at Hyper-K from the systematic uncertainties assumed in this study.}
\centering
\begin{tabular}{cccccc}  \hline \hline
&  & ~~Flux \& ND-constrained~~ & ~~ND-independent~~  & \multirow{2}{*}{~~Far detector~~}  & \multirow{2}{*}{Total} \\
 & &   cross section &  cross section \\ \hline
\multirow{2}{*}{$\nu$ mode~~} 			& Appearance 	& 3.0\% & 0.5\% & 0.7\% & 3.2\% \\
									& Disappearance 	& 3.3\% & 0.9\% & 1.0\% & 3.6\% \\ \hline
\multirow{2}{*}{$\overline{\nu}$ mode~~}	& Appearance 	& 3.2\% & 1.5\% & 1.5\% & 3.9\% \\
									& Disappearance 	& 3.3\% & 0.9\% & 1.1\% & 3.6\% \\
\hline \hline
\end{tabular}

\label{tab:sens:systsummary}
\end{table}%

\subsubsection{Measurement of $CP$ asymmetry}

\begin{figure}[tbp]
\centering
\includegraphics[width=0.48\textwidth]{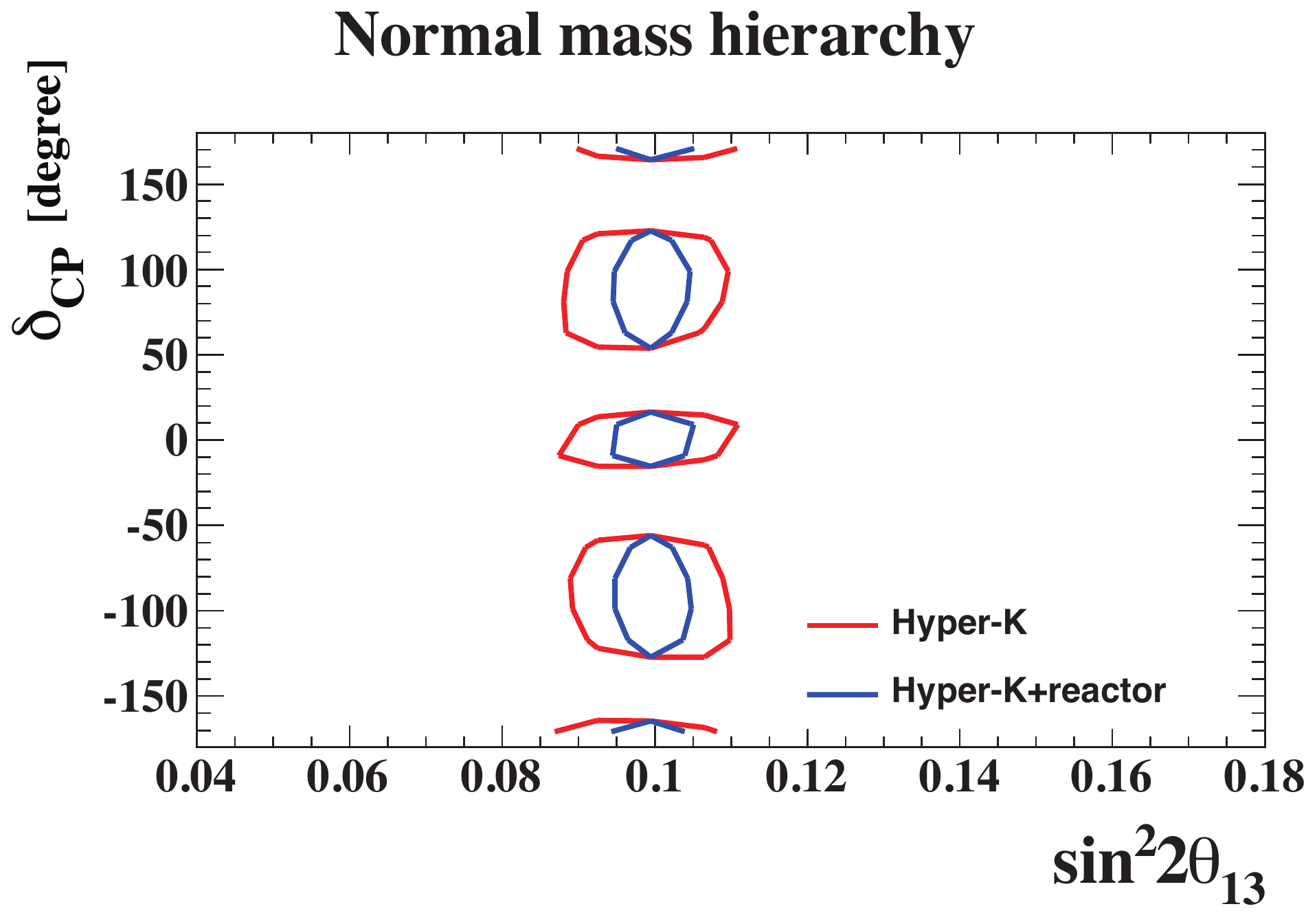}
\includegraphics[width=0.48\textwidth]{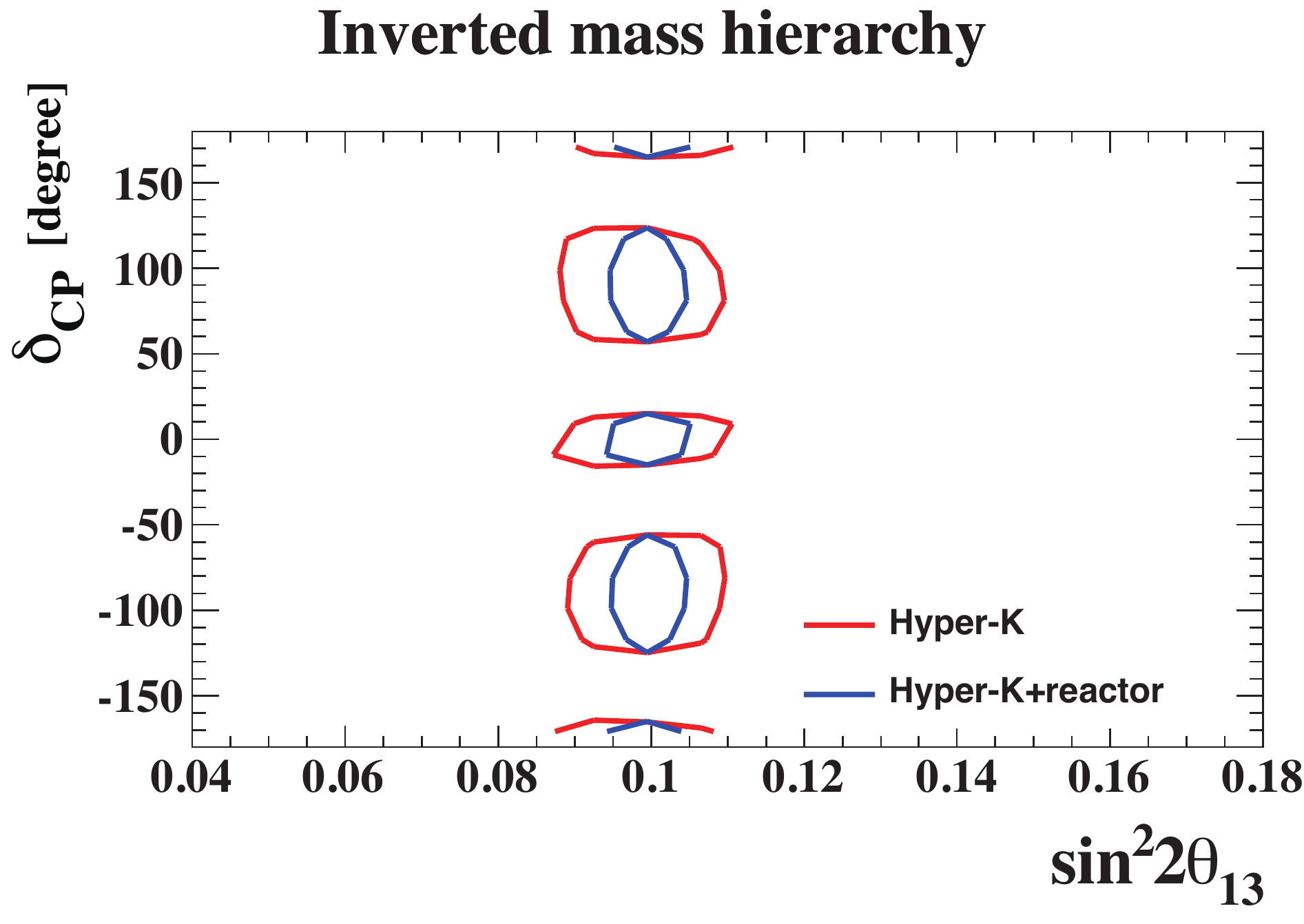}
\caption{The expected 90\% CL allowed regions in the $\sin^22\theta_{13}$-$\deltacp$ plane.
The results for the true values of $\deltacp = (-90^\circ, 0, 90^\circ, 180^\circ)$ are shown.
Left: normal hierarchy case. Right: inverted hierarchy case.
Red (blue) lines show the result with Hyper-K only (with $\sin^22\theta_{13}$ constraint from reactor experiments). 
\label{fig:CP-contour}}
\end{figure}

\begin{figure}[tbp]
\centering
\includegraphics[width=0.5\textwidth]{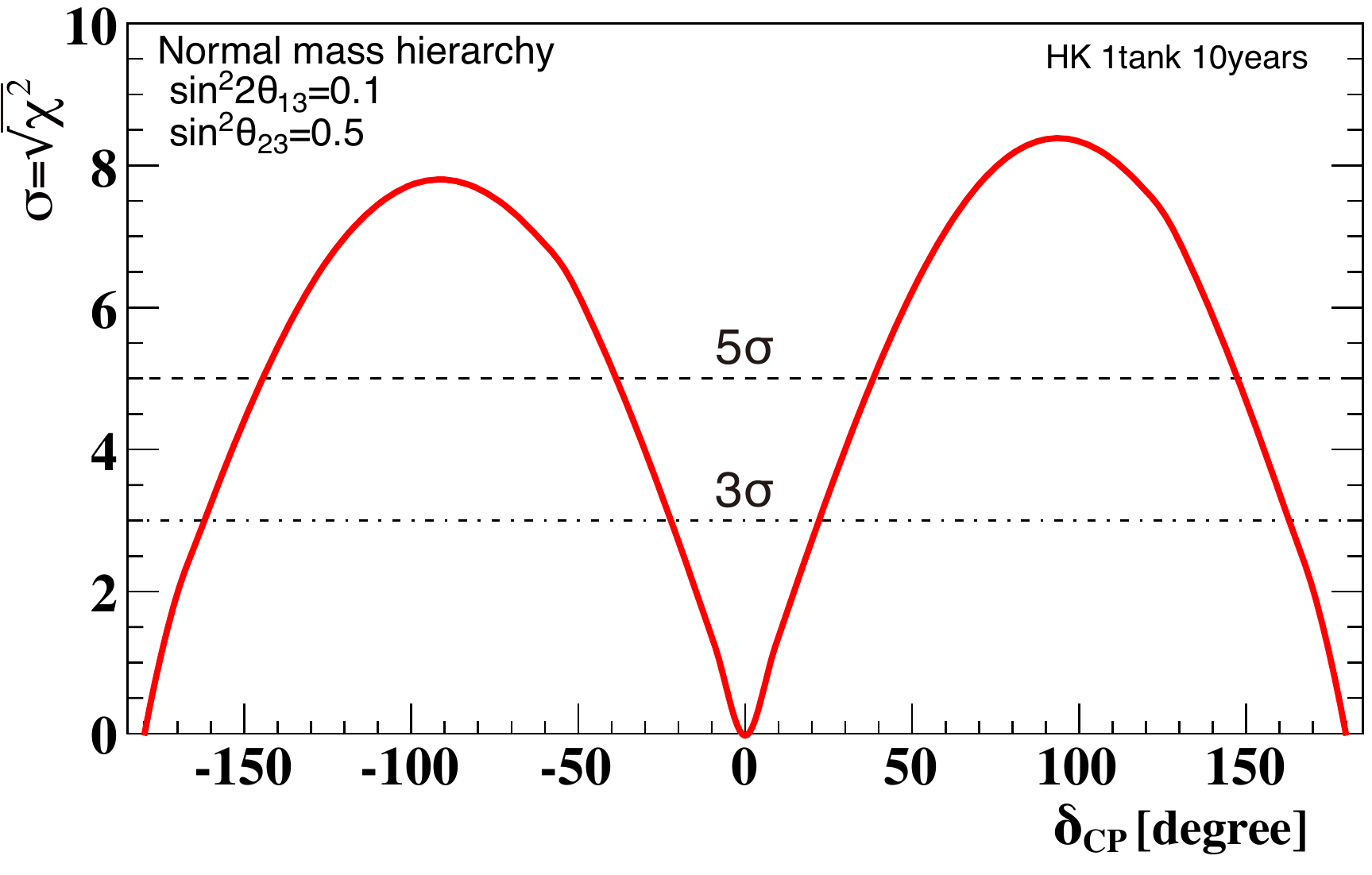}
\caption{Expected significance to exclude $\sin\deltacp = 0$ in case of normal hierarchy.  Mass hierarchy is assumed to be known.
\label{fig:CP-chi2}}
\end{figure}

\begin{figure}[tbp]
\centering
\includegraphics[width=0.6\textwidth]{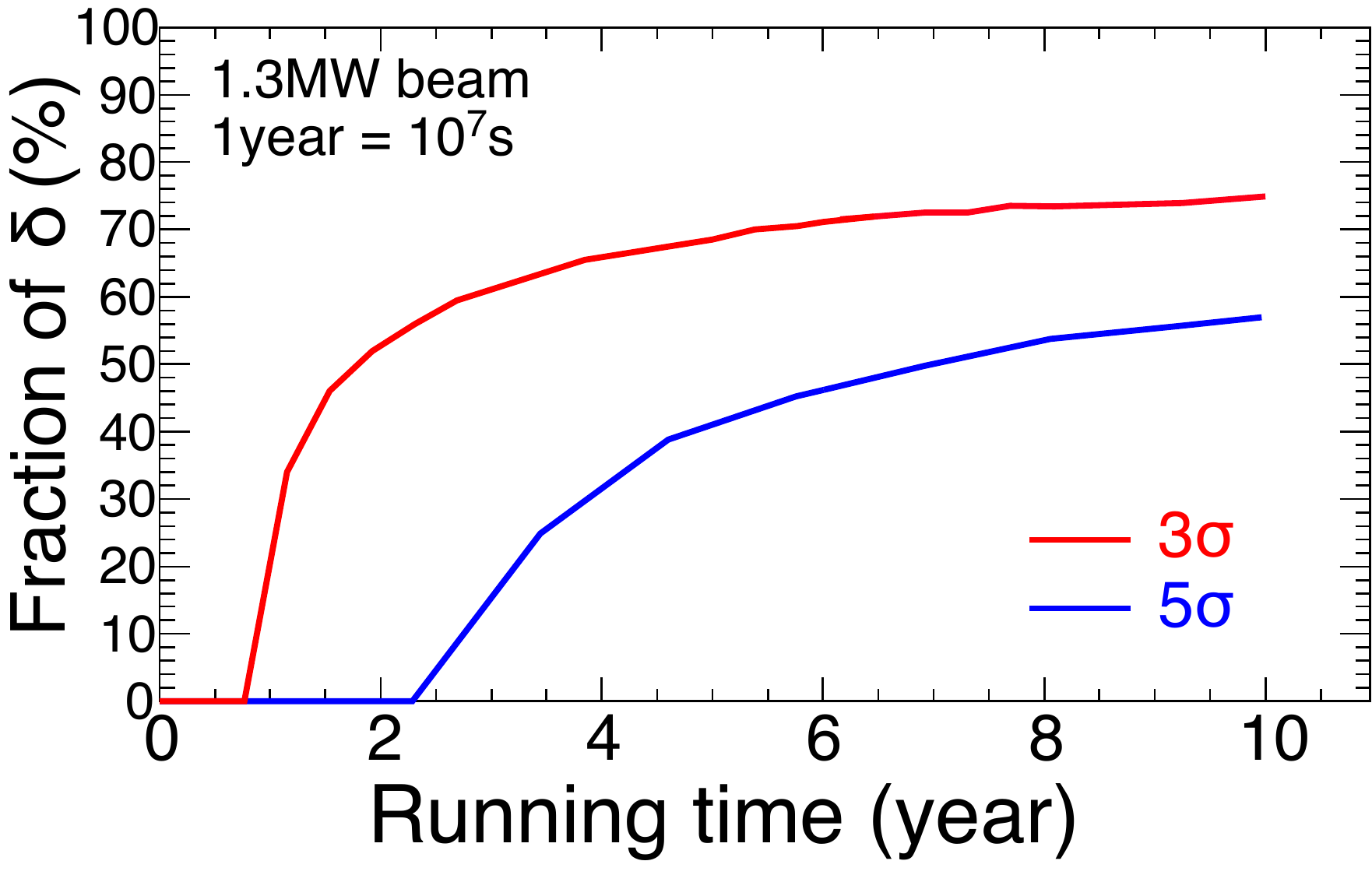}
\caption{Fraction of $\deltacp$ for which $\sin\deltacp= 0$ can be excluded with  more than 3\,$\sigma$ (red) and 5\,$\sigma$ (blue) significance as a function of the running time. 
For the normal hierarchy case, and mass hierarchy is assumed to be known.
The ratio of neutrino and anti-neutrino mode is fixed to 1:3. 
\label{fig:delta-sens-time}}
\end{figure}

Figure~\ref{fig:CP-contour} shows examples of the 90\% CL allowed
regions on the $\sin^22\theta_{13}$--$\deltacp$ plane resulting from
the true values of $\deltacp = (-90^\circ, 0, 90^\circ, 180^\circ)$.
The left (right) plot shows the case for the normal (inverted) mass
hierarchy.  Also shown are the allowed regions when we include a
constraint from the reactor experiments,
$\sin^22\theta_{13}=0.100 \pm0.005$.  With reactor constraints,
although the contour becomes narrower in the direction of
$\sin^22\theta_{13}$, the sensitivity to $\deltacp$ does not
significantly change because $\delta_{CP}$ is constrained by the
comparison of neutrino and anti-neutrino oscillation probabilities by
Hyper-K and not limited by the uncertainty of $\theta_{13}$.

Figure~\ref{fig:CP-chi2} shows the expected significance to exclude
$\sin\deltacp = 0$ (the $CP$ conserved case).  The significance is
calculated as $\sqrt{\Delta \chi^2}$, where $\Delta \chi^2$ is the
difference of $\chi^2$ for the \textit{trial} value of \deltacp\ and
for $\deltacp = 0^\circ$ or 180$^\circ$ (the smaller value of
difference is taken).  We have also studied the case with a reactor
constraint, but the result changes only slightly.
Figure~\ref{fig:delta-sens-time} shows the fraction of $\deltacp$ for
which $\sin\deltacp= 0$ is excluded with more than 3\,$\sigma$ and
5\,$\sigma$ of significance as a function of the integrated beam
power.  The ratio of integrated beam power for the neutrino and
anti-neutrino mode is fixed to 1:3.  The normal mass hierarchy is
assumed.  The results for the inverted hierarchy are almost the same.
$CP$ violation in the lepton sector can be observed with more than
3(5)\,$\sigma$ significance for 76(57)\% of the possible values of $\deltacp$.

\begin{figure}[tbp]
\centering
\includegraphics[width=0.6\textwidth]{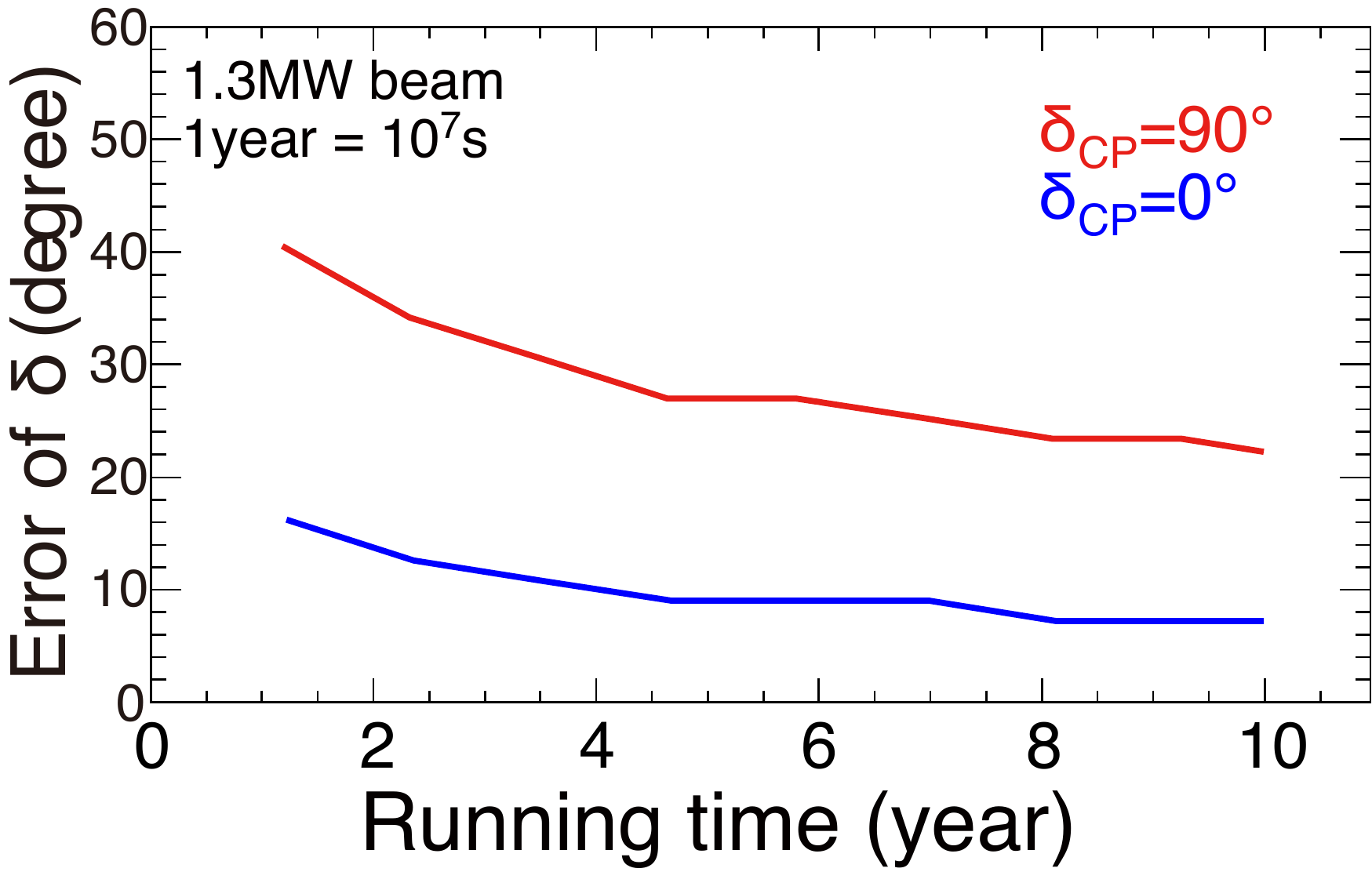}
\caption{Expected 68\% CL uncertainty of $\deltacp$ as a function of running time.  
For the normal hierarchy case, and mass hierarchy is assumed to be known. 
\label{fig:delta-error-time}}
\end{figure}

\begin{figure}[tbp]
\centering
\includegraphics[width=0.6\textwidth]{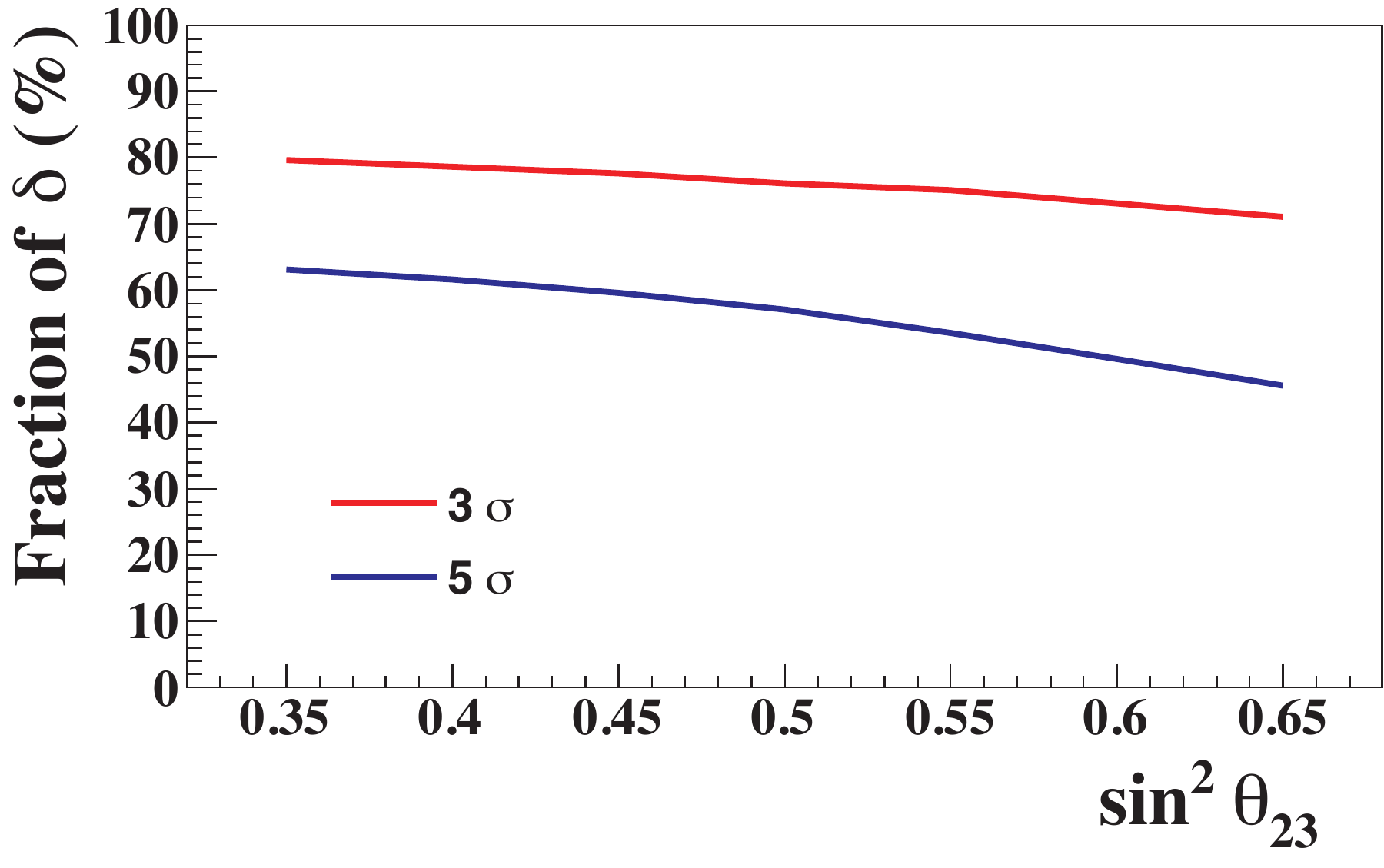}
\caption{Fraction of $\deltacp$ for which $\sin\deltacp= 0$ can be excluded with more than 3\,$\sigma$ (red) and 5\,$\sigma$ (blue) significance as a function of the true value of $\sin^2\theta_{23}$, for the normal hierarchy case.
\label{fig:delta-theta23}}
\end{figure}

Figure~\ref{fig:delta-error-time} shows the 68\% CL uncertainty of
$\deltacp$ as a function of the integrated beam power.
The value of $\deltacp$ can be determined with an
uncertainty of 7.2$^\circ$ for $\deltacp=0^\circ$ or $180^\circ$, and
23$^\circ$ for $\deltacp=\pm90^\circ$.

As the nominal value we use $\sin^2\theta_{23}=0.5$, but the
sensitivity to $CP$ violation depends on the value of $\theta_{23}$.
Figure~\ref{fig:delta-theta23} shows the fraction of $\deltacp$ for
which $\sin\deltacp= 0$ is excluded with more than 3\,$\sigma$ and
5\,$\sigma$ of significance as a function of the true value of
$\sin^2\theta_{23}$.
T2K collaboration reported $\sin^{2}\theta_{23}=0.55^{+0.05}_{-0.09}$ 
in case of the normal hierarchy~\cite{Abe:2017vif}.

\begin{table}[htbp]
\caption{Comparison of CP sensitivity with different configurations. As a reference, the former results published in PTEP~\cite{Abe:2015zbg} are also shown, where 560kton fiducial volume, 7.5~MW$\times 10^7$s integrated beam power, and an old estimate of systematic uncertainty with larger anti-neutrino errors are assumed.}
\begin{center}
\begin{tabular}{l|cc|cc}
                     & \multicolumn{2}{c|}{($\sin\deltacp=0$) exclusion} & \multicolumn{2}{c}{68\% uncertainty of $\deltacp$}\\
Configuration & ~~$>3\sigma$~~ & ~~$>5\sigma$~~  & ~~$\deltacp=0^\circ$~~ & ~~$\deltacp=90^\circ$~~\\ \hline \hline
1 tank & 76\% & 57\% & 7.2$^\circ$ & 23$^\circ$\\ \hline
Staging & 78\% & 62\% & 7.2$^\circ$ & 21$^\circ$ \\ \hline
PTEP~\cite{Abe:2015zbg} & 76\% & 58\% & 7$^\circ$ & 19$^\circ$\\ \hline
\end{tabular}
\end{center}
\label{tab:cp-tank-comparison}
\end{table}%

Table~\ref{tab:cp-tank-comparison} shows a comparison of several configurations for CP violation sensitivities.

\subsubsection{Precise measurements of $\Delta m^2_{32}$ and $\sin^2\theta_{23}$}
\begin{figure}[tbp]
\centering
\includegraphics[width=0.6\textwidth]{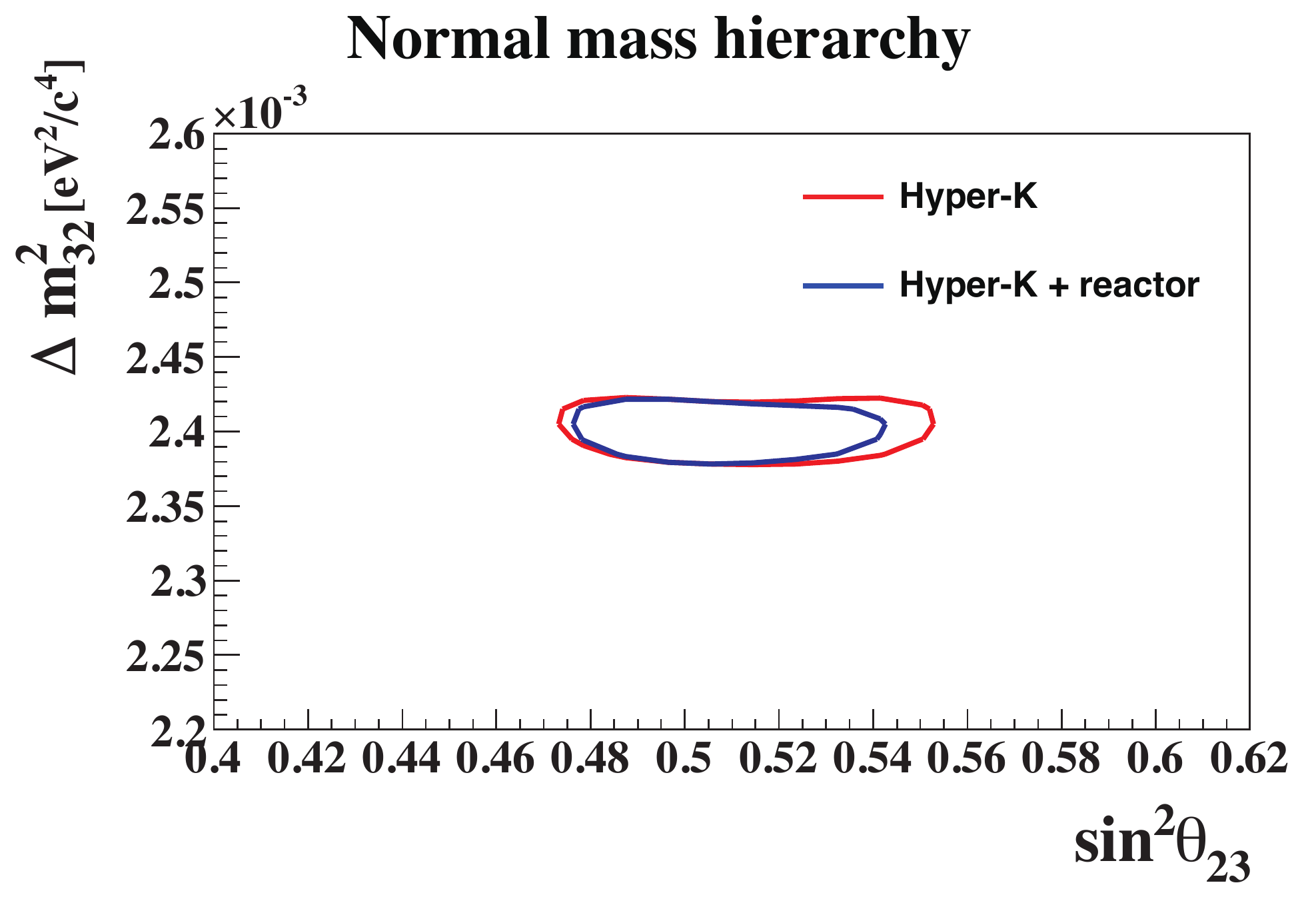}
\caption{ The 90\% CL allowed regions in the $\sin^2\theta_{23}$--$\Delta m^2_{32}$ plane.
The true values are $\sin^2\theta_{23}=0.5$ and $\Delta m^2_{32} = 2.4 \times 10^{-3}$~eV$^2$.
Effect of systematic uncertainties is included. The red (blue) line corresponds to the result with Hyper-K alone (with a reactor constraint on $\sin^22\theta_{13}$).
\label{fig:theta23-0.50}}
\end{figure}

\begin{figure}[tbp]
\centering
\includegraphics[width=0.48\textwidth]{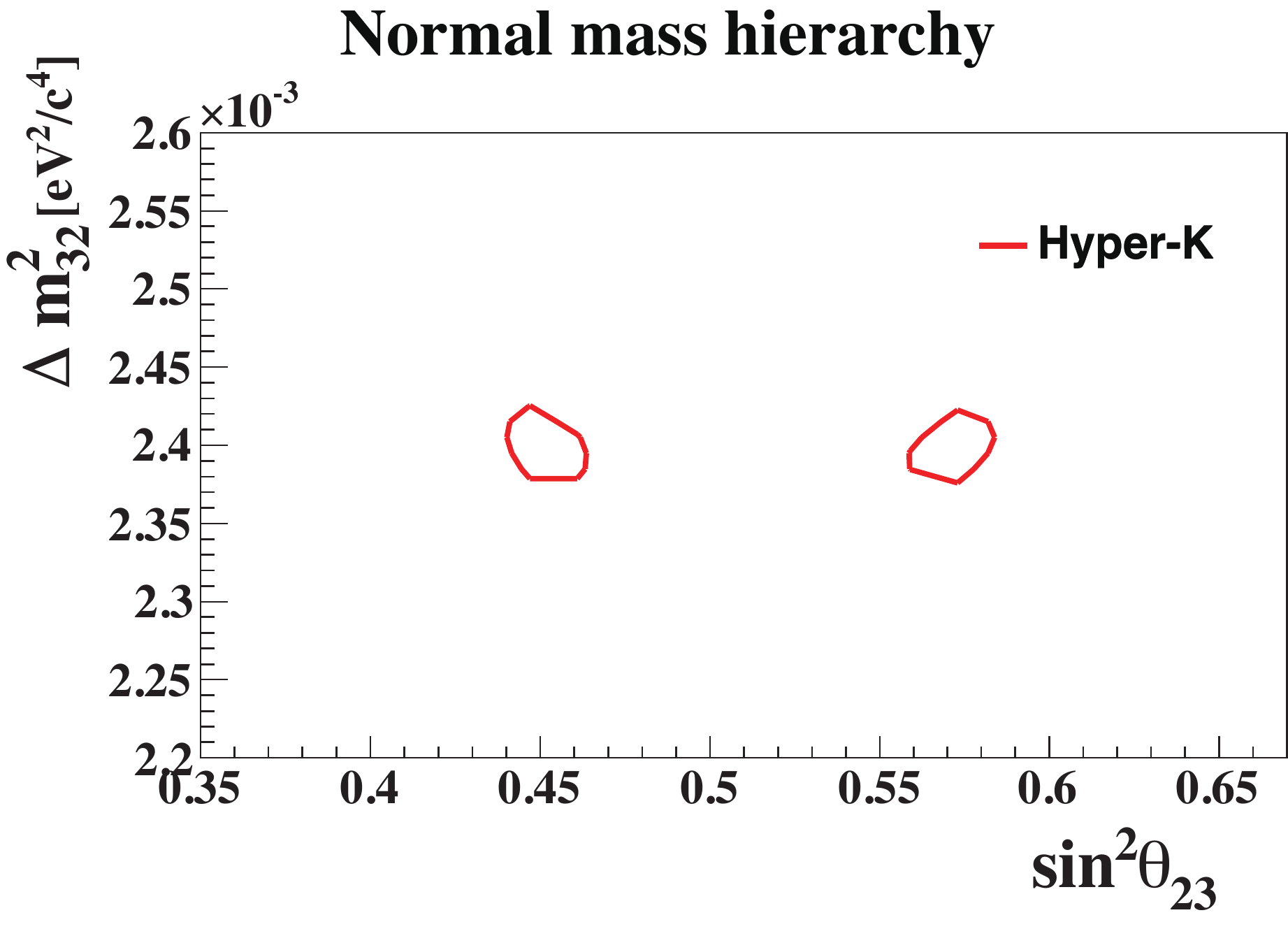}
\includegraphics[width=0.48\textwidth]{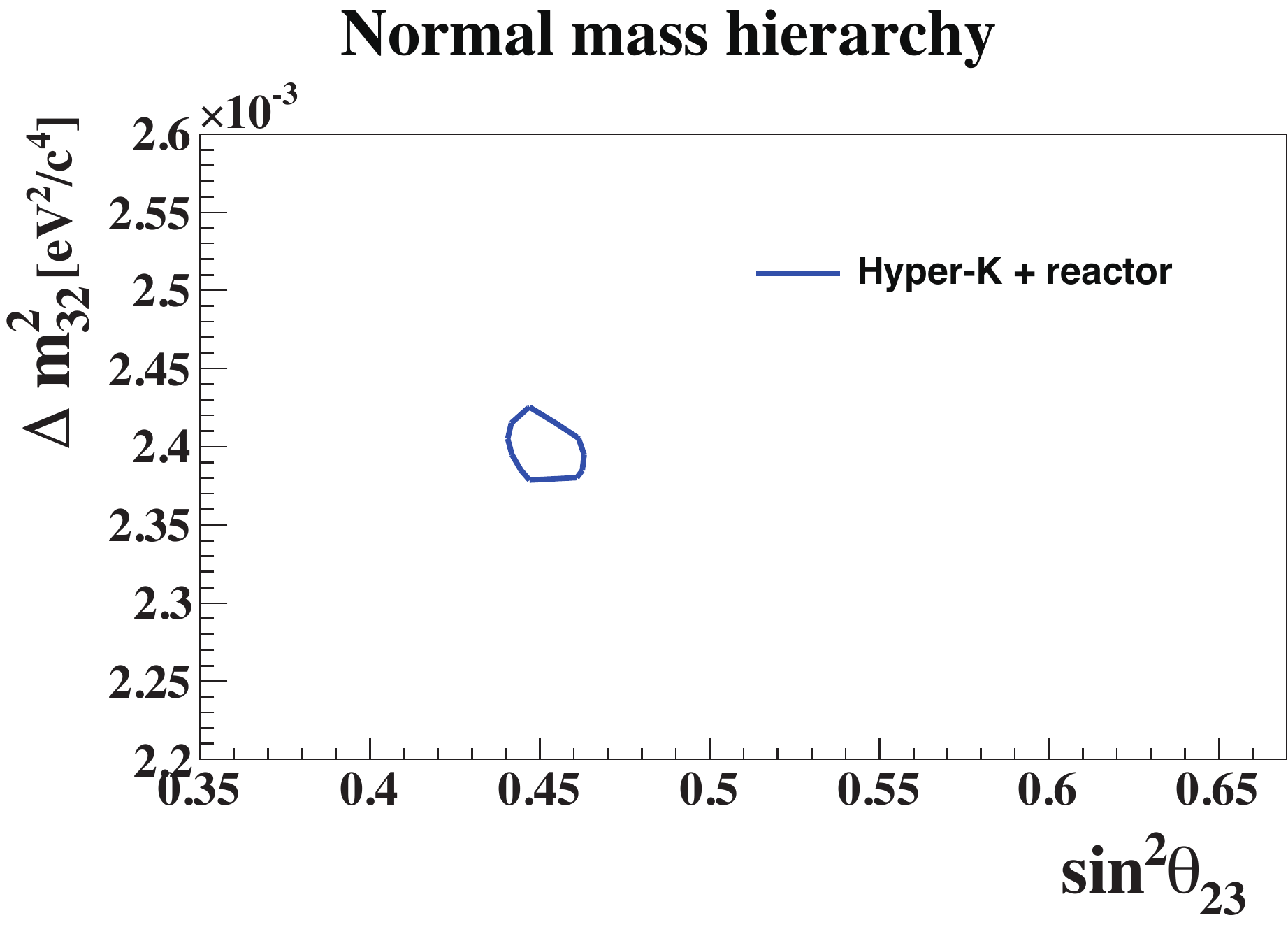}
\caption{ 90\% CL allowed regions in the $\sin^2\theta_{23}$--$\Delta m^2_{32}$ plane.
The true values are $\sin^2\theta_{23}=0.45$ and $\Delta m^2_{32} = 2.4 \times 10^{-3}$~eV$^2$.
Effect of systematic uncertainties is included.
Left: Hyper-K only. Right: With a reactor constraint. 
\label{fig:theta23-0.45}}
\end{figure}

\begin{table}[tbp]
\caption{Expected 1$\sigma$ uncertainty of $\Delta m^2_{32}$ and $\sin^2\theta_{23}$  for true $\sin^2\theta_{23}=0.45, 0.50, 0.55$. 
Reactor constraint on $\sin^22\theta_{13}=0.1\pm 0.005$ is imposed.}
\begin{center}
\begin{tabular}{ccccccc} \hline \hline
True $\sin^2\theta_{23}$	& \multicolumn{2}{c}{$0.45$}  		& \multicolumn{2}{c}{$0.50$} 		& \multicolumn{2}{c}{$0.55$}\\ 
Parameter  				& $\Delta m^2_{32}$ (eV$^2$) 	& $\sin^2\theta_{23}$  & $\Delta m^2_{32}$	(eV$^2$)	& $\sin^2\theta_{23}$ 	& $\Delta m^2_{32}$ (eV$^2$)	& $\sin^2\theta_{23}$\\ \hline
 NH	& $1.4\times10^{-5}$		& 0.006 			& $1.4\times10^{-5}$		& 0.017				& $1.5\times10^{-5}$			& 0.009\\
 IH & $1.5\times10^{-5}$		& 0.006 			& $1.4\times10^{-5}$		& 0.017				& $1.5\times10^{-5}$			& 0.009\\
\hline \hline
\end{tabular}
\end{center}
\label{tab:23sensitivity}
\end{table}%

A joint fit of the $\numu$ and $\nue$ samples enables us to also precisely measure $\sin^2\theta_{23}$ and $\Delta m^2_{32}$.
Figure~\ref{fig:theta23-0.50} shows the 90\% CL allowed regions for the true value of $\sin^2\theta_{23}=0.5$.
The expected precision of $\Delta m^2_{32}$ and $\sin^2\theta_{23}$ for true $\sin^2\theta_{23}=0.45, 0.50, 0.55$ with reactor constraint on $\sin^22\theta_{13}$ is summarized in Table~\ref{tab:23sensitivity}.

\begin{figure}[tbp]
\centering
\includegraphics[width=0.6\textwidth]{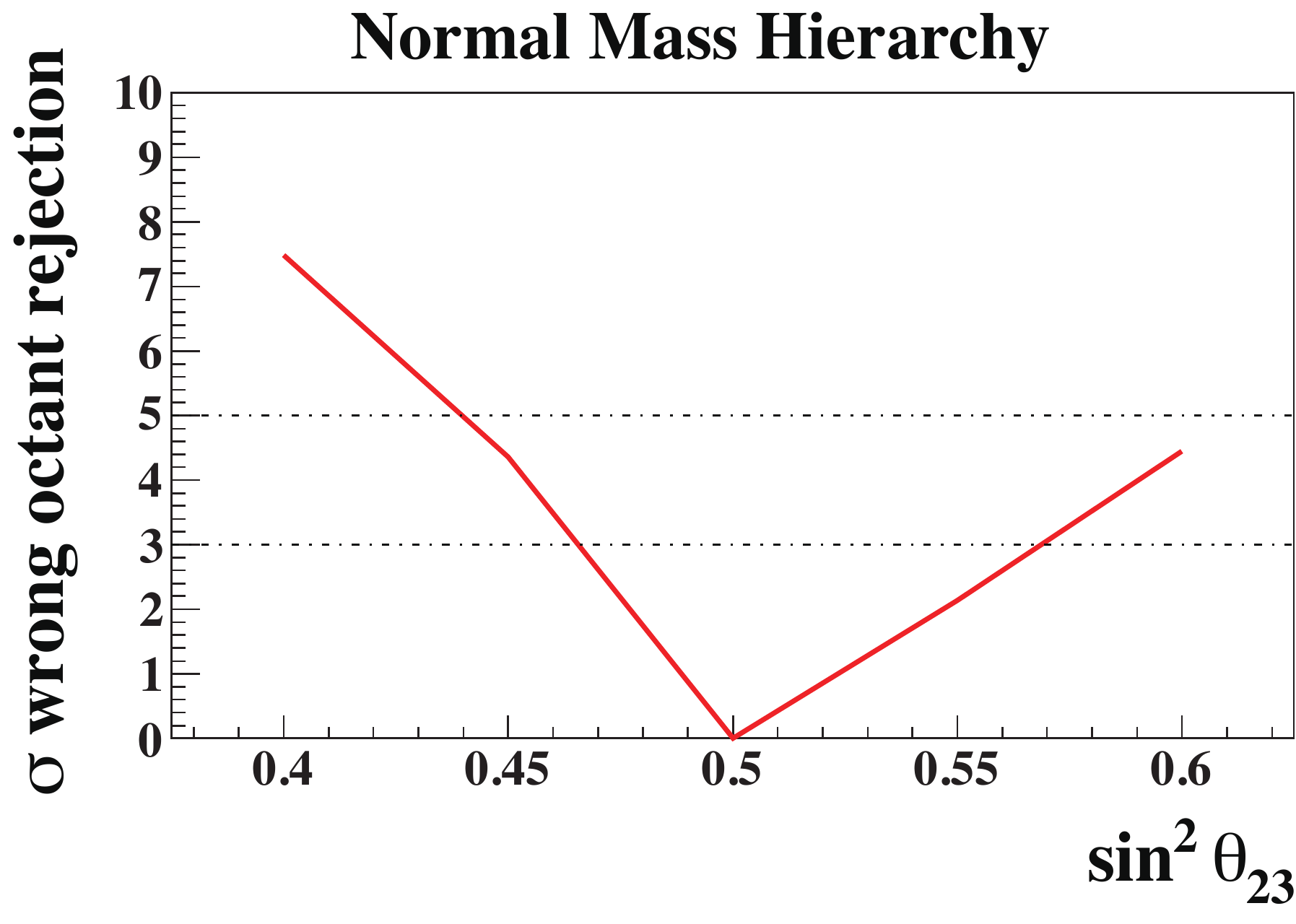}
\caption{The expected significance $(\sigma \equiv \sqrt{\Delta \chi^2})$ for wrong octant rejection, by beam neutrino measurement with reactor constraint, as a function of true $\sin^2\theta_{23}$ in the normal hierarchy case.
\label{fig:lbl-octant}}
\end{figure}

Figure~\ref{fig:theta23-0.45} shows the 90\% CL allowed regions on the
$\sin^2\theta_{23}$--$\Delta m^2_{32}$ plane, for the true values of
$\sin^2\theta_{23}=0.45$ and $\Delta m^2_{32} = 2.4 \times
10^{-3}$~eV$^2$.  With a constraint on $\sin^22\theta_{13}$ from the
reactor experiments, Hyper-K measurements can resolve the octant
degeneracy and precisely determine $\sin^2\theta_{23}$.
Figure~\ref{fig:lbl-octant} shows the expected significance $(\sigma \equiv \sqrt{\Delta \chi^2})$ for wrong octant rejection with beam neutrino measurement alone as a function of true value of $\sin^2\theta_{23}$ in the normal hierarchy case.

As discussed earlier, a precision measurement of $\Delta m^2_{32}$,
compared with reactor measurements of $\Delta m^2_{ee}$, will enable a
consistency check of the mixing matrix framework.  The difference
expected from the current knowledge of oscillation parameters is a
few \%.  The uncertainty of $\Delta m^2_{32}$ by Hyper-K is expected
to reach $0.6\%$, while measurements by future reactor experiments are
expected to achieve $<1\%$ precision.  Thus, the comparison will yield
a significant consistency check.

\subsubsection{Neutrino cross section measurements}

\begin{figure}[htbp]
\begin{center}
\includegraphics[width=0.85\textwidth]{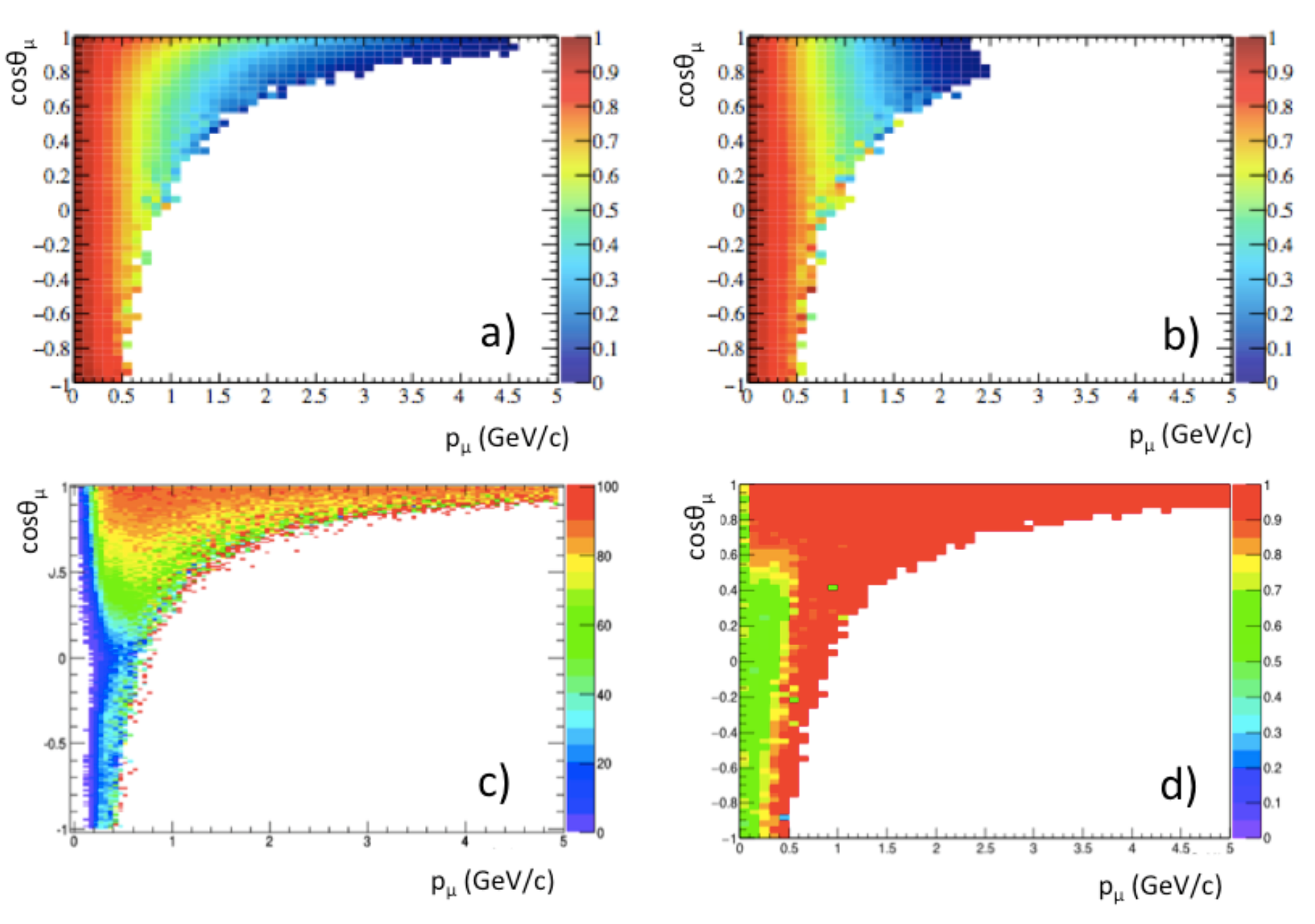}
\caption{
Example detector efficiency in muon momentum and direction. a) A
horizontally oriented 2\,kton cylindrical intermediate WC detector at
2km, b) the same detector vertically oriented with respect to the
beam, c) the current ND280 near detector and d) the WAGASCI
detector. Good coverage of this phase space helps to constrain
uncertainties in cross-section models. }
\label{fig:lbl-ndcrosseff}
\end{center}
\end{figure}

With a set of highly capable neutrino detectors envisioned for Hyper-K
project, a variety of neutrino interaction cross section measurements
will become possible. The near detector suite offers a range of
capabilities to probe different theoretical models for neutrino
interactions: in particular data across different momenta ranges and a
range of lepton emission angles. Figure~\ref{fig:lbl-ndcrosseff} shows
the efficiency of different detectors as a function of angle and muon
momentum. The ability to measure exclusive hadronic final states,
using techniques such as high pressure gas TPCs or emulsion detectors,
provides valuable additional information for exclusive
cross-sections. In table
\ref{tab:lbl-ndcross} we estimate the sensitivity of each proposed near detector for key selections 
based on a flux of $10^{21}$POT.

 \begin{table}[htbp]
\begin{center}
\begin{tabular}{|c|c|c|p{5.5cm}|} \hline
Detector	& Selection & Nevents & Selection Characteristics \\ \hline\hline
ND280 detector, 280m	& $\nu_{\mu}$CC0$\pi$ & 20k	& FGD1 (1--3\,GeV), $P\approx $72\%  \cite{Abe:2015awa}	 \\ \hline
ND280 detector, 280m	& $\nu_{\mu}$CC1$\pi$ & 6k	& FGD1 (1--3\,GeV), $P\approx $50\%\ \cite{Abe:2015awa}	 \\ \hline
ND280 detector, 280m	& $\nu_{\mu}$CC inclusive & 40k	& FGD1 (1--3\,GeV), $P\approx $90\%\cite{Abe:2015awa}	 \\ \hline
\hline
INGRID &  $\nu_{\mu}$CC inclusive & 17.6$\times 10^6$ & $\epsilon>$70\%  (1--3\,GeV), $P = $  97\% \cite{Abe:2015biq}\\ \hline
\hline
HPTPC, 8\,m$^3$, 10\,bar Ne (CF$_4$) & $\nu_{\mu}$CC inclusive & 4.2k (18.4k) & $\epsilon \approx $70\%, protons $>$ 5\,MeV detected \\ \hline
HPTPC, 8\,m$^3$, 10\,bar Ne (CF$_4$) & $\nu_{e}$CC inclusive & 80 (450)       & $\epsilon \approx $70\%, protons $>$ 5\,MeV detected \\ \hline
\hline
WAGASCI	 &  $\nu_{\mu}$CC0$\pi$  & 63k &P=75\%, proton reconstruction: $\epsilon \approx 15$\% at  p=500\,MeV/c, water in; $\epsilon \approx 27$\% at  p=250\,MeV/c, water out (15\% @ 150MeV/c) \\ \hline
WAGASCI	 &  $\nu_{\mu}$CC1$\pi$  & 10k &P=50\% (protons as above)\\ \hline
WAGASCI	 &  $\nu_{\mu}$CC inclusive  & 75k &P=96\% (protons as above)\\ \hline

\hline
200kg Water target 		& $\nu_\mu$ CC+NC inclusive & 10k-20k & 4$\pi$ automated readout \\
emulsion off-axis, 280m    & & & proton $>$ 10-30\,MeV detected \\ \hline
200kg Water target 	& $\nu_e$ CC inclusive & 1k & 4$\pi$ automated readout \\
emulsion off-axis, 280m    & & & proton $>$ 10-30\,MeV detected \\ \hline

\hline
1kton WC \@ 1\,km & $\nu_{\mu}$CC0$\pi$ (1-2$^{\circ}$,2-3$^{\circ}$,3-4$^{\circ}$) & 1682k,1060k,519k  &  $P\approx$92\%,95\%,95\% \\ \hline 
1kton WC \@ 1\,km & $\bar{\nu}_{\mu}$CC0$\pi$ (1-2$^{\circ}$,2-3$^{\circ}$,3-4$^{\circ}$) & 519k,331k,186k  &  $P\approx$74\%,77\%,76\% \\ \hline 
1kton WC \@ 1\,km & $\nu_{\mu}$CC1$\pi$ (1-2$^{\circ}$,2-3$^{\circ}$,3-4$^{\circ}$) & 208k,65k,27k  &  $P\approx$46\%,44\%,31\% \\ \hline 
1kton WC \@ 1\,km & $\nu_{e}$CC0$\pi$ (1-2$^{\circ}$,2-3$^{\circ}$,3-4$^{\circ}$) & 11.2k,6.9k,4.6k  &  $P\approx$54\%,71\%,80\% \\ \hline 
1kton WC \@ 1\,km & $\nu$NC$\pi^{0}$ (1-2$^{\circ}$,2-3$^{\circ}$,3-4$^{\circ}$) & 300k,111k,45k  &  $P\approx$58\%,63\%,60\% \\ \hline 
\end{tabular}

\caption{
Some of the primary cross-section measurements accessible with
different elements of the Near Detector Suite (see chapter 2 for
details). The predicted number of events or measurement precision have
been evaluated for $10^{21}$POT.  $\epsilon$ = efficiency = number
selected / total events for given topology, $P$ = purity = number of
given topology / total events selected. For the ND280 measurements
only events for a single fine grained detector (FGD1) are projected,
the second FGD plus the use of other detector components as targets
increases the statistics. Numbers are obtained either from independent
Monte Carlo studies, or extrapolated from the cited references.}

\label{tab:lbl-ndcross}
\end{center}
\end{table}%

\subsubsection{Searches for new physics}
In addition to the study of standard neutrino oscillation, the
combination of intense beam and high performance detectors enables us
to search for new physics in various ways.  Examples of possible
searches for new physics with Hyper-K and accelerator beam are listed
below.

\paragraph{Search for sterile neutrinos}
Sterile neutrinos can be searched for in both disappearance and
appearance channels in near and intermediate detectors.  With neutrino
energy of 0.1--few~GeV and baseline of 0.3~km and 1--2~km, it will be
sensitive to $\Delta m^2$ of $\mathcal{O}(1)$~eV$^2$, which is
the interesting region in light of several anomalies reported by recent
experiments.  T2K has searched for sterile neutrino in $\nue$
appearance and $\numu$ disappearance with near
detectors~\cite{Abe:2014nuo, Dewhurst:2015aba}.  With more statistics,
improved detectors, and possible two detector configuration with near
and intermediate detectors, Hyper-K near detectors will have chance to
improve the sensitivity for sterile neutrino searches.

Neutral current measurements at the far detector will be also sensitive to the sterile neutrino,
because the neutral current channel measures the total active flavor content.
By selecting two electron-like ring events with no decay electron and invariant mass consistent with $\pi^0$, neutral current events with 96\% purity can be obtained.
With $1.56\times10^{22}$ protons on target and 560~kton fiducial mass, more than 4,000 
NC $\pi^0$ events are expected after selection.
Normalization of the NC $\pi^{0}$ production can be strongly
constrained by the high statistics data at the intermediate detector
as shown in Table~\ref{tab:lbl-ndcross}.

\paragraph{Test of Lorentz and CPT invariance}
Lorentz Violation arises when the behavior of a particle depends on
its direction or boost velocity and is predicted to occur at the
Planck scale (10$^{19}$~GeV).  Searches for Lorentz Violation have
been performed by various experiments, including T2K, by looking for a
sidereal time dependence of the neutrino event rate.  Similar searches
can be carried out with larger statistics and improved detectors.

\paragraph{Heavy neutrino search}
The existence of heavy neutral leptons (heavy neutrinos) is predicted
in many extensions of the Standard Model.  Such heavy neutrinos may be
produced in decays of kaons and pions from the target.  Then, decays
of heavy neutrinos can be detected in the near detector.  The
feasibility of search for heavy neutrinos in accelerator neutrino
experiment, in particular with T2K, is studied in \cite{Asaka:2012bb}
and the sensitivity is expected to be better than previous searches.
Because interactions of ordinary neutrinos produce background to this
search, having a low density detector such as a gas TPC inside a
magnetic field like ND280 is an advantage for this search.  The
sensitivity will be further enhanced if a larger volume of gas
detector is employed.

\subsubsection{Summary}
The sensitivity to leptonic $CP$ asymmetry of a long baseline
experiment using a neutrino beam directed from J-PARC to the
Hyper-Kamiokande detector has been studied based on a full simulation
of beamline and detector.  
With $1\times$10$^8$~sec of running time with 1.3~MW beam power,
the value of $\deltacp$ can be determined with 7.2$^\circ$ for $\deltacp=0^\circ$ 
or $180^\circ$, and 23$^\circ$ for $\deltacp=\pm90^\circ$.  
$CP$ violation in the lepton sector can be observed with more than
3~$\sigma$(5~$\sigma$) significance for 76\%(57\%) of the possible
values of $\deltacp$.
Using both $\nu_e$ appearance and $\nu_\mu$ disappearance data,
precise measurements of $\sin^2\theta_{23}$ and $\Delta m^2_{32}$ will
be possible.  The expected 1$\sigma$ uncertainty of
$\sin^2\theta_{23}$ is 0.017 (0.006) for $\sin^2\theta_{23}=0.5 (0.45)$.
The uncertainty of $\Delta m^2_{32}$ is expected to reach $<1\%$.

There will be also a variety of measurements possible with both near
and far detectors, such as neutrino-nucleus interaction cross section
measurements and search for exotic physics, using the well-understood
neutrino beam.

\newpage
\graphicspath{{physics-atmnu/figures}}

\subsection{Atmospheric neutrinos}\label{section:atmnu}

  Primary cosmic ray interactions with nuclei in the atmosphere produce
charged hadrons whose decays further create a continuous flux of
neutrinos known as atmospheric neutrinos.  Since the primary cosmic
ray flux is known to be nearly isotropic about the earth, the
resulting neutrino flux is present at all zenith angles observed by a
terrestrial detector.  Further, these neutrinos are less energetic
than their cosmic ray parents by roughly an order of magnitude meaning
that the range of available neutrino energies for observation spans
several orders of magnitude starting near O(100)~MeV.  This
diversity of both energy and pathlength, which ranges from O(10) to
O($10^{4}$)~km, makes atmospheric neutrinos a particularly versatile
tool for studying neutrino oscillations.  However, these neutrinos
represent the most serious background to nucleon decay searches
(discussed in Section~\ref{section:pdecay}).  At the same time they
form the basis for searches for exotic particles, such as dark matter
(discussed in Section~\ref{section:darkmatter}), whose interactions
may produce neutrinos that would subsequently appear atop the
atmospheric neutrino spectrum.  For these reasons a precise
characterization of the atmospheric neutrino flux is key to future
discoveries at Hyper-Kamiokande.

Though atmospheric neutrinos were used to make the first discovery of
the neutrino oscillation phenomenon, for very large detectors like
Hyper-Kamiokande, they provide excellent sensitivity to many of the
remaining open questions in oscillation physics.  Indeed, current
neutrino telescopes have demonstrated constraints on the atmospheric
mixing parameters comparable to beam measurements using atmospheric
neutrinos alone.  Additionally, future experiments seek to use these
neutrinos to study the mass hierarchy.  While both of these
measurements and more are available to Hyper-Kamiokande, it offers two
distinct advantages over current and planned projects.  First, with its
exquisite ability to distinguish between charged current $\nu_{e}$ and
$\nu_{\mu}$ interactions, it will have improved access to the
oscillation modes with the most hierarchy sensitivity,
$\nu_{\mu} \rightarrow \nu_{e}$ and
$\bar \nu_{\mu} \rightarrow \bar \nu_{e}$.  It is indeed the asymmetry
in these two probabilities for few GeV neutrinos traversing the earth
that provides the cleanest signature of the mass hierarchy.
Additionally, Hyper-Kamiokande will make combined beam and atmospheric
neutrino oscillation measurements to yield increased sensitivity.
Details of Hyper-K's physics potential using atmospheric neutrinos by
themselves and in conjunction with beam neutrinos are presented in the
next two subsections.

  \subsubsection{Neutrino oscillation studies (MH, \(\theta_{23}\) octant, $CP$ phase)}
      
     \graphicspath{{physics-atmnu/figures}}
As atmospheric neutrinos span both low and high energies as well as
long and short path lengths, they are in principal sensitive to all
parameters in the PMNS mixing paradigm.  That being said, the most
apparent oscillation features are driven by the so-called atmospheric
mixing parameters, $\theta_{23}$ and $\Delta m^{2}_{32}$, and they
induce a deficit of observed upward-going $\nu_{\mu}$ interactions at
predominantly multi-GeV energies as these neutrinos oscillate into
primarily unobserved $\nu_{\tau}$.  However, now that the value of
$\theta_{13}$ is known to be non-zero the presence of matter effects
on atmospheric neutrinos that traverse the earth makes
important contributions to this picture.  Matter-induced parametric
oscillations in the energy range 2-10 GeV lead to significant
enhancement of either the $\nu_{\mu} \rightarrow \nu_{e}$ or the $\bar \nu_{\mu} \rightarrow \bar \nu_{e}$ 
appearance probability for upward-going neutrinos depending upon the mass hierarchy.
For the normal (inverted) hierarchy neutrino (antinuetrino) oscillations are enhanced.
This enhancement leads to appearance probabilities around 50\% for both hierarhies.
The separation of atmospheric neutrino data
into neutrino-like and antineutrino-like subsets can therefore be used to
extract the hierarchy signal.  Importantly, the features of these
oscillations are also a strong function of $\theta_{23}$ and, to a lesser extent, the value of $\delta_{CP}$.
It should be noted that these matter effects affect both the appearance and
disappearance, $\nu_{\mu}\rightarrow \nu_{\mu}$ , channels, enabling
mass hierarchy sensitivity in both the $\nu_{e}$-like and
$\nu_{\mu}$-like data at Hyper-K.

Including these effects the flux of atmospheric $\nu_{e}$ at the detector may be written roughly as, 
\begin{eqnarray}
 \frac{\Psi(\nu_e)}{\Psi_0(\nu_e)} -1 & \approx & P_2\cdot(r\cdot \cos^2\theta_{23}-1) \nonumber \\
&& -r\cdot \sin\tilde{\theta}_{13}\cdot \cos^2\tilde{\theta}_{13}\cdot \sin{2\theta_{23}}\cdot(\cos{\delta}\cdot R_2-\sin{\delta}\cdot I_2) \nonumber \\
&& +2\sin^2\tilde{\theta}_{13}\cdot(r\cdot \sin^2\theta_{23}-1),
\label{eqn:atm-nue-flux}
\end{eqnarray}
\noindent where these three terms are identified as the ``solar'', ``interference,'' and ``parametric (resonance)'' terms, respectively.
Here $P_{2}$ is the two neutrino oscillation probability
$\nu_{e} \rightarrow \nu_{\mu,\tau}$ driven by the solar mass
splitting $\Delta m^{2}_{21}$.  Note that the effect of $\delta_{CP}$
appears in the interference term, which is controlled by an effective
mixing angle in matter, $\tilde \theta_{13}$, and where $R_{2}$ and
$I_{2}$ denote amplitudes for CP-even and CP-odd oscillations.
Antineutrino oscillations are described by changes to these amplitudes
under an inversion of the sign of both the matter potential and of
$\delta_{CP}$.  At sub-GeV energies the flux ratio
$\nu_{\mu}/\nu_{e}$, $r$, is $\sim 2$, and increases above 1~GeV until
reaching $\sim 3$ at 10~GeV.

\begin{figure}[thb]
  \begin{center}
    \includegraphics[scale=1.0]{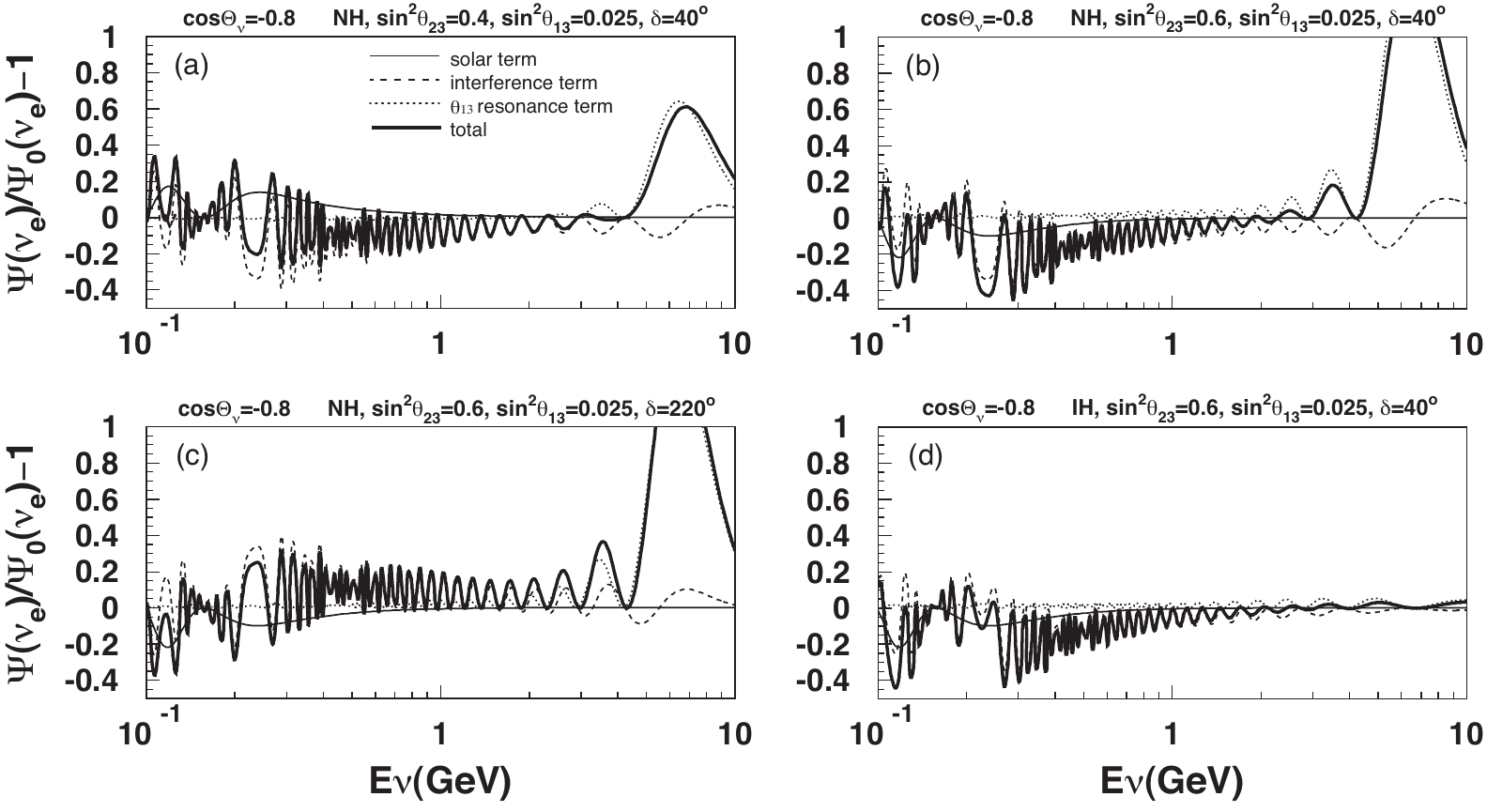}
  \end{center}
  \caption{Oscillated \(\nu_e\) flux relative to the non-oscillated flux
           as a function of neutrino energy
           for the upward-going neutrinos
           with zenith angle $\cos\Theta_\nu=-0.8$.
           $\bar \nu_e$ is not included in the plots.
           Thin solid lines, dashed
           lines, and dotted lines correspond to the solar
           term, the interference term, and the
           \(\theta_{13}\) resonance term,
           respectively
           (see Eq. \ref{eqn:atm-nue-flux}).  Thick solid lines
           are total fluxes.
           Parameters are set as
           \((\sin^2\theta_{21},\sin^2\theta_{13},\sin^2\theta_{23},\delta,
           \Delta m^2_{21},\Delta m^2_{32})\)
           unless otherwise noted.
           The $\theta_{23}$ octant effect can be seen by comparing
           (a) (\(\sin^2\theta_{23}=0.4\)) and
           (b) (\(\sin^2\theta_{23}=0.6\)).
           $\delta$ value is changed to $220^\circ$ in (c) to be compared
           with $40^\circ$ in (b).
           The mass hierarchy is inverted only in (d) so
           \(\theta_{13}\) resonance (MSW) effect disappears in this plot.
           For the inverted hierarchy the MSW
           effect should appear in the $\bar \nu_e$ flux,
           which is not shown in the plot.
           }
  \label{fig:4nue-flux}
\end{figure}

Figure~\ref{fig:4nue-flux} shows the expected $\nu_{e}$ flux at the
detector normalized to the unoscillated prediction under this
approximation for four configurations of the oscillation parameters
and neutrinos with $\cos(\theta_{zenith}) = -0.8$.  At energies between
5-10~GeV the most prominent feature of the figure is the parametric
resonance driven by $\tilde \theta_{13}$, whose amplitude increases
with $\mbox{sin}^{2} \theta_{23}$ ( c.f. panels a. and b.).  Further,
this resonance becomes suppressed in the neutrino channel when the
hierarchy is switched from normal to inverted (compare panels a. and
d.).  Though some change in the resonance can be seen via the
interference term as $\delta_{CP}$ is varied, the dominant effect
appears below 1~GeV (panels a. and c.).  For these reasons the
atmospheric neutrino oscillation analysis has been designed to
maximize each of these potential effects.

Hyper-Kamiokande's reconstruction performance is expected to meet or
exceed that of its predecessor, Super-Kamiokande.  Nominally the size
and configuration of the two detectors are similar enough that event
selections and systematic errors are not expected to differ largely.
While the larger statistics afforded by Hyper-K may result in improved detector 
systematic uncertainties, such improvements are not assumed here.
Similarly, improvements in the uncertainty of the atmospheric neutrino flux,
the leading systematic for studies of $\delta_{CP}$ with atmospheric 
neutrinos, or the cross section model, which largely affects the 
mass hierarchy sensitivity, are not assumed in the studies below.
The analyses proceed following
those at Super-K using simulation and reconstruction tools tuned and
validated for that experiment.

\begin{figure}[thb]
  \begin{center}
    \includegraphics[scale=0.7]{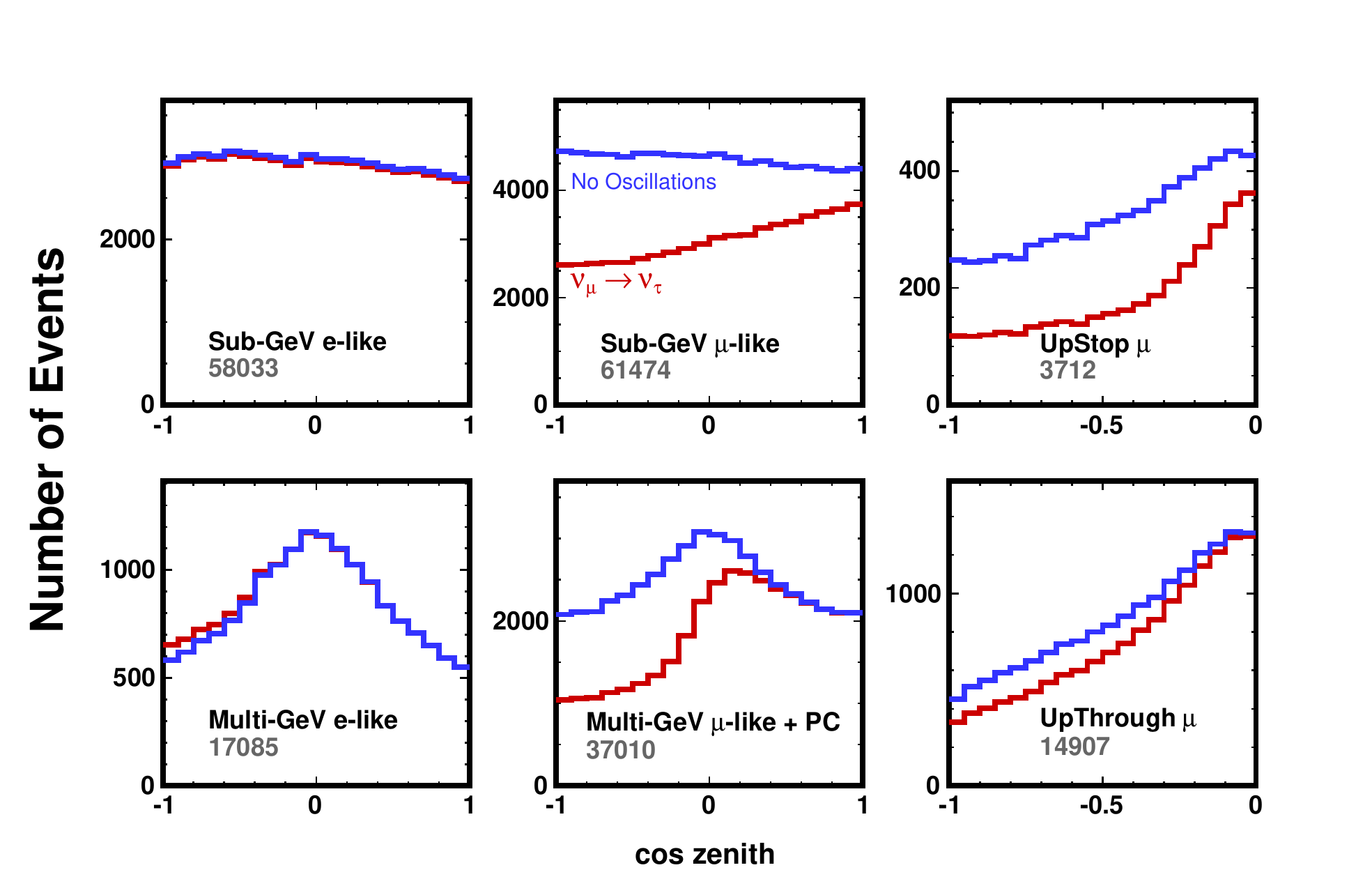}
  \end{center}
  \caption{Zenith angle distributions of a subset of the atmospheric neutrino analysis samples.
           The blue (red) line shows the expectation for a normal (inverted) hierarchy 
           after a 1.9~Mton$\cdot$year exposure. Gray numbers in each panel represent 
           the size of the event sample. }
\label{fig:atmnu_hier_zenith}
\end{figure}

Atmospheric neutrino MC corresponding to a 75 year exposure of the
186~kton Hyper-K detector has been generated based on the flux model
presented in~\cite{Honda:2006qj} and using the NEUT interaction
generator.  The analysis is based on 19 samples optimized for
sensitivity to potential oscillation signals at both high (multi-GeV)
and low (sub-GeV) energies.  Interactions are divided into sub-samples
based upon the particle ID of their most energetic reconstructed
Cherenkov ring ($e$-like or $\mu$-like) and the number of such rings.
Additional selections are made to the multi-GeV $e$-like samples to
separate them into antineutrino-like and neutrino-like
subsamples~\cite{Pik:2012qsy}.  Event topologies with particles
exiting the inner detector and depositing light in the outer veto as
well as muons from neutrino interactions in the rock surrounding the
detector are also included in the analysis.  Further details of the
sample selection are presented elsewhere~\cite{Abe:2014gda}.  Zenith
angle distributions for six of the analysis samples 
are shown for both assumptions in figure~\ref{fig:atmnu_hier_zenith}.

\begin{figure}[thb]
  \begin{center}
    \includegraphics[scale=0.35]{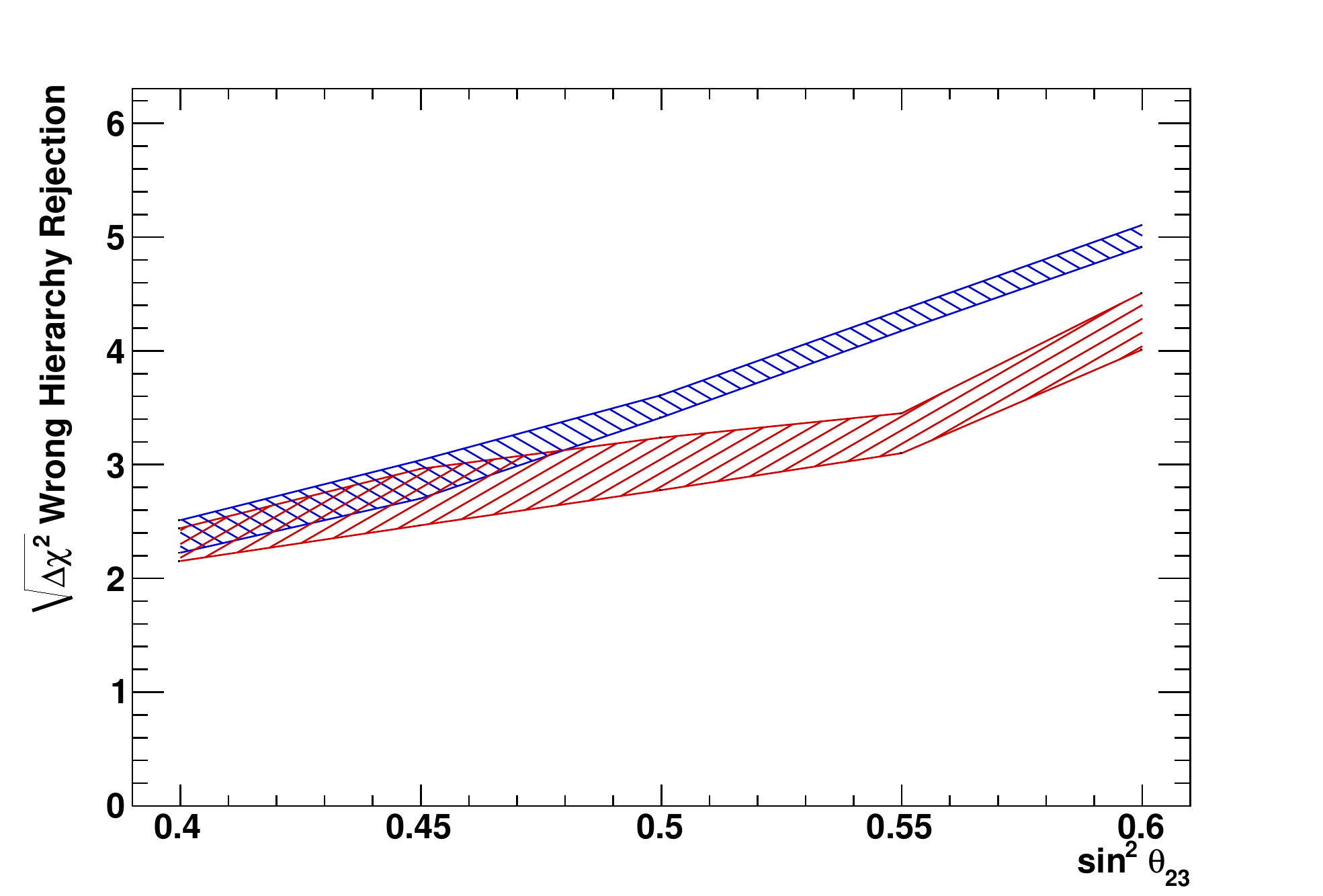}
    \includegraphics[scale=0.35]{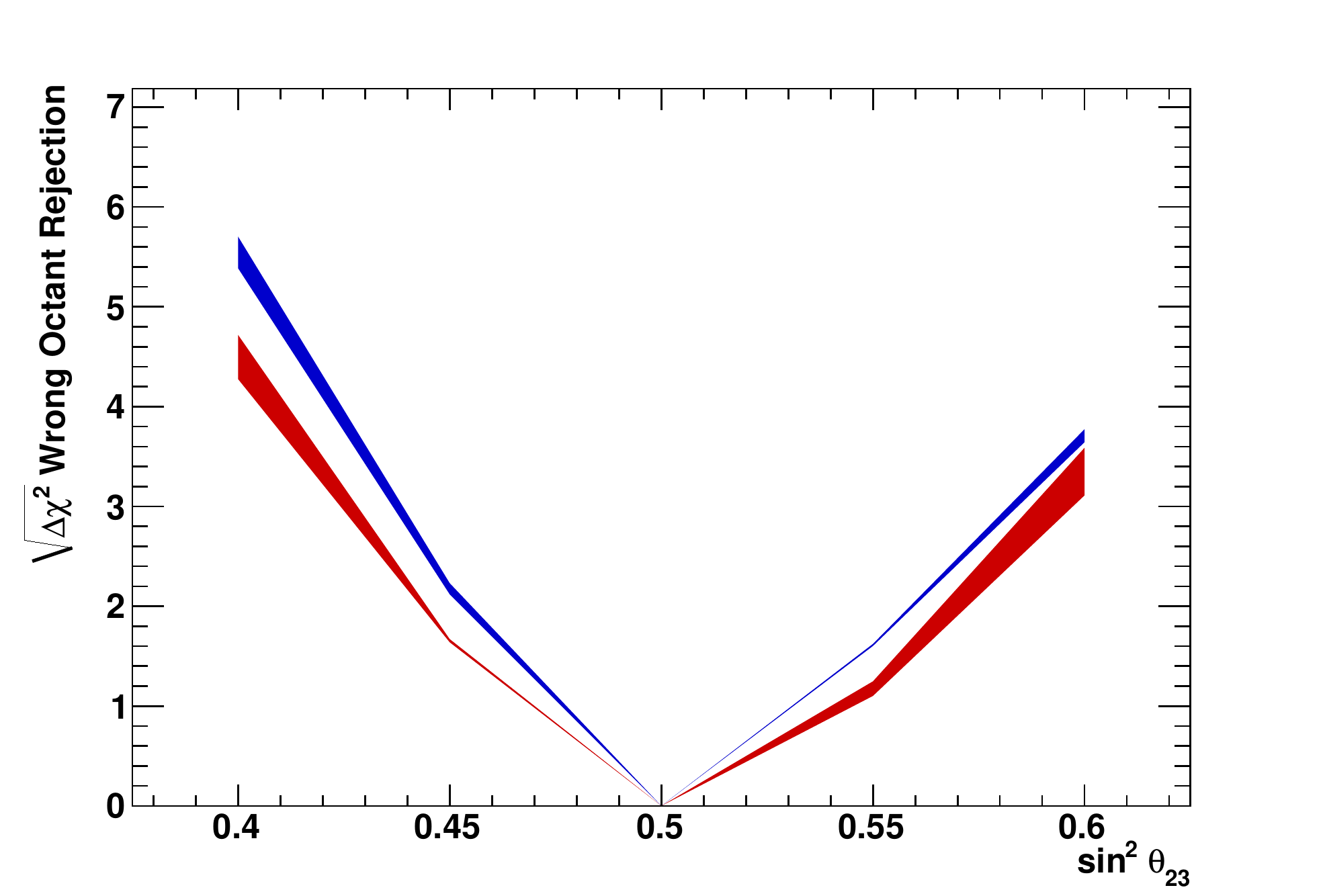}
  \end{center}
  \caption{Neutrino mass hierarchy sensitivity (left) and octant sensitivity (right) as a function of the 
           true value of $\mbox{sin}^{2}\theta_{23}$ for a single detector after 10 years. 
           (a 1.9 Mton$\cdot$year exposure). 
           In both figures the blue (red) band denotes the normal (inverted) hierarchy
           and the uncertainty from $\delta_{CP}$ is shown by the width of the band. }
\label{fig:atmnu_hier_sens}
\end{figure}

Figure~\ref{fig:atmnu_hier_sens} shows the Hyper-K's expected
sensitivity to resolving the neutrino mass hierarchy and the octant of
$\theta_{23}$ assuming a single detectors after 10 years.
Both panels of the figure
are shown as a function of the true value of $\mbox{sin}^{2}\theta_{23}$
for the range of values allowed by recent measurements from the T2K
experiment~\cite{Abe:2015awa} and the width of the bands in the
figures illustrates the uncertainty from $\delta_{CP}$.  In each panel
the sensitivity is defined as $\sqrt{\Delta \chi^{2}}$, which for mass
hierarchy resolution corresponds to
$\Delta \chi^{2} \equiv \chi^{2}_{AH} - \chi^{2}_{TH}$, where $TH$ and
$AH$ refer to the true hierarchy and the alternate hierarchy
hypotheses, respectively.  The octant sensitivity is defined
similarly.  After 10 years (a 1.9 Mton$\cdot$year exposure) Hyper-K is expected to
resolve the mass hierarchy at $\sqrt{\Delta \chi^{2}} > 3$ for both
hierarchy assumptions and when
$\mbox{sin}^{2}\theta_{23} > 0.53$.  Similarly the atmospheric neutrino data
alone can be used to determine the $\theta_{23}$ octant at
$\sqrt{\Delta \chi^{2}} > 2$ when $|\theta_{23} - 45| > 4^{\circ}.$
Note further that atmospheric neutrinos can be used to measure the
chemical composition of the Earth's interior.  Details of Hyper-K's
expected sensitivity are presented in
section~\ref{section:radiography}.

  \subsubsection{Combination with Beam Neutrinos}

\begin{figure}[thb]
\begin{center}
    \includegraphics[width=0.7\textwidth]{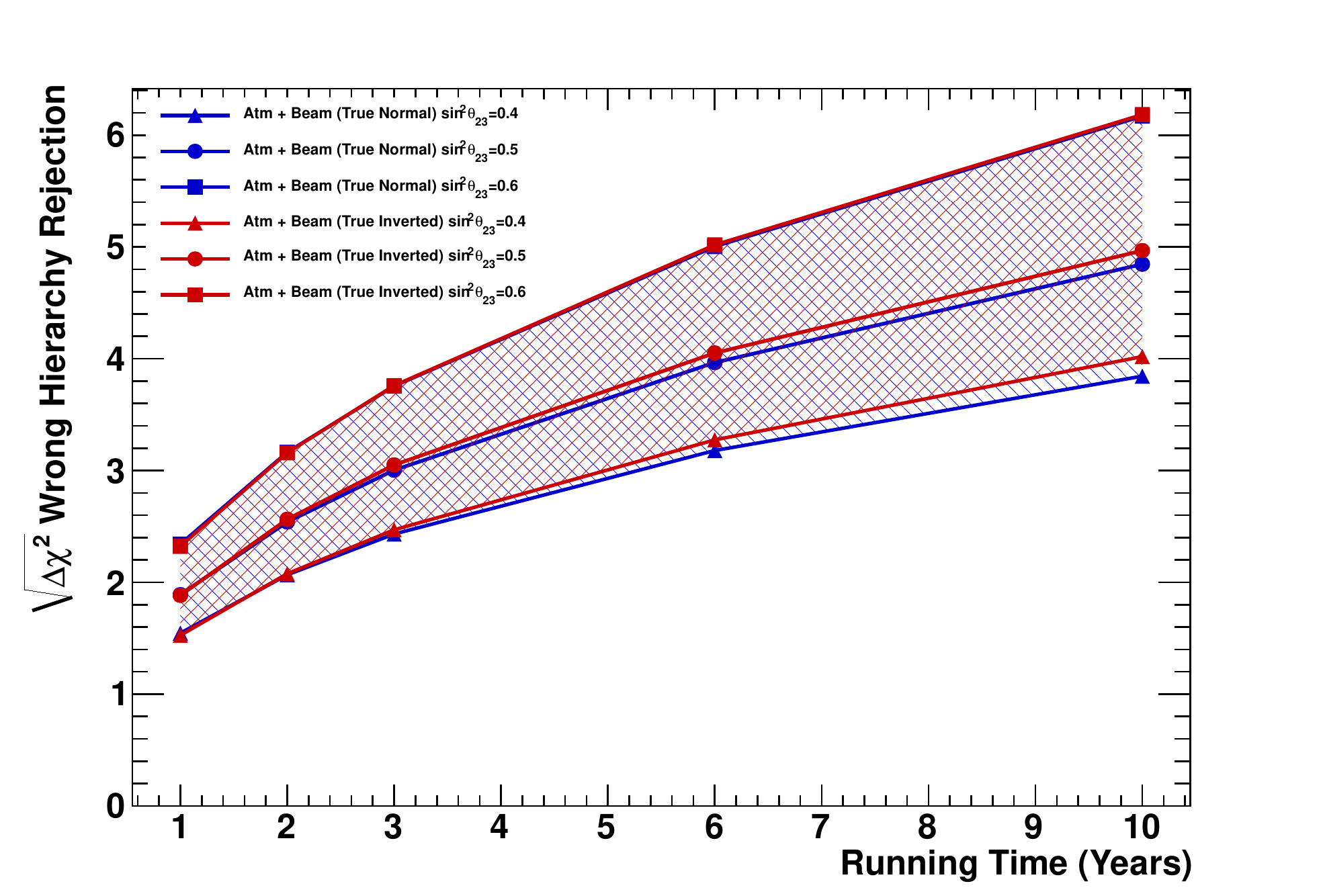}
  \caption{ Expected sensitivity to the mass hierarchy as a function 
           of time assuming $\sin^{2}\theta_{23} =$ 0.4 (triangle), 0.5 (circle), and 0.6 (square) 
           from a combined analysis of atmospheric and accelerator neutrinos data at Hyper-K.
           Blue (red) colors denote the normal (inverted) hierarchy. }
\label{fig:atmnu_beam_sens_hier}
\end{center}
\end{figure}

\begin{figure}[thb]
\begin{center}
    \includegraphics[width=0.7\textwidth]{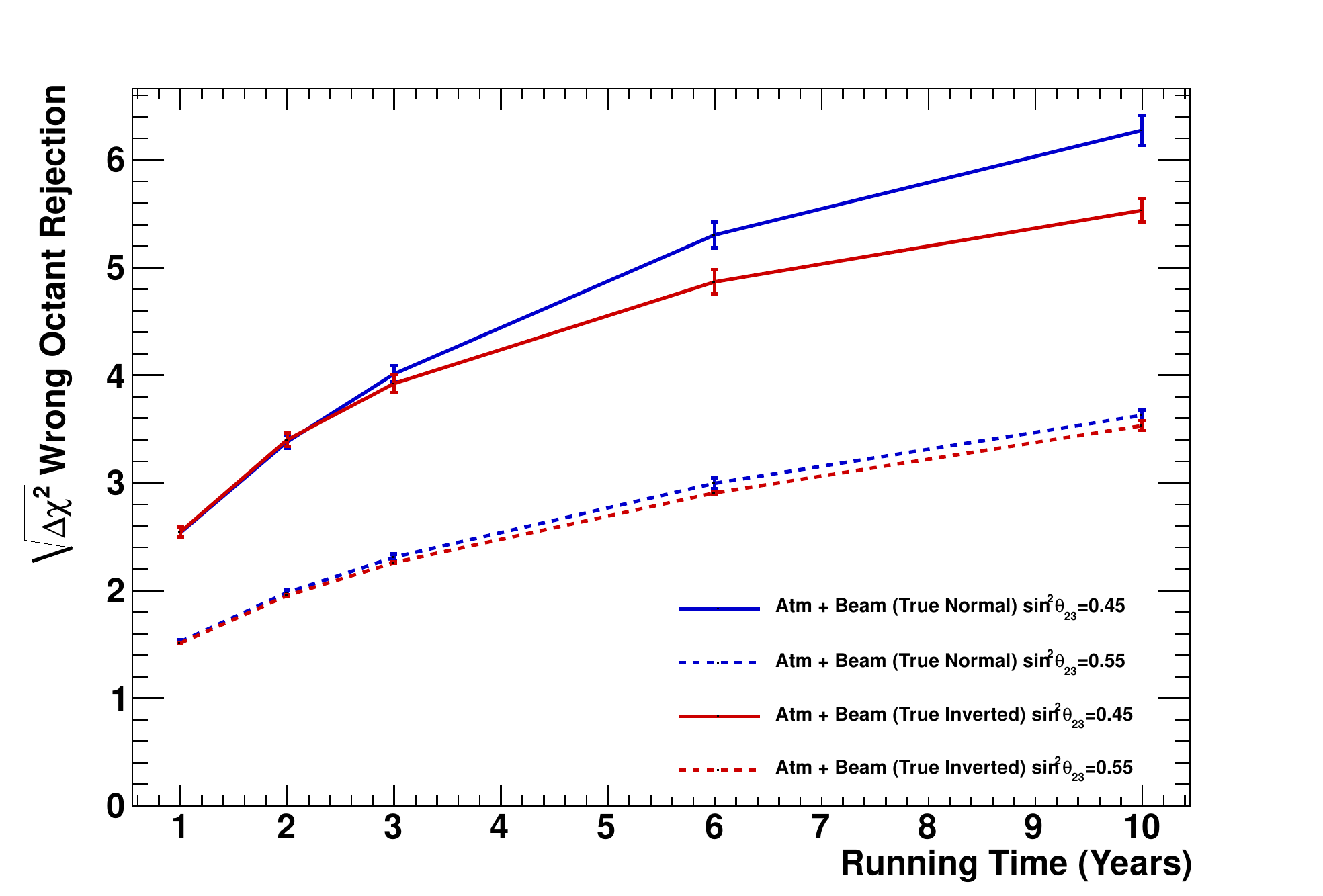}
  \caption{Expected octant resolution sensitivity for $\sin^{2}\theta_{23} = 0.45$ (solid) and  $ 0.55 $ (dashed)
           as a function of time. Error bars indicate the uncertainty from $\delta_{CP}$.
           Blue (red) colors denote the normal (inverted) hierarchy. }
\label{fig:atmnu_beam_sens_oct}
\end{center}
\end{figure}

     Hyper-K will have improved sensitivity to neutrino oscillations by
joint analysis of its atmospheric and accelerator (long-baseline beam)
neutrino data sets.  A particularly striking example comes in the form
of mass hierarchy resolution.  Since matter effects are small for the
295~km baseline Tokai-to-HK baseline, the beam neutrino data will have limited
sensitivity to the mass hierarchy.  At the same time, though matter
effects are strong in the resonance-enhanced oscillation region of the
atmospheric neutrino energy spectrum, lacking precise knowledge of
these neutrinos' true baseline limits their ability to constrain the
atmospheric neutrino mixing parameters which govern the size of the
expected enhancement namely, $\theta_{23}$.  Further, approximate
degeneracies between this parameter and the sign of $\Delta m^{2}_{32}$
weaken the mass hierarchy sensitivity.  Note, for instance, the
dramatic influence $\theta_{23}$ has on Hyper-K's atmospheric
neutrino-only sensitivity in Figure~\ref{fig:atmnu_hier_sens}.  The
off-axis angle of the beam measurement, on the other hand, provides a
clean measurement of the atmospheric mixing parameters and therefore
provides for a precise prediction of the expected amount of
$\nu_{\mu} \rightarrow \nu_{e}$ appearance expected in the resonance
region.  Fitting the two data sets together in turn improves the
overall mass hierarchy sensitivity.
Figure~\ref{fig:atmnu_beam_sens_hier} shows the evolution 
of the combined sensitivity over time.
With five years of data with the \hksingletank\ detector 
the combined atmospheric neutrino and beam samples show better than 
$3\sigma$ ability to reject the incorrect mass hierarchy, 
assuming either the normal or inverted hierarchy is true.
Similarly, the ability to resolve the $\theta_{23}$ octant improves 
with the combination as shown in Figure~\ref{fig:atmnu_beam_sens_oct}. 
While atmospheric neutrinos alone can resolve the octant at 3~$\sigma$ if 
$|\theta_{23} - 45| > 4^{\circ},$ in the combined analysis 
it can be resolved when this difference is only $2.3^{\circ}$ in ten years.

\begin{figure}[thb]
  \begin{center}
    \includegraphics[width=0.47\textwidth]{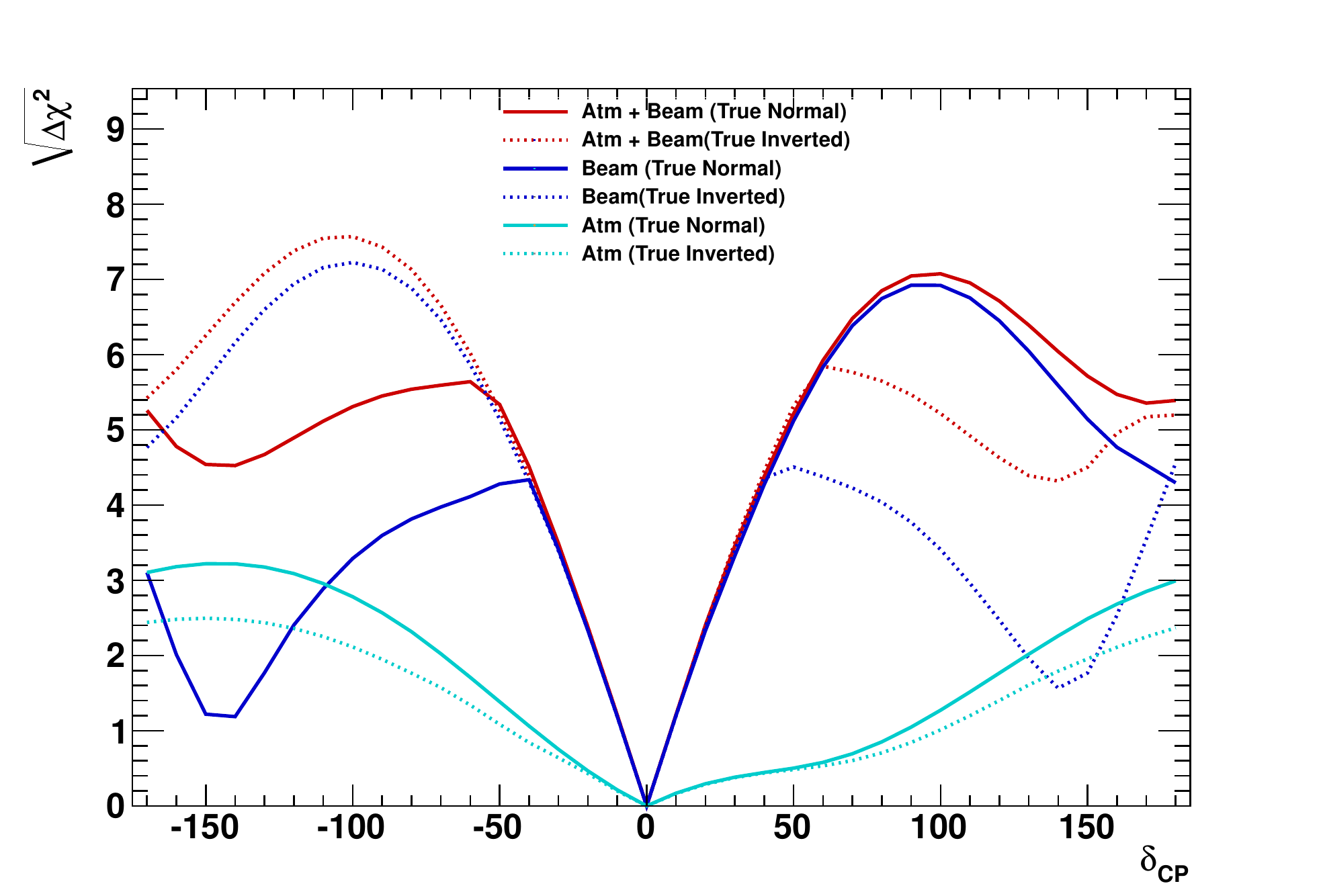}
    \includegraphics[width=0.47\textwidth]{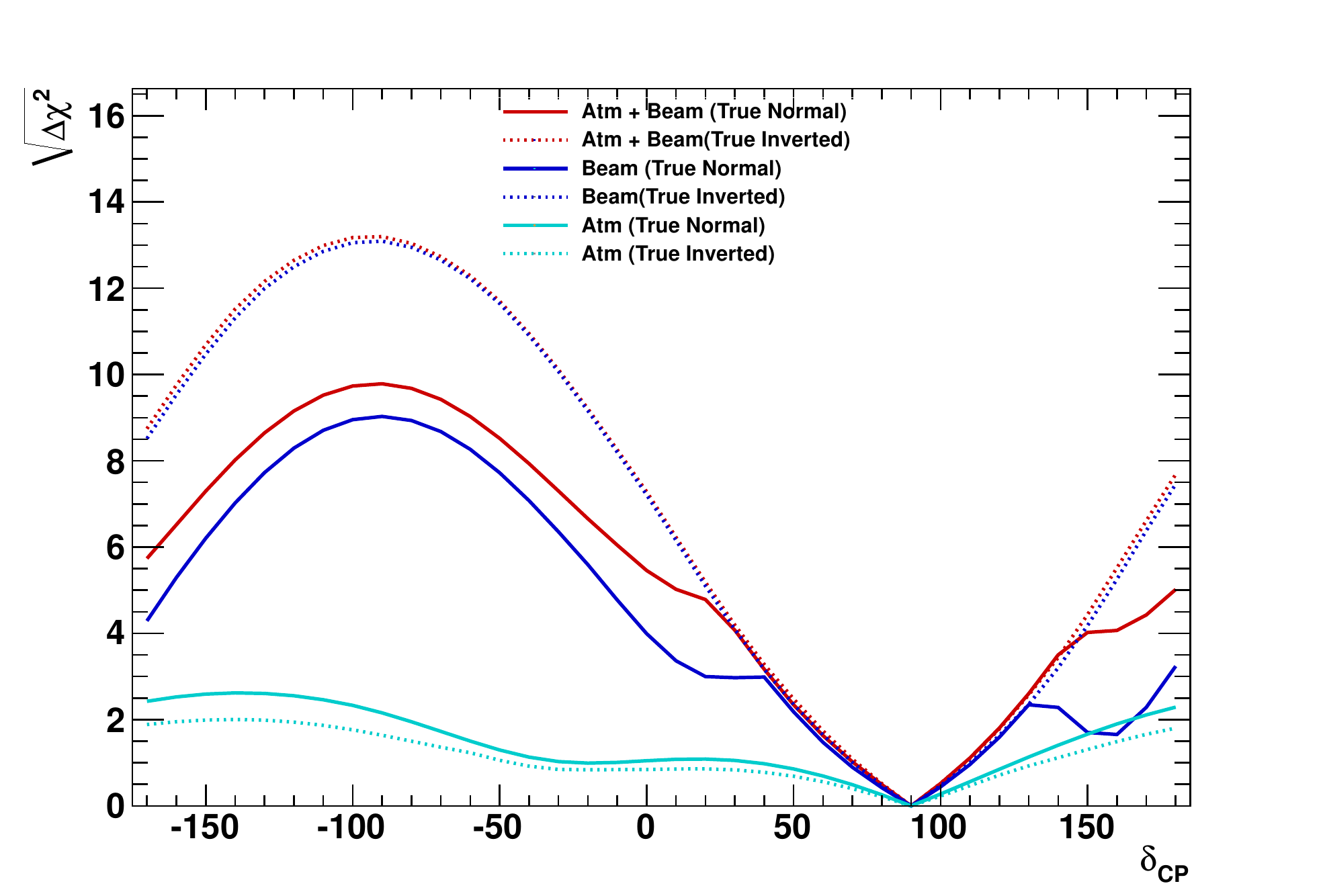}
  \end{center}
  \caption{Constraints on $\delta_{CP}$ after a 10 year exposure of Hyper-K assuming the normal 
           mass hierarchy. Cyan and blue lines show the constraint from the atmospheric  
           neutrino sample and beam neutrino sample individually, whereas the constraint from 
           their combination appears in the red line.  The left (right) figure assumes the 
           true value of $\delta_{CP}$ is $0^{\circ}$ ($90^{\circ}$). Solid and dashed lines 
           denote the normal and inverted hierarchies, respectively. }
\label{fig:comb_cp}
\end{figure}

\begin{figure}[thb]
\begin{center}
    \includegraphics[width=0.7\textwidth]{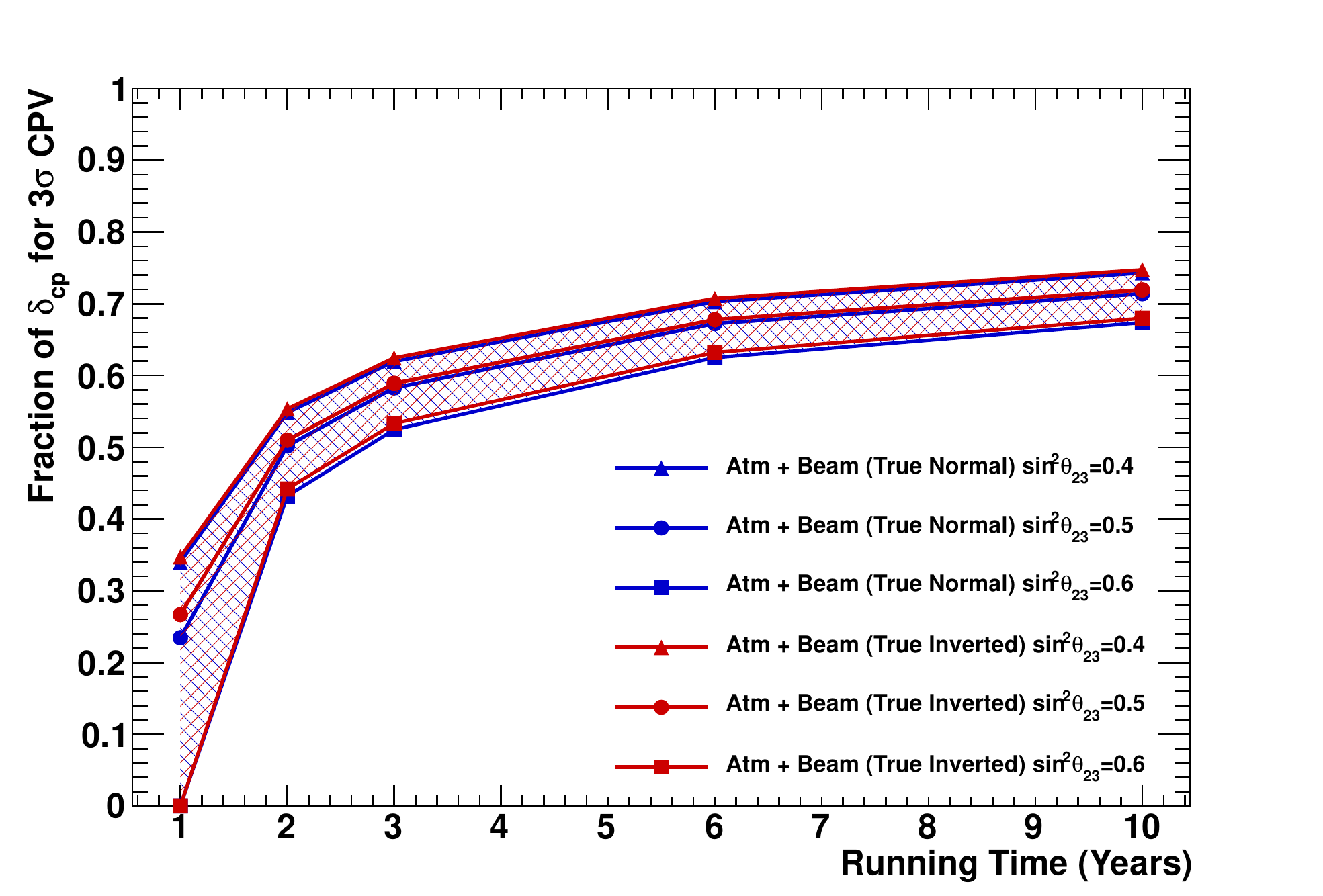}
  \caption{ Fraction of $\delta_{CP}$ phase space at which a $3\sigma$
           observation of CP violation can be made as a function 
           of time assuming $\sin^{2}\theta_{23} =$ 0.4 (triangle) , 0.5 (circle) , and 0.6 (square) 
           from a combined analysis of atmospheric and accelerator neutrinos data at Hyper-K.
           Blue (red) colors denote the normal (inverted) hierarchy. }
\label{fig:atmnu_beam_sens_cp}
\end{center}
\end{figure}

However, it is not just the atmospheric neutrinos that benefit from
combined measurements.  Indeed, for a fixed baseline uncertainty in
the mass hierarchy leads to parameter degeneracies in the the beam
neutrino measurement of $\delta_{CP}$.  The atmospheric neutrino data,
on the other hand, can be used not only to resolve the mass hierarchy
and subsequently these degeneracies, but they also provide
complementary sensitivity to $\delta_{CP}$.  The left panel of
Figure~\ref{fig:comb_cp} shows the expected constraint on this
parameter assuming its true value is $\delta_{CP} = 0$ for separate
beam and atmospheric neutrino measurements.  While the beam
measurement (gray) shows excellent precision near the true parameter
value, there is a strong degeneracy near $\delta_{CP} = \pi$ when the
mass hierarchy is unknown.  This false solution is not present in the
atmospheric neutrino measurement (blue), so the combination of the two (red)
measurements yields a very precise measurement. 
The situation is similar in the right panel, which shows the sensitivity 
when  $\delta_{CP} = \pi / 2.$ 
Hyper-K's ability observe CP violation at $3\sigma$ or better 
is shown as a function of time for the combined measurement in 
Figure~\ref{fig:atmnu_beam_sens_cp}.
Table~\ref{tbl:atm_summary} summarizes Hyper-K's
expected sensitivity to various parameters when using atmospheric neutrinos only and when
combining them with the beam data.
\begin{table*}[htbp]
\begin{center}
\begin{tabular}{l|c|c|c}
\hline
\multicolumn{2}{l|}{Metric}  & \multicolumn{2}{c}{\hksingletank }                     \\
           & $\mbox{sin}^{2} (\theta_{23})$      & Atmospheric  $\nu$  &  Atm $+$ Beam           \\
\hline
\hline
 \multirow{2}{*}{Hierarchy}  & $0.40$  & 2.2 $\sigma$  & 3.8 $\sigma$    \\ 
                             & $0.60$  & 4.9 $\sigma$  & 6.2 $\sigma$    \\ 
\hline                                                  
 \multirow{2}{*}{Octant}     & $0.45$  & 2.2 $\sigma$  & 6.2 $\sigma$    \\ 
                             & $0.55$  & 1.6 $\sigma$  & 3.6 $\sigma$    \\ 
\hline
\end{tabular}
\caption{ Summary of Hyper-K's sensitivity in various metrics with atmospheric neutrinos only (Atmospheric) and 
          with the combination of atmospheric neutrino and beam data (Atm $+$ Beam ) for  the 
          staged \hksingletank design. 
          These numbers assume a normal hierarchy, 
          $\Delta m_{23}^{2} = 2.5\times 10^{-3} \mbox{eV}^{2}$, 
          $\mbox{sin}^{2} \theta_{13} = 0.0219$, 
          and the value of $\delta_{CP}$ that minimizes the sensitivity. 
          Entries in the table are in units of $\sqrt{\Delta \chi^{2}}$. See text for details.\label{tbl:atm_summary} }
\end{center}
\end{table*}

  \subsubsection{Exotic Oscillations And Other Topics}

     \newcommand{\dm}{\ensuremath{\Delta m^{2}}\xspace}
\newcommand{\dmsq}[1]{\ensuremath{\dm_{#1}}\xspace}
\newcommand{\dmnew}{\dm}
\renewcommand{\th}[1]{\ensuremath{\theta_{#1}}\xspace}
\newcommand{\sn}[1]{\ensuremath{ \sin^{2}(\theta_{#1}) }\xspace }
\newcommand{\snt}[1]{\ensuremath{ \sin^{2}(2\theta_{#1}) }\xspace }
\newcommand{\Hlv}{\ensuremath{H_{LV}}\xspace}

\newcommand{\at}{\ensuremath{a^{T}}\xspace}
\newcommand{\ctt}{\ensuremath{c^{TT}}\xspace}
\newcommand{\ats}[1]{\ensuremath{\at_{#1}}\xspace}
\newcommand{\ctts}[1]{\ensuremath{\ctt_{#1}}\xspace}
\newcommand{\as}[1]{\ensuremath{\left(a^{T}_{#1}\right)^*}\xspace}
\newcommand{\cs}[1]{\ensuremath{\left(c^{TT}_{#1}\right)^*}\xspace}

Though the standard paradigm of neutrino oscillations driven by two
mass differences has been well established, hints for a third mass
difference with $\Delta m^{2}_{s} \sim 1 \mbox{eV}^{2}$ have been seen
in a variety of short-baseline experiments
(c.f.~\cite{Aguilar:2001ty,Aguilar-Arevalo:2013pmq,Huber:2011wv,Mention:2011rk}).
Measurements of the $Z^{0}$ decay width, however, indicate that there
are only three neutrinos that participate in the weak interaction and
therefore explaining the short-baseline data with an additional
neutrino means it cannot couple to the $Z$ and therefore cannot
participate in ordinary weak interactions.  Such a state is referred
to as ``sterile.''  Even without weak interactions the existence of
such sterile neutrinos can make imprints on the atmospheric neutrino
spectrum visible at Hyper-K.  In addition to sterile neutrinos, other
sub-dominant contributions to the standard oscillation picture, such
as effects of Lorentz-invariance violating (LV) processes are expected
to influence the oscillations of atmospheric neutrinos.  Positive
observation of LV would provide access to physics at the Planck scale,
an energy regime far beyond the reach of current accelerator
technology.  
Though Hyper-K will have an atmospheric neutrino sample
of unprecedented size, its increased sensitivity to sterile and LV
oscillations relative to existing measurements is hampered by our current
understanding of atmospheric neutrino flux and interaction
uncertainties.
In particular, reductions in the uncertainty on the absolute neutrino flux, 
the $\nu_{\mu}/\nu_{e}$ ratio below 10~GeV, and the $\nu_{\mu}/\nu_{e}$ cross section 
ratio will improve Hyper-K's sensitivity to these oscillations. 
\begin{figure}[thb]
  \begin{center}
    \includegraphics[scale=0.35]{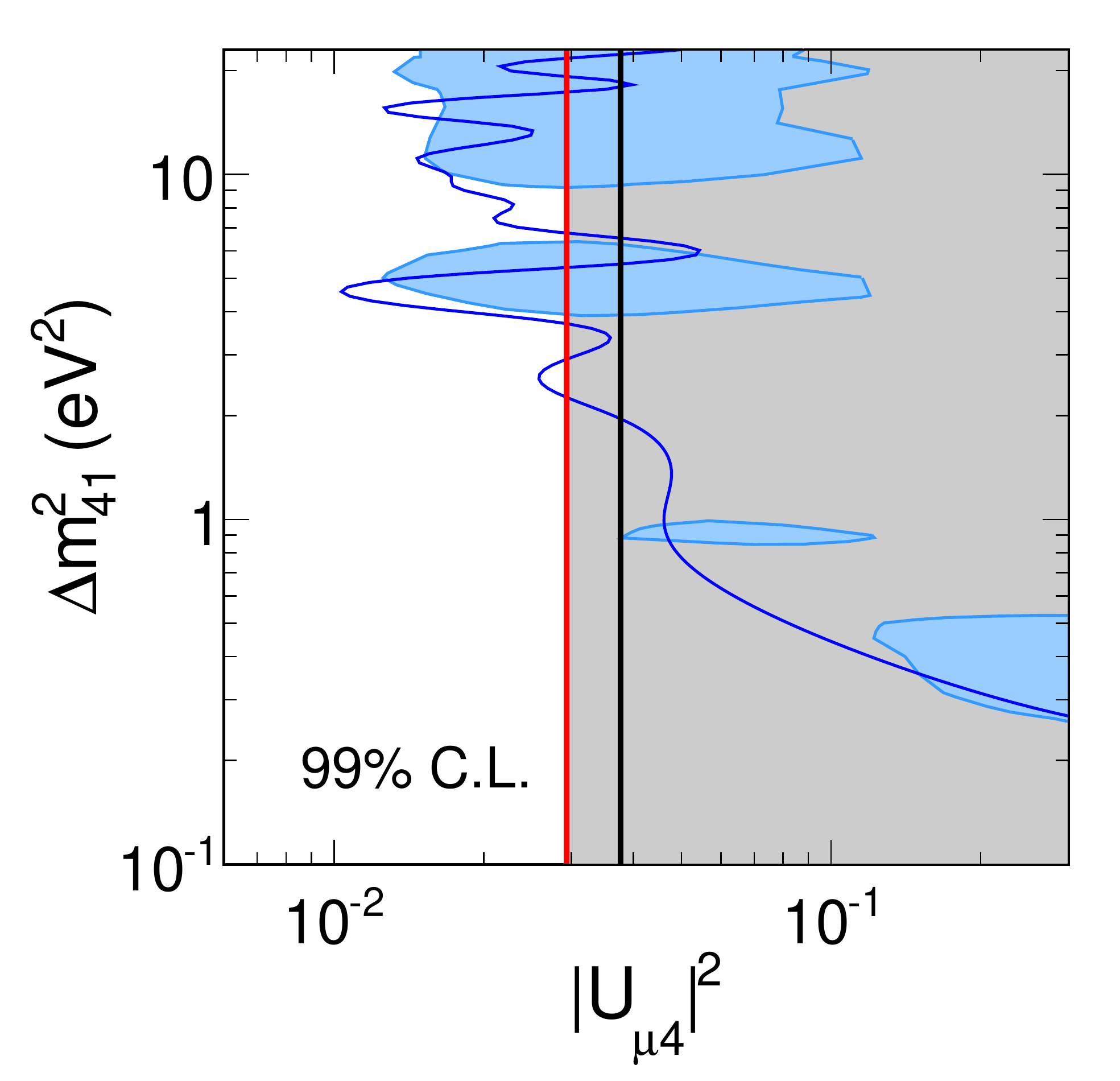}
    \includegraphics[scale=0.35]{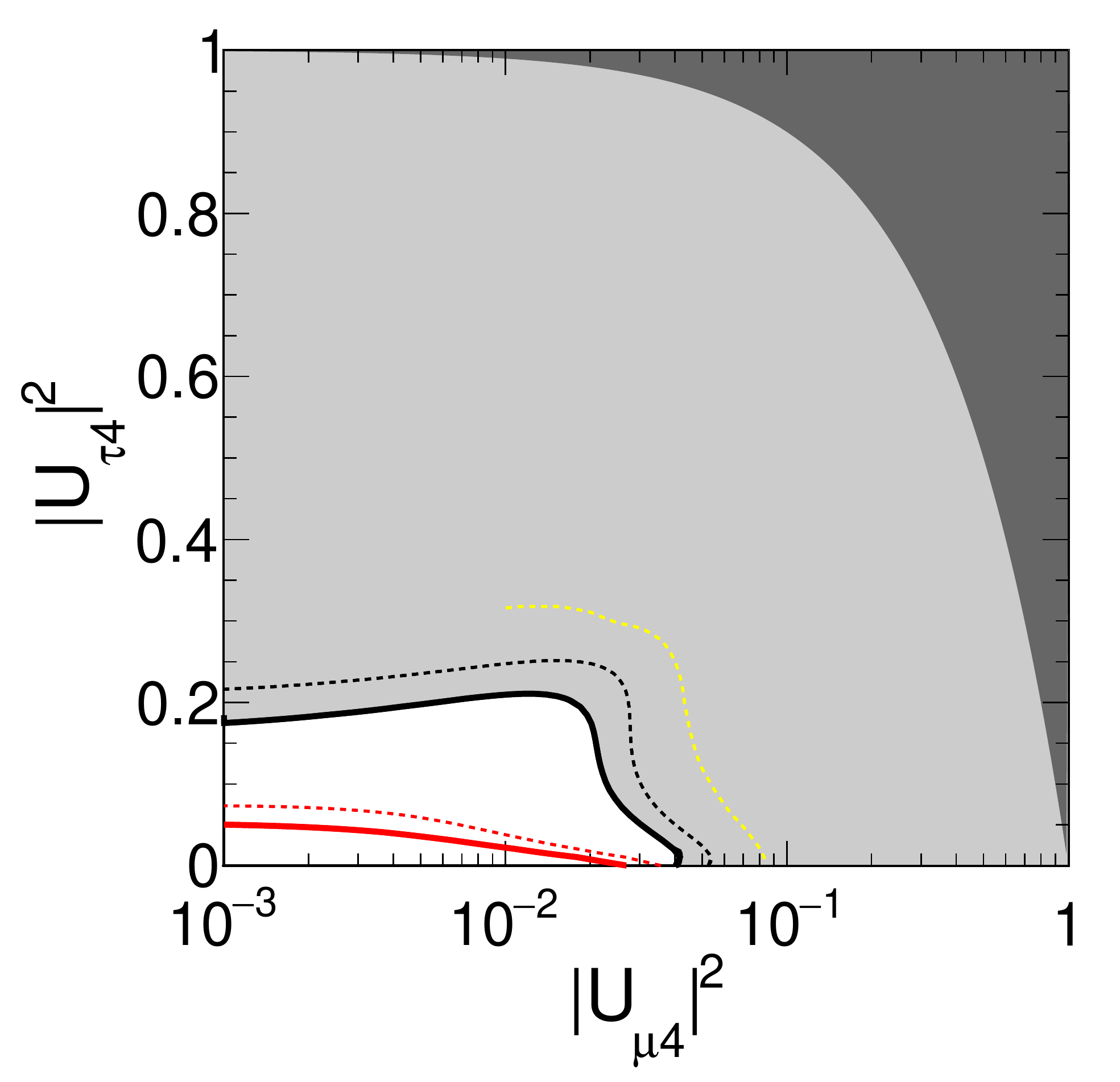}
  \end{center}
  \caption{ Hyper-K's expected 90\% C.L. upper limits on
    $|U_{\mu4}|^2$ appear as red lines in the left figure. 
    The right figure shows 90\% (solid) and 99\% (dashed) C.L. limits on $|U_{\mu4}|^2$ vs $|U_{\tau4}|^2$
    for a 5.6~Mton$\cdot$year exposure (red) in comparison with
    recent limits from Super-K (black)~\cite{Abe:2014gda}.
    Limits at
    90\% C.L. from a joint analysis of MiniBooNE and SciBooNE
    data~\cite{Cheng:2012yy} and light blue filled areas show the
    allowed regions from a joint fit to global $\nu_{e}$ appearance
    and disappearance data from~\cite{Kopp:2013vaa}. 
    A yellow dashed line shows the limit from~\cite{Kopp:2013vaa}.  }
\label{fig:sterile_limits}
\end{figure}

Extensions of the standard (PMNS) oscillation framework to include
sterile states expand the mixing matrix with additional rows and
columns that include terms describing mixing between the active and
sterile neutrinos, such as $U_{e4}$, $U_{\mu4}$, and $U_{\tau4}.$
Interestingly, for mass differences $\Delta m^{2}_{s} > 0.1
\mbox{eV}^{2}$, as suggested by short-baseline measurements,
atmospheric neutrinos are not sensitive to the exact value of the
splitting and further, they are essentially insensitive to the exact
number of sterile neutrinos.  Nonetheless, non-zero mixing of the
active and sterile neutrinos is expected to produce spectral
distortions of the atmospheric neutrino flux or suppress it.  Sterile
neutrinos lack NC interactions, which makes them subject to an
additional effective potential, $V_{s} = \pm (G_F/ \sqrt{2})N_n $
\noindent when traveling through matter. 
Here, $N_n$ is the local neutron density and $G_F$ the Fermi constant.

Due to the complications of simultaneously describing sterile matter
effects and oscillations that include $\nu_{e}$ for atmospheric
neutrinos two searches for sterile neutrinos are performed following
the formalism of~\cite{Maltoni:2007zf}.  For the first analysis both
$\theta_{13}$ and $\theta_{12}$ are set to zero, decoupling $\nu_{e}$
oscillations from the other active neutrinos.  In so doing,
atmospheric neutrinos are sensitive to the effects of $|U_{\mu 4}|$,
which is expected to decrease the overall $\nu_{\mu}$ survival
probability, and $|U_{\tau 4}|$, which causes an energy dependent
distortion of the $\nu_{\mu}$ flux through the sterile matter
potential.  However, this comes at the expense of a slight bias in the
measurement of $|U_{\mu 4}|$.  Constraints on $|U_{\tau 4}|$ are
possible via its effect on the partially contained (PC) and upward-going $\mu$ (Up $\mu$)samples, both of which
are enriched in $\nu_{\mu}$ interactions.  
In the second analysis,
sterile matter effects are assumed to be negligible and $\nu_{e}$
oscillations are reintroduced.  This approximation allows for an
unbiased measure of $|U_{\mu 4}|$.  The results of these analyses are
shown in comparison to limits from Super-K are shown in
Figure~\ref{fig:sterile_limits}.  It should be noted that the
relatively modest improvement in Hyper-K's measurement of $|U_{\mu
  4}|$ is due to uncertainties in the $(\nu_{\mu} + \bar
\nu_{\mu})/(\nu_{e} + \bar \nu_{e})$ flux and CCQE cross section below
10 GeV.
\begin{table*}[htbp]
\begin{center}
\begin{tabular}{l|ccc|ccc}
\hline
\hline
Analysis    & \multicolumn{3}{c|}{ $\alpha^{T}_{XY}$ [GeV]}            &         \multicolumn{3}{c}{ $C^{TT}_{XY}$ }         \\ 
            & $e \mu$          & $e \tau$         &  $\mu \tau$        &  $e \mu$         & $e \tau$          &  $\mu \tau$  \\ 
\hline
\hline
Super-K     & $2\times10^{-23}$& $4\times10^{-23}$ & $6\times10^{-24}$ & $2\times10^{-26}$& $1\times10^{-24}$ & $5\times10^{-27}$ \\ 
Hyper-K     & $7\times10^{-24}$& $2\times10^{-23}$ & $2\times10^{-24}$ & $6\times10^{-27}$& $7\times10^{-25}$ & $2\times10^{-27}$ \\ 
\hline
\hline
\end{tabular}
\caption{ Comparison of Hyper-K and Super-K~\cite{Abe:2014wla}
          90\% C.L. limits on Lorentz-invariance violation within the context of the SME.  \label{tbl:liv_limits}
         }
\end{center}
\end{table*}
 
Atmospheric neutrino oscillations are particularly sensitive to
effects of LV processes due to their interferometric nature.  Though
such effects may manifest in atmospheric neutrinos as either sidereal
variations in their oscillations or as distortions in the oscillated
spectra, the present analysis considers only the latter.  The search
is performed within the context of an effective field theory which
contains the standard model, general relativity, and all possible LV
operators known as the standard model extension
(SME)~\cite{Kostelecky:2003fs}.  Much like the sterile oscillations
described above the SME extends neutrino Hamiltonian by introducing a
LV component,
\begin{align}
\Hlv =& 
\left(\begin{array}{ccc}
0          & \ats{e\mu}      & \ats{e\tau} \\
\as{e\mu}  & 0               & \ats{\mu\tau} \\
\as{e\tau} & \as{\mu\tau}    & 0  \\
\end{array}\right) 
  -E 
\left(\begin{array}{ccc}
0          & \ctts{e\mu}     & \ctts{e\tau} \\
\cs{e\mu}  & 0               & \ctts{\mu\tau} \\
\cs{e\tau} & \cs{\mu\tau}    & 0  \\
\end{array}\right),
\end{align}
\noindent where $\mbox{a}_{\alpha\beta}^{T}$ and
$\mbox{c}_{\alpha\beta}^{TT}$ are complex coefficients for isotropic
LV operators.  In general the $\mbox{a}_{\alpha\beta}^{T}$ parameters
produce oscillation effects proportional to the neutrino propagation
distance, $L$, while the $\mbox{c}_{\alpha\beta}^{TT}$ induce effects
that depend on $LE$, where $E$ is the neutrino energy.  While
atmospheric neutrinos are effective probes of such exotic oscillations
due to their the large variety of pathlengths and energies, this same
feature prohibits perturbative solutions to SME Hamiltonian and
instead the problem must be fully diagonalized.

Event modest amounts of LV can have large effects on atmospheric
neutrino oscillations, but generally they are expected to appear
primarily in the multi-GeV (both $e$-like and $\mu$-like), PC, and
Up$\mu$ samples.  Both the real and imaginary components of each of
the LV coefficients found in the Hamiltonian above are then fit for in
the analysis, with each coefficient considered individually while all
others are held at zero.  Limits from a 5.6~Mton$\cdot$year exposure of Hyper-K
within this framework appear in conjunction with limits from Super-K
in Table~\ref{tbl:liv_limits}.  For each of the considered
coefficients the expected Hyper-K limit is roughly three or four times
more stringent.  Additional sensitivity may be gained by incorporating
more detailed shape information into the Hyper-K analysis through
finer binning.

\begin{figure}[thb]
  \begin{center}
    \includegraphics[scale=0.35]{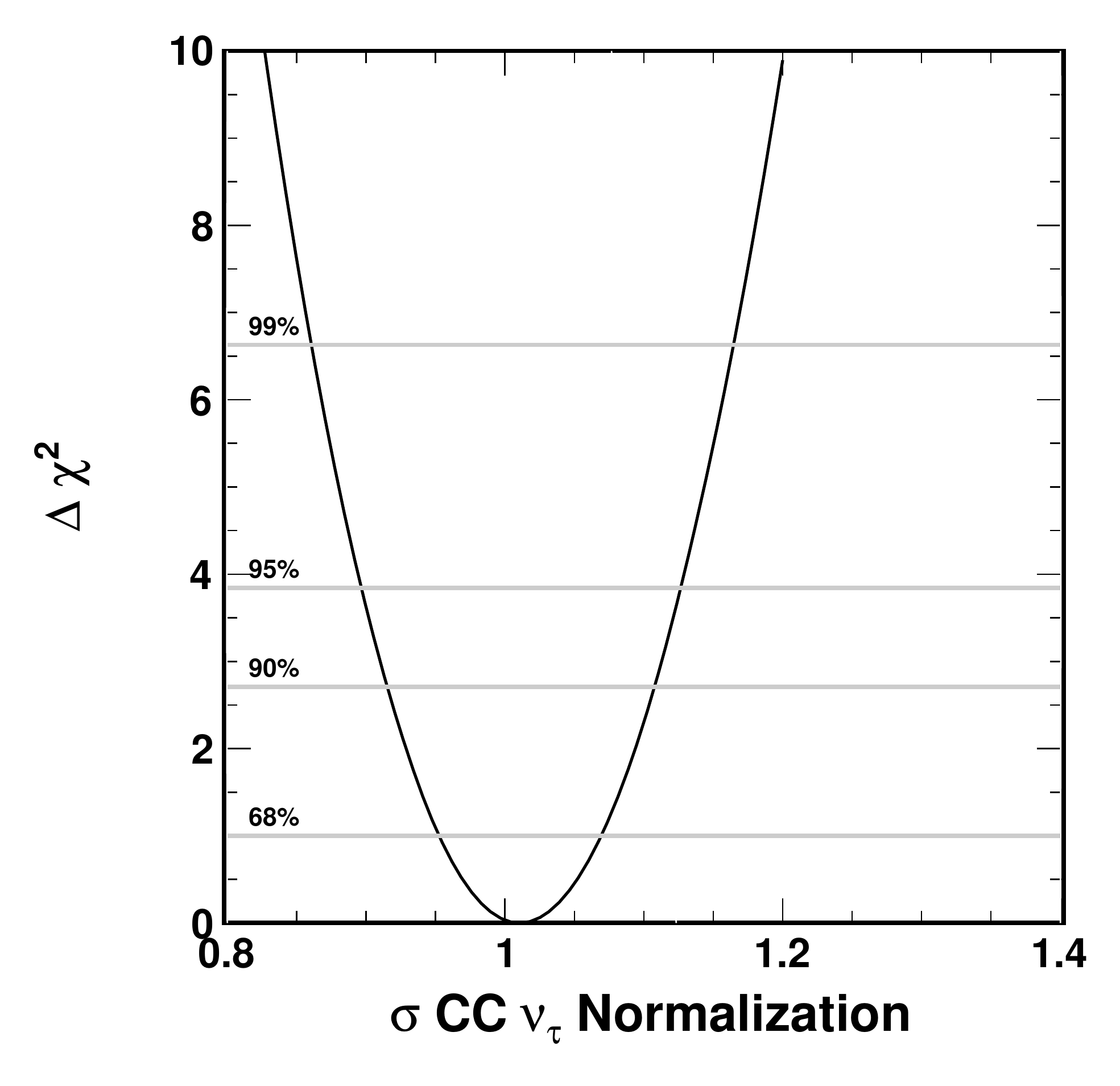}
  \end{center}
  \caption{ Expected constraint on the CC $\nu_{\tau}$ cross section normalization 
            for a 5.6~Mton$\cdot$year exposure of Hyper-K's to atmospheric neutrinos. }
\label{fig:taucc_norm}
\end{figure}

In addition to the studies presented above, Hyper-K's atmospheric
neutrino sample is expected to enable measurements of several other
physical phenomena, including the CC $\nu_{\tau}$ cross section,
non-standard neutrino interactions, the primary atmospheric neutrino
flux, and others.  The study of $\nu_{\tau}$ interactions is a topic
of particular importance, since oscillation-induced $\nu_{\tau}$
events within the detector are often reconstructed as multi-GeV e-like
interactions in the upward-going direction making them a significant
background to Hyper-K's mass hierarchy sensitivity.  In addition,
since creating and subsequently observing $\nu_{\tau}$ in accelerator
experiments is difficult the world data is comprised of only a handful
of events (nine events from DONuT~\cite{Kodama:2007aa}, and five from
OPERA~\cite{Agafonova:2015jxn}).  For this reason the uncertainty in
the interaction cross section is large, with
reference~\cite{Abe:2014gda} assigning a 25\% systematic error based
on a survey of theoretical models.  Super-Kamiokande has developed a
method to extract oscillation-induced $\nu_{\tau}$ events from the
atmospheric neutrino background based on a neural network procedure
trained to select CC interactions in which the $\tau$ lepton has
decayed hadronically.  This technique was used to successfully
identify an oscillation-induced $\nu_{\tau}$ signal of $\sim 40$
events atop of a background of $\sim 448$ events per 100 kton
years~\cite{Abe:2012jj}.  Extrapolating to a 10 year exposure of Hyper-K,
this would correspond to more than 700 CC interactions.  Further, by
incorporating the neural network variable into the oscillation
analysis described in Section~\ref{section:atmnu} it is possible to
isolate the $\tau$-like events and use them to measure their cross
section normalization.  Hyper-K's expected sensitivity to this
normalization is shown in Figure~\ref{fig:taucc_norm}.

\newpage
\graphicspath{{physics-solarnu/figures}}

\subsection{Solar neutrinos}

\label{section:solar}

\begin{figure}[htb]
 \begin{center} \includegraphics[height=9.5cm]{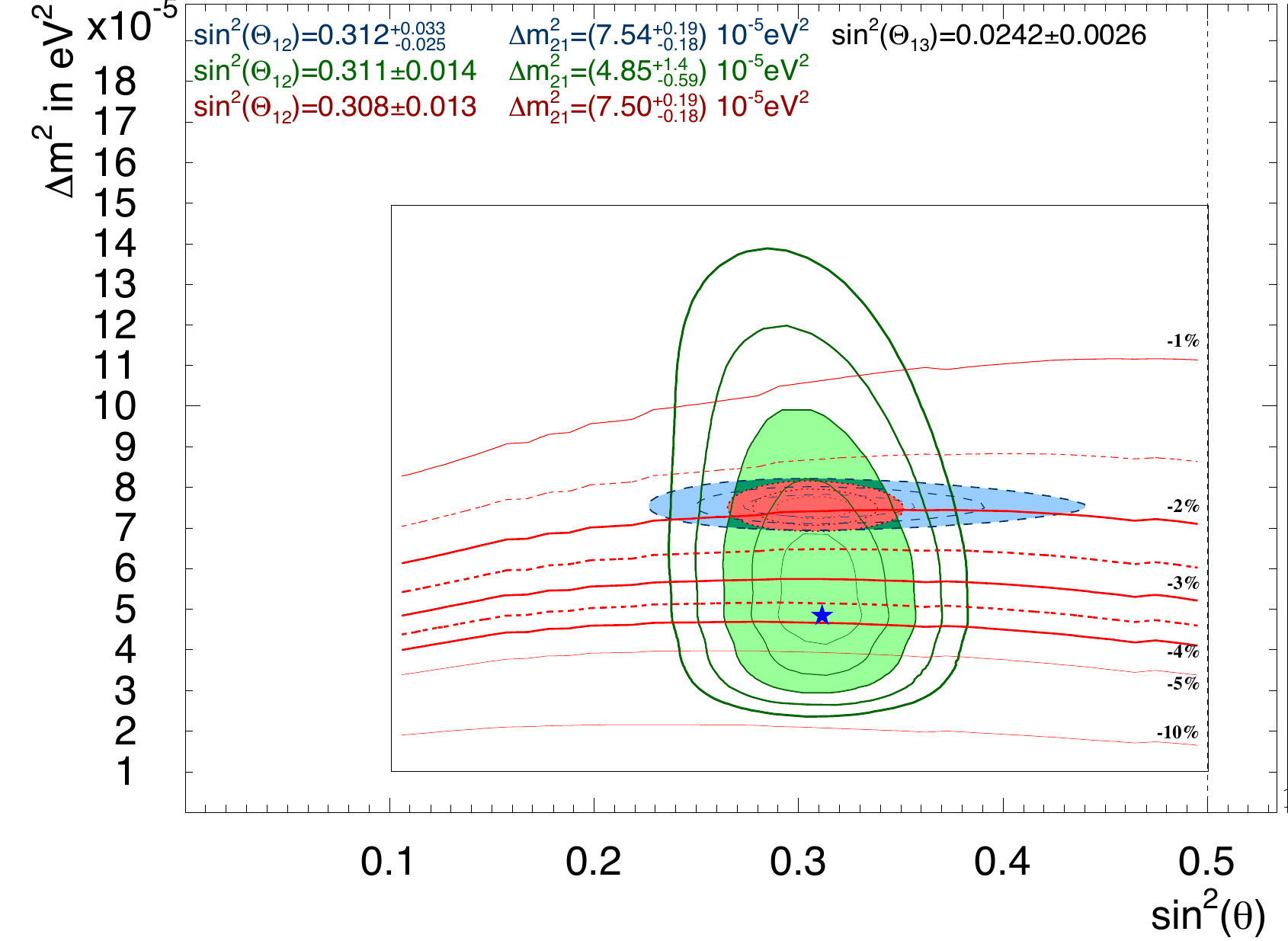} \\ \end{center} \caption{Allowed
  neutrino oscillation parameter region from all the solar neutrino
  experiments (green), reactor neutrino from KamLAND (blue) and
  combined (red) from one to five sigma lines and three sigma filled
  area.  The star shows the best fit parameter from the solar
  neutrinos.  The contour of the expected day-night asymmetry with
  6.5\,MeV (in kinetic energy) energy threshold is overlaid.}  \label{fig:sol-osc}
\end{figure}

The solar neutrino measurements are capable of determining the neutrino
oscillation parameters between mass eigenstate $\nu_1$ and $\nu_2$ in
the equation (\ref{eq:mixing}).  Figure~\ref{fig:sol-osc} shows the
latest combined results of the allowed neutrino oscillation
parameters, $\theta_{12}$ and $\Delta m^2_{21}$ from all the solar
neutrino experiments, as well as the reactor neutrino experiment
KamLAND~\cite{sol-neutrino2014}. The mixing angle is consistent between
solar and reactor neutrinos, while there is about 2$\sigma$ tension in
$\Delta m^2_{21}$.  This mainly comes from the recent result of the solar
neutrino day-night asymmetry and energy spectrum shape observed in Super-K.
In solar neutrino oscillations, regeneration of the electron neutrinos through the MSW
matter effect in the Earth is expected.  According to the MSW model,
the observed solar neutrino event rate in water Cherenkov detectors in
the nighttime is expected to be higher -- by about a few percent in
the current solar neutrino oscillation parameter region -- than that
in the daytime as shown in the Figure~\ref{fig:sol-osc}.  Super-K
found the first indication of this day-night flux asymmetry at the
3$\sigma$ level~\cite{sk4-daynight}, but conclusive evidence is
expected in Hyper-K.  If the 2$\sigma$ tension of $\Delta m^2_{21}$
between solar ($\nu_e$) and reactor ($\bar{\nu_e}$) neutrinos is a
real effect, new physics must be introduced.

In addition to that, the observation of the upturn in the solar
neutrino survival probability might be possible. The spectrum upturn is produced by
the transition of the survival probability in $\nu_e$ from the matter
dominant energy region to the vacuum dominant energy region in the
solar neutrino oscillation, and has been observed by the comparison
between $^8$B solar neutrino flux in Super-K and SNO and $^7$Be solar
neutrino flux in BOREXINO~\cite{borexino-7be}. However, the precise
measurement of the spectrum shape can distinguish the usual neutrino
oscillation scenario from several exotic models such as non standard
interaction~\cite{sol-nsi}, MaVaN~\cite{sol-mavan}, and sterile
neutrino~\cite{sol-sterile}, for example.  
Due to the high photo-coverage of 40~\%, the
lower energy threshold required to measure the upturn is possible because of better energy
resolution and reduction of the radio active background.~\footnote{
Though the energy of the radioactive events are lower than the energy
threshold, they can be misreconstructed such that they contaminate the
region above the energy threshold. 
If the energy resolution is better, the background will be reduced more.
See more in Section~\ref{sec:lowe_bg}.
}

In the following sections, the sensitivity of the day-night flux
asymmetry and spectrum upturn in Hyper-K are described.

\if 0
In solar neutrino oscillations, regeneration of the electron neutrinos
through the Mikheyev-Smirnov-Wolfenstein (MSW) matter effect
\cite{Wolfenstein:1977ue,Mikheyev:1985zz,Mikheyev:1986zz} 
in the Earth is expected. 
Regeneration of the solar electron neutrinos in the Earth would constitute 
concrete evidence of the MSW matter effect, and so it is important 
to experimentally observe this phenomenon.
However, the matter effect acting on solar neutrinos passing through the Earth has not been 
directly confirmed yet, since the sensitivities of the current 
solar neutrino experiments are not sufficient.
According to the MSW model, the observed solar neutrino event rate 
in water Cherenkov detectors in the nighttime is expected to be higher 
-- by about a few percent in the current solar
neutrino oscillation parameter region --
than that in the daytime. 
We would like to measure this difference in Hyper-Kamiokande. 
\fi

Hyper-K also could be used for variability analyses of the Sun.  For
example, the $^8$B solar neutrino flux highly depends on the Sun's
present core temperature~\cite{1996PhRvD..53.4202B}.
Unlike multiply scattered, random-walking
photons or slow-moving helioseismic waves, free streaming solar
neutrinos are the only available messengers with which to precisely
investigate ongoing conditions in the core region of the Sun.
Hyper-K, with its unprecedented statistical power, could measure the
solar neutrino flux over short time periods. Therefore, short time
variability of the temperature in the solar core could be monitored by
the solar neutrinos in Hyper-K.

\if 0
In order to achieve these precision measurements, background event
levels must be sufficiently small.  Here, we have estimated the basic
performance of Hyper-K for low energy events assuming some typical
background levels.  In this study, the current analysis tools and the
detector simulation for the low energy analysis~\cite{Abe:2011xx} in
Super-K were used.  The dark rate of the PMTs and the water
transparency were assumed to be similar to those in the current
Super-K detector.  A brief summary of the low energy event
reconstruction performance in Hyper-K is listed in
Table~\ref{tab:performance}.

The analysis threshold of the total energy of the recoil electrons in
Hyper-K will be 7.0\,MeV or lower, since a 7.0\,MeV threshold was
previously achieved in the SK-II solar neutrino
analysis~\cite{sk2-solar}.  The current analysis tools will work all
the way down to 4.5\,MeV in Hyper-K with a vertex resolution of 3.0~m.
Not surprisingly, higher energy events will be reconstructed with even
better vertex resolution.
\fi

\subsubsection{Background estimation}
\label{section:solar_bg}
The major background sources for the $^8$B solar neutrino measurements
are the radioactive spallation products created by cosmic-ray muons~\cite{sk-spa} and
the radioactive daughter isotopes of ${}^{222}\mbox{Rn}$ in water.
The spallation products is discussed in detail in the paragraph~\ref{par:spallation_reduction},
and the rate of spallation which result in relevant backgrounds is 2.7 times higher in Hyper-K compared to Super-K
because of its shallow depth.
\if 0
${}^{222}\mbox{Rn}$  will be reduced to a similar (or lower) level as that currently
in the Super-K detector, since Hyper-K will employ a similar water
purification system and design improvements may well occur over the
next several years.  However, the spallation products will be
increased in Hyper-K.
\fi
As the radioactive daughter isotopes, ${}^{222}\mbox{Rn}$  is an important background source for the spectrum upturn
measurement.  First of all, the water purification system must achieve ${}^{222}\mbox{Rn}$ levels similar to that achieved at Super-K.
Furthermore, this background level must be achieved across the full fiducial volume, unlike at Super-K, where only a limited volume can be used for events with less than 5\,MeV of energy.
It is a challenging task but we believe that this should be
possible by design improvements over the next several years.
Therefore, the same ${}^{222}\mbox{Rn}$ background level as Super-K in full fiducial volume is assumed in the following calculation.

\subsubsection{Oscillation studies}

In order to calculate the neutrino oscillation sensitivity, the signal
and background rates in each option have to be estimated.  As for the
day-night asymmetry analysis, the energy threshold is set to 6.5\,MeV
(in kinetic energy)
since its effect is larger at higher energy region.  In this
energy region, only spallation backgrounds should be considered.  The
remaining spallation background rate in Super-K phase IV (40\%
photo-coverage) has been reduced by a factor three comparing to Super-K
phase II (20\% photo-coverage) because of the better energy resolution
and better vertex resolution.
From this experience in Super-K, the spallation background in Hyper-K
will be reduced by a factor three because of the higher photon detection
efficiency than Super-K.

\begin{figure}[htb]
 \begin{center}
  \includegraphics[height=9.5cm]{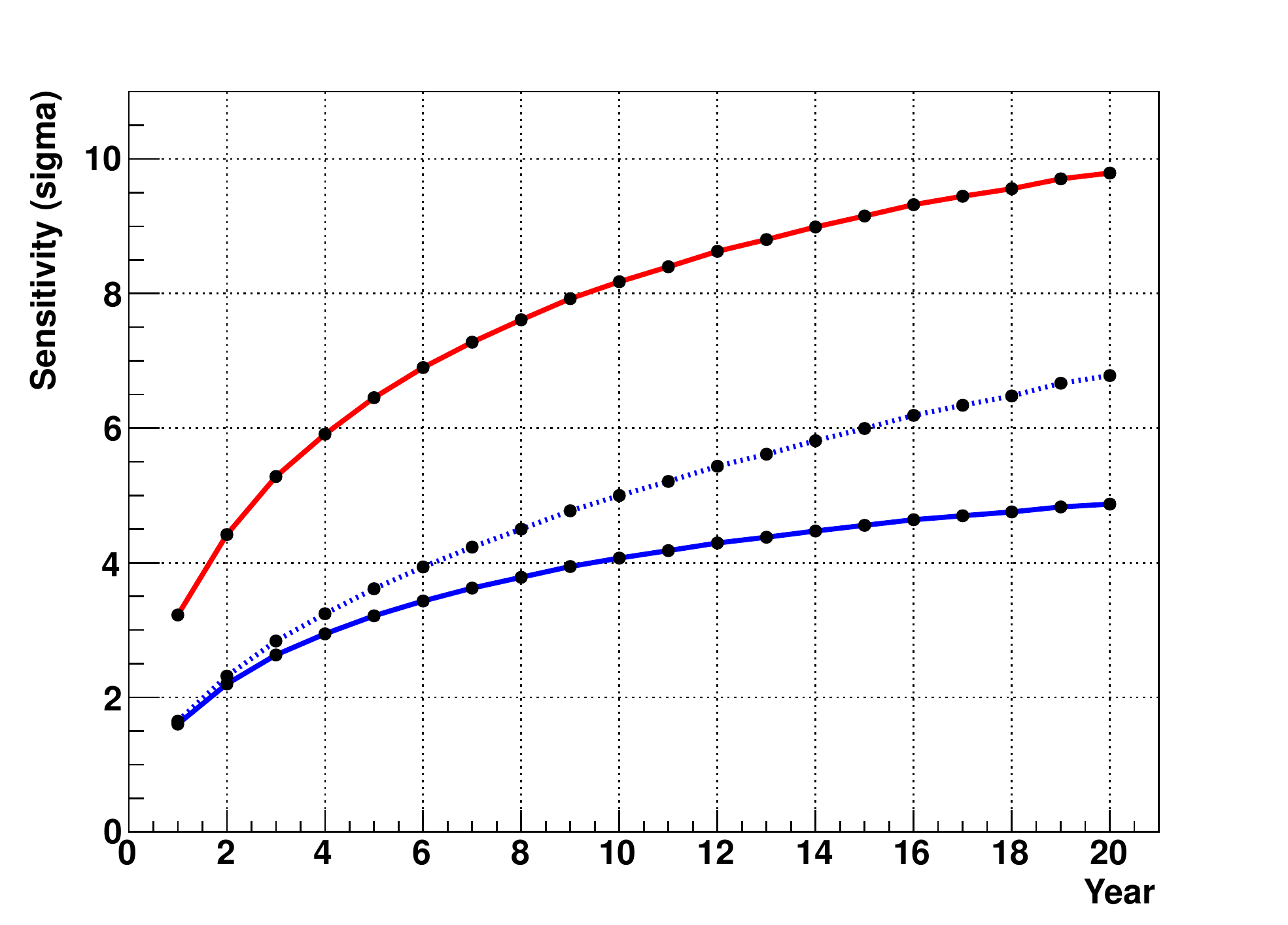} \\
 \end{center}
 \caption{Day-night asymmetry observation sensitivity as a function of observation time. The red line shows the sensitivity from the no asymmetry, while the blue line shows from the asymmetry expected by the reactor neutrino oscillation. The solid line shows that the systematic uncertainty which comes from the remaining background direction is 0.3\%, while the dotted line shows the 0.1\% case.}
 \label{fig:sol-dn}
\end{figure}

Figure~\ref{fig:sol-dn} shows the sensitivity of the day-night asymmetry
as a function of the observation time.
The $\Delta m^2_{21}$ separation ability between solar neutrino (HK) and
reactor anti-electron neutrino (KamLAND) is expected 4$\sim$5$\sigma$ level in ten years observation, though it depends on the systematic uncertainty.

\begin{figure}[htb]
 \begin{center}
  \includegraphics[height=9.5cm]{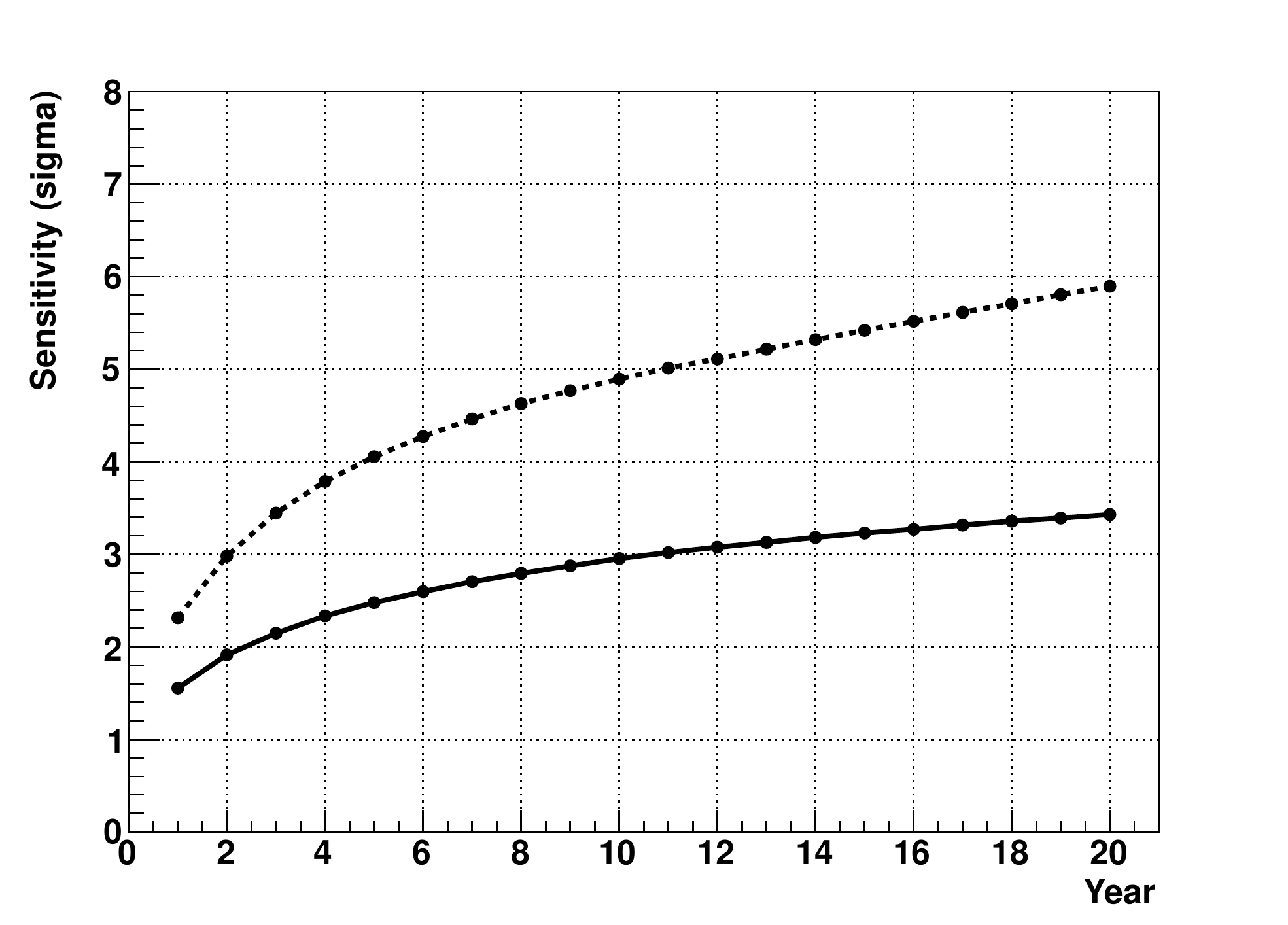} \\
 \end{center}
 \caption{Spectrum upturn discovery sensitivity as a function of observation time. The solid line shows that the energy threshold is 4.5MeV, while the dotted line shows the 3.5MeV}
 \label{fig:sol-upturn}
\end{figure}

In the measurement of the spectrum upturn, the ${}^{222}\mbox{Rn}$  background is
critical because the $^{214}$Bi beta decay events (3.27\,MeV end point
energy) will come above the energy threshold due to the energy
resolution.  The Hyper-K detector, which has better energy resolution
because of the higher photon detection efficiency than Super-K,
is strong to reduce such kind of radioactive background.
Furthermore, the precise energy calibration has to be considered.
Here, it is assumed that the same background level with full fiducial
volume and the same precision of the energy calibration as Super-K are
achieved in Hyper-K.
Figure~\ref{fig:sol-upturn} shows the sensitivity of the spectrum upturn
discovery as a function of the observation time.
It is about 3$\sigma$ level in ten years observation with 4.5\,MeV energy threshold.

\subsubsection{Hep solar neutrino}

Hep solar neutrino produced by the $^{3}$He + $p$ fusion reaction 
has the highest energy in solar neutrinos.
But, most of the hep energy spectrum is overlapped with $^8$B solar neutrinos.
The expected $^8$B solar neutrino flux is more than 100 times larger than that of hep solar
neutrino in Standard Solar Model (SSM).  
So far, only upper limits were reported from SNO and SK group~\cite{hep-sno,hep-sk}, 
but a recent improved analysis of the SNO charged-current data shows hints of 
a hep solar neutrino signal~\cite{hep-sno-2016}, and indicates a higher hep 
flux than the SSM prediction.
 
The measurement of the hep solar neutrino could provide new
information on solar physics.  The production regions of the $^8$B and
hep neutrinos are different in the Sun. 
The energy production peak of hep neutrinos is located at the outermost 
radius in the solar core region
among all the solar neutrinos in pp-chain~\cite{bahcall-textbook}.
So, they could be used as a new probe of the solar interior around core region.
Non-standard solar models, originally motivated by the solar neutrino problem, 
predict the potential enhancement of the hep neutrino flux~\cite{Bahcall:1988}.
This is realized through the mixing of $^{3}$He into the inner core on a time scale 
shorter than the $^{3}$He burning time. Significant mixing is 
already ruled out by helioseismology, however, the hep neutrino observation
can be a sensitive probe of the degree of the mixing in the solar core.
There is also the solar abundance problem. $^8$B and hep solar neutrino
fluxes show different behavior with GS98 and AGSS09 chemical
compositions~\cite{solar-abundance}.
Theoretical calculation of hep solar neutrinos is a difficult 
challenge~\cite{solar-fusion}. The measurement of the hep solar neutrino
flux will provide a better understanding of SSM.
Hep solar neutrinos could be also used to test non-standard neutrino physics
in the energy range ($\sim 18$~MeV)~\cite{kubodera-hep}. 

\begin{figure}[htb]
 \begin{center}
  \includegraphics[height=9.5cm]{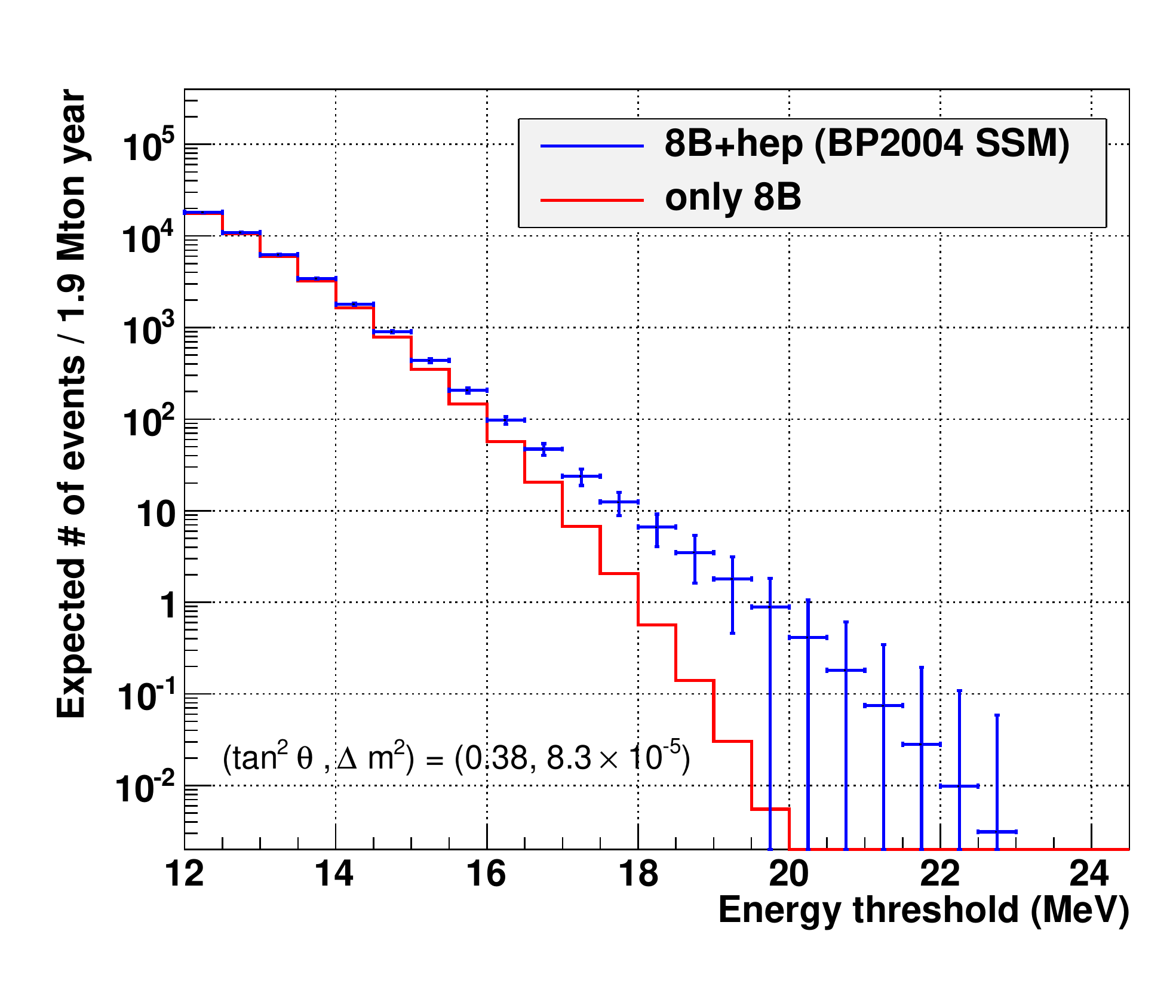} \\
 \end{center}
 \caption{Expected solar neutrino fluxes with neutrino oscillation 
  in Hyper-K. 
 The horizontal axis is the energy threshold in electron total energy 
 and the vertical axis is expected event rate in the energy range 
 from the threshold up to 25\,MeV in 10-year observation in Hyper-K.
 BP2004 SSM fluxes are assumed. 
 The effect of background events, reduction efficiencies, systematic
 uncertainties are not considered.}
 \label{fig:sol-hep}
\end{figure}

In Hyper-K, a high sensitivity measurements of hep solar neutrino flux
would be possible, since the detector has a good energy resolution.
Figure~\ref{fig:sol-hep} shows the expected solar neutrino fluxes
in 1.9 Mton year in Hyper-K detector.
The separation between $^8$B and hep solar neutrinos highly depends
on the energy resolution of the detector. 
Table~\ref{tab:sol-hep} shows a list of expected numbers of solar
neutrino events in typical energy regions.
\begin{table}[htb]
 \caption{Expected solar neutrino event rates in water Cherenkov detectors. 
The assumptions are same as Fig.~\ref{fig:sol-hep}.}
\begin{center}
\begin{tabular}{ccccc}
\hline \hline
Energy resolution & Energy range & $^8$B        & hep          & hep / $^8$B \\
                   & [MeV]        & [/1.9 Mton/year] & [/1.9 Mton/year] &  \\
\hline
\hline
 
 SK-III/IV &  19.5--25.0 & 0.77 & 3.03 &  3.9 \\
 Hyper-K   &  18.0--25.0 & 0.56 & 6.04 & 10.6 \\
\hline \hline
\end{tabular}
\end{center}
\label{tab:sol-hep}
\end{table}
Hyper-K has a better separation between hep and $^8$B solar neutrinos
comparing to SK-III/IV.\par

\begin{figure}[htb]
 \begin{center}
  \includegraphics[height=6.5cm]{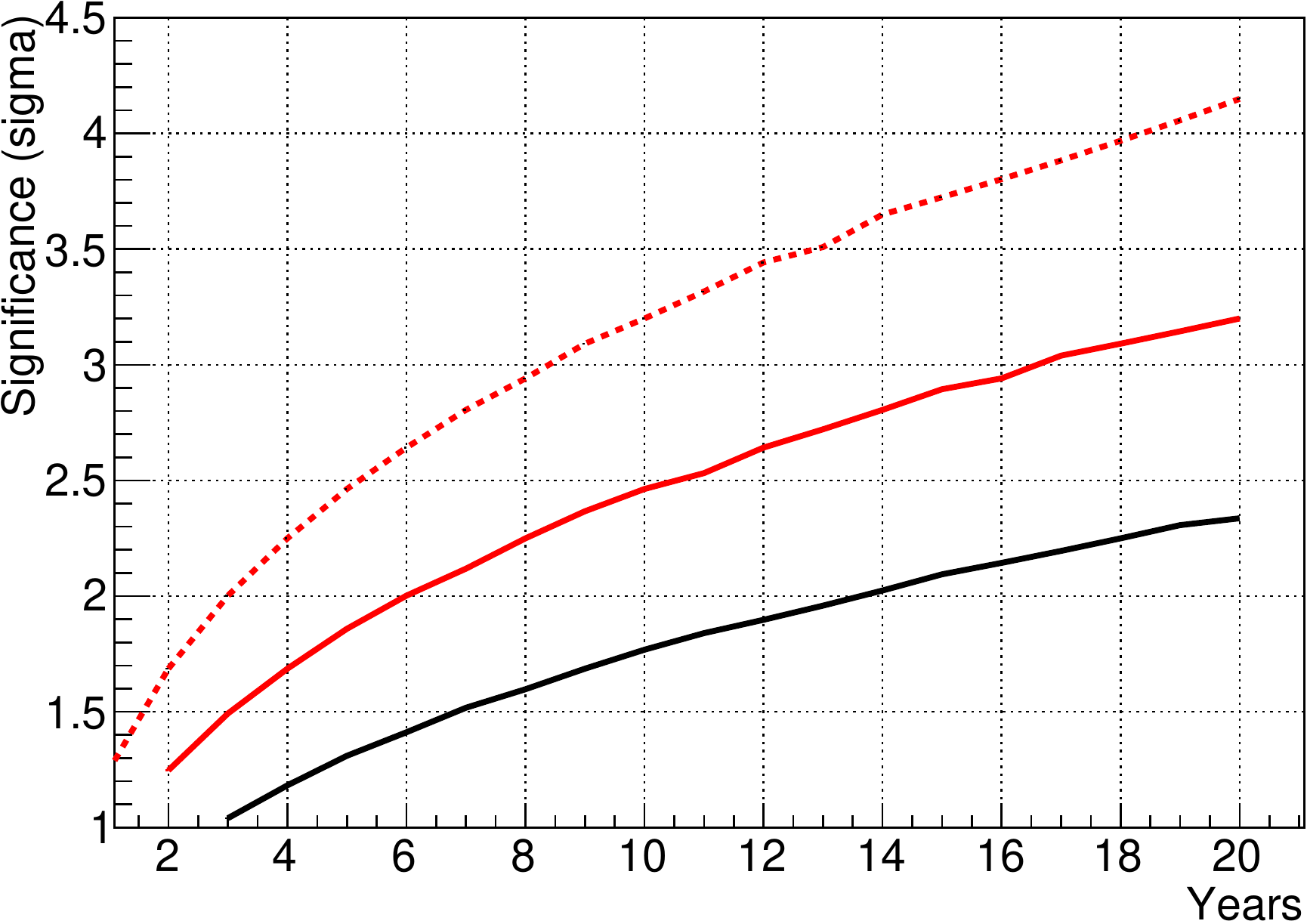} \\
 \end{center}
 \caption{
	 Expected soler hep neutrino sensitivity with 187\,kt fiducial volume.
 The horizontal axis is the observation time and the vertical axis is non-zero significance of hep neutrino signal expected from a energy spectrum analysis.
 BP2004 SSM fluxes are assumed. 
 The effects of remaining spallation background, detection efficiency and systematic uncertainties of the energy scale and resolution estimated from SK-IV data are considered.
 The black solid line shows the expected sensitivity in Tochibora site.
 The red solid and dashed lines show the cases in Mozumi site and no spallation background, respectively.
 }
 \label{fig:sol-hep-sensitivity}
\end{figure}

Figure~\ref{fig:sol-hep-sensitivity} shows an estimation of hep neutrino detection sensitivity.
A spectrum fit analysis is performed here,
considering the spallation background, detection efficiency and systematic uncertainties of the energy scale and resolution.
The statistical error due to remaining spallation background is the dominant source of ambiguity.
When we simply scale the current remaining spallation background level in SK-IV solar analysis, with the cosmic muon rate at Tochibora,
the uncertainty of the hep neutrino flux will be $\sim$60\% ($\sim$40\%) and the non-zero significance will be 1.8$\sigma$ (2.3$\sigma$) in ten (twenty) years observation in Hyper-K.
Due to the higher energy resolution of Hyper-K, there is still chance to improve the sensitivity.
If we can reduce the remaining spallation background to the SK-IV level, the uncertainty of hep neutrino flux will be $\sim$40\% ($\sim$30\%) and non-zero significance will be improved to 2.5$\sigma$ (3.2$\sigma$) in ten (twenty) years observation.
Here the same systematic uncertainties of detector energy scale (0.5~\%) and resolution (0.6~\%) as SK-IV are considered.

\subsubsection{Summary}

In this section, estimates of potential solar neutrino measurements
are reported.  The solar neutrino analysis is sensitive to the
detector resolutions and background levels.  We have estimated
expected sensitivities in 10 years of Hyper-K observation based on the
current Super-K knowledge.

As a result of its shallower site, the increase of the spallation
background level in Hyper-K will be up to a factor of 2.7 larger as compared
to Super-K.  However -- due to its much greater size and high energy
resolution-- the statistical uncertainties on solar
neutrino measurements would actually be reduced in Hyper-K as compared
to Super-K on an equal time basis.

The sensitivity to the difference in neutrino oscillation
parameters between solar and reactor neutrinos due to the day-night asymmetry
is estimated to be 4$\sim$5$\sigma$.
The possibility of spectrum upturn observation is estimated to be at
the 3$\sigma$ level.

The solar hep neutrino could be measured in 
Hyper-K 
with a few Mton year data at the level of $2\sim3 \sigma$.

\if 0
In this section, rough estimates of potential solar neutrino
measurements are reported.  The solar neutrino analysis is sensitive
to the detector resolutions and background levels.  We have estimated
the expected sensitivities based on the current Super-K analysis tools.

As a result of its shallower site, the increase of the background
level in Hyper-K will be up to a factor of 20 as compared to Super-K.
However -- due to its much greater size -- the statistical
uncertainties on solar neutrino measurements would actually be reduced
by a factor of at least two in Hyper-K as compared to Super-K on an
equal time basis, assuming similar detector resolutions.

The day/night asymmetry of the solar neutrino flux -- concrete
evidence of the matter effect on oscillations -- could be discovered
and then precisely measured in Hyper-K, given that the detector
up-down response is understood to better than about 1\%.  Good
calibration tools will be a must for this physics.

Hyper-K will provide short time and high precision variability
analyses of the solar core activity.  The solar core temperature could
be monitored within a few percent accuracy day by day, and to a tenth
of a percent over the period of several months.

The solar hep neutrino could be measured in a \hksingletank{} 
detector with a few Mton year data.
\fi

\newpage
\section{Nucleon Decays}
\graphicspath{{physics-pdecay/figures}}

\subsection{Nucleon decays }\label{section:pdecay}
Optimizing Hyper-Kamiokande for the observation and discovery of a
nucleon decay signal is one if its primary design drivers.  In order
to significantly extend sensitivity beyond existing limits, many of
which have been set by Super-Kamiokande, Hyper-K needs both a much
larger number of nucleons than its predecessor and sufficient
reconstruction ability to extract signals and suppress backgrounds.
While it is possible to target specific decay channels, one of the
strengths of water Cherenkov technology is its sensitivity to a wide variety
of modes.  Using MC and analysis techniques originally developed for
Super-Kamiokande, this section details Hyper-K's expected sensitivity
to both the flagship proton decay modes, $p \rightarrow e^{+} \pi^{0}$
and $p \rightarrow \overline{\nu} K^{+}$, as well as other $\Delta
(B-L)$ conserving, $\Delta B =2$ dinucleon, and $\Delta (B-L)=2$ decays.  
%%%%%%
\begin{figure}[htbp]
  \begin{center}
    \includegraphics[scale=0.4]{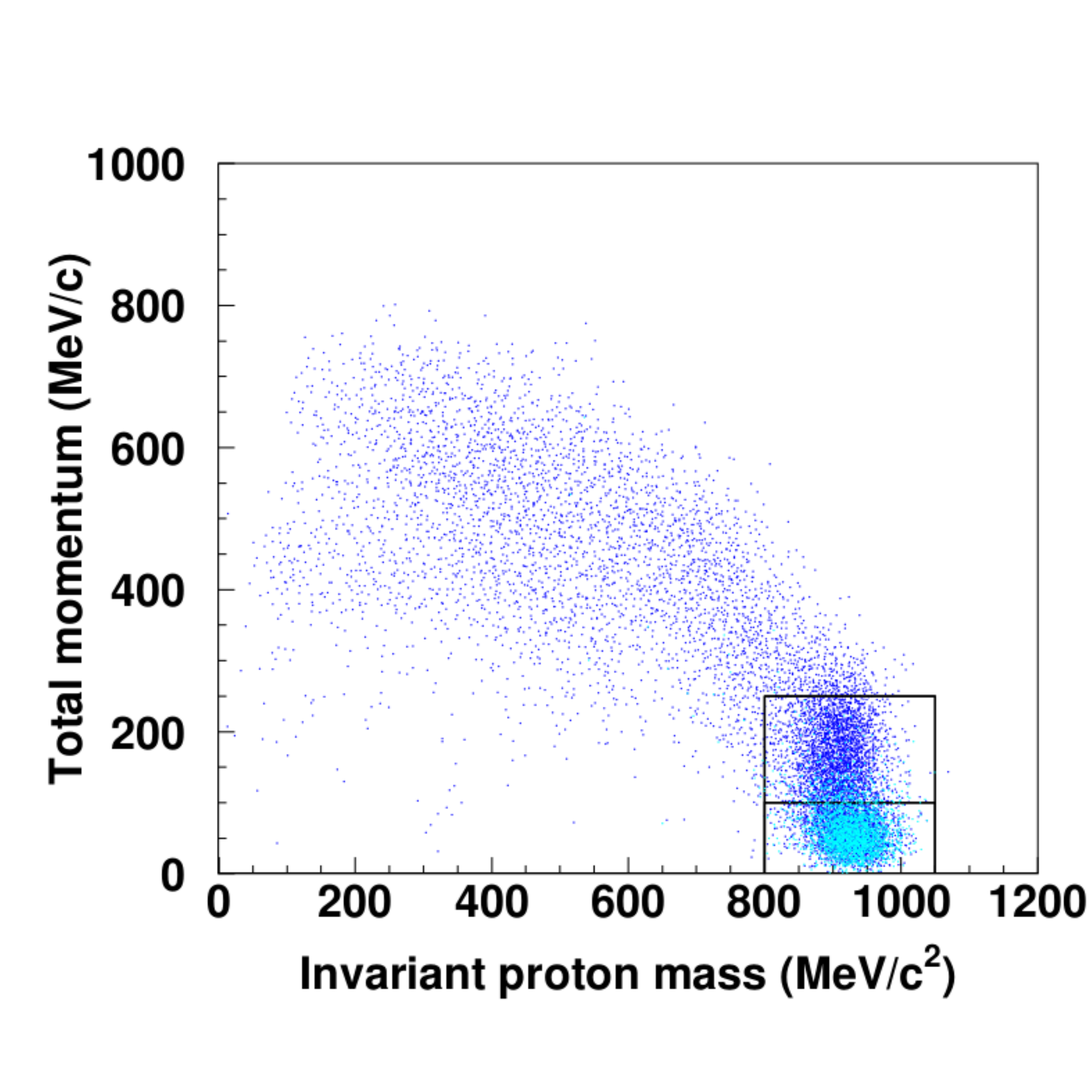}
    \includegraphics[scale=0.4]{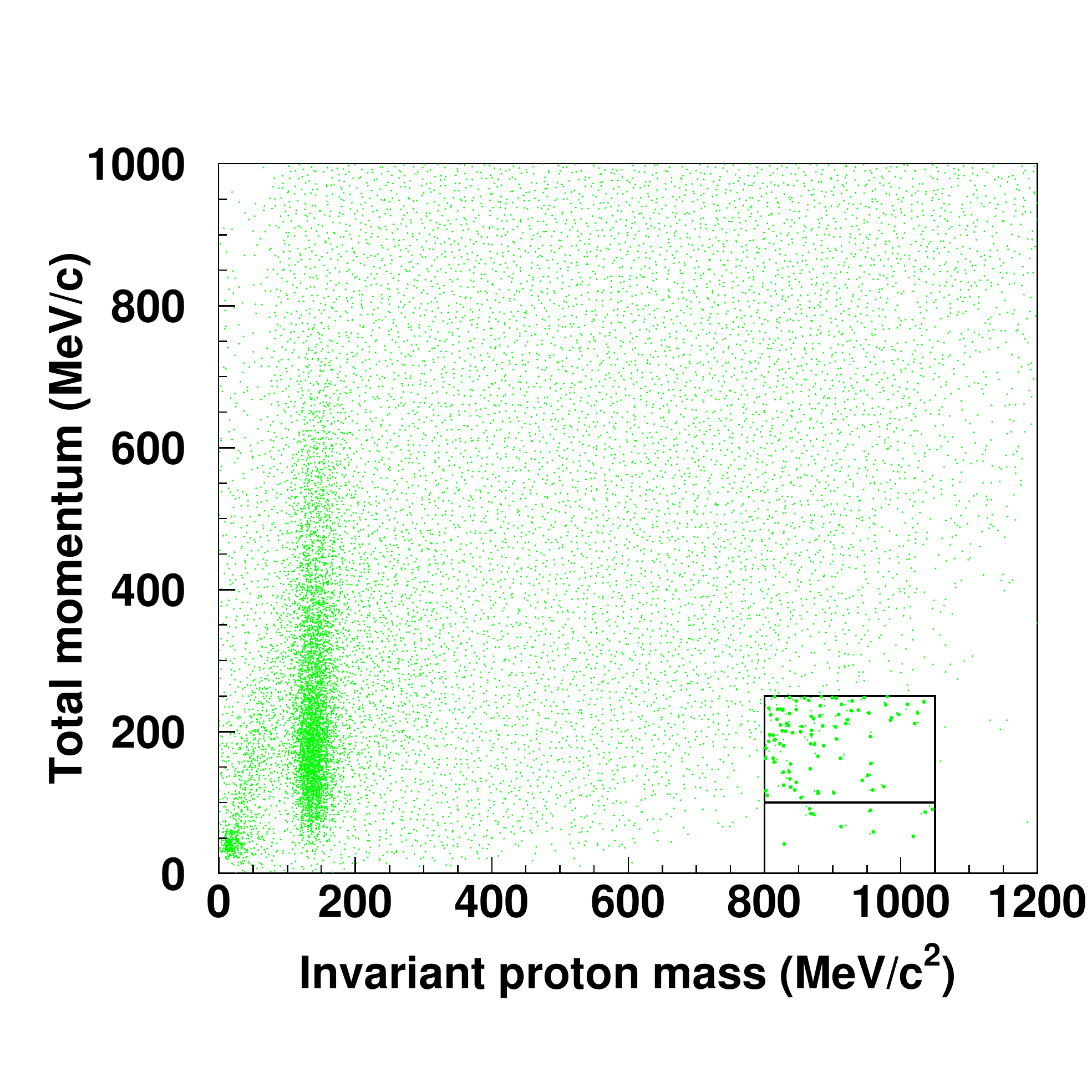}
  \end{center}
\caption {Reconstructed invariant mass and total momentum distributions for the $p \rightarrow e^{+} \pi^{0}$ 
          MC (left) and atmospheric neutrino MC (right) after all
          event selections except the cuts on these variables.  The
          final signal regions are shown by two black boxes in the
          plane.  In the signal plot decays from bound and free
          protons have been separated by color, dark blue and cyan
          respectively.  Background events have been generated for a
          45~Mton$\cdot$year exposure and those falling in the signal
          regions have been enlarged for
          visibility. } \label{fig:epi0_plane}
\end{figure}

%%%%%
\begin{figure}[htbp]
  \begin{center}
     \includegraphics[width=0.47\textwidth,trim={0 4.5cm 0 4cm},clip]{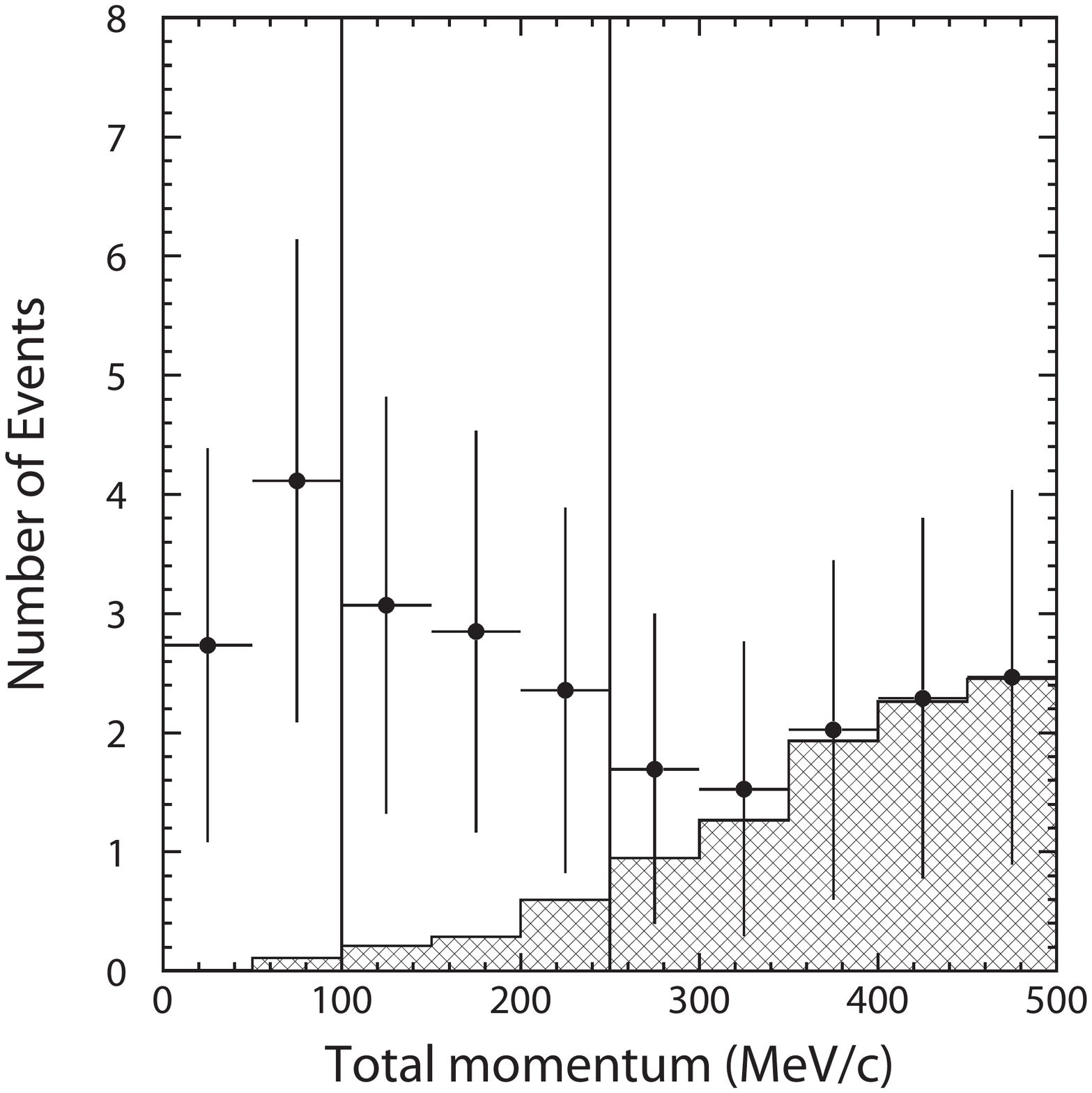}
     \includegraphics[width=0.45\textwidth]{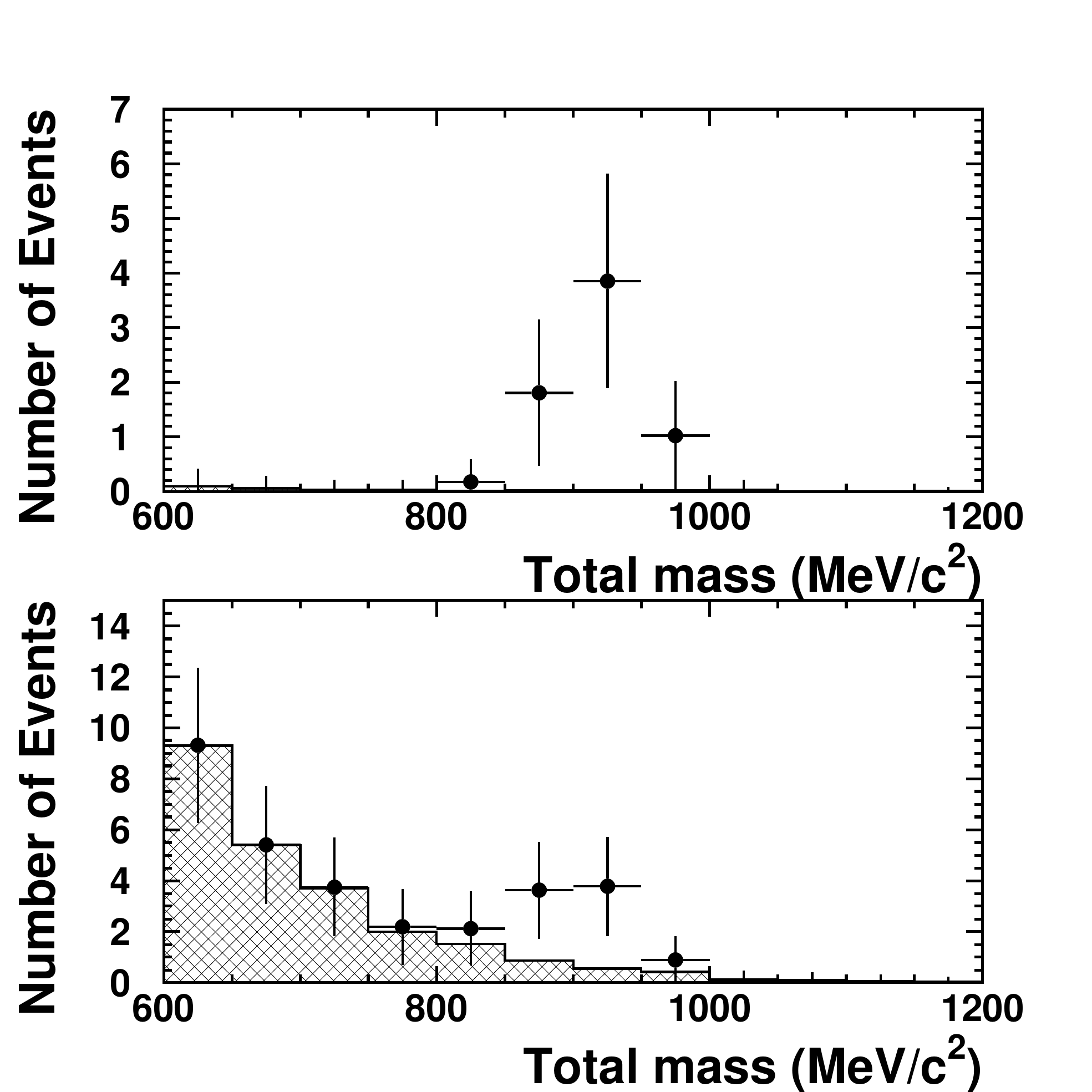}
  \end{center}
\caption { Total momentum distribution of events passing all steps of the $p \rightarrow e^{+} \pi^{0}$ event selection 
          except the momentum cut after a 10 year exposure of a single Hyper-K tank (left).
          Reconstructed invariant mass distribution of events passing all steps of the $p \rightarrow e^{+} \pi^{0}$ event selection 
          except the invariant mass cut after a 10 year exposure of a single Hyper-K tank (right).
          The hatched histograms show
          the atmospheric neutrino background and the solid crosses
          denote the sum of the background and proton decay
          signal. Here the proton lifetime is assumed to be,
          $1.7 \times 10^{34}$\,years, just beyond current Super-K
          limits.   
          The free and bound proton-enhanced bins are shown by the lines in the left plot, 
          and are the upper and lower panels of the right plot. } \label{fig:pmass_epi0}
\end{figure} 
%%%%%%%%%%%%%%%%%%%%%%%%

\subsubsection{Sensitivity to $p \rightarrow e^{+} + \pi^{0}$ Decay}
Proton decay into a positron and neutral pion is a favored mode of
many GUT models.
Experimentally this decay has a very clean
event topology, with no invisible particles in the final state.  As a
result it is possible fully reconstruct the proton's mass from its
decay products and as a two body process the total momentum of
the recoiling system should be small.  The event selection focuses on
identifying fully contained events within the Hyper-K fiducial volume
with two or three electron-like Cherenkov rings.  Though the decay of
the pion is expected to produce two visible gamma rays, for
forward-boosted decays the two photons may be close enough in space to
be reconstructed as a single ring.  Atmospheric neutrino events with
a muon below threshold are removed by requiring there are no Michel
electrons in the event.  For those events with three rings, the two
rings whose invariant mass is closest to the $\pi^{0}$ mass are
labeled the $\pi^{0}$ candidate.  An additional cut on the mass of
those candidates, $ 85 < m_{\pi} < 185$~MeV/$c^{2}$, is applied.
The signal sample is selected by requiring the total invariant mass of
the event be near the proton mass, $800 < m_{inv} <
1050$~MeV/$c^{2}$ and that the total momentum, $p_{tot}$, be less
than 250~MeV/$c$.  Water is an excellent molecule for studying proton
decay because in addition to providing 10 protons per molecule, two of
those are unbound free protons.  Decays from those protons are not
subject to nuclear effects and result in final state particles with
very low total momentum.  At the same time, very few atmospheric
neutrino interactions are reconstructed with both a proton-like invariant
mass and low total momentum.  To optimize the analysis sensitivity,
proton-decay candidates are divided into two signal regions after the
invariant mass cut, a free proton decay enriched region ($0 < p_{tot}
< 100$~MeV/$c$) and a bound proton decay region ($100 < p_{tot}
< 250$~MeV/$c$).  
Finally, neutron tagging (described in a later section)
is used to reject background events by requiring events have no neutron candidates
in the final state.
Figure~\ref{fig:epi0_plane} shows the signal
and background MC in the invariant mass and total momentum plane
before the final signal region is defined.  Signal efficiencies and
background rates with corresponding systematic errors after all
selections are listed in Table~\ref{tbl:pdkepi0}.

\begin{table}[htb]
  \begin{center}
  \begin{tabular}{c|c||c|c}
\hline
\hline
\multicolumn{2}{c||}{ $0 < p_{tot} < 100$~MeV/$c$ } &  \multicolumn{2}{c}{ $100 < p_{tot} < 250$~MeV/$c$ } \\
$\epsilon_{sig}$ [\%] & Bkg [/Mton$\cdot$yr] & $\epsilon_{sig}$ [\%]   &  Bkg [/Mton$\cdot$yr] \\
\hline
\hline
$18.7 \pm 1.2$   &  $0.06 \pm  0.02$    &     $19.4 \pm 2.9$    &  $0.62 \pm 0.20$           \\
\hline
\hline
  \end{tabular}
  \end{center}
  \caption{Signal efficiency and background rates as well as estimated systematic uncertainties 
           for the $p \rightarrow e^{+} \pi^{0}$ analysis at Hyper-K. }
  \label{tbl:pdkepi0}
\end{table}

\begin{figure}[p]
  \begin{center}
    \includegraphics[width=0.7\textwidth,trim={0 0.5cm 0 0},clip]{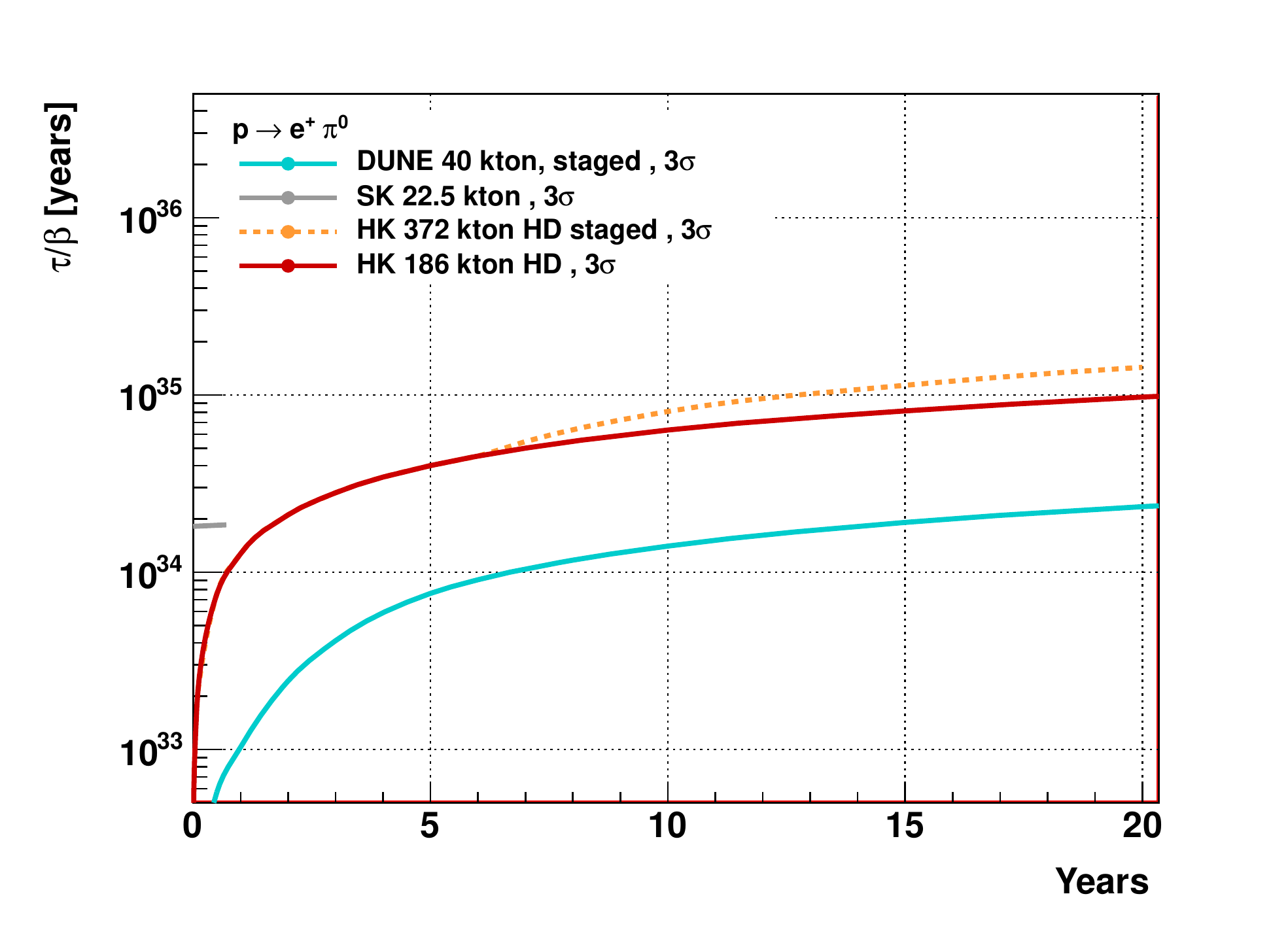}
  \end{center}
  \caption{Comparison of the 3~$\sigma$ $p \rightarrow e^{+}\pi^{0}$ discovery potential as a function of year
           Hyper-K (red solid) assuming a single tank 
           as well as that of the 40~kton liquid argon detector DUNE (cyan solid) following~\cite{Acciarri:2015uup}. 
           In the orange dashed line an additional Hyper-K tank is assumed to
           come online six years after the start of the experiment.
           Super-K's discovery potential in 2026 assuming 23 years of data is also shown.
          }
  \label{fig:epi_discovery}
  \begin{center}
    \includegraphics[width=0.7\textwidth,trim={0 0.5cm 0 0},clip]{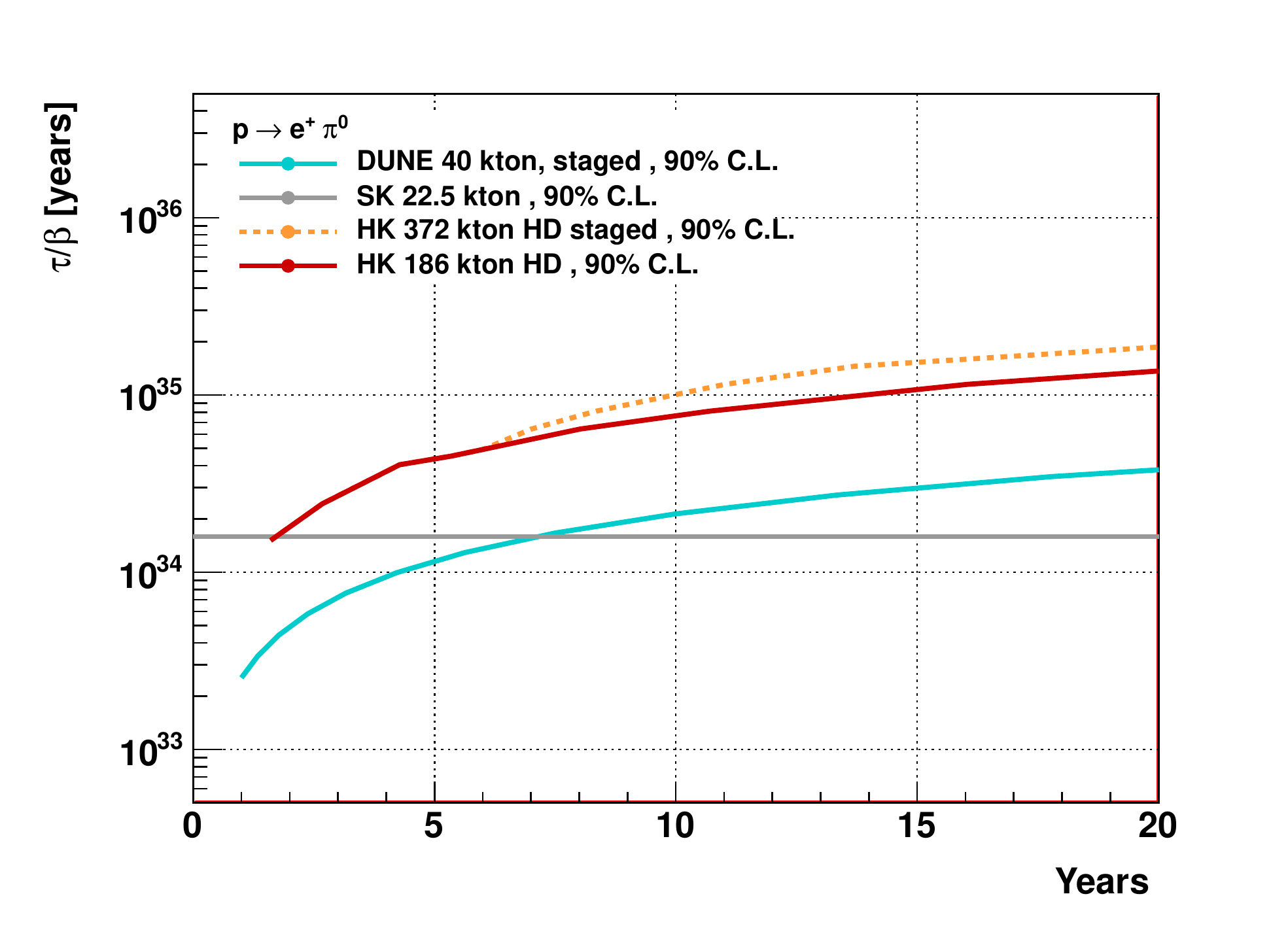}
  \end{center}
\caption {Hyper-K's sensitivity to the $p \rightarrow e^{+} \pi^{0}$ decay mode 
          at 90\% C.L. as a function of run time appears in red assuming one detector in comparison with other experiments (see caption of Figure~\ref{fig:epi_discovery}).
          Super-K's current limit is shown by a horizontal line.
         }
  \label{fig:sens_epi0}
\end{figure}
%%%%

Monte Carlo simulation of these decays includes the effects of the
nucleon binding energy, Fermi motion, and interactions within the
$^{16}O$ nucleus.  The latter represents a significant, but
unavoidable, source of inefficiency as the signal $\pi^{0}$ may be
lost to absorption or charge exchange prior to exiting the nucleus.
Although the signal efficiency of free proton decays is roughly 87\% 
these nuclear effects reduce the
efficiency of bound proton decays such that the overall efficiency is
only $\sim$ 40\% when all decays are considered.

Atmospheric neutrino interactions are the main background to proton
decay searches.  Not only can CC $\nu_{e}$ single-pion production
processes produce the same event topology expected in the
$p \rightarrow e^{+} \pi^{0}$ decay, but recoiling nucleons from
quasi-elastic processes can produce pion final states through hadronic
scatters that mimic the signal.  After the event selection above the
expected background rate is 0.06 events in the free proton
enhanced bin and 0.62 events in the bound proton enhanced bin
per Mton$\cdot$year.
These background expectations have been experimentally verified using
beam neutrino measurements with the K2K experiment's one kiloton water
Cherenkov detector, which found an expected atmospheric neutrino
background contamination to $p \rightarrow e^{+} \pi^{0}$ searches of
$1.63^{+0.42}_{-0.33}(stat)^{+0.45}_{-0.51}(sys)$ events per
Megaton$\cdot$year~\cite{Mine:2008rt} without neutron tagging.

The ability to reconstruct the proton's invariant mass is a powerful
feature of this decay mode.  The left panel of Figure~\ref{fig:pmass_epi0} shows the
expected distribution of this variable for both signal events and
atmospheric neutrino backgrounds after a 10 year exposure assuming the
proton lifetime is $\tau_{x} = 1.7 \times 10^{34}$ years.  
Similarly, the left panel shows the total momentum distribution
of candidate events prior to the momentum selection.
Both plots illustrate 
the marked difference in the signal and background expectations 
for the free proton and bound proton analysis bins.

Hyper-Kamiokande's proton decay discovery potential has been estimated 
based on a likelihood ratio method. 
A likelihood function is constructed from a Poisson probability density 
function for the event rate in each total momentum bin of the $p \rightarrow e^{+} \pi^{0}$
analysis with systematic errors on the selection efficiency and background 
rate represented by Gaussian nuisance parameters.
The experiment's expected sensitivity to a proton decay signal at a given 
confidence level, $\alpha$, is calculated as the fastest proton decay rate, $\Gamma$, 
whose median likelihood ratio value assuming a proton decay signal
yields a p-value not larger than $\alpha$ from the likelihood ratio distribution
under the background-only hypothesis.
That is, 
\begin{equation}
\Gamma_{disc} = \max_{\Gamma} \left[ \alpha = \int_{ M(\Gamma) }^{\infty} f( q_{0} | b ) d q_{0}  \right],  
\end{equation}
\noindent where $f$ is the distribution function of the likelihood ratio, $q_{\Gamma}$, and 
$ M(\Gamma) = \mbox{median}[ f( q_{\Gamma} | \Gamma \epsilon \lambda + b ) ]$. 
Here $\epsilon$ is the selection efficiency, $\lambda$ the detector exposure, 
and $b$ is the expected number of background events.
For the calculated significance the corresponding proton lifetime is then $\tau_{disc} = 1/\Gamma_{disc}$.
Under this definition Hyper-K's $3 \sigma$ (one-sided) discovery potential as a function of run
time is shown in Figure~\ref{fig:epi_discovery}.  Note that if the
proton lifetime is as short as $\tau_{x}$ its decay into
$e^{+}\pi^{0}$ will be seen at this significance within two years of
Hyper-K running. 

Even in the absence of a signal Hyper-K is expected
to extend existing limits considerably.
Hyper-K's sensitivity to this decay mode is computed as
\begin{equation}
\tau_{limit} = \sum_{n=0}^{\infty} O(n|b) / \Gamma_{n},
\end{equation}
\noindent where
\begin{equation}
\Gamma_{n} = \left[ \Gamma_{l} : 1 - \alpha = \int_{0}^{\Gamma_{l}} P( \Gamma | n ) d\Gamma \right]
\end{equation}
\noindent and $P( \Gamma | n )$ is the probability that the proton decay rate is $\Gamma$ given 
an observation of $n$ events.  
Similarly, $O(n|b)$ is the Poisson probability to observe $n$ events given a mean of $b$.
The function $P( \Gamma | n )$ is obtained using Bayes' theorem and the likelihood function 
outlined above. 
Hyper-K's expected sensitivity to $p \rightarrow e^{+} \pi^{0}$ using this metric is shown 
in Figure~\ref{fig:sens_epi0}.
In both this and the discovery potential figure, systematic errors on the signal and background have been
included based on the Super-K analysis; The errors are listed in Table~\ref{tbl:pdkepi0}.  
A detailed description of the systematic error estimation and the limit calculation may be found
in~\cite{Nishino:2012bnw}.

\subsubsection{Sensitivity study for the $p \rightarrow \overline{\nu} K^{+}$ mode} 

Proton decays into a lepton and a kaon are a feature of supersymmetric grand unified theories, of
which $p \rightarrow \overline{\nu} K^{+}$ is among the most
prominent.  Searching for the $K^{+}$ in a water Cherenkov detector is
complicated by the fact it emerges from the decay with a momentum of
only 340\,MeV/$c$, well below the threshold for light production,
749\,MeV/$c$.  As a result the $K^{+}$ must be identified by its decay
products: $K^+ \to \mu^{+}+\nu$ (64\% branching fraction) and
$K^+ \to \pi^{+}+\pi^{0}$ (21\% branching fraction).  Since both of
these modes are two body decays the outgoing particles have
monochromatic momenta.  Furthermore, the 12~ns lifetime of the kaon
makes it possible to observe prompt $\gamma$ ray emission produced
when the proton hole leftover from a bound proton decay is filled by the
de-excitation of another proton.  For $^{16}O$ nuclei the probability
of producing a 6\,MeV $\gamma$ from such a hole is roughly 40\%, making
this a powerful tool for identifying the $K^{+}$ decay products and rejecting 
atmospheric neutrino backgrounds.

\begin{figure}[htbp]
  \begin{center}
    \includegraphics[height=19pc]{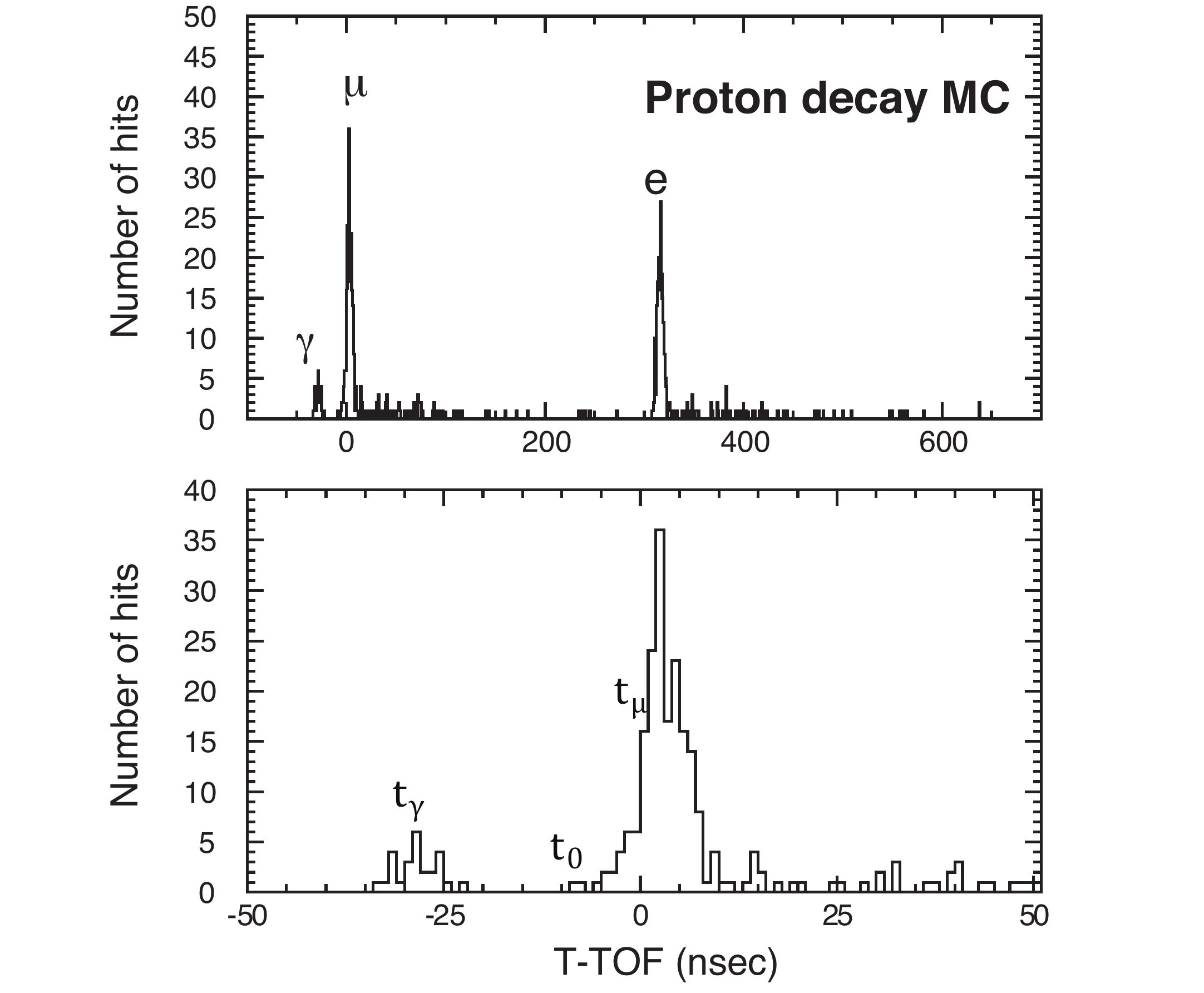}
  \end{center}
\caption {Schematic view of the expected timing distribution of events in the 
          prompt $\gamma$ search method for $p \rightarrow \overline{\nu} K^{+}$ decays.
          The upper panel shows the full event time window with the $\mu$, Michel, 
          and $\gamma$ candidate clusters. The bottom panel shows the time of flight 
          subtracted timing distribution around the $\mu$ candidate.
         }
  \label{fig:prompt-gamma}
\end{figure}

Three methods, each targeting different aspects of the $K{+}$ decay,
are used to search for $p \rightarrow \overline{\nu} K^{+}$
events~\cite{Abe:2014mwa}.  The ``prompt $\gamma$'' method searches
for a prompt nuclear de-excitation $\gamma$ ray (6.3\,MeV) occurring
prior to and separated in time from a 236\,MeV/$c$ muon.  A schematic
of this process appears in Figure~\ref{fig:prompt-gamma}.  In the
second method the same monochromatic muon is searched for but without
the $\gamma$ tag.  Finally, the third method, or ``$\pi^{+}\pi^{0}$
method'', searches for a monochromatic $\pi^{0}$ with light from a
backward $\pi^{+}$.  Here the charged pion is only slightly above
Cherenkov threshold, making it difficult to reconstruct a full 
ring.

%%%%%%

\begin{table}[htb]
  \begin{center}
  \begin{tabular}{c|c||c|c||c|c||c}
\hline
\hline
\multicolumn{2}{c||}{ Prompt $\gamma$ } &  \multicolumn{2}{c||}{ $\pi^{+}\pi^{0}$  } & \multicolumn{3}{c}{ $p_{\mu}$ Spectrum } \\
$\epsilon_{sig}$ [\%] & Bkg  [/Mton$\cdot$yr] &  $\epsilon_{sig}$ [\%]   &  Bkg  [/Mton$\cdot$yr]  &  $\epsilon_{sig}$ [\%]   & Bkg  [/Mton$\cdot$yr] & $\sigma_{fit}$ [\%] \\
\hline
\hline
$12.7 \pm 2.4$  &  $0.9 \pm 0.2$  &  $10.8 \pm 1.1$  &  $0.7 \pm 0.2$  & 31.0  &   1916.0  &  8.0             \\
\hline
\hline
  \end{tabular}
  \end{center}
  \caption{Signal efficiency and background rates as well as estimated systematic uncertainties 
           for the $p \rightarrow \bar \nu K^{+}$ analysis at Hyper-K.}
  \label{tbl:pdknuk}
\end{table}

%%%%%%%%%%
\begin{figure}[htbp]
  \begin{center}
     \includegraphics[width=0.48\textwidth,trim={0 4.5cm 0 4cm},clip,trim={0 4.5cm 0 4cm},clip]{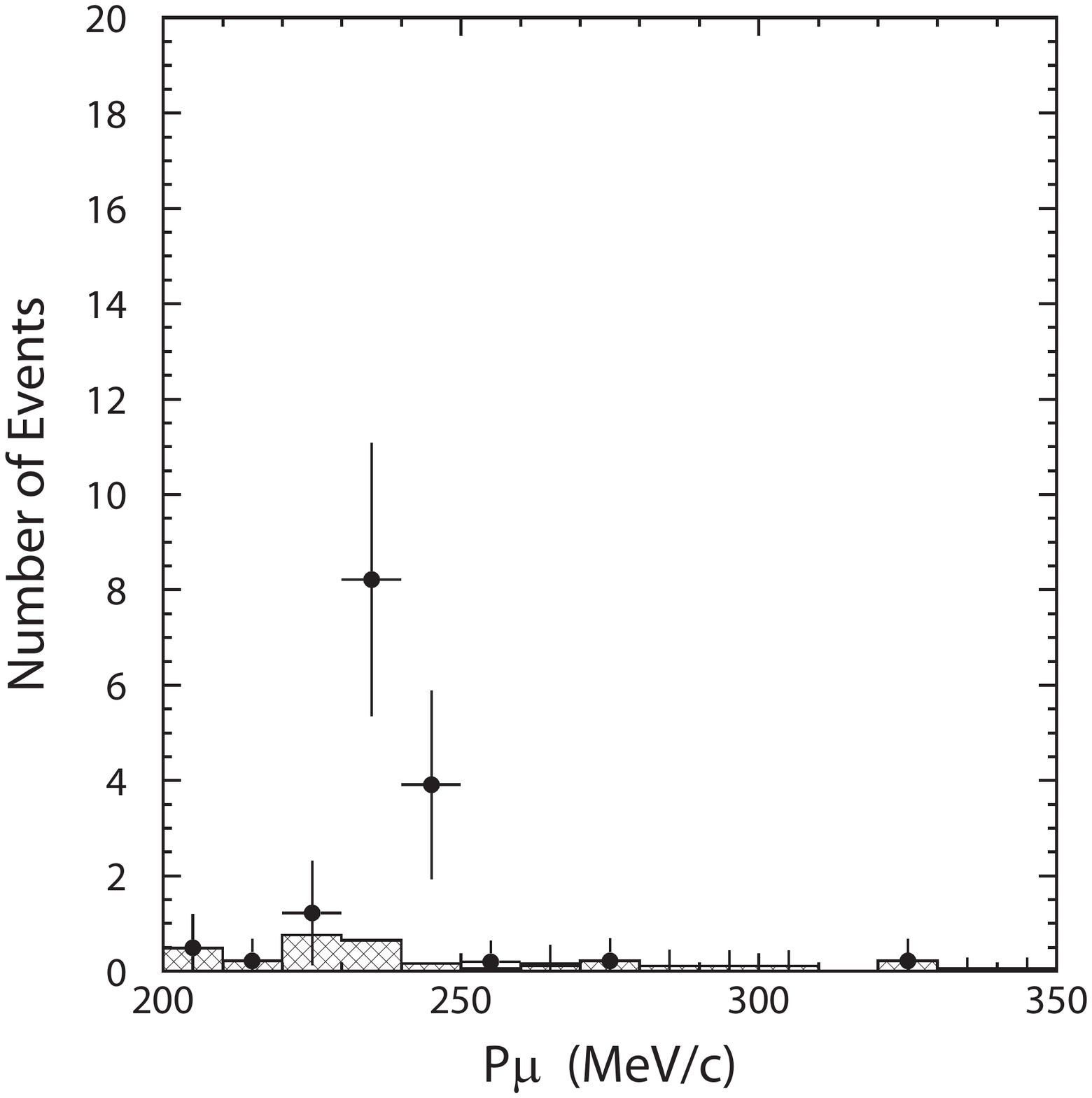}
     \includegraphics[width=0.48\textwidth,trim={0 4.5cm 0 4cm},clip]{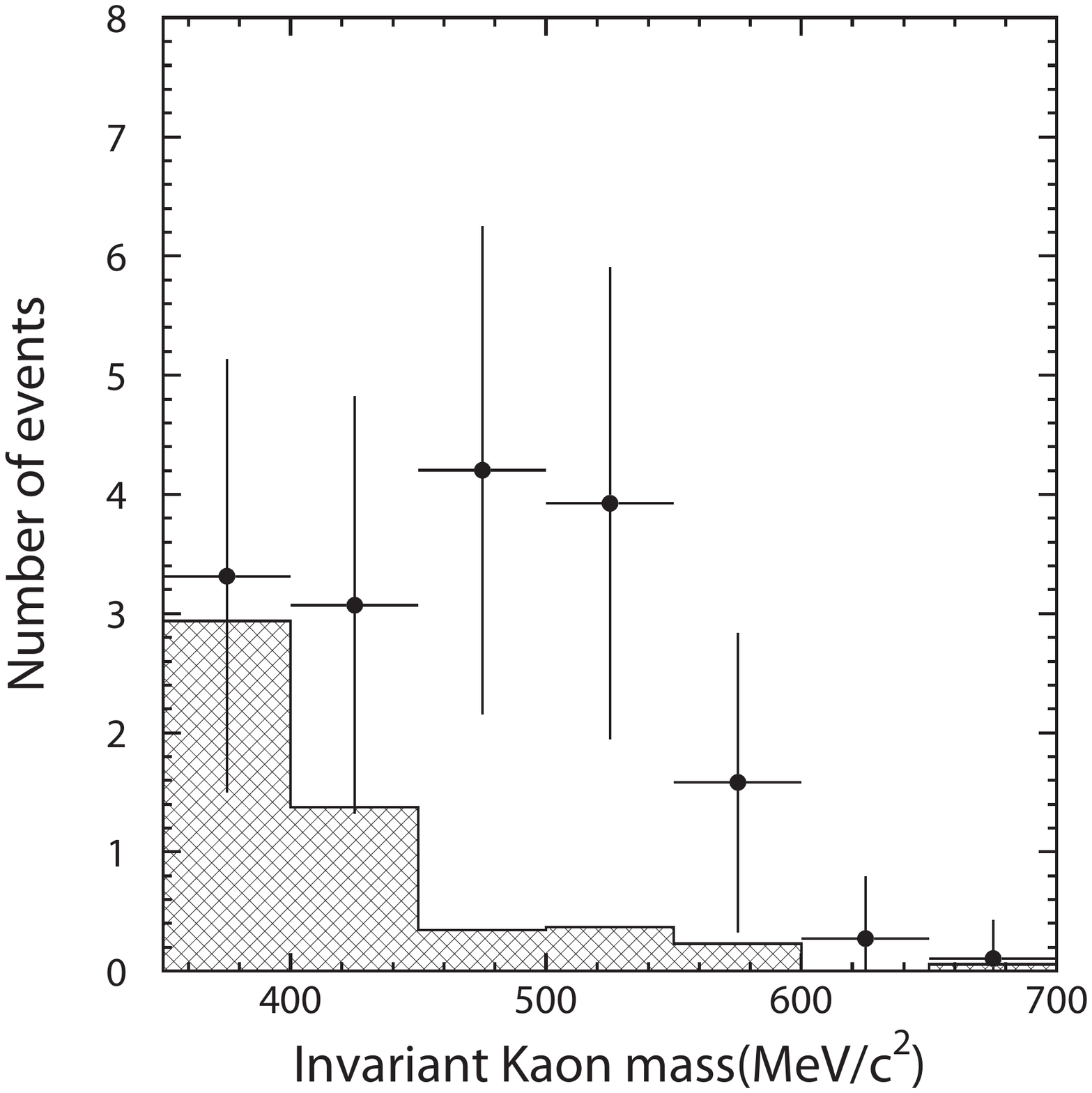}
  \end{center}
\caption { Reconstructed muon momentum distribution for muons found in the prompt $\gamma$ search 
          of the $p \rightarrow \bar \nu K^{+}$ analysis after a  10 year exposure of a single Hyper-K tank (left).
          The right figure shows the reconstructed kaon mass based on the reconstructed final state in candidates from the  
          $\pi^{+}\pi^{0}$. 
          The hatched histograms show the atmospheric neutrino background and the solid crosses
          denote the sum of the background and proton decay signal. 
          Here the proton lifetime is assumed to be,
          $6.6 \times 10^{33}$ years, just beyond current Super-K limits.
          All cuts except for the cut on visible energy opposite the $\pi^{0}$ candidate have been applied in the right plot.  }
  \label{fig:nuk_pmu}
\end{figure}

The prompt $\gamma$ search method proceeds by identifying fully
contained fiducial volume interactions with a single $\mu$-like ring
accompanied by a Michel electron.  In order to suppress atmospheric
neutrino backgrounds, particularly those with an invisible muon, the
muon momentum is required to be $215< p_{\mu} < 260$\,MeV/$c$ and the
distance between its vertex and that of the Michel electron cannot
exceed 200~cm.  Events with high momentum protons that create a
$\mu$-like Cherenkov ring are rejected using a dedicated likelihood
designed to select in favor of genuine muons based on the PMT hit
pattern and the Cherenkov opening angle.  Searching backward in time
from the muon candidate, de-excitation $\gamma$ ray candidates are
identified as the largest cluster of PMT hits within a 12~ns sliding
window around the time-of-flight-subtracted time distribution.  
The time difference between the center of the time window
containing the $\gamma$ candidate, $t_{\gamma}$, and the muon time,
$t_{\mu}$ is then required be consistent with decay of a kaon,
$t_{\mu}- t_{\gamma} < 75$~ns ($\sim 6\tau_{K^+}$).  Finally, only
events whose number of hits in this window, $N_{\gamma}$, is
consistent with 6\,MeV of energy deposition ($8 < N_{\gamma} < 120$) are
kept.  The left panel of Figure~\ref{fig:nuk_pmu} shows the expected $p_{\mu}$
distribution after all selection cuts have been applied assuming a
proton lifetime of $6.6\times 10^{33}$ years, which is slightly less than the current 
Super-K limit.

The second search ($p_{\mu}$ spectrum) method also focuses on
identifying the monochromatic muon but with relaxed search criterion.
Only those cuts applied before the proton likelihood cut
are used.  A fit is then applied to the resulting muon momentum
distribution to identify any proton decay-induced excess of muon
events over the considerable atmospheric neutrino background.

%%%%
%%%%
\begin{figure}[p]
  \begin{center}
    \includegraphics[width=0.7\textwidth,trim={0 0.5cm 0 0},clip]{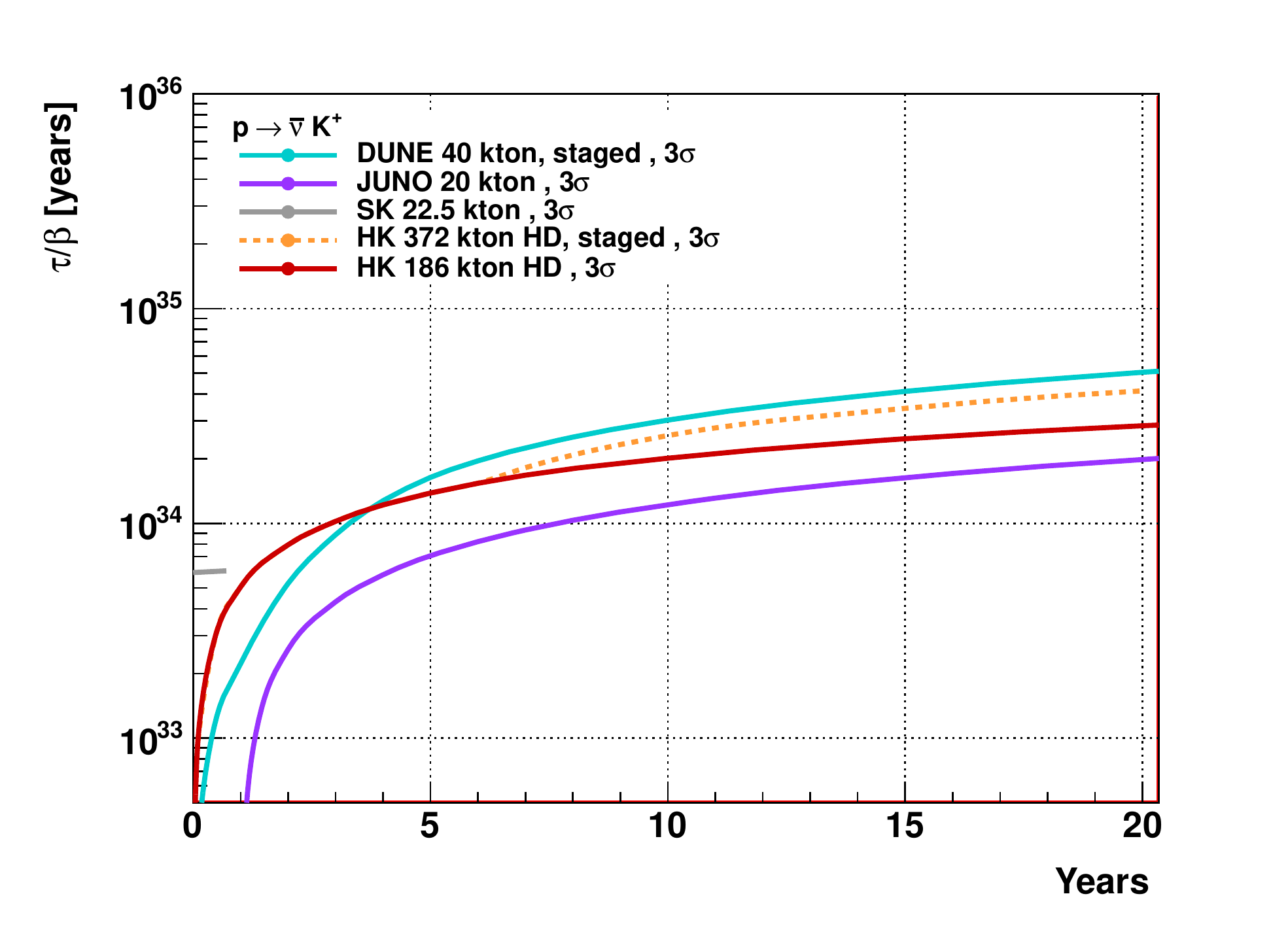}
  \end{center}
  \caption{Comparison of the 3~$\sigma$ $p \rightarrow \bar \nu K^{+}$ discovery potential as a function of year for 
           the Hyper-K as well as that of the 40~kton DUNE detector (cyan solid) based on~\cite{Acciarri:2015uup} and 
           the 20~kton JUNO detector based on~\cite{An:2015jdp}. 
           The red line denotes a single Hyper-K tank, while the orange line shows the expectation when a second 
           tank comes online after six years.
           The expected discovery potential for Super-K by 2026 assuming 23 years of data is also shown.
          }
  \label{fig:nuK_discovery}
  \begin{center}
    \includegraphics[width=0.7\textwidth,trim={0 0.5cm 0 0},clip]{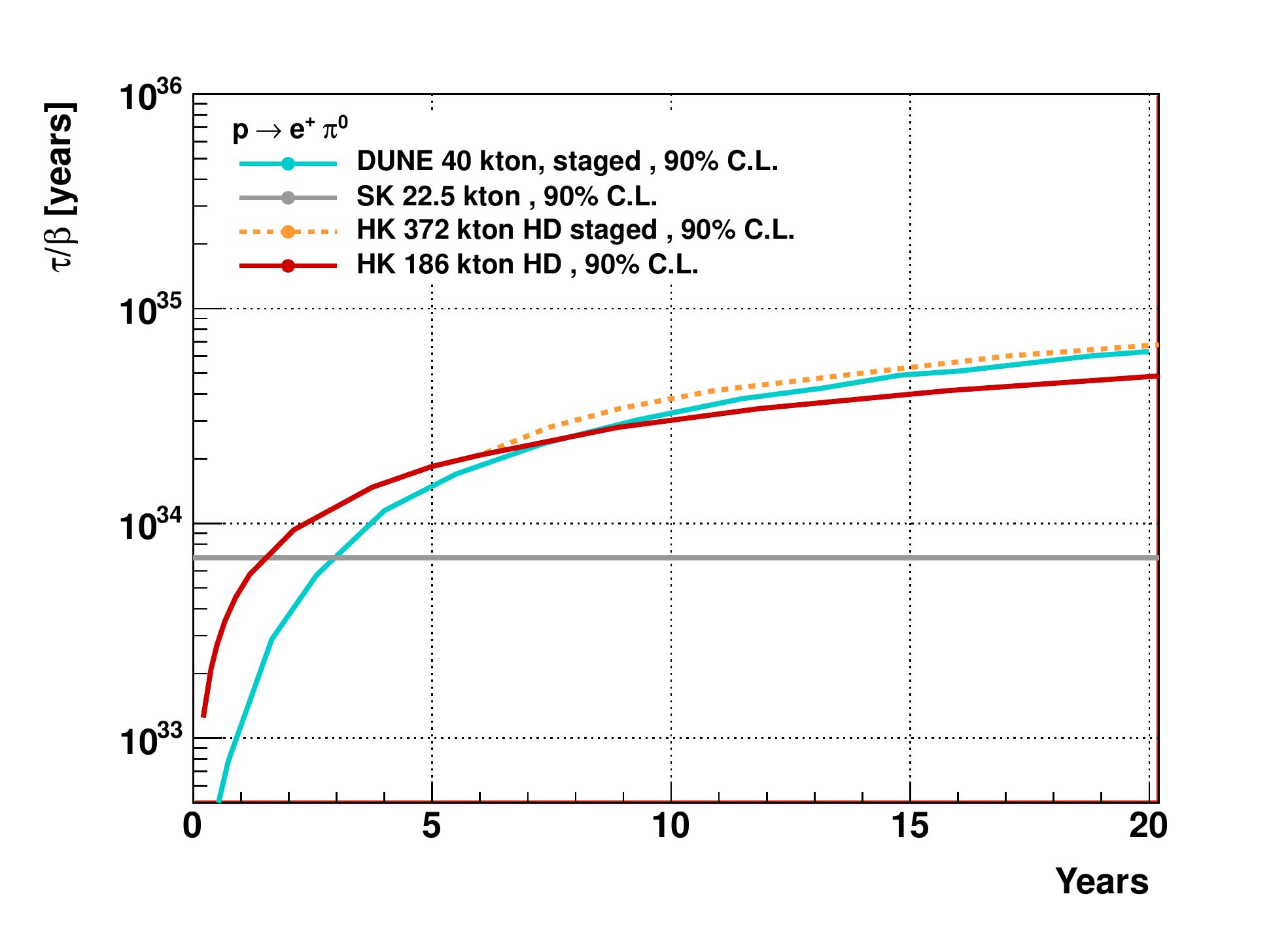}
  \end{center}
\caption {Hyper-K's sensitivity to the $p \rightarrow \bar \nu K^{+}$ decay mode 
          at 90\% C.L. as a function of run time is shown in red against other experiments (see caption of Figure~\ref{fig:nuK_discovery}).
          Super-K's current limit is also shown.
         }
  \label{fig:sens_nuk}
\end{figure}

Like the prompt $\gamma$ search, the $\pi^{+}\pi^{0}$ search relies on
more sophisticated event selections to identify the signal.  While the
momentum of both the $\pi^{+}$ and $\pi^{0}$ from the $K^{+}$ decay
will be 205\,MeV/$c$, the former does not deposit enough light to be
reconstructed fully.  To compensate for this the search method focuses
on PMT hits opposite the direction of a $\pi^{0}$ at the correct
momentum.  Fully contained events with one or two reconstructed rings,
all of which are $e$-like, are selected.  In addition there should be
one Michel electron from the charged pion decay chain.  For two-ring
events the invariant mass is required to be consistent with that of a
neutral pion, $ 85 < m_{\pi^{0}} < 185$\,MeV/$c^2$.  Further, the total
momentum should be between 175 and 250\,MeV/$c$.  At this stage the
$\pi^{+}$ light is identified as electron-equivalent visible energy between 7 and 17\,MeV
located at an angular separation between 140 and 180\,degrees from the
$\pi^{0}$ candidate's direction.  Finally a likelihood method is used
to evaluate the consistency of this light pattern with that produced
by a $\pi^{+}$.  Details of the full search method are documented
elsewhere~\cite{Abe:2014mwa}.
In addition, as in the $p\rightarrow e^{+}\pi^{0}$ search, the number 
of tagged neutron candidates is required to be zero.

Table~\ref{tbl:pdknuk} summarizes the expected signal efficiency and background rates 
for each of the search methods. 
Accompanying systematic errors have been calculated based on Super-K methods~\cite{Abe:2014mwa} and are included in 
the table. 
Assuming a proton lifetime close to current Super-K limits, 
Figures~\ref{fig:nuk_pmu} show the reconstructed muon momentum from 
the prompt gamma search and the reconstructed kaon mass spectrum from the $\pi^{+}\pi^{0}$ search.
Note that while the latter is unused in the analysis itself it provides a demonstration 
of the signal reconstruction since the proton mass for this decay mode cannot be reconstructed. 
The expected discovery potential and sensitivity of Hyper-K in comparison with other experiments 
appears in Figures~\ref{fig:nuK_discovery} and~\ref{fig:sens_nuk}.

\subsubsection{Sensitivity study for other nucleon decay modes}
Although the $p \rightarrow e^{+} \pi^{0}$ decay mode is predicted to be dominant 
in many GUT models, a variety of other decay modes are possible, each with a sizable 
branching ratio.
Table~\ref{tab:branch} shows the branching ratio distribution and ratio of neutron to proton 
lifetimes as predicted by several GUT models.
The diversity in these predictions suggests that in order to make a discovery and to subsequently 
constrain proton decay models, it is critical to probe as many nucleon decay modes as possible.
Fortunately, Hyper-Kamiokande is expected to be sensitive 
to many decay modes beyond the two standard decays discussed above. 

\begin{table}[htb]
\caption{Branching ratios for various proton decay modes together with the
ratio of the neutron to proton lifetimes as predicted by SU(5) and SO(10) models.
.\label{tab:branch}}
\vspace{0.4cm}
\begin{center}
\begin{tabular}{l|rrrrr}
\hline\hline
        & \multicolumn{5}{c}{Br.($\%$)} \\ 
\hline
        & \multicolumn{4}{c}{SU(5)}& SO(10)  \\ 
\hline
References& ~\cite{mach} & ~\cite{gav} & ~\cite{dono} & ~\cite{bucc} & ~\cite{bucc}   \\ 
\hline
$p \rightarrow e^{+} \pi^{0}$ & 33   & 37  & 9  & 35 & 30 \\ 
\hline
$p \rightarrow e^{+} \eta^{0}$ & 12  & 7 & 3 & 15 & 13 \\
\hline 
$p \rightarrow e^{+} \rho^{0}$ & 17    & 2   & 21   & 2 & 2   \\ 
\hline
$p \rightarrow e^{+} \omega^{0}$ & 22  & 18 & 56 & 17 & 14 \\
\hline 
Others & 17   & 35   & 11   & 31 & 31  \\ 
\hline
\hline
$\tau_{p}/\tau_{n}$ & 0.8 &1.0 & 1.3 &  & \\
\hline  \hline  
\end{tabular}
\end{center}
\end{table}

Hyper-Kamiokande's sensitivity to other nucleon decay modes has been 
estimated based in part on efficiencies and background rates
from Super-Kamiokande~\cite{:2009gd}. 
Table~\ref{tab:other_mode} shows the 90~$\%$ CL
sensitivities with a 1.9\,Megaton$\cdot$year exposure in the \hksingletank\ configuration. 
Under these conditions Table~\ref{tab:disc_other} shows the $3\sigma$ discovery potential of Hyper-K 
for a selected number of decay modes after a 1.9~Mton$\cdot$year exposure.

\begin{table}[htp]
\caption{Summary of Hyper-K's sensitivity to various $|B-L|$ conserving nucleon decay modes after a 1.9\,Megaton$\cdot$year exposure of the \hksingletank\  design compared in comparison with existing lifetime limits. The current limits for $p \rightarrow e^{+} \pi^{0}$, $p \rightarrow \mu^{+} \pi^{0}$ are from a 0.316\'Megaton$\cdot$year exposure of Super-Kamiokande~\cite{Miura:2016krn}, $p \rightarrow \overline{\nu} K^{+}$ is from a 0.26\,Megaton$\cdot$year exposure~\cite{Abe:2014mwa}, and the other modes are from a 0.316\,Megaton$\cdot$year exposure~\cite{TheSuper-Kamiokande:2017tit}.
\label{tab:other_mode}}
\vspace{0.4cm}
\begin{center}
\begin{tabular}{l|c|c}
\hline
Mode        & Sensitivity (90$\%$ CL) [years]  & Current limit [years] \\
\hline
\hline
\textcolor{blue}{$p \rightarrow e^{+} \pi^{0}$}        & 7.8 $\times 10^{34}$ & 1.6$\times 10^{34}$ \\
\textcolor{blue}{$p \rightarrow \overline{\nu} K^{+}$} & 3.2 $\times 10^{34}$   & 0.7$\times 10^{34}$  \\
\hline
\hline
$p \rightarrow \mu^{+} \pi^{0}$ & 7.7$\times 10^{34}$  & 0.77$\times 10^{34}$  \\
\hline
$p \rightarrow e^{+} \eta^{0}$& 4.3$\times 10^{34}$  & 1.0$\times 10^{34}$ \\
\hline
$p \rightarrow \mu^{+} \eta^{0}$& 4.9$\times 10^{34}$  & 0.47$\times 10^{34}$ \\
\hline
$p \rightarrow e^{+} \rho^{0}$& 0.63$\times 10^{34}$  & 0.07$\times 10^{34}$ \\
\hline
$p \rightarrow \mu^{+} \rho^{0}$& 0.22$\times 10^{34}$  & 0.06$\times 10^{34}$ \\
\hline
$p \rightarrow e^{+} \omega^{0}$& 0.86$\times 10^{34}$  & 0.16$\times 10^{34}$ \\
\hline
$p \rightarrow \mu^{+} \omega^{0}$& 1.3$\times 10^{34}$  & 0.28$\times 10^{34}$ \\
\hline
$n \rightarrow e^{+} \pi^{-}$ & 2.0$\times 10^{34}$   & 0.53$\times 10^{34}$ \\
\hline
$n \rightarrow \mu^{+} \pi^{-}$ & 1.8$\times 10^{34}$   & 0.35$\times 10^{34}$ \\
\hline
\hline
\end{tabular}
\end{center}
\end{table}

The decay modes in Table~\ref{tab:other_mode} all conserve baryon
number minus lepton number, $(B-L)$.  
However,  the $(B+L)$
conserving mode, $n \rightarrow e^{-} K^{+}$, was also given attention
and searched for by Super-Kamiokande.  In $n \rightarrow e^{-} K^{+}$,
the $K^{+}$ stops in the water and decays into $\mu^{+} + \nu$.  The
final state particles observed in $n \rightarrow e^{-} K^{+},K^+ \to \mu^+ \nu$ are $e^{-}$ and $\mu^{+}$.
 Both $e^{-}$ and
$\mu^{+}$ have monochromatic momenta as a result of originating from
two-body decays.  Furthermore, the timing of the $\mu^{+}$ rings are 
delayed with respect to the $e^{-}$ rings because of the $K^{+}$
lifetime.  In SK-II, the estimated efficiencies and the background
rate are 8.4\% and 1.1\,events/Megaton$\cdot$year, respectively. From
those numbers, the sensitivity to the $n \rightarrow e^{-} K^{+}$ mode
with a 1.9\,Megaton$\cdot$year exposure is estimated to be 1.0$\times
10^{34}$ years.

The possibility of $n \overline{n}$ oscillation is another interesting
phenomenon; it violates baryon number $(B)$ by $|\Delta B|$ = 2.
These $n \overline{n}$ oscillations have been searched for in
Super-Kamiokande with an exposure of 0.09\,Megaton$\cdot$year~\cite{jang}. 
Further improvement of the $n \overline{n}$
oscillation search is expected in Hyper-Kamiokande.

\begin{table}[htp]
\caption{Summary of Hyper-K's $3 \sigma$ discovery potential for several nucleon decay modes in the \hksingletank\  configuration.
         A 10 year exposure of a single detector  
         has been assumed. Numbers in brackets denote the potential assuming a second detector comes online six years 
         after the start of the experiment. Current limits are summarized in Table~\ref{tab:other_mode}. }
\label{tab:disc_other}
\vspace{0.4cm}
\begin{center}
\begin{minipage}[b][][b]{.45\linewidth}
\begin{tabular}{>{\raggedright}p{3cm}|>{\centering}p{3cm}}
\hline
Mode  &  $\tau_{disc}$ $3\sigma$ [years]   \tabularnewline
\hline
\hline
\textcolor{blue}{$p \rightarrow e^{+} \pi^{0}$}        & 6.3 (8.0)$\times 10^{34}$  \tabularnewline
\textcolor{blue}{$p \rightarrow \overline{\nu} K^{+}$} & 2.0 (2.5)$\times 10^{34}$  \tabularnewline
\hline
\hline
$p \rightarrow \mu^{+} \pi^{0}$  & 6.9 (8.7)$\times 10^{34}$ \tabularnewline
\hline
$p \rightarrow e^{+} \eta^{0}$   & 3.0 (3.9)$\times 10^{34}$  \tabularnewline 
\hline
$p \rightarrow \mu^{+} \eta^{0}$ & 3.4 (4.7)$\times 10^{34}$  \tabularnewline 
\hline
$p \rightarrow e^{+} \rho^{0}   $   & 3.4 (5.0)$\times 10^{33}$ \tabularnewline 
\hline
$p \rightarrow \mu^{+} \rho^{0}$     & 1.3 (1.6) $\times 10^{33}$ \tabularnewline 
\hline
$p \rightarrow e^{+} \omega    $ & 5.4 (6.9)$\times 10^{33}$  \tabularnewline
\hline
$p \rightarrow \mu^{+} \omega  $ & 0.78 (1.0)$\times 10^{34}$  \tabularnewline
\hline
\hline
\end{tabular}
\end{minipage}
\begin{minipage}[b][][b]{.45\linewidth}
\begin{tabular}{>{\raggedright}p{3cm}|>{\centering}p{3cm}}
\hline
Mode  &  $\tau_{disc}$ $3\sigma$ [years]   \tabularnewline
\hline
\hline
$n \rightarrow e^{+} \pi^{-}$    & 1.3 (1.6)$\times 10^{34}$   \tabularnewline
\hline
$n \rightarrow \mu^{+} \pi^{-}$  & 1.1 (1.5)$\times 10^{34}$  \tabularnewline
\hline
$n \rightarrow e^{+} \rho^{-}$    & 1.1 (1.5)$\times 10^{33}$   \tabularnewline
\hline
$n \rightarrow \mu^{+} \rho^{-}$  & 6.2 (8.1)$\times 10^{32}$  \tabularnewline
\hline
\hline
\end{tabular}
\end{minipage}
\end{center}
\end{table}

\begin{table}[htp]
\caption{Summary of Hyper-K's sensitivity to various $| \Delta (B-L) | = 2 $ and $| \Delta B | = 2$ proton and dinucleon decay modes after a 5.6\,Megaton$\cdot$year exposure compared with existing lifetime limits. 
The current limits are taken from a 0.273~Mton$\cdot$year exposure of Super-K.
Limits on the dinucleon decay modes are reported per $^{16}O$ nucleus, whereas the single nucleon decays are displayed 
as limits per nucleon.
}
\label{tab:dbl_eq_2}
\begin{center}
\begin{tabular}{l|c|c}
\hline
Mode                               & Sensitivity (90$\%$ CL) [years]  & Current limit [years] \\
\hline
$p \rightarrow e^{+}   \nu\nu$       & 10.2 $\times 10^{32}$             &  1.7  $\times 10^{32}$  \\
$p \rightarrow \mu^{+} \nu\nu$       & 10.7 $\times 10^{32}$             &  2.2  $\times 10^{32}$  \\
\hline
\hline
$p \rightarrow e{+} X        $       & 31.1 $\times 10^{32}$             &  7.9  $\times 10^{32}$  \\
\hline
$p \rightarrow \mu^{+} X     $       & 33.8 $\times 10^{32}$             &  4.1  $\times 10^{32}$  \\
\hline
$n \rightarrow \nu  \gamma   $       & 23.4 $\times 10^{32}$             &  5.5  $\times 10^{32}$  \\
\hline
\hline
$np \rightarrow e^{+} \nu    $       &  6.2 $\times 10^{32}$             &  2.6  $\times 10^{32}$  \\
\hline
$np \rightarrow \mu^{+} \nu  $       &  4.2 $\times 10^{32}$             &  2.0  $\times 10^{32}$  \\
\hline
$np \rightarrow \tau^{+} \nu $       &  6.0 $\times 10^{32}$             &  3.0  $\times 10^{32}$  \\
\hline
\hline
\end{tabular}
\end{center}
\end{table}

In addition to $(B-L)$ conserving modes several other nucleon channels
are available.  Unification schemes invoking left-right symmetry
(c.f.~\cite{Pati:1974yy}) predict trilepton decays such as
$p \rightarrow e^{+} (\mu^{+}) \nu\nu$, which violate $(B-L)$ by two units
($| \Delta (B-L) |= 2 $).
Though there are two particles in the final state that are invisible to Hyper-Kamiokande,
the presence of a positron or muon from such decays can, in principal, be detected.
While this type of single charged lepton signature would naturally be subject to large 
atmospheric backgrounds, with a sufficiently large decay rate, spectral information can be 
used to separate the two. 
Going one step farther, then, spectral analysis makes it possible to search for generic decay 
modes as well, such as $p \rightarrow e^{+} (\mu^{+}) X$, where $X$ is an unknown and unseen particle.

Though only single nucleon decays have been considered up until this
point, it is worth noting that dinucleon processes, in which a
neutron-proton (or proton-proton) pair decays into a pair of leptons,
can also be studied at Hyper-Kamiokande.
Modes where $\Delta B =2$ , such as $np \rightarrow l^{+}\nu$,
appear in models with extended Higgs sectors~\cite{Perez:2013osa} and
have connections to Baryogenesis.  Interestingly, these models have the
additional property that the single nucleon decay modes are suppressed
relative to the dinucleon decays.  This wealth of predictions for
possible channels further emphasizes the need to search for as many
nucleon decay signatures as possible in the quest for grand
unification.  Table~\ref{tab:dbl_eq_2} lists Hyper-K's expected
sensitivity to both $| \Delta (B-L) | =2 $ and $\Delta B =2$ decays
for a 5.6~Mton$\cdot$year exposure.  While searches for these modes
are dominated by the atmospheric neutrino background, Hyper-K can be
expected to extend existing limits by a factor of three to ten if 
no signal is observed.

\subsection{Impact of Photocathode Coverage and Improved Photosensors}\label{section:pdecay-coverage}
\begin{figure}[htbp]
  \begin{center}
    \includegraphics[scale=0.35]{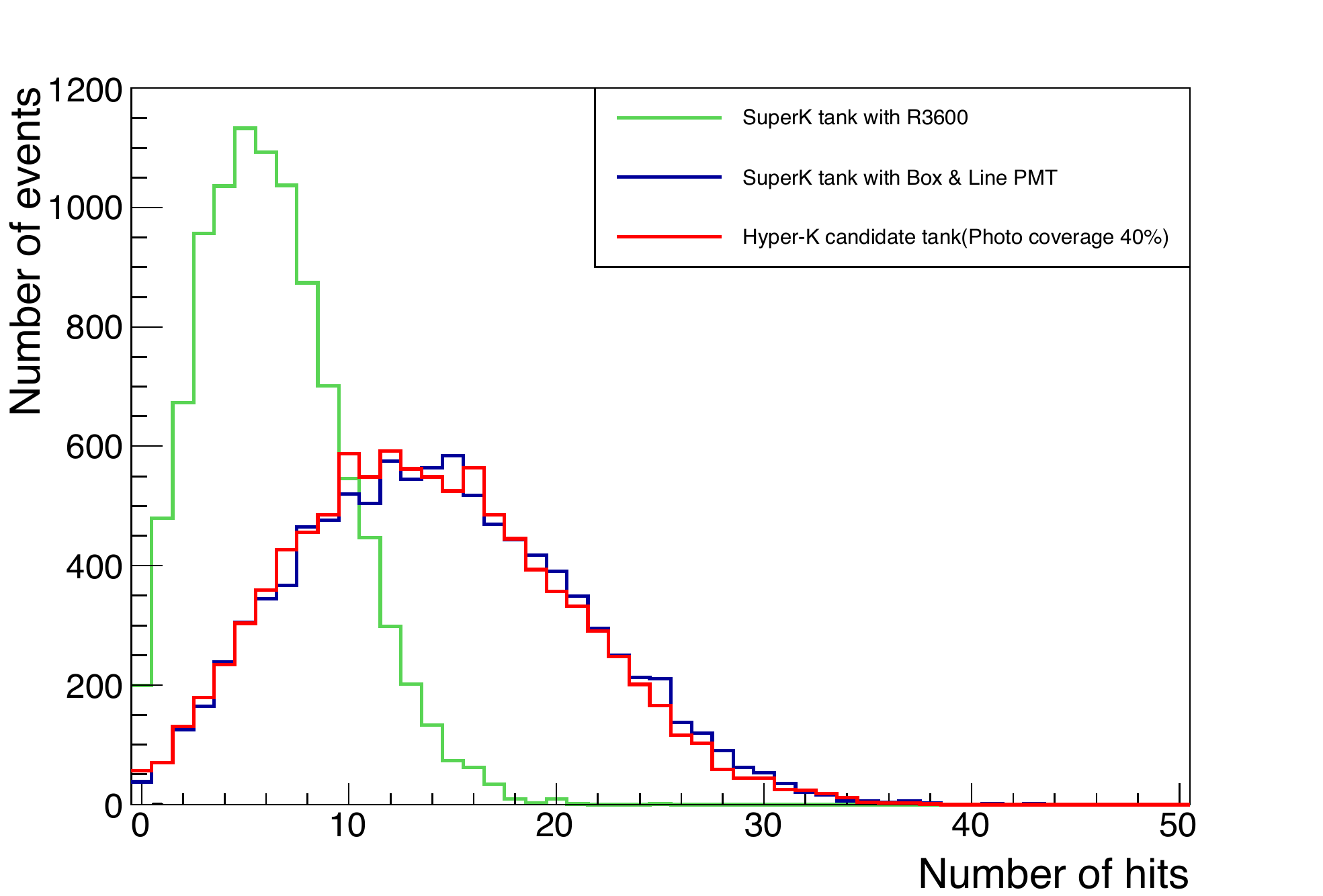}
  \end{center}
  \caption{Number of hit PMTs for a toy MC simulation of the 2.2\,MeV $\gamma$ rays 
           emitted following neutron capture on hydrogen. 
           Cutting at more than nine hits in the Super-K distribution yields an 
           estimated 18\% tagging efficiency.}
  \label{fig:n_dg_p_nhit}
\end{figure} 

\begin{figure}[htbp]
  \begin{center}
    \includegraphics[scale=0.35]{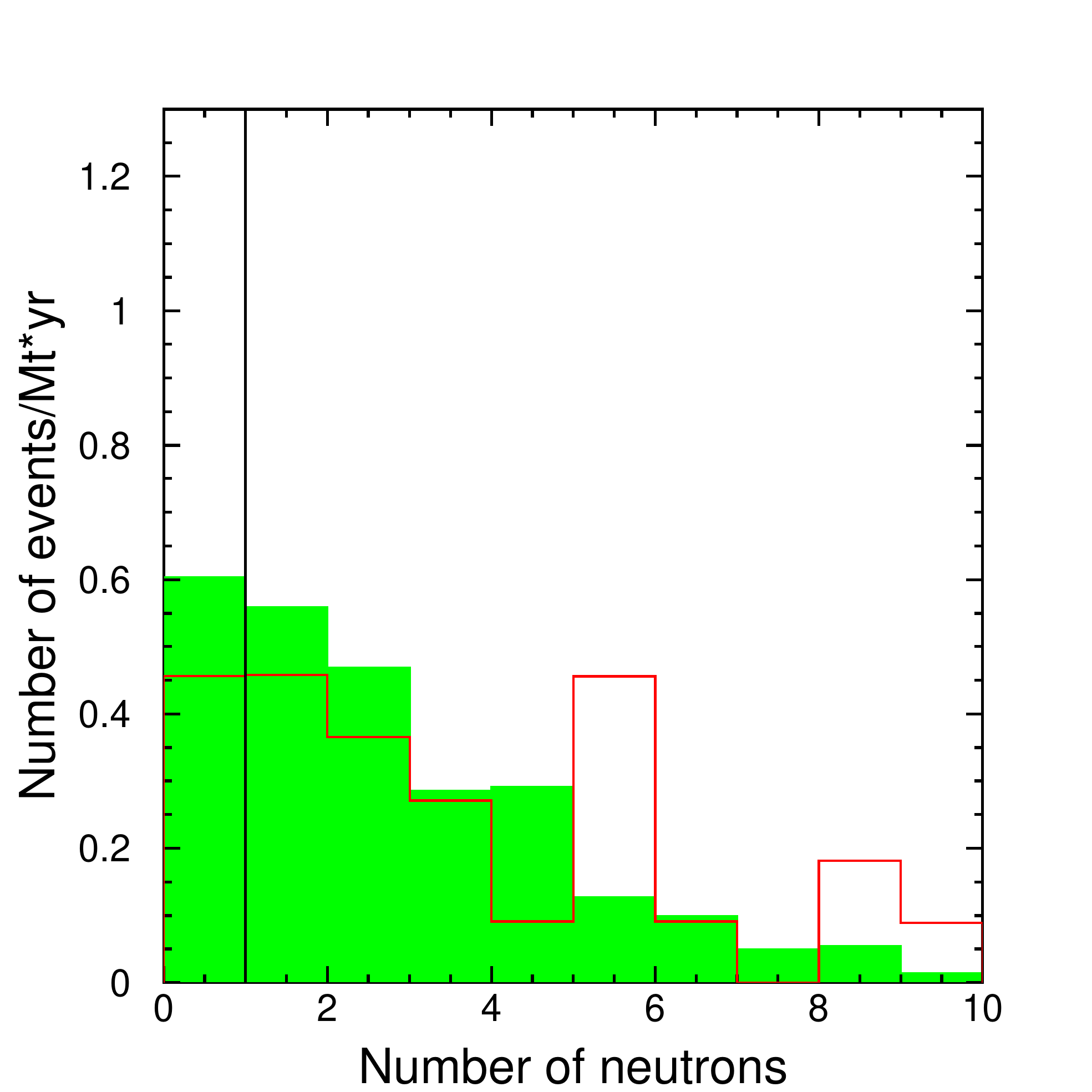}
  \end{center}
  \caption{Neutron multiplicity distribution for background events in the $p \rightarrow e^{+}\pi^{0}$
           search at Hyper-K. The red histogram shows the distribution prior to 
           neutron tagging and the green histogram shows the result of 
           applying a tagging algorithm with 70\% efficiency. }
  \label{fig:ep0_n_mult}
\end{figure} 

Improved photon collection with larger photocathode coverages, higher
quantum efficiency photosensors, and their combination have
a dramatic effect on the physics sensitivity of Hyper-K.  
Nucleon decay searches, in particular, are expected to benefit significantly
from enhanced ability to detect low levels of Cherenkov light.  With
the large exposures Hyper-K will provide, the atmospheric neutrino
background to these searches becomes sizable and
can inhibit the discovery potential of the experiment.  However, these
same backgrounds are often expected to produce neutrons either
directly through the CC interaction of antineutrinos or indirectly via
the secondary interaction of hadrons in the interaction.  Proton decay
events, in contrast, are only rarely expected to be accompanied by
neutrons.  Though such neutrons are ordinarily transparent to water
Cherenkov detectors, Super-Kamiokande has demonstrated the ability to
tag the 2.2\,MeV photon emerging from neutron capture on hydrogen,
$n(p,d)\gamma$.  Naturally this channel will be available to
Hyper-K.
In this section we compare the performance of Hyper-K \hksingletank\,   configuration with its 40\% photocathode 
coverage and high quantum efficiency photosensors
against a design with 20\% coverage and the same PMTs used in Super-K.

Since 2.2\,MeV is only barely visible in Super-K, their tagging algorithm
makes use of a neural network to isolate signal neutrons, which are
correlated spatially and temporally with the primary neutrino
interaction, from background sources.  Though the method achieved
20.5\% tagging efficiency with a 1.8\% false tag
probability~\cite{Wendell:2014dka}  it is worth noting that so far it
has only been successful during the SK-IV phase of the experiment.
Despite 40\% photocathode coverage with Hamamatsu R3600 50~cm PMTs, on
average the neutron capture signal produces only 7~hits in the
detector~\cite{Wendell:2014dka}.  Since the average photon travel
length in Hyper-K will be larger than that of SK-IV, in order to take
advantage of the neutron signal it is essential to augment Hyper-K's
photon yield wherever possible.

While a full analysis of Hyper-K's neutron tagging capability is in
development a rough estimation for the \hksingletank\,   design 
(c.f. Section~\ref{section:photosensors}) has been determined via
simulation.  Figure~\ref{fig:n_dg_p_nhit} shows the number of hit
photosensors for a simulated sample of the 2.2\,MeV $\gamma$ ray
emitted when a neutron captures on a proton for three water Cherenkov
detector configurations.  The green distribution shows the response of
a Super-K-sized detector with 40\% photocathode coverage by Hamamatsu
R3600 PMTs, while the blue line shows the effect of implementing the HQE
Box and Line photosensors. 
Using the same photocathode coverage and the
latter photosensors but with an enlarged tank representative of a
the Hyper-K design (60~m diameter and 74~m height results)
results in the red curve.  Using these distributions the number of hit
photosensors which reproduces the tagging efficiency realized in the
Super-K analysis is chosen as the metric to represent the expected
tagging efficiency at Hyper-K.  For the Super-K distribution gamma
events with more than nine PMT hits result in an 18\% tagging efficiency.
Assuming the same 10 hit threshold can be used at Hyper-K, the
equivalent distribution (red curve in the figure) suggests 73\%
efficiency can be achieved.  The sensitivity studies presented in this 
document assume this number for Hyper-K when neutron tagging is employed. 
In what follows sensitivity estimates are
presented assuming both this and the Super-K tagging efficiency for comparison.

\begin{figure}[htbp]
\begin{center}
  \includegraphics[scale=0.35]{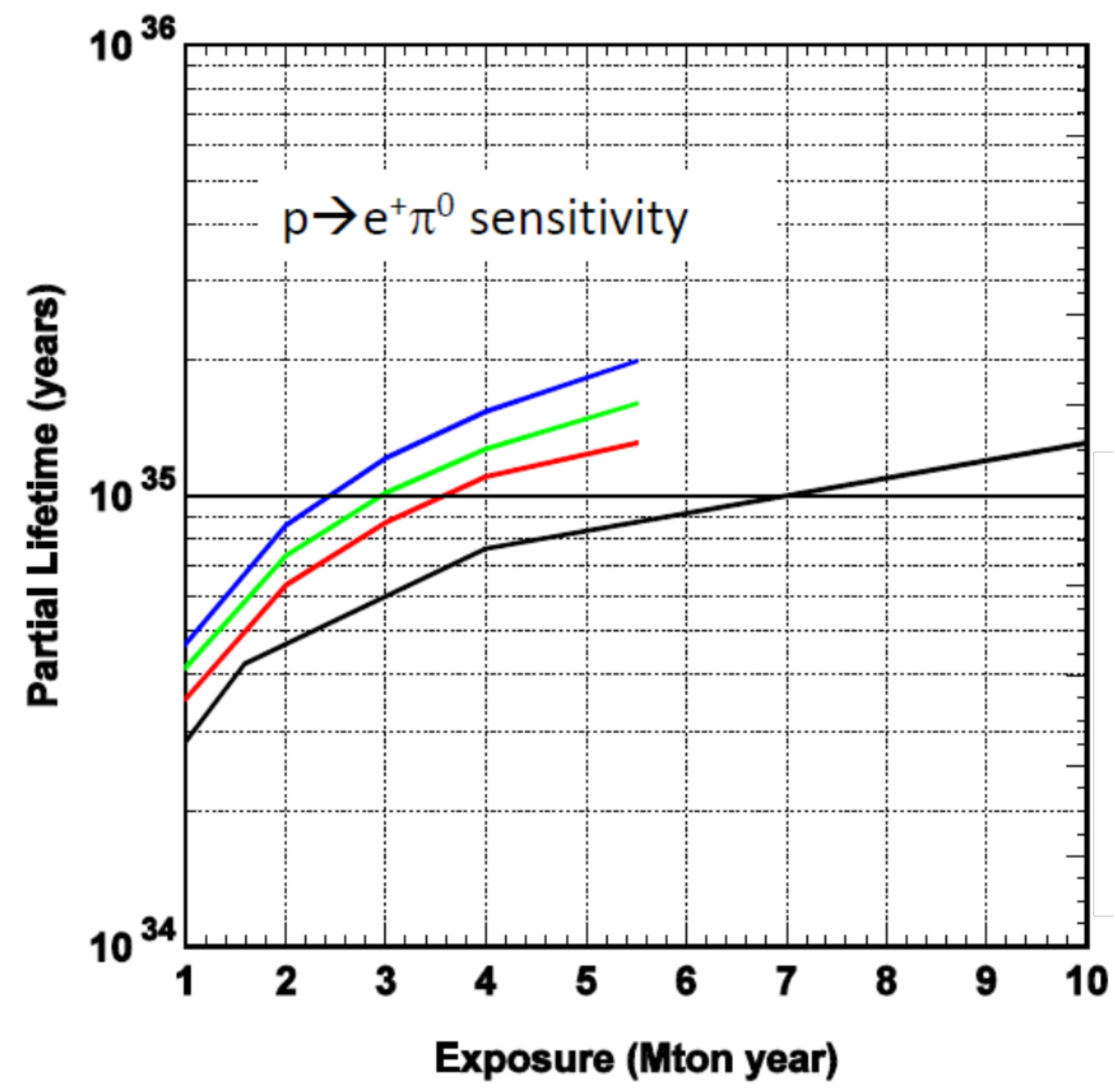}
\end{center}
\caption {Hyper-K sensitivity to proton decay for the $p \rightarrow e^{+}\pi^{0}$ mode 
          as a function of exposure. The black curve illustrates the sensitivity 
          assuming equivalent performance to Super-K.
          Improved sensitivity curves as a result of a finer binned signal region 
          (red) in conjunction with a 50\% reduction in background (green) 
          or a 70\% reduction (blue) are also shown.
          The latter is the analysis presented in Section~\ref{section:pdecay}.  }
\label{fig:ep0_bkg_exp}
\end{figure}

Based on MC studies, the fraction of atmospheric neutrino backgrounds
to the $p \rightarrow e^{+}\pi^{0}$ search which are accompanied by at
least one final state neutron is $> 80\%$.  At the same time, only 4\%
of signal MC events have such a neutron.  The multiplicity
distribution for the background sample is shown in
Figure~\ref{fig:ep0_n_mult}.  Due to the presence of high multiplicity
events, assuming 70\% neutron tagging efficiency, the
neutron background can be reduced by approximately 70\% (green
histogram) by rejecting events with one or more neutron tags.
Reducing the background in this manner will have a large impact on the
sensitivity to this decay mode, as shown in
Figure~\ref{fig:ep0_bkg_exp}.  In the figure the black curve represents
the sensitivity of the analysis without neutron tagging and cuts defining 
only a single signal region.
If that signal region is divided into a piece where free proton
decays are enhanced ($p_{tot} < 100 \mbox{MeV/c}^{2}$) and a region
with predominantly bound decays ($p_{tot} > 100 \mbox{MeV/c}^{2}$), as is assumed 
in the analysis above, the
resulting sensitivity is shown by the red line.  Further, if the total
background is then reduced by 50\% (70\%), by the introduction of
neutron tagging, the sensitivity improves as shown in
the green (blue) lines.  
With these background reductions Hyper-K will require exposures of 
3.0 and 2.4 Mton $\cdot$years to reach lifetime limits of $10^{35}$ years if no signal is observed.
Without any background reduction 7.0~Mton$\cdot$years are required.

It should be noted that $p \rightarrow e^{+}\pi^{0}$ is not the only
mode that is expected to benefit from a higher photon yield; Most
modes are similarly expected to have reduced atmospheric neutrino
backgrounds with neutron tagging.  However, the $p
\rightarrow \bar \nu K^{+}$ search can also benefit from enhanced
light collection to improve the signal efficiency.  Its two lowest
efficiency, but most sensitive search modes, one in which the decay is
accompanied by a prompt 6.3 MeV de-excitation $\gamma$ from the
recoiling nucleus and the other in which the $K^{+}$ decays into
$\pi^{+}\pi^{0}$, both have components that are looking for small
amounts of light.  Higher photon yields are thus connected directly to
efficiency improvements in these channels.

\begin{figure}[htbp]
\begin{center}
  \includegraphics[scale=0.35]{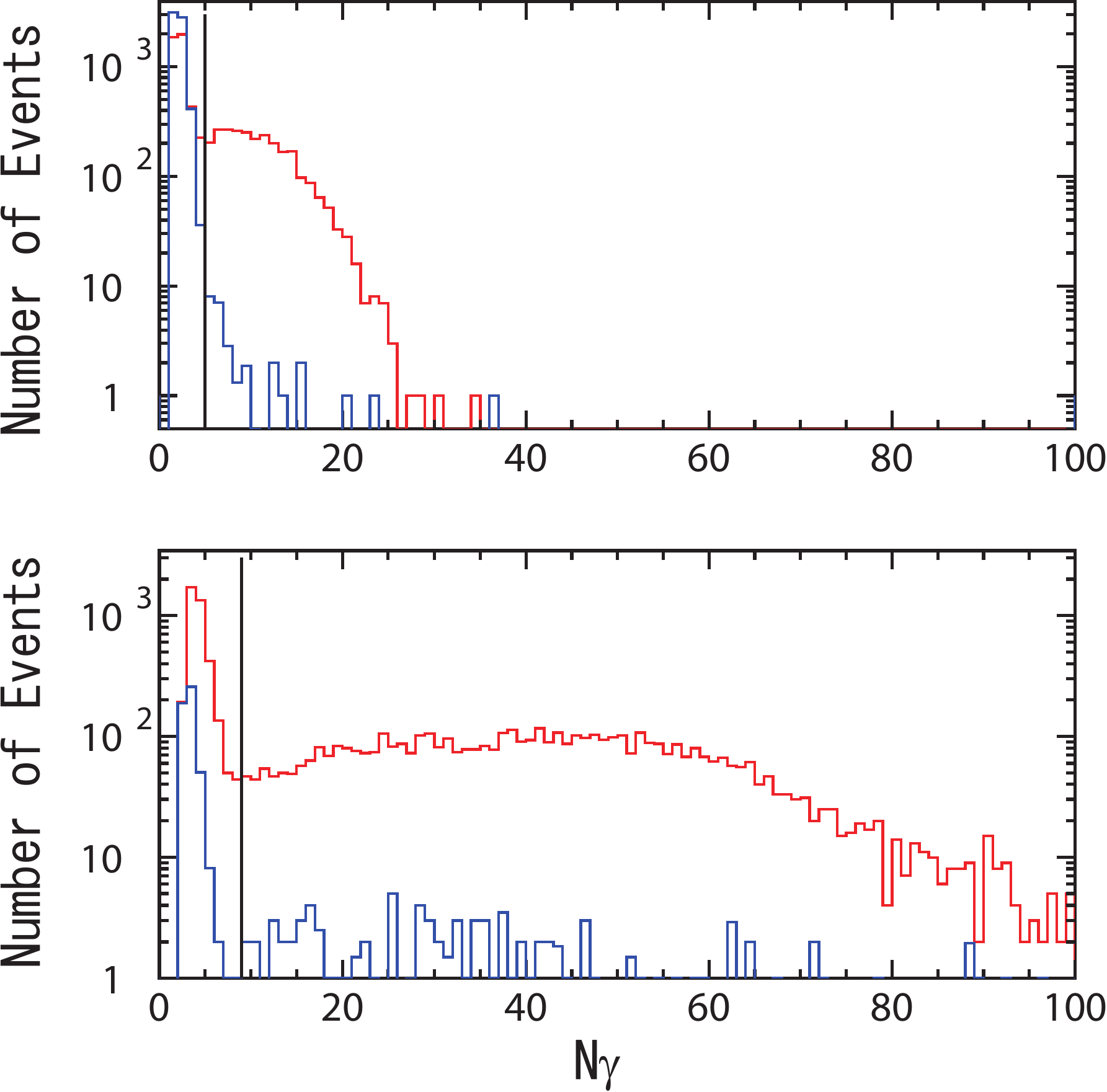}
\end{center}
\caption {Distribution of the number of hits within a 12~nsec timing window used to search for the 6.3 MeV $\gamma$ 
          from $p \rightarrow \bar \nu K^{+}$ events. 
          The upper panel corresponds to a Hyper-K design with 20\% photocathode coverage 
          and the same PMTs used in SK (Hamamatsu R3600).
          In the lower panel the result for the \hksingletank\,   design is shown. 
          Red (blue) lines show the proton decay signal (atmospheric neutrino background) distribution. 
          The proton decay signal region is defined as $4 < N_{\gamma} < 30$ hits and $8 < N_{\gamma} < 120$ 
          for the 20\% coverage and \hksingletank\,   coverage designs, respectively.
          Vertical lines in the plots show the lower bound of these signal regions.  }
\label{fig:ngam-bl}
\end{figure} 

\begin{figure}[htbp]
\begin{center}
  \includegraphics[scale=0.35]{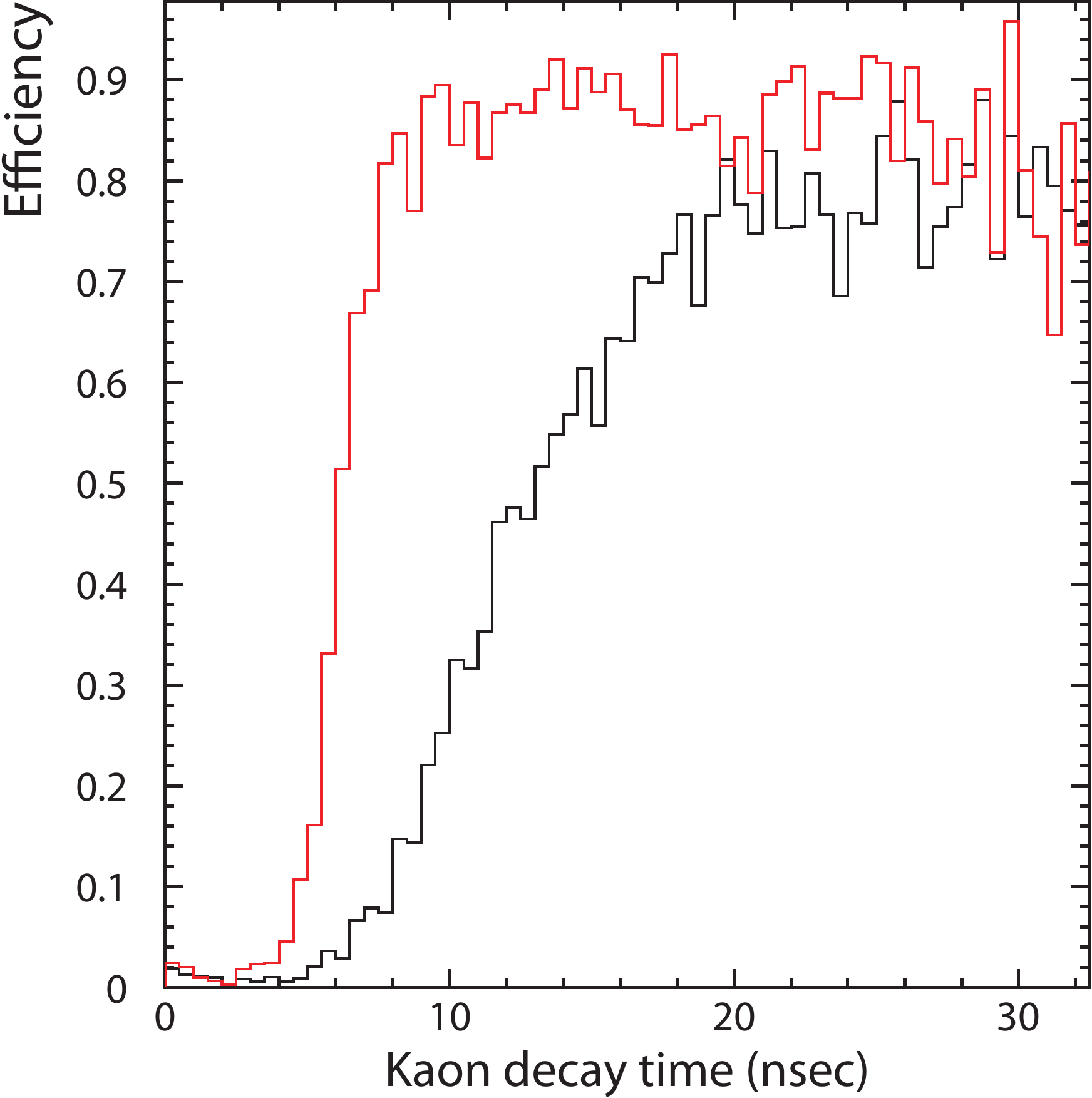}
\end{center}
\caption {Prompt $\gamma$ tagging efficiency as a function of the 
          $K^{+}$ decay time for the $p \rightarrow \bar \nu K^{+}$ decay mode. 
          A Hyper-K design with only 20\% photocathode coverage is shown in black and 
          the design with 40\% photocathode coverage and photosensors with 1.9 times 
          higher photon yield (\hksingletank\,  ) is shown in red. 
          }
\label{fig:gam-eff}
\end{figure}

As discussed above, the search for the prompt $\gamma$ ray is done
using the number of hit PMTs within a 12~nsec wide timing widow prior
to the muon candidate from the $K^{+} \rightarrow \mu^{+} \nu$
(c.f. Figure~\ref{fig:prompt-gamma}).  Figure~\ref{fig:ngam-bl} shows
this distribution for the proton decay signal (red) and background
from atmospheric neutrinos (blue) for a Hyper-K design with
20\% photocathode coverage (upper panel) and the \hksingletank\,   design. 
In the latter study the photon yield is assumed to be 1.9
times greater than that of the PMT used in the 20\% coverage design and
with half the intrinsic timing resolution of the photosensor
(1~nsec at 1~p.e.).  Additionally, these sensors are assumed to have a
higher dark rate of 8.4~kHz.  For both detector configurations the peak
near zero corresponds to events in which a $\gamma$ was not found; 
these hits are dominated by dark noise.  Though
this peak is not well separated from the feature at higher hits in the
20\% coverage configuration, with improved photosensors and 40\% coverage a clear distinction
between the two can be seen.  Further, since the $\gamma$ search is
designed to avoid hit contamination from the $\mu^{+}$, the narrower
timing resolution of the improved sensors means that the search can
occur closer in time to the muon.  Practically speaking, this means $K^{+}$
with earlier decay times can be used in the analysis as shown in
Figure~\ref{fig:gam-eff}.  Both of these effects lead to an overall
increase in the signal efficiency of the $p \rightarrow \bar \nu K^{+}$
mode.  Based on this distribution the proton decay signal region is
defined as $4 < N_{\gamma} < 30$ hits for the 20\% coverage design 
and $8 < N_{\gamma} < 120$ for the \hksingletank\,   case.

\begin{figure}[htbp]
\begin{center}
  \includegraphics[scale=0.35]{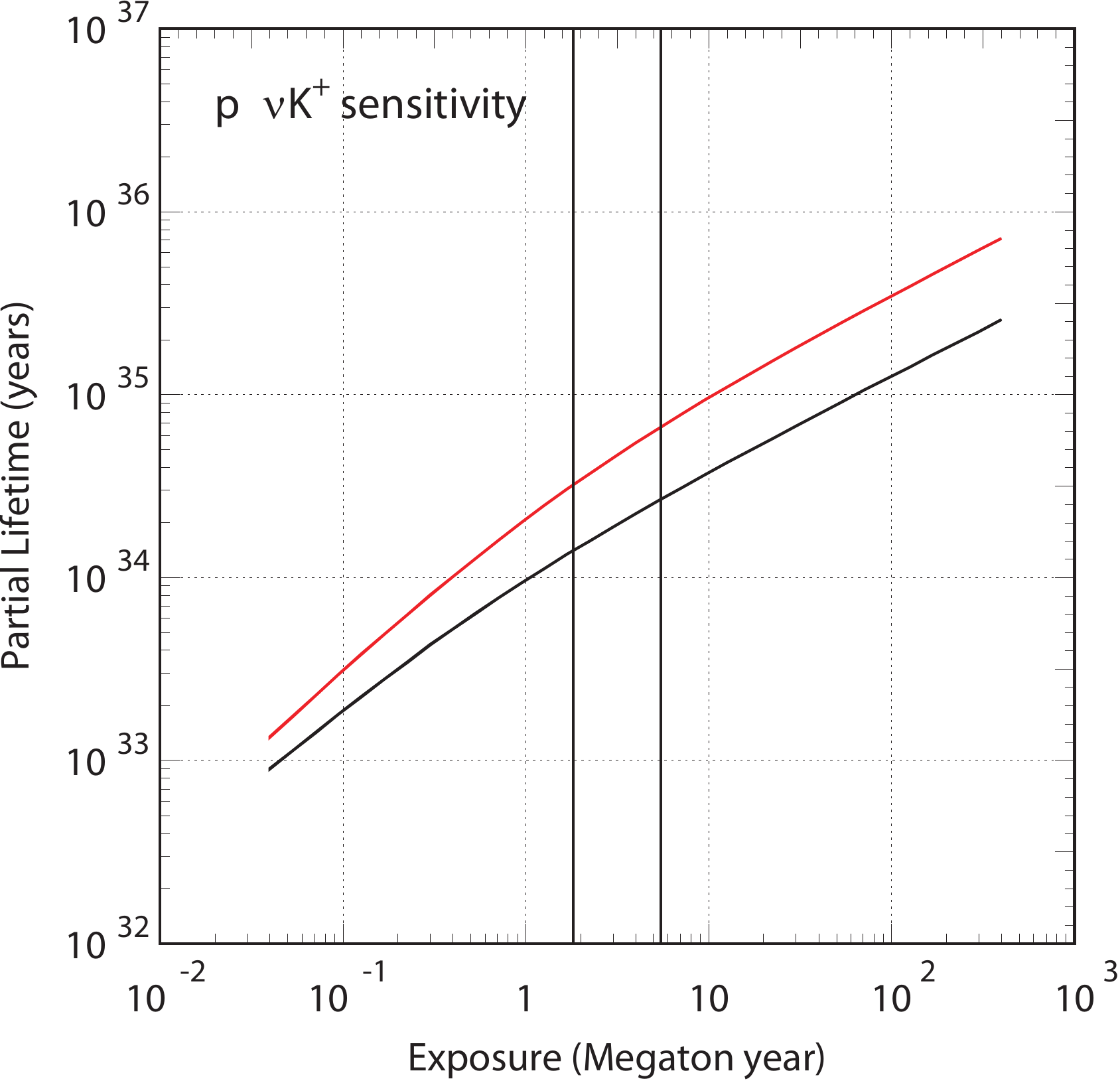}
\end{center}
\caption {Sensitivity to $p \rightarrow \bar \nu K^{+}$ at 90\% C.L. for the \hksingletank\,   design is shown in red. 
          The black curve shows the corresponding sensitivity for a design with 20\% coverage and Super-K PMTs. }
\label{fig:bl-sens}
\end{figure}

Accordingly, the signal efficiency for the prompt $\gamma$ tag method
to search for $p \rightarrow \bar \nu K^{+}$ increases from 6.7\% in
the design with 20\% photocathode coverage to 12.7\% in the \hksingletank\,   configuration.
Though not detailed here, improved tagging of the $\pi^{+}$ from the
$K \rightarrow \pi^{+} \pi^{0}$ component of this search yields 10.2\%
signal efficiency in the \hksingletank\,   design compared to 6.7\% in the
20\% coverage case.
As in the $p \rightarrow e^{+} \pi^{0}$ study
above, if neutron tagging with 70\% efficiency is assumed
(c.f. Figure~\ref{fig:ep0_bkg_exp}) the background to these $p
\rightarrow \bar \nu K^{+}$ search methods can be reduced to 0.87 and
0.71~events/Mton$\cdot$year, respectively.  For comparison the
background rates are 2.8 and
3.4~events/Mton$\cdot$year without neutron tagging.  
Figure~\ref{fig:bl-sens} shows the
corresponding sensitivity curves for both the design with 20\% coverage (black) and
for the \hksingletank\,   design (red).  Two vertical lines at exposures of 1.9 and
5.6~Mton$\cdot$year in the figure denote positions of comparable
sensitivity between the two designs.  

It should be noted that since backgrounds for the Super-K neutron
tagging algorithm are taken from its data, it is difficult to provide
a realistic estimate of the potential of a similar algorithm at
Hyper-K at this time.  
However assuming similar performance, Hyper-K's dramatic
improvement in proton decay sensitivity makes this topic among the
most fundamental to the development of its future program.  It is clear
that the potential for a discovery is connected to
Hyper-K's background reduction and efficiency enhancement
capabilities, both of which are realizable with higher photon yields.

\section{Neutrino Astrophysics and Geophysics}

%\newpage
\graphicspath{{physics-supernova/figures}}
\subsection{Supernova}\label{sec:supernova}
\subsubsection{Supernova burst neutrinos}\label{sec:supernova-burst}
\paragraph{Introduction}

Core-collapse supernova explosions are the last process in the
evolution of massive ($>8$M$_{\rm sun}$) stars.  Working their way
successively through periods of predominantly hydrogen fusion, helium
fusion, and so on, eventually silicon fusion starts making iron.  Once
an iron core has formed, no more energy can be released via its fusion
into still-heavier elements, and the hydrodynamic balance between
gravity and stellar burning is finally and catastrophically disrupted.
The sudden gravitational collapse of their iron cores --
which will go on to form either a neutron star or a black hole -- is
the main source of energy from this type of supernova explosion.  The
energy released by a supernova is estimated to be $\sim 3 \times
10^{53}$\,ergs, making it one of the most energetic phenomena in the
universe.  Since neutrinos interact weakly with matter, almost 99\% of
the released energy from the exploding star is carried out by
neutrinos.  As a result, the detection of supernova neutrinos gives
direct information of energy flow during the explosion.  The neutrino
emission from a core collapse supernova starts with a short
($\sim$10\,millisecond) burst phase of electron captures ($p +
e^- \rightarrow n + \nu_e$) called the neutronization burst, which
releases about $10^{51}$\,ergs.  Following that, the majority of the
burst energy is released by an accretion phase ($< \sim$1\,second) and
a cooling phase (several seconds) in which all three flavours of
neutrinos (as well as anti-neutrinos) are created.\par
The observation of a handful (25 in total) of supernova burst
neutrinos from SN1987a by the Kamiokande, IMB, and Baksan experiments
proved that the basic scenario of the supernova explosion was correct.
However, more than three decades later the detailed mechanism of
explosions is still not known.
Achieving the necessary conditions
for a supernova explosion in computer simulations has been a
long-standing challenge.
Recently, successful supernova explosions in two-dimensional and
three-dimensional simulations have been reported, the details of which will be
described later\cite{Tamborra:13,SASIBlondin,SASIScheck,Takiwaki:2017tpe}.
Though several models have produced successful
supernova explosions in simulations,
detecting supernova neutrinos will provide further input to improve the physics accuracy of the models.
Hyper-K can detect neutrinos with energy down to $\sim$3\,MeV and can point the supernova, due to its event-by-event directional sensitivity.
Because the supernova neutrinos are detected as a burst in a short time period, we can neglect the low energy radioactive backgrounds.
The localization in time also makes it possible to utilize most of full inner volume for our analysis, $i.e.$ 220\,kt for each tanks.
Compared with the current or planned experiments, $e.g.$ Super-K, the ice Cherenkov detector IceCube/PINGU~\cite{Aartsen:2014oha} and large liquid Ar TPC detectors like DUNE,
Hyper-K has several advantages for the supernova measurement.
The first advantage is the large volume and statistics.  
Hyper-K will have a FV of 8\,times to 16\,times larger than the Super-K detector, resulting in a commensurate increase in the number of detected supernova neutrinos and sensitivity to supernovae occurring in nearby galaxies.
Likewise, Hyper-K will be significantly larger than DUNE
with mass in the tens of kt scale.
Furthermore, Hyper-K primarily detects anti-electron-neutrino from the
supernova explosion using the inverse beta decay reaction
($\bar{\nu}_e + p \rightarrow e^+ + n$), unlike LArTPC detectors which
primarily detect electron neutrinos.
Large neutrino telescopes like IceCube/PINGU has capability of huge
statistics detection, however, they would detect only single PMT hits from such low energy neutrino events, allowing them to separate supernova neutrinos from their dark noise only on a statistical basis.
In contrast, every single event will be reconstructable with Hyper-K down to an energy analysis threshold of $\sim$3\,MeV.
Our precise event-by-event measurement will be essential for the comprehensive study of supernova neutrinos, especially with the detailed and time-dependent energy spectrum.
With these advantages, Hyper-K is able to perform unique measurements to reveal the mechanism of supernova explosions.\\

\paragraph{Expected observation in Hyper-Kamiokande}

In order to correctly estimate the expected number of neutrino events detected
in Hyper-K, we must consider the neutrino oscillation due to the MSW matter
effect through the stellar medium.
The flux of each neutrino type emitted from a supernova is related to the originally produced fluxes ($F^0_{\nu_e}$, $F^0_{\bar{\nu}_e}$ and $F^0_{\nu_x}$, where $\nu_x$ is $\nu_{\mu,\tau}$ and $\bar{\nu}_{\mu,\tau}$) by the following formulas~\cite{Dighe:1999id, Fogli:2004ff} :

\noindent
For normal hierarchy,
\begin{eqnarray}
F_{\bar{\nu}_e}  &\simeq&  \cos^2\theta_{12} F^0_{\bar{\nu}_e} + \sin^2\theta_{12}F^0_{\nu_x}\ , \nonumber \\
F_{\nu_e} &\simeq& \sin^2\theta_{12} P_{H} F^0_{\nu_e} + (1-\sin^2\theta_{12} P_{H}) F^0_{\nu_x}, \nonumber \\
F_{\nu_\mu} + F_{\nu_\tau} &\simeq&  (1-\sin^2\theta_{12} P_{H}) F^0_{\nu_e} + (1 + \sin^2\theta_{12} P_{H}) F^0_{\nu_x}, \nonumber \\
F_{\bar{\nu}_\mu} + F_{\bar{\nu}_\tau}  &\simeq&  (1-\cos^2\theta_{12}) F^0_{\bar{\nu}_e} + (1 + \cos^2\theta_{12}) F^0_{\nu_x} , \nonumber
\end{eqnarray}

\noindent
and, for inverted hierarchy,
\begin{eqnarray}
F_{\bar{\nu}_e} &\simeq& \cos^2\theta_{12} P_{H} F^0_{\bar{\nu}_e} + (1-\cos^2\theta_{12} P_{H}) F^0_{\nu_x}\ , \nonumber \\
F_{\nu_e} &\simeq& \sin^2\theta_{12} F^0_{\nu_e}+ \cos^2\theta_{12}F^0_{\nu_x} , \nonumber \\
F_{\nu_\mu} + F_{\nu_\tau} &\simeq& (1-\sin^2\theta_{12}) F^0_{\nu_e} + (1 + \sin^2\theta_{12}) F^0_{\nu_x} , \nonumber \\
F_{\bar{\nu}_\mu} + F_{\bar{\nu}_\tau}  &\simeq&  (1-\cos^2\theta_{12}P_H) F^0_{\bar{\nu}_e} + (1 + \sin^2\theta_{12} P_H) F^0_{\nu_x} , \nonumber
\end{eqnarray}

\noindent
where $P_H$ is the crossing probability through the matter resonant
layer corresponding to $\Delta m^2_{32}$. $P_H = 0$ ($P_H = 1$) for
adiabatic (non-adiabatic) transition.
Recent measurement of $\theta_{13}$ indicates the adiabatic transition ($P_H=0$) for the matter transition in the supernova envelope.
The supernova neutrino spectrum is
affected not only by stellar matter but also by other neutrinos and
anti-neutrinos at the high density core (so-called collective
effects).  These collective effects, which swap the $\nu_{e}$ and
$\bar{\nu}_e$ spectra with those of $\nu_x$ in certain energy
intervals bounded by sharp spectral splits, were first discussed
in~\cite{Duan:2006an,Duan:2006jv}.  This has become an active field of
study whose recent investigations include taking into account the
possibility of multiple splits~\cite{Dasgupta:2009dd}, computation with
three neutrino flavors~\cite{Friedland:2010sc}, and utilizing the full
multi-angle framework~\cite{Duan:2010bf}.
So, in the following description of the performance of the
Hyper-Kamiokande detector, three cases are considered in order to
fully cover the possible variation of expectations: (1)
no~oscillations, (2) normal~hierarchy (N.H.) with $P_H=0$, and (3)
inverted~hierarchy (I.H.) with $P_H=0$.  The process depends critically on
$\theta_{12}$; in what follows we assume $\sin^2\theta_{12}=0.31$.
Concerning the neutrino fluxes and energy spectra at the production
site, we used results obtained by the Livermore
simulation~\cite{Totani:1997vj}.

\begin{figure}[tbp]
  \begin{center}
    \includegraphics[width=16cm]{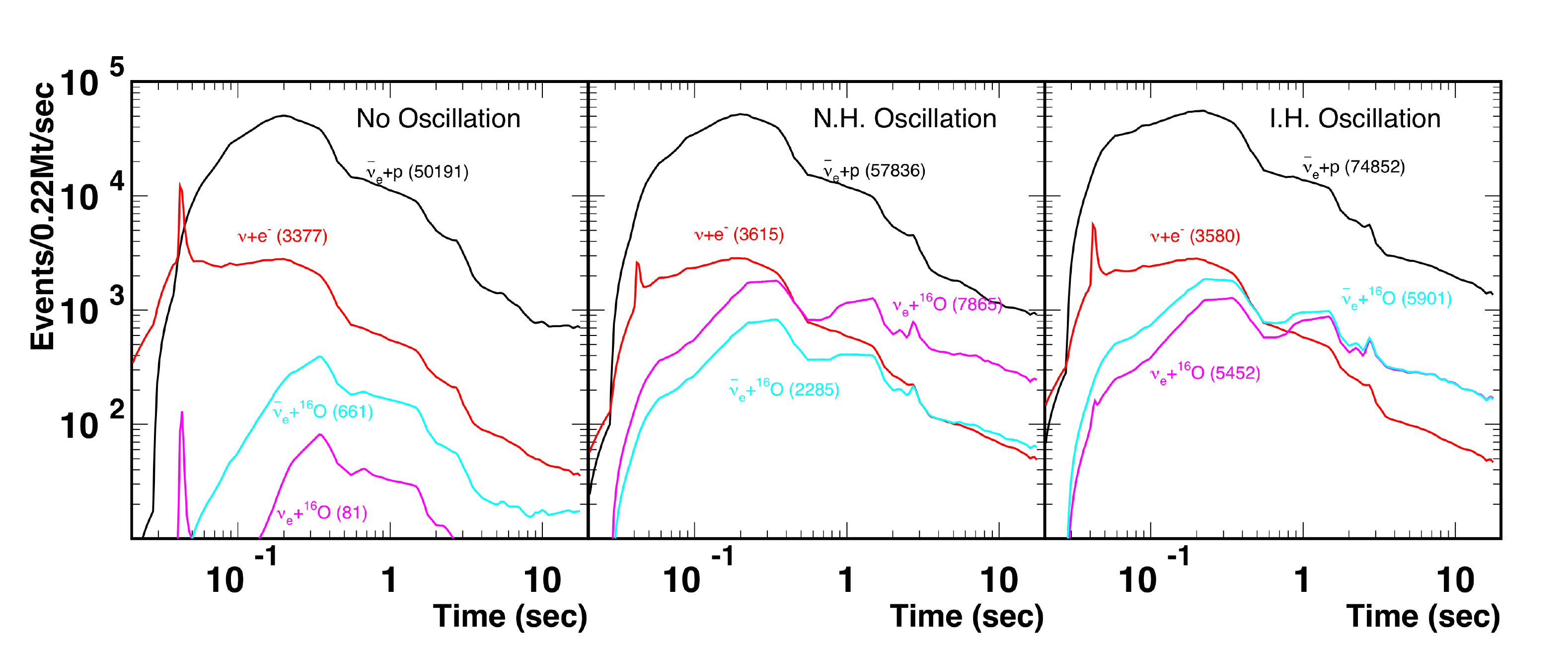}
  \end{center}
\vspace{-1cm}
\caption{Expected time profile of a supernova at 10\,kpc with 1 tank. 
Left, center, and right figures show profiles for no oscillation, normal 
hierarchy, and inverted hierarchy, respectively.
Black, red, purple, and light blue curves show event rates for
interactions of inverse beta decay ($\bar{\nu}_e + p \rightarrow e^+ + n$), 
$\nu e$-scattering ($\nu + e^- \rightarrow \nu + e^-$), 
$\nu_e~^{16}$O CC ($\nu_e + {\rm ^{16}O} \rightarrow e^- + {\rm^{16}F^{(*)}}$),
and 
$\bar{\nu}_e~^{16}$O CC
($\bar{\nu}_e + {\rm ^{16}O} \rightarrow e^+ + {\rm ^{16}N^{(*)}}$), respectively.
The numbers in parentheses are integrated
number of events over the burst. 
The fluxes and energy spectra are from the Livermore 
simulation~\cite{Totani:1997vj}
  \label{fig:sn-rate}
  }
\end{figure}

Figure~\ref{fig:sn-rate} shows time profiles for various interactions
expected at the Hyper-Kamiokande detector for a supernova at a
distance of 10\,kiloparsecs~(kpc).  This distance is a bit farther than
the center of the Milky Way galaxy at 8.5\,kpc; it is chosen as being
representative of what we might expect since a volume with a radius of
10\,kpc centered at Earth includes about half the stars in the galaxy.
\begin{figure}[tbp]
  \begin{center}
    \includegraphics[width=8cm]{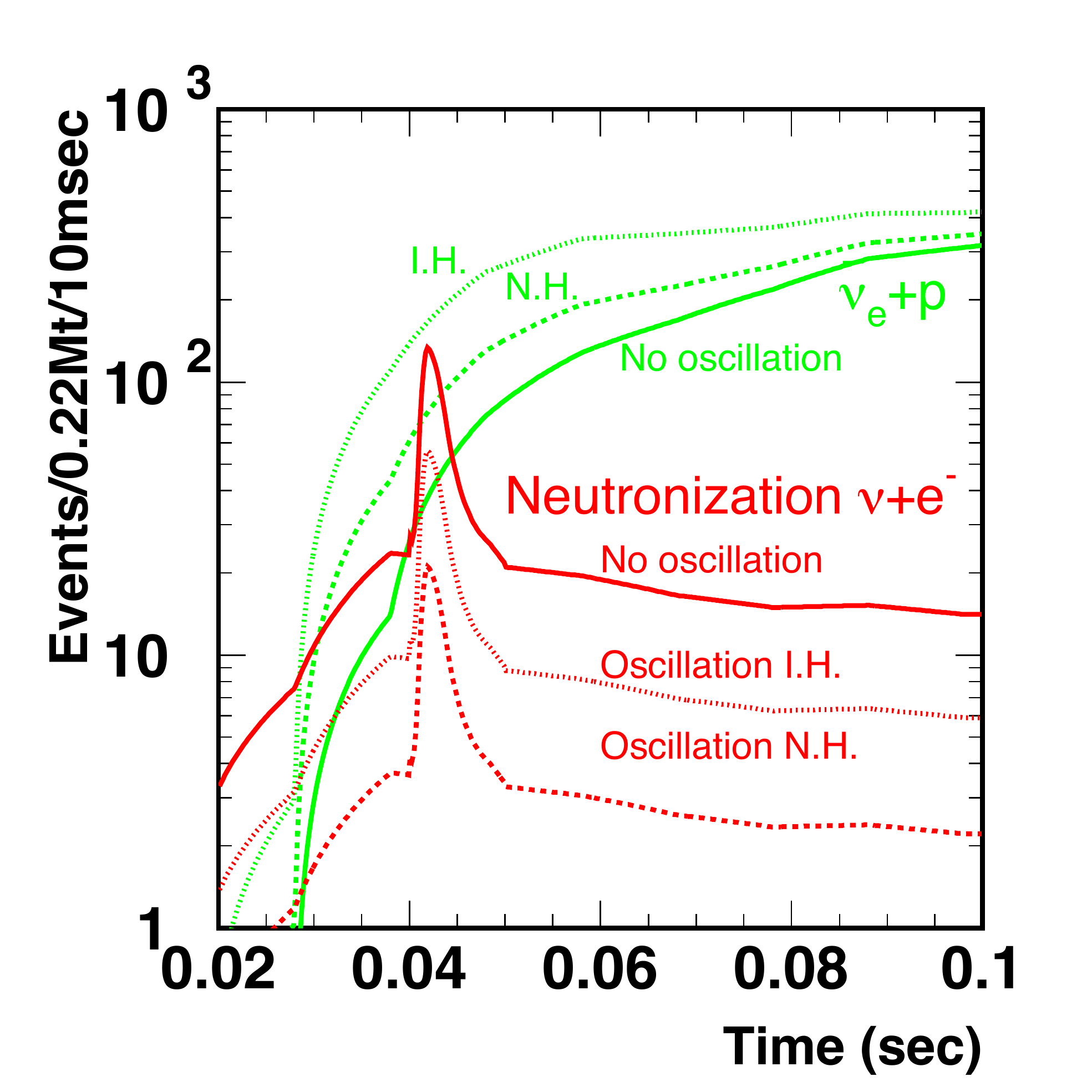}
  \end{center}
\vspace{-1cm}
  \caption{Expected event rate at the time of neutronization burst for
	  a supernova at 10\,kpc with 1 tank.  Red and green
	  show event rates for $\nu e$-scattering events originated with neutronization neutrino and inverse beta
	  events, respectively.  Solid, dotted, and dashed curved
	  indicate the neutrino oscillation scenarios of no
	  oscillation, N.H., and I.H.,
	  respectively.
  \label{fig:sn-neutronization}
  }
\end{figure}
\begin{figure}[tbp]
  \begin{center}
    \includegraphics[width=8cm]{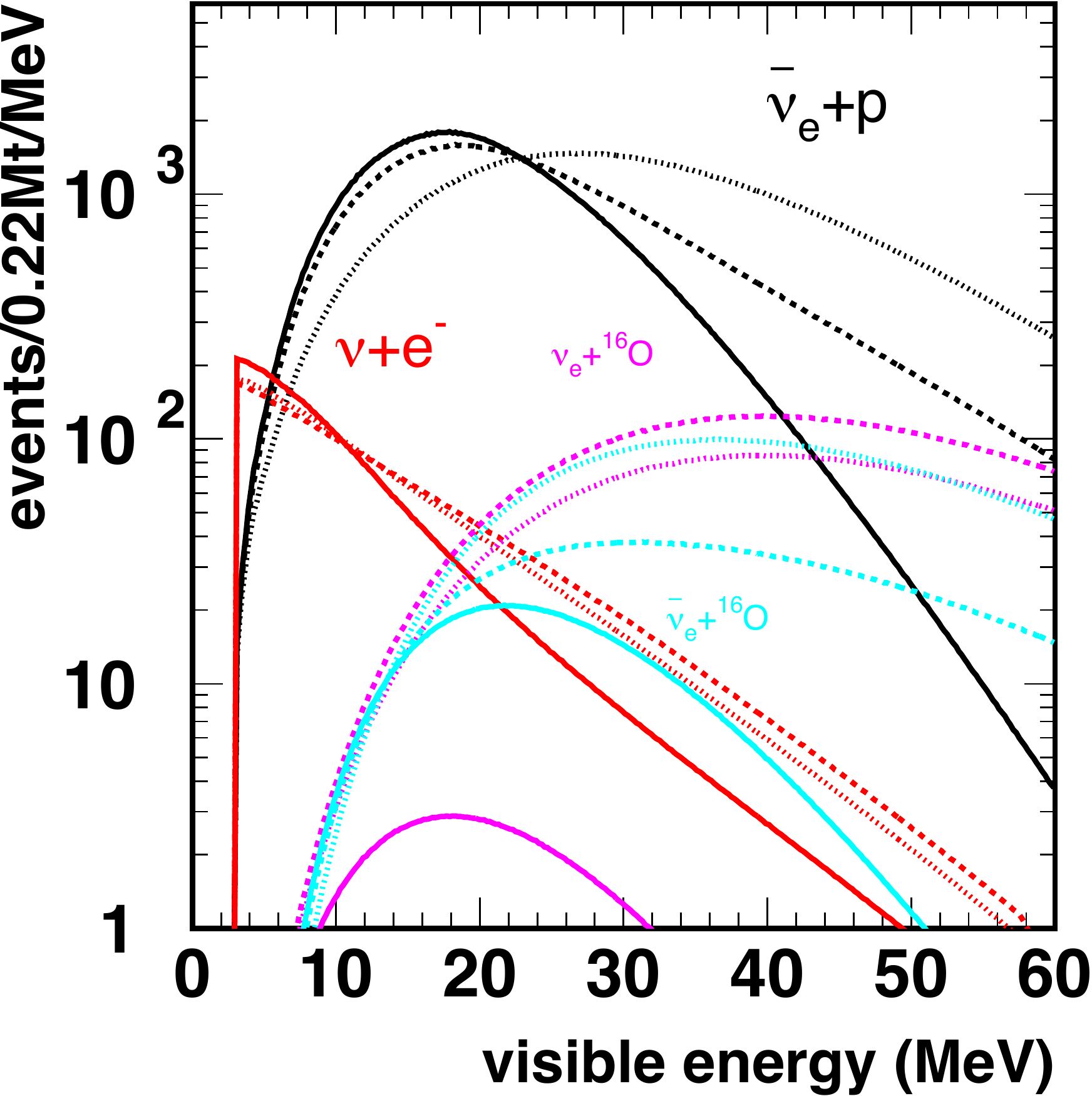}
  \end{center}
\vspace{-1cm}
  \caption{Total energy spectrum for each interaction
	  for a supernova at 10\,kpc with 1 tank. 
Black, red, purple, and light blue curves show event rates for
interactions of inverse beta decay, $\nu e$-scattering, 
$\nu_e+^{16}$O CC, and $\bar{\nu}_e+^{16}$O CC, respectively.
Solid, dashed, and dotted curves correspond to no oscillation, N.H., and
I.H., respectively.
  \label{fig:sn-spec}
}
\end{figure}
The three graphs in the figure show the cases of no oscillation, normal hierarchy and inverted hierarchy, respectively.
Colored curves in the figure show event rates for inverse beta decay
($\bar{\nu}_e + p \rightarrow e^+ + n$), 
$\nu e$-scattering($\nu + e^- \rightarrow \nu + e^-$), 
$\nu_e+^{16}$O CC($\nu_e + {\rm ^{16}O} \rightarrow e^- + {\rm ^{16}F^{(*)}}$),
and 
$\bar{\nu}_e+^{16}$O CC
($\bar{\nu}_e + {\rm ^{16}O} \rightarrow e^+ + {\rm ^{16}N^{(*)}}$).
The burst time period is about 10\,s and the peak event rate of inverse beta decay events reaches about 50\,kHz at 10 kpc.
The DAQ and its buffering system of Hyper-K will be designed to accept the broad range of rates, for a galactic SN closer than 10 kpc.
A sharp timing spike is expected for $\nu e$-scattering events at the time of neutronization.
Fig.~\ref{fig:sn-neutronization} shows the expanded plot around the neutronization burst peak region.
We expect $\sim$9, $\sim$23 and $\sim$55 $\nu e$-scattering events in this region for a supernova at 10\,kpc, for N.H., I.H., and no oscillation respectively.
Although the number of inverse beta events is $\sim$100 (N.H.), $\sim$210 (I.H.), and $\sim$60 (no oscillation) in the 10\,ms bin of the neutronization burst, 
the number of events in the direction of the supernova is typically 1/10 of the total events. 
So, the ratio of signal events ($\nu e$-scattering) to other events (inverse beta) is expected to be about 9/10~(N.H.),
23/21~(I.H.) and 55/6~(no oscillation).

The energy distributions of each interaction are shown in Fig.~\ref{fig:sn-spec}, where the energy is the electron-equivalent total energy measured by a Cherenkov detector.
The energy spectrum of $\bar{\nu}_e$ can be extracted from the distribution.

\begin{figure}[htbp]
  \begin{center}
    \includegraphics[width=8cm]{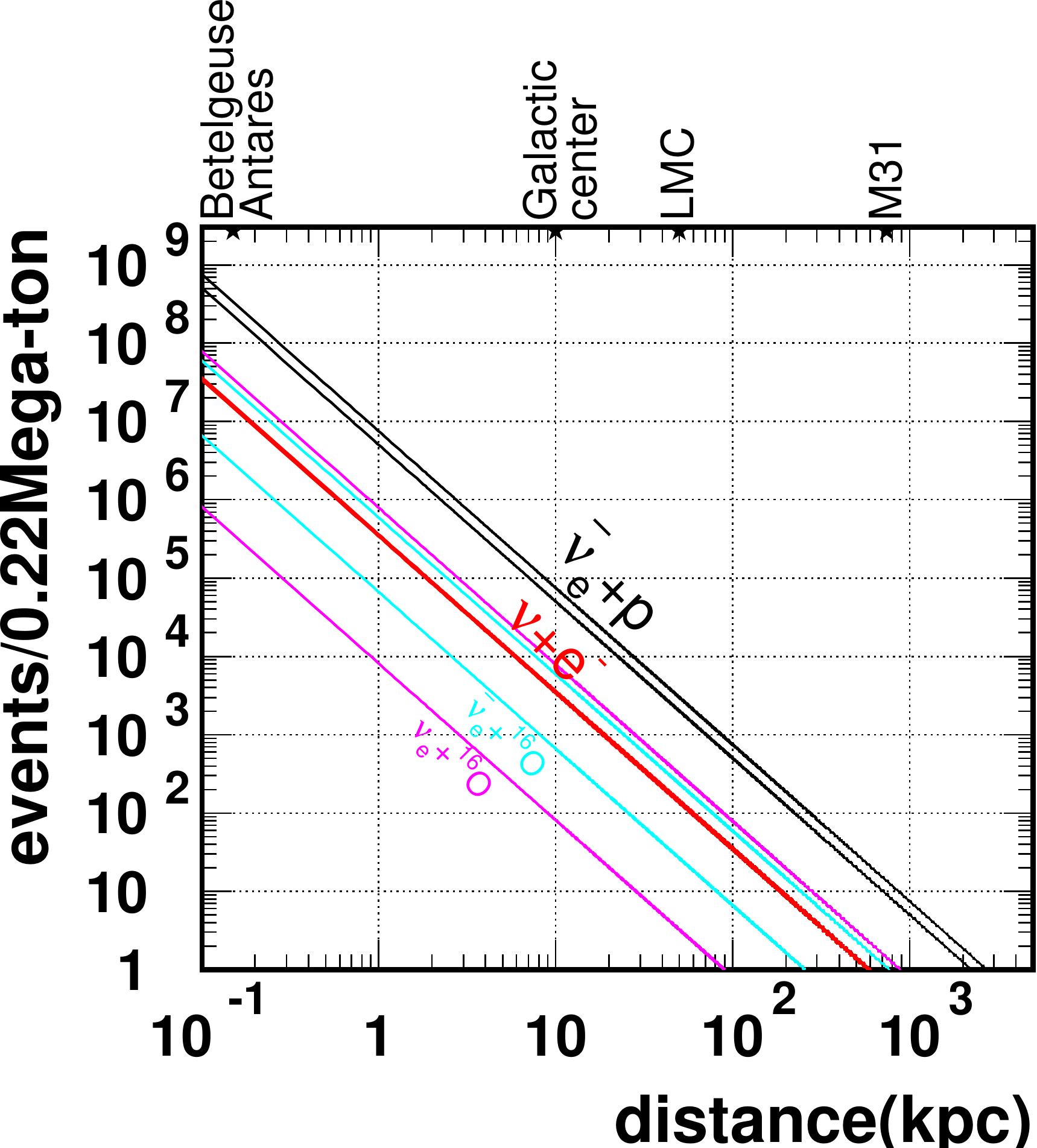}
  \end{center}
\vspace{-1cm}
  \caption{Expected number of supernova burst events for each
	  interaction as a function of the distance to a supernova with 1 tank.
The band of each line shows the possible variation due to the assumption
of neutrino oscillations.
   \label{fig:sn-evtvsdist}
}
\end{figure}

Figure~\ref{fig:sn-evtvsdist} shows the expected number of supernova neutrino events at Hyper-K versus the distance to a supernova.
At the Hyper-Kamiokande detector, we expect to see about 50,000 to 75,000 inverse beta decay events, 3,400 to 3,600 $\nu e$-scattering events, 80 to 7,900 $\nu_e+^{16}$O CC events, and 660 to 5,900 $\bar{\nu}_e+~^{16}$O CC events, in total 54,000 to 90,000 events, for a 10\,kpc supernova.
The range of each of these numbers covers possible variations due to the neutrino oscillation scenario (no oscillation, N.H., or I.H.).
Even for a supernova at M31 (Andromeda Galaxy), about 10 to 16 events are expected at Hyper-K.
In the case of the Large Magellanic Cloud (LMC) where SN1987a was located, about 2,200 to 3,600 events are expected.

\begin{figure}[tbp]
  \begin{center}
    \includegraphics[width=10cm]{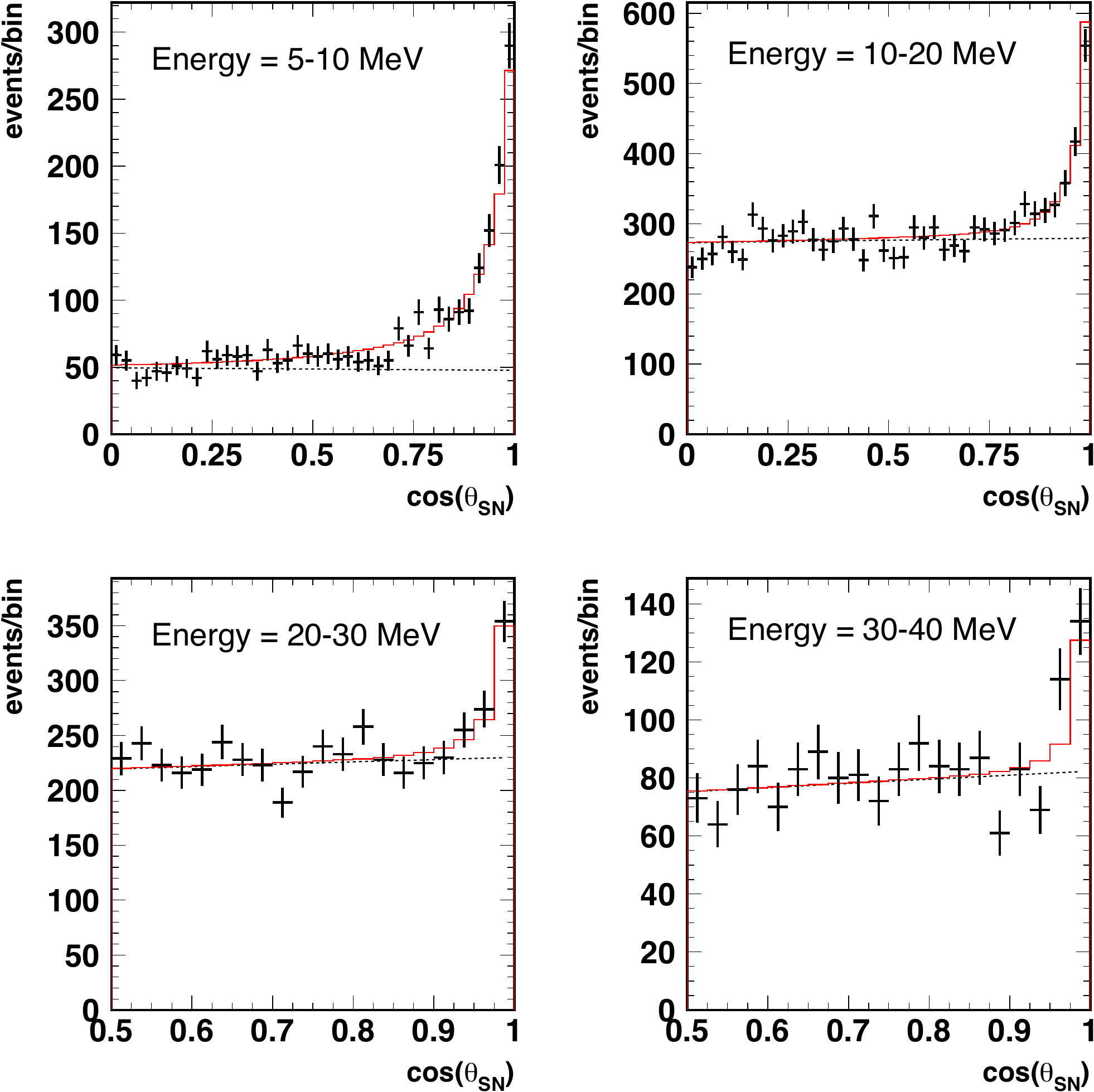}
  \end{center}
\vspace{-1cm}
  \caption{Angular distributions of a simulation of
	  a 10\,kpc supernova with 1 tank. The plots show a visible energy range of 
5-10\,MeV (left-top), 
10-20\,MeV (right-top), 20-30\,MeV (left-bottom), and 30-40\,MeV (right-bottom).
The black dotted line and the red solid histogram (above the black dotted line)
are fitted contributions of inverse beta decay and $\nu e$-scattering events.
Concerning the neutrino oscillation scenario, the $no~oscillation$ case is shown
here.
  \label{fig:sn-cossn}
}
\end{figure}

The observation of supernova burst neutrino and the directional
information can provide an early warning for electromagnetic observation
experiments, e.g. optical and x-ray telescopes. 
Figure~\ref{fig:sn-cossn} shows expected angular
distributions with respect to the direction of the supernova for four
visible energy ranges.  The inverse beta decay events have a nearly
isotropic angular distribution.  On the other hand, $\nu e$-scattering
events have a strong peak in the direction coming from the supernova.
Since the visible energy of $\nu e$-scattering events are lower than
the inverse beta decay events, the angular distributions for lower energy
events show more enhanced peaks.  The direction of a supernova at
10\,kpc can be reconstructed with an accuracy of about 1 to 1.3 degrees with Hyper-K,
assuming the performance of event direction reconstruction similar to Super-K~\cite{Abe:2016waf}.
This pointing accuracy will be precise enough for the multi-messenger measurement of supernova at the center of our galaxy,
with the world's largest class telescopes, $i.e.$ Subaru HSC and future LSST telescope~\cite{Nakamura:2016kkl}.\\

\paragraph{Physics impacts}

\begin{figure}[tbp]
  \begin{center}
    \includegraphics[width=7cm]{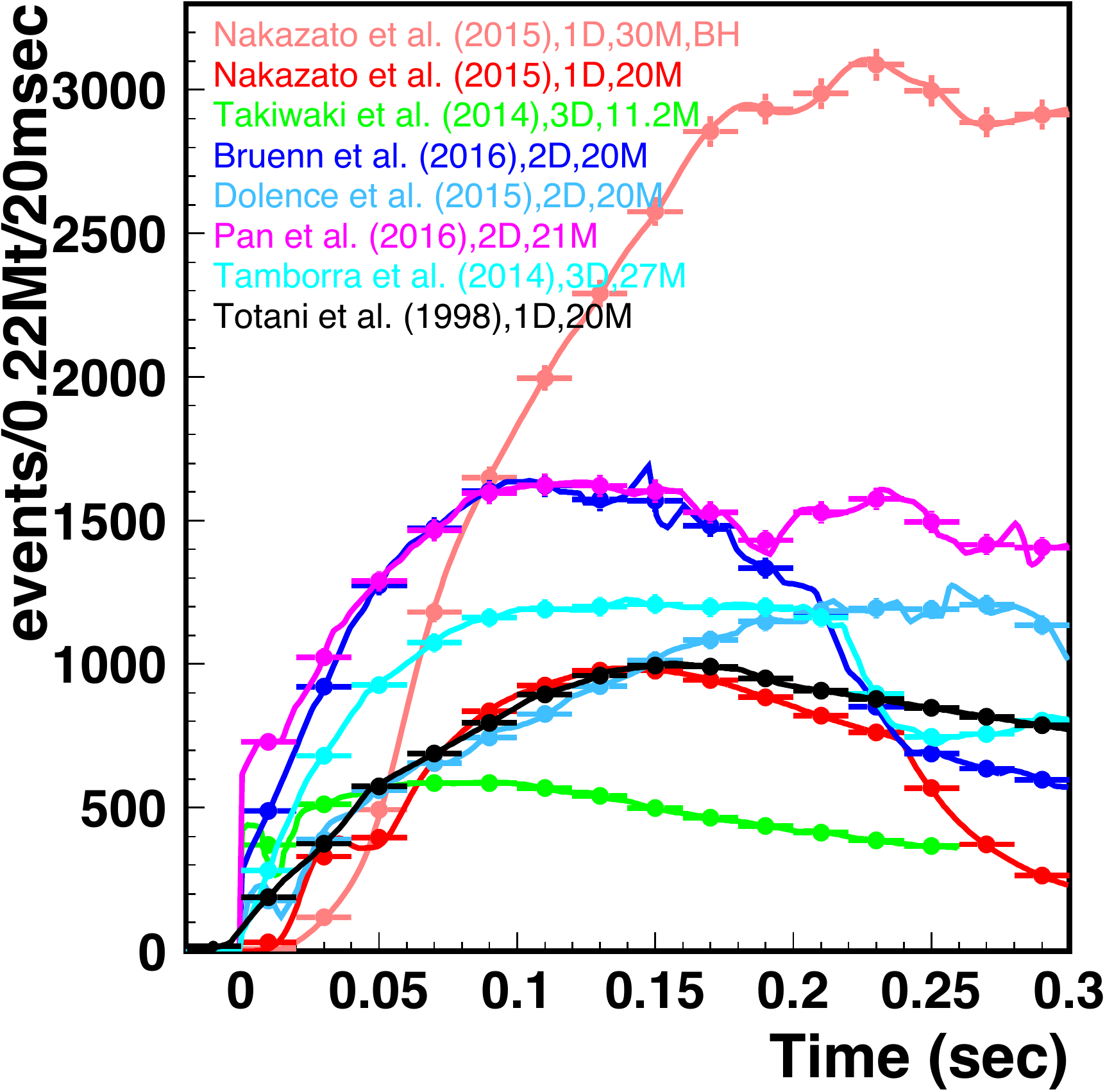}
    \includegraphics[width=7cm]{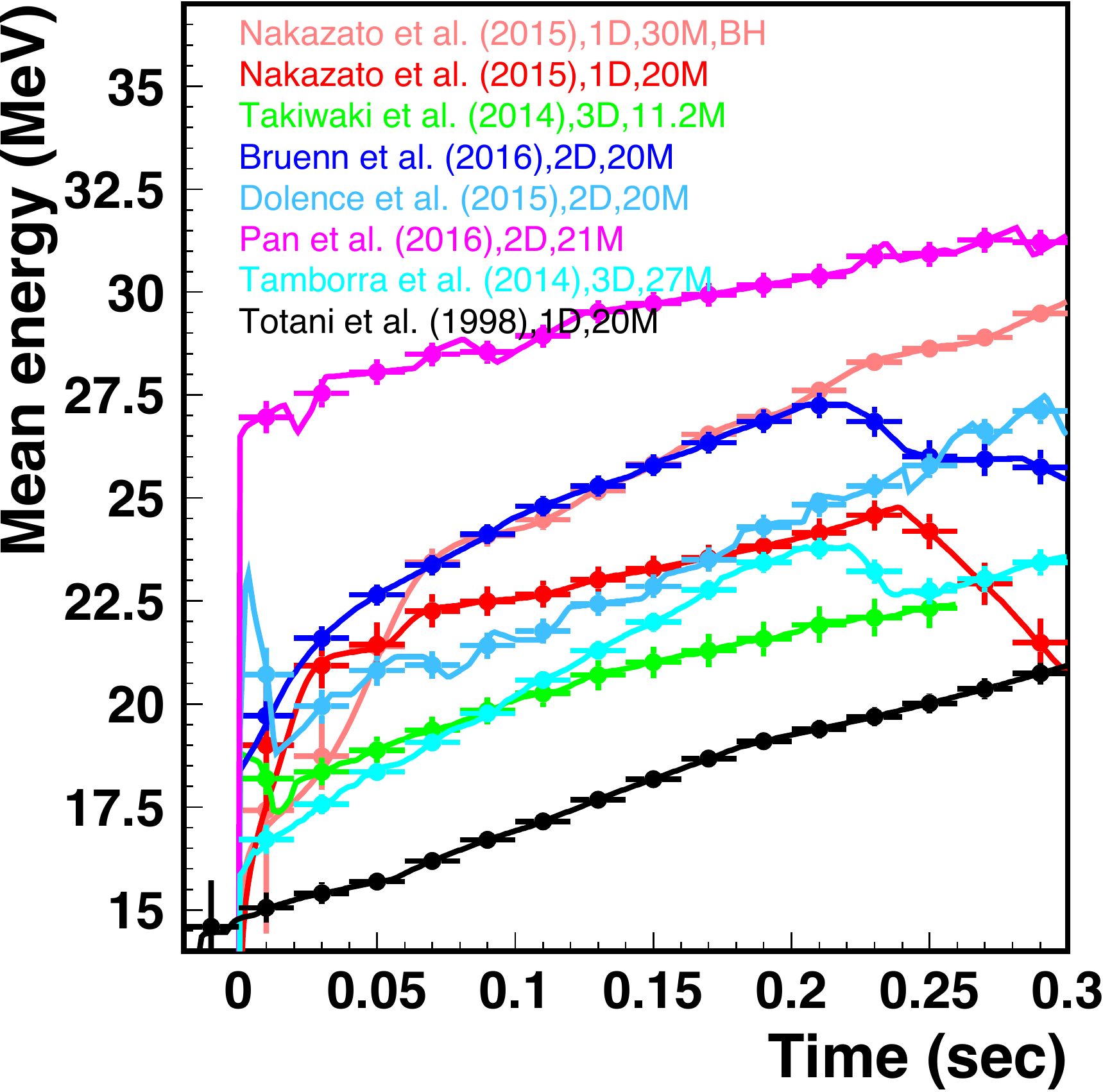}
  \end{center}
\vspace{-1cm}
  \caption{
	  Time profiles of the observed inverse beta decay event rate (left) and mean energy of these events, 
predicted by supernova simulations~\cite{Totani:1997vj,Takiwaki:2013cqa,Tamborra:2014hga,Dolence:2014rwa,Pan:2015sga,Nakazato:2015rya,Bruenn:2014qea} for the first 0.3\,seconds after the onset 
of a 10\,kpc distant burst with Hyper-K 1 tank.
  \label{fig:sn-model} 
}
\end{figure}

The shape of the rising time of supernova neutrino flux and energy strongly depends on the model.
Figure~\ref{fig:sn-model} shows inverse beta decay event rates and mean $\bar{\nu}_e$ energy distributions predicted by various models
\cite{Totani:1997vj,Takiwaki:2013cqa,Tamborra:2014hga,Dolence:2014rwa,Pan:2015sga,Nakazato:2015rya,Bruenn:2014qea} for the first 0.3\,sec after the onset of a burst.
The statistical error is much smaller than the difference between the models, and so Hyper-K should give crucial data for comparing model predictions.
The left plot in Fig.~\ref{fig:sn-model} shows that about 150-500
events are expected in the first 20 millisecond bin.  This means that
the onset time can be determined with an accuracy of about 1 ms.  This
is precise enough to allow examination of the infall of the core in
conjunction with the signals of neutronization as well as
possible data from future gravitational wave detectors.
Our measurement will also provide an opportunity to observe black hole formation directly, as a sharp drop of the neutrino flux~\cite{Sekiguchi:2010ja}.

We can use the sharp rise of the burst to make a measurement of
the absolute mass of neutrinos.  Because of the finite mass of
neutrinos, their arrival times will depend on their energies.  This
relation is expressed as \\
\begin{eqnarray}
	\Delta t = 5.15~{\rm msec} (\frac{D}{10~{\rm kpc}})
	(\frac{m}{1~{\rm eV}})^2 (\frac{E_\nu}{10~{\rm MeV}}) ^{-2}
\end{eqnarray}
where $\Delta t$ is the time delay with respect to that assuming zero
neutrino mass, $D$ is the distance to the supernova, $m$ is the
absolute mass of a neutrino, and $E_\nu$ is the neutrino energy.
Totani~\cite{Totani:1998nf} discussed Super-Kamiokande's sensitivity
to neutrino mass using the energy dependence of the rise time; scaling
these results to the much larger statistics provided by Hyper-K, we
expect a sensitivity of 0.5 to 1.3\,eV for the absolute neutrino
mass~\cite{Totani:2005pv}.  Note that this measurement of the absolute
neutrino mass does not depend on whether the neutrino is a Dirac or
Majorana particle.

Hyper-K can also statistically extract an energy distribution of
$\nu_e + \nu_X (X = \mu, \tau)$ events using the angular distributions
in much the same way as solar neutrino signals are separated from
background in Super-K.  Although the effect of neutrino oscillations
must be taken into account, the $\nu_e + \nu_X$ spectrum gives another
handle on the temperature of neutrinos.
Hyper-K will be able to evaluate the temperature
difference between $\bar{\nu}_e$ and $\nu_e + \nu_X$.  This would be a
valuable input to model builders. For example, the prediction from
many of the 
models that the energy of $\nu_e$ is less than $\nu_X$ can be confirmed.
The temperature is also critical for the nucleosynthesis via supernova explosion~\cite{2016inpc.confE.249H}.

From recent computer simulation studies, new characteristic
modulations of the supernova neutrino flux are proposed.
These modulations are due to the dynamic motions in the supernovae such as convection.
The stall of shock wave after core bounce has been an issue in supernova computer simulations, which was not able to achieve successful explosions.
These dynamic motions enable the inner materials to be heated more efficiently by the neutrinos from collapsed core, and realize the shock wave revival.
Such modulations can be detected as a variance of the neutrino event rate in Hyper-K.
It will be the clear evidence that neutrino is the driver of supernova burst process.
One source of such modulation is the Standing Accretion Shock
Instability (SASI)~\cite{Tamborra:13,SASIBlondin,SASIScheck}.
Figure~\ref{fig:sn_tamborra_sasi} shows the detection rate modulation
in Hyper-K, induced by SASI.
The modulation can be observed as a variance of $\sim$10\% of the number of supernova events in Hyper-K detector
\cite{Tamborra:13}.
This flux variance caused by SASI has a characteristic peak in the frequency space.
When we assume a 3\% flux modulation on the total supernova flux,
though the amount of the flux modulation depends on several variables, e.g. progenitor mass or equation of states,
it is possible to see the existence of SASI for the supernovae within about 15\,kpc distance.\cite{Migenda:16}
Under this assumption of a 3\% flux modulation, we will have chances to prove SASI effects for $\sim$90\% of galactic supernovae with Hyper-K, compared with only $\sim$15\% with Super-K.
\\
Another source of modulation is the rotation of supernova~\cite{Takiwaki2014,Takiwaki:2017tpe}.
The size of variation depends on the angle between the direction of
earth and the rotational axis of supernova.  When the supernova
rotational axis is orthogonal with the direction to the earth, the
detectable variance will be the maximum. In that case, Hyper-K will
detect a variance of $\sim$50\% in the number of observed signals
as shown in Figure~\ref{fig:sn_tamborra_sasi}.
The observation of these modulation with Hyper-K
will be a good test of such simulations and also provide important
information for understanding the dynamics in supernovae.

\if0
\begin{figure}[tbp]
  \begin{center}
    \includegraphics[width=8cm]{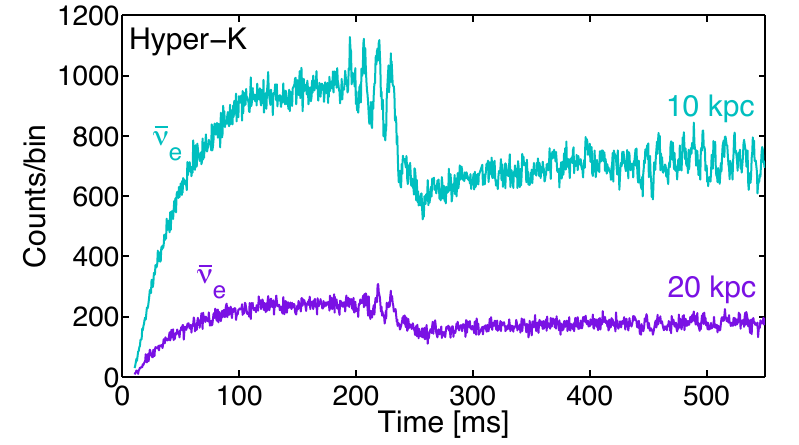}
  \end{center}
\vspace{-1cm}
  \caption{
	  Detection rate modulation induced by SASI in 5\,ms bins (0.56 Mt).
	  The SN progenitor mass is 27 solar mass.
	  The direction to the detector is chosen for strong signal modulation.
	  This figure is adopted from~\cite{Tamborra:13}.
  \label{fig:sn_tamborra_sasi}
  }
\end{figure}
\fi
\begin{figure}[tbp]
  \begin{center}
    \includegraphics[width=8cm]{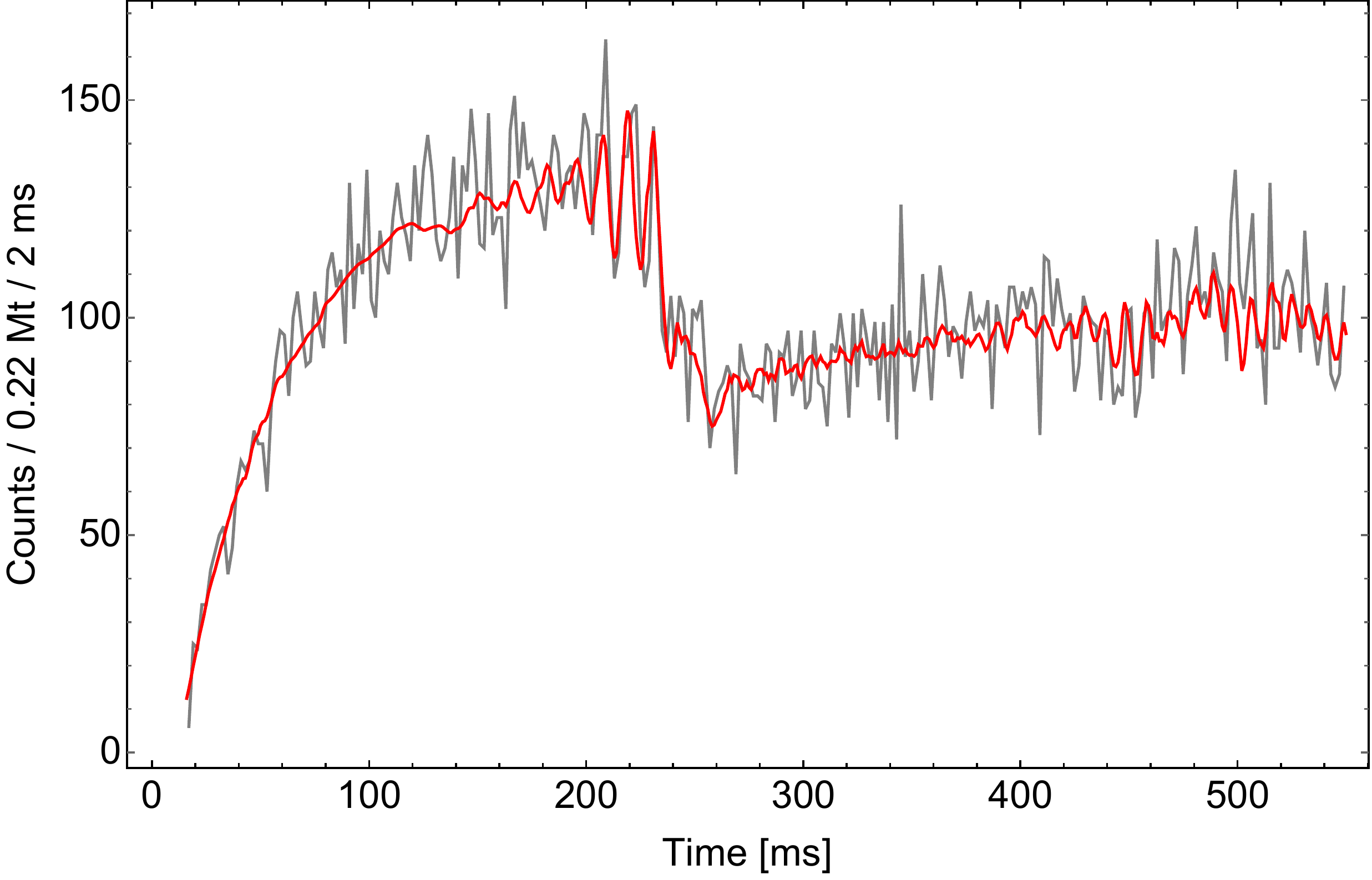}
  \end{center}
\vspace{-1cm}
  \caption{
	  Detection rate modulation induced by SASI in Hyper-K 1 tank.
	  Red line shows the theoretical event rate estimation for the inverse beta decay reaction.
	  Gray line shows a simulated event rate taking into account statistical fluctuation.
	  The SN progenitor mass is 27 solar mass.
	  The direction to the detector is chosen for strong signal modulation.
	  This neutrino flux is adopted from~\cite{Tamborra:13}.
  \label{fig:sn_tamborra_sasi}
  }
\end{figure}

\begin{figure}[tbp]
  \begin{center}
    \includegraphics[width=10cm]{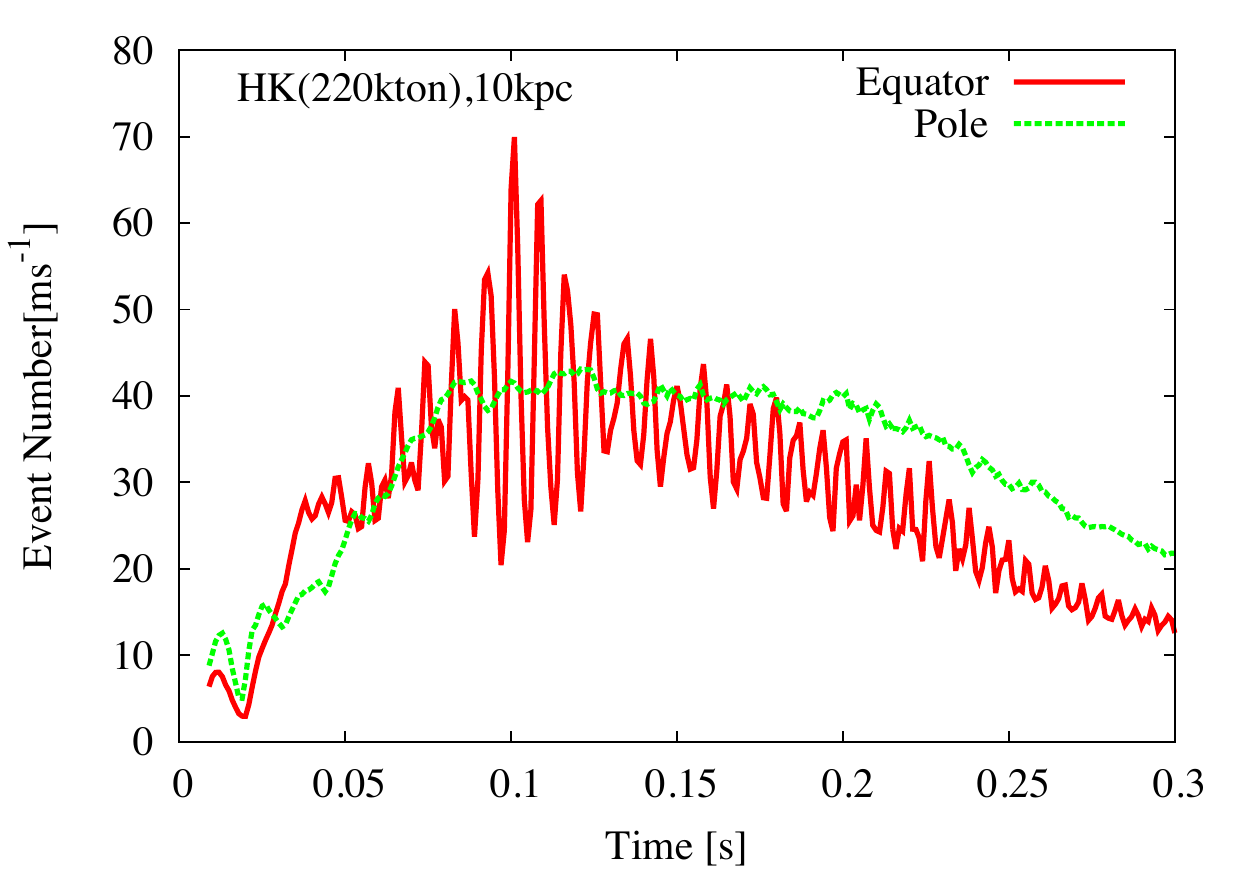}
  \end{center}
\vspace{-1cm}
  \caption{
	  Detection rate modulation produced by a rotating SN model in Hyper-K 1 tank.
	  The SN progenitor mass is 27 solar mass.
	  The supernova rotational axis is orthogonal (red) and parallel (green) with the direction to the earth.
	  This figure is adopted from~\cite{Takiwaki:2017tpe}.
  \label{fig:sn_tatewaki_rotation}
  }
\end{figure}

Neutrino oscillations could be studied using supernova neutrino events.
There are many papers which discuss the possibility of extracting signatures of neutrino oscillations free from uncertainties of supernova models \cite{Totani:1997vj,Takiwaki:2013cqa,Tamborra:2014hga,Dolence:2014rwa,Pan:2015sga,Nakazato:2015rya,Bruenn:2014qea}.
One big advantage of supernova neutrinos over other neutrino sources (solar, atmospheric, accelerator neutrinos) is that they inevitably pass through very high density matter on their way to the detector.
This gives a sizeable effect in the time variation of the energy spectrum~\cite{Schirato:2002tg, Fogli:2004ff, Tomas:2004gr}.
Though the combination of MSW stellar matter effects and collective effects makes the prediction quite difficult~\cite{Duan:2006an,Duan:2006jv,Dasgupta:2009dd,Friedland:2010sc,Duan:2010bf},
we will still have opportunities to determine the neutrino mass hierarchy from the supernova burst.
The first chance is the neutronization burst, where mostly pure $\nu_e$ will be emitted from the proto-neutron star.
Since, the collective effect through $\nu_e\bar{\nu}_e\rightarrow\nu_X\bar{\nu}_X$ interaction can be negligible.
The multi-angle effect can also be ignored, because these neutrinos are emitted only from the very center of the supernova.
The flux is well predicted and hardly affected by the physics
modelling of the EOS or the progenitor mass~\cite{Kachelriess:2004ds, Mirizzi:2015eza}.
The number of event will be about 50\% larger in IH case comparing to NH, after 20\,ms from the core bounce.
In the succeeding accretion phase, we will have another chance by observing the rise-time of neutrino event rate.
The mixing of $\bar{\nu}_X$ to $\bar{\nu}_e$, will result in a 100\,ms faster rise time for the inverted hierarchy compared to the normal hierarchy case~\cite{Serpico:2011ir}.
We will have fair chance to investigate it for a supernova at the galactic center, see Fig.~\ref{fig:sn-model}.

\begin{figure}[tbp]
  \begin{center}
    \includegraphics[width=16cm]{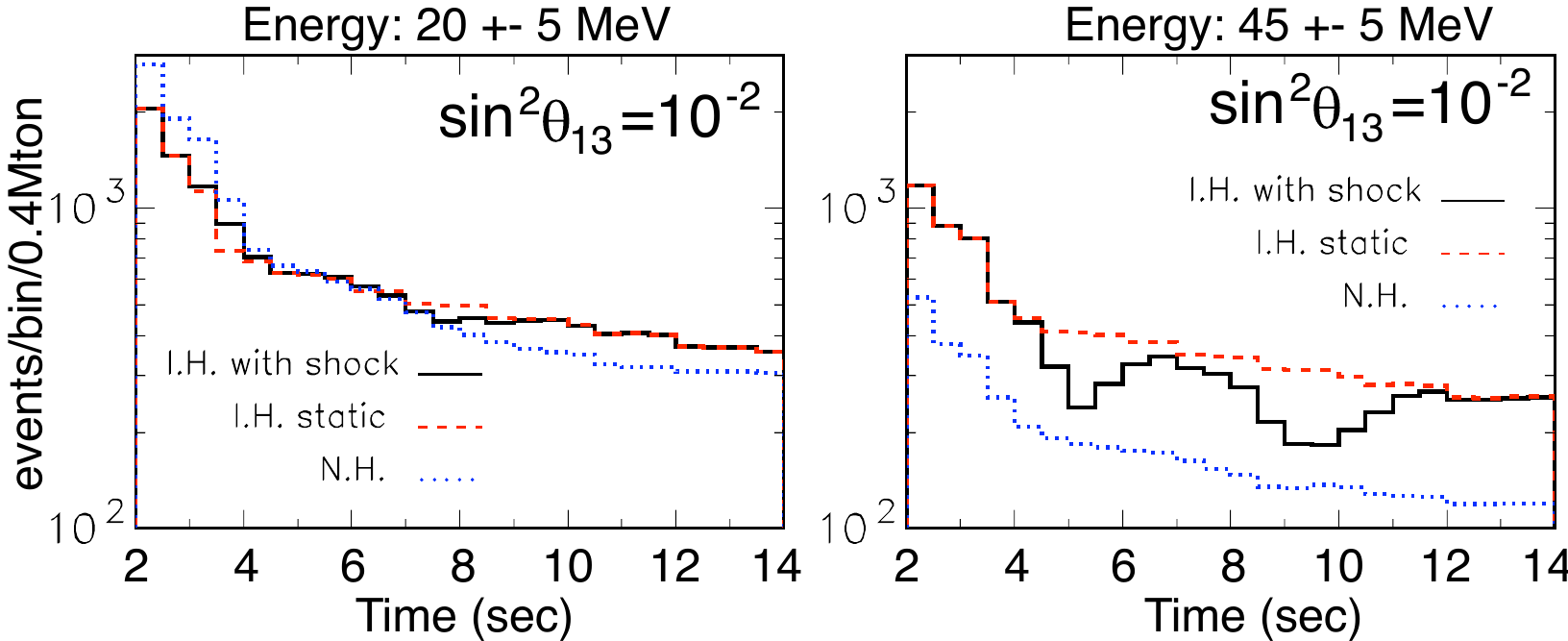}
  \end{center}
\vspace{-1cm}
  \caption{Time variation of the neutrino event rate affected by
	  neutrino conversion by matter due to shock wave propagation
	  (reproduced from~\cite{Fogli:2004ff}).
	  Left (right) plot shows the time variation of inverse beta
	  events for the energy range of 20$\pm$5\,MeV (45$\pm$5\,MeV).
	  Solid black, dashed red, and blue dotted histograms show the
	  event rates for I.H. with shock wave propagation, I.H. with
	  static matter density profile, and N.H., respectively.  It
	  has been assumed that
	  $\sin^2\theta_{13}=10^{-2}$.  \label{fig:sn-nuosc}}
\end{figure}

\if0
An example is shown in Figure~\ref{fig:sn-nuosc}~\cite{Fogli:2004ff}.
The propagation of the supernova shock wave causes time variations in
the matter density profile through which the neutrinos must travel.
Because of neutrino conversion by matter, there may be a bump in the
time variation of the inverse beta event rate for a particular energy
range ($i.e.$, 45$\pm$5\,MeV as shown in
Fig.~\ref{fig:sn-nuosc}(right)) while no change is observed in the
event rate near the spectrum peak ($i.e.$, 20$\pm$5\,MeV as shown in
Fig.~\ref{fig:sn-nuosc}(left)).  This effect is observed only in the
case of inverted mass hierarchy; this is one way in which the mass
hierarchy could be determined by a supernova burst.
\fi

\begin{figure}[tbp]
  \begin{center}
    \includegraphics[width=7cm]{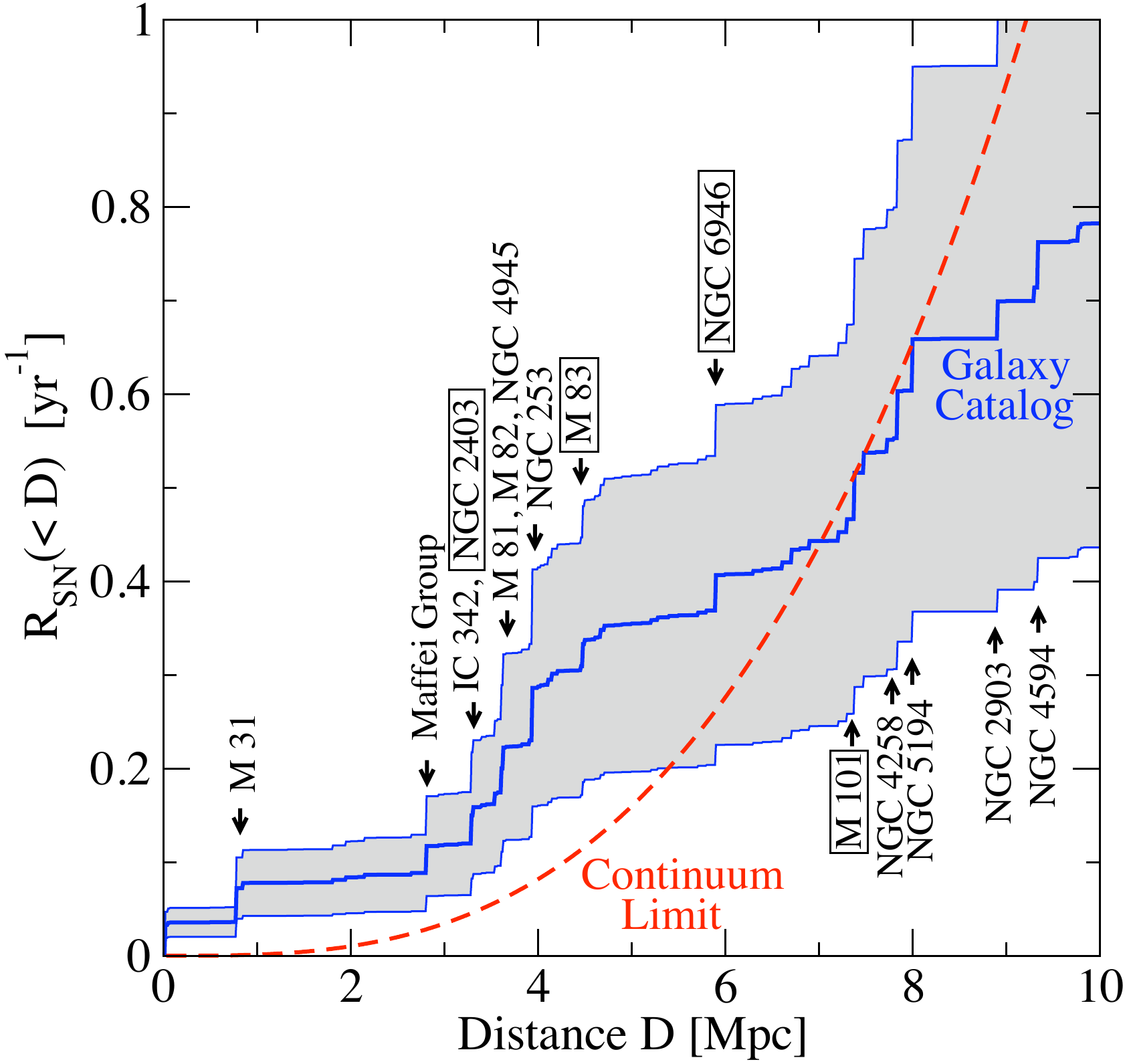}
  \end{center}
\vspace{-1cm}
  \caption{Cumulative calculated supernova rate versus distance for 
	  supernovae in nearby galaxies. The dashed line is core-collapse supernova rate expectation, using the $z = 0$ limit of star formation rate measured by GALEX.  The figure is reproduced from ref.~\cite{Ando:2005ka}.
  \label{fig:sn-nearby}
}
\end{figure}

\begin{figure}[tbp]
  \begin{center}
    \includegraphics[width=7cm]{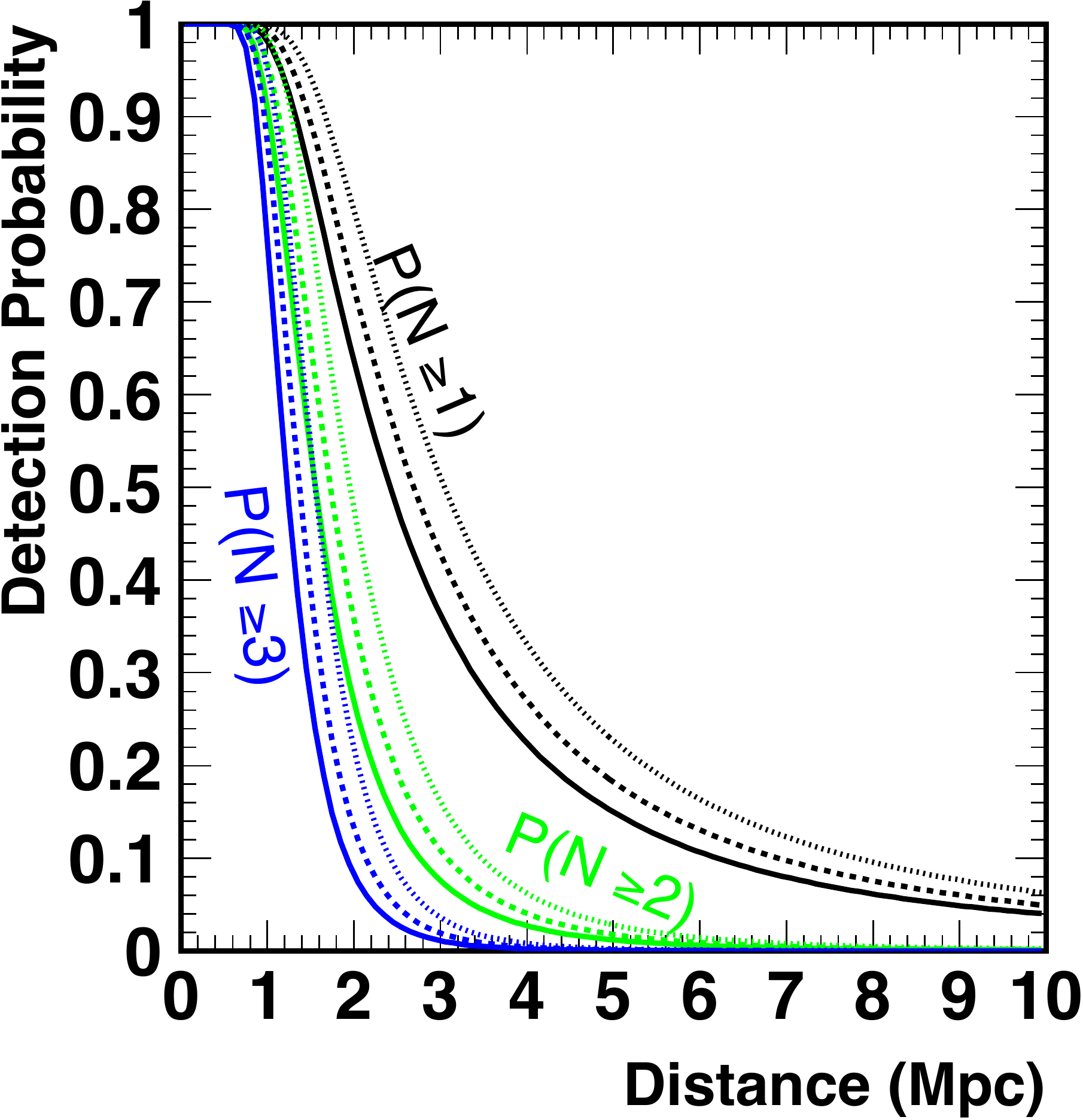}
\hspace{3mm}
    \includegraphics[width=7cm]{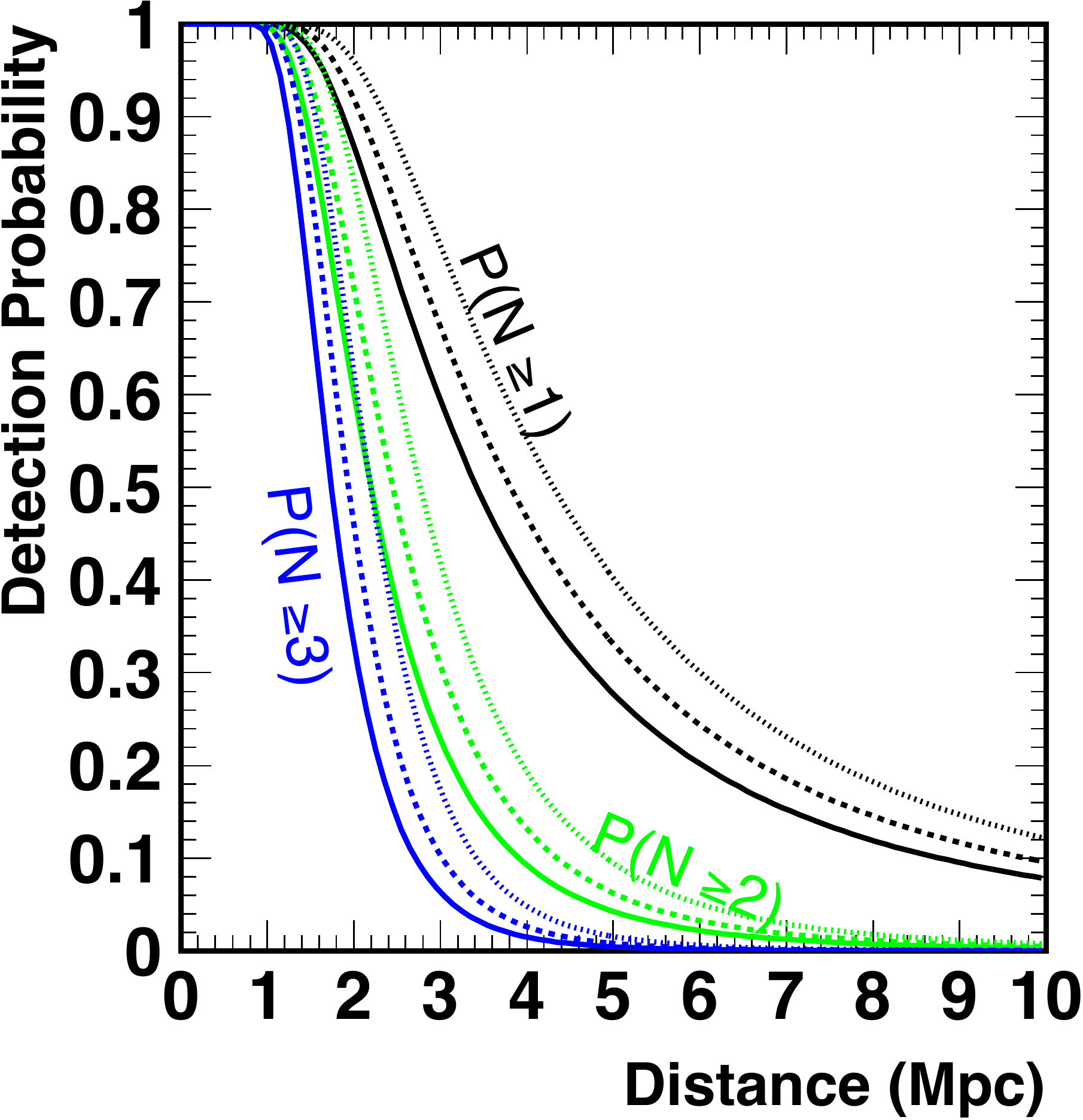}
  \end{center}
\vspace{-1cm}
  \caption{(Left) Detection probability of supernova neutrinos versus distance at Hyper-K assuming a 187\,kiloton fiducial volume and 10\,MeV threshold for this analysis.
Black, green, and blue curves show the detection efficiency resulting in requiring more than or equal to one, two, and three events per burst, respectively.
Solid, dotted, and dashed curves are for neutrino oscillation scenarios of no oscillation, N.H., and I.H., respectively.
	(Right) Detection probability of supernova neutrinos with Hyper-K 2 tanks.
  \label{fig:sn-nearbyprob}
}
\end{figure}

In Hyper-K, it could be possible to detect burst neutrinos from
supernovae in nearby galaxies.  As described above, we expect to
observe a very large number of neutrino events from a galactic
supernova.  However, galactic supernovae are expected to happen once
per 30\,-\,50 years.  So, we cannot count on seeing many galactic
supernova bursts.  In order to examine a variety of supernova bursts,
supernovae from nearby galaxies are useful even though the expected
number of detected events from any single such burst is small.
Furthermore, the merged energy spectrum from these supernovae will be highly useful for understanding that of supernova relic neutrinos (see next sub-section) for the absence of red-shift.
The supernova events from nearby galaxies provide a reference energy spectrum for this purpose.
The supernovae rate in nearby galaxies was discussed in~\cite{Ando:2005ka} and a figure from the
paper is shown in Fig.~\ref{fig:sn-nearby}.
It shows the cumulative supernova rate versus distance and indicates that if Hyper-K can see signals out to 4\,Mpc then we could expect a supernova about every three years.
It should be noted that recent astronomical observations indicate about 3 times higher nearby supernova rate~\cite{Horiuchi:2013}, compared to the conservative calculation.
It is also valuable to mention that two strange supernovae have been found at $\sim$2\,Mpc distance in the past 11 years observation, which are called dim supernovae~\cite{Horiuchi:2013}.
The detections of supernova neutrinos from these dim supernovae will prove their explosion mechanism is core-collapse.
Figure~\ref{fig:sn-nearbyprob} shows detection probability versus distance for the Hyper-K detector 1 tank (left) and 2 tanks configuration (right).
In this estimate, we required the neutrino energy be greater than 10
MeV in order to reduce background.
Requiring the number of neutrino events to be more than or or equal to two\,(one), the detection probability is 27 to 48\% (64 to 80\%) with 1 tank, for a supernova at 2\,Mpc.
The probability will be 3 to 6\% (22 to 33\%) for the supernovae at 4\,Mpc.
With 2 tank configuration and 375 kt fiducial volume, the detection probability will be increased to 10 to 18\% (40 to 57\%) for a supernova at 4\,Mpc.
If we can use a tight timing coincidence with other types of supernova sensors (e.g. future gravitational wave detectors), we should be able to identify even single supernova neutrinos.
We expect to observe 2.4 to 4.6 or 0.6 to 1.4 supernovae with and without dim supernovae within 10\,Mpc respectively, during 20 years of Hyper-K one tank operation.
Here we required two or more neutrino events for each supernovae, and referred to the nearby galactic supernova rate given by CCSNe counting in ref.~\cite{Horiuchi:2013}.
The number of observations can be increased to 5.0 to 8.2 and 1.6 to 3.3 supernovae with and without dim supernova respectively, with Hyper-K staging two tank scenario.

\paragraph{Summary}

The expected number of supernova neutrino events in the Hyper-Kamiokande detector is summarized in Table~\ref{tab:phys_supernova_summary}.
These events will be enough to provide detailed information about the time profile and the energy spectrum for inspecting supernova explosion mechanism, including black hole formation.
They will also provide an opportunity for further physics topics, $i.e.$ the neutrino mass, the mass hierarchy and the neutrino oscillations in the supernova.
Physics beyond the standard model also can be tested, $e.g.$
sterile neutrinos~\cite{Hidaka:2006sg, Loveridge:2003fy, Hidaka:2007se, Tamborra:2011is},
the neutrino magnetic moment~\cite{Lattimer:1988mf, Barbieri:1988nh}
the neutrino non-standard interaction~\cite{EstebanPretel:2007yu, Ohlsson:2012kf} and
the axion~\cite{Turner:1987by, Burrows:1988ah, Keil:1996ju, Raffelt:2006cw, Fischer:2016cyd}.
The better energy resolution from high photo-coverage will help these analyses.
The lower energy threshold will also help to understand the whole picture of supernova explosion.
The detection of supernovae in nearby galaxies will be possible even with a single tank, though the number of events will be small.
The energy spectrum of these events will still be useful to understand the nature of the supernova explosion.
\begin{table}[t]
		\caption{Summary table of expected supernova neutrino events in the Hyper-Kamiokande detector, with E$_\nu$ upto 100\,MeV and the detection threshold of 3\,MeV. A supernova at Galactic center (10\,kpc) is assumed. Our references for each cross-section are also shown.
		\label{tab:phys_supernova_summary}
		}
	\begin{ruledtabular}
		\begin{tabular}{lccc}
			Neutrino source  & Single Tank (220\,kt Full Volume) & 2 Tanks (440\,kt Full Volume) & Ref.\\
			\hline
			$\bar{\nu}_e + p$ & 50,000 - 75,000\,events& 100,000 - 150,000\,events&\cite{Vogel:1999zy}\\
			${\nu} + e^-$ & 3,400 - 3,600~events& 6,800 - 7,200~events&\cite{tHooft:1971ucy} \\
			${\nu}_e + ^{16}O$ CC     & 80 - 7,900~events& 160 - 11,000~events&\cite{Tomas:2003xn} \\
			$\bar{\nu}_e + ^{16}O$ CC     & 660 - 5,900~events& 1,300 - 12,000~events& \cite{Kolbe:2002gk}\\
			${\nu} + e^-$ (Neutronization) & 9 - 55~events& 17 - 110~events &\cite{tHooft:1971ucy}\\
			\hline
			Total & 54,000 - 90,000~events& 109,000 - 180,000~events \\
		\end{tabular}
	\end{ruledtabular}
\end{table}

\subsubsection{High-energy neutrinos from supernovae with interactions with circumstellar material}
Core-collapse supernovae are promising sources of high-energy ($\gtrsim$ GeV) neutrinos as well as multi-MeV neutrinos.
The supernova shock
propagates in the stellar material and experiences a shock breakout,
which can be observed at ultraviolet or X-ray wavelengths.  Before the
shock breakout, the supernova shock is mediated by radiation since the
photon diffusion time is longer than the expansion time.  During this
time, the conventional cosmic-ray (CR) acceleration is inefficient, so
associated neutrino production is not promising.  However, as the
shock becomes collisionless after the breakout, the CR acceleration
starts to be effective~\cite{wl01,mur+11}. The situation is expected
to be analogous to supernova remnants, which are almost established as
CR accelerators and widely believed as the origin of Galactic CRs.

In the early phase just after the breakout, the matter density is still high, so that accelerated CRs are efficiently used for neutrino production via inelastic $pp$ scatterings.
For type II supernovae, which are associated with red
super-giants, the released energy of high-energy neutrinos is
typically ${\mathcal E}_{\nu}\sim{10}^{47}$\,erg~\cite{wl01}.
One to two events of GeV neutrinos are expected in a timescale of hours after the core-collapse for a Galactic supernova at 10kpc in Hyper-K 1 tank.

About 10\% of core-collapse supernovae show strong interactions with
ambient circumstellar material, which are often called
interaction-powered supernovae.  If the circumstellar material mass is
$\sim0.1$-$1~M_\odot$, the released high-energy neutrino energy
reaches ${\mathcal
E}_{\nu}\sim{10}^{49}$-${10}^{50}$~erg~\cite{mur+11}.  For example,
Eta Carinae at 2.3\,kpc is an interesting candidate that showed violent
mass eruptions in the past.  If a real supernova occurs and the ejecta
collides with the circumstellar material shell with $\sim10\,M_\odot$,
one may expect $\sim300$ events with Hyper-K.  However, because of
the long duration (from months to years), the signal can overwhelm the
background only at early times and sufficiently high energies.

High-energy neutrinos from supernovae are detectable hours to months
after the core-collapse, and detecting the signals will give us new
insights into supernova physics, such as how collisionless shocks are
formed and CR acceleration starts, as well as the connection to
supernova remnants as the origin of Galactic CRs.  We may be able to
see the time evolution of multi-energy neutrino emission from the
core-collapse to shock breakout and following interactions with the
circumstellar material.
       
\subsubsection{Supernova relic neutrinos}\label{sec:supernova-relic}
The neutrinos produced by all of the
supernova explosions since the beginning of the universe are called
supernova relic neutrinos (SRN) or diffuse supernova neutrino background (DSNB).  They must fill the present universe
and their flux is estimated to be a few tens/cm$^2$/sec.  If we can
detect these neutrinos, it is possible to explore the history of how
heavy elements have been synthesized since the onset of stellar
formation.
\par
\begin{figure}[tbp]
  \begin{center}
    \includegraphics[width=8cm]{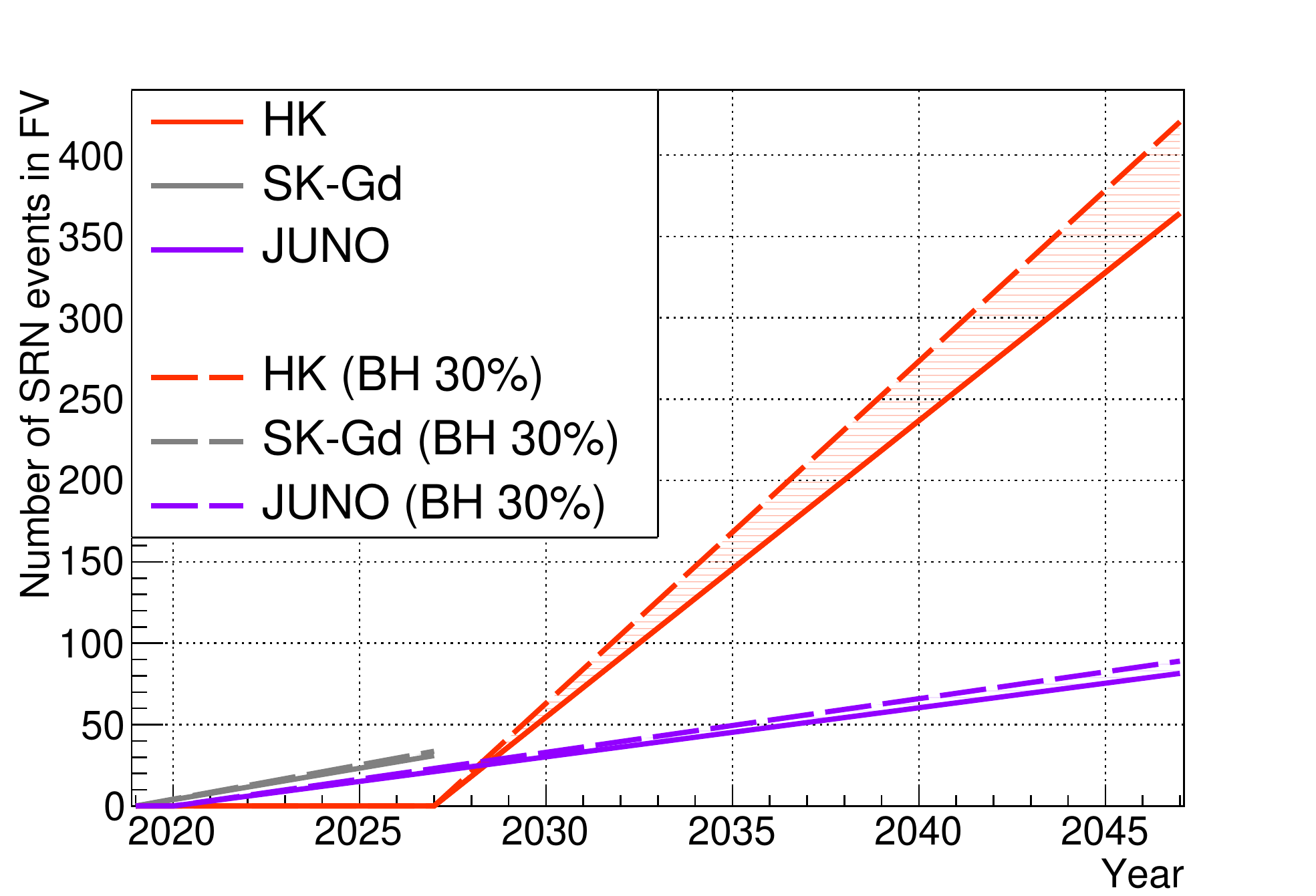}
  \end{center}
\vspace{-1cm}
  \caption{Expected number of inverse beta decay reactions due to supernova relic neutrinos in several experiments as a function of year.
Red, gray and purple line shows Hyper-Kamiokande, SK-Gd, and JUNO, respectively.
The sizes of their fiducial volume and analysis energy thresholds were considered.
The neutrino temperature is assumed to be 6MeV.
Solid line corresponds to the case, in which all the core-collapse supernovae emits neutrinos with the particular energy.
Dashed line corresponds to the case, in which 30\% of the supernovae form black hole and emits higher energy neutrinos corresponding to the neutrino temperature of 8\,MeV.  \label{fig:srn-comp} }
\end{figure}
\begin{figure}[tbp]
  \begin{center}
    \includegraphics[width=8cm]{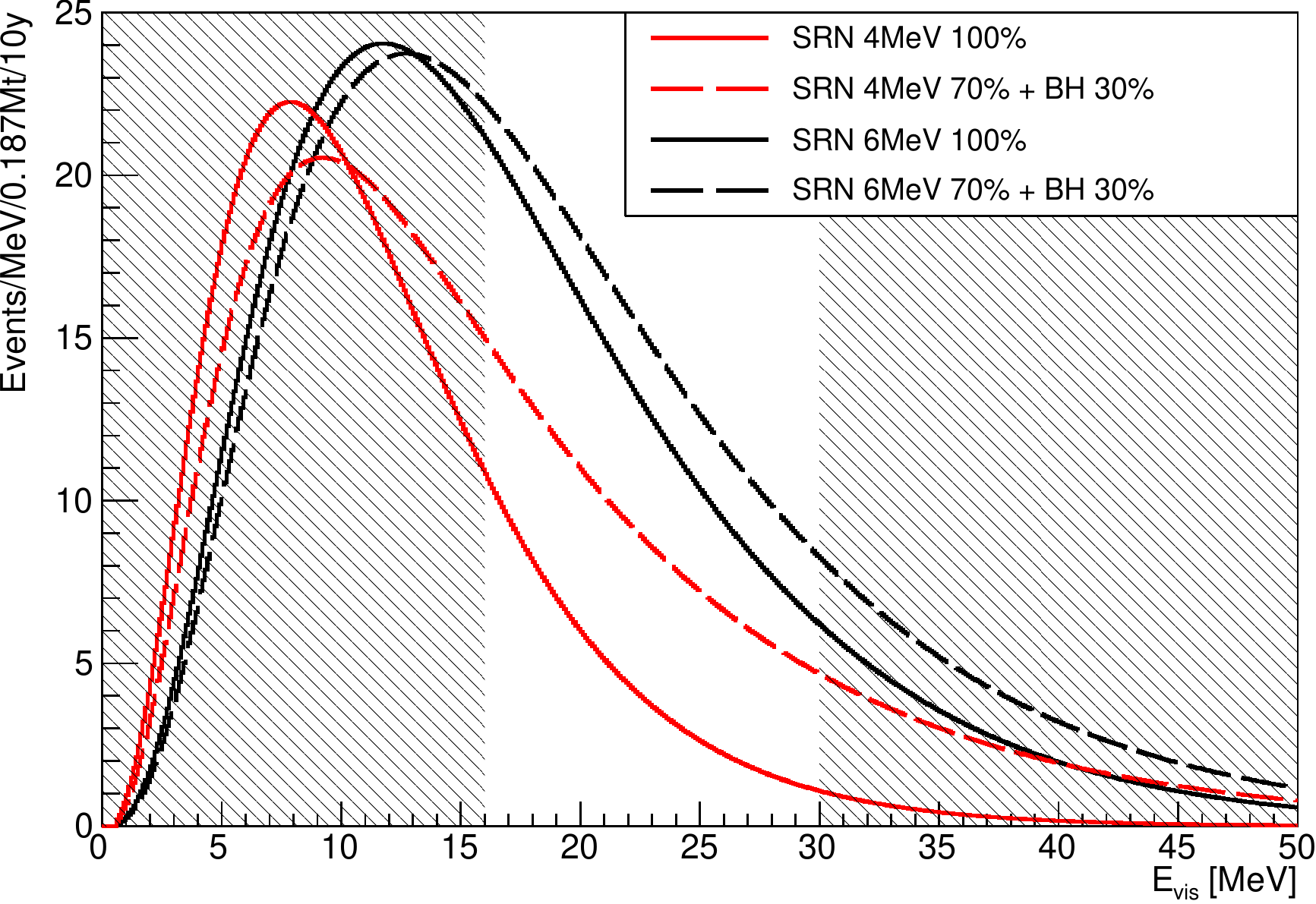}
  \end{center}
\vspace{-1cm}
  \caption{
	  The SRN signal expectations in Hyper-K 1 tank fiducial volume and 10 years measurement.
	  Black line shows the case of neutrino temperature in supernova of 6\,MeV, and red shows the case of 4\,MeV~\cite{Horiuchi:2008jz,Horiuchi:2017pv}.
	  Solid line corresponds to the case, in which all the core-collapse supernovae emits neutrinos with the particular energy.
	  Dashed line corresponds to the case, in which 30\% of the supernovae form black hole and emits higher energy neutrinos corresponding to the neutrino temperature of 8\,MeV.
	  Shaded energy region shows the range out of SRN search window at Hyper-K.
  \label{fig:srn-BH} }
\end{figure}

\if0
\begin{figure}[tbp]
  \begin{center}
    \includegraphics[width=8cm]{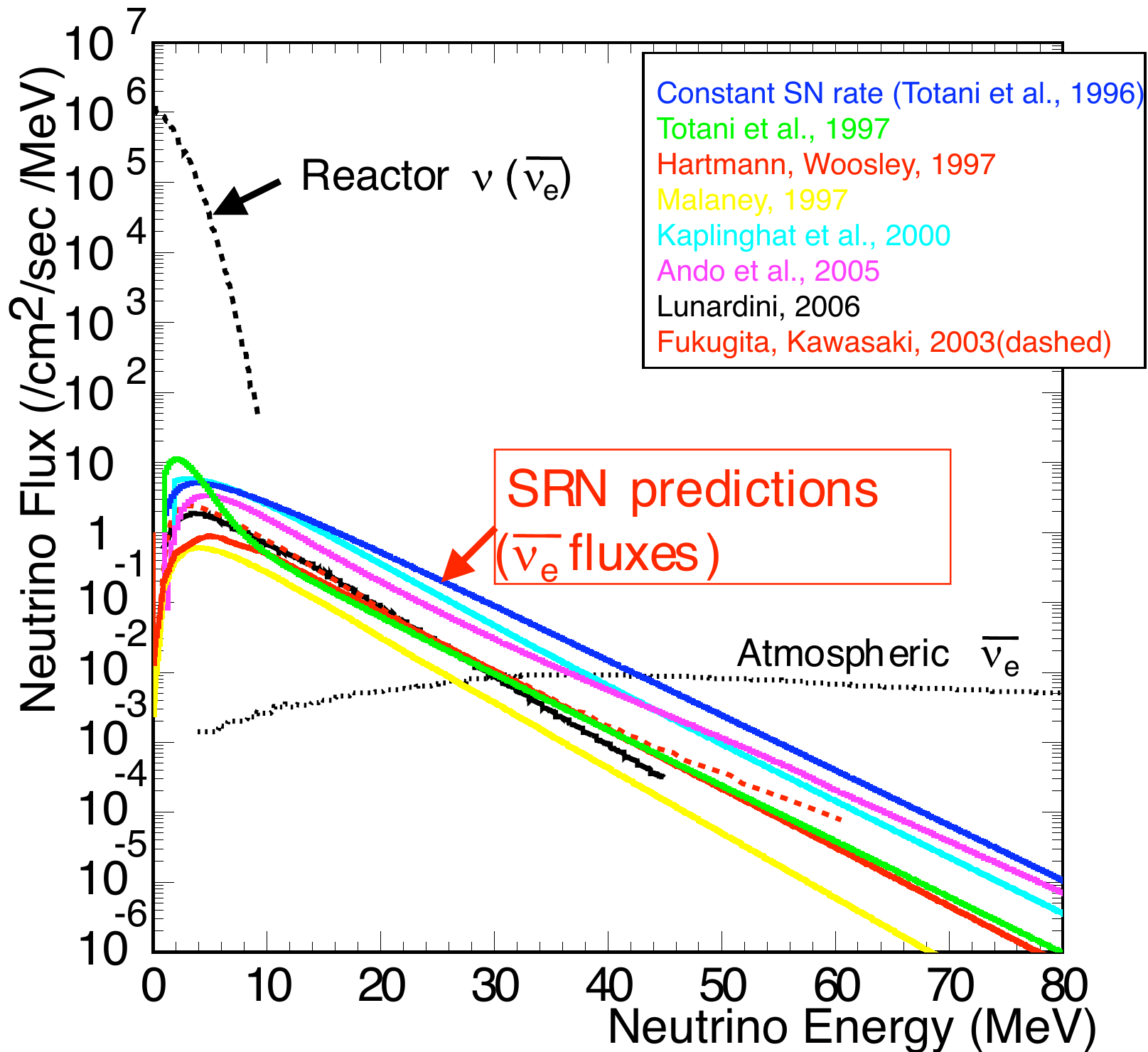}
  \end{center}
\vspace{-1cm}
  \caption{Predictions of the supernova relic neutrino (SRN) spectrum.
Fluxes of reactor neutrinos and atmospheric neutrinos are also shown.
	\cite{Totani:1995rg,Totani:1995dw,Hartmann:1997qe,Malaney:1996ar,Kaplinghat:1999xi,Ando:2002ky,Lunardini:2006pd,Fukugita:2002qw}
  \label{fig:sn-srn-prediction} }
\end{figure}
\fi
\begin{figure}[tbp]
  \begin{center}
    \includegraphics[width=8cm]{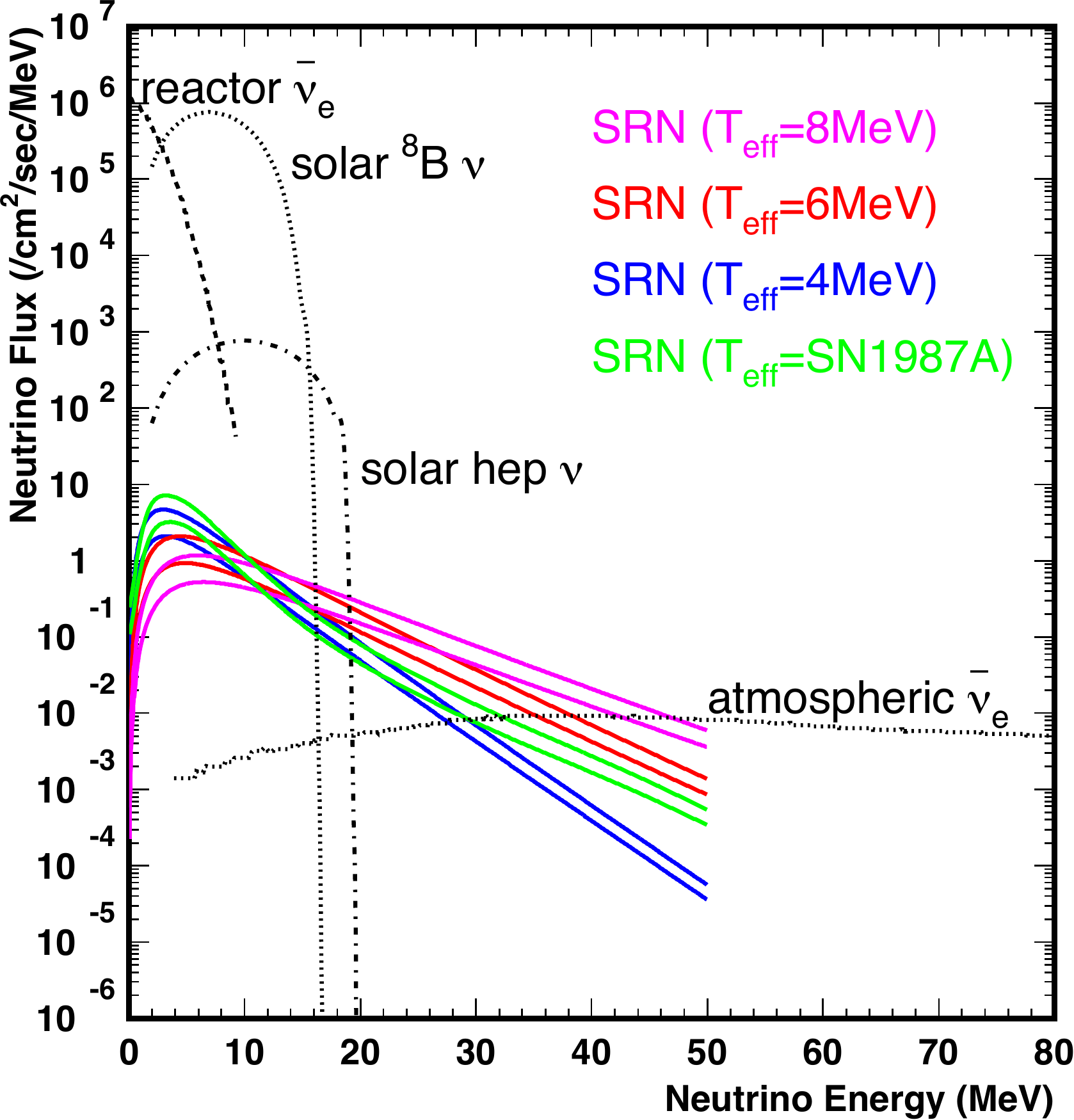}
  \end{center}
\vspace{-1cm}
  \caption{Predictions of the supernova relic neutrino (SRN) spectrum.
Fluxes of reactor neutrinos and atmospheric neutrinos are also shown
	\cite{Horiuchi:2008jz}.
  \label{fig:sn-srn-prediction} }
\end{figure}
Figure~\ref{fig:sn-srn-prediction} shows the SRN spectra predicted by
various models.  Although searches for SRN have been conducted at
large underground detectors, no evidence of SRN signals has yet been
obtained because of the small flux of SRN.  The expected inverse beta
($\bar{\nu}_e+p \rightarrow e^++n$) event rate at Super-Kamiokande is
0.8-5 events/year above 10\,MeV, but because of the large number of
spallation products and the low energy atmospheric neutrino background
(decay electrons from muons below Cherenkov threshold produced by
atmospheric muon neutrinos, the so-called invisible muon background),
SRN signals have not yet been observed at Super-Kamiokande.  In order
to reduce background, lower the energy threshold, individually
identify true inverse beta events by tagging their neutrons, and
thereby positively detect SRN signals at Super-Kamiokande, a project
to add 0.1\% gadolinium (Gd) to tank (the SK-Gd project, called
GADZOOKS! project previously) was proposed by J.F. Beacom and
M.R. Vagins~\cite{Beacom:2003nk}.
As the result of very active R\&D works, the detector upgrade to SK-Gd is planned in 2018.
The first observation of the SRN could be made by the SK-Gd project.
However, a megaton-scale detector is still desired to measure the spectrum of the SRN and to investigate the history of the universe because of its huge statistics as shown in Fig.~\ref{fig:srn-comp}.
Furthermore, Hyper-K could measure the SRN neutrinos at E =
16-30\,MeV, while the SK-Gd project concentrates on the energy of 10-20\,MeV.
These observation at a different energy region can measure the contribution of extraordinary supernova bursts on the SRN, e.g. black hole formation~\cite{Lunardini:2009,Horiuchi:2015HK}.
Figure~\ref{fig:srn-BH} shows the SRN signals in Hyper-K fiducial volume, with a different fraction of black hole formations.
Because the successful formation of black hole depends on the initial mass and metallicity of the progenitor,
the rate will provide information of the history about the formation of stars and their metallicity.

\begin{figure}[tbp]
  \begin{center}
    \includegraphics[width=7cm]{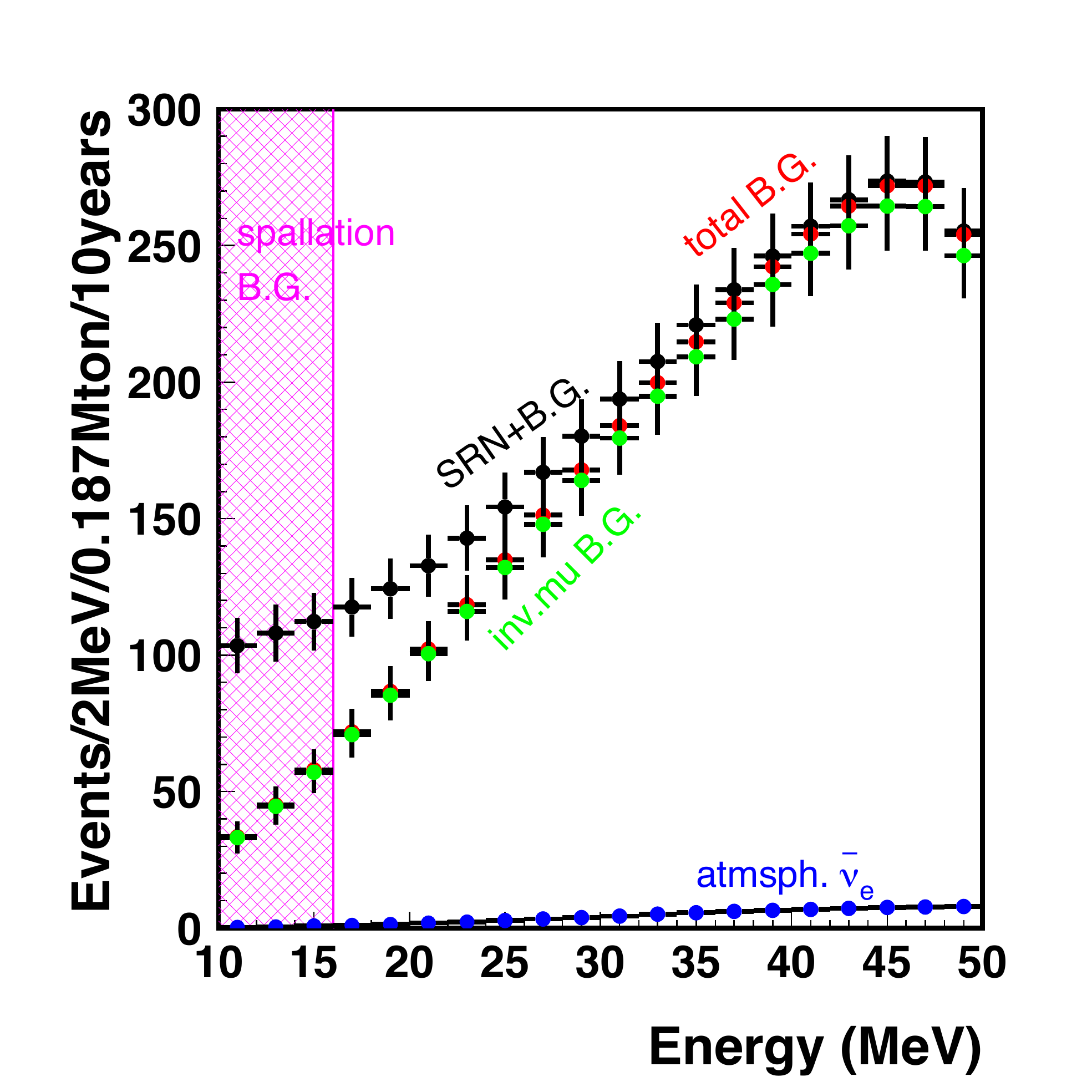}
	  \hspace{3mm}
    \includegraphics[width=7cm]{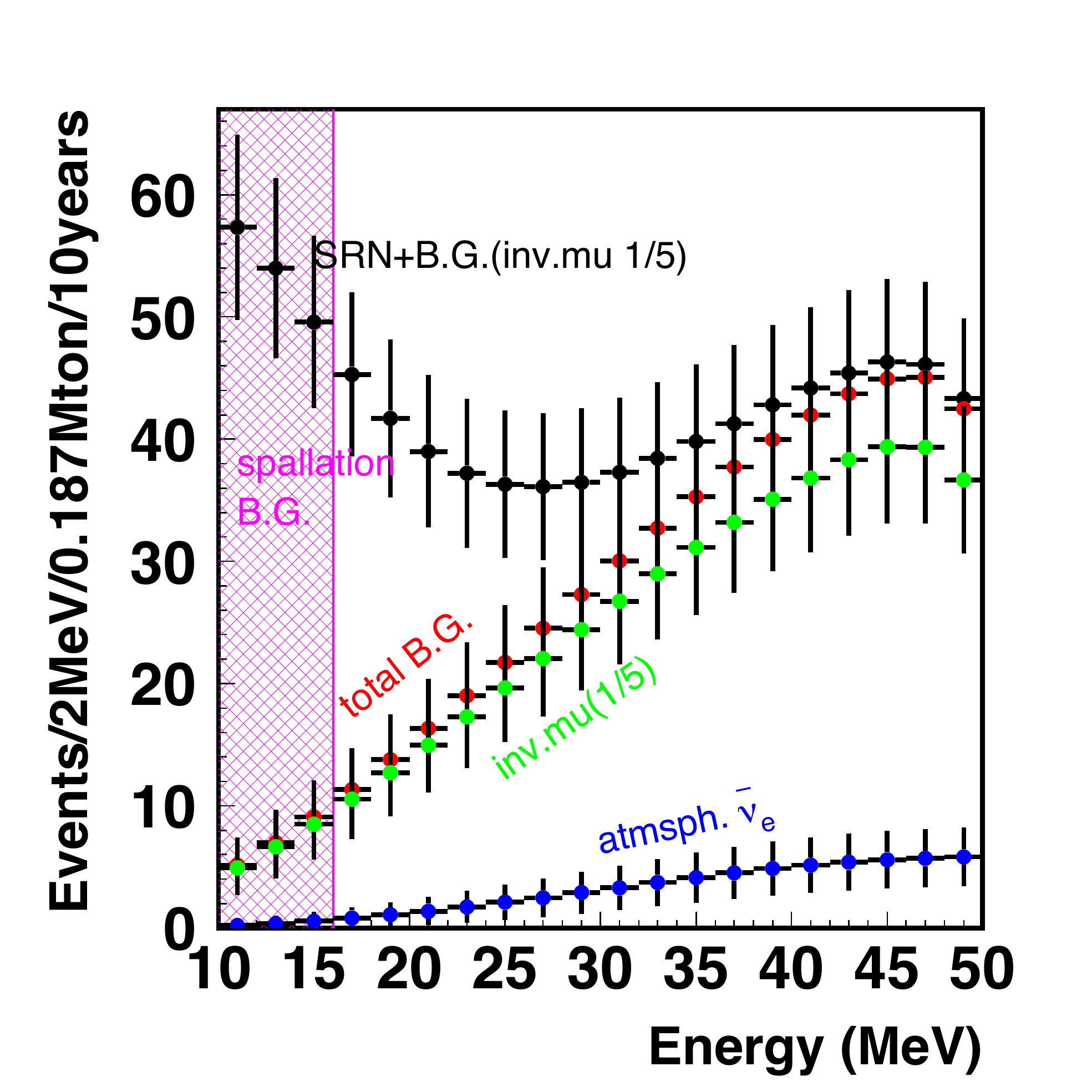}
  \end{center}
\vspace{-1cm}
  \caption{
	  Expected spectrum of SRN signals at Hyper-K with 10 years of livetime without tagging neutrons.
	  Left figure shows the case without tagging neutrons, assuming a signal selection efficiency of 90\%.
	  Neutron tagging were applied for right figure, with the tagging efficiency of 67\% and the pre-gamma cut for invisible muon background reduction.
	The black dots show the sum of the signal and the total background, while the red shows the total background.
Green and blue show background contributions from the invisible muon and 
$\nu_e$ components of atmospheric neutrinos.
The SRN flux prediction in~\cite{Ando:2003aa} is applied.
  \label{fig:srn-no-n-tag} }
\end{figure}

\begin{figure}[tbp]
	  \begin{center}
		      \includegraphics[width=8cm]{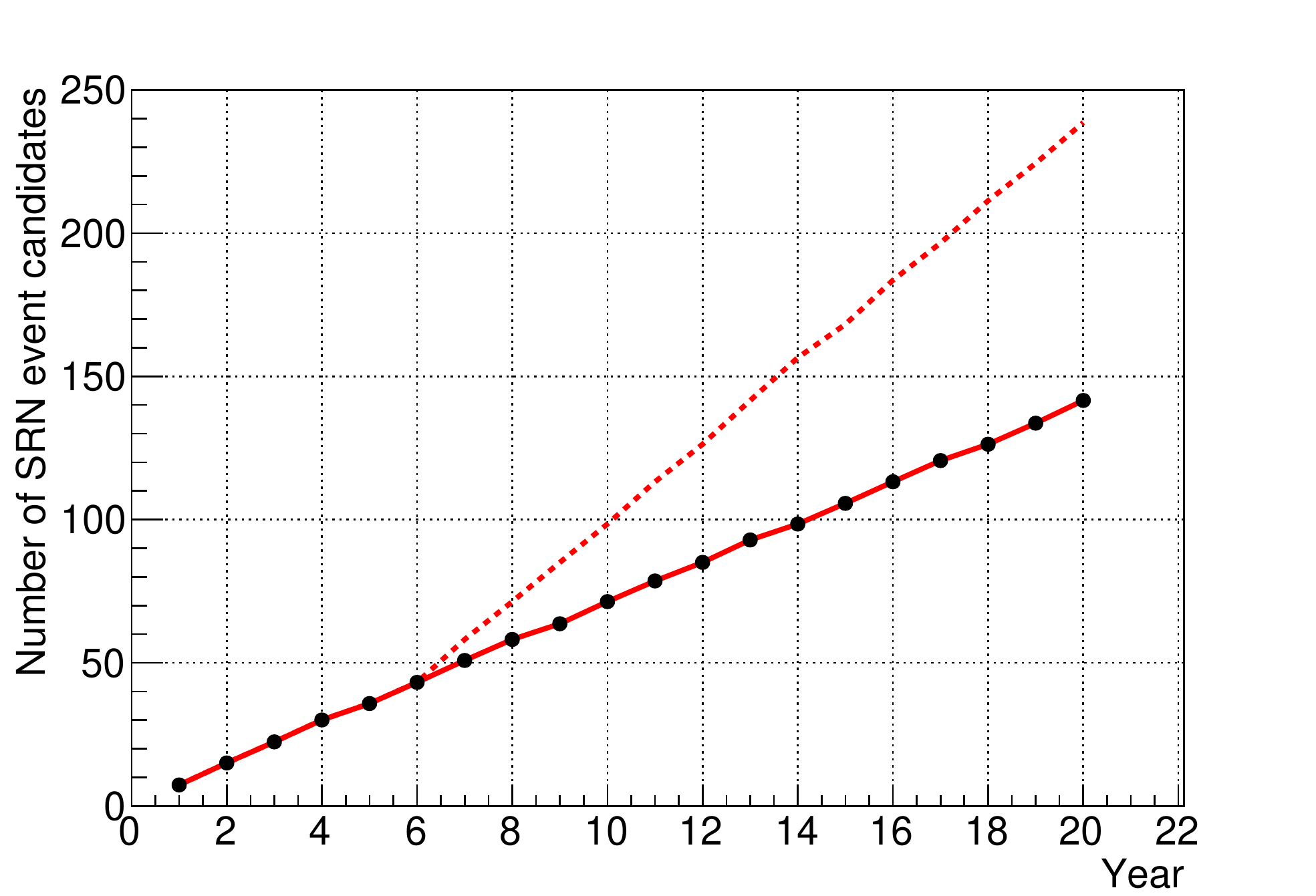}
			      \includegraphics[width=8cm]{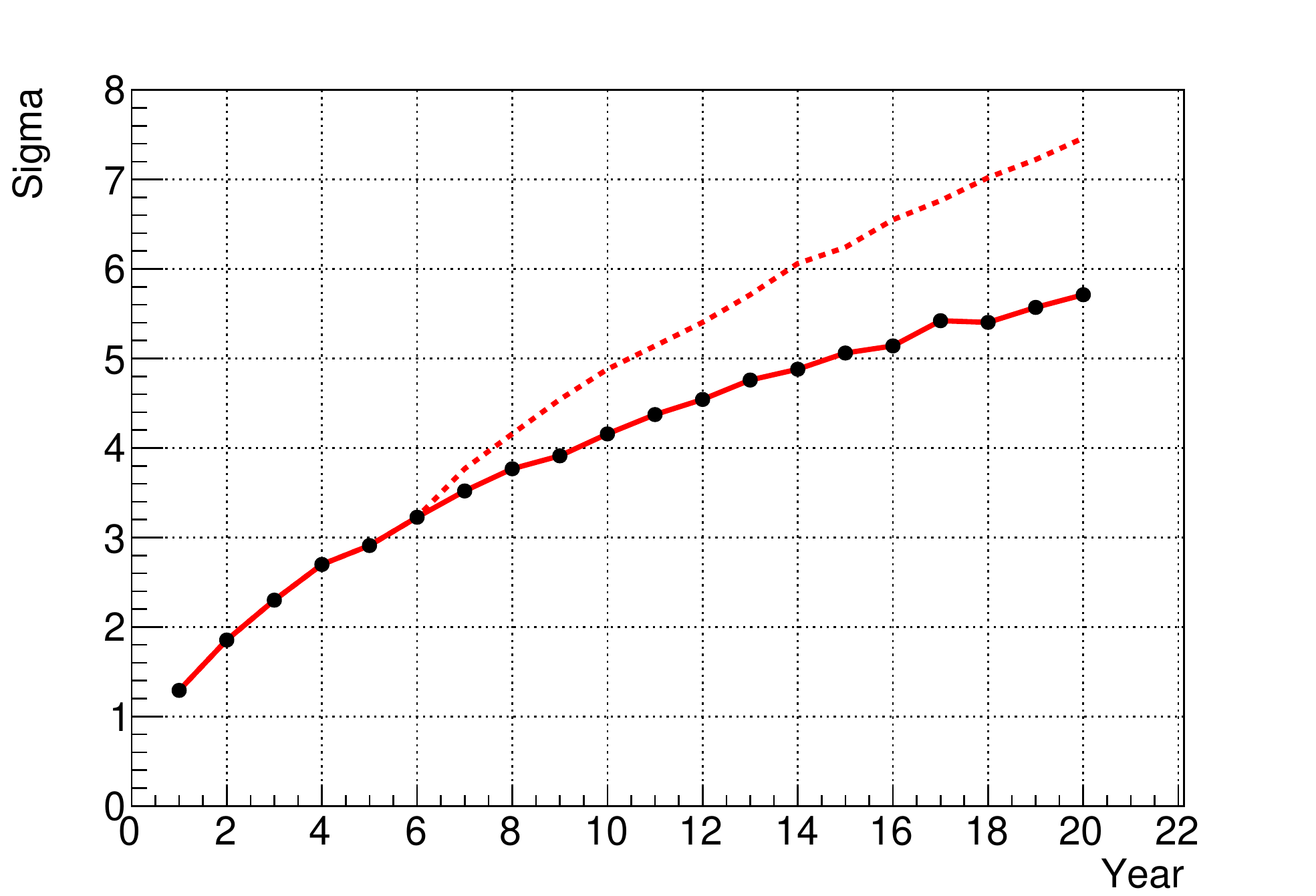}
				    \end{center}
					\vspace{-1cm}
					  \caption{The left (right) plot shows the number of observed SRN events (the discovery sensitivity) as a function of observation period.
						  Red solid line shows the continuous measurement with 1 tank and red dashed line shows the staging scenario, respectively.
						    \label{fig:srn-det} }
						\end{figure}

Figure~\ref{fig:srn-no-n-tag} shows expected SRN signals at Hyper-K with 10\,years' livetime.
Because of the high background rate below 20\,MeV from spallation products, the detection of SRN signals is limited to above $\sim$16\,MeV,
while above 30\,MeV the atmospheric neutrino backgrounds completely overwhelm the signal.
Considering the event selection efficiency after spallation product background reduction,
the expected number of SRN events in E = 16 to 30\,MeV is about 70 (140) after 10 (20) years observation with Hyper-K 1 tank.
The statistical error will be 17 (25) events, corresponding to an observation of SRN in the energy range 16 to 30\,MeV with 4.2 (5.7) $\sigma$ significance (fig.~\ref{fig:srn-det}). 
Here, we assumed the flux prediction described in ref.~\cite{Ando:2003aa} and neutron tagging using $n + p \rightarrow d + \gamma\,(2.2\,$MeV$)$ with the tagging efficiency of 70\%.

\begin{figure}[tbp]
  \begin{center}
    \includegraphics[width=8cm]{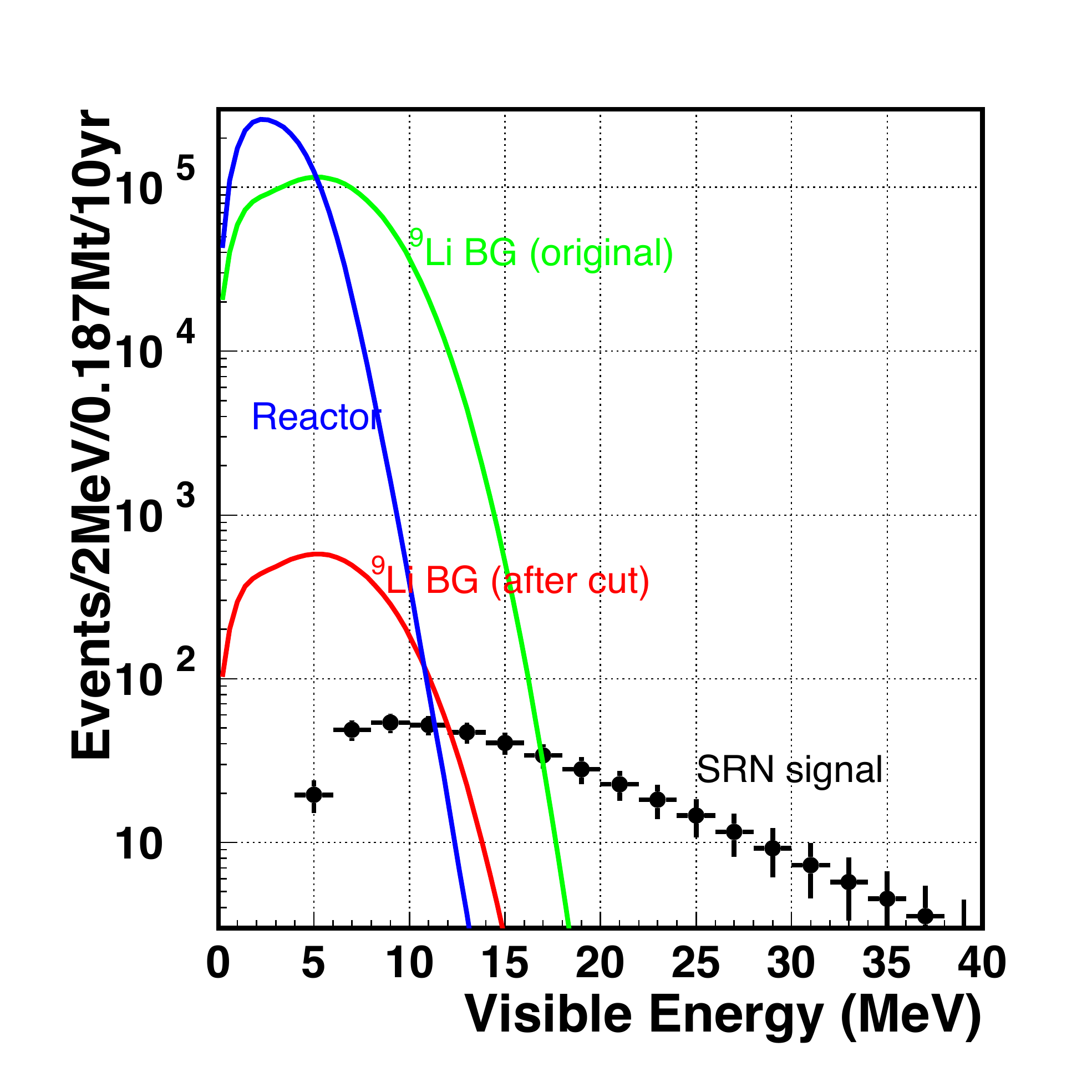}
  \end{center}
\vspace{-1cm}
  \caption{Green (red) curve shows the estimated $^9$Li production rate
before (after) applying cuts based on a correlation with cosmic ray muons.
Blue shows estimated background from reactor neutrinos at full intensity.
Black data points show expected SRN signal based on the
flux prediction in~\cite{Ando:2003aa}.
\label{fig:srn-bg}}
\end{figure}

It is still important to measure the SRN spectrum down to $\sim$ 10\,MeV in order to explore the history of supernova bursts back to the epoch of red shift (z) $\sim$1.
Therefore, in the following discussion of the expected SRN signal with
gadolinium neutron tagging, we assume that an
analysis with a lower energy threshold of $\sim$ 10\.MeV is possible.
Inverse beta reactions can be identified by coincident detection of both positron and delayed neutron signals, and requiring tight spatial and temporal correlations between them.
With 0.1\% by mass of gadolinium dissolved in the water, neutrons are captured on gadolinium with about 90\% capture efficiency; the excited Gd nuclei then de-excite by emitting 8\,MeV gamma cascades.
The time correlation of about 30\,$\mu$sec between the positron and the Gd(n,$\gamma$)Gd cascade signals, and the vertex correlation within about 50\,cm are strong indicators of a real inverse beta event.
Requiring both correlations (as well as requiring the prompt event to be Cherenkov-like and the delayed event to be isotropic) can be used to reduce background of spallation products by many orders of magnitude while also reducing invisible muon backgrounds by about a factor of 5.
The expected number of SRN events in the energy range of 10-30\,MeV is about 280 (390) with 10 years of live time with Gd-loaded Hyper-K 1 tank (staging 2 tanks).
\if0
\begin{figure}[tbp]
  \begin{center}
    \includegraphics[width=8cm]{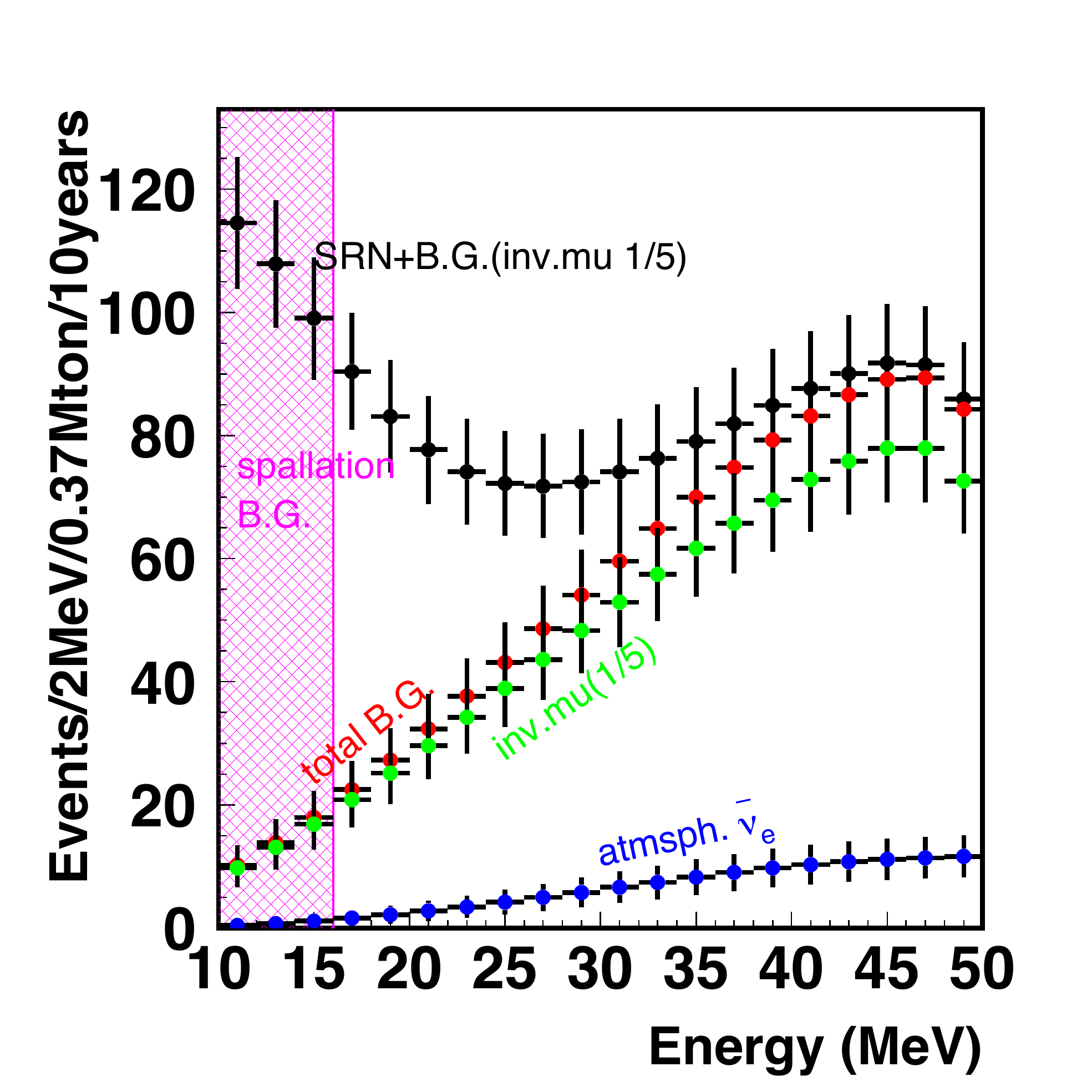}
  \end{center}
\vspace{-1cm}
  \caption{Expected spectrum of the SRN signals at Hyper-K with 10 years
of livetime. The black dots show signal+background (red component).
Green and blue show background contributions from the invisible muon and 
$\nu_e$ components of atmospheric neutrinos.
A SRN flux prediction~\cite{Ando:2003aa} was used, and 
a 67\% detection efficiency of 8\,MeV gamma cascades and
a factor of 5 reduction in the invisible muon background were assumed.
\label{fig:srn-megaton}}
\end{figure}
\fi
In addition, by comparing the
total SRN flux with optical supernova rate observations, a
determination of the fraction of failed (optically dark) supernova
explosions, currently unknown but thought to occur in not less than
5\% and perhaps as many as 50\% of all explosions, will be possible.
Possible backgrounds to the SRN search down to $\sim$ 10\,MeV are (1) accidental
coincidences with the spallation event, (2) spallation products with accompanying neutrons, and
(3) the resolution tail of the reactor neutrinos.  For (1) accidental
coincidences, the possible source of the prompt event is the
spallation products. By requiring time coincidence, vertex correlation
and energy and pattern of the delayed event, the accidental
coincidence rate can be reduced by a factor about 5 and can be below
the level of the expected SRN signal.
For (2) spallation products with accompanying neutrons, the only
possible spallation product is $^9$Li and an estimation by a Geant4
simulation is shown in Fig.~\ref{fig:srn-bg}.  Because of the short
half-life of $^9$Li ($\tau_{\frac{1}{2}}=0.18$~sec), a high rejection
efficiency of $\sim$99.5\% is expected.  With this expectation, the
$^9$Li background is less than the signal level above 12~MeV; this
could be lowered by further development of the background reduction
technique. For (3) the resolution tail of the reactor neutrinos, the
estimated background rate is about 100(20)/10\,years above 10\,MeV
(11\,MeV) as shown in Fig.~\ref{fig:srn-bg} with full reactor
intensity.

\newpage
\graphicspath{{physics-darkmatter/figures}}

    \subsection{Dark matter searches}\label{section:darkmatter}
    
       Despite overwhelming evidence for the existence of dark matter in the
universe, it has yet to been definitively identified or detected in
terrestrial experiments.  Based on astronomical observations it is
known to have gravitational interactions and lacks electromagnetic
interactions, but whether or not it carries the weak, strong, or some
other interactions remains an open question.  Assuming that dark
matter is a particle with weak interactions, several direct detection
experiments have looked for evidence of so-called WIMP elastic
scattering off target nuclei.  It is thought though
that the self-interaction or decay of dark matter particles bound in
strong gravitational potentials, such as those of the milky way galaxy
itself or even our sun, may produce standard model particles.
In particular, neutrinos may be produced either through direct annihilation or decay 
of dark matter particles or through the decays of heavier particles produced in these 
processes, and can be observed at Hyper-Kamiokande.
In this case the atmospheric neutrino sample described in
Section~\ref{section:atmnu} becomes an overwhelming background.
However, a potential dark matter signal is expected to have an angular
distribution peaked sharply near the center of the binding potential,
which at Hyper-Kamiokande would manifest as the direction towards the
galactic center, the sun, or the Earth.  By studying the angular distance to these
sources it is possible to extract the dark matter signal since the
atmospheric neutrino background is expected to be uniformly
distributed in this coordinate, particularly for neutrinos from the 
sun or galactic center.  It should be noted that the momentum
distribution of the signal is important for extracting the mass of the
dark matter candidate producing any observed neutrino event excess.

In the analyses below dark matter is assumed to produce standard model
particles such as $\chi\chi \rightarrow W^{+}W^{-}$,
$\tau^{+}\tau^{-}$, $b \bar b$, $\mu^{+} \mu^{-}$, and $\nu \bar \nu$
each with 100\% branching fraction.  The expected distribution of
signal neutrinos is simulated and searched for using the same analysis
samples described in previous sections.  Unlike other indirect
detection experiments, such as the neutrino telescopes,
Hyper-Kamiokande is expected to have superior sensitivity to lower
mass (below 100 GeV/$c^{2}$) WIMPS and the ability to resolve a signal
with both $\nu_{e}$ and $\nu_{\mu}$ components.  Hyper-K's expected
sensitivity to WIMP annihilation in the galactic center and the Earth
after a 1.9~Mton$\cdot$ year exposure is presented below.

       \subsubsection{Search for WIMPs at the Galactic Center}

Dark matter trapped in the gravitational potential of galaxies is said
to form a halo.  Halo models predict dark matter density distributions
that peak sharply near the center of a given galaxy and drop with
radial distance.  For example, for the Milky Way galaxy the expected
density distribution at the position of our solar system, $r =
8.5$~kpc, is roughly 1000 times smaller than that at the galactic
center in the NFW model~\cite{Navarro:1995iw}.  When simulating the
expected signal distribution expected at Hyper-Kamiokande a diffuse
dark matter profile following the full NFW density distribution is
assumed and accordingly the signal is expected to arise primarily from
the galactic center.

\begin{figure}[thb]
  \begin{center}
    \includegraphics[scale=0.55]{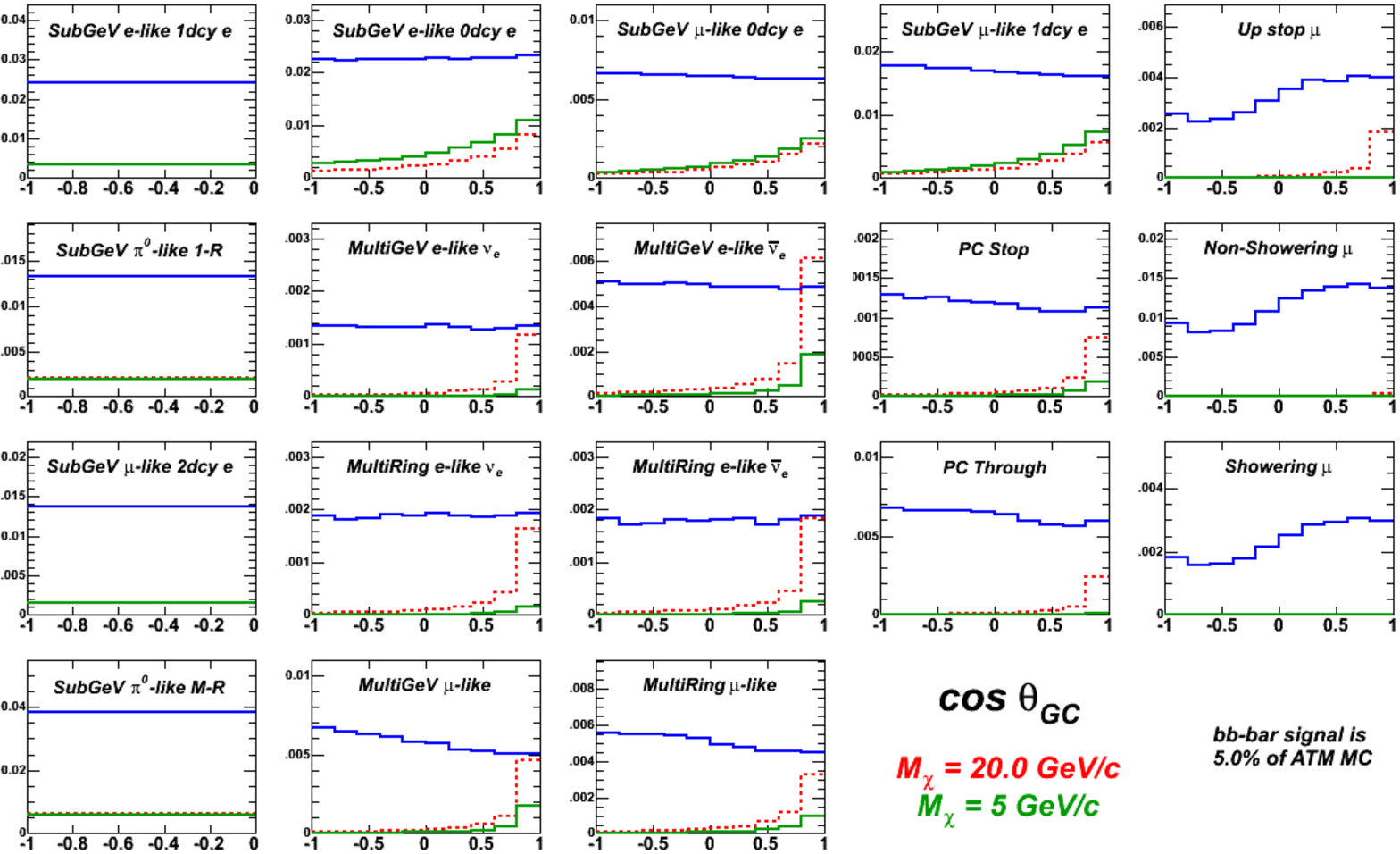}
  \end{center}
  \caption{Signal and background (blue) distributions used in the Hyper-K sensitivity study of dark matter 
           annihilating via $\chi\chi \rightarrow b \bar b$ at the galactic center. 
           Analysis samples are binned in $\mbox{cos}\theta_{gc}$, the direction to the galactic 
           center.
           Two WIMP hypotheses are shown: $m_{\chi} = 5 \mbox{GeV/c}^{2}$ in green and $m_{\chi} = 20 \mbox{GeV/c}^{2}$
           in red. 
           All distributions have been area normalized with the WIMP normalization taken to 5\% of the 
           background MC.
           }
\label{fig:gc_signal_demo}
\end{figure}

The differential neutrino flux arising from WIMP annihilation into the
standard model particles listed above is simulated using the DarkSUSY
package~\cite{Gondolo:2005we}, and the resulting spectrum adjusted to account for
oscillations on the way to the detector.  An independent set of
atmospheric neutrino MC is reweighted to this distribution to give a
reconstructed signal MC at Hyper-Kamiokande.  In computing the
sensitivity to an additional neutrino source, the analysis samples are
rebinned in momentum and $\mbox{cos}\theta_{gc}$, where $\theta_{gc}$
is the angle between the galactic center position (RA = 266$^{\circ}$,
Dec = -28$^{\circ}$) and the reconstructed direction.
Figure~\ref{fig:gc_signal_demo} shows the $\mbox{cos}\theta_{gc}$
distributions of the atmospheric neutrino background and
two WIMP hypotheses for each of the analysis samples.  During the fit
MC data sets without a WIMP signal are fit against a PDF built from
the atmospheric background MC plus a WIMP signal modified by a
normalization parameter, $\beta$.  Here the maximum value of $\beta$
that is consistent with the background-only model within errors is
used to compute the upper limit on the amount of additional neutrinos
from the galactic center allowed after a 3.8~Mton$\cdot$year exposure
of Hyper-K.

\begin{figure}[thb]
  \begin{center}
    \includegraphics[width=0.90\linewidth]{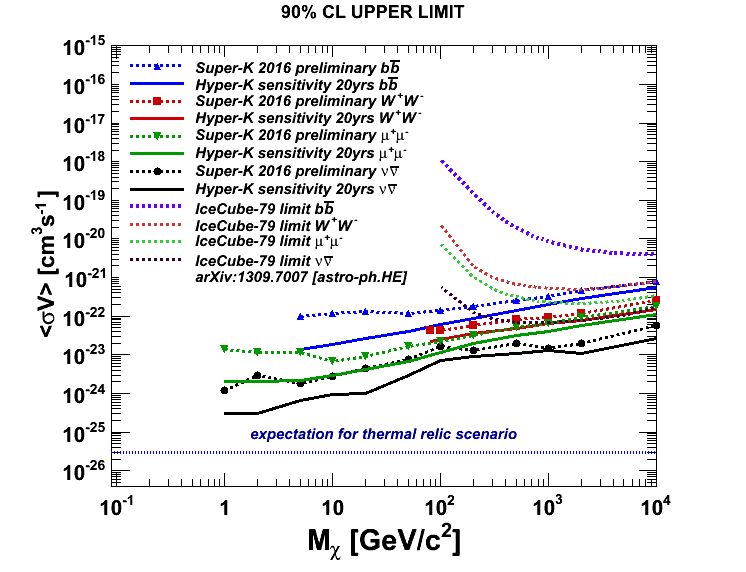}
  \end{center}
  \caption{Hyper-K's expected 90\% C.L. limit on the WIMP velocity averaged 
           annihilation cross section for several annihilation modes after 
           a 3.8~Mton$\cdot$year exposure overlaid with 
            limits from several experiments. 
            Limits are shown as a function of the dark matter mass. }
\label{fig:gc_wimps}
\end{figure}

Unlike direct detection experiments this search method is insensitive
to the WIMP-nucleon interaction cross section.  Instead limits can be
placed on the velocity averaged self-annihilation cross section,
$< \sigma \times v > $, where $v$ is the assumed velocity distribution
of WIMPs in the halo.  Figure~\ref{fig:gc_wimps} shows the expected
sensitivity of Hyper-K to WIMP annihilations at the galactic center.
The analysis makes use of potential signals
in both $\nu_{e}$- and $\nu_{\mu}$-enriched samples across the entire
energy range of atmospheric neutrinos and their energy and directional distributions
contribute to the sensitivity.  
Hyper-Kamiokande's sensitivity to the WIMP velocity
averaged self-annihilation cross section is expected to exceed that of
Super-Kamiokande's limits by factors of three to ten, depending on 
the assumed WIMP mass and annihilation channel.
Unlike other experiments,
Hyper-K's ability to reconstruct down to $O(100)$~MeV neutrino
interactions gives it unparalleled sensitivity to WIMPs with masses
less than $\sim$ 100 GeV/$c^{2}$.

\subsubsection{Search for WIMPs from the Earth}

\begin{figure}[thb]
  \begin{center}
    \includegraphics[width=0.9\textwidth]{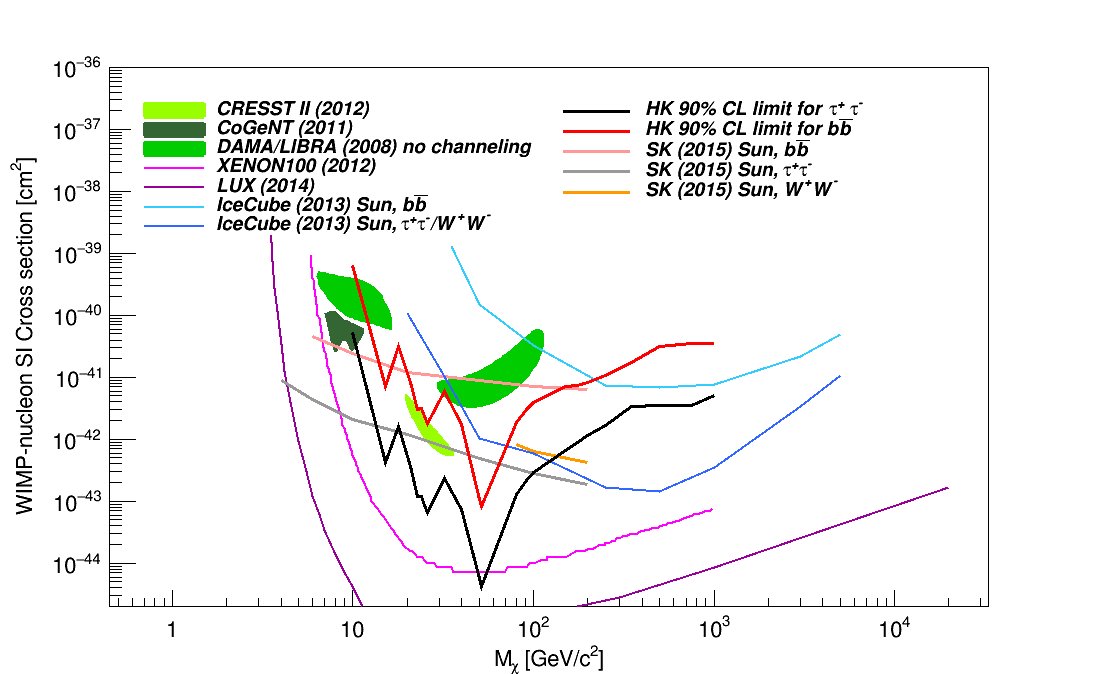}
  \end{center}
  \caption{The 90\% C.L. upper limits on the
 spin-independent WIMP-nucleon scattering cross section based on a search from
WIMP-induced neutrinos coming from the center of the earth for several annihilation channels.
Limits (lines) and allowed regions (hatched regions) from other experiments are also shown.
Results from Super-Kamiokande assuming annihilations in the sun are taken from~\cite{Choi:2015ara}.}
\label{fig:earth_wimps}
\end{figure}

WIMPs bound in the halo of the galaxy may also become gravitationally
trapped within the Earth (or sun) after losing energy via scattering processes
with nuclei in its interior.  If these then pair annihilate and
produce neutrinos, they will escape the core of the Earth and be detectable at
Hyper-Kamiokande.  Assuming that the rate of WIMP capture within the
sun is in equilibrium with the annihilation rate, measurements of the
WIMP-induced neutrino flux can be directly translated into
measurements of the WIMP-nucleon scattering cross section without the
need to measure the self-annihilation cross section.  
Since the Earth is composed of heavy nuclei (relative to hydrogen) 
it is further possible to study WIMP interactions that are 
not coupled to the nuclear spin (spin independent, SI).

In the analysis below the local dark matter density is assumed to be
0.3 GeV/$\mbox{cm}^{3}$ with an RMS velocity of 270~km/s.  The
rotation of the solar system through the halo is taken to be 220~km/s.
Signal MC has been generated by reweighting atmospheric neutrino MC
events to spectra produced by the WIMPSIM package~\cite{Blennow:2007tw}, which accounts for
the passage of particles through terrestrial matter.  Oscillation between
flavors as the neutrinos travel from the Earth core to the detector are
included.  An independent set of atmospheric neutrino MC is used to
model the background.

The search for WIMPs bound and annihilating at the center of the Earth 
proceeds along similar lines as the search for events from the
galactic center, though events are now binned in momentum and
$\mbox{cos}\theta_{zenith}$, the zenith angle of the reconstructed 
lepton direction relative to Hyper-K. 
In these coordinates the atmospheric neutrino background takes its 
characteristic shape while the WIMP signal MC is peaked sharply in the direction 
of the Earth core; the most upward-going bin.
Limits on the WIMP-induced neutrino flux are translated into
limits on the WIMP-nucleon SI cross sections using the DarkSUSY
simulation.  Hyper-K's sensitivity with a 1.9 Mton$\cdot$year exposure
shown in Figure~\ref{fig:earth_wimps}.  
The plot shows the sensitivity to the WIMP-nucleon SI cross section
for masses $m_{\chi} > 4 \mbox{GeV/c}^{2}$ compared to allowed regions
(shown as hatched spaces) and limits (shown as lines) from current
experiments.  These limits have been produced assuming WIMPs have only
SI interactions and have been estimated for
$\chi\chi \rightarrow W^{+}W^{-}, b \bar b,$ and $\tau^{+}\tau^{-}.$
Hyper-K's is expected to produce limits a factor of $3\sim 4$ times
stringent than Super-K if no WIMP signal is seen.
Further, current hints for a positive SI
dark matter signal~\cite{Bernabei:2008yi,Savage:2009mk,Agnese:2013rvf,Angloher:2011uu,Aalseth:2010vx}, 
can be probed completely by Hyper-K's $\tau\tau$ channel.

It is worth noting that in both the Earth and galactic center analyses
no improvement in systematic errors beyond Super-K's current
understanding has been assumed.  At Hyper-K the statistical
uncertainty in the data is small enough that increases in sensitivity
relative to Super-K is limited by systematic errors in the atmospheric
neutrino flux and cross section model.  Due to the relatively high energy
of the signal events and the expected directional resolution of the
detector, systematic errors in the detector response, while currently
less well known, are expected to be less significant.  While it is
uncertain how the flux and cross section model will evolve in the
future, improved modeling will translate directly into better
sensitivity to WIMP-induced neutrinos at Hyper-K.

\newpage
\graphicspath{{physics-astronu/figures}}

    \subsection{Other astrophysical neutrino sources}\label{section:astro}
   
\subsubsection{Solar flare}
Solar flares are the most energetic bursts which occur in the solar
surface.  Explosive release of energy stored in solar magnetic fields
is caused by magnetic reconnections, resulting in plasma heating,
particle accelerations, and emission of synchrotron X-rays or charged
particles from the solar surface.  In a large flare, an energy of
10$^{33}$ ergs is emitted over 10's of minutes, and the accelerated
protons can reach energies greater than 10 GeV.  Such high energy
protons can produce pions by nuclear interactions in the solar
atmosphere. Evidence of such nuclear interactions in the solar
atmosphere are obtained via observations of solar neutrons, 2.2 MeV
gamma rays from neutron captures on protons, nuclear de-excitation
gamma rays, and possible $>100$ MeV gamma rays from neutral pion
decays.  Thus, it is likely that neutrinos are also emitted by the
decay of mesons following interactions of accelerated particles.
Detection of neutrinos from a solar flare was first discussed in
1970's by R.Davis~\cite{dav, bacall_sf}, but no significant signal has
yet been found~\cite{aglietta,hirata}. There have been some estimates
of the number of neutrinos which could be observed by large water
Cherenkov detectors~\cite{fargion,kocharov}. According
to \cite{fargion}, about 6-7 neutrinos per tank will be observed at
Hyper-Kamiokande during a solar flare as large as the one in 20
January 2005, although the expected numbers have large uncertainties.
Therefore, regarding solar flares our first astrophysics goal is to
discover solar flare neutrinos with Hyper-K.  This will give us
important information about the mechanism of the particle acceleration
at work in solar flares.

\subsubsection{Gamma-Ray Burst Jets and Newborn Pulsar Winds}
Gamma-ray bursts (GRBs) are the most luminous astrophysical phenomena
with the isotropically-equivalent gamma-ray luminosity,
$L_\gamma\sim{10}^{52}~{\rm erg}~{\rm s}^{-1}$, which typically occur
at cosmological distance.  Prompt gamma rays are observed in the MeV
range, and their spectra can be fitted by a smoothened broken power
law~\cite{grbrev}.  The prompt emission comes from a relativistic jet
with the Lorentz factor of $\Gamma\sim{10}^{2}-{10}^{3}$, which is
presumably caused by a blackhole with an accretion disk or
a fast-spinning, strongly-magnetized neutron star.  Observed gamma-ray
light curves are highly variable down to $\sim1$~ms, suggesting
unsteady outflows.  However, GRB central engines and their radiation
mechanism are still unknown, and GRBs have been one of the biggest
mysteries in high-energy astrophysics.

Internal shocks are naturally expected for such unsteady jets, and the
jet kinetic energy can be converted into radiation via shock
dissipation.  In the ``classical'' internal shock
scenario~\cite{rm94}, observed gamma rays are attributed to
synchrotron emission from non-thermal electrons accelerated at
internal shocks.  It has been suggested that GRBs may also be
responsible for ultrahigh-energy cosmic rays (CRs), and TeV-PeV
neutrinos have been predicted as a smoking gun of CR acceleration in
GRBs~\cite{grbnu1}.  One of the key advantages in GRB neutrino
searches is that atmospheric backgrounds can significantly be reduced
thanks to space- and time-coincidence, but no high-energy neutrino
signals correlated with GRBs have been found in any neutrino detector
including Super-K~\cite{skgrb} and IceCube~\cite{icecubegrb}.

On the other hand, the recent theoretical and observational progress
has suggested that the above classical scenario has troubles in
explaining observational features such as the low-energy photon
spectrum.  Alternatively, the photospheric scenario, where prompt
gamma rays are generated around or under the ``photosphere'' (where
the Compton scattering optical depth is unity), has become more
popular~\cite{photosphere1,photosphere2,photosphere3}.  Indeed,
observations have indicated a thermal-like component in GRB
spectra~\cite{fermigrb,fermigrb2}.  Energy dissipation may be caused
by inelastic nucleon-neutron
collisions~\cite{neutron1,neutron2,neutron3}, and neutrons can
naturally be loaded by GRB central engines either
blackhole-accretion-disk system or strongly-magnetized neutron star.
Then, quasi-thermal GeV-TeV neutrino emission is an inevitable consequence of such inelastic nucleon-neutron collisions~\cite{mur+13}.  Since neutrinos easily leave the flow, predictions for these neutrinos are insensitive to details of gamma-ray spectra.  
Hyper-K will enable us to search these quasi-thermal GeV-TeV neutrinos
from GRB jets, and it also has an advantage over 
IceCube (that is suitable for higher-energy $>10$-$100$~GeV
neutrinos).  The GeV-TeV neutrino detection is feasible if a GRB
happens at $\lesssim100$~Mpc, and successful detections should allow
us to discriminate among prompt emission mechanisms and probe the jet
composition, leading to a breakthrough in understanding GRB physics.
However, GRBs are rare astrophysical phenomena, so we have little
chance to expect such a nearby bright burst in the next 10-100 years.
Much more promising targets as high-energy neutrino sources would be
energetic supernovae driven by relativistic
outflows~\cite{mw01,rmw04,ab05,mi13}.  A significant fraction of GRB
jets may fail to puncture their progenitor star, and photon emission
from the jets can easily be hidden.  Indeed, theoretical studies
revealed the existence of conditions for a jet to make a successful
GRB.  ``Choked jets'' or ``failed GRBs'' are naturally predicted when
the jet luminosity is not sufficient or the jet duration is too short
or the progenitor is too big.  The choked jets can explain
trans-relativistic supernovae or low-luminosity GRBs, which show
intermediate features between GRBs and supernovae~\cite{sod+06}.

The neutrino event rate expected in Hyper-K depends on the
isotropically-equivalent dissipation energy ${\mathcal E}_{\rm
diss}^{\rm iso}$, Lorentz factor $\Gamma$, and distance $d$.  For
${\mathcal E}_{\rm diss}^{\rm iso}={10}^{53}~{\rm erg}~{\rm s}^{-1}$,
$\Gamma=10$ and $d=10$~Mpc, the characteristic energy of quasi-thermal
GeV-TeV neutrinos is $E_\nu^{\rm qt}\sim3$~GeV~\cite{mur+13}.  The
neutrino-nucleon cross section for the charged-current interaction at
1~GeV is $\sim0.6\times {10}^{-38}~{\rm cm}^2$ (averaged over $\nu$
and $\bar{\nu}$), so the effective area of Hyper-K is
$\sim2\times{10}^{-3}~{\rm cm}^2$ at 1~GeV.  Then, Hyper-K will be
able to detect $\sim5$ signal events from such a jet-driven supernova.
Successful detections enable us to probe jet physics that cannot be
directly studied by electromagnetic observations.  Neutrinos enable us
to understand how jets are accelerated and what the jet composition
is, and will give us crucial keys to the mysterious connection between
GRBs and energetic supernovae~\cite{tho+04}.  Also, in principle,
matter effects in neutrino oscillation could be
investigated~\cite{fs08,rs10}.  Moreover, we will able to study how CR
acceleration operates in dense radiation environments inside a GRB
progenitor star.  Whether CRs are accelerated or not depends on
properties of shocks. The conventional shock acceleration mechanism
can effectively operate only if the shock is radiation-unmediated
collisionless~\cite{mi13}.  On the other hand, when the shock is
mediated by radiation, the so-called neutron-proton-converter
acceleration mechanism can work efficiently~\cite{npc,kas+13}, which
boosts the energy of quasi-thermal neutrinos produced by
nucleon-neutron collisions~\cite{mur+13}.

As discussed above, relativistic outflows containing neutrons should
naturally lead to GeV-TeV neutrino production, but the outflows do not
have to be jets.  Another interesting case may be realized when a
supernova explosion leaves a fast-spinning neutron star.  Neutrons are
loaded in the proto-neutron star wind via neutrino heating.  Around
the base of the outflow, the particle density is so high that neutrons
and ions are tightly coupled via elastic collisions.  Neutrons should
be accelerated together with ions as the Poynting-dominated pulsar
wind is accelerated.
Once the outflow becomes relativistic enough to exceed the pion-production threshold, inelastic collisions naturally occur as the main dissipation process of relativistic neutrons.  
The neutrons then interact with the material decelerated by the shock
and possibly with the overlying stellar material, producing 0.1-1~GeV
neutrinos~\cite{mur+14}.  Detecting this signal would probe the
otherwise completely obscured process of jet acceleration and the
physics of rotating and magnetized proto-neutron star birth during the
core collapse of massive stars.  Hyper-K may expect $\sim20-30~({\mathcal
E}_{\nu}^{\rm iso}/{10}^{48}~{\rm erg})$ events for a core-collapse
supernova at 10~kpc.  In addition, Hyper-K could also allow us to see
$\sim10-100$~MeV neutrinos through the $\bar{\nu}_ep\rightarrow e^+n$
channel.  However, detection of these lower energy neutrinos would be
more difficult because of the smaller cross sections at lower energies
and because the signal may be buried in the exponential tail of
thermal MeV neutrinos from the proto-neutron star.

To detect high-energy neutrino signals from hidden GRB jets or newborn
pulsar winds embedded in supernovae, it is crucial to reduce
atmospheric backgrounds using space and time coincidence, so
information at other wavelengths is relevant.  The atmospheric
neutrino background at GeV energies is $\approx1.3\times{10}^{-2}~{\rm
GeV}~{\rm cm}^{-2}~{\rm s}^{-1}~{\rm sr}^{-1}$ for $\nu_e+\bar{\nu}_e$
and $\approx2.6\times{10}^{-2}~{\rm GeV}~{\rm cm}^{-2}~{\rm
s}^{-1}~{\rm sr}^{-1}$ for $\nu_\mu+\bar{\nu}_\mu$,
respectively~\cite{hon+11}.  We may take the time window of $t_{\rm
thin}\sim10-100$~s after the explosion time that is measurable with
MeV neutrinos or possibly gravitational waves.  The localization is
possible by follow-up observations at x-ray, optical, and infrared
bands.  The atmospheric background flux in the typical angular and
time window is $\sim2\times{10}^{-3}~{\rm erg}~{\rm cm}^{-2}$, which
can be low enough for a nearby supernova.

Note that it is critical to have large volume detectors for the
purpose of detecting GeV-TeV neutrinos.  The present Super-K and
liquid scintillator detectors such as JUNO and RENO-50 are too small
to detect high-energy signals from astrophysical objects especially if
extragalactic, and much bigger detectors such as Hyper-K and PINGU are
necessary to have a good chance to hunt high-energy neutrinos from
GRBs and energetic supernovae.  Because of the atmospheric background,
sensitivities above GeV energies are typically essential but searches
for neutrinos below $\sim1$~GeV could also be useful for nearby
events.

\subsubsection{Neutrinos from gravitational-wave sources}

100 years after the prediction by Einstein, gravitational waves (GWs)
have been detected by advanced-LIGO in 2015~\cite{YS-ref1}. This
has allowed us to conduct multi-messenger observations of astrophysical
objects via multiple signals, i.e., electromagnetic (EM) waves (from
radio to gamma-rays), neutrinos (from MeV to PeV) and GWs
(kHz). Some strong GW emitters are also expected to be strong sources of
neutrinos, e.g. supernovae and gamma-ray bursts. The searches
for neutrino counterparts to GW sources have been
performed~\cite{YS-ref2}, including a search done by Super-Kamiokande
collaboration~\cite{YS-ref3}, but there has been no clear counterpart
found.
This is consistent with the theoretical prediction because these GW
sources (GW150914 and GW151226) are binary black-hole mergers.
Although binary black-hole mergers as GW sources are not expected to
generate other detectable signals, the mergers that contain at least
one neutron star (i.e. black hole-neutron star binary or neutron
star-neutron star binary) could emit other signals, includeing $\sim
10^{53}$ erg of neutrinos~\cite{YS-ref4}. Indeed, a number of EM
counterparts associated with GW170817 (neutron-star binary merger)
were detected, including GRB170817A~\cite{YS-ref5,YS-ref6}. A search
for neutrinos by Super-Kamiokande was also reported~\cite{YS-ref7}, which did
not detect any coincident neutrino. Hyper-Kamiokande has the potential to
detect thermal neutrinos from nearby ($\lesssim 10$Mpc) neutron
star meger events.

Depending on the event rates, these objects would contribute to the
relic neutrino spectrum. The central engine of gamma-ray bursts are
also candidates of strong emitters of neutrinos and GWs.  It is
evident that there are ultra-relativistic jets which are driven by the
central engine. However, the mechanism that generates the jet is
still unclear. If this jet is driven by neutrino annihilation, which
is one of the promising scenarios, concurrent observations of
neutrinos and GWs will be important probe of the very central part of
the violent cosmic explosions at Hyper-Kamiokande era~\cite{YS-ref8}.

\newpage
\graphicspath{{physics-geophys/figures}}

    \subsection{Neutrino geophysics}\label{section:radiography}

The chemical composition of the Earth's core is one of the most
important properties of the planet's interior, because it is deeply
connected to not only the formation and evolution of the
Earth~\cite{Allegre1995} itself but also to the origin of the
geomagnetic field~\cite{Fearn1981}.  While paleomagnetic evidence
suggests that the geomagnetic field has existed for roughly three
billion years, it is known that a core composed of iron alone could
not sustain this magnetic field for more than 20,000 years.
Explaining the continued generation of the geomagnetic field as well
as its other properties requires knowledge of composition of the core
matter.  Based on seismic wave velocity measurements and the
composition of primordial meteorites the composition of the core is
presumed to be an iron-nickel alloy that additionally includes light
elements, such as oxygen, sulfur, or silicon~\cite{McDonough1995}.
However, since no sample of the Earth's mantle has even been acquired,
let alone a sample of the core, the composition of the latter,
particularly its light element a abundance, remains highly uncertain.
Since the deepest wellbore to date has a depth of
14~km~\cite{Popov1999}, and the depth to the outer core is 2900~km it
is unlikely that a core sample can be obtained within this century.
As a result, addition methods of determining the chemical composition
of the core are essential to understanding the Earth and its magnetic
field.

\begin{figure}[thb]
  \begin{center}
    \includegraphics{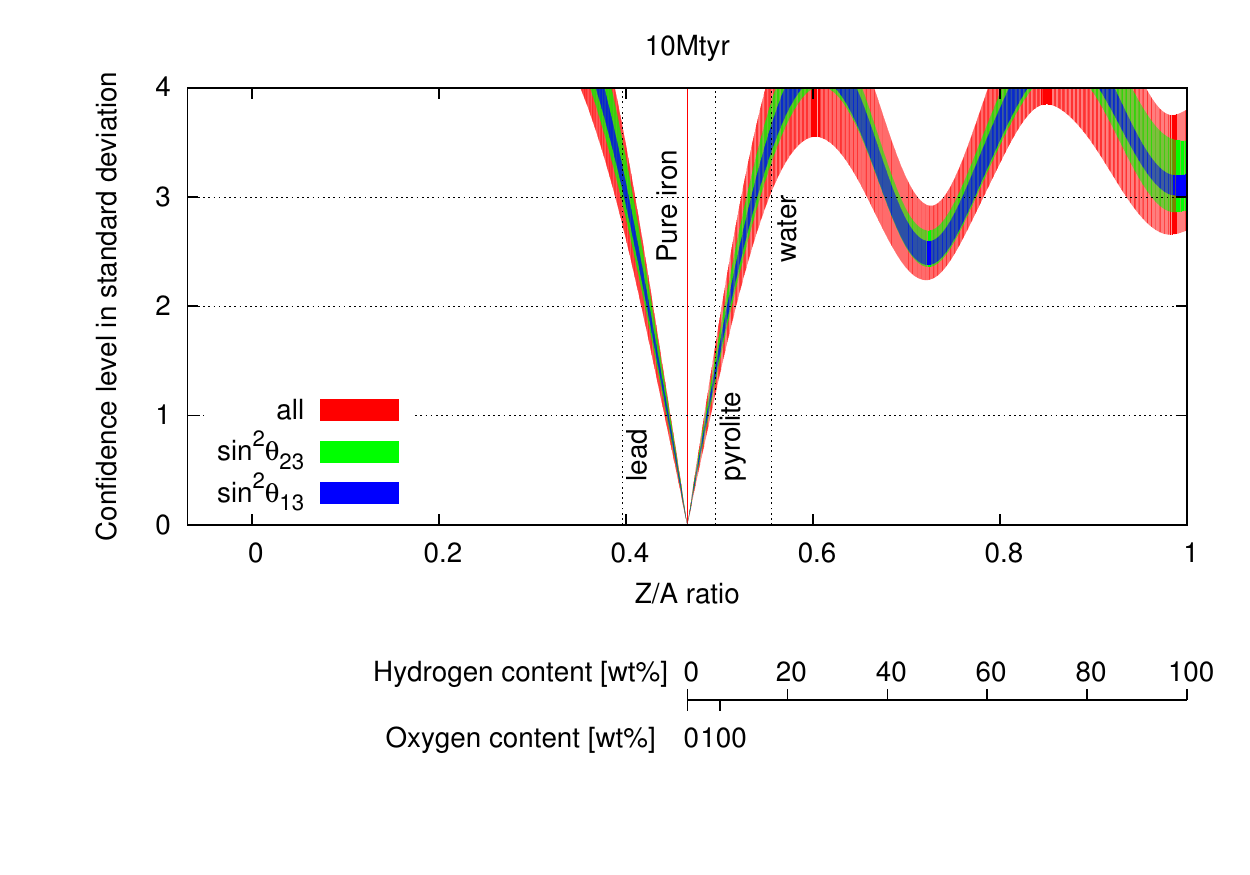}
  \end{center}
  \caption{Constraints on the proton to nucleon ratio of the Earth's outer core 
           for a 10 Mton year exposure of Hyper-K to atmospheric neutrinos. 
           Colored bands indicate the effect of present uncertainties in the 
           neutrino mixing parameters.}
\label{fig:geophys_za}
\end{figure}

As discussed in Section~\ref{section:atmnu}, the oscillation
probability of atmospheric neutrinos depends on not only the various
mixing angles, the neutrino mass differences, and CP-violating phase,
$\delta_{CP}$, but also on the electron density of the media they
traverse.  This last property makes atmospheric neutrinos an ideal
probe for measuring the electron density distribution of the Earth
presuming the other oscillation parameters are well measured.  Since
accelerator neutrino measurements at Hyper-K itself are expected to
dramatically improve on the precision of these parameters
(c.f. Section~\ref{sec:cp}), Hyper-K may be able to make the first
measurement of the core's chemical composition using its atmospheric
neutrino sample.

Hyper-K's sensitivity in this regard has been studied in the context
of atmospheric neutrino spectrum's dependence upon the ratio of the
proton to nucleon ratio (Z/A) of material in the outer core.
Constraints from the combination of measurements of the Earth's
geodetic-astronomical parameters, such as its precession and nutation,
with its low frequency seismic oscillation modes (free oscillations),
and seismic wave velocity measurements have yielded precise knowledge
of the planet's density profile~\cite{Dziewonski1981}.  Using this
information the inner core and mantle layers of the Earth are fixed to
pure iron (Z/A = 0.467) and pyrolite (Z/A = 0.496) and the Z/A value
of the outer core is varied.  The analysis uses the same analysis
samples presented in Section~\ref{section:atmnu} and focuses on
upward-going events between 1 and 10 GeV.  Assuming the outer and
inner core chemical compositions are the same,
figure~\ref{fig:geophys_za} shows the expected constraint on the Z/A
parameter.  After a 10 Mton year exposure Hyper-K can exclude lead and
water (pyrolite) outer core hypotheses by approximately $\sim 3\sigma
(1\sigma)$.  
While geophysics models will ultimately require even
greater precision in such measurements, Hyper-K has the potential to
make the spectroscopic measurements of the Earth's core.
It is worth noting that other proposed experiments with the ability 
to make similar geochemical measurements, such as the next generation of neutrino 
telescopes, rely primarily on the muon disappearance channel. 
Hyper-Kamiokande's, on the other hand, is unique in that its sensitivity 
is derived from the electron appearance channel.

\newpage

\clearpage
\part{Second Detector in Korea}
\label{section:secon-detector-korea}
\section{Second Detector in Korea}
\graphicspath{{second-tank-korea/figures}}

%=====================================================================
% **********  Motivations ****************
%=====================================================================
\subsection{Motivations}\label{sec:secondtankkorea-motivations}

\begin{figure}
\begin{center}
\includegraphics[width=1.0\textwidth]{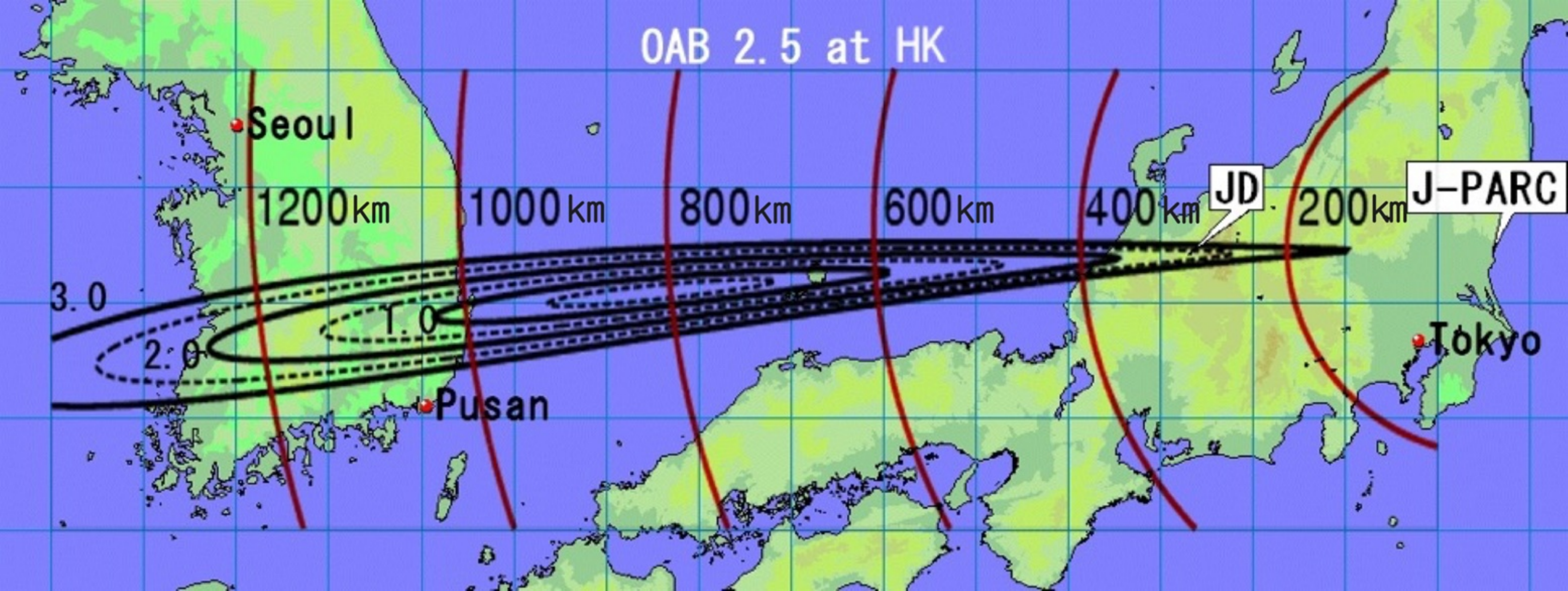}
\end{center}
\caption{ Contour map of the J-PARC off-axis angle beam to Korea~\cite{Hagiwara2006,Hagiwara2006e,Hagiwara2007}. }
\label{f:OAB}
\end{figure}

Strategy of the future Hyper-K experiment is to build two identical water-Cherenkov detectors in stage with 260 kton of purified water per detector.
The first detector will be built at Tochibora mine in Japan, and this Hyper-K design report is dedicated to describe the design of the first detector in Japan.
Locations in Korea are currently investigated for the second detector where the J-PARC neutrino beam passes through, and this section focuses on the benefits of building the second detector in Korea. 
Figure~\ref{f:OAB} shows the J-PARC neutrino beam aiming at the Japanese site with 2.5 degree off-axis angle (OAA) and also passing through Korea
with an 1$\sim$3 degree OAA range.

In fact, having a 2$^\text{nd}$ detector regardless of its location, will improve sensitivities of all areas of physics covered by the Hyper-K 1$^\text{st}$ detector.
However, building a 2$^\text{nd}$ detector in Korea rather than in Japan further enhances physics capabilities on neutrino oscillation physics
(leptonic CP violation, determination of neutrino mass ordering, non-standard neutrino interaction etc.)
using beam neutrinos thanks to the longer baseline ($\sim$1,100~km).

\begin{figure}[ht]
\centering
\includegraphics[width=1.0\textwidth]{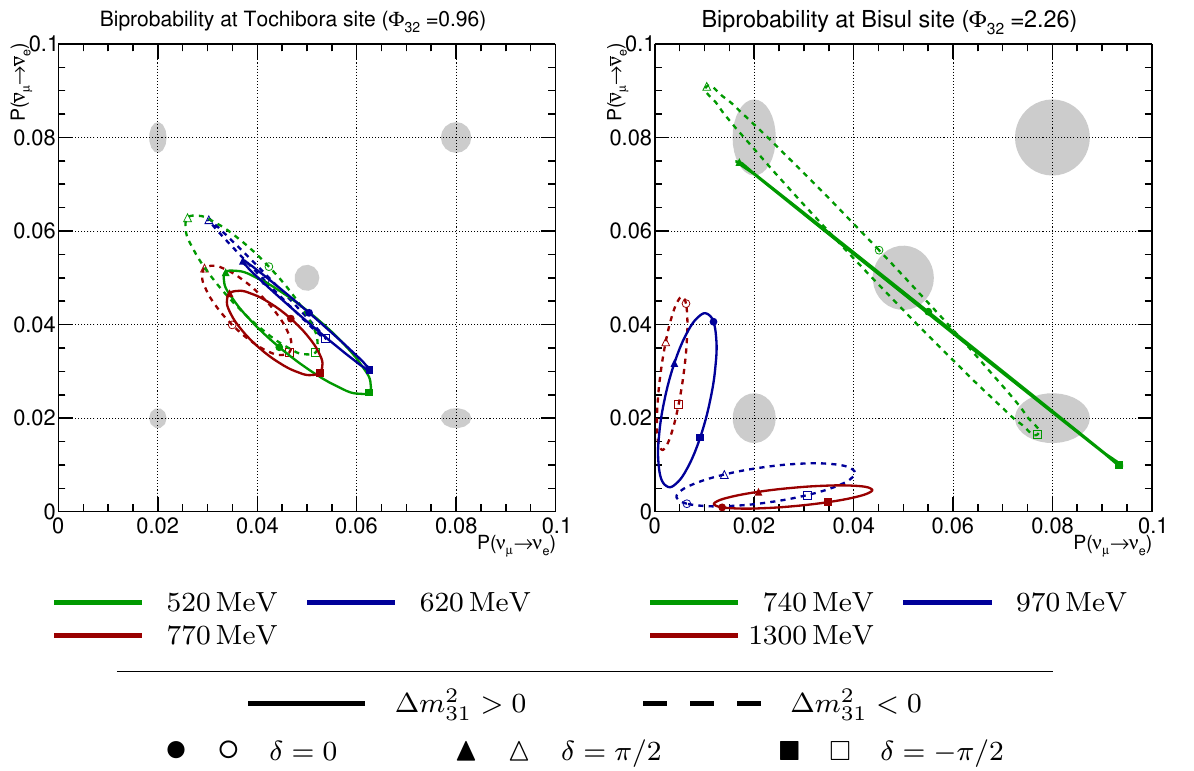}
\caption{Appearance bi-probability plots at Hyper-K sites in Tochibora (left) at $2.5^{\circ}$ OAA and in Mt. Bisul (right) at $1.3^{\circ}$ OAA. 
These plots show the principle by which Hyper-K determines the mass ordering and measures the CP phase in three representing energies
drawn with green, blue and red colors. 
Normal (inverted) ordering for different $\delta_{CP}$ values is represented by a solid (dotted) curves. 
Grey ellipses represent the sizes of statistical error for a ten-year exposure on one Hyper-K detector. 
The $\Phi_{32}$ is defined as $\frac{2}{\pi}\frac{\left|\Delta m^2_{32}\right|L}{E}$ 
and is close to odd-integer values for oscillation maxima, and close to even-integer values for oscillation minima.
}
\label{f:bp}
\end{figure}

The benefits of Korean site for beam neutrino physics are well expressed by an appearance bi-probability plot: 
$P(\nu_{\mu}\rightarrow\nu_{e})$ vs. $P(\bar{\nu}_{\mu}\rightarrow\bar{\nu}_{e})$. 
Figure~\ref{f:bp} shows appearance bi-probability plots at Hyper-K sites in Tochibora, Japan (left) and in Mt. Bisul, Korea (right). 
Each point in colored ellipses represents different $\delta_{CP}$ value for normal (inverted) neutrino mass ordering 
in solid (dotted) lines.
Three different colors correspond to three representing neutrino energies for each site.
The blue ellipses correspond to the peak energy at each OAA and the green and red ones represent median
energy after splitting the events into two parts below and above the peak. 
Note that there are high energy ($>$ 1.25 GeV) appearance events observed in Korean site due to the longer baseline,
and this results in an important difference in physics sensitivities between Japan and Korean sites. 
The gray ellipses indicates the sizes of the statistical
uncertainties given by $\sqrt N$ from the number of events around peak energy in $\nu_{e}$ and $\overline{\nu_{e}}$ appearance signals. 
It is clear from the bi-probability plots that the ellipses in Mt. Bisul site are well separated compare to those at Tochibora site
although the statistics is lower with the longer distance. 
This will allow us to resolve degeneracies (overlaps of ellipses) in neutrino mass ordering and $\delta_{CP}$ parameters.

Thanks to the longer baselines in Korean sites the 1$^\text{st}$ and 2$^\text{nd}$ oscillation maxima of the appearance probability are reachable
with higher ($>$ 1.25 GeV) neutrino energy, and this allows much better sensitivities especially on the neutrino mass ordering determination and non-standard neutrino interaction (NSI) in matter. 
By having the 2$^\text{nd}$ detector in Korea the fraction of $\delta_{CP}$ coverage is also increased and more precise measurement of $\delta_{CP}$ is achieved. 
According to the sensitivity studies performed, described later, smaller OAA in Korean site gives best sensitivities in most physics cases with beam neutrinos. 
Reaching higher energy with smaller OAA, however, introduces more $\pi^{\circ}$ background but a good news is that T2K has recently reduced 
the $\pi^{\circ}$ background and its systematic uncertainty~\cite{Abe:2015awa}. 
Thanks to the larger overburdens in Korean sites the sensitivities on solar neutrino and Supernova relic neutrinos (SRN) are also further improved.

%=====================================================================
% ********** Location and Detector **********
%=====================================================================
\subsection{Location and Detector}\label{sec:secondtankkorea-location}

There are several candidate sites to locate the 2$^\text{nd}$ detector in Korea and they are listed in Table~\ref{t:six_sites}.
Among the six candidate sites Mt. Bisul with the smallest OAA seems to be the most favorite ones according to our sensitivity study described later. 

\begin{table}[hbt]
\captionsetup{justification=raggedright,singlelinecheck=false}
\small
 \caption{Candidate sites with the off-axis angles between 1 and 2.5 degrees for the 2$^\text{nd}$ Hyper-K detector in Korea. The baseline is the distance from the production point of the J-PARC\
 neutrino beam~\cite{Abe:2016t2hkk}.}
 \centering
 \begin{tabular*}{0.95\textwidth}{@{\extracolsep{\fill}} l c c c l}

  \hline \hline
Site                 &  Height &  Baseline  &  Off-axis angle   &  Composition of rock \\[-1.4ex]
     & (m) & (km) & (degree) & \\
\hline
 Mt. Bisul           &   1084    &  1088    &  1.3$^{\circ}$      &   Granite porphyry,\\[-1.4ex] %1.30
  & & & & andesitic breccia \\[0.4ex]
 Mt. Hwangmae  &   1113    &  1141    &  1.9$^{\circ}$      &   Flake granite, \\[-1.4ex] %1.94
& & & & porphyritic gneiss  \\[0.4ex]
 Mt. Sambong    &   1186    &  1169    &  2.1$^{\circ}$     &   Porphyritic granite, \\[-1.4ex] %2.06
  & & & & biotite gneiss \\[0.4ex]
 Mt. Bohyun       &   1124    &  1043    &  2.3$^{\circ}$    &   Granite, volcanic rocks, \\[-1.4ex] %2.29
  & & & & volcanic breccia \\[0.4ex]
 Mt. Minjuji         &   1242    &  1145    &  2.4$^{\circ}$    &   Granite, biotite gneiss \\[0.4ex] %2.38
 Mt. Unjang        &   1125    &  1190    &  2.2$^{\circ}$     &   Rhyolite, granite porphyry, \\[-1.4ex] %2.21
  & & & & quartz porphyry  \\
  \hline \hline
%\end{center}
  \end{tabular*}
 \label{t:six_sites}
 \end{table}

\begin{figure}[tb]
\centering
\includegraphics[width=0.7\textwidth]{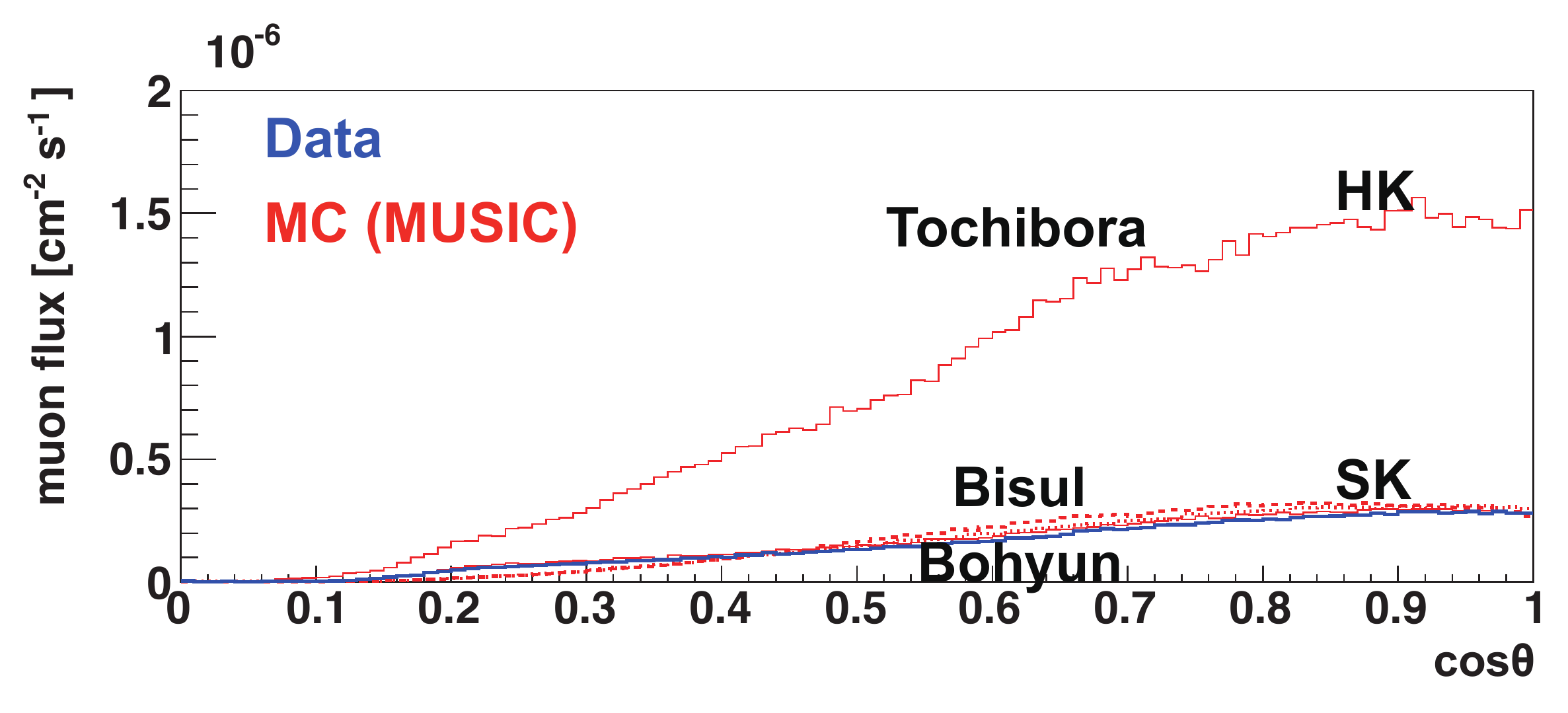}
\includegraphics[width=0.7\textwidth]{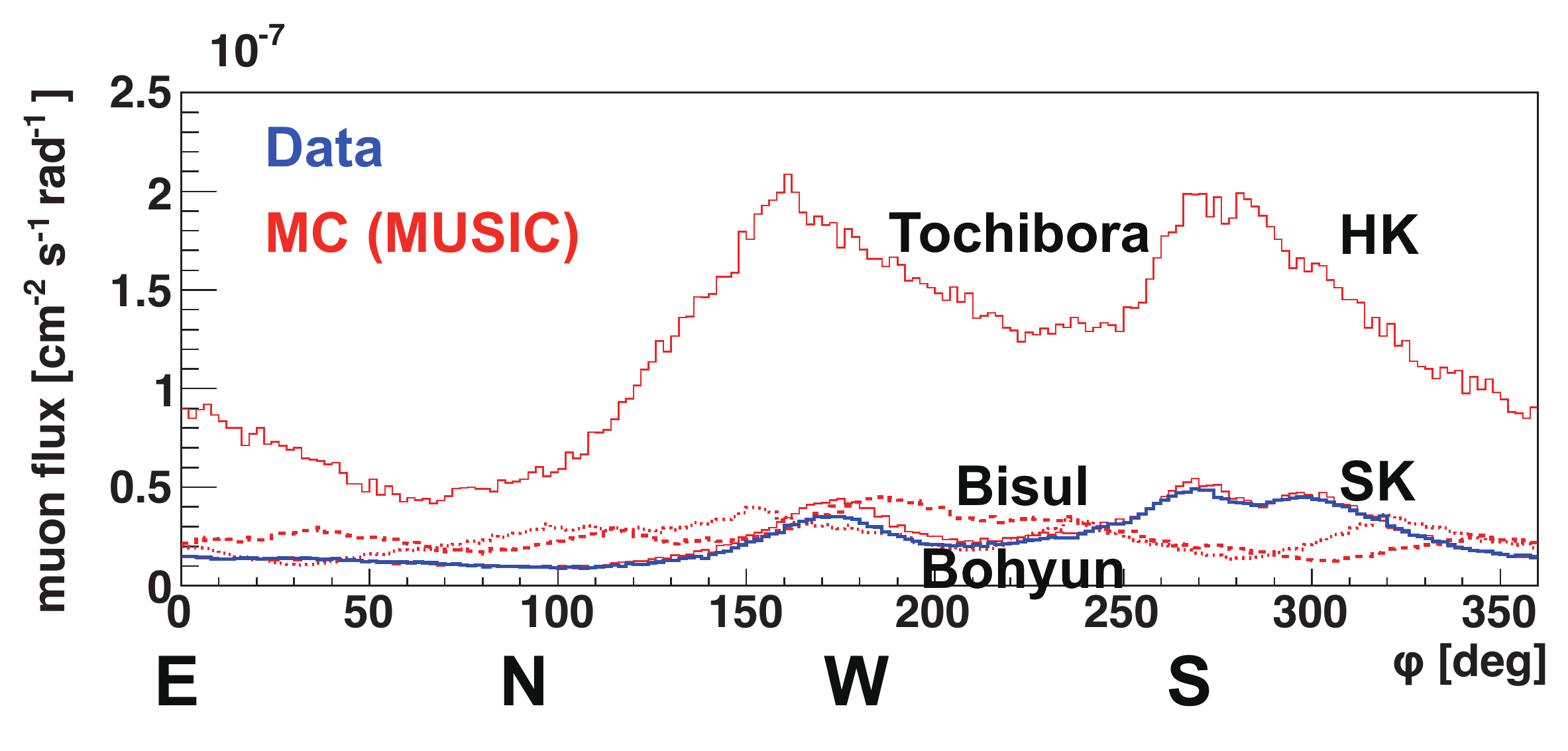}
\caption{Muon flux as a function of cosine of zenith angle $\cos\theta$ (upper) and azimuth angle $\phi$ (lower) for Hyper-K (Tochibora), 
Mt.\ Bisul (1,000~m overburden), Mt.\ Bohyun (1,000~m overburden), and Super-K.
The east corresponds to the azimuth angle of zero degree.
The blue lines show the data for Super-K, and the red lines show the MC predictions based on the \texttt{MUSIC} simulation. }
\label{fig:muon-direction}
\end{figure}

The larger overburdens of the Korean candidate sites allow much less muon flux and spallation isotopes. 
With a flat tunnel the overburden is $\sim$820~m for both Mt. Bisul and Mt. Bohyun, the two favorable sites in a more favorable order.
By making a sloped tunnel the overburden becomes $\sim$1000~m for both mountains, and this is the default tunnel construction plan.
Figure~\ref{fig:muon-direction} shows muon flux as a function of zenith (top) and azimuth (bottom) angles
for Mt. Bisul and Mt. Bohyun as well as Super-K and Hyper-K (Tochibora) sites for comparison.
The muon flux at the two Korean sites with $\sim$1,000~m overburden are similar to that of Super-K site
and about four times smaller than Hyper-K Tochibora site. 

%=====================================================================
% ********** Physics Sensitivities ***********
%=====================================================================
\subsection{Physics Sensitivities}\label{sec:secondtankkorea-physics}

Physics sensitivity studies are performed 
for three different off-axis angles (1.5$^{\circ}$, 2.0$^{\circ}$, and 2.5$^{\circ}$) at a baseline distance of $L=1,000$~km.
Sensitivity study results are obtained for the determination of the mass ordering, the discovery of CP violation by excluding of the sin($\delta_{cp}$) = 0
hypothesis, the precision measurement of $\delta_{cp}$, fraction of $\delta_{cp}$ as a function of CPV significance and exposure, $\theta_{23}$ octant, 
atmospheric parameters and NSI.
Only part of these studies are shown here due to a limited space, but the full studies will be published in a separate paper. 
The results are obtained by assuming 2.7$\times$10$^{22}$ proton-on-target with $\nu : \bar{\nu} = 1 : 3$
which corresponds to 10-year operation with 1.3 MW beam power for 187 kton fiducial volume mass per detector. 
In this section relatively simplistic systematic uncertainty model~\cite{Abe:2016t2hkk} is used in the sensitivity studies 
while the size of the systematic uncertainties are mostly based on the current T2K analyses. 
Note that the same simplistic systematic uncertainty model is used for the Japanese detector, 
which is a little different from the systematic uncertainties used in other sections of this document. 
In sensitivity studies other oscillation parameter values are from the Particle Data Group (PDG) 2015 Review of Particle Physics~\cite{Olive:2016xmw}
except for $\theta_{23}$ and $\Delta m^{2}_{32}$, where $\sin^{2}\theta_{23}=0.5$ and $\Delta m^{2}_{32} = 2.5 \times 10^{-3}$ eV$^{2}$ are used 
due to their large uncertainties in their absolute values, and no CP violation ($\delta$ = 0) is assumed unless it is specified.
Two detector configuration, either JD (Japanese Detector) + KD (Korean Detector) or JD $\times$ 2, assumes no staging in the sensitivity studies shown in this section. 

\begin {figure}[htbp]
\captionsetup{justification=raggedright,singlelinecheck=false}
  \begin{center}
    \includegraphics[width=0.80\textwidth]{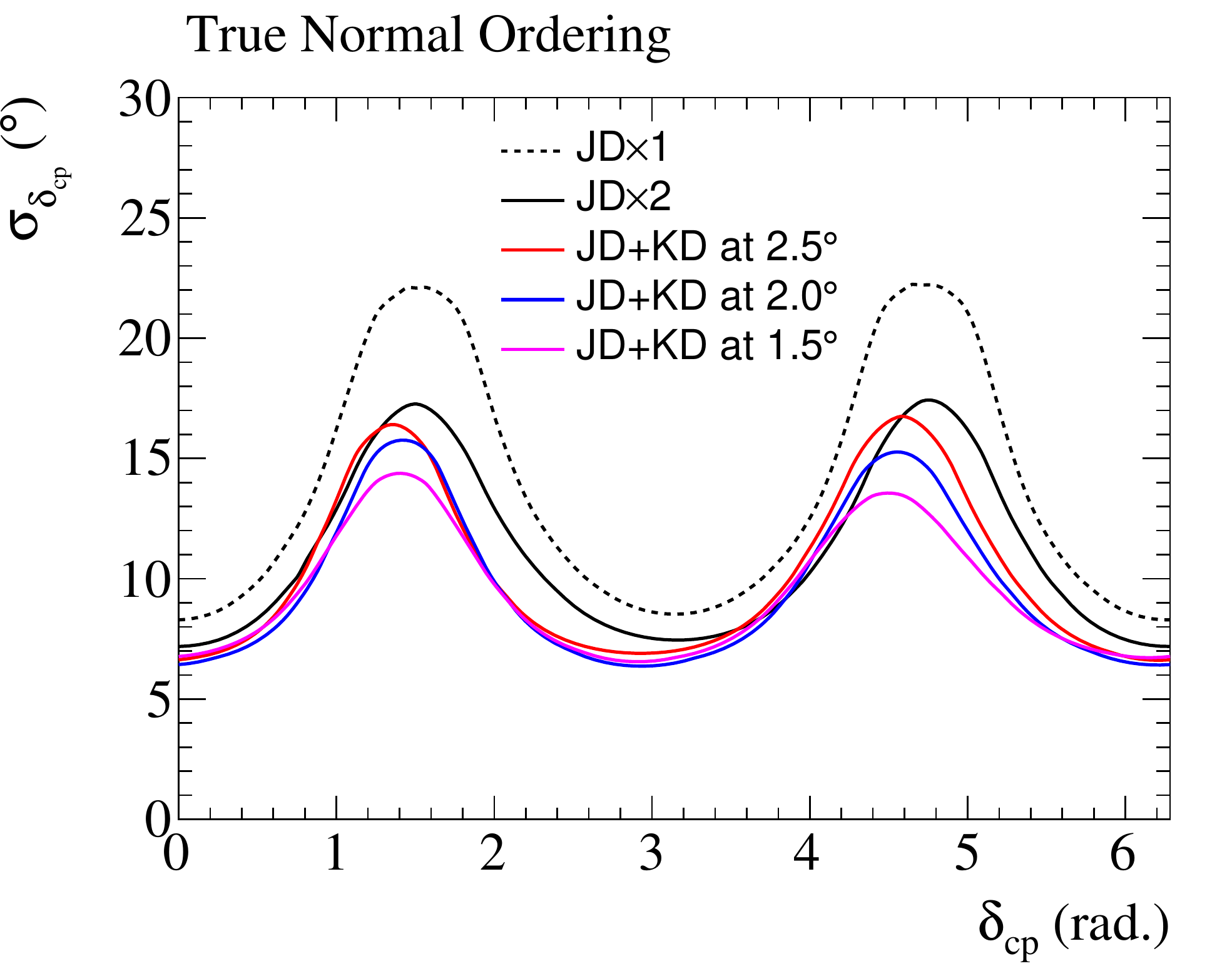}\\
    \includegraphics[width=0.80\textwidth]{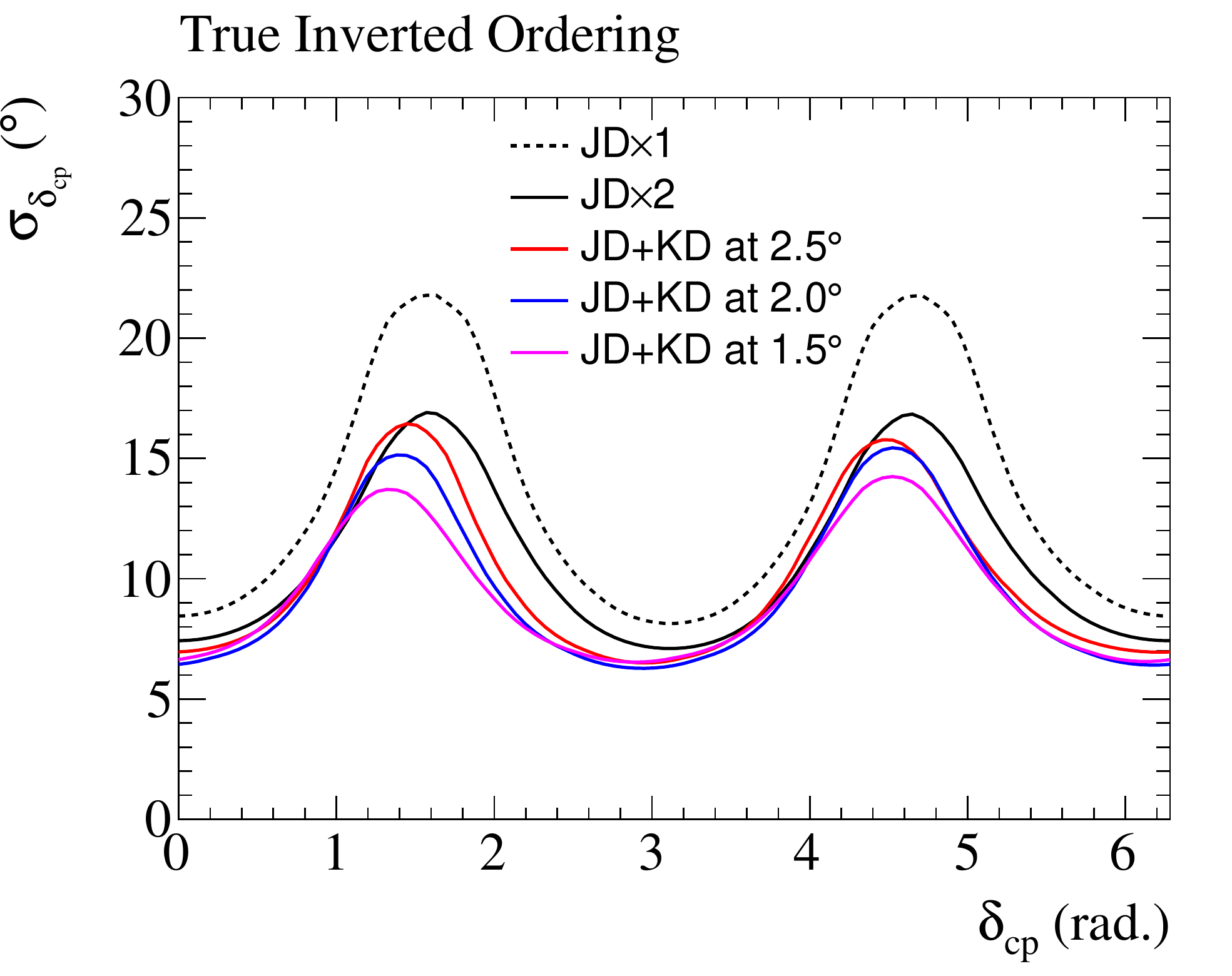}
    \caption{The 1$\sigma$ precision of the $\delta_{cp}$ measurement as a function of the true $\delta_{cp}$ value.  Here, it is assumed there is no prior knowledge of the
mass ordering.}
    \label{fig:cp_precision}
  \end{center}
\end {figure}

\begin {figure}[htbp]
\captionsetup{justification=raggedright,singlelinecheck=false}
  \begin{center}
    \includegraphics[width=0.8\textwidth]{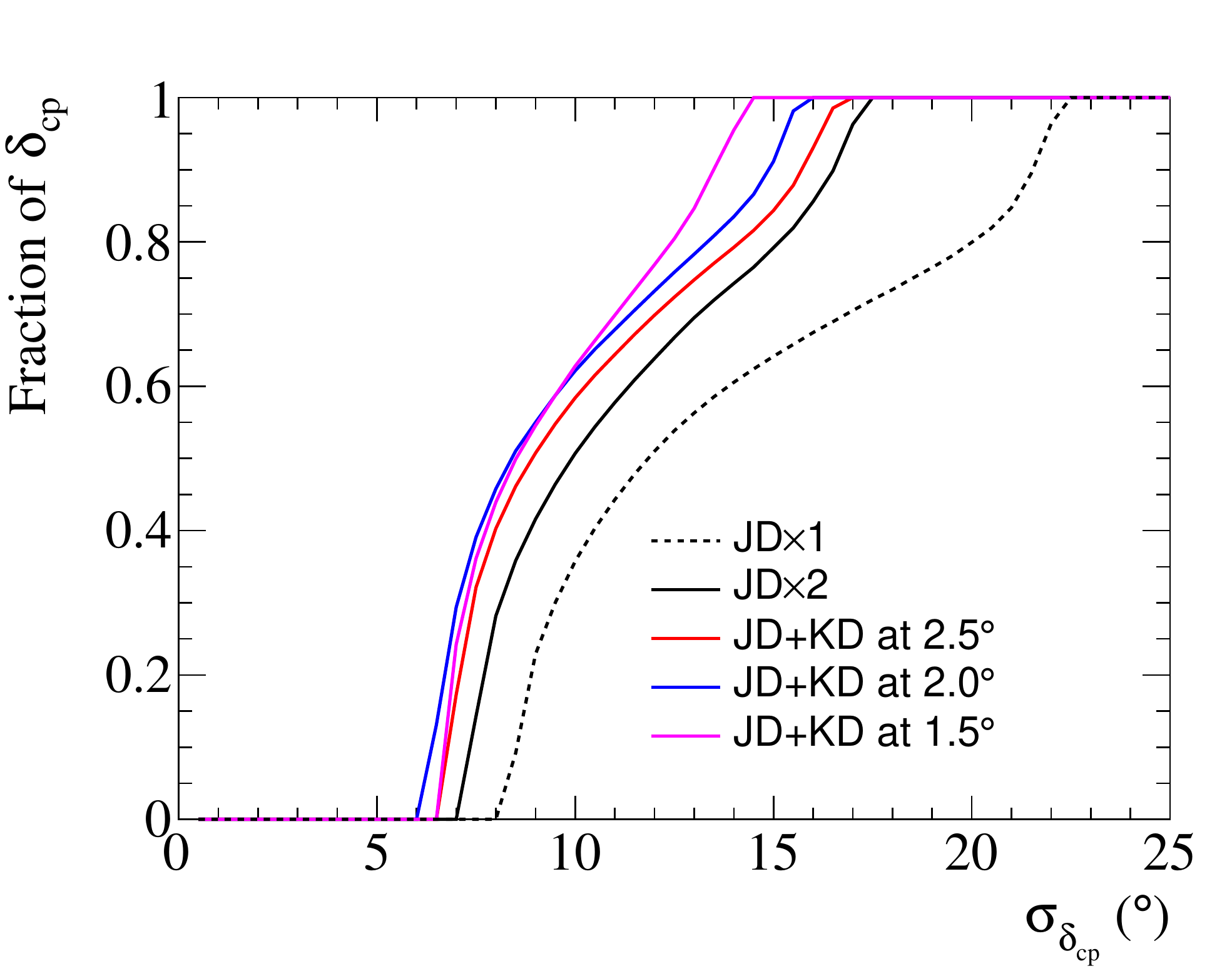}
    \caption{The fraction of $\delta_{cp}$ values (averaging over the true mass ordering) for which a given precision or better on $\delta_{cp}$ can be achieved.}
    \label{fig:cp_prec_cp_frac}
  \end{center}
\end {figure}

Figure~\ref{fig:cp_precision} shows the sensitivity on the 1$\sigma$ precision of the $\delta_{cp}$ measurement assuming no prior knowledge on neutrino mass ordering. 
For most of the true $\delta_{cp}$ the best sensitivity is from JD + KD with 1.5$^{\circ}$ OAA. 
Especially when the CP is maximally violated the sensitivity difference is largest between the JD + KD with 1.5$^{\circ}$ OAA and JD $\times$ 2 or JD $\times$ 1. 
Figure~\ref{fig:cp_prec_cp_frac} shows the sensitivity on the fraction of $\delta_{cp}$ for the 1$\sigma$ precision of the $\delta_{cp}$ measurement. 
The results with JD + KD with any off-axis angle are always better than JD $\times$ 2 and JD $\times$ 1. 
The sensitivity with 2.0$^{\circ}$ OAA is slightly better than that of 1.5$^{\circ}$ OAA if the $\delta_{cp}$ precision is less than 10 degree
but otherwise the 1.5$^{\circ}$ OAA gives the best sensitivity. 
More details on the physics sensitivity studies are found in Ref~\cite{Abe:2016t2hkk}.
According to our sensitivity study CP violation sensitivity improves a little in JD + KD compared to JD $\times$ 2. 
The $\theta_{23}$ octant sensitivity and atmospheric parameter sensitivities get slightly worse in JD + KD configuration compared to JD $\times$ 2. 

The sensitivity on NSI of neutrinos in matter (not in production nor in detection) 
is also greatly enhanced by having an additional  2$^\text{nd}$ detector in Korea,
and the details on NSI sensitivity studies are found in Ref~\cite{Fukasawa:2016lew,Liao2017}.

Thanks to the similar overburden to Super-K in Korean candidate sites (see Fig.~\ref{fig:muon-direction}) 
but with a larger detection volume, 
low energy astrophysics such as solar neutrinos and SRN sensitivities are expected to be good. 
According to our sensitivity study on the SRN we expect about 5 (4) sigma sensitivity for 10 years of data taking 
in Mt. Bisul or Mt. Bohyun (Tochibora) sites~\cite{Yeom2017_SRN_sens}. 
Moreover we expect to observe spectral distribution of SRN due to large statistics from 8.4 times larger detection volume than Super-K,
and this might lead to solving SN burst rate problem. 

%=====================================================================
% **********  Conclusion ****************
%=====================================================================
\subsection{Conclusion}\label{sec:secondtankkorea-conclusion}

Having a 2$^\text{nd}$ Hyper-K detector in addition to the 1$^\text{st}$ one will enhance physics sensitivities
from beam neutrino physics to astroparticle physics due to the increased detection volume.
According to our sensitivity studies, physics capabilities of the Hyper-K project are further improved by locating the 2$^\text{nd}$ detector in Korea, 
such as determination of neutrino mass ordering, precision of $\delta_{CP}$ measurement and test of NSI.
With the longer baseline in Korean site both the 1$^\text{st}$ and the 2$^\text{nd}$ oscillation maximum of the appearance neutrino probability are reachable,
and this is a unique opportunity since no other past and current experiment (MINOS, T2K, NOvA) can reach the 2$^\text{nd}$ oscillation maxima. 
The longer baseline to the detector in Korea allows to resolve the degeneracy between the mass ordering and the value of delta CP that would happen 
for only certain values of the parameters with a detector(s) in Japan. The bi-probability plots can intuitively show this feature even before 
performing any sensitivity studies.
NSI sensitivities are also improved especially with the smaller OAA site in Korea. 
According to our sensitivity studies the best Korean candidate site seems to be the Mt. Bisul. 
With $\sim$1000~m overburden sensitivities on solar neutrino and SRN physics are further enhanced in the Korean candidate sites
than those in the Tochibora mine ($\sim$650~m overburden). 

\clearpage
\appendix
\part{Appendix}
\label{section:appendices}
\color{black}
\section{\label{sec:linertests} Liner Sheet Tests}
\graphicspath{{appendixB/figs/}}

\begin {figure}[htbp]
  \begin{center}
    \includegraphics[width=0.8\textwidth]{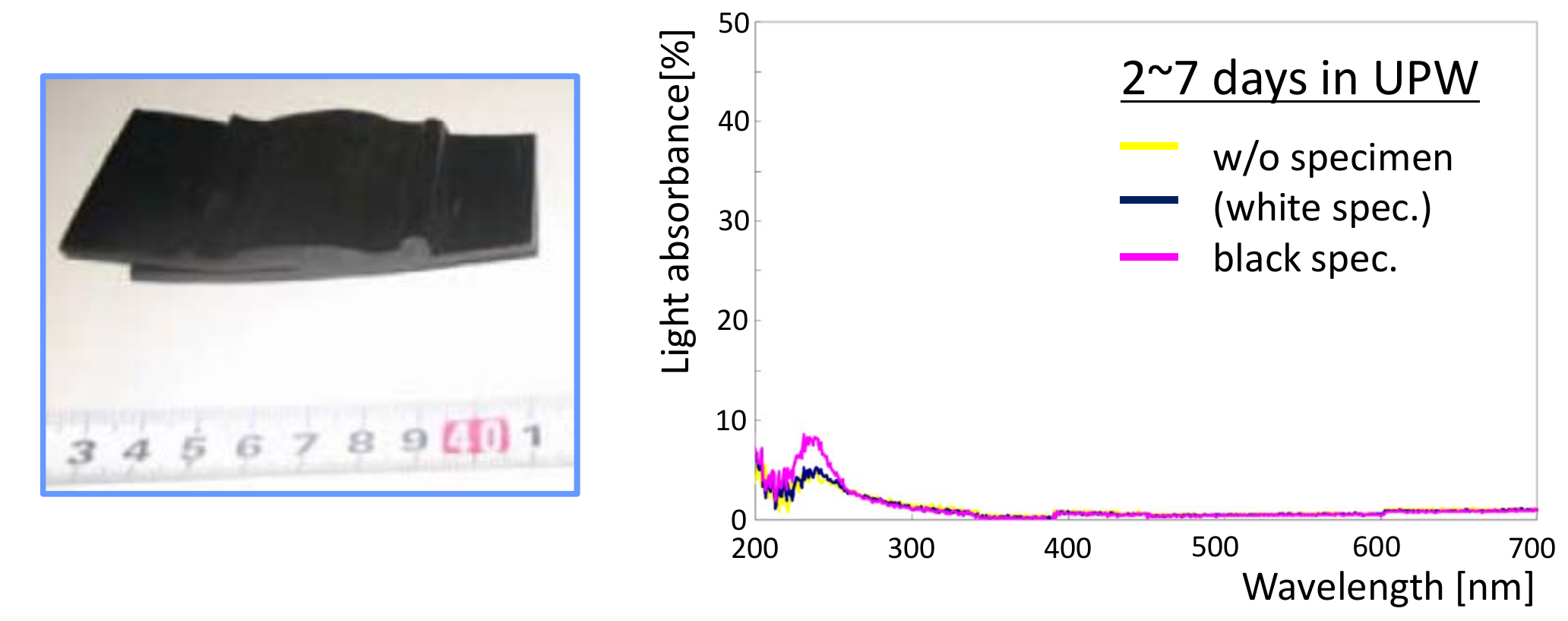}
\caption{A typical specimen (GSE Gundle sheet) for immersion test and typical results of light absorbance of ultra-purified water after several days with specimens soaked in it (pink line), which is compared to a control sample without specimen (yellow line).}
    \label{fig:liner-immersion} \end{center}
\end {figure}

\subsection{Immersion test}
Specimens of HDPE lining sheet (GSE Gundle sheet, whose material is
identical to that for the CPL), with artificial extrusion-welded seam,
were immersed into the ultra-purified water (UPW) for certain periods
(1, 2 to 7, and 8 to 31 days), and absorbance was compared to a
control sample without specimens. Amount of eluted materials into UPW,
i.e. total organic carbon (TOC), anions and metals, were also
measured.
Figure \ref{fig:liner-immersion} show the specimen and an example of
measurements, where increase of the light absorbance were observed
between the wavelength range of 200$\sim$300\,nm. Some amount of
material elution were observed, where eluted amounts per unit area and
time were significantly less for later periods. Although relation
between the increase of light absorbance and the material elusion
should be studied, it is noted that range of PMT-sensitive wavelength
is somewhat higher (300$\sim$650\,nm), so the effect to the
experiment may be limited. Similar results were obtained for
Gadolinium sulfate solutions.

\subsection{Measurements on material strength}
Tension tests were carried out for the CPL to estimate yield strength,
tensile strength, and Young's modulus. Since HDPE has large elongation
before breaking ($\sim$500\%), 1.0\% proof stress was used as the
yield strength (instead of 0.2\% proof stress which is generally used
for other materials). Here, varying measurement conditions were
examined for tensioning velocity (0.05\,mm/min and 0.5\,mm/min) and for temperature (typical room temperature 23.5$^\circ$
Celsius and 15$^\circ$ Celsius simulating water temperature).
It was found that measured yield strengths were smaller by a few to
several tens of percent than the specification value (15.2\,MPa as
listed in Table~\ref{tab:liner-spec}). In general, lower tensioning
speed gives lower strength, due to large plasticity of HDPE. Strength
also depends significantly on temperature: HDPE becomes harder with
lowering temperature, as common properties in high-polymer
materials. The measurement in 15$^\circ$ Celsius gave about 20\%
higher strengths than those at 23.5$^\circ$ Celsius.
The tension tests were also repeated on the samples with an
extrusion-welded seam. For the most cases, none of peeling, fracture,
nor other troubles were observed on welded seams, but the deformation
occurred at the base material. The yield/tensile strength were almost
identical to the values for base material.

\subsection{Creep test}
Creep tests were performed with various tensile loads: 1/4$\times$M,
3/8$\times$M, or 1/2$\times$M, where M = 18.2\,MPa is the observed
yield strength as 1.0\% proof stress with tensioning speed of 5.0\,cm/min in 20$^\circ$ Celsius. For the tests with load of 1/4$\times$M
({\it i.e.} 4.6 MPa), creep was not observed for about 30 days. Meanwhile, for tests with load of 3/8$\times$M (6.8\,MPa) and
1/2$\times$M (9.1\,MPa), clear generation of creep was observed with 5
and 10 days, respectively.

\subsection{Resistivity to localized water pressure}
\begin {figure}[htbp]
  \begin{center}
    \includegraphics[width=0.8\textwidth]{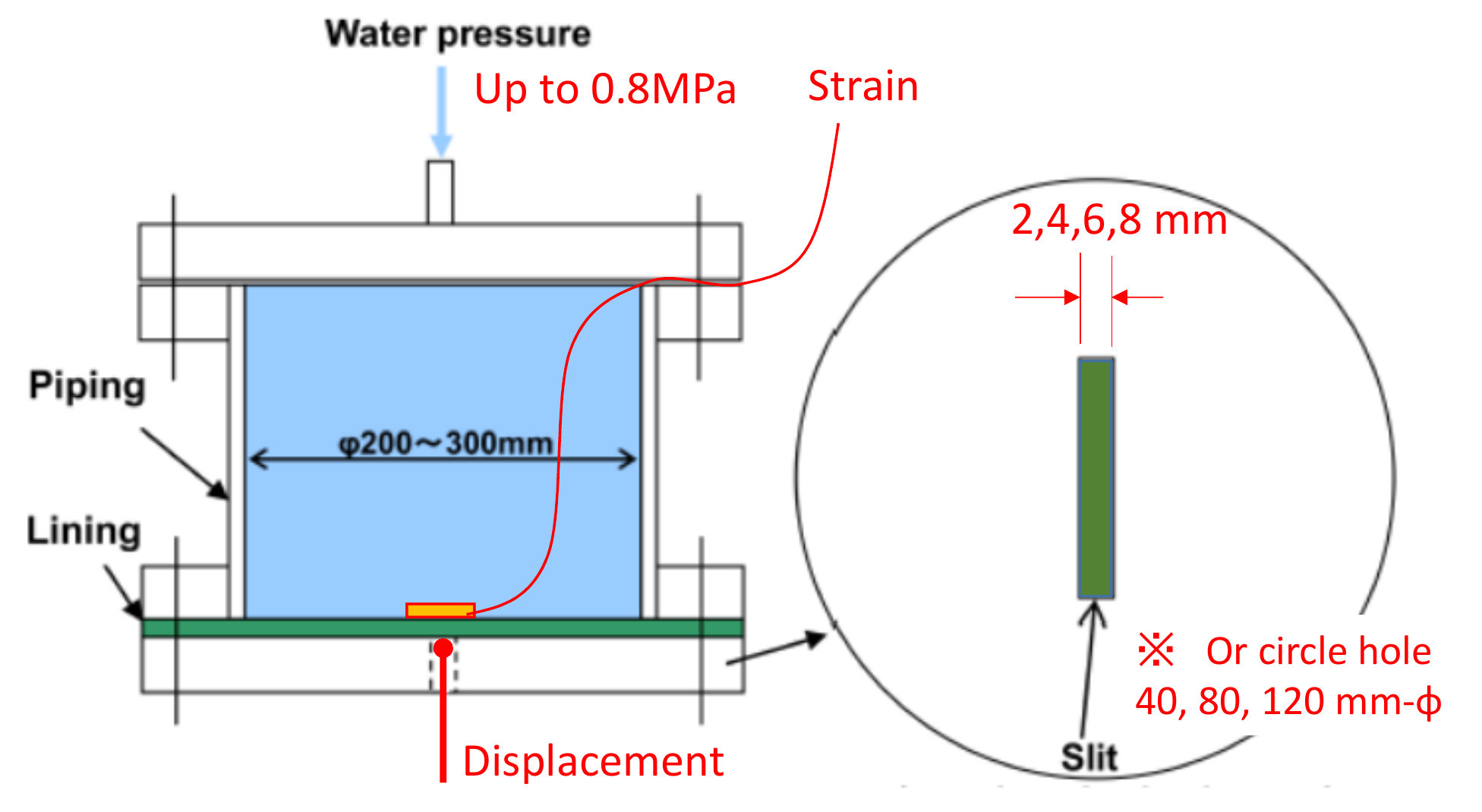}
\caption{Setup for the test applying localized water pressure to the lining material.}
    \label{fig:liner-pressure}
  \end{center}
\end {figure}
Ideally, the CPL sheet should be closely attached to the surface of
flatly-backfilled concrete walls and thus firmly supported by
them. However, it is probable that cracks or rough holes exist or
happen in the backfill concrete wall, and thus the liner should
locally stand for water pressure by itself. To simulate the situation,
tests to apply localized water pressure for the lining were performed
with variety of slit widths (2 to 8\,mm) and hole diameters (40
to 120\,mm-$\phi$), as illustrated in
Fig.~\ref{fig:liner-pressure}. It is found that the liner survived for
0.8\,MPa water pressure without breaking for all cases. Although the
test load was applied only in a short period and durability for longer
period should be examined, it is probable that the expected water
pressure is enough lower than the critical pressure causing creep.
\subsection{Water permeability}
Generally, plastic materials have a property to pass water as moisture
vapor through molecules. It is referred as moisture permeability, or
water vapor transmission rate (WVTR), being represented with
transferred mass through unit area and time (g/m$^2$/24-hours). The
permeability was studied for the GSE geomembrane (Gundle sheet, whose
material is quite simmilar to that for CPL
Studliner)~\cite{liner:JAEA-Tech-2013-036}. For the sheet with
thickness $t$= 1.5\,mm, WVTR ($P_{a1}$) was obtained to be
1\,g/m$^2$/24-hours at most for standard testing temperature
(40$^\circ$\,Celsius). The water permeability coefficient ($k$) was
then deduced to be
\[
k = P_{a1} \times t \times \frac{g}{\Delta P_v} =  2.5 \times 10^{-12} cm/s,
\]
where $g$ is standard gravitational acceleration (9.8\,m$^2$/s) and
$\Delta P_v$ is difference of the water-vapor pressures of both sides
of the geomembrane (90\% of saturated vapor pressure at
40$^\circ$\,Celsius, 75.22\,hPa). Since the CPL is 5\,mm thick, time
until water permeates to backside of the CPL is:
\[
0.5 {\rm [cm]} \times \frac{1}{2.5 \times 10^{-12} {\rm [cm/s]}} = 2 \times 10^{11} {\rm [s]},
\]
{\it i.e.} about 6,300 years. Since the total inner surface area of
the tank is 18,259 (54,750)\,m$^2$ for 1 (3) tank options, amount of
water permeation through the entire liner surface will be:
\begin{eqnarray}
&~& 18,250 (54,750) \times 10^4\ {\rm [cm^2]} \times 2.5 \times 10^{-12}\ {\rm [cm/s]} \nonumber \\
&=& 4.6 \times 10^{-4}(1.4 \times 10^{-3}){\rm [cm^3/s]} = 40 (120) \ {\rm [cm^3/day]}, \nonumber
\end{eqnarray} 
thus being negligible amount.

%----------------------------%
\subsection{Penetration structure}
%----------------------------%

\begin {figure}[htbp]
  \begin{center}
    \includegraphics[width=0.8\textwidth]{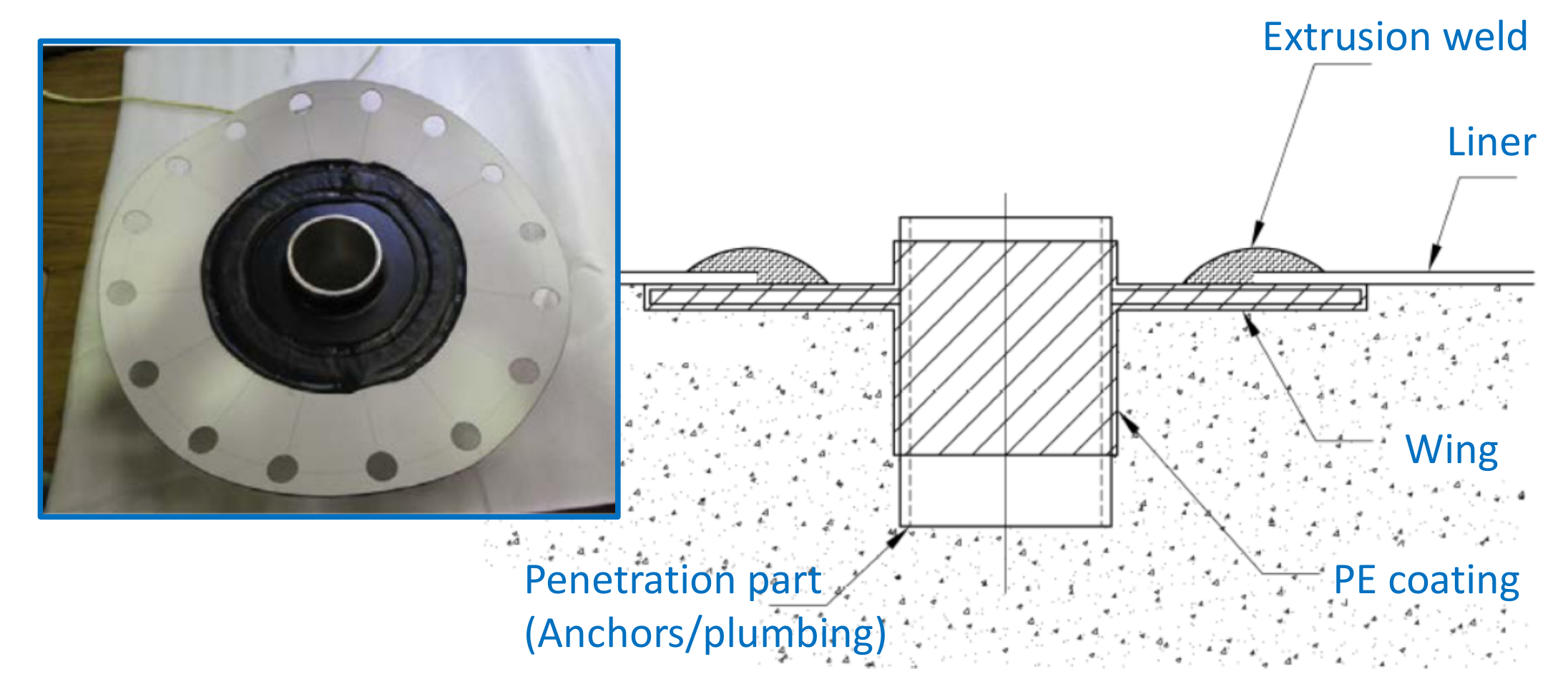}
\caption{Schematic drawing for the penetration structure of a water pipe and 
         a photo of its prototype.}
    \label{fig:liner-penetration}
  \end{center}
\end {figure}
The leak can happen around the components which penetrate the water
tank lining, such as anchors to support PMT framework columns, water
supply/return pipes, and so on. A possible design of the penetration
structure for the water pipes is illustrated in
Fig.~\ref{fig:liner-penetration}. A metal pipe with a flange is coated
together with PE resin of about 1.5\,mm thickness, which can be
extrusion-welded to adjacent CPLs.  A prototype was made for the
design as shown together in the figure, and tested with pressure up to
1 MPa for 30 minutes. Tests with cyclic pressure (0.5\,MPa, repeating
on/off in a day for 5 days) and with continuous pressure (3 months
with applying 0.5\,MPa were also performed. For all of the cases, no
water leak was observed.

\clearpage
\section{\label{sec:hakamagoshi} Hakamagoshi Option}
\graphicspath{{appendixC/figs/}}

The area around Mount Hakamagoshi in Toyama Prefucture's Minami-Tonami is being considered as a possible candidate for the second Hyper-K tank.
Lying nearly along an extension of the line between Tokai-mura and Super-Kamiokande Mount Hakamagoshi can observe neutrinos from the J-PARC beam 
in the same way as the detector in Kamioka.
Though a detailed survey of the rock quality in the area is still needed, the 1000~m height of the peaks suggest that high-quality stable rock 
can be expected. 
A detector at Hakamagoshi would have an overburden of more than 1~km and would therefore be well suited to the observation of 
low energy neutrinos such as those from supernovae and the sun. 
Further, since the Tokai-Hokuriku expressway passes through the mountain, access to the experimental site is straight-forward.

\subsection{Rock Quality Information for Hakamagoshi and its Surroundings} 

The Hakamagoshi area is characterized by Mount Hakamagoshi (elev. 1,159~m), Mount Mikata (elev. 1,142~m),
and Mount Sarugayama (elev. 1,448~m), which form a mountain chain running from northeast to southwest. 
Mountain streams along the north side of these peaks form part of the Oyabegawa river system, while 
similar streams on the southern side running to the east and south form the network of waterways feeding the Shogawa
river. 
In short, the area around Hakamagoshi is the watershed for the Oyabegawa and Shogawa rivers, which is a 
strong indicator of abundant underground water.
Indeed, measurements have yielded spring water flows in excess of 100~tons per hour. 
Hakamagoshi and its surroundings are formed from the 
primary Futomiyama formation (from the Paloegene-Paleocene), sporadically covered with so-called Hida bedrock 
(a mixture of Shirakawa granite and Nohi ryolite). 
The rock types in the Futomiyama formation as it is distributed throughout Hakamagoshi are primarily
ryholitic tuff, ryholitic welded tuff, and ryholitic lava. 
In addition, quartz porphyry, porphyrite, dorelite and other rock intrusions of no known era are found 
within the formation. 
The upper layers of the formation are inconsistently covered by Tori conglomerate from the Neogene-Miocene eras.
In particular this conglomerate is found above the Hakamagoshi tunnel at elevations around 900~m.
The upper most layers are composed of andesite and andesitic pyroclast from the same eras. 
On site investigations of the lithic fragments from the Futomiyama formation indicate large amounts of CH$\sim$CM class
hardness rock.  

\begin {figure}[htbp]
  \begin{center}
    \includegraphics[width=0.8\textwidth]{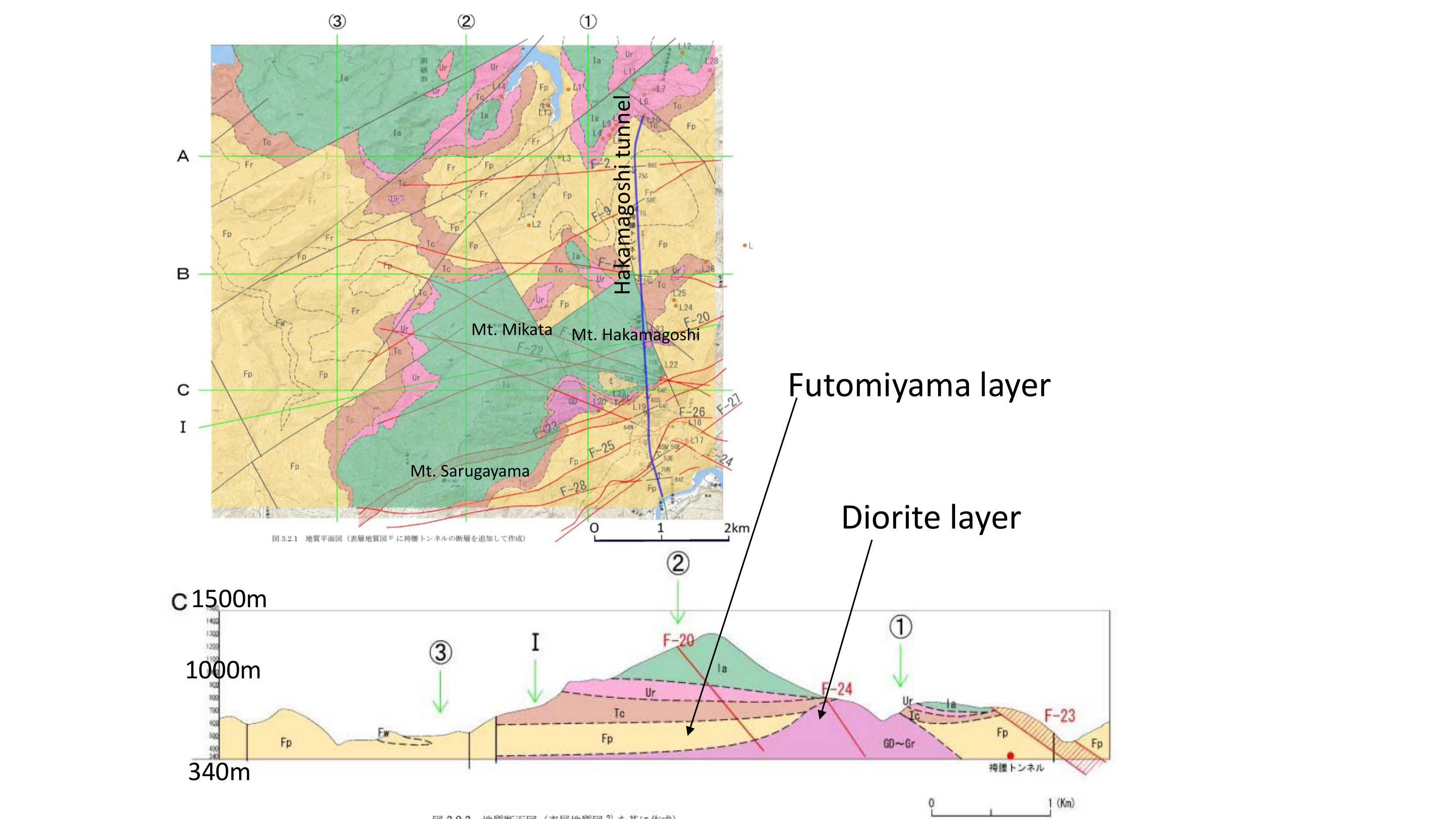}
\caption{ Overhead and cross sectional views of the rock quality in the Hakamagoshi area appear in the upper and lower panels, respectively. 
The dark pink color shows the diorite layer which is thought to extend to 
to areas where the rock overburden is 1000~m. 
}
    \label{fig:hakamagoshi_layer} \end{center}
\end {figure}

Figure~\ref{fig:hakamagoshi_layer} shows the rock quality distribution around Hakamagoshi in hardness classes.
In the lower layers of the Futomiyama formation Shogawa granodiorite is widespread, and in 
addition to this formation, at 1000~m depths other distributions of Shogawa granodiorite and 
Kose diorite may be found.
Since it is generally thought that diorite layers are higher quality in comparison to the Futomiyama formation,
the key to realizing the Hakamagoshi option lies in determining whether or not such layers exist at the site. 

\subsection{Physics sensitivities}
\subsubsection{Muon rate at Mt. Hakamagoshi} 

\begin {figure}[htbp]
  \begin{center}
    \includegraphics[width=0.8\textwidth]{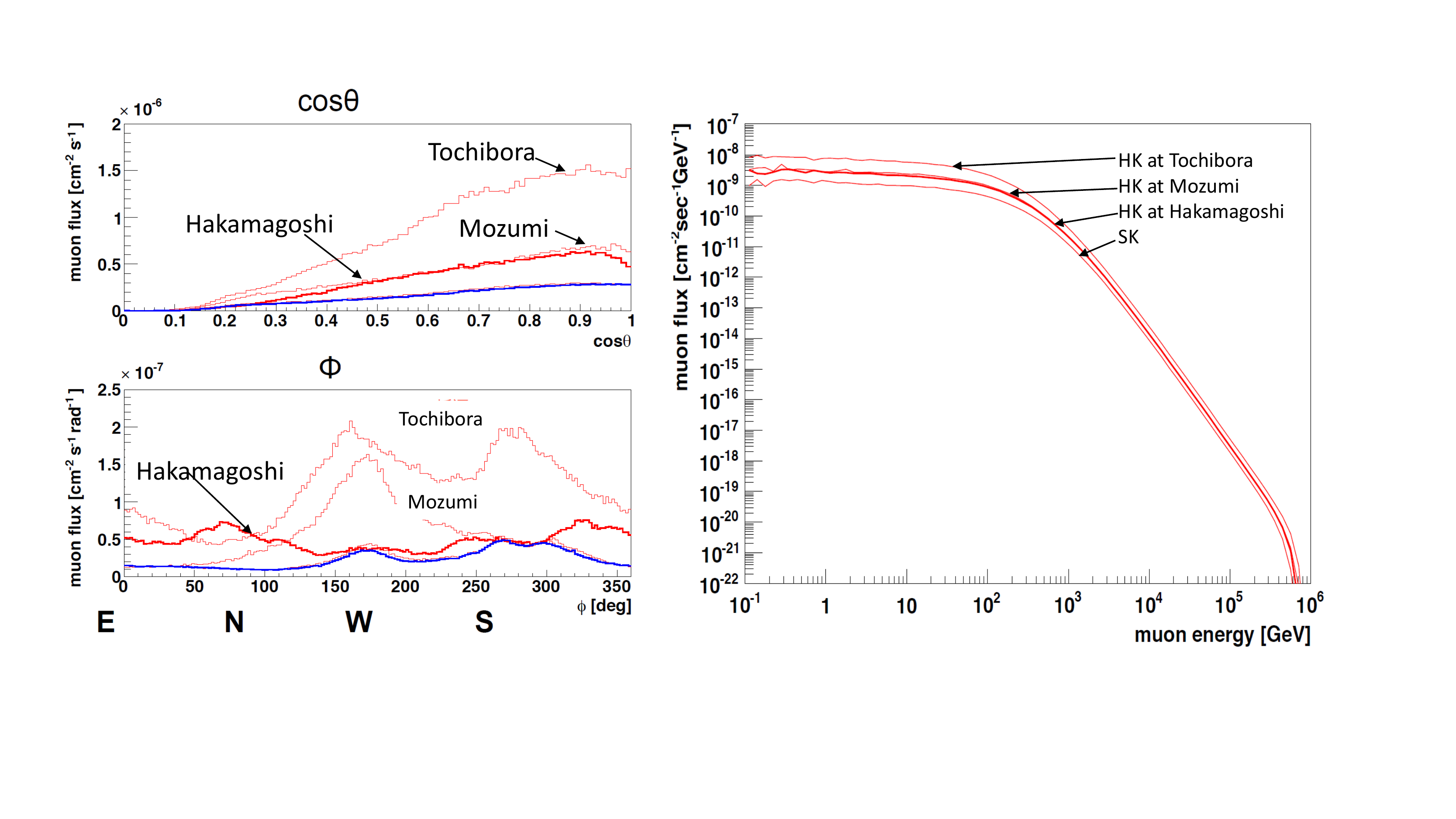}
\caption{ Muon zenith (left top) and azimuthal (left bottom) angular distributions as well as the energy distributions (right) for the Hakamagoshi site assuming an 850~m rock overburden are shown in thick red lines. Results for the Hyper-K Tochibora and 
Mozumi sites, as well as the Super-K site are shown in thin red lines. 
Blue lines show measured values. 
}
    \label{fig:hg_mu_rate} \end{center}
\end {figure}

For the moment we assume that a detector is placed directly beneath the peak of Mount Hakamagoshi at the same 
level as the expressway tunnel  and calculate the expected muon rate.
In order to study the reliability of these estimates the calculation is additionally performed for 
the area inside of the tunnel and then compared with measurements made with a plastic scintillator-based detector.
The estimation uses 30~m elevation data from the ALOS database in the same way as calculations performed for the 
Tochibora site. 
MUSIC and FLUKA are used for the muon simulation. 
With a peak elevation of 1,159~m and a tunnel elevation of approximately 300~m, here the rock overburden is taken to be 
roughly 850~m. 
Similarly the rock density is assumed to be the same as in Kamioka, 2.7~g/$\mbox{cm}^{3}$.
Figure~\ref{fig:hg_mu_rate} shows zenith, azimuthal, and energy distributions from the simulation.
Comparing to the Tochibora site, the muon rate at the Hakamagoshi site is roughly a factor of two times smaller 
and the mean muon energy is reduced by about 10\%.
The following sensitivity studies have been performed based on these results.

To validate the results of these simulations the muon rate inside the Hakamagoshi tunnel was measured 
in cooperation with NEXCO Central Nippon in an air conditioning chamber from January 18-20, 2016. 
This chamber is not located beneath the mountain peak and has an overburden of only 450~m.
Two plastic scintillators of dimension $1000\times200\times45\mbox{mm}^{3}$ and separated by 
100~mm were used for coincident muon identification. 
For comparison the same measurement was repeated at the Super-Kamiokande experimental site.
Using this apparatus the muon rates were measured to be $1.9\pm 0.1 \times 10^{-3}$~Hz 
and $1.9\pm 0.2\times 10^{-4}$~Hz at the Hakamagoshi and Super-K sites, respectively.
That is, the air conditioning chamber's muon rate is $10.0\pm1.1$ times larger.
Based on the simulations outlined above the expected difference is a factor of 8.9, indicating 
that the data and simulation are consistent within errors.
For this reason the simulation of the Hakamagoshi site below the mountain peak is taken to be reasonable.

\subsubsection{Neutrino beam from J-PARC} 

\begin {figure}[htbp]
  \begin{center}
    \includegraphics[width=0.8\textwidth]{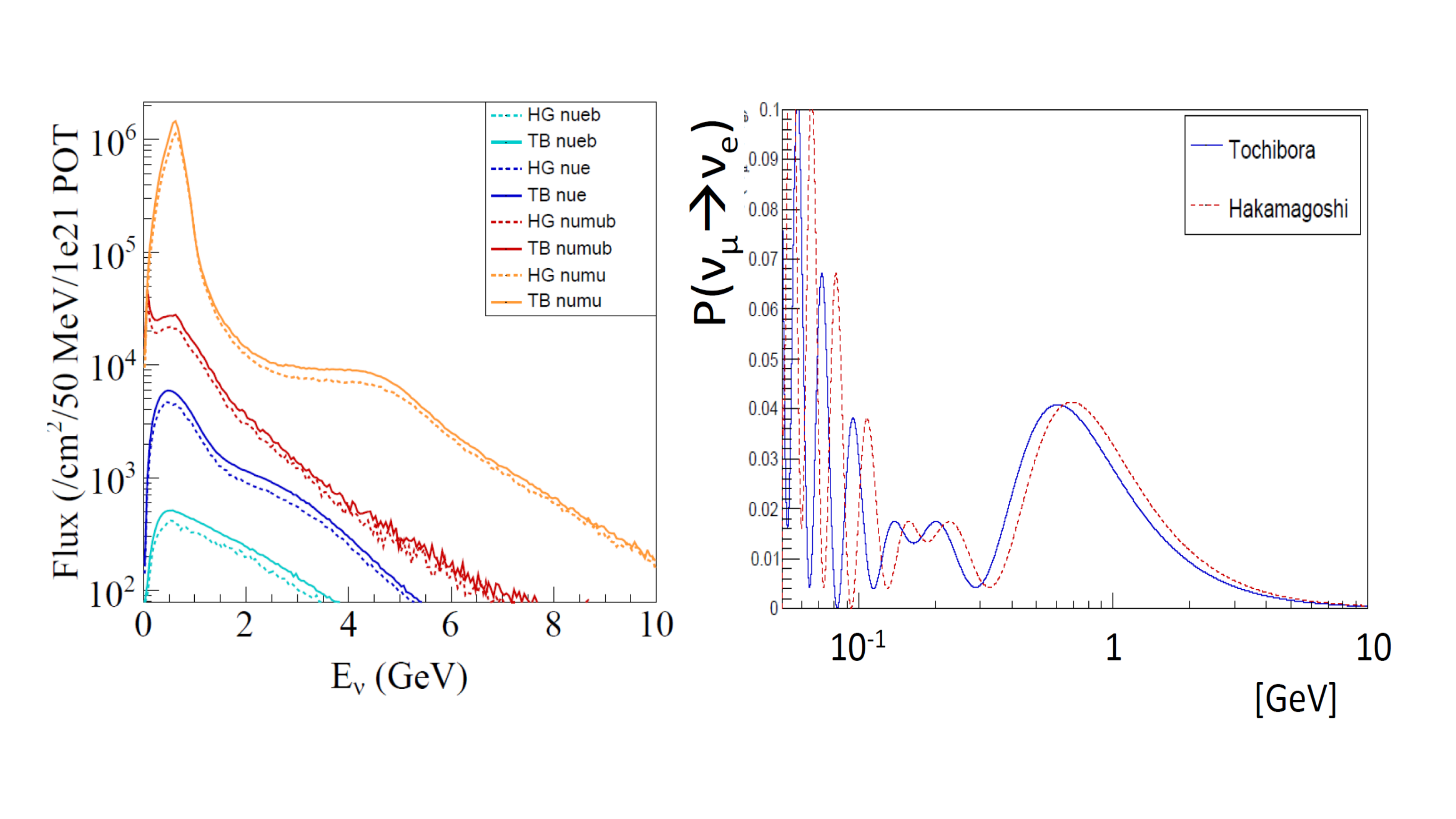}
\caption{ The neutrino flux at the Tochibora (TB) and Hakamagoshi (HG) sites from the J-PARC beam is
shown in the left figure.
In the right figure the electron neutrino appearance probability for both sites is shown 
as a function of energy.}
    \label{fig:hg_flux_osc} \end{center}
\end {figure}

\begin {figure}[htbp]
  \begin{center}
    \includegraphics[width=0.8\textwidth]{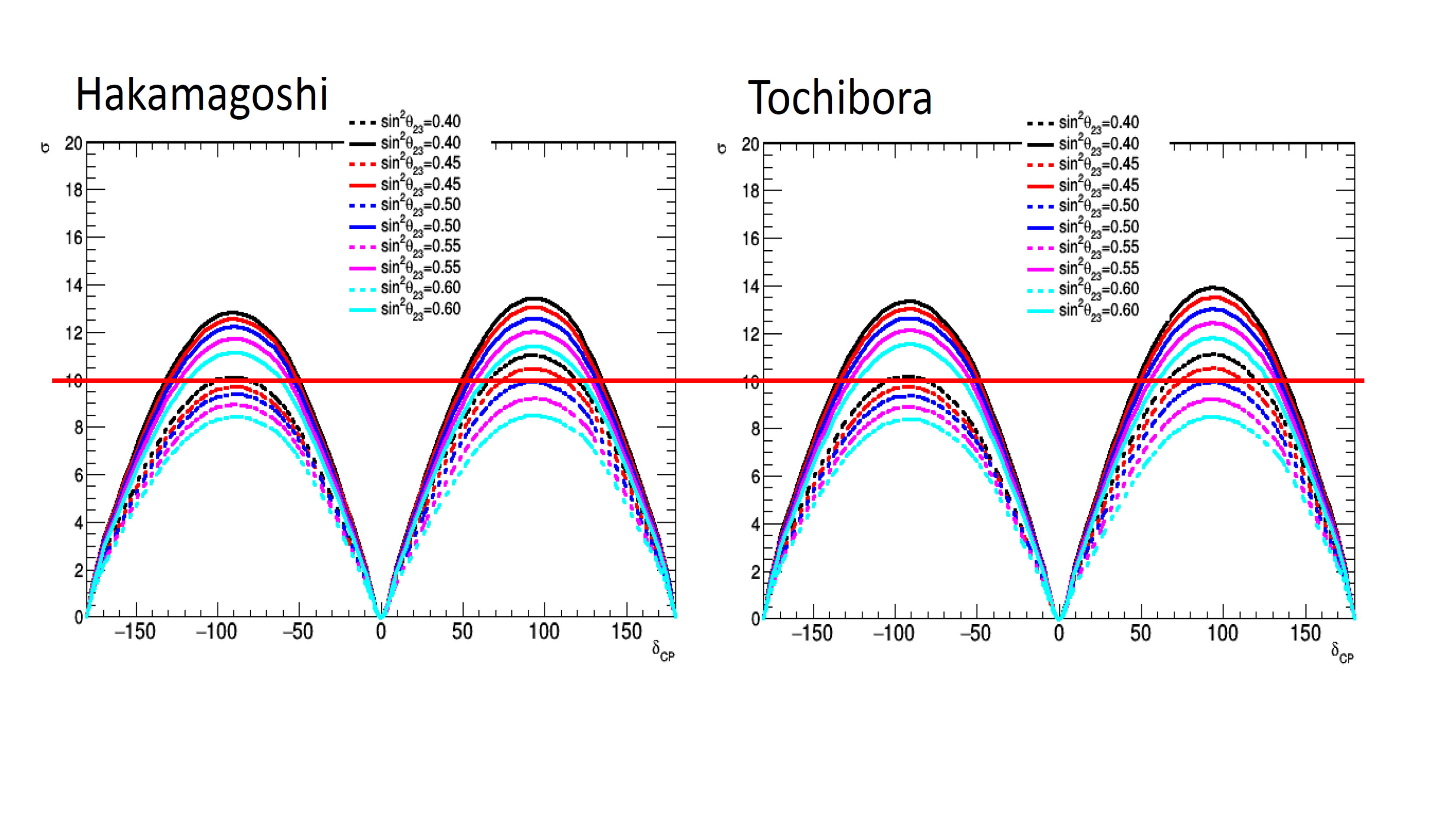}
\caption{ Sensitivity to CP violation as a function of the true value of $\delta_{cp}$ and for 
various assumed values of $\mbox{sin}^{2}\theta_{23}$.  
Both plots assume a single detector operating for five years with a second detector 
beginning operations in the sixth year. 
The left (right) plot shows the result assuming the second detector is placed in Hakamagoshi (Tochibora).
Solid lines show the sensitivity assuming only statistical errors and dashed lines 
include both systematic and statistical uncertainties.
}
    \label{fig:hg_cpv} \end{center}
\end {figure}

In the following the Hakamakoshi detector is assumed to be beneath the peak of the mountain, as in the simulations above, 
and the sensitivity neutrino oscillations using the J-PARC neutrino beam is estimated.
Hakamagoshi has an off-axis angle of $2.4^\circ$ and a baseline of 335~km.
The corresponding flux and $\nu_{\mu} \rightarrow \nu_{e}$ oscillation probabilities are shown in Figure~\ref{fig:hg_flux_osc}.
Since the off-axis angle is sightly smaller than that for the Tochibora site, the flux peak has shifted to slightly 
higher energies, but since the baseline is correspondingly larger the oscillation maximum is found at nearly the same energy.
Using this information the sensitivity to CP violation has been estimated as shown in Figure~\ref{fig:hg_cpv}.
Here the first tank is assumed to be located in Tochibora with the second tank beginning operations six years later.
A comparison is made between a second tank in Tochibora and one in Hakamagoshi.
Due to the larger baseline to Hakamagoshi there is a corresponding decrease in event rate, resulting in a slightly larger 
statistical error, though the systematic errors are taken to be the same.

\subsubsection{Low Energy Neutrino Observations} 

\begin {figure}[htbp]
  \begin{center}
    \includegraphics[width=0.8\textwidth]{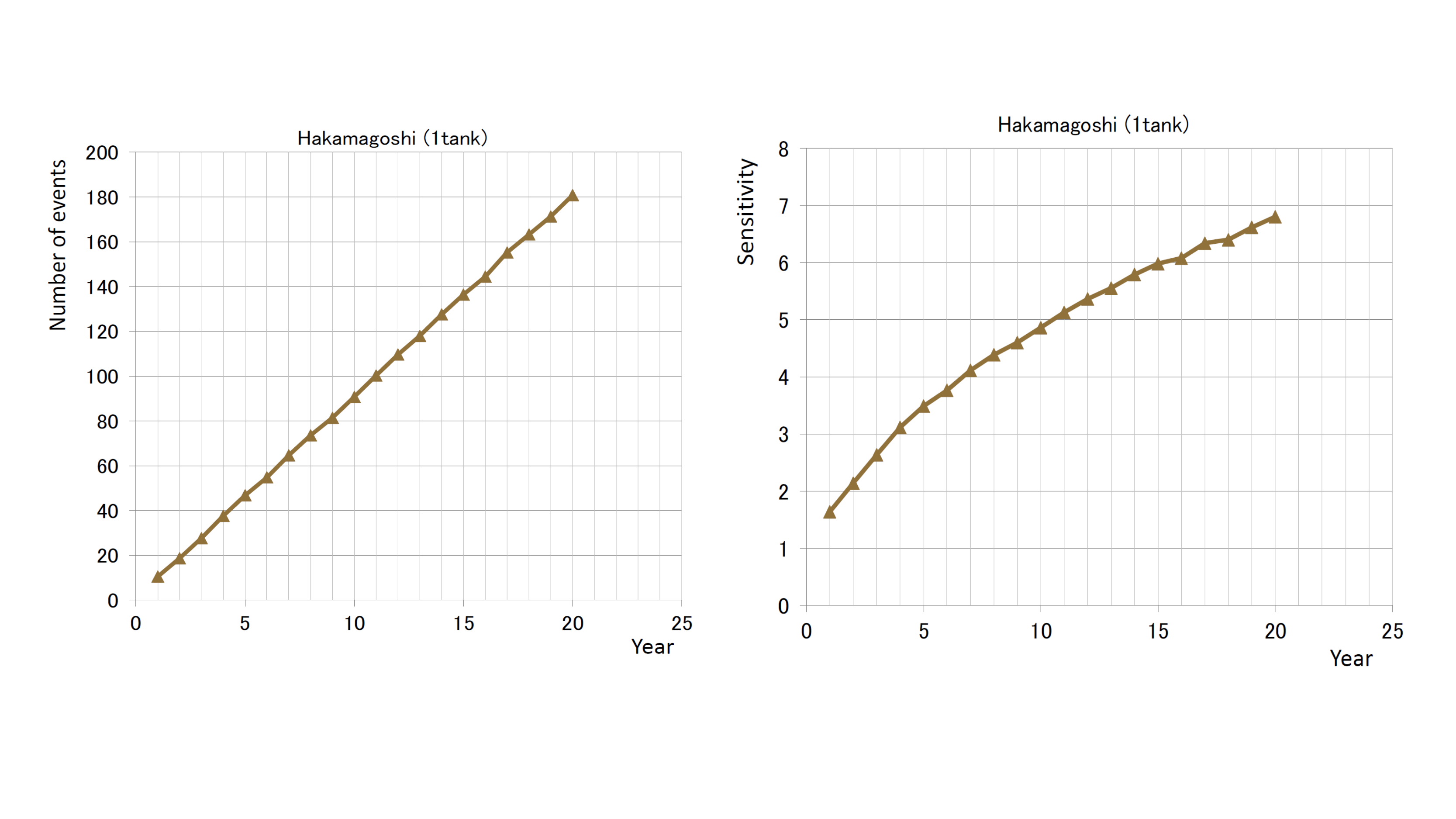}
\caption{ Sensitivity to supernova relic neutrinos as a function of operation time.
The left figure shows the number of relic neutrino candidate events and 
the right figure shows the ability to discern this flux from the 
background in units of $\sigma$.
}
    \label{fig:hg_srn} \end{center}
\end {figure}

Spallation products from cosmic ray muons form the main background to low energy neutrino physics in Hyper-K, including 
supernova and solar neutrino measurements.
Since the muon background at Hakamagoshi is lower than that at Tochibora, the former is expected to have improved sensitivity 
to these neutrinos.
Using the muon simulation results presented above to derive the spallation backgrounds, Hyper-K's sensitivity 
to the observation of supernova relic neutrinos has been estimated.
In this analysis an analysis sample selected using neutron tagging (70\% tagging efficiency) 
to identify the inverse beta decay signal in an energy window of 16 to 30~MeV has been assumed.
Figure~\ref{fig:hg_srn} shows the expected sensitivity assuming only one tank at at the Hakamagoshi site.
Assuming a standard model of the relic neutrino flux and the first detector in Tochibora, the signal can 
be observed with $3\sigma$ ($5\sigma$) significance after 4 (11)~years.  

\clearpage

\clearpage
\bibliography{reference}

\end{document}